%% file: thesis_dissertation_template.tex
\documentclass[PHD]{macro/neu_msthesis}

\title{Cross-Layer Optimization and System-Level Design of Next-Generation Wireless Networks via Intelligent RAN Control}

\author{Maria Tsampazi}

\nuid{002989668}

\dept{Electrical and Computer Engineering}

\degreename{Electrical and Computer Engineering}

\field{Electrical and Computer Engineering}

\submitdate{June 2026}

\numberofmembers{3}
\principaladviser{Prof. Tommaso Melodia}
\firstreader{Prof. Tommaso Melodia}
\secondreader{Prof. Josep-Miquel Jornet}
\thirdreader{Prof. Athanasios D. Panagopoulos}
\chairman{Prof. Edmund Yeh}
\dean{Dr. Waleed Meleis}
\input{macro/macro}

\input{tex/acronyms.tex}

\begin{document}

% add a pdf bookmark to the cover page
\pdfbookmark[1]{Cover}{cover}

% --- title page ---
\titlepage

% --- front matter ---
\begin{frontmatter}
% %%%%%% NO SIGNATURE PAGE WILL BE GENERATED, PLEASE DOWNLOAD FROM COE WEB SITE (READ file Readme.pdf in this folder!)
%\signaturepage
% dedication

\input{tex/dedication.tex}

% table of content (add bookmark for convenience)

\pdfbookmark[1]{Table of Contents}{contents}
\tableofcontents
\listoffigures
\newpage\ssp
\listoftables

% include a list of Acronyms (only referenced ones appear automatically)
\printnoidxglossary[type=acronym,title={List of Acronyms},sort=word]

%% include any of the front matter files that contain text
%% attention the input does cause a page break, the include on
%% the other hand does not
% \input{tex/acknowledgments.tex}
\input{tex/abstract.tex}

\end{frontmatter}

% --- body of the document ---
\pagestyle{headings}

%% include each chapter like below
\input{tex/intro.tex}

%% Enabling AI at the Edge in O-RAN
\input{tex/oran_edge.tex}

%% Optimizing Wireless Propagation with RIS
\input{tex/ris_propagation.tex}
%% AI-RAN Link Adaptation Leveraging Digital Twins
\input{tex/ai_ran_link_adaptation.tex}

%% Enabling Spectrum Sharing in FRli3
\input{tex/spectrum_sharing.tex}

%% Conclusions
\input{tex/conclusion.tex}

% --- Bibliography ----
\bibliographystyle{IEEEtran}

%% include bibliography definition
\bibliography{bib/thesis}

% --- Index ----
\printindex
\markboth{}{}
\input{tex/acknowledgments.tex}

% --- that's it ---
\end{document}

%% file: macro/macro.tex
\usepackage{lmodern}        % scalable fonts at any size (suppresses font size warnings)
\usepackage{amsfonts,amssymb,amsmath}			% special math characters
\usepackage{amsthm}         % theorem/definition environments
\usepackage{times}		% use Postscript fonts
\usepackage{bm}
\usepackage{xspace}

\newcommand{\oran}{O-RAN\xspace}

\newcommand{\ifno}[1]{}

\usepackage{multirow}
\makeindex
\usepackage{makeidx}
\usepackage[acronym,toc,nonumberlist,nopostdot]{glossaries}
\makenoidxglossaries
\usepackage{url}
\ifnum\pdfoutput>0
\usepackage[pdftex]{graphicx}
\usepackage{epstopdf}
\else
\usepackage{graphicx}
\fi

\ifnum\pdfoutput>0
\usepackage[pdftex,%                    % hyper-references for ps2pdf
bookmarks=true,%                        % generate bookmarks ...
bookmarksnumbered=true,%                % ... with numbers
hypertexnames=false,%                   % needed for correct links to figures
breaklinks=true           %Allows link text to break across lines;
]{hyperref}
\else
\usepackage[hypertex,%                  % hyper-references for ps2pdf
bookmarks=true,%                        % generate bookmarks ...
bookmarksnumbered=true,%                % ... with numbers
hypertexnames=false,%                   % needed for correct links to figures
breaklinks=true           %Allows link text to break across lines;
]{hyperref}
\fi

\hypersetup{				% setup PDF information fields
pdfauthor = {\authorRef},
pdftitle = {\titleRef},
pdfsubject = {\expandafter{\degreeRef} thesis submitted to Northeastern University},
pdfkeywords = {add keywords here}
pdfcreator = {LaTeX with hyperref package},
pdfproducer = {dvips + ps2pdf}}

\usepackage{microtype}

\usepackage[noadjust]{cite}

\usepackage{algorithm}
\usepackage{algorithmic}
\makeatletter
\def\endfigure{\end@float}
\def\endtable{\end@float}
\makeatother
\usepackage{subfig}
\usepackage{listings}
\usepackage{xcolor}
\usepackage{pgfplots}
\pgfplotsset{compat=1.18}
\usepackage{tikz}
\usetikzlibrary{plotmarks,patterns,decorations.pathreplacing,backgrounds,calc,arrows,arrows.meta,spy,matrix,scopes}
\usepgfplotslibrary{patchplots,groupplots}
\usepackage{newunicodechar}
\newunicodechar{λ}{$\lambda$}
\usepackage{booktabs}
\usepackage{tabularx}
\usepackage{makecell}
\usepackage{adjustbox}
\usepackage{verbatim} % provides \begin{comment} environment used in integrated paper sources

%% file: tex/acronyms.tex
\newacronym{3gpp}{3GPP}{3rd Generation Partnership Project}
\newacronym{4g}{4G}{4th generation}
\newacronym{5g}{5G}{5th generation}
\newacronym{5gc}{5GC}{5G Core}
\newacronym{6g}{6G}{6th generation}
\newacronym{adc}{ADC}{Analog to Digital Converter}
\newacronym{aerpaw}{AERPAW}{Aerial Experimentation and Research Platform for Advanced Wireless}
\newacronym{awgn}{AWGN}{Additive White Gaussian Noise}
\newacronym{ai}{AI}{Artificial Intelligence}
\newacronym{aimd}{AIMD}{Additive Increase Multiplicative Decrease}
\newacronym{am}{AM}{Acknowledged Mode}
\newacronym{amc}{AMC}{Adaptive Modulation and Coding}
\newacronym{amf}{AMF}{Access and Mobility Management Function}
\newacronym{aoa}{AoA}{Angle of Arrival}
\newacronym{aod}{AoD}{Angle of Departure}
\newacronym{aops}{AOPS}{Adaptive Order Prediction Scheduling}
\newacronym{ap}{AP}{Access Point}
\newacronym{api}{API}{Application Programming Interface}
\newacronym{apn}{APN}{Access Point Name}
\newacronym{dt}{DT}{Digital Twin}
\newacronym{aqm}{AQM}{Active Queue Management}
\newacronym{ariadne}{ARIADNE}{AI-RAN Informed Link Adaptation in Digital Twin Network Environments}
\newacronym{ausf}{AUSF}{Authentication Server Function}
\newacronym{avc}{AVC}{Advanced Video Coding}
\newacronym{balia}{BALIA}{Balanced Link Adaptation Algorithm}
\newacronym{bbu}{BBU}{Base Band Unit}
\newacronym{bdp}{BDP}{Bandwidth-Delay Product}
\newacronym{ber}{BER}{Bit Error Rate}
\newacronym{bf}{BF}{Beamforming}
\newacronym{bler}{BLER}{Block Error Rate}
\newacronym{brr}{BRR}{Bayesian Ridge Regressor}
\newacronym{bs}{BS}{Base Station}
\newacronym{bsr}{BSR}{Buffer Status Report}
\newacronym{bss}{BSS}{Business Support System}
\newacronym{cv2x}{C-V2X}{Cellular Vehicle-to-Everything}
\newacronym{ca}{CA}{Carrier Aggregation}
\newacronym{caas}{CaaS}{Connectivity-as-a-Service}
\newacronym{cb}{CB}{Code Block}
\newacronym{cc}{CC}{Congestion Control}
\newacronym{ccid}{CCID}{Congestion Control ID}
\newacronym{cdd}{CDD}{Cyclic Delay Diversity}
\newacronym{cdf}{CDF}{Cumulative Distribution Function}
\newacronym{cdn}{CDN}{Content Distribution Network}
\newacronym{cir}{CIR}{Channel Impulse Response}
\newacronym{cn}{CN}{Core Network}
\newacronym{codel}{CoDel}{Controlled Delay Management}
\newacronym{comac}{COMAC}{Converged Multi-Access and Core}
\newacronym{cord}{CORD}{Central Office Re-architected as a Datacenter}
\newacronym{cornet}{CORNET}{COgnitive Radio NETwork}
\newacronym{cosmos}{COSMOS}{Cloud Enhanced Open Software Defined Mobile Wireless Testbed for City-Scale Deployment}
\newacronym{cots}{COTS}{Commercial Off-the-Shelf}
\newacronym{cp}{CP}{Control Plane}
\newacronym{cpu}{CPU}{Central Processing Unit}
\newacronym{cqi}{CQI}{Channel Quality Information}
\newacronym{cr}{CR}{Cognitive Radio}
\newacronym{cran}{CRAN}{Cloud \acrshort{ran}}
\newacronym{crs}{CRS}{Cell Reference Signal}
\newacronym{csi}{CSI}{Channel State Information}
\newacronym{csirs}{CSI-RS}{Channel State Information - Reference Signal}
\newacronym{cu}{CU}{Central Unit}
\newacronym{d2tcp}{D$^2$TCP}{Deadline-aware Data center TCP}
\newacronym{d3}{D$^3$}{Deadline-Driven Delivery}
\newacronym{d2d}{D2D}{Device-to-Device}
\newacronym{dac}{DAC}{Digital to Analog Converter}
\newacronym{dag}{DAG}{Directed Acyclic Graph}
\newacronym{das}{DAS}{Distributed Antenna System}
\newacronym{dash}{DASH}{Dynamic Adaptive Streaming over HTTP}
\newacronym{dc}{DC}{Dual Connectivity}
\newacronym{dccp}{DCCP}{Datagram Congestion Control Protocol}
\newacronym{dce}{DCE}{Direct Code Execution}
\newacronym{dci}{DCI}{Downlink Control Information}
\newacronym{dcqi}{dCQI}{delayed CQI}
\newacronym{dctcp}{DCTCP}{Data Center TCP}
\newacronym{dl}{DL}{Downlink}
\newacronym{dmr}{DMR}{Deadline Miss Ratio}
\newacronym{dmrs}{DMRS}{DeModulation Reference Signal}
\newacronym{dqn}{DQN}{Deep Q-Network}
\newacronym{drl}{DRL}{Deep Reinforcement Learning}
\newacronym{drlcc}{DRL-CC}{Deep Reinforcement Learning Congestion Control}
\newacronym{drs}{DRS}{Discovery Reference Signal}
\newacronym{dtr}{DTR}{Decision Tree Regressor}
\newacronym{dectree}{DTree}{Decision Tree}
\newacronym{du}{DU}{Distributed Unit}
\newacronym{eutran}{E-UTRAN}{Evolved Universal Terrestrial Access Network}
\newacronym{e2e}{E2E}{end-to-end}
\newacronym{ecaas}{ECaaS}{Edge-Cloud-as-a-Service}
\newacronym{ecn}{ECN}{Explicit Congestion Notification}
\newacronym{edf}{EDF}{Earliest Deadline First}
\newacronym{embb}{eMBB}{Enhanced Mobile Broadband}
\newacronym{empower}{EMPOWER}{EMpowering transatlantic PlatfOrms for advanced WirEless Research}
\newacronym{endc}{EN-DC}{E-UTRAN-\acrshort{nr} \acrshort{dc}}
\newacronym{enb}{eNB}{evolved Node Base}
\newacronym{epc}{EPC}{Evolved Packet Core}
\newacronym{eps}{EPS}{Evolved Packet System}
\newacronym{es}{ES}{Edge Server}
\newacronym{eta}{ETA}{Estimated Time of Arrival}
\newacronym{etsi}{ETSI}{European Telecommunications Standards Institute}
\newacronym{faas}{FaaS}{Function-as-a-Service}
\newacronym{fapi}{FAPI}{Functional Application Platform Interface}
\newacronym{fbs}{FBS}{First Bounce Scatterer}
\newacronym{fdd}{FDD}{Frequency Division Duplexing}
\newacronym{fdm}{FDM}{Frequency Division Multiplexing}
\newacronym{fdma}{FDMA}{Frequency Division Multiple Access}
\newacronym{fed4fire}{FED4FIRE+}{Federation 4 Future Internet Research and Experimentation Plus}
\newacronym{fir}{FIR}{Finite Impulse Response}
\newacronym{fit}{FIT}{Future \acrlong{iot}}
\newacronym{fpga}{FPGA}{Field Programmable Gate Array}
\newacronym{fqi}{FQI}{Fitted Q-Iteration}
\newacronym{fr1}{FR1}{Frequency Range 1}
\newacronym{fr2}{FR2}{Frequency Range 2}
\newacronym{fr3}{FR3}{Frequency Range 3}
\newacronym{fs}{FS}{Fast Switching}
\newacronym{fscc}{FSCC}{Flow Sharing Congestion Control}
\newacronym{ftp}{FTP}{File Transfer Protocol}
\newacronym{fw}{FW}{Flow Window}
\newacronym{gbsm}{GBSM}{Geometry-Based Stochastic Model}
\newacronym{gbs}{GBSMb}{Geometry-Based Stochastic}
\newacronym{ge}{GE}{Gaussian Elimination}
\newacronym{gnb}{gNB}{Next Generation Node Base}
\newacronym{gop}{GOP}{Group of Pictures}
\newacronym{gpr}{GPR}{Gaussian Process Regressor}
\newacronym{gps}{GPS}{Global Positioning System}
\newacronym{gpu}{GPU}{Graphics Processing Unit}
\newacronym{gtp}{GTP}{GPRS Tunneling Protocol}
\newacronym{gtpc}{GTP-C}{GPRS Tunnelling Protocol Control Plane}
\newacronym{gtpu}{GTP-U}{GPRS Tunnelling Protocol User Plane}
\newacronym{gtpv2c}{GTPv2-C}{\acrshort{gtp} v2 - Control}
\newacronym{gw}{GW}{Gateway}
\newacronym{harq}{HARQ}{Hybrid Automatic Repeat reQuest}
\newacronym{hetnet}{HetNet}{Heterogeneous Network}
\newacronym{hh}{HH}{Hard Handover}
\newacronym{hol}{HOL}{Head-of-Line}
\newacronym{hqf}{HQF}{Highest-quality-first}
\newacronym{hss}{HSS}{Home Subscription Server}
\newacronym{http}{HTTP}{HyperText Transfer Protocol}
\newacronym{ia}{IA}{Initial Access}
\newacronym{iab}{IAB}{Integrated Access and Backhaul}
\newacronym{ic}{IC}{Incident Command}
\newacronym{ids}{IDS}{Independent Diffusive Scatterer-based}
\newacronym{ietf}{IETF}{Internet Engineering Task Force}
\newacronym{imsi}{IMSI}{International Mobile Subscriber Identity}
\newacronym{imt}{IMT}{International Mobile Telecommunication}
\newacronym{inr}{INR}{interference-to-noise ratio}
\newacronym{inw}{INW}{Impedance Network-based}
\newacronym{iot}{IoT}{Internet of Things}
\newacronym{ip}{IP}{Internet Protocol}
\newacronym{itu}{ITU}{International Telecommunication Union}
\newacronym{kf}{KF}{Kalman Filter}
\newacronym{kkt}{KKT}{Karush-Kuhn-Tucker}
\newacronym{kpi}{KPI}{Key Performance Indicator}
\newacronym{kpm}{KPM}{Key Performance Measurement}
\newacronym{kvm}{KVM}{Kernel-based Virtual Machine}
\newacronym{lbs}{LBS}{Last Bounce Scatterer}
\newacronym{leo}{LEO}{Low Earth Orbit}
\newacronym{lmmse}{LMMSE}{Linear Minimum Mean Square Error}
\newacronym{los}{LOS}{Line-of-Sight}
\newacronym{lsm}{LSM}{Link-to-System Mapping}
\newacronym{lsp}{LSP}{Large Scale Parameter}
\newacronym{lstm}{LSTM}{Long Short Term Memory}
\newacronym{lte}{LTE}{Long Term Evolution}
\newacronym{lxc}{LXC}{Linux Container}
\newacronym{m2m}{M2M}{Machine to Machine}
\newacronym{mab}{MAB}{Multi-Armed Bandit}
\newacronym{mac}{MAC}{Medium Access Control}
\newacronym{manet}{MANET}{Mobile Ad Hoc Network}
\newacronym{mano}{MANO}{Management and Orchestration}
\newacronym{mc}{MC}{Multi-Connectivity}
\newacronym{mcc}{MCC}{Mobile Cloud Computing}
\newacronym{mchem}{MCHEM}{Massive Channel Emulator}
\newacronym{mcs}{MCS}{Modulation and Coding Scheme}
\newacronym{mdp}{MDP}{Markov Decision Process}
\newacronym{mec}{MEC}{Multi-access Edge Computing}
\newacronym{mecn}{MECN}{Mobile Edge Cloud Network}
\newacronym{mfc}{MFC}{Mobile Fog Computing}
\newacronym{mgen}{MGEN}{Multi-Generator}
\newacronym{mi}{MI}{Mutual Information}
\newacronym{mib}{MIB}{Master Information Block}
\newacronym{miesm}{MIESM}{Mutual Information Based Effective SINR}
\newacronym{mimo}{MIMO}{Multiple Input, Multiple Output}
\newacronym{minlp}{MINLP}{Mixed Integer Non-Linear Programming}
\newacronym{ml}{ML}{Machine Learning}
\newacronym{mlp}{MLP}{Multilayer Perceptron}
\newacronym{mlr}{MLR}{Maximum-local-rate}
\newacronym{mme}{MME}{Mobility Management Entity}
\newacronym{mmtc}{mMTC}{Massive Machine-Type Communications}
\newacronym{mmwave}{mmWave}{millimeter wave}
\newacronym{mpdccp}{MP-DCCP}{Multipath Datagram Congestion Control Protocol}
\newacronym{mpc}{MPC}{Multipath Component}
\newacronym{mptcp}{MPTCP}{Multipath TCP}
\newacronym{mr}{MR}{Maximum Rate}
\newacronym{mrdc}{MR-DC}{Multi \acrshort{rat} \acrshort{dc}}
\newacronym{mse}{MSE}{Mean Square Error}
\newacronym{mss}{MSS}{Maximum Segment Size}
\newacronym{mt}{MT}{Mobile Termination}
\newacronym{mtc}{MTC}{Machine-type Communications}
\newacronym{mtd}{MTD}{Machine-Type Device}
\newacronym{mtu}{MTU}{Maximum Transmission Unit}
\newacronym{mumimo}{MU-MIMO}{Multi-user \acrshort{mimo}}
\newacronym{mvno}{MVNO}{Mobile Virtual Network Operator}
\newacronym{nalu}{NALU}{Network Abstraction Layer Unit}
\newacronym{nas}{NAS}{Non-Access Stratum}
\newacronym{nbiot}{NB-IoT}{Narrow Band IoT}
\newacronym{nextg}{NextG}{Next Generation}
\newacronym{nfv}{NFV}{Network Function Virtualization}
\newacronym{nfvi}{NFVI}{Network Function Virtualization Infrastructure}
\newacronym{nic}{NIC}{Network Interface Card}
\newacronym{nlos}{NLOS}{Non-Line-of-Sight}
\newacronym{noma}{NOMA}{Non-Orthogonal Multiple Access}
\newacronym{now}{NOW}{Non Overlapping Window}
\newacronym{nr}{NR}{New Radio}
\newacronym{nrf}{NRF}{Network Repository Function}
\newacronym{nsa}{NSA}{Non Stand Alone}
\newacronym{nse}{NSE}{Network Slicing Engine}
\newacronym{nsm}{NSM}{Network Service Mesh}
\newacronym{nssf}{NSSF}{Network Slice Selection Function}
\newacronym{ntn}{NTN}{Non-Terrestrial Network}
\newacronym{o2i}{O2I}{Outdoor to Indoor}
\newacronym{oai}{OAI}{OpenAirInterface}
\newacronym{oaicn}{OAI-CN}{\acrshort{oai} \acrlong{cn}}
\newacronym{oairan}{OAI-RAN}{\acrlong{oai} \acrlong{ran}}
\newacronym{oam}{OAM}{Operations, Administration and Maintenance}
\newacronym{oco}{OCO}{Online Convex Optimization}
\newacronym{ofdm}{OFDM}{Orthogonal Frequency Division Multiplexing}
\newacronym{olia}{OLIA}{Opportunistic Linked Increase Algorithm}
\newacronym{olla}{OLLA}{Outer Loop Link Adaptation}
\newacronym{omec}{OMEC}{Open Mobile Evolved Core}
\newacronym{onap}{ONAP}{Open Network Automation Platform}
\newacronym{onf}{ONF}{Open Networking Foundation}
\newacronym{onos}{ONOS}{Open Networking Operating System}
\newacronym{oom}{OOM}{\acrshort{onap} Operations Manager}
\newacronym{opnfv}{OPNFV}{Open Platform for \acrshort{nfv}}
\newacronym{orbit}{ORBIT}{Open-Access Research Testbed for Next-Generation Wireless Networks}
\newacronym{os}{OS}{Operating System}
\newacronym{osc}{OSC}{O-RAN Software Community}
\newacronym{osm}{OSM}{Open Source \acrshort{nfv} Management and Orchestration}
\newacronym{oss}{OSS}{Operations Support System}
\newacronym{ota}{OTA}{over-the-air}
\newacronym{pa}{PA}{Position-aware}
\newacronym{pase}{PASE}{Prioritization, Arbitration, and Self-adjusting Endpoints}
\newacronym{pawr}{PAWR}{Platforms for Advanced Wireless Research}
\newacronym{pbch}{PBCH}{Physical Broadcast Channel}
\newacronym{pcef}{PCEF}{Policy and Charging Enforcement Function}
\newacronym{pcfich}{PCFICH}{Physical Control Format Indicator Channel}
\newacronym{pcrf}{PCRF}{Policy and Charging Rules Function}
\newacronym{pdcch}{PDCCH}{Physical Downlink Control Channel}
\newacronym{pdcp}{PDCP}{Packet Data Convergence Protocol}
\newacronym{pdsch}{PDSCH}{Physical Downlink Shared Channel}
\newacronym{pdu}{PDU}{Packet Data Unit}
\newacronym{pf}{PF}{Proportionally Fair}
\newacronym{pgw}{PGW}{Packet Gateway}
\newacronym{phich}{PHICH}{Physical Hybrid ARQ Indicator Channel}
\newacronym{phy}{PHY}{Physical}
\newacronym{pl}{PL}{Path-Loss}
\newacronym{pmch}{PMCH}{Physical Multicast Channel}
\newacronym{pmi}{PMI}{Precoding Matrix Indicators}
\newacronym{pnf}{PNF}{Physical Network Function}
\newacronym{powder}{POWDER}{Platform for Open Wireless Data-driven Experimental Research}
\newacronym{ppo}{PPO}{Proximal Policy Optimization}
\newacronym{ppp}{PPP}{Poisson Point Process}
\newacronym{prach}{PRACH}{Physical Random Access Channel}
\newacronym{prb}{PRB}{Physical Resource Block}
\newacronym{psnr}{PSNR}{Peak Signal to Noise Ratio}
\newacronym{pss}{PSS}{Primary Synchronization Signal}
\newacronym{pucch}{PUCCH}{Physical Uplink Control Channel}
\newacronym{pusch}{PUSCH}{Physical Uplink Shared Channel}
\newacronym{qam}{QAM}{Quadrature Amplitude Modulation}
\newacronym{qci}{QCI}{\acrshort{qos} Class Identifier}
\newacronym{qoe}{QoE}{Quality of Experience}
\newacronym{qos}{QoS}{Quality of Service}
\newacronym{quadriga}{QuaDRiGa}{QUAsi Deterministic RadIo channel GenerAtor}
\newacronym{quic}{QUIC}{Quick UDP Internet Connections}
\newacronym{ra}{RA}{Resource Allocation}
\newacronym{rach}{RACH}{Random Access Channel}
\newacronym{ran}{RAN}{Radio Access Network}
\newacronym{rat}{RAT}{Radio Access Technology}
\newacronym{rbg}{RBG}{Resource Block Group}
\newacronym{rc}{RC}{RAN Control}
\newacronym{rcn}{RCN}{Research Coordination Network}
\newacronym{rec}{REC}{Radio Edge Cloud}
\newacronym{red}{RED}{Random Early Detection}
\newacronym{relu}{ReLU}{Rectified Linear Unit}
\newacronym{renew}{RENEW}{Reconfigurable Eco-system for Next-generation End-to-end Wireless}
\newacronym{rf}{RF}{Radio Frequency}
\newacronym{rfr}{RFR}{Random Forest Regressor}
\newacronym{rfc}{RFC}{Request for Comments}
\newacronym{ric}{RIC}{\glsentryshort{ran} Intelligent Controller}
\newacronym{ris}{RIS}{Reconfigurable Intelligent Surface}
\newacronym{rl}{RL}{Reinforcement Learning}
\newacronym{rlc}{RLC}{Radio Link Control}
\newacronym{rlf}{RLF}{Radio Link Failure}
\newacronym{rlnc}{RLNC}{Random Linear Network Coding}
\newacronym{rmr}{RMR}{RIC Message Router}
\newacronym{rmse}{RMSE}{Root Mean Squared Error}
\newacronym{rnis}{RNIS}{Radio Network Information Service}
\newacronym{rr}{RR}{Round Robin}
\newacronym{rrc}{RRC}{Radio Resource Control}
\newacronym{rrm}{RRM}{Radio Resource Management}
\newacronym{rru}{RRU}{Remote Radio Unit}
\newacronym{rs}{RS}{Remote Server}
\newacronym{rsrp}{RSRP}{Reference Signal Received Power}
\newacronym{rsrq}{RSRQ}{Reference Signal Received Quality}
\newacronym{rss}{RSS}{Received Signal Strength}
\newacronym{rssi}{RSSI}{Received Signal Strength Indicator}
\newacronym{rsu}{RSU}{Road Side Unit}
\newacronym{rtt}{RTT}{Round Trip Time}
\newacronym{ru}{RU}{Radio Unit}
\newacronym{rw}{RW}{Receive Window}
\newacronym{rx}{RX}{Receiver}
\newacronym{rzf}{RZF}{Regularized Zero-Forcing}
\newacronym{s1ap}{S1AP}{S1 Application Protocol}
\newacronym{sa}{SA}{standalone}
\newacronym{sack}{SACK}{Selective Acknowledgment}
\newacronym{salad}{SALAD}{Self-Adaptive Link Adaptation}
\newacronym{sap}{SAP}{Service Access Point}
\newacronym{sc2}{SC2}{Spectrum Collaboration Challenge}
\newacronym{sca}{SCA}{Successive Convex Approximation}
\newacronym{scef}{SCEF}{Service Capability Exposure Function}
\newacronym{sch}{SCH}{Secondary Cell Handover}
\newacronym{scoot}{SCOOT}{Split Cycle Offset Optimization Technique}
\newacronym{sctp}{SCTP}{Stream Control Transmission Protocol}
\newacronym{sdap}{SDAP}{Service Data Adaptation Protocol}
\newacronym{sdk}{SDK}{Software Development Kit}
\newacronym{sdm}{SDM}{Space Division Multiplexing}
\newacronym{sdma}{SDMA}{Spatial Division Multiple Access}
\newacronym{sdn}{SDN}{Software-defined Networking}
\newacronym{sdr}{SDR}{Software-defined Radio}
\newacronym{se}{SE}{Spectral Efficiency}
\newacronym{seba}{SEBA}{SDN-Enabled Broadband Access}
\newacronym{sgsn}{SGSN}{Serving GPRS Support Node}
\newacronym{sgw}{SGW}{Service Gateway}
\newacronym{si}{SI}{Study Item}
\newacronym{sib}{SIB}{Secondary Information Block}
\newacronym{sic}{SIC}{Successive Interference Cancellation}
\newacronym{sinr}{SINR}{Signal to Interference plus Noise Ratio}
\newacronym{sip}{SIP}{Session Initiation Protocol}
\newacronym{siso}{SISO}{Single Input, Single Output}
\newacronym{sla}{SLA}{Service Level Agreement}
\newacronym{sm}{SM}{Service Model}
\newacronym{smf}{SMF}{Session Management Function}
\newacronym{smo}{SMO}{Service Management and Orchestration}
\newacronym{sms}{SMS}{Short Message Service}
\newacronym{smsgmsc}{SMS-GMSC}{\acrshort{sms}-Gateway}
\newacronym{snr}{SNR}{Signal-to-Noise-Ratio}
\newacronym{son}{SON}{Self-Organizing Network}
\newacronym{sptcp}{SPTCP}{Single Path TCP}
\newacronym{srb}{SRB}{Service Radio Bearer}
\newacronym{srn}{SRN}{Standard Radio Node}
\newacronym{srs}{SRS}{Sounding Reference Signal}
\newacronym{ss}{SS}{Synchronization Signal}
\newacronym{ssp}{SSP}{Small Scale Parameter}
\newacronym{sss}{SSS}{Secondary Synchronization Signal}
\newacronym{st}{ST}{Spanning Tree}
\newacronym{star}{STAR}{Simultaneous Transmitting And Reflecting}
\newacronym{svc}{SVC}{Scalable Video Coding}
\newacronym{svd}{SVD}{Singular Value Decomposition}
\newacronym{tb}{TB}{Transport Block}
\newacronym{tcp}{TCP}{Transmission Control Protocol}
\newacronym{tdd}{TDD}{Time Division Duplexing}
\newacronym{tdm}{TDM}{Time Division Multiplexing}
\newacronym{tdma}{TDMA}{Time Division Multiple Access}
\newacronym{tfl}{TfL}{Transport for London}
\newacronym{tfrc}{TFRC}{TCP-Friendly Rate Control}
\newacronym{tft}{TFT}{Traffic Flow Template}
\newacronym{tgen}{TGEN}{Traffic Generator}
\newacronym{tip}{TIP}{Telecom Infra Project}
\newacronym{tm}{TM}{Transparent Mode}
\newacronym{to}{TO}{Telco Operator}
\newacronym{toa}{ToA}{Time-of-Arrival}
\newacronym{tr}{TR}{Technical Report}
\newacronym{trp}{TRP}{Transmitter Receiver Pair}
\newacronym{ts}{TS}{Technical Specification}
\newacronym{tti}{TTI}{Transmission Time Interval}
\newacronym{ttt}{TTT}{Time-to-Trigger}
\newacronym{tx}{TX}{Transmitter}
\newacronym{usa}{U.S.}{United States}
\newacronym{uas}{UAS}{Unmanned Aerial System}
\newacronym{uav}{UAV}{Unmanned Aerial Vehicle}
\newacronym{udm}{UDM}{Unified Data Management}
\newacronym{udp}{UDP}{User Datagram Protocol}
\newacronym{udr}{UDR}{Unified Data Repository}
\newacronym{ue}{UE}{User Equipment}
\newacronym{uhd}{UHD}{\acrshort{usrp} Hardware Driver}
\newacronym{ul}{UL}{Uplink}
\newacronym{um}{UM}{Unacknowledged Mode}
\newacronym{uml}{UML}{Unified Modeling Language}
\newacronym{upa}{UPA}{Uniform Planar Array}
\newacronym{upf}{UPF}{User Plane Function}
\newacronym{urllc}{URLLC}{Ultra Reliable and Low Latency Communications}
\newacronym{usim}{USIM}{Universal Subscriber Identity Module}
\newacronym{usrp}{USRP}{Universal Software Radio Peripheral}
\newacronym{utc}{UTC}{Urban Traffic Control}
\newacronym{v2x}{V2X}{Vehicle-to-everything}
\newacronym{vim}{VIM}{Virtualization Infrastructure Manager}
\newacronym{vm}{VM}{Virtual Machine}
\newacronym{vnf}{VNF}{Virtual Network Function}
\newacronym{volte}{VoLTE}{Voice over \acrshort{lte}}
\newacronym{voltha}{VOLTHA}{Virtual OLT HArdware Abstraction}
\newacronym{vr}{VR}{Virtual Reality}
\newacronym{vran}{vRAN}{Virtualized \acrshort{ran}}
\newacronym{vss}{VSS}{Video Streaming Server}
\newacronym{wbf}{WBF}{Wired Bias Function}
\newacronym{wf}{WF}{Waterfilling}
\newacronym{wlan}{WLAN}{Wireless Local Area Network}
\newacronym{xapp}{xApp}{Intelligent Application}
\newacronym{xpr}{XPR}{Cross-polarization Ratio}

\newacronym{pdf}{PDF}{Probability Density Function}
\newacronym{ez}{EZ}{Exclusion Zone}
\newacronym{wsr}{WSR}{Weighted Sum Rate}
\newacronym{ttd}{TTD}{True-Time-Delay}
\newacronym{sqp}{SQP}{Sequential Quadratic Programming}
\newacronym{jfi}{JFI}{Jain's Fairness Index}
\newacronym{dof}{DOF}{Degrees of Freedom}
\newacronym{mu}{MU}{Multi-User}
\newacronym{epfd}{EPFD}{Equivalent Power Flux-Density}
\newacronym{fss}{FSS}{Fixed-Satellite Service}
\newacronym{cpi}{CPI}{Conditional Policy Improvement}
\newacronym{ddpg}{DDPG}{Deep Deterministic Policy Gradient}
\newacronym{marl}{MARL}{Multi-Agent Reinforcement Learning}
\newacronym{mrat}{Multi-RAT}{Multi-Radio Access Technology}
\newacronym{nn}{NN}{Neural Network}
\newacronym{sgd}{SGD}{Stochastic Gradient Descent}
\newacronym{td3}{TD3}{Twin Delayed Deep Deterministic Policy Gradient}

\newacronym{vpn}{VPN}{Virtual Private Network}
\newacronym{tdl}{TDL}{Tapped Delay Line}
\newacronym{sba}{SBA}{Service-Based Architecture}
\newacronym{cbrs}{CBRS}{Citizen Broadband Radio Service}
\newacronym{esc}{ESC}{Environmental Sensing Capability}
\newacronym{sas}{SAS}{Spectrum Access System}
\newacronym{otic}{OTIC}{Open Testing and Integration Center}
\newacronym{rt}{RT}{Real-time}
\newacronym{xai}{XAI}{Explainable AI}
\newacronym{ci}{CI}{Continuous Integration}
\newacronym{cd}{CD}{Continuous Deployment}
\newacronym{ct}{CT}{Continuous Testing}
\newacronym{cucp}{CU-CP}{Central Unit Control Plane}
\newacronym{cuup}{CU-UP}{Central Unit User Plane}
\newacronym{cast}{CaST}{Channel Emulation Scenario Generator and Sounder Toolchain}
\newacronym{wi}{WI}{Wireless InSite}
\newacronym{osm2}{OSM}{Open Street Map}

\newacronym{ks}{K-S}{Kolmogorov-Smirnov}
\newacronym{int}{INT}{Integral Area}
\newacronym{ecdf}{ECDF}{Empirical Cumulative Distribution Function}
\newacronym{pr}{PR}{Proportional Fair}

\newacronym{FR3}{FR3}{Frequency Range 3}
\newacronym{NTN}{NTN}{Non-Terrestrial Network}

%% file: tex/dedication.tex
% !TEX root = ../thesis_dissertation template.tex
% dedication.tex:

\begin{dedication}
\begin{center}
\vspace{-1.2cm}
To my family, my professors, and my friends who always believed in me.

% \vspace{0.25cm}
\vspace{1.5cm}

\textit{\textbf{Ithaka}}, by Constantine P. Cavafy (1911)

\vspace{0.1cm}

As you set out for Ithaka, \\
hope your road is a long one, \\
full of adventure, full of discovery. \\

Laistrygonians, Cyclops, \\
angry Poseidon—don’t be afraid of them: \\
you’ll never find things like that on your way \\
as long as you keep your thoughts raised high, \\
as long as a rare excitement \\
stirs your spirit and your body. \\

Laistrygonians, Cyclops, \\
wild Poseidon—you won’t encounter them \\
unless you bring them along inside your soul, \\
unless your soul sets them up in front of you. \\

\smallskip

[...] \\

\smallskip

Keep Ithaka always in your mind. \\
Arriving there is what you’re destined for. \\
But don’t hurry the journey at all. \\

Better if it lasts for years, \\
so you’re old by the time you reach the island, \\
wealthy with all you’ve gained on the way, \\
not expecting Ithaka to make you rich. \\

\smallskip

Ithaka gave you the marvelous journey. \\
Without her you wouldn't have set out. \\
She has nothing left to give you now. \\

\smallskip

And if you find her poor, Ithaka won’t have fooled you. \\
Wise as you will have become, so full of experience, \\
you’ll have understood by then what these Ithakas mean. \\

\vspace{0.1cm}

\footnotesize{Translated by Edmund Keeley and Philip Sherrard.}

\end{center}

\end{dedication}

%% file: tex/abstract.tex
% !TEX root = ../thesis_dissertation template.tex
% abstract.tex:

\begin{abstract}
\glsresetall
Recent years have seen the evolution of the traditional \gls{ran} toward more open, programmable, softwarized, disaggregated, and intelligent architectures, a concept referred to as Open \gls{ran}. Future \gls{nextg} networks are envisioned to be \gls{ai}-native, enabling on-demand, data-driven closed-loop optimization of \gls{bs} resources. At the same time, key enablers such as \glspl{ris} are emerging as transformative technologies for optimizing wireless propagation and improving spectral efficiency. However, the efficient utilization of the already crowded spectrum is becoming a critical issue for the successful commercial deployment of \gls{6g} and beyond networks. Therefore, efficient optimization of \gls{bs} scheduling functionalities is essential for enhancing performance and resource orchestration in cellular networks.

This Ph.D. dissertation focuses on the design, optimization, and experimental evaluation of \gls{nextg} \glspl{ran} that integrate Open~\gls{ran} principles, data-driven control loops, and intelligent resource allocation mechanisms. The work emphasizes cross-layer optimization across the full protocol stack, including system-level experimental analysis for energy-efficient power control, and explores the use of \gls{ai} in the \gls{ran} for network slicing, scheduling, and link adaptation. The results demonstrate the feasibility of \gls{nextg} \glspl{ran} that can be reconfigured in real time to meet the requirements of \gls{6g} and beyond use cases. First, the dissertation analyzes the architectural enablers, algorithms, and modeling frameworks that underpin these technologies. Next, it prototypes and evaluates the proposed solutions on experimental platforms and \glspl{dt}.

The main contributions include: (i) the design of \gls{drl} solutions for network slicing and scheduling; (ii) the development of the PandORA framework for the automatic design, training, and deployment of \gls{drl}-based solutions, deployed at the network edge for real-time control of Open~\gls{ran} applications within the Colosseum wireless network emulator; (iii) accurate \gls{phy} \gls{ris}-enabled channel modeling and optimized resource allocation through comprehensive studies of energy efficiency and power control across different spectrum bands; (iv) full-stack, system-level emulation and performance evaluation of \gls{ris}-assisted channels for \gls{embb} and \gls{urllc}; (v) the integration of \gls{ris} within the Open~\gls{ran} ecosystem; (vi) the design of online \gls{rl} solutions for link adaptation; and (vii) the study of spectrum sharing mechanisms between cellular and \gls{ntn} links through power control and beamforming.

Overall, this dissertation addresses the ever-growing needs of both industry and academia, which stress the importance of designing novel algorithmic solutions and tools for advanced wireless research. In this work, we provide algorithmic designs for slicing, scheduling, link adaptation, energy-efficient power control, and beamforming; develop system-level frameworks; and provide insights through validation ranging from simulation and emulation in \glspl{dt} and hardware-in-the-loop channel emulators to experimental evaluation in realistic \gls{ota} \gls{5g} testbeds.

\end{abstract}

%% file: tex/intro.tex
% !TEX root = ../thesis_dissertation template.tex
% intro.tex

\chapter{Introduction}
\label{chap:intro}

\section{Motivation \& Context}
\label{sec:intro:motivation}

\subsection{How Standards and Industry Shape \texorpdfstring{\gls{nextg}}{NextG} Research}
\label{sec:intro:standardsguide}
% \glsreset{ran}
% \glsreset{ric}

\begin{figure}[ht]
    \centering
    \includegraphics[width=0.99\textwidth]{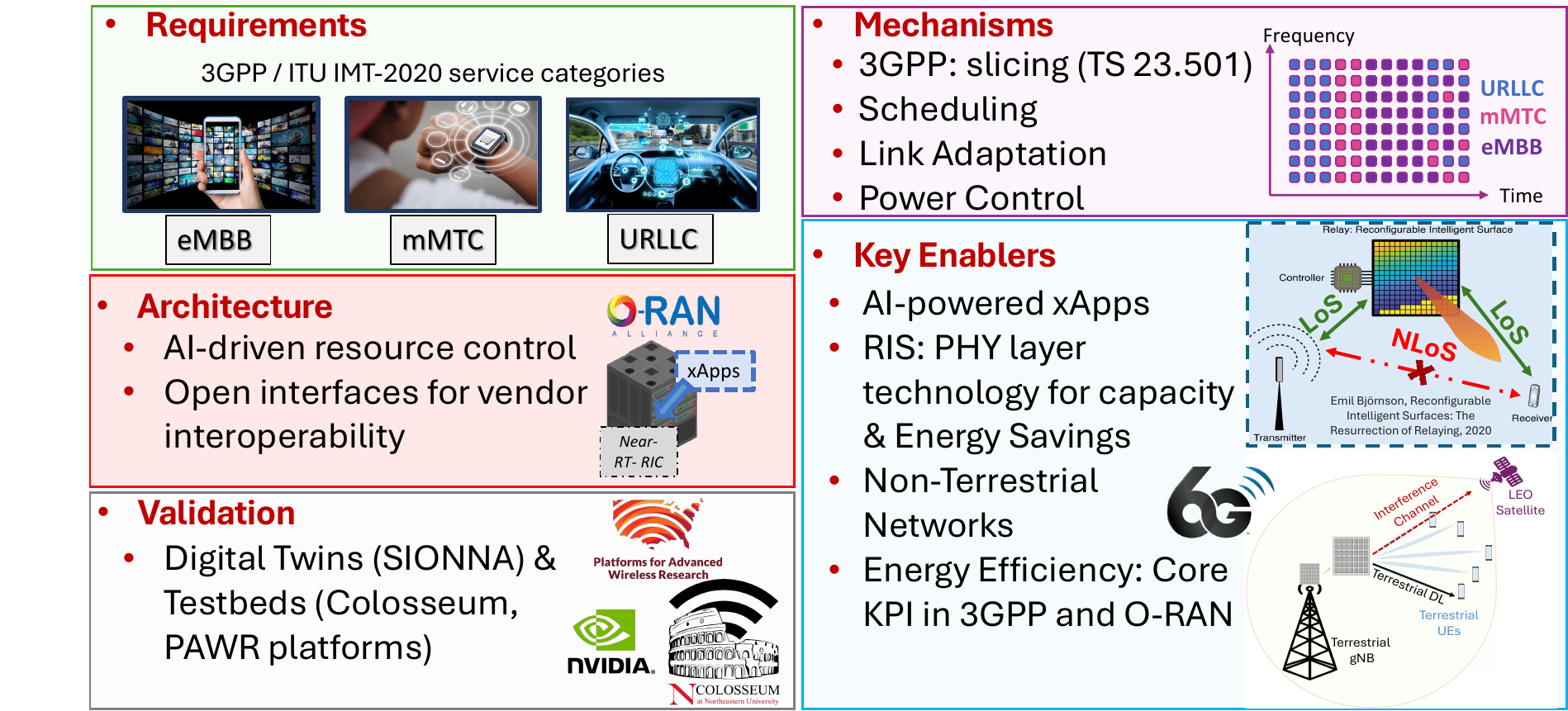}
    \caption{NextG Research Framework: Requirements, Mechanisms, Architecture, Enablers, and Experimental Validation.}
    \label{fig:standards}
\end{figure}

 Standards bodies and industry fundamentally shape the field of \gls{5g} and beyond cellular networks. The first requirement that needs to be addressed, as seen in Fig.~\ref{fig:standards}, pertains to the types of services the network shall support. The \gls{5g} spectrum is shared among \glspl{ue} with diverse \gls{qos} requirements, where video streaming applications require high throughput, \gls{iot} devices require massive data transmission, and \gls{v2x} applications require ultra-fast and ultra-reliable communication links. These become the high-level \glspl{kpi} that the system must satisfy, with operators targeting gigabit speeds for \gls{embb} applications, five-nines reliability and ultra-low latency for \gls{urllc} \glspl{ue}, and massive device connectivity for \gls{mmtc} devices~\cite{3gpp22261}.

Once the requirements are specified, mechanisms are defined to meet these demands. \gls{3gpp} defines network slicing for resource isolation, while scheduling is the operational lever used to meet service \glspl{kpi}. In detail, slicing refers to the allocation of available spectrum resources (e.g., \glspl{prb}) across different network slices (e.g., \gls{embb}, \gls{mmtc}, \gls{urllc}) according to their requirements and respective \glspl{sla}~\cite{3gpp23501}. Scheduling defines the orchestration of radio resources depending on channel and traffic conditions, ranging from round-robin (\gls{rr}) to more fairness-oriented approaches such as \gls{wf} and \gls{pf}~\cite{bonati2021scope}.

Important network functionalities also include link adaptation and power control. Link adaptation refers to how efficiently the spectrum is used by selecting appropriate modulation and coding schemes based on the current channel state, while power control enables efficient use of network power resources through open- and closed-loop mechanisms~\cite{3gpp38214, 3gpp38213}.
The \oran Alliance defines the \glspl{ric} for \gls{ai}-driven closed-loop control and open interfaces for interoperability~\cite{polese2022understanding, oran-wg3-ricarch}. This architecture serves as an industry playground for innovation, with vendors developing \oran-compliant solutions. Since standards require performance validation and industry trials, research relies on testbeds and \glspl{dt} as a bridge between specification and deployment~\cite{breen2020powder, bonati2023openran}.

Evaluating cutting-edge applications and studying coexistence techniques is necessary for successful commercial deployment of the proposed key enabling technologies. These span \gls{ai}-powered xApps for on-demand network optimization~\cite{polese2021colo} to \glspl{ris} for improved resource utilization and energy efficiency, representing a cross-cutting \gls{kpi} dimension emphasized by both \gls{3gpp} and \oran~\cite{oran-wg1-use-cases,3gpp22261}. Finally, the study of coexistence mechanisms and strategies between terrestrial networks and \glspl{ntn} is essential for the successful deployment of future cellular architectures~\cite{3GPP38821, kang2024terrestrial, agarwal2023coexistence}, especially for \gls{6g}, which introduces \gls{6g}-ready \gls{ris}, \gls{ai}-native \gls{ran}~\cite{AIRANAlliance2024}, \glspl{dt}, and dense \gls{leo} constellations for improved resource efficiency.

\subsection{Open~\texorpdfstring{\gls{ran}}{RAN}: Enabling Resilient, Reconfigurable, and Autonomous Networks}
\label{sec:intro:openranintro}
 
Recent years have seen traditional \glspl{ran} transitioning toward more open, programmable, virtualized, and disaggregated architectures~\cite{polese2022understanding}. These concepts are at the foundation of the Open~\gls{ran} paradigm, which has revolutionized how \gls{nextg} networks are designed, operated, and deployed. The traditional \gls{bs}, which was once a closed, monolithic opaque system, now encompasses a distributed functional unit split among the \gls{cu}, \gls{du}, and \gls{ru} components. Each unit is responsible for different network functionalities, with open interfaces enabling these units to communicate seamlessly via standardized protocols. Through softwarization, various network functionalities (e.g., \gls{prb} allocation and scheduling decisions) can now be run as network functions on general-purpose or accelerated multi-vendor hardware within virtualized environments, facilitating network scalability and continuous integration and testing.

\begin{figure}[ht]
    \centering
    \includegraphics[width=0.99\textwidth]{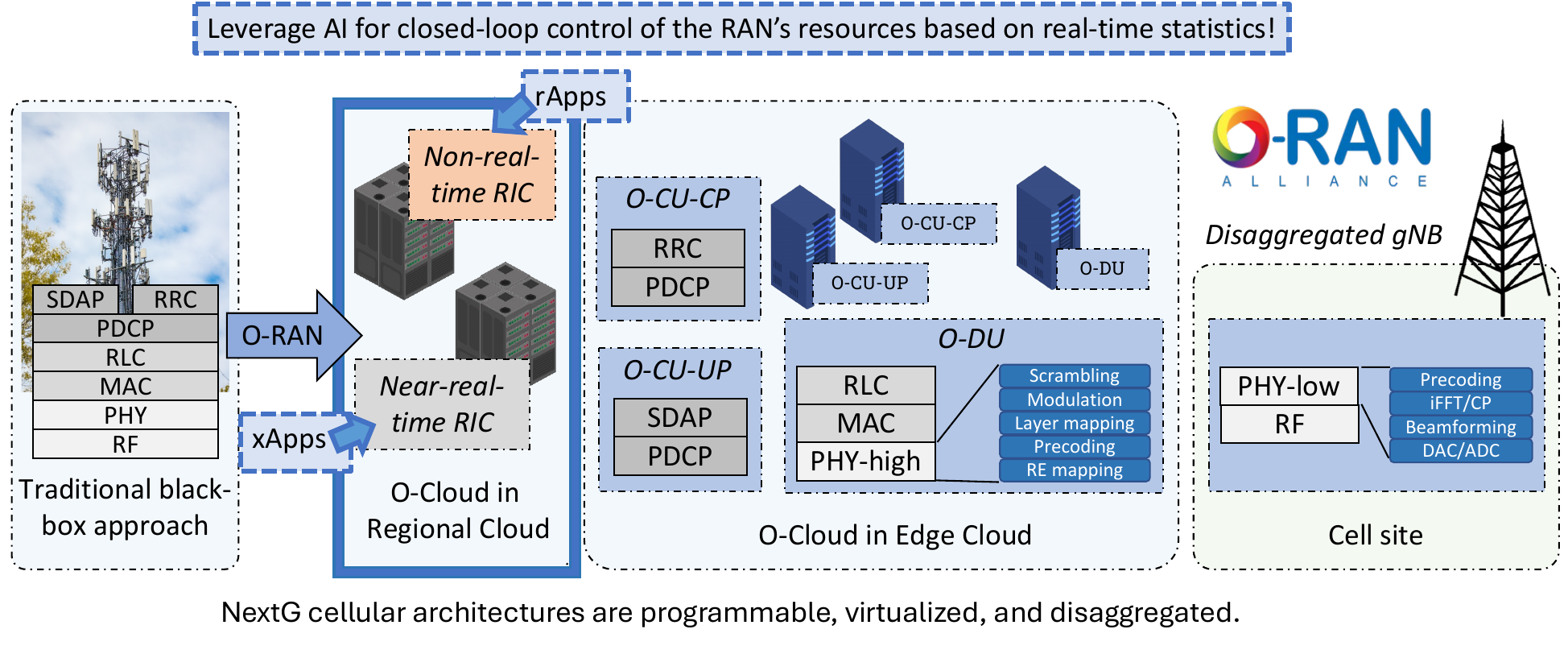}
    \caption{The Open~\gls{ran} Paradigm: programmable, virtualized, and disaggregated \gls{nextg} cellular architectures embedding intelligent control at the network edge, adapted from~\cite{polese2022understanding}.}
    \label{fig:oran}
\end{figure}

Furthermore, network intelligence can be added to the infrastructure at the network edge through \gls{ran} \glspl{ric}, enabling on-demand fine-tuning of network resources. Specifically, through the implementation of rApps, xApps, and dApps, network resources can be controlled at various timescales—from coarse to fine granularity—depending on the application. Various optimization solutions, ranging from classic optimization to \gls{ai}/\gls{ml} algorithms, can now be embedded directly into the network. Open~\gls{ran} further reinforces the notion of a truly \gls{ai}-native \gls{ran}, where multi-vendor solutions for intelligent \gls{ran} control can coexist seamlessly within the network, enhancing network management and orchestration. In this way, novel algorithm designs can be tested and validated in terms of achieved resource allocation efficiency and compared to standard research approaches that have been leveraged for years and instructed by standards bodies such as \gls{3gpp}. Indeed, the Open~\gls{ai}~\gls{ran} offers unprecedented opportunities that are poised to revolutionize the future of telecommunications and spectrum management. Finally, Fig.~\ref{fig:oran} summarizes the \oran architecture and the evolution of \gls{nextg} cellular \glspl{ran} into more open and distributed architectures which can now be optimized on demand through \gls{ran} intelligent control.

 \subsection{The Importance of \texorpdfstring{\gls{ris}}{RIS} technology for the \texorpdfstring{\gls{5g}}{5G} and Beyond era}
  \label{sec:intro:risintro}   

Another emerging question pertains to the optimization of \gls{rf} propagation, where the wireless environment is no longer treated as a fixed input to the optimization problem~\cite{renzo2019smart}. Thanks to \gls{ris} technology, the wireless environment is instead treated as an entity that can be controlled and optimized to enhance radio propagation, by intelligently steering the \gls{ris} elements’ phase shifts through dedicated controllers. \gls{ris} technology aligns with the requirements of the \gls{5g} and beyond era, as it has been shown to improve connectivity and coverage, enhance system capacity, reduce latency, and improve both spectral and energy efficiency across diverse deployments~\cite{ETSI_RIS001}.

 \subsection{The \texorpdfstring{\gls{5g}}{5G}-Satellite Spectrum Clash in the Upper Mid-Band}
  \label{sec:intro:spectrumsharingchallenge}  

\gls{fr3} is an emerging band for \gls{6g} applications. This mid-band spectrum, typically spanning frequencies between $7.125$--$24.25$~GHz, balances coverage and capacity by offering better propagation characteristics than mmWave frequencies and larger bandwidths than the sub-$6$~GHz band. However, this frequency range is already used by satellite systems (e.g., Starlink). Consequently, spectrum-sharing conflicts arise when \gls{5g} and beyond cellular applications attempt to access the already scarce spectrum resources~\cite{bazzi2025upper}.

Indeed, a major challenge for the successful commercial deployment of \gls{6g} stems from the operation of cellular networks that may disrupt incumbent satellite service links. Since cellular \glspl{bs} employ \gls{mimo} transmissions and high transmit power, the sidelobes generated by beamformed transmissions toward terrestrial \glspl{ue} can result in significant interference toward satellite \glspl{ul}. In particular, satellites operating at low elevation angles are especially vulnerable to such interference~\cite{kang2024terrestrial}.

Consequently, given the shared spectrum, devising coexistence mechanisms and spectrum-sharing strategies that enable the harmonious operation of terrestrial and satellite networks is essential. Such solutions must ensure satisfactory \gls{qos} for \gls{5g} and beyond cellular networks without disrupting satellite operations, ultimately enabling reliable service for terrestrial \glspl{ue} while protecting passing satellites along their predefined trajectories.

 \subsection{\texorpdfstring{\glspl{dt}}{DTs}: A Primer for \texorpdfstring{\gls{6g}}{6G} Research}
  \label{sec:intro:6gdt}   
\gls{6g} research is expected to emerge from simulation~\cite{nokia_ran_dt2026}, with \glspl{dt} playing a key role in the successful testing and evaluation of \gls{ai}-native \gls{ran} solutions before commercial deployment. Indeed, \gls{ai} for \gls{6g} and beyond \gls{ran} requires adaptive data-driven control for applications such as link adaptation and power control across various network slices. \glspl{dt} offer a cost-effective playground for \gls{ai} research that accelerates the \gls{ai}-\gls{ran} development cycle.

As shown in Fig.~\ref{fig:standards}, popular \gls{dt} platforms include Colosseum, the Open~\gls{ran} \gls{dt}~\cite{bonati2021colosseum} from \gls{pawr}~\cite{kochan2017ppo}, and SIONNA developed by NVIDIA~\cite{sionna2022}. Colosseum, hosted at Northeastern University, is the world’s largest wireless network emulator with hardware-in-the-loop capabilities. It leverages protocol stacks for the deployment of end-to-end \gls{ran} systems, supports large-scale data collection and \gls{ran} fine-tuning, and provides an \oran-compliant control architecture for \gls{ai}/\gls{ml}-based xApp execution in the near-real-time \gls{ric}.
% It enables experimentation under diverse \gls{rf} and network conditions through a channel emulator, which uses \gls{fpga}-based \gls{fir} filters to reproduce realistic wireless environments.
NVIDIA’s SIONNA provides an all-in-one solution spanning ray tracing to system-level experimental evaluation on ray-traced channels, while also enabling seamless plug-in integration of \gls{ai}/\gls{ml} solutions.

\section{Contributions \& Dissertation Outline}
\label{sec:intro:contributions}

It is evident that both research and industry emphasize the need for testbeds, novel algorithms, system-level frameworks, design tools, \glspl{dt}, openness, and \gls{ai}-native architectures. In this context, the present work contributes algorithmic designs for slicing, scheduling, link adaptation, energy-efficient power control, and beamforming, and discusses the integration of \glspl{ris} into Open~\gls{ran} architectures through extensive system-level experimental evaluation. Finally, the proposed approaches are validated using experimental platforms, including the Colosseum Testbed, \glspl{dt}, and real \gls{ota} \gls{5g} data, thereby providing a bridge between theoretical design and practical deployment.

  % Chapter~\ref{chap:intro}  introduces \gls{nextg} use cases and the necessity of experimental platforms for framework prototyping. 
  
The remainder of this dissertation is organized as follows. Chapter~\ref{chap:oran_edge} (based on~\cite{10643670,10766614,10437367,11007770}) explores the Colosseum testbed as a \gls{dt} for \oran, focusing on \gls{drl}-based xApps for slicing and scheduling. Chapter~\ref{chap:ris_propagation} (based on~\cite{10973151,11152828}) examines the cross-layer evaluation of \gls{ris} and its integration with Open~\gls{ran}. Chapter~\ref{chap:ai_ran_link_adaptation} (based on~\cite{tsampazi2026ariadne}) details \gls{ai}-\gls{ran} link adaptation using \glspl{dt} and compares it against industry standards and state-of-the-art methods. Chapter~\ref{chap:spectrum_sharing} (based on~\cite{tsampazi2026satellite}) discusses spectrum sharing in \gls{fr3}, specifically power control and beamforming strategies to protect \gls{ntn} systems while maintaining terrestrial service quality. Chapter~\ref{chap:conclusion} provides the final remarks and conclusions.

% --- EOF ---

%% file: tex/oran_edge.tex
% !TEX root = ../thesis_dissertation template.tex
% oran_edge.tex

\chapter[Enabling AI at the Edge in O-RAN]{Enabling \texorpdfstring{\gls{ai}}{AI} at the Edge in \texorpdfstring{\oran}{O-RAN}}
\label{chap:oran_edge}

% Local macro definitions for this chapter
\newcommand{\pandora}{PandORA\xspace}
\newcommand{\name}{PACIFISTA\xspace}
\newcommand{\kpms}{\glspl{kpm}\xspace}

% Macros required for the Colosseum DT integration (from colosseum-openran source)
\newcommand{\coloran}{ColO-RAN\xspace}
\newcommand{\openrangym}{OpenRAN Gym\xspace}
\newcommand{\scope}{SCOPE\xspace}
\newcommand{\neutran}{NeutRAN\xspace}
\newcommand{\nearrt}{Near-\gls{rt}\xspace}
\newcommand{\nonrt}{Non-\gls{rt}\xspace}
\newcommand{\ric}{\gls{ric}\xspace}
\newcommand{\rics}{\glspl{ric}\xspace}
\newcommand{\ran}{\gls{ran}\xspace}
\newcommand{\ru}{\gls{ru}\xspace}
\newcommand{\rus}{\glspl{ru}\xspace}
\newcommand{\du}{\gls{du}\xspace}
\newcommand{\dus}{\glspl{du}\xspace}
\newcommand{\prb}{\gls{prb}\xspace}
\newcommand{\prbs}{\glspl{prb}\xspace}
\newcommand{\ai}{\gls{ai}\xspace}
\newcommand{\xai}{\gls{xai}\xspace}
\newcommand{\cicd}{\gls{ci}/\gls{cd}\xspace}
\newlength\fheight
\newlength\fwidth
\tikzstyle{llarrow} = [line width=0.1mm,->,>=stealth]

% PACIFISTA math macros
\newcommand{\apps}{\mathcal{A}}
\newcommand{\params}{\mathcal{P}}
\newcommand{\metrics}{\mathcal{K}}
\newcommand{\gp}{G^\mathrm{P}}
\newcommand{\gk}{G^\mathrm{K}}
\newcommand{\ops}{\mathcal{C}}
\newcommand{\confp}{\Pi}
\newcommand{\confa}{\Theta}
\theoremstyle{definition}
\newtheorem{definition}{Definition}[chapter]

% Define JSON language for listings (not built-in)
\lstdefinelanguage{json}{
  basicstyle=\ttfamily\small,
  numbers=left,
  numberstyle=\tiny,
  stepnumber=1,
  showstringspaces=false,
  breaklines=true,
  string=[s]{"}{"},
  comment=[l]{//},
  morecomment=[s]{/*}{*/},
  literate=
     *{0}{{{\color{blue}0}}}{1}
      {1}{{{\color{blue}1}}}{1}
      {2}{{{\color{blue}2}}}{1}
      {3}{{{\color{blue}3}}}{1}
      {4}{{{\color{blue}4}}}{1}
      {5}{{{\color{blue}5}}}{1}
      {6}{{{\color{blue}6}}}{1}
      {7}{{{\color{blue}7}}}{1}
      {8}{{{\color{blue}8}}}{1}
      {9}{{{\color{blue}9}}}{1}
      {:}{{{\color{red}{:}}}}{1}
      {,}{{{\color{red}{,}}}}{1}
}

\section[Colosseum Testbed as an Open RAN Digital Twin Playground]{Colosseum Testbed as an Open \texorpdfstring{\gls{ran}}{RAN} \texorpdfstring{\gls{dt}}{Digital Twin} Playground}
\label{sec:oran_edge:colosseum}

\begin{figure}[t]
  \centering
  \includegraphics[width=\textwidth]{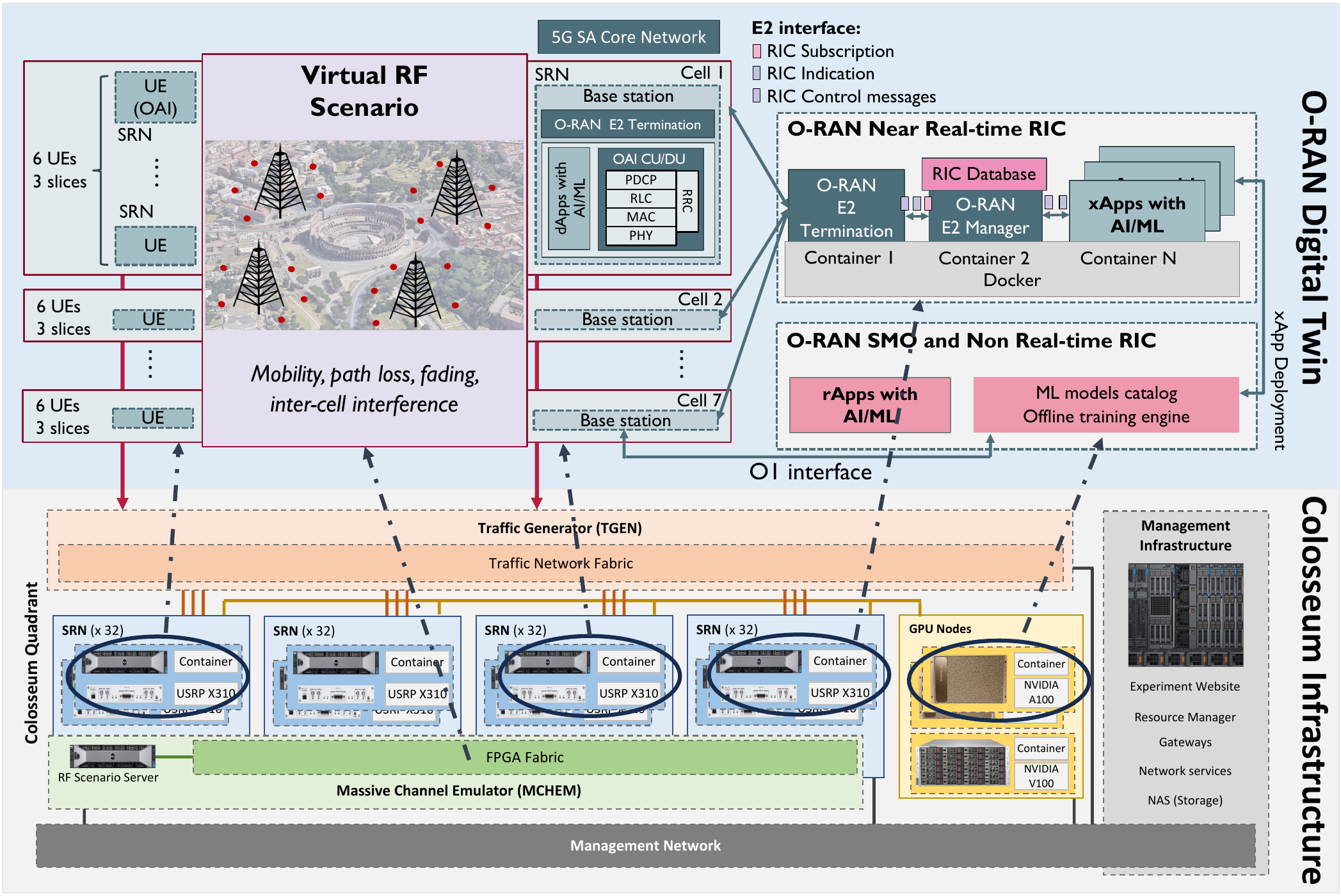}
  \caption{Open RAN twinning capabilities in Colosseum.}
  \label{fig:oran_edge:colosseum:overview}
\end{figure}

\subsection{Introduction}
\label{sec:oran_edge:colosseum:introduction}
Recent years have witnessed the Open \gls{ran} paradigm transforming the fundamental ways cellular systems are deployed, managed, and optimized.
This shift is driven by concepts such as openness, softwarization, programmability, interoperability, and network intelligence, which have already emerged in wired networks through \gls{sdn} but still lag behind in cellular systems.
The realization of the Open \gls{ran} vision in practical architectures, intelligent data-driven control loops, and efficient software implementations, however, is a multifaceted challenge that requires (i)~datasets to train \gls{ai} and \gls{ml} models; (ii)~facilities to test models without disrupting production networks; (iii)~continuous and automated validation of the \gls{ran} software; and (iv)~significant testing and integration efforts.
This chapter describes how Colosseum---the world's largest wireless network emulator with hardware in the loop---can provide the research infrastructure and tools to bridge the gap between the Open \gls{ran} vision and the deployment and commercialization of open and programmable networks.
We describe how Colosseum implements an Open \gls{ran} \gls{dt} through a high-fidelity \gls{rf} channel emulator and end-to-end softwarized O-RAN and \gls{5g}-compliant protocol stacks, thus allowing users to reproduce and experiment with topologies representative of real-world cellular deployments.
Finally, we showcase the development, prototyping, and testing of \gls{ai}/\gls{ml} solutions for Open \gls{ran}.

Wireless networks and systems are evolving toward more programmable, open, and virtualized solutions. The transition to software-defined networking~\cite{mckeown2008openflow}, which has transformed management operations in the wired networking industry over the past decade, is now extending to cellular networks.
While introducing software and programmability in radio systems is more challenging than in switching and routing, it also creates opportunities to deploy and configure bespoke networks that can be dynamically tuned to optimally support a variety of verticals.
The Open \gls{ran} paradigm, its embodiment in the technical specifications of the O-RAN ALLIANCE, and its evolution in the 3GPP \gls{ran} and core network designs are converging toward a network architecture that is open, disaggregated, programmable, and intelligent~\cite{polese2022understanding,abdalla2022toward}.
Openness, embedded in interfaces that expose \gls{ran} telemetry and control, enables the programmability and optimization of \gls{ran} functions through intelligent closed-loop control driven by the so-called \glspl{ric}~\cite{oran-wg3-ricarch,balasubramanian2021ric}. Disaggregation, implemented by splitting \gls{ran} functionalities into self-contained units across user and control planes, fosters innovation and introduces multi-vendor interoperability.
Finally, programmability, realized through open \glspl{api} of software solutions, enables swift reconfiguration of the white-box components of cellular networks to adapt to ever-changing network conditions and demand.
Overall, these ingredients have the potential to transform how cellular networks are deployed, managed, and optimized.

However, advancing the Open \gls{ran} vision toward a fully developed architecture with robust and reliable algorithmic components and seamless multi-vendor integration requires addressing several challenges at the architectural, algorithmic, and system levels, including the following.

\textbf{Need for Datasets.} To develop robust and scalable \gls{ai} and \gls{ml} solutions that generalize well across a variety of real-world deployment scenarios, it is necessary to leverage rich datasets of \gls{ran} telemetry, data, and performance indicators~\cite{challita2020when}. While network operators are in a unique position to collect such datasets, it is often impractical or impossible to use them for research and development due to privacy and security concerns. Wireless testbeds represent a feasible path to overcome this limitation~\cite{villa2024dt,breen2020powder,raychaudhuri2020cosmos,marojevic2020advanced,zhang2021ara}. However, they are often limited by the \gls{rf} characteristics and topology of their deployment areas.

\textbf{End-to-end \gls{ai} and \gls{ml} Testing.} Once trained, \gls{ai}/\gls{ml}-based control solutions need to be validated and tested in controlled environments to avoid disruption to production networks. At the same time, testing conditions must be realistic to obtain meaningful results that account for, for instance, user load, traffic patterns, and \gls{rf} characteristics of the real-world deployments the models are intended to control.

\textbf{Automated Integration and Testing of Disaggregated Components.} Disaggregation enables a more robust supply chain but also introduces the need to validate interoperability across vendors and devices. This is a labor-intensive and often manual process that calls for the development of automated techniques~\cite{dellOroRAN}.

In this section, we describe how Colosseum---the world's largest wireless network emulator with hardware in the loop~\cite{bonati2021colosseum}---can be leveraged as a \gls{dt} platform to address these challenges and to develop end-to-end, fully integrated, and reliable solutions for Open \gls{ran}, as shown in Figure~\ref{fig:oran_edge:colosseum:overview}.
Through its channel and traffic emulation capabilities, Colosseum can replicate a wide range of real-world scenarios representative of cellular deployments and generate datasets that can be used to train \gls{ai}/\gls{ml} models robust to network changes~\cite{polese2021colo}.
A combination of generic compute nodes, \glspl{sdr}, and the possibility of integrating \gls{cots} devices allows the digital replication of Open \gls{ran} and \gls{5g}-and-beyond protocol stacks~\cite{kaltenberger2020openairinterface,gomez2016srslte}.
This enables repeatable experiments in which different network configurations and protocol stacks can be tested under the same channel and traffic conditions, while also providing a safe environment for testing \gls{ai}/\gls{ml} solutions.

\subsection{Related Work}
\label{sec:oran_edge:colosseum:soa}

The development of \glspl{dt} and testbeds for cellular systems, and specifically for Open \gls{ran}, has recently received significant attention, owing to the importance of addressing the challenges described in Section~\ref{sec:oran_edge:colosseum:introduction}. The O-RAN ALLIANCE itself has dedicated a study area to \glspl{dt} for Open \gls{ran} within its next Generation Research Group (nGRG).

\gls{dt} frameworks for cellular networks are discussed in~\cite{jagannath2022digital,masaracchia2023digital,corici2023digital,nguyen2021digital,lin20236g,alkhateeb2023real,MCMANUS2023110000,mirzaei2023network}, with analyses of the challenges and opportunities of using \glspl{dt} for Open \gls{ran} and beyond-\gls{5g} design.
Paper~\cite{vila2023design} proposes a network \gls{dt} framework for \gls{5g} and develops a use case on \gls{ran} control with reinforcement learning, using a simulation-based twin that replicates a real-world scenario in Barcelona, Spain. The authors of~\cite{li2022rlops} discuss the opportunity of integrating \glspl{dt} into automated operations for the development of reinforcement learning algorithms for Open \gls{ran}. Reference~\cite{ndikumana2023digital} presents a simulation study framed in the context of \glspl{dt} to optimize fixed wireless access use cases for cellular networks. The studies in~\cite{deng2021digital,ren2023end} leverage \glspl{dt} to create virtual environments for testing network updates driven by \gls{ai}/\gls{ml} components. These works, however, are primarily based on simulation-driven twins, which fail to capture the complexity of real-world \gls{rf} equipment. In addition, they provide limited evaluation of the \gls{dt} in real-world settings, e.g., through experiments in operational networks or testbeds. The Colosseum \gls{dt} described in this chapter is based on real-time channel emulation and end-to-end Open \gls{ran} protocol stacks that can be either automatically deployed and tested in the real world or used within the \gls{dt} as-is, thus achieving a high level of fidelity, as demonstrated by the experiments in~\cite{villa2024dt, bonati2023openran}.

Implementations of \glspl{dt} or components of \glspl{dt} for networked systems are presented in~\cite{moorthy2022middleware,HU2023342,baranda2021demo}, focusing on middleware for \glspl{dt} for aerial networks and on optimizing network services for the replication of real-world setups (e.g., for Industry 4.0) within a \gls{dt}. However, these works do not focus on end-to-end Open \gls{ran} systems or on the development of intelligent control solutions. Testbeds and frameworks for the evaluation of Open \gls{ran} systems are discussed in~\cite{bahl2023accelerating,villa2025x5g,upadhyaya2023open,upadhyaya2022prototyping,koumaras20185genesis}, and Open \gls{ran} evaluation is also supported by the \gls{pawr} platforms~\cite{breen2020powder,raychaudhuri2020cosmos,marojevic2020advanced,zhang2021ara}. The PAWR platforms are the closest in terms of capabilities and programmability to Colosseum, being mostly based on \glspl{sdr}, but they lack the emulated environment that Colosseum supports.

\subsection{Colosseum Wireless Network Emulator}
\label{sec:oran_edge:colosseum:emulator}

The architecture of Colosseum is shown in the bottom part of Figure~\ref{fig:oran_edge:colosseum:overview}.
Its main components are: (i)~128~pairs of generic compute servers and \glspl{sdr}, named \glspl{srn}; (ii)~a channel emulation system; and (iii)~a state-of-the-art \gls{ai}/\gls{ml} infrastructure.

Each \gls{sdr} is paired with a Dell PowerEdge R750 server connected to an NI/Ettus \gls{usrp} X310 \gls{sdr} through a dedicated $10$\:Gbps link.
Users of Colosseum can reserve these \glspl{srn} and deploy either custom or pre-configured protocol stacks through \gls{lxc}.
Container images are stored in a dedicated storage infrastructure, which is part of a larger management infrastructure that includes services such as web interfaces, a \gls{vpn} for external connectivity, and networking functionalities.

The Colosseum channel emulator, namely \gls{mchem}, is responsible for reproducing digital representations of real-world \gls{rf} propagation scenarios, which represent the state of the \gls{rf} channel at given time instants.
Specifically, each pair of transceivers, i.e., those of the \glspl{sdr}, is modeled through a \gls{tdl} with up to four non-zero taps that represent the multipath components of the \gls{cir} for each millisecond of the captured scenario.
At its core, \gls{mchem} includes four quadrants with 64~\glspl{fpga} and 128~\glspl{sdr} that up-/down-convert the signal between \gls{rf} and baseband.
The \glspl{fpga} implement real-time convolution operations on the digital baseband signal, modeling the \gls{rf} channel effects of real-world environments.
These taps, which have a maximum propagation delay of $5.12\:\mathrm{\mu}$s~\cite{chaudhari2018scalable}, are streamed to the \glspl{fpga} from a dedicated \gls{rf} scenario server.
We refer readers to~\cite{bonati2021colosseum} for a comprehensive overview of the Colosseum architecture.

The \gls{ai}/\gls{ml} infrastructure of Colosseum, shown in yellow at the bottom of Figure~\ref{fig:oran_edge:colosseum:overview}, consists of two NVIDIA DGX A100 systems with 8~\glspl{gpu} each, providing $10$\:petaFLOPS of compute power, and one Supermicro Superserver 8049U-E1CR4T with 6 NVIDIA V100 \glspl{gpu} and $3$\:TB of RAM.
Compared to the virtualization system of the \glspl{srn}, the \gls{gpu} nodes leverage Docker containers.

\subsection{End-to-End Open RAN Twinning in Colosseum}
\label{sec:oran_edge:colosseum:twinning}

The emulation capabilities of Colosseum can be leveraged to deploy, study, and profile wireless protocol stacks in controlled and repeatable environments.
As an example, Figure~\ref{fig:oran_edge:colosseum:overview} illustrates the deployment of an end-to-end, fully programmable Open \gls{ran} cellular system in Colosseum.
The top-left portion of the figure depicts the \gls{ran}, which consists of a core network and cellular base stations---either monolithic or disaggregated into \glspl{cu} and \glspl{du}---deployed over an emulated \gls{rf} scenario of interest and serving mobile \glspl{ue}.
The top-right portion of the figure shows the \gls{smo} and \oran \glspl{ric}, which interface with the \gls{ran} base stations through open interfaces and augment their operation using \gls{ai}/\gls{ml} applications deployed therein, namely xApps and rApps.
We refer readers to~\cite{polese2022understanding} for a detailed overview of the O-RAN architecture and its functionalities.

In Colosseum, these O-RAN-aligned solutions are provided to users through ready-to-use \gls{lxc} images that implement functionalities exposed via the \openrangym framework, which is described next.
\subsubsection{Open-source Cellular Protocol Stacks}
\label{sec:oran_edge:colosseum:stacks}

Colosseum enables experimentation with different open-source \gls{5g} softwarized protocol stacks, such as \gls{oai}~\cite{kaltenberger2020openairinterface} and srsRAN~\cite{gomez2016srslte}. These are provided as ready-to-use images~\cite{colosseumImages}—the same images that can be used for \gls{ota} experiments with minor variations in the configuration parameters. These open-source projects provide reference implementations of \gls{3gpp} cellular protocol stacks, including that of \gls{5g} \gls{nr}.
The \gls{nr} system consists of a \gls{5g} \gls{ran} and a core network that run on Linux-based general-purpose compute platforms and can control \glspl{sdr}, e.g., the NI/Ettus \glspl{usrp} deployed in Colosseum. The \gls{ran} applications support the instantiation of \glspl{gnb} and \glspl{ue}.
Compared to \gls{ota} environments, however, the additional components in the Colosseum transmit and receive chains (e.g., those that are part of \gls{mchem}) require adjusting the timing advance during the establishment of the initial connection between \gls{gnb} and \glspl{ue}.
This can be done using the \texttt{ta} input argument when starting the base station application.

\subsubsection{OpenRAN Gym}
\label{sec:oran_edge:colosseum:openrangym}

\openrangym is an open-source framework for O-RAN experimentation that enables users to instantiate softwarized \glspl{ran} based on the \gls{oai} and srsRAN protocol stacks, and to control them through xApps deployed on an O-RAN \nearrt \gls{ric}~\cite{ bonati2023openran}.
Its components include (i)~\scope~\cite{bonati2021scope}, a wrapper for \gls{ran} software that simplifies the user experiment workflow and enables data collection at scale, and (ii)~\coloran~\cite{polese2021colo}, a simplified version of the O-RAN Software Community \nearrt \gls{ric} that provides an \gls{sdk} for xApp design, as well as pipelines for data collection and training of \gls{ai}/\gls{ml} models for \gls{ran} inference and control to be embedded into xApps.
Overall, \openrangym allows users to instantiate a fully compliant \nearrt \gls{ric} using Docker containers, as well as xApps that can interact with the \gls{ran} (e.g., to receive \glspl{kpi} and enforce control policies) via the E2 interface.

Examples of developed xApps include \gls{drl} agents that optimize \gls{ran} performance in real time by controlling scheduling and slicing configurations~\cite{polese2021colo,10437367}.

\subsection{Radio-Frequency Twinning}
\label{sec:oran_edge:colosseum:rf-twinning}

The channel and traffic emulation capabilities, together with the cellular protocol stacks that can be deployed on Colosseum, provide key building blocks toward the creation of high-fidelity digital replicas of Open \gls{ran} systems with \gls{rf} hardware in the loop.
This section describes a set of toolchains used to twin real-world \gls{rf} scenarios in Colosseum. \gls{rf} scenarios are generated offline and then replayed in real time by the scenario server.
To enable this process, we rely on the \gls{rf} twinning methodology of the \gls{cast} framework~\cite{villa2022cast}. The framework comprises two main components: (i)~a streamlined framework for creating realistic mobile wireless scenarios; and (ii)~an \gls{sdr}-based channel sounder for characterizing emulated \gls{rf} channels. The combination of these components enables the generation and validation of \gls{dt} \gls{rf} scenarios, ensuring that the digital replicas closely match their real-world counterparts.

The main steps of the scenario creation workflow are as follows:
(i)~identify the wireless environment to be captured, i.e., the physical location to be twinned, which may vary in size and type, e.g., indoor, outdoor, urban, or rural;
(ii)~obtain a 3D model of the environment by leveraging online databases such as \gls{osm2};
(iii)~load the model into a ray-tracing tool, e.g., MATLAB ray tracer, \gls{wi}, or SIONNA;
(iv)~define the location and trajectories of nodes;
(v)~sample the channel using the ray tracer, capturing effects such as node mobility;
(vi)~parse the ray-tracer output to extract the channel for each pair of nodes at each millisecond of the emulated scenario;
(vii)~approximate the resulting channel representation into a format suitable for the Colosseum channel emulator, i.e., $512$ channel taps with at most $4$ non-zero taps, spaced at $10$\:ns intervals and with a maximum delay spread of $5.12\:\mathrm{\mu s}$;
(viii)~install the scenario in Colosseum by converting it into an \gls{fpga}-based format.
This scenario creation toolchain is modular and allows users to modify or provide inputs at any of the above steps.

After a scenario is installed in Colosseum, it is validated to ensure that it closely matches the expected real-world behavior.
This validation relies on a channel sounder implemented in GNU Radio~\cite{gnuradio} and consists of the following steps:
(i)~transmission of a known code sequence used as a reference signal via the GNU Radio transmitter;
(ii)~reception at the GNU Radio receiver of the signal processed by \gls{mchem}, which applies the corresponding channel characteristics;
(iii)~post-processing of the received signal using the known code sequence to extract the \gls{cir} and \gls{pl} of the waveform;
(iv)~validation of the results against the original modeled channel taps.

\subsection{Use Cases of Colosseum as Open RAN Digital Twin}
\label{sec:oran_edge:colosseum:research}

Colosseum has been extensively used to study open problems in the algorithmic, architectural, and system-level design of Open \gls{ran} systems, leveraging the realism of its \gls{rf} scenario emulation and the ability to run state-of-the-art software-defined protocol stacks that replicate real-world deployments. In this section, we review recent results as examples of Open \gls{ran} studies that can be conducted on Colosseum.

\subsubsection{Slicing and Scheduling Optimization}

\gls{ran} slicing allows operators to dynamically partition the available spectrum bandwidth to accommodate \glspl{ue} with heterogeneous \gls{qos}, performance, and \gls{sla} requirements. A common classification of services and \glspl{ue} in \gls{5g} includes \gls{embb}, \gls{mtc}, and \gls{urllc}. To achieve flexibility in \gls{ran} control, \gls{ai}/\gls{ml} agents can be designed and embedded in xApps, then deployed on the \nearrt \gls{ric} and used to adapt the \gls{ran} to varying network conditions and demand. Specifically, \gls{drl} can be leveraged in the design of such control solutions for Open~\gls{ran}, as it can be trained a priori and does not require prior knowledge of run-time network dynamics~\cite{10437367}. Since these solutions require large datasets for training prior to deployment in production \gls{ran} environments (to avoid disrupting network operations), the scale and controlled nature of \glspl{dt} provide a suitable environment for their development. Through the \openrangym framework (described in Section~\ref{sec:oran_edge:colosseum:openrangym}), Colosseum provides the tools and infrastructure for data collection, design, training, testing, and evaluation of \gls{ai}/\gls{ml} xApps prior to their deployment in over-the-air systems.

\subsection{Colosseum for O-RAN Testing and Integration}
\label{sec:oran_edge:colosseum:otic}

The Colosseum Open \gls{ran} \gls{dt} is also an \gls{otic} center officially recognized by the O-RAN ALLIANCE for testing and integrating Open \gls{ran} components from multiple vendors. Current \gls{otic} activities primarily focus on achieving interoperability in disaggregated Open \gls{ran} architectures, including conformance testing of O-RAN interfaces and equipment with respect to O-RAN specifications, as well as interoperability certification of component pairs and end-to-end systems. The Colosseum data center already supports such testing; its digital twinning capabilities, together with its controlled and repeatable experimental environment and \gls{ai}/\gls{ml} infrastructure, make it a key enabler for the testing and eventual certification of \gls{ai}-based solutions for Open \gls{ran}.

\subsection{Conclusions}
\label{sec:oran_edge:colosseum:conclusions}

This chapter reviewed how Colosseum, the world's largest wireless network emulator with hardware in the loop, can be leveraged as a \gls{dt} for end-to-end Open \gls{ran} systems. We introduced the Colosseum architecture and described how it enables a high-fidelity digital replica of real-world \gls{rf} environments, as well as the deployment of real-world protocol stacks. We then presented use cases where Colosseum’s \gls{dt} capabilities have been used to advance algorithms, architectures, and testing of Open \gls{ran} systems. We now showcase how Colosseum, through its emulation capabilities and support for softwarized stacks, enables the development of effective solutions for the control and optimization of the \gls{ran}, such as network slicing and scheduling.

\section[DRL-based xApps for Slicing and Scheduling Optimization]{\texorpdfstring{\gls{drl}}{DRL}-based xApps for Slicing and Scheduling Optimization}
\label{sec:oran_edge:drl_xapps}

\subsection{Introduction}\label{sec:oran_edge:pandora:intro}

The highly heterogeneous ecosystem of \gls{nextg} wireless communication systems calls for novel networking paradigms where functionalities and operations can be dynamically and optimally reconfigured in real time to adapt to changing traffic conditions and satisfy stringent and diverse \gls{qos} demands.
Open \gls{ran} technologies, and specifically those being standardized by the \oran Alliance, make it possible to integrate network intelligence into the once monolithic \gls{ran} via intelligent applications, namely, xApps and rApps.
These applications enable flexible control of the network resources and functionalities, network management, and orchestration
through data-driven intelligent control loops. 
Recent work has shown how \gls{drl} is effective in dynamically controlling \oran systems. However, how to
design these solutions in a way that manages heterogeneous optimization goals and prevents unfair resource allocation is still an open challenge, with the logic within \gls{drl} agents often considered as an opaque system.
In this chapter, we introduce \pandora~\cite{10766614}, a framework to automatically design and train \gls{drl} agents for Open RAN applications, package them as xApps and evaluate them in the Colosseum wireless network emulator.
We benchmark $23$ xApps that embed \gls{drl} agents trained using different architectures, reward design, action spaces, and decision-making timescales, and with the ability to hierarchically control different network parameters. We test these agents on the Colosseum testbed under diverse traffic and channel conditions, in static and mobile setups.
Our experimental results indicate how suitable fine-tuning of the \gls{ran} control timers, as well as proper selection of reward designs and \gls{drl} architectures can boost network performance according to the network conditions and demand. Notably, finer decision-making granularities can improve \gls{mmtc}'s performance by $\sim56\%$ and even increase \gls{embb} Throughput by $\sim99\%$.

Programmable, virtualized, and disaggregated architectures are seen as key enablers of \gls{nextg} cellular networks. Indeed, the flexibility offered through softwarization, virtualization, and open standardized interfaces provides new self-optimization capabilities through \gls{ai}. These concepts are at the foundation of the Open \gls{ran} paradigm, which is being specified by the \oran Alliance. Thanks to the \glspl{ric} proposed by \oran (i.e., the near- and non-real-time \glspl{ric}), intelligence can be embedded into the network and leveraged for on-demand closed-loop control of its resources and functionalities~\cite{polese2022understanding}. This is achieved via intelligent applications, called xApps and rApps, which execute on the near- or non-real-time \glspl{ric}, respectively. Through the \glspl{ric}, these applications interface with the network nodes and implement data-driven closed-loop control based on real-time statistics received from the \gls{ran}, thus realizing the vision of resilient, reconfigurable and autonomous networks. Since they do not require prior knowledge of the underlying network dynamics~\cite{sutton2018reinforcement}, \gls{drl} techniques are usually preferred in the design of such control solutions for the Open \gls{ran}~\cite{wang2022self,polese2021colo}.

\subsubsection{Related Work}\label{sec:oran_edge:pandora:related}

Intelligent control in \oran through xApps has widely attracted the interest of the research community. For example, \cite{johnson2022nexran} proposes the NexRAN xApp to control and balance the throughput of different \gls{ran} slices. In~\cite{kak2023hexran}, an \oran-compliant \gls{ran} platform is introduced for \gls{mrat} environments, enabling network slicing and slice-specific scheduling. The FlexSlice framework in~\cite{EURECOM+7416} addresses \gls{ran} slicing and scheduling with finer control loop granularity for real-time efficiency. In~\cite{wiebusch2023towards}, the authors discuss an \oran-based framework for predictive \gls{ul} network slicing, leveraging a Deep Learning-based xApp for dynamic reconfiguration of \gls{ran} scheduling for \gls{urllc} services. Similarly, in~\cite{yeh2023deep}, the authors present a Deep Learning-based approach to develop a \gls{ran} slicing xApp.
The authors of~\cite{kouchaki2022actor} develop a \gls{rl} xApp to assign resource blocks to certain users according to their \gls{csi} and with the goal of maximizing the aggregated data rate of the network. A deep Q-learning-based xApp for controlling slicing policies to minimize latency for \gls{urllc} slices is presented in~\cite{filali2023communication}, while a cooperative multi-agent \gls{rl} algorithm for Open RAN slicing and radio resource management, taking into account various \gls{sla} constraints, is discussed in~\cite{10329913}. The authors of~\cite{polese2021colo} experimentally evaluate and demonstrate three \gls{drl}-based xApps under a variety of traffic and channel conditions, and investigate how different action space configurations impact the network performance. Finally, other research efforts focus on coordinating multiple xApps to control different parameters via a combination of federated learning and team learning~\cite{iturria2022multi,zhang2022team,zhang2022federated,9933014}.

\gls{drl} has been widely used for resource allocation~\cite{luong2019applications,alwarafy2021deep} in the wireless communications and networking field. In~\cite{suh2022deep}, the authors propose a \gls{qos}-oriented resource allocation scheme for network slicing and throughput maximization in \gls{5g} and beyond networks. DeepSlicing in~\cite{liu2020deepslicing} relies on a \gls{drl} approach to determine the number of resources required by the \glspl{ue} in each slice to ensure their \gls{qos} demands are met. In~\cite{sun2019dynamic}, the authors present a \gls{drl}-based framework for \gls{nextg} \gls{ran} slicing to improve resource utilization and meet the slices' \gls{qos} requirements by leveraging feedback generated by the \gls{dqn} agents. In~\cite{yan2023deep}, the authors discuss a two-level scheduling framework to jointly maximize the \gls{qoe} and \gls{se}. Specifically, the higher layer controller manages bandwidth allocation, while the lower level controller is responsible for scheduling the \gls{prb} and power allocation at a smaller timescale, utilizing \gls{mimo} antenna technology. In~\cite{naderializadeh2021resource} a \gls{marl} approach is introduced for joint optimization of \gls{ue} scheduling and power control in a wireless environment characterized by the coexistence of multiple \glspl{tx} with multiple associated \glspl{ue}. In~\cite{azimi2021energy}, an energy-efficient \gls{ran} slicing scheme utilizing \gls{drl}-assisted learning for resource allocation in \gls{5g} networks is introduced. Specifically, a combination of Deep Learning and \gls{drl} is employed for resource allocation on large and small time-scales, respectively. In~\cite{rahimi2022novel} a hierarchical \gls{drl}-based approach is employed for resource allocation, coupled with a load balancing strategy to enhance energy efficiency while ensuring user fairness from a \gls{qos} perspective.

\begin{figure}[htbp]
  \centering
  \includegraphics[width=.99\textwidth]{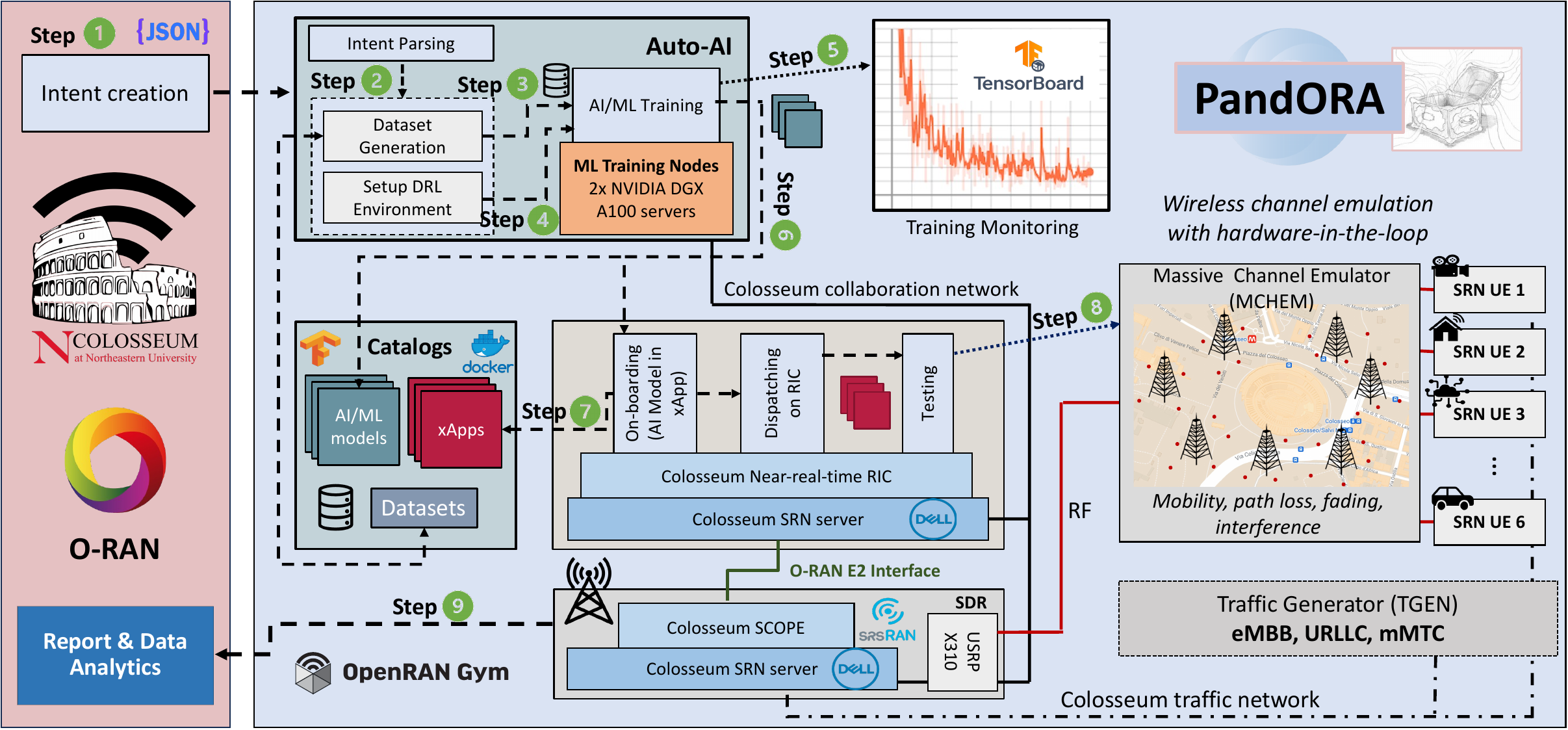}
  \caption{\pandora framework for intent-driven DRL training, xApp on-boarding, and testing with Open RAN in Colosseum.}
  \label{fig:pandora-toolbox}
\end{figure}

\subsubsection{Contributions and Outline}\label{sec:oran_edge:pandora:contributions}

The above works clearly show that \gls{drl} and \gls{ai} are catalysts in the design and development of intelligent control solutions for the Open \gls{ran}. However, despite early results showing their success and effectiveness, designing \gls{drl} agents which are effective at controlling and optimizing complex Open \gls{ran} scenarios---characterized by the coexistence of diverse traffic profiles and potentially conflicting \gls{qos} demands---is still an open challenge that, as we describe below, we aim at addressing in this chapter. Specifically, our goal is to go beyond merely using \gls{ai}, and specifically \gls{drl}, in a black-box manner. Instead, we try to address some fundamental design questions that are key for the success of intelligence in Open \gls{ran} systems.

We consider an Open \gls{ran} delivering services to \gls{urllc}, \gls{mmtc} and \gls{embb} network slices. Specifically, we use OpenRAN Gym~\cite{bonati2023openran}---an open-source framework for \gls{ml} experimentation in \oran---to deploy such Open \gls{ran} on the Colosseum wireless network emulator~\cite{bonati2021colosseum}, and control it through xApps using $23$ different \gls{drl} agent designs. These xApps have been trained to perform slice-based resource allocation (i.e., scheduling profile selection and \gls{ran} slicing control) and to meet the diverse requirements of each slice. We investigate the trade-off between long-term and short-term rewards, we discuss and compare different design choices of action set space and \gls{drl} architecture, hierarchical decision-making policies and action-taking timescales. Finally, we show how these choices greatly impact network performance and affect each slice differently.

To the best of our knowledge, in~\cite{10437367}, we conducted the first experimental study that comprehensively evaluated the design choices for \gls{drl}-based xApps to provide insights on the design of xApps for \gls{nextg} Open \glspl{ran}. In this chapter, we extend our previous work~\cite{10437367} and present \pandora, a large-scale evaluation and profiling of \gls{drl} agents for Open RAN, leveraging a framework to automate the training of \gls{drl} agents and their on-boarding as xApps to be executed in the near-real-time \gls{ric}. This all-in-one solution, spanning from the automated end-to-end streamlining of \gls{drl} model training, to testing on an Open \gls{ran} network, is enabled by the Colosseum testbed. Moreover, we extend the analysis in~\cite{10437367} by considering two new directions that affect \gls{drl}-based xApp design. Specifically, (i) we investigate the trade-offs between using a global \gls{drl} agent that takes decisions for all slices against the case of training and deploying per-slice dedicated xApps; and (ii) we analyze the effect of \gls{ran} control timing (i.e., the temporal granularity we use to compute and enforce new control actions). We explore the aforementioned directions by training a variety of \gls{drl} agents, which we then onboard on xApps. We test these agents both under network conditions similar to those encountered in the training dataset (i.e., in-sample experimental evaluation), as well as under previously unseen network conditions not encountered during the training process (i.e., out-of-sample experimental evaluation).

The remainder of this chapter is organized as follows. Section~\ref{sec:oran_edge:pandora:sysmodel} describes the \pandora System Model and Evaluation Framework. Section~\ref{sec:oran_edge:pandora:strategies} presents the different \gls{drl} optimization strategies considered in this work, while Section~\ref{sec:oran_edge:pandora:setup} details our experimental setup and training methodology, through \pandora. Experimental results are discussed in Sections~\ref{sec:oran_edge:pandora:insample},~\ref{sec:oran_edge:pandora:outofsample} and~\ref{sec:oran_edge:pandora:broader}. Finally, Section~\ref{sec:oran_edge:pandora:conclusions} draws our conclusions.

\subsection{The \texorpdfstring{\pandora}{PandORA} System Model and Evaluation Framework}\label{sec:oran_edge:pandora:sysmodel}

Before providing thorough implementation details of the \pandora system design and its operational procedures (Section~\ref{sec:oran_edge:pandora:overview}), we first briefly introduce the \oran system model considered in this work (Section~\ref{sec:oran_edge:pandora:system_model}).

\subsubsection{System Model and Reference Use-Case Scenario}\label{sec:oran_edge:pandora:system_model}

In this work, we consider an Open \gls{ran} multi-slice scenario where \glspl{ue} generate traffic with diverse profiles and \gls{qos} demands. Without loss of generality, we assume that traffic generated by \glspl{ue} can be classified into \gls{embb}, \gls{urllc}, or \gls{mmtc} slices.

We focus on the case where a set of xApps execute in the near-real-time \gls{ric} to intelligently control the resource allocation process in a way that satisfies the diverse \gls{qos} demands required by each slice. These xApps take these control decisions by leveraging \gls{ai}/\gls{ml} algorithms that can control \gls{bs} parameters and functionalities such as \gls{ran} slicing (i.e., the portion of available \glspl{prb} that are allocated to each slice at any given time) and \gls{mac} layer scheduling policies (i.e., how such \glspl{prb} are internally allocated to users belonging to each slice). It is worth mentioning that xApps can assign a certain scheduler profile (to be selected among \gls{rr}, \gls{wf} and \gls{pf}) to each slice, but do not take \gls{tti}-level scheduling decisions, which is instead left to the \gls{enb}. xApps make decisions based on \gls{ue} traffic demand, load, performance and network conditions that are given by \glspl{kpm} periodically reported by the \gls{ran}.

\subsubsection{\texorpdfstring{\pandora}{PandORA} Overview and Procedures}\label{sec:oran_edge:pandora:overview}

The architecture and building blocks of \pandora are illustrated in Fig.~\ref{fig:pandora-toolbox}. These are key to our extensive evaluation campaign, as they streamline the \gls{drl} design and xApp evaluation. \pandora leverages automated pipelines that cover the end-to-end lifecycle of \gls{ai} development for \oran systems. Specifically, it embeds the following components and functionalities:
\begin{itemize}
    \item \textbf{Catalogs.} These include \gls{ai}/\gls{ml} models trained offline, xApps, and datasets that can be used to train and test data-driven solutions.
    \item \textbf{Intent-driven Auto-\gls{ai} Engine.} This is a fully-customizable software component that parses training intents defined in JSON format and automatically extracts the necessary data from datasets in the catalog. Once the training process is instantiated, \gls{ai} solutions are trained based on the intent, and the \gls{ai} training is configured to match the desired inputs, outputs, and goals. It also offers web-based dashboards (e.g., TensorFlow's TensorBoard~\cite{tensorflow2015-whitepaper}) to monitor the training phase.
    \item \textbf{xApp On-boarding Module.} Templates and pipelines are provided to convert trained \gls{ai} solutions into xApps that are subsequently published to the catalog. These xApps are composed of two interconnected units, namely a service model connector and a data-driven logic unit. The former is responsible for handling the communication with the near-real-time \gls{ric} (e.g., extracts live \glspl{kpi} from the \gls{ran} nodes or sends control actions, via the E2 termination), while the latter is tasked with embedding trained \gls{ai} solutions~\cite{bonati2023openran}.
    \item \textbf{xApp Dispatcher Module.} This module automates the instantiation of xApps in order to facilitate their \emph{testing} process. This is done through scripts that automatically create Docker containers of the xApps to dispatch and deploy them on the near-real-time \gls{ric}.
    \item \textbf{Integration with OpenRAN Gym and Colosseum.} \pandora is seamlessly integrated with OpenRAN Gym~\cite{bonati2023openran} and Colosseum~\cite{bonati2021colosseum} to enable end-to-end \oran-compliant testing of xApps in a reproducible RF environment. Via OpenRAN Gym, \pandora also offers data collection and analytics functionalities that are useful to validate xApp behavior and evaluate their performance under diverse conditions and deployments.
\end{itemize}

In a nutshell, \pandora users (e.g., telco operators, xApp developers) can generate an intent in JSON format (Step~$1$ in Fig.~\ref{fig:pandora-toolbox}) by specifying the control objective (e.g., maximizing a certain set of \glspl{kpm} for each slice), desired control parameters and observable inputs (e.g., throughput measurements, \gls{cqi}, to name a few). \pandora parses the intent (Step $2$) and: (i) produces a dataset to be used to train the \gls{ai} algorithms (Step $3$); (ii) configures the \gls{ai} algorithm to be trained, e.g., a \gls{drl} agent based on the \gls{ppo} architecture controlling slicing policies to maximize a certain reward function (Step $4$); and (iii) trains the \gls{ai} solution while offering a dashboard to monitor the training phase (Step $5$). Once the training phase is complete, the \gls{ai}/\gls{ml} models are published to the Catalog, and \pandora integrates with OpenRAN Gym~\cite{bonati2023openran} to convert the trained \gls{ai} solution into an xApp (Step $6$), which then gets published to the respective xApp Catalog (Step $7$). Once the publishing is complete, the xApp can be tested on Colosseum under one or more Colosseum scenarios (Step $8$), and testing data is collected and stored to enable performance evaluation, xApp validation and data analytics (Step $9$).

\pandora is developed in Python using Tensorflow libraries~\cite{tensorflow2015-whitepaper}. Although it supports the training of any type of \gls{ai}/\gls{ml} model, for the scope of this work, we primarily focus on training \gls{drl} agents which are the state-of-the-art for intelligent decision making in \oran~\cite{wang2023resource, filali2023communication, brik2022deep}.

Following \oran specifications and requirements~\cite{polese2022understanding}, \pandora trains \gls{ai} solutions offline using an extended and customized version of the \texttt{tf\_agents} library~\cite{TFAgents} where the configuration of the training environment, the \gls{drl} agents and the datasets are generated at run-time starting from a JSON-based intent file.

A simplified example of a minimal intent document is shown in Listing~\ref{lst:intent}. The example shows how it is possible to specify the slices to be controlled by the agent, the actions that can be controlled, the \glspl{kpi} for each slice that constitute the observation of the state, and which should be considered to compute the reward. \pandora \emph{intents} use a modular approach where \pandora users can specify slice-specific rewards and set a global reward to combine them. In Listing~\ref{lst:intent}, we illustrate a case where a \gls{drl} agent is trained to maximize both the average throughput and number of transmitted packets in the \gls{dl} for the \gls{embb} slice (Lines $4$--$9$), while maximizing the average number of \gls{dl} transmitted packets for the \gls{mmtc} slice (Lines $10$--$15$) and minimizing the maximum \gls{dl} buffer occupancy for the \gls{urllc} slice (Lines $16$--$21$).\footnote{Note that the weight associated to the \gls{urllc} slice is configured in Line $25$ and has negative value. Therefore, the $\texttt{MaxElemReward}$ function in Line $18$ effectively results in minimizing the maximum buffer occupancy. It is worth mentioning that the \gls{enb} does not have direct access to end-to-end latency measurements, therefore we use the buffer occupancy as a proxy for latency~\cite{10437367}.}

These three slice-specific rewards are then combined using a global reward that aims at maximizing the weighted sum of the individual reward contributions for each slice (weights are defined at Line $25$).

Users can specify a general control action space (e.g., controlling \gls{ran} slicing policies and scheduling profile as shown in Line $23$) without the need to specify the explicit values of these control actions. At run-time, \pandora parses the JSON-formatted intent and generates a dataset that fulfills the intent. This dataset includes all necessary \glspl{kpm} required to compute observations and rewards. Additionally, it generates the proper action space from the available datasets. For instance, \pandora processes these datasets, automatically determining the available scheduling profiles (e.g., \gls{rr}, \gls{wf}, and \gls{pf} as described in Section~\ref{sec:oran_edge:pandora:system_model}), along with the \gls{ran} slicing policies suitable for the given slices. More advanced \pandora users can also specify the subset of actions that must be considered by the agent. For example, they can force the agent to only consider PF and RR schedulers and exclude \gls{wf}.

Apart from offering modules to generate custom reward functions, \pandora provides a set of pre-defined reward functions that can be selected and combined to generate a global reward. This includes rewards aimed at maximizing or minimizing specific metrics through average, maximum, minimum, and median values. It enables the prioritization of certain \glspl{kpm} by configuring weights, as demonstrated in Line $25$. \pandora also offers pre-configured slice configurations with tailored rewards and observable \glspl{kpm} that users can utilize. These configurations serve as default values for each slice, and they can be overridden by the user.

It is also worth noting that intents, slice configurations, rewards, actions, and observation configurations are modular and can be combined to generate new intents. In this way, new intents specific to each slice are created, which are later combined, providing further flexibility to \pandora and its users.

Via \texttt{tf\_agents}~\cite{TFAgents}, \pandora provides access to a range of \gls{drl} agents, including \gls{dqn}, \gls{ppo}, \gls{ddpg}, \gls{td3}, and various others. Users can utilize \pandora to select the type of agents they want to use, along with specifying hyperparameters. Once the selection is complete, \pandora configures the agents at run-time, adjusting action and observation spaces as well as rewards based on the specified intent. Similarly to intents, users have the flexibility to customize their \gls{drl} architecture and hyperparameters or use default values pre-configured in \pandora.

\begin{lstlisting}[language=json,
  caption={A simplified example of a JSON intent for \protect\pandora.},
  label={lst:intent}]
{
  "intent": {
    "slices": [
      {
        "name": "embb",
        "reward": "MaxAverageReward",
        "reward_KPIs": ["dl_brate", "dl_tx_pkts"],
        "observation_KPIs": ["dl_buffer", "dl_tx_pkts"]
      },
      {
        "name": "mmtc",
        "reward": "MaxAverageReward",
        "reward_KPIs": ["dl_tx_pkts"],
        "observation_KPIs": ["dl_brate", "dl_tx_pkts"]
      },
      {
        "name": "urllc",
        "reward": "MaxElemReward",
        "reward_KPIs": ["dl_buffer"],
        "observation_KPIs": ["dl_buffer", "dl_brate"]
      }
    ],
    "actions": ["scheduling", "ran_slicing"],
    "global_reward_type": "NestedSumWeightedReward",
    "global_reward_weights": [0.5, 0.25, -1]
  }
}
\end{lstlisting}

\subsubsection{\texorpdfstring{\gls{drl}}{DRL} Agent Architectures tested in this work}\label{sec:oran_edge:pandora:architectures}

To control \gls{ran} slicing, and scheduling, we focus on data-driven approaches that rely on \gls{rl}. In \gls{rl}, an \emph{agent} iteratively interacts with the \emph{environment} by performing \emph{observations}, in order to learn the optimal control \emph{policy} that maximizes the desired cumulative \emph{reward}. More specifically, the \emph{agent} explores the \emph{environment} and takes \emph{actions} in several environmental \emph{states}, without knowing a priori which actions are more beneficial, and eventually learns the best policy through experience. The \emph{reward} is a metric that defines the effectiveness of an \emph{action}, while a sequence of \emph{states}, \emph{actions}, and \emph{rewards} that result in a terminal \emph{state} is called an \emph{episode}. \emph{State Space $S$}, represents all the possible states of the environment $s \in S$, while \emph{Action Space $A$} defines all the feasible actions $a \in A$ that can be taken by the \emph{agent}. Finally, in this work, we focus on \gls{drl} agents due to their ability to learn directly from experience, without relying on pre-existing models or explicit knowledge of the wireless environment~\cite{mnih2015human, kaloxylos2021ai, yang2019application} and therefore of its complex and dynamically changing conditions. Moreover, we focus on discrete actions and, for this reason, we limit our analysis to \gls{ppo}~\cite{schulman2017proximal} and \gls{dqn}~\cite{mnih2013playing} architectures which are two state-of-the-art algorithms respectively for on-policy and off-policy \gls{drl} for discrete action spaces.

\textbf{\gls{ppo}} has been demonstrated several times to outperform other architectures~\cite{polese2021colo, kouchaki2022actor}. It is based on an actor-critic network architectural approach, where the actor and critic network ``work'' cooperatively to learn a policy that selects actions that deliver the highest reward possible for each state. While the actor's task is to take actions based on current network states, the critic's target is to evaluate actions taken by the actor network and provide feedback that reflects how effective the action taken by the actor is. In this way, the critic helps the actor in taking actions that lead to the highest rewards for each given state. The Clipped Surrogate Objective function used by \gls{ppo} is defined in Eq.~\eqref{eq:clip-surog-fun}:
\begin{equation}\label{eq:clip-surog-fun}
\begin{aligned}
   L^{\text{CLIP}}(\theta)=\hat{\mathbb{E}}_t\left[\min \left(q_t(\theta) \hat{A}_t, \operatorname{clip}\left(q_t(\theta), 1-\epsilon, 1+\epsilon\right) \hat{A}_t\right)\right],
\end{aligned}
\end{equation}

\noindent
where $\hat{\mathbb{E}}_t$ represents the empirical average; $\hat{A}_t$ is the estimator of the advantage function denoted as $A_t$, which assesses how well an action performed compared to the expected performance under the current policy; and $q_t(\theta)=\frac{\pi_\theta\left(a_t \mid s_t\right)}{\pi_{\theta_{\text {old }}}\left(a_t \mid s_t\right)}$. This ratio represents the probability of taking action $a_t$ at state $s_t$ following the current policy $\pi_{\theta}$, divided by the respective probability when following the previous policy (i.e., $\pi_{\theta_{\text {old }}}$). When multiplied by the estimator of the advantage function at time-step $t$ (i.e., $\hat{A}_t$), the unclipped part of Eq.~\eqref{eq:clip-surog-fun} is obtained, as shown in Eq.~\eqref{eq:unclipped-ratio}, where \gls{cpi} stands for the technique leveraged to avoid large policy updates~\cite{kakade2002approximately}:
\begin{equation}\label{eq:unclipped-ratio}
\begin{aligned}
   L^{\text{CPI}}(\theta)=\hat{\mathbb{E}}_t\left[\frac{\pi_\theta\left(a_t \mid s_t\right)}{\pi_{\theta_{\text {old }}}\left(a_t \mid s_t\right)} \hat{A}_t\right]=\hat{\mathbb{E}}_t\left[q_t(\theta) \hat{A}_t\right].
\end{aligned}
\end{equation}

\noindent
Since maximizing Eq.~\eqref{eq:unclipped-ratio} directly would result in large policy updates, by \emph{clipping} the probability ratio, we constrain the surrogate objective function (i.e., Eq.~\eqref{eq:clip-surog-fun}) and remove the incentive to move the probability ratio (i.e., $q_t(\theta)$) outside the interval $[1-\epsilon, 1+\epsilon]$. $\epsilon$ is a hyperparameter that determines the clip range and is chosen in a way to ensure that the policy update will not be too large, indicating a significant divergence between the old and the current policy. Lastly, we take the minimum of the clipped and non-clipped objectives. Therefore, the final objective serves as a lower pessimistic bound of the unclipped objective. This indicates that we select either the clipped or the non-clipped objective based on $q_t(\theta)$ and the advantage. The final form of the Clipped Surrogate Objective Loss for the actor-critic implementation of \gls{ppo} is given in Eq.~\eqref{eq:final-ppo-obj}. It is a combination of the Clipped Surrogate Objective function (i.e., $L_t^{\text{CLIP}}$), the squared-error value loss function (i.e., $L_t^{\text{VF}}(\theta)=\left(V_\theta\left(s_t\right)-V_t^{\operatorname{targ}}\right)^2$) which is the difference between predicted and target cumulative reward estimations, and the entropy bonus (i.e., $S''$), with the latter ensuring sufficient exploration, while $c_1$, $c_2$ are control parameters:
\begin{equation}\label{eq:final-ppo-obj}
\begin{aligned}
  L_t^{\text{CLIP+VF+S}}(\theta)=\hat{\mathbb{E}}_t\left[L_t^{\text{CLIP}}(\theta)-c_1 L_t^{\text{VF}}(\theta)+c_2 S''\left[\pi_\theta\right]\left(s_t\right)\right].
\end{aligned}
\end{equation}

\noindent
In addition, based on a comparative analysis between models with diverse number of layers and neurons, the actor and critic networks we select for this work both consist of fully-connected neural networks with 3 layers of 30 neurons each. The hyperbolic tangent serves as the activation function while the learning rate is set to $10^{-3}$.

The \textbf{\gls{dqn}} algorithm leverages Q-learning principles to estimate the Q-function. The Q-function represents the expected discounted cumulative reward obtained by taking action $a$ in state $s$ and then following the policy $\pi$. The optimal state-action pair is computed using the Bellman Equation, defined in Eq.~\eqref{eq:bellm-eq}:
\begin{equation}\label{eq:bellm-eq}
   Q^*(s, a)=\left[r+\gamma \max _{a^{\prime}} Q^*\left(s^{\prime}, a^{\prime}\right)\right].
\end{equation}

\noindent
This equation is approximated by a \gls{dqn} that calculates the cumulative future reward, denoted as $r_t$ at time $t$, and discounted by a factor $\gamma \in[0,1]$, while $s'$ and $a'$ represent the state and the action taken in the next time-step. In addition, the algorithm uses a replay buffer to store experience and to cope with instabilities caused by correlation between consecutive episodes. The set of experiences is denoted as $D_t=\left\{e_1, \ldots, e_t\right\}$, where each $e=\left(s, a, r, s'\right)$ is a tuple that represents the state, action, reward, and the next state. The system transitions to the next state $s'$ after the agent takes an action $a$ at each time-step $t$ during the training phase, while the experience vector is stored in the replay buffer. In order to reduce the correlation between the Q-function value and the optimal $Q^*$ value, a second Q-Network is employed, namely the target Q-Network. Both the main and target Q-Network share the same structure, while the weights of the latter are periodically updated to match those of the former. During the training phase, the replay buffer is used to randomly select batches of previous experiences. These experiences are then utilized to calculate the \gls{dqn} weights, denoted as $\theta_t$ in Eq.~\eqref{eq:dqn-loss}. This calculation involves minimizing the loss function using \gls{sgd} method. Concludingly, the loss equation leveraged during the Q-Learning update at iteration $t$ is given by Eq.~\eqref{eq:dqn-loss}:
\begin{equation}\label{eq:dqn-loss}
\begin{aligned}
L_t\left(\theta_t\right)=\mathbb{E}_{s, a, r, s^{\prime}}\left[\left(r+\gamma \max _{a^{\prime}} Q\left(s^{\prime}, a^{\prime};\theta_t^{-}\right)-Q\left(s,a; \theta_t\right)\right)^2\right],
\end{aligned}
\end{equation}

\noindent
where $\theta_t^{-}$ are the weights of the target Q-Network, and $\gamma$ is the discount factor that prioritizes instantaneous rewards against long-term rewards. Additionally, the $\epsilon_{\text{DQN}}$-greedy is employed, where the \gls{drl} agent's choice of action depends on the parameter $\epsilon_{\text{DQN}}\in[0,1]$. In detail, with a probability of $1-\epsilon_{\text{DQN}}$, an action computed by the \gls{dqn} is selected. Conversely, with a probability of $\epsilon_{\text{DQN}}$, the agent selects a random action from the action space $A$. This strategy effectively balances the exploration of new actions, randomly selected, and the exploitation of the agent's learned knowledge, based on actions selected by the \gls{dqn}. In our analysis, the latter is implemented as a feed-forward multi-layer \gls{nn}, consisting of $5$~hidden layers of $50$~neurons each. The learning rate is set to $10^{-3}$, while the discount factor is $\gamma=0.95$. Finally, the replay buffer memory size is set to $10{,}000$, while $\epsilon_{\text{DQN}}=0.1$.

We follow the same approach as in~\cite{polese2021colo}, where the input (e.g., \glspl{kpm}) of the \gls{drl} agent is first processed by the encoding part of an autoencoder for dimensionality reduction. This also synthetically reduces the state space and makes training more efficient in terms of time and generalization. In detail, the autoencoder converts an input matrix of $K=10$ individual measurements of $M=3$ \gls{kpm} metrics (i.e., \gls{dl} throughput, buffer occupancy, and number of transmitted packets) into a single $M$-dimensional vector. The \gls{relu} activation function and four fully-connected layers of $256$, $128$, $32$ and $3$ neurons are also used in the encoder. Although the channel state information (e.g., \gls{cqi}, \gls{mcs}) is not directly fed to the AI/ML agents, it is worth mentioning that its effect on performance is indirectly captured by the \glspl{kpi} used to feed the \gls{drl} agents (e.g., Throughput, Buffer Occupancy, Transmitted Packets), which has been shown to be sufficient to implicitly capture channel effects on network performance~\cite{polese2021colo}.

\begin{figure}[htbp]
  \centering
  \includegraphics[width=0.99\textwidth]{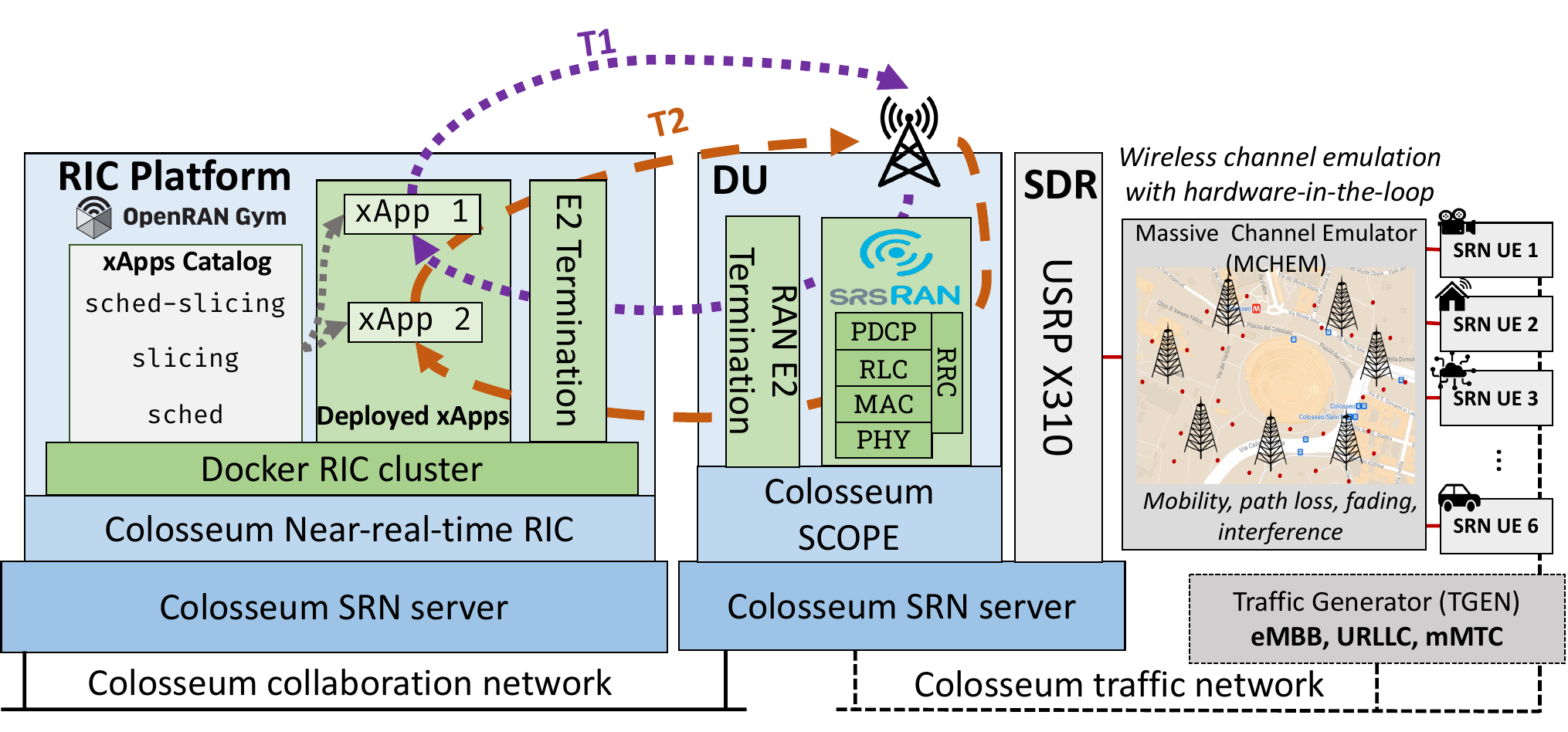}
  \caption{Reference \oran testing architecture with focus on the case of two xApps operating at different time scales, $T_{i}$, as described in Section~\ref{sec:oran_edge:pandora:hierarchical}.}
  \label{fig:ext-arch}
\end{figure}

In the case of the joint-slice optimization, the cumulative average reward function of the \gls{drl} agent is designed to jointly satisfy the \gls{qos} demand of the three slices with respect to their \gls{kpm} requirements. For instance, \gls{embb} users aim to maximize throughput, while \gls{mmtc} users aim at maximizing the number of transmitted packets. Finally, the goal of \gls{urllc} users is to deliver packets with minimum latency. Since the base station cannot measure end-to-end application-layer latency (which is instead measured at the receiver side), we measure latency in terms of number of bytes in the transmission buffer, the smaller the buffer, the smaller the latency. The reward is formulated as the weighted sum in Eq.~\eqref{eq:weighted_reward}:
\begin{equation}\label{eq:weighted_reward}
   R = \sum_{t=0}^{\infty} \gamma^t \left( \sum_{j=1}^{N} w_{j} \cdot r_{j,t} \right),
\end{equation}

\noindent
where $t$ represents the training step, and $N=3$ is the total number of slices, $w_{j}$ represents the weight associated to slice $j$, considered for reward maximization in the three corresponding slices. Finally, $\gamma$ is the discount factor and $r_{j,t}$ describes the slice-specific reward obtained at each training step $t$. In our case, $r_{j,t}$ represents the average value of the \gls{kpm} measured by all users of slice $j$ at time $t$ (e.g., throughput for the \gls{embb} slice). Note that the weight $w_{j}$ for the \gls{urllc} slice is negative to model the minimization of the buffer occupancy. The models that we have designed and trained are deployed as xApps on the near-real-time \gls{ric}, as illustrated in Fig.~\ref{fig:ext-arch}.

In the case of the per-slice optimization, the respective weighted average reward function of the \gls{drl} agent is given as follows in Eq.~\eqref{eq:per_slice_reward}:
\begin{equation}\label{eq:per_slice_reward}
   R = \sum_{t=0}^{\infty} \gamma^t \left( w_{j} \cdot r_{j,t} \right).
\end{equation}

\begin{figure}[htbp]
  \centering
  \includegraphics[width=0.99\textwidth]{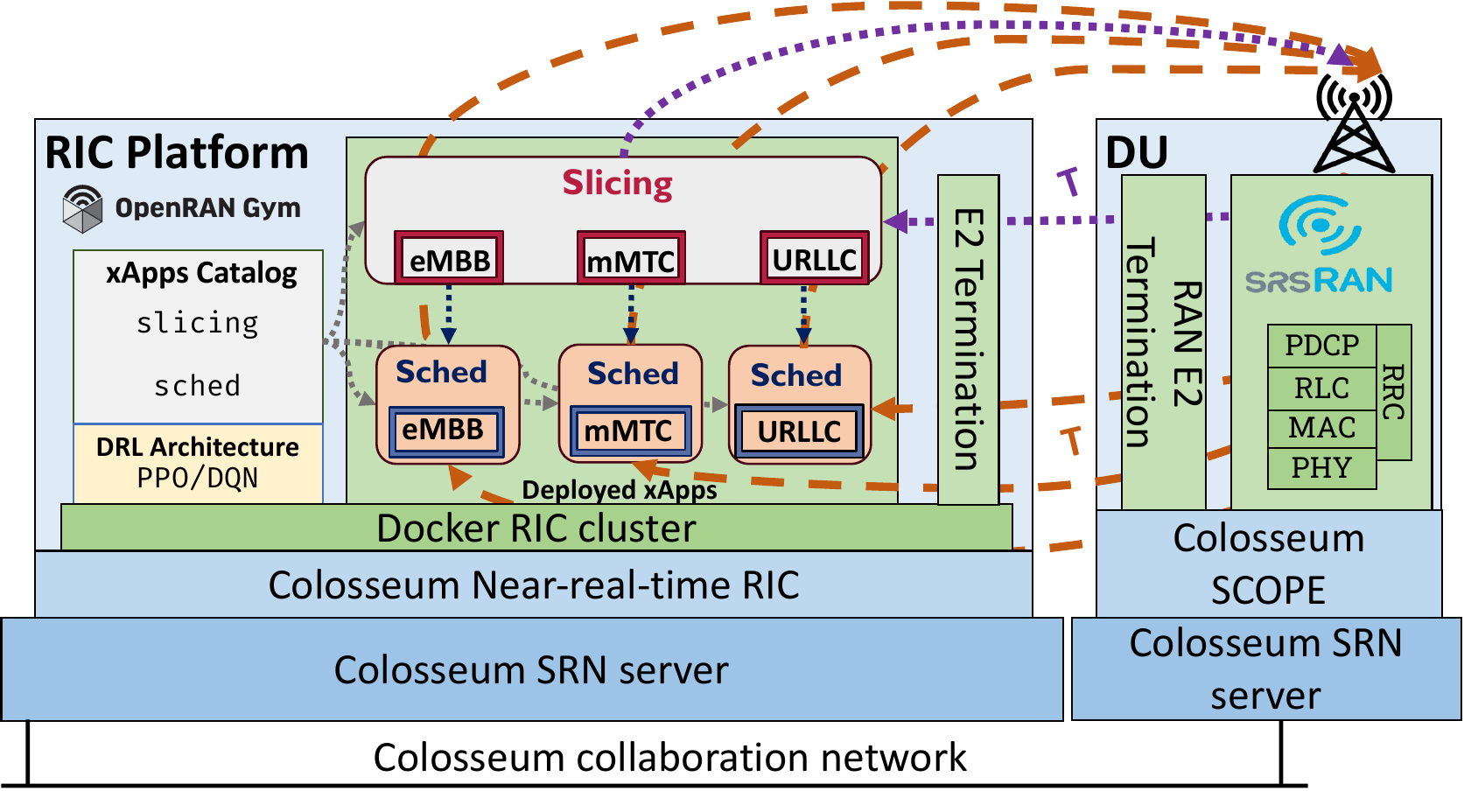}
  \caption{Reference \oran testing architecture with focus on the case of four xApps operating at time scale, $T$, as described in Section~\ref{sec:oran_edge:pandora:perslice}.}
  \label{fig:4xapps-arch}
\end{figure}

\noindent
The models that we have designed and trained are deployed as xApps on the near-real-time \gls{ric}, as illustrated in Fig.~\ref{fig:4xapps-arch}.

\subsection{\texorpdfstring{\gls{drl}}{DRL} Optimization Strategies}\label{sec:oran_edge:pandora:strategies}

Our goal is to investigate how different design choices affect the effectiveness and decision-making of the \gls{drl}-based xApps. For this reason, we consider the following design choices, which resulted in the generation and testing of a total of $23$ xApps.

\textbf{Short-term vs. Long-term Rewards.} We train \gls{drl} agents with different values of the discount factor $\gamma$. The \gls{ppo} discount factor weights instantaneous rewards against long-term rewards. A higher value prioritizes long-term rewards, while a lower $\gamma$ prioritizes short-term rewards. Results of this exploration are provided in Section~\ref{sec:oran_edge:pandora:discount}.

\textbf{Hierarchical Decision-Making.} We investigate the case of two xApps configuring different parameters in parallel but at different timescales. In this way, we investigate how multiple xApps with different optimization goals and operating timescales impact the overall network performance. The findings of this investigation are provided in Section~\ref{sec:oran_edge:pandora:hierarchical}, and a practical example is illustrated in Fig.~\ref{fig:ext-arch}.

\textbf{Per-Slice Scheduling Profile Selection.} We explore a hierarchical decision-making configuration of four xApps operating simultaneously at the same time granularity and reconfiguring the \gls{bs}'s control parameters. In detail, one slicing xApp simultaneously allocates \glspl{prb} to all slices. The remaining three xApps are exclusively assigned one slice each and select a dedicated scheduling profile for each slice. In this way, we aim to examine the effects of multiple xApps operating at the same timescale, and to compare a widely used state-of-the-art off-policy architecture (i.e., the \gls{dqn}), with its on-policy counterpart (i.e., the \gls{ppo} algorithm). A practical example is depicted in Fig.~\ref{fig:4xapps-arch}, while the results of this investigation are presented in Section~\ref{sec:oran_edge:pandora:perslice}.

\textbf{Impact of Reward's Weights.} We test different values for the weights $w_i$ of the slices in Eq.~\eqref{eq:weighted_reward}. A different weight configuration affects how \gls{drl} agents prioritize each slice. The results of this analysis are reported in Section~\ref{sec:oran_edge:pandora:weights}, where we show how weights significantly impact the overall performance and can result in inefficient control strategies.

\textbf{Effect of \gls{ran} Control Timers.} Finally, we aim at understanding how different \gls{ran} control timers (i.e., different periodicities for \gls{ran} telemetry, report and control) will have an impact on the decision-making process and how different control sets will prioritize one slice over the other. Specifically, we look into three control timers namely the \gls{kpm} log time, the \gls{du} Report Timer, and the action update time, i.e., how frequently the actions sent back by the \gls{ric} get updated and enforced by the \gls{bs}. The reported findings of this strategy are provided in Section~\ref{sec:oran_edge:pandora:timers}.

It is reminded that the action set $A$ consists of both scheduling and \gls{ran} slicing policies. In our analysis, we consider the cases where an agent can control either scheduling or slicing decisions individually, or control both jointly. The state $S$ is represented by the output of the autoencoder which is used to compress input \glspl{kpi} collected over the E2 interface and convert them into latent representations. Finally, although in our analysis we evaluate diverse reward designs that consider long-term and short-term goals, and combine diverse target \glspl{kpi} for each slice, the general form of the reward $R$ considered in all of our \gls{drl} agents is defined in Eq.~\eqref{eq:weighted_reward}.

\subsection{Experimental Setup and \texorpdfstring{\gls{drl}}{DRL} Training}\label{sec:oran_edge:pandora:setup}

\begin{table}[htbp]
\centering
\small
\caption{Traffic Profiles}
\begin{tabular}{cccc}
\toprule
\textbf{Profile Name} & \textbf{eMBB [Mbps]} & \textbf{mMTC [kbps]} & \textbf{URLLC [kbps]} \\
\midrule
Profile 1 & $1$ & $30$ & $10$ \\
Profile 2 & $4$ & $44.6$ & $89.3$ \\
\bottomrule
\end{tabular}
\label{table:traffic-profiles}
\end{table}

To experimentally evaluate the \gls{drl} agents, we leverage the capabilities of OpenRAN Gym~\cite{bonati2023openran}, an open-source experimental toolbox for end-to-end design, implementation, and testing of \gls{ai}/\gls{ml} applications in \oran. It features:
\begin{itemize}
   \item End-to-end \gls{ran} and core network deployments through the srsRAN~\cite{gomez2016srslte} softwarized open-source protocol stack;
    \item Large-scale data collection, testing and fine-tuning of \gls{ran} functionalities through the SCOPE framework~\cite{bonati2021scope}, which adds open \glspl{api} to srsRAN for the control of slicing and scheduling functionalities, as well as for \glspl{kpm} collection;
    \item An \oran-compliant control architecture to execute \gls{ai}/\gls{ml}-based xApps via the ColO-RAN near-real-time \gls{ric}~\cite{polese2021colo}. The E2 interface between \gls{ran} and the \gls{ric} and its \glspl{sm}~\cite{polese2022understanding} manage streaming of \glspl{kpm} from the RAN and control actions from the xApps.
\end{itemize}

\begin{table}[htbp]
\centering
\small
\caption{Weight Configurations}
\begin{tabular}{cccc}
\toprule
\textbf{Weights} & \textbf{eMBB} & \textbf{mMTC} & \textbf{URLLC} \\
\midrule
\texttt{Default} & $72.0440333$ & $0.229357798$ & $0.00005$ \\
\texttt{Alternative} & $72.0440333$ & $1.5$ & $0.00005$ \\
\bottomrule
\end{tabular}
\label{table:weight-confs-list}
\end{table}

We deploy OpenRAN Gym on Colosseum~\cite{bonati2021colosseum}, a publicly available testbed with $128$~\glspl{srn}, i.e., pairs of Dell PowerEdge R730 servers and NI \gls{usrp} X310 \glspl{sdr}. Colosseum enables large-scale experimentation in diverse \gls{rf} environments and network deployments. The channel emulation is done through the \gls{mchem} component, which leverages \gls{fpga}-based \gls{fir} filters to reproduce different conditions of the wireless environment (e.g., path loss, fading, attenuation, interference of signals) modeled a priori through ray-tracing software, analytical models, or real-world measurements. Similarly, the Colosseum \gls{tgen}, built on top of the \gls{mgen} TCP/UDP traffic generator~\cite{mgen}, emulates different traffic types, demand, and distributions (e.g., Poisson, periodic).

We deploy a 3GPP-compliant cellular network with one base station and $6$~\glspl{ue} uniformly distributed across $3$~different slices. These are: (i) \gls{embb} that concerns high traffic modeling of high-quality multimedia content and streaming applications; (ii) \gls{urllc} for time-critical applications, such as autonomous driving in \gls{v2x} scenarios; and (iii) \gls{mmtc} for \gls{iot} devices with low data rate requirements but with high need for consistent information exchange. In terms of physical deployment, we consider two different topologies both located in the urban environment of Rome, Italy~\cite{bonati2021scope}. Specifically, \textbf{Location 1} concerns \glspl{ue} uniformly distributed within a $50$\:m radius from the \gls{bs}, while in \textbf{Location 2}, the corresponding \glspl{ue} are placed within a $20$\:m radius from the \gls{bs}.

The bandwidth of the \gls{bs} is set to $10$\:MHz (i.e., $50$~\glspl{prb}) and is divided among the $3$~slices, with $2$ users statically assigned to each slice. Slice-based traffic is generated following the specifications reported in Table~\ref{table:traffic-profiles}. Specifically, we consider two different traffic profile configurations (i.e., Profile $1$ and Profile $2$) with different source bitrates. However, for both profiles, we consider a constant bitrate traffic for \gls{embb} users, while \gls{urllc} and \gls{mmtc} \glspl{ue} generate traffic based on a Poisson distribution.

\begin{table}[htbp]
\centering
\small
\caption{Per-Slice Top performing xApps under the Default Weight Configuration}
\begin{tabular}{lll}
\toprule
 & \textbf{eMBB} & \textbf{mMTC} \\
\midrule
\texttt{1)} & \texttt{Sched \& Slicing 0.5} & \texttt{Slicing 0.99} \\
\texttt{2)} & \texttt{Sched \& Slicing 0.99} & \texttt{Slicing 0.5} \\
\texttt{3)} & \texttt{Slicing 0.5} & \texttt{Sched 0.99} \\
\texttt{4)} & \texttt{Slicing 0.99} & \texttt{Sched \& Slicing 0.99} \\
\texttt{5)} & \texttt{Sched 0.99} & \texttt{Sched 0.5} \\
\texttt{6)} & \texttt{Sched 0.5} & \texttt{Sched \& Slicing 0.5} \\
\bottomrule
\end{tabular}
\label{table:default-comp-analysis}
\end{table}

To train the \gls{drl} agents, we used the publicly available dataset described in~\cite{polese2021colo}. This dataset contains about $8$\:GB of \glspl{kpm} collected by using OpenRAN Gym and the Colosseum network emulator over $89$~hours of experiments, and concerns setups with up to 7~base stations and 42~\glspl{ue} belonging to different \gls{qos} classes, and served with heterogeneous scheduling policies. Each \gls{drl} model evaluated in the following sections takes as input \gls{ran} \glspl{kpm} such as throughput, buffer occupancy, number of \glspl{prb}, and outputs resource allocation policies (e.g., RAN slicing and/or scheduling) for \gls{ran} control.

Abiding by the \oran specifications~\cite{alliance2021ran}, which do not permit the deployment of untrained data-driven solutions, we train our \gls{ml} models \emph{offline} on Colosseum's GPU-accelerated environment, which includes two NVIDIA DGX A100 servers with $8$~GPUs each. Notably, \gls{ai}/\gls{ml} solutions should be trained offline to avoid actions that can potentially lead to performance degradation in the network~\cite{polese2022understanding}. Then, the trained \gls{drl}-agents are onboarded on xApps inside softwarized containers implemented via Docker and deployed on the ColO-RAN near-real-time \gls{ric}.

\begin{table}[htbp]
\centering
\small
\caption{Hierarchical Reporting Setup}
\begin{tabular}{ccc}
\toprule
\textbf{Setup ID} & \textbf{\texttt{Slicing 0.5}} & \textbf{\texttt{Sched 0.99}} \\
\midrule
\textbf{1} & $1$\:s & $10$\:s \\
\textbf{2} & $1$\:s & $5$\:s \\
\textbf{3} & $10$\:s & $1$\:s \\
\textbf{4} & $5$\:s & $1$\:s \\
\bottomrule
\end{tabular}
\label{table:hier-setups}
\end{table}

\subsection{In-Sample Experimental Evaluation}\label{sec:oran_edge:pandora:insample}

In this section, we present the results of an extensive performance evaluation campaign, with more than $38$~hours of experiments in Colosseum, to profile the impact of the strategies discussed in Section~\ref{sec:oran_edge:pandora:strategies}. These results were produced by taking the median as the most representative statistical value of a dataset, and averaged over multiple repetitions of experiments in the \gls{dl} direction of the communication system. Last, all the experiments of the in-sample experimental evaluation campaign concern \textbf{Location 1} which is the configuration used to collect training data, and all the \glspl{ue} are assumed static. Out-of-sample evaluation, i.e., testing the \gls{drl} agents against data collected in an entirely different network deployment, will be the focus of Section~\ref{sec:oran_edge:pandora:outofsample}.

\subsubsection{Impact of Discount Factor on the Action Space}\label{sec:oran_edge:pandora:discount}

We explore how \gls{ran} slicing, \gls{mac} scheduling, and joint slicing and scheduling control are affected by training procedures that favor short-term against long-term rewards. All the reported results were obtained with $\gamma\in\{0.5,0.99\}$, while the reward's weight configuration is shown in Table~\ref{table:weight-confs-list} and identified as \texttt{Default}.

\begin{figure}[htbp]
\centering
\subfloat[\gls{embb} Throughput]{\includegraphics[height=4cm]{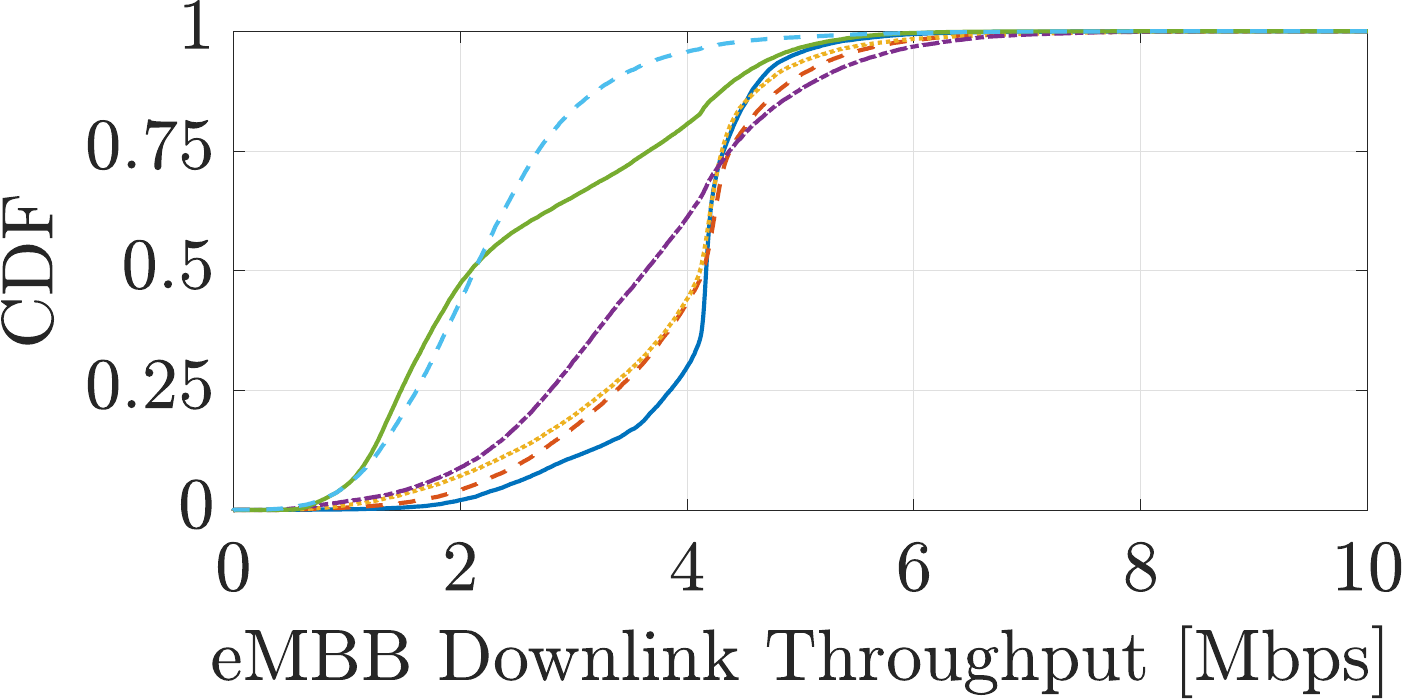}\label{fig:Figure4a}}
% \hfil
\subfloat[\gls{mmtc} Packets]{\includegraphics[height=4cm]{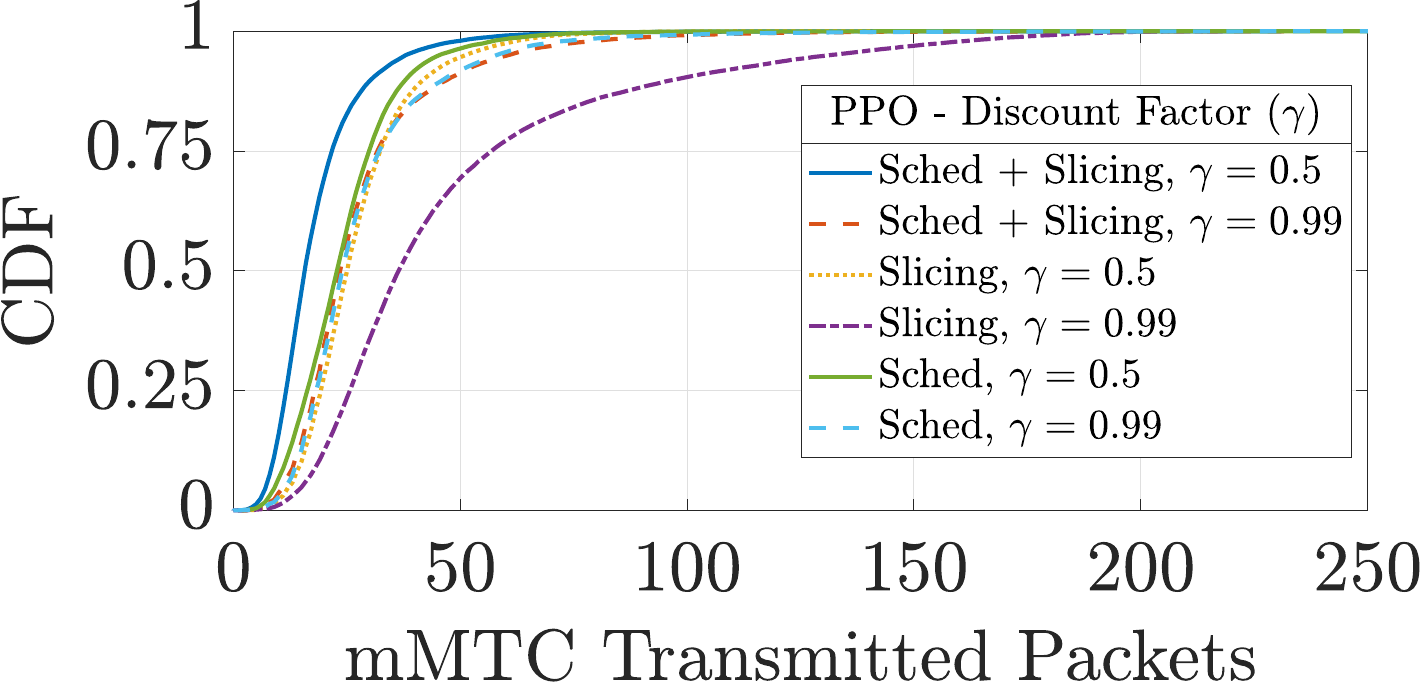}\label{fig:Figure4b}}
\hfil
\subfloat[\gls{urllc} Buffer Occupancy]{\includegraphics[height=4cm]{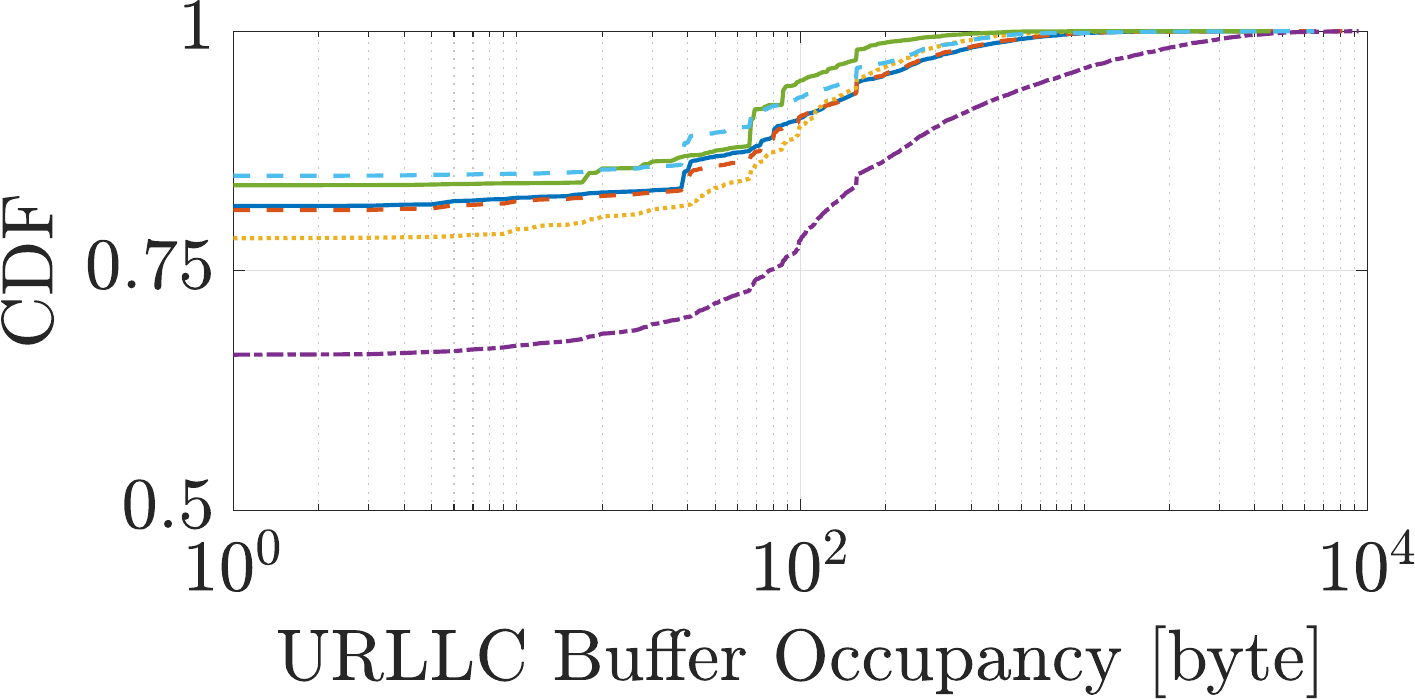}\label{fig:Figure4c}}
\caption{Performance evaluation under different action spaces and values of the $\gamma$ parameter with the PPO DRL Architecture.}
\label{Figure4-1a}
\end{figure}

\begin{figure}[htbp]
\centering
\subfloat[\gls{embb} DL Throughput]{\includegraphics[width=3in]{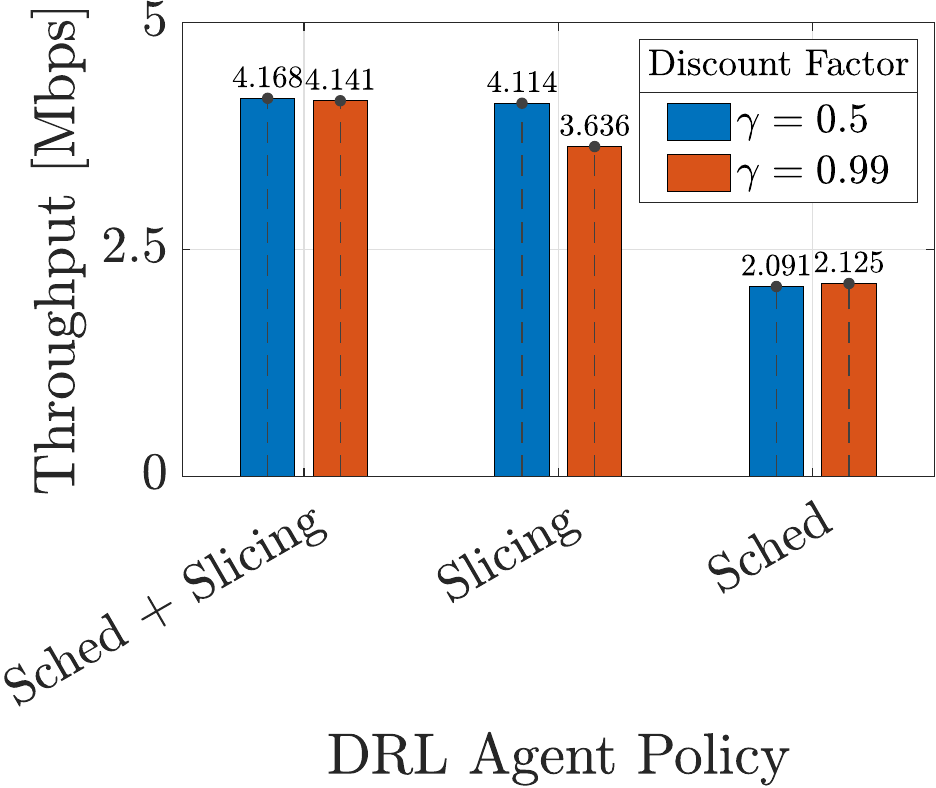}\label{fig:Figure4d}}
\hfil
\subfloat[\gls{mmtc} Packets]{\includegraphics[width=3in]{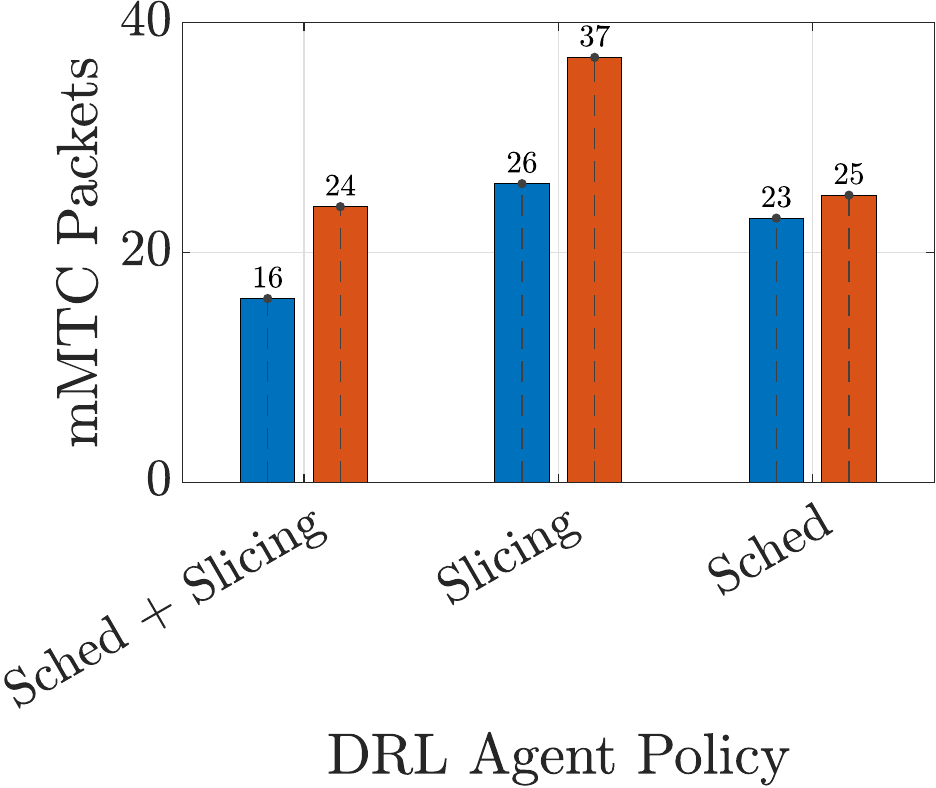}\label{fig:Figure4e}}
\caption{Median values under different action spaces and values of $\gamma$ with the PPO DRL Architecture.}
\label{Figure4-1b}
\end{figure}

In Fig.~\ref{Figure4-1a}, we report the \gls{cdf} of individual \glspl{kpm} for each slice and for different xApps trained to control different sets of actions and using different values of $\gamma$. The median of such measurements for the \gls{embb} and \gls{mmtc} slices is instead reported in Fig.~\ref{Figure4-1b}. The median for the \gls{urllc} slice is not reported, as this value is zero in all configurations. The best performing configurations for the \gls{embb} and \gls{mmtc} slices are instead listed in numerical order in Table~\ref{table:default-comp-analysis} from best to worst performing.

Our results show that \texttt{Sched \& Slicing} and \texttt{Slicing 0.5} favor \gls{embb} the most, with \texttt{Sched \& Slicing 0.5} being the best configuration among the ones considered. Moreover, slicing is essential to ensure high throughput values (the four top-performing xApps for \gls{embb} include slicing as a control action). We also notice that prioritizing immediate rewards (i.e., $\gamma=0.5$) results in higher throughput values if compared to xApps embedding agents trained to maximize long-term rewards. This design option, when combined with a bigger action space (e.g., scheduling \& slicing) ultimately yields a higher throughput.

For the \gls{mmtc} slice, the \texttt{Slicing 0.99} xApp always yields the best performance. However, we notice that \texttt{Sched \& Slicing 0.5}, which is the best-performing xApp for \gls{embb}, yields the worst performance for \gls{mmtc}. Although a larger action space and a short-term reward design are ideal for \gls{embb} (e.g., \texttt{Sched \& Slicing 0.5}), we notice that this performance gain comes at the expense of the \gls{mmtc} slice. Indeed, in Figs.~\ref{fig:Figure4d} and~\ref{fig:Figure4e}, we observe that the higher the \gls{embb} performance, the lower the \gls{mmtc}'s. This is clearly illustrated when we compare the ``best'' per-slice policies, respectively \texttt{Sched \& Slicing 0.5} (\gls{embb}) and \texttt{Slicing 0.99} (\gls{mmtc}). The former delivers the highest reported \gls{embb} throughput ($4.168$\:Mbps) but the lowest number of \gls{mmtc} packets ($16$\:packets), while the latter delivers the highest number of \gls{mmtc} packets ($37$\:packets) and one of the lowest measured \gls{embb} throughput values (i.e., $3.636$\:Mbps).

Hence, \gls{embb}-\gls{mmtc} slices indicate a competitive behavior, since we cannot optimally satisfy both of them without loss in their respective rewards, as they compete for the amount of packets required for transmission. Our results show that, in general, controlling scheduling only is not ideal as it strongly penalizes \gls{embb} performance with a modest improvement in terms of number of transmitted \gls{mmtc} packets.

\subsubsection{Impact of Hierarchical Decision-Making}\label{sec:oran_edge:pandora:hierarchical}

In this analysis, we evaluate the effectiveness of making disjoint decisions to control scheduling and slicing policies. We select the best performing single-action xApps from Table~\ref{table:default-comp-analysis}, i.e., \texttt{Slicing 0.5} and \texttt{Sched 0.99}, and we compare their execution at different timescales. The former, provides a good balance in terms of \gls{embb} throughput ($\sim 4$\:Mbps) and number of \gls{mmtc} packets, while the latter, provides the best performance for the \gls{mmtc} slice. With this design choice, we expect to maintain high performance for both \gls{embb} and \gls{mmtc}.

\begin{table}[H]
\centering
\small
\caption{Feasible \gls{prb} Allocation}
\label{tab:feasible_prb_allocation}
\begin{tabular}{ccc}
\toprule
\textbf{\gls{embb}} & \textbf{\gls{mmtc}} & \textbf{\gls{urllc}} \\
\midrule
 30 & 9 & 11 \\
 30 & 15 & 5 \\
 36 & 9 & 5 \\
 24 & 21 & 5 \\
 24 & 15 & 11 \\
 18 & 15 & 17 \\
 18 & 9 & 23 \\
 18 & 21 & 11 \\
 12 & 27 & 11 \\
 12 & 15 & 23 \\
 12 & 9 & 29 \\
 6 & 27 & 17 \\
 6 & 39 & 5 \\
 6 & 15 & 29 \\
 6 & 9 & 35 \\
 36 & 3 & 11 \\
\bottomrule
\end{tabular}
\end{table}

\begin{figure}[htbp]
\centering
\subfloat[\gls{embb} Throughput]{\includegraphics[height=4cm]{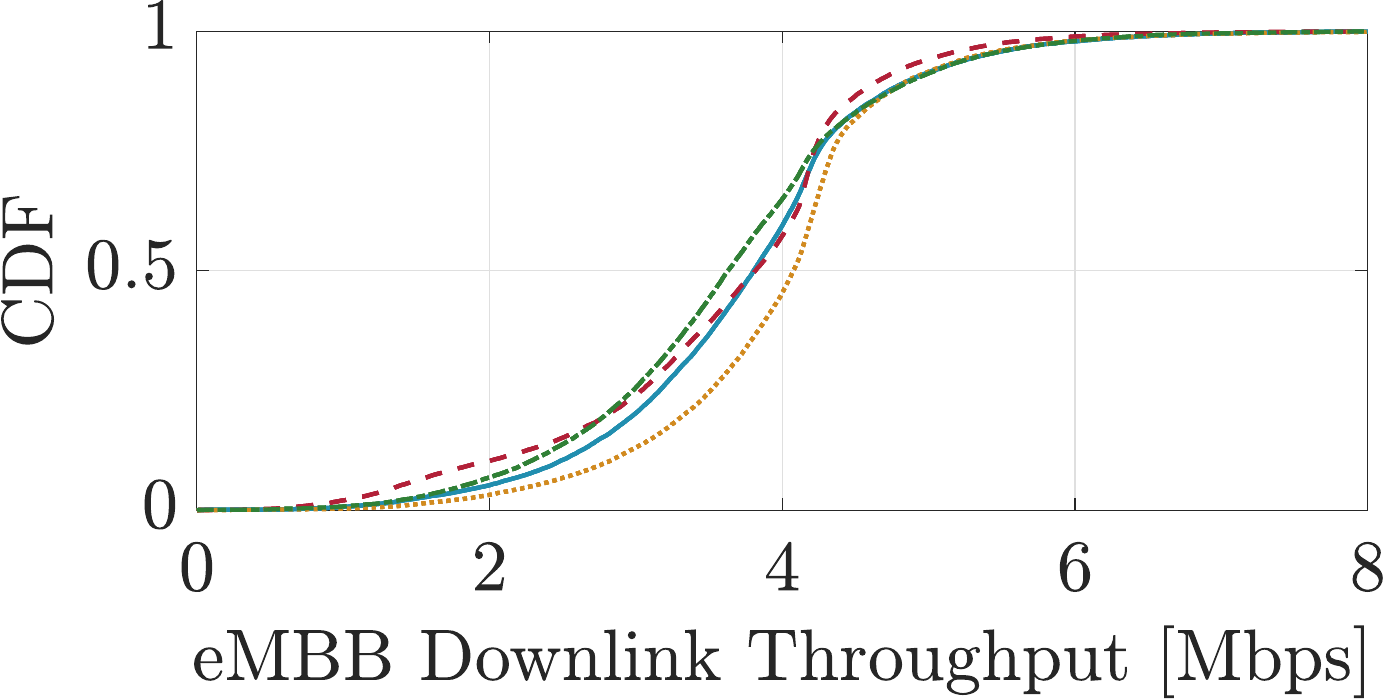}\label{Figure4a2}}
% \hfil
\subfloat[\gls{mmtc} Packets]{\includegraphics[height=4cm]{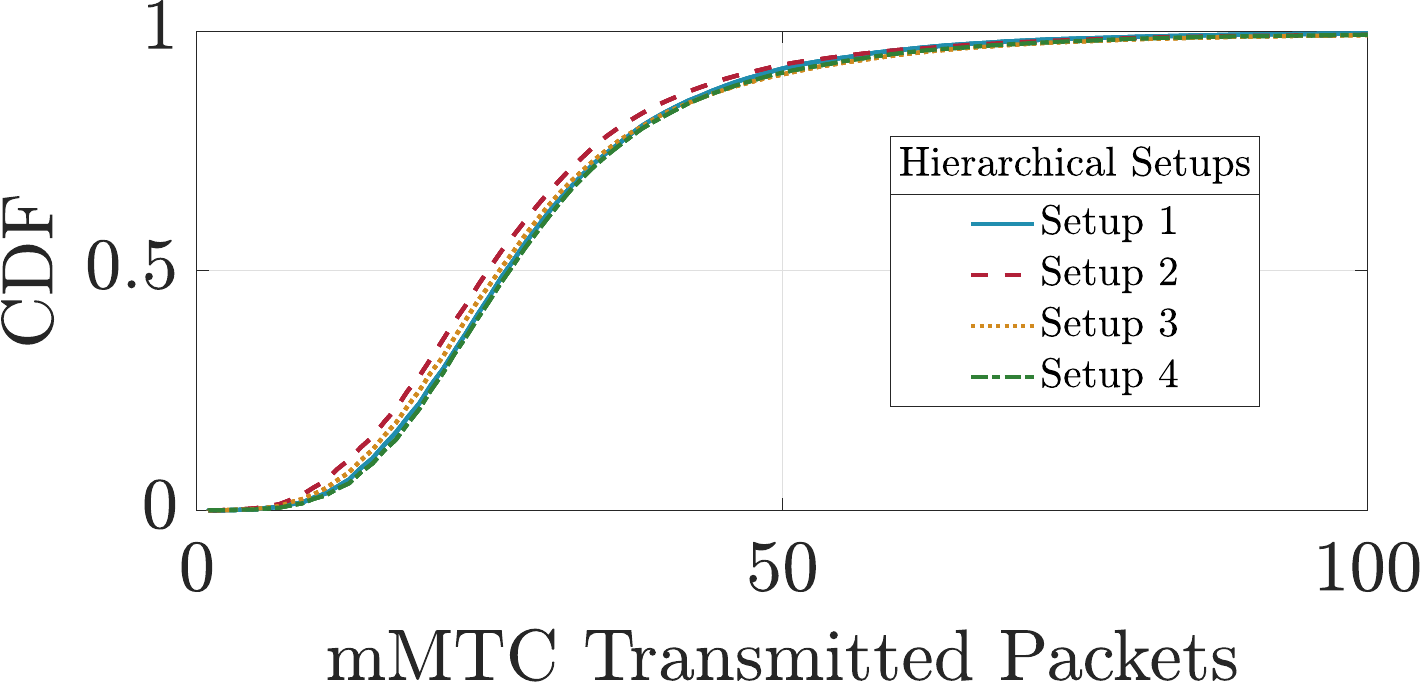}\label{Figure4b2}}
\hfil
\subfloat[\gls{urllc} Buffer Occupancy]{\includegraphics[height=4cm]{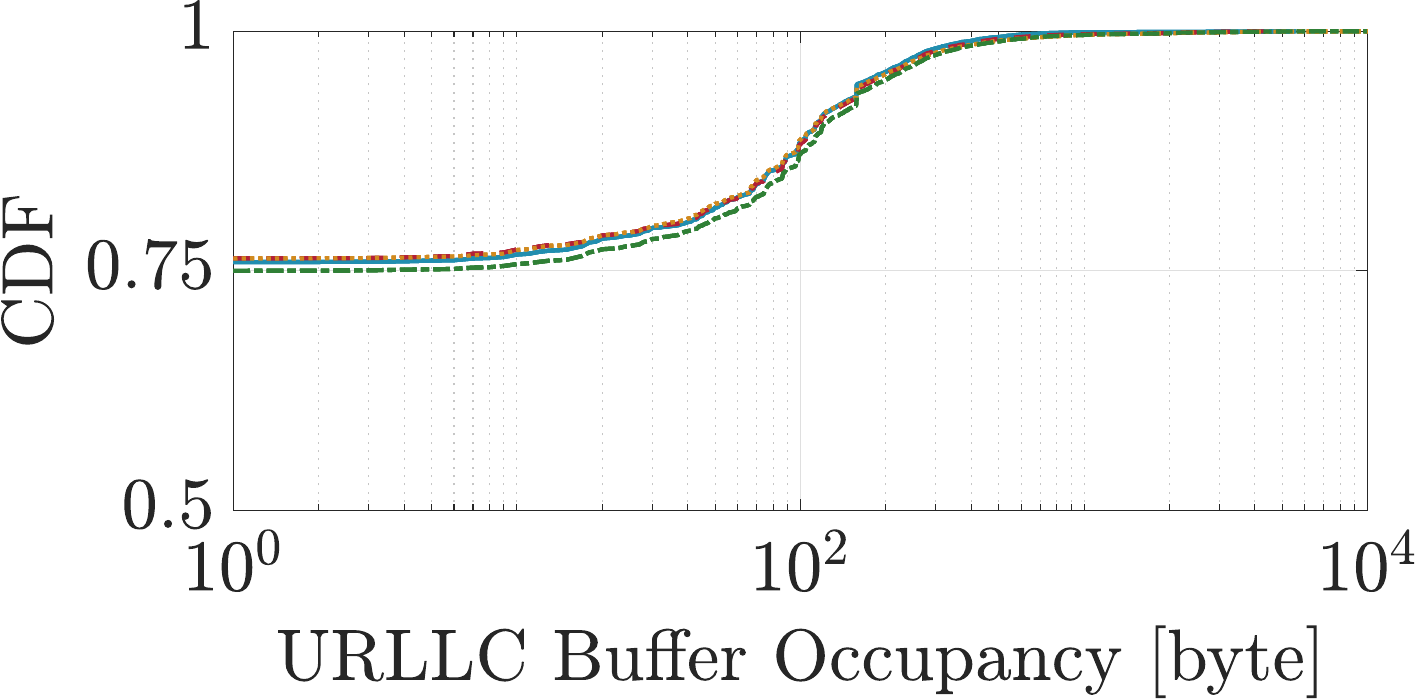}\label{Figure4c2}}
\caption{Performance evaluation under different hierarchical configurations with the PPO DRL Architecture.}
\label{Figure4-2a}
\end{figure}

\begin{figure}[htbp]
\centering
\subfloat[\gls{embb} DL Throughput]{\includegraphics[width=2.85in]{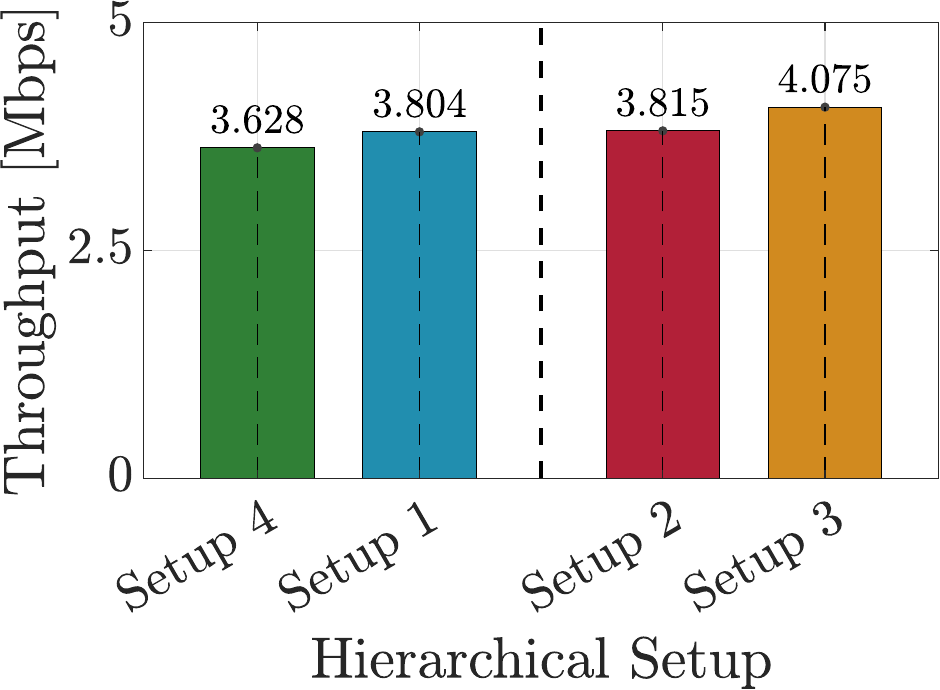}\label{Figure4d2}}
\hfil
\subfloat[\gls{mmtc} Packets]{\includegraphics[width=2.85in]{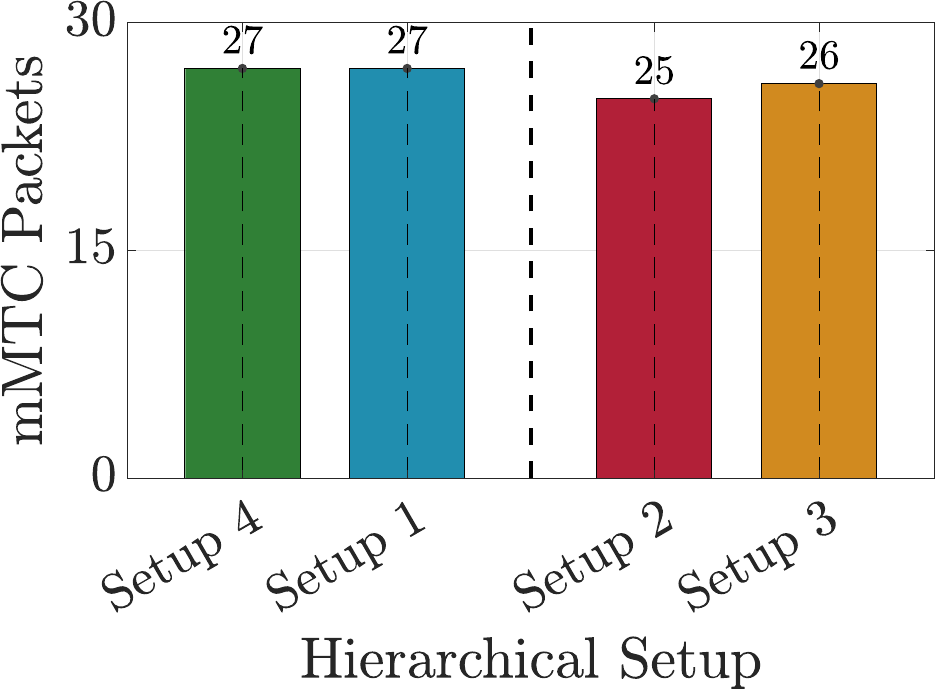}\label{Figure4e2}}
\caption{Median values under different hierarchical configurations with the PPO DRL Architecture.}
\label{Figure4-2b}
\end{figure}

We consider four setups, summarized in Table~\ref{table:hier-setups}. Each entry describes how frequently the \gls{bs} reports \glspl{kpm} to the \gls{ric}. For instance, in Setup~1, the xApp for slicing control receives data from the \gls{bs} every $1$\:s, while the scheduling agent receives the respective metrics every $10$\:s. Despite taking into account \gls{ran} telemetry reported every $1, 5$ or $10$\:s, the \gls{drl} decision-making process and the enforcement of a control policy on the \gls{bs} occur within a granularity of tens of milliseconds, and hence the intelligent control loops are still in compliance with the timescale requirements of the near-real-time \gls{ric}.

Results of this analysis are presented in Figs.~\ref{Figure4-2a} and~\ref{Figure4-2b}. From Fig.~\ref{Figure4a2}, \textit{Setup 3} delivers the best \gls{embb} performance, \textit{Setups} $1$ and $2$ perform almost equally, while \textit{Setup 4} performs the worst. For \gls{mmtc}, in Fig.~\ref{Figure4b2} we notice that all combinations perform similarly and deliver approximately $26$ packets, with \textit{Setup 1} and \textit{Setup 4} delivering an additional packet. From Fig.~\ref{Figure4c2}, we notice that all setups deliver the same performance for the \gls{urllc} slice and, despite not being reported in the figures, they all yield a median buffer occupancy of $0$\:byte, i.e., they maintain an empty buffer to ensure low latency values. In Fig.~\ref{Figure4d2}, we notice that \textit{Setups} $2$ and $3$ deliver the highest \gls{embb} throughput. In Fig.~\ref{Figure4e2}, instead, we notice that \textit{Setups} $1$ and $4$ deliver the highest number of transmitted \gls{mmtc} packets.

Our findings on hierarchical control verify \gls{embb}'s and \gls{mmtc}'s competitive behavior for individual reward maximization. Our results show that the rewards of \gls{embb} and \gls{mmtc} slices are competing with one another, as the best configuration for \gls{embb} corresponds to the worst configuration for \gls{mmtc}, and vice versa. Among all considered configurations, \textit{Setup 3} offers the best trade-off, as it delivers the highest throughput at the expense of a single \gls{mmtc} packet less being transmitted.

\subsubsection{Impact of Per-Slice Scheduling Profile Selection}\label{sec:oran_edge:pandora:perslice}

We extend our prior work~\cite{10437367} by evaluating the coexistence of a multitude of xApps delivering services to an Open \gls{ran} multi-slice scenario. Specifically, we consider the case where a single xApp distributes the available \glspl{prb} to all three slices, while three xApps are each tasked with selecting a dedicated scheduling profile for a single slice. All four xApps operate at the same timescale as given in Set~$1$ of Table~\ref{table:control-time}, where \glspl{kpm} and actions are both logged and updated every $250$~ms respectively, while fresh \gls{ran} data are reported by the \gls{du} within a granularity of $1$~s.

\begin{figure}[htbp]
\centering
\subfloat[\gls{embb} Throughput]{\includegraphics[height=4cm]{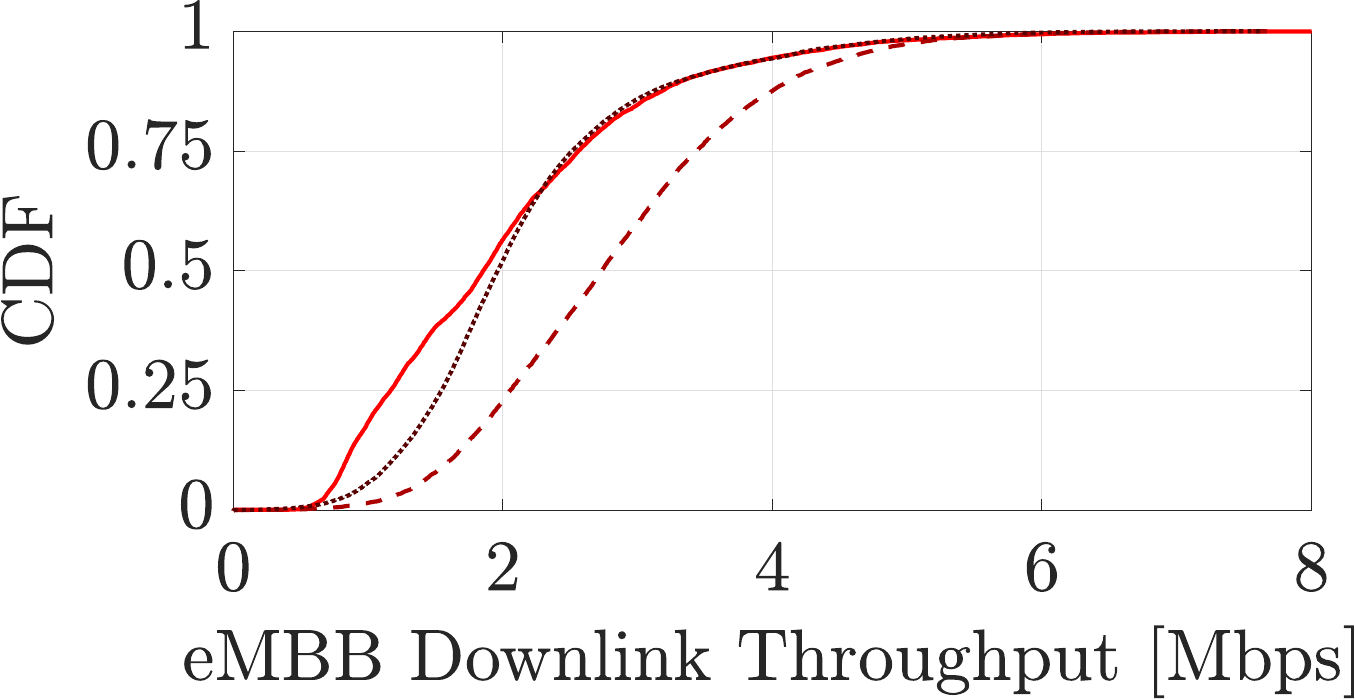}\label{fig:Figure9a}}
% \hfil
\subfloat[\gls{mmtc} Packets]{\includegraphics[height=4cm]{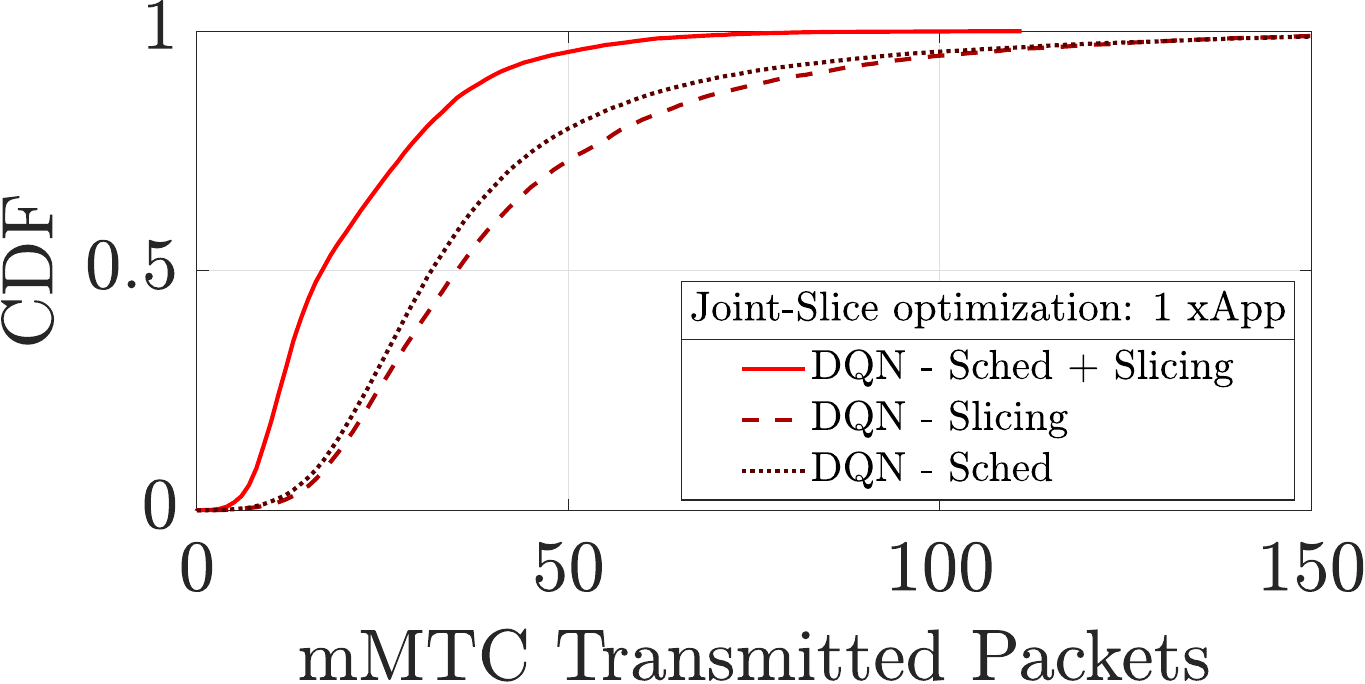}\label{fig:Figure9b}}
\hfil
\subfloat[\gls{urllc} Buffer Occupancy]{\includegraphics[height=4cm]{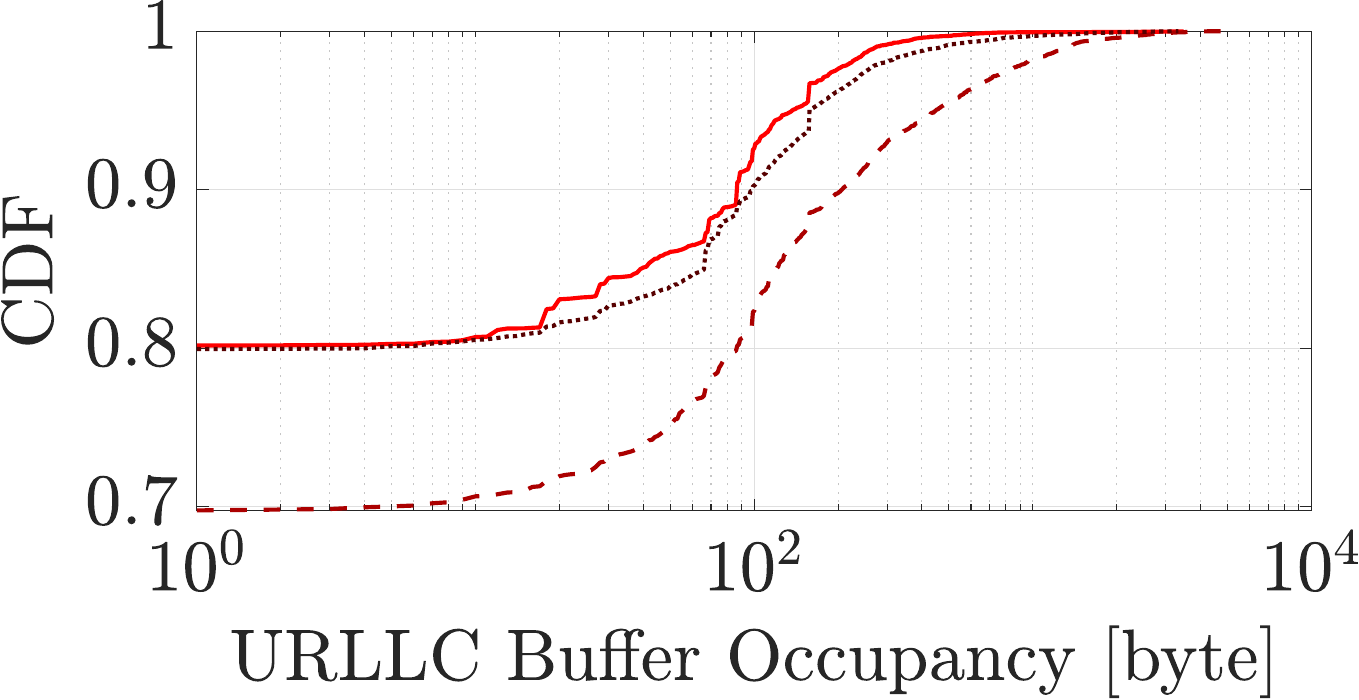}\label{fig:Figure9c}}
\caption{Performance evaluation under the \textit{Default} weight configuration for different action spaces and discount factor $\gamma=0.95$ with the DQN DRL Architecture.}
\label{Figure9-1a}
\end{figure}

\begin{figure}[htbp]
\centering
\subfloat[\gls{embb} DL Throughput]{\includegraphics[width=2.68in]{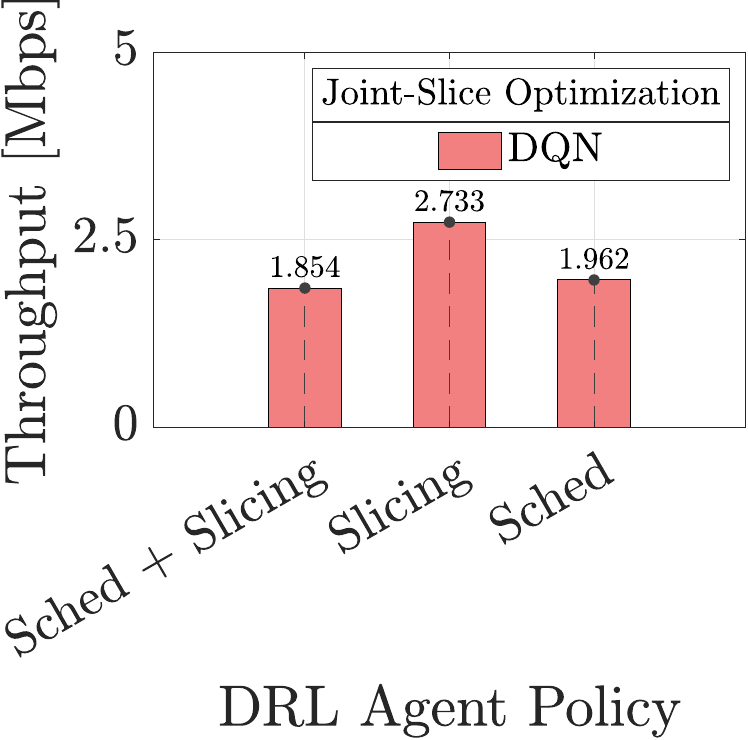}\label{fig:Figure9d}}
\hfil
\subfloat[\gls{mmtc} Packets]{\includegraphics[width=2.68in]{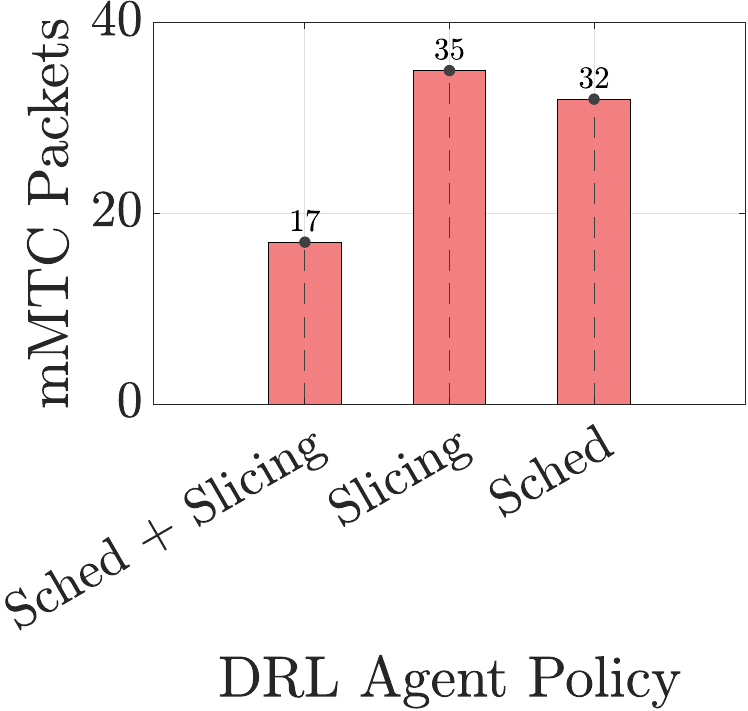}\label{fig:barplot_mmtc_dqn}}
\caption{Median values under the \textit{Default} weight configuration for different action spaces and discount factor $\gamma=0.95$ with the DQN DRL Architecture.}
\label{fig:Figure9e}
\end{figure}

\begin{figure}[htbp]
\centering
\subfloat[\gls{embb} Throughput]{\includegraphics[height=4cm]{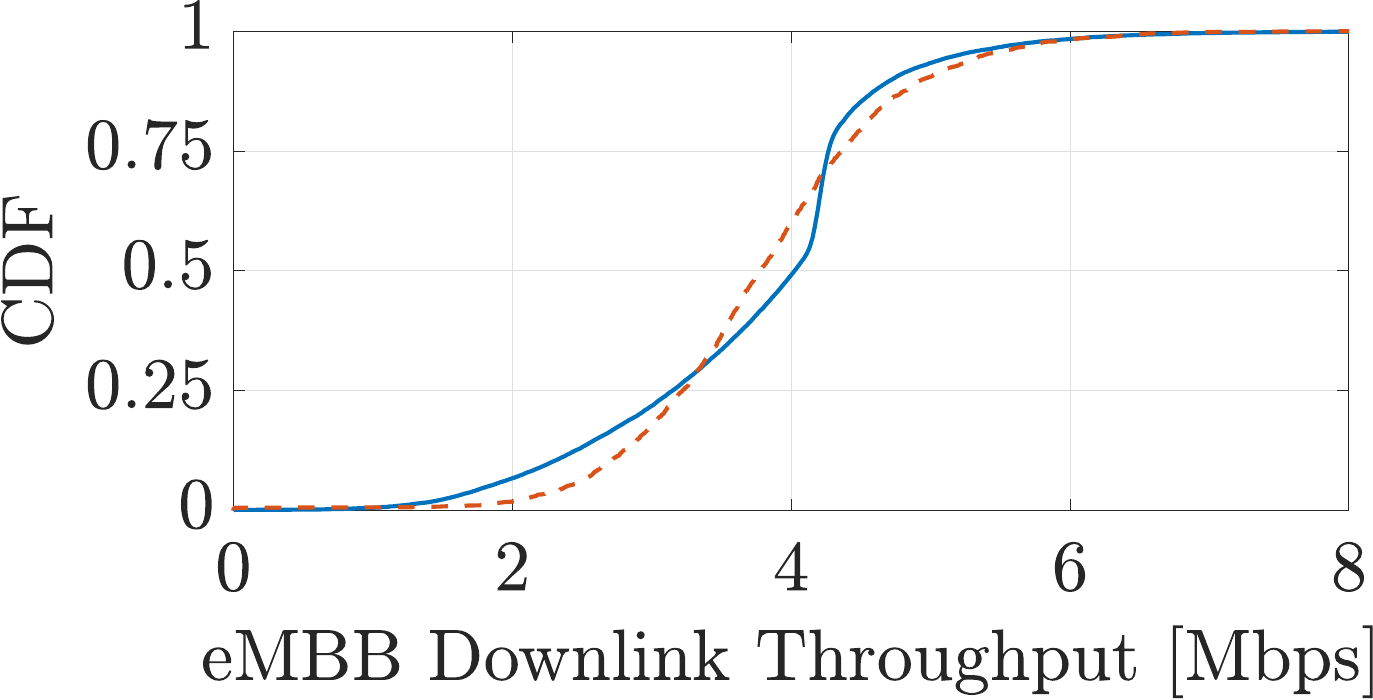}\label{fig:Figure8a}}
% \hfil
\subfloat[\gls{mmtc} Packets]{\includegraphics[height=4cm]{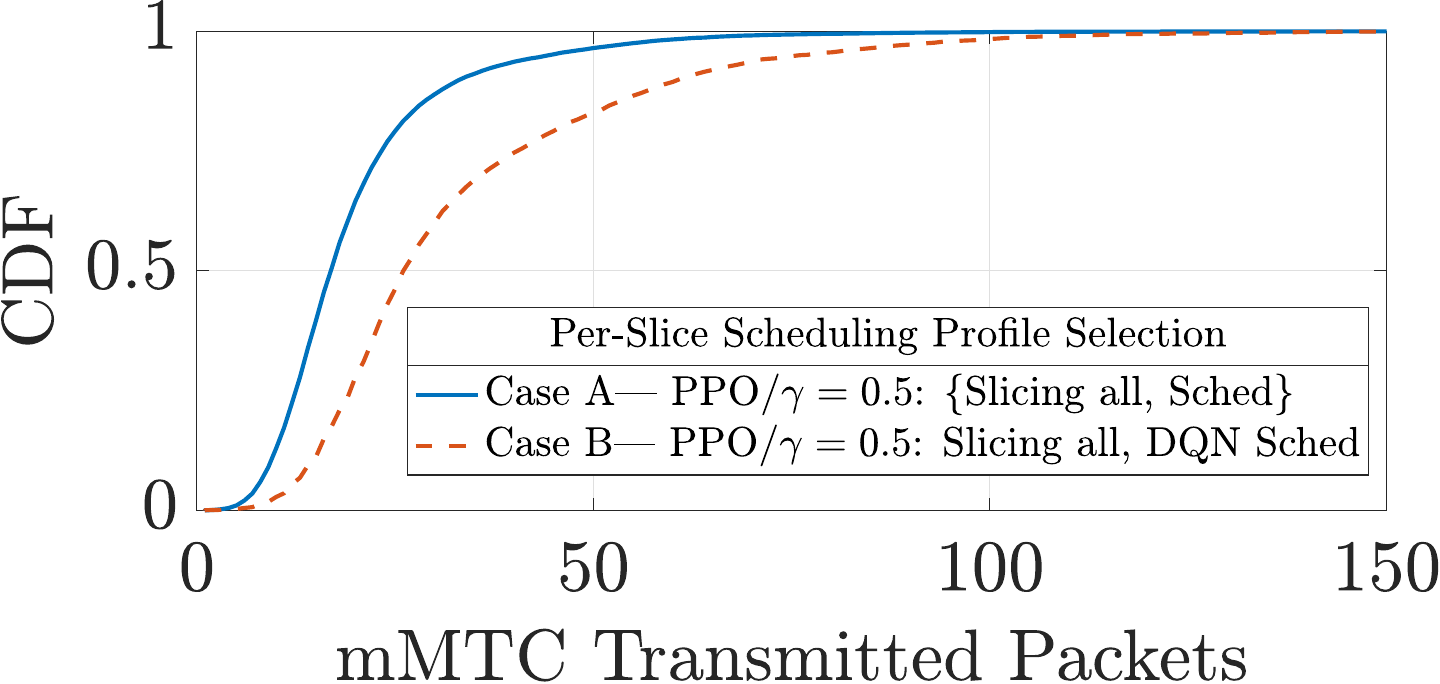}\label{fig:Figure8b}}
\hfil
\subfloat[\gls{urllc} Buffer Occupancy]{\includegraphics[height=4cm]{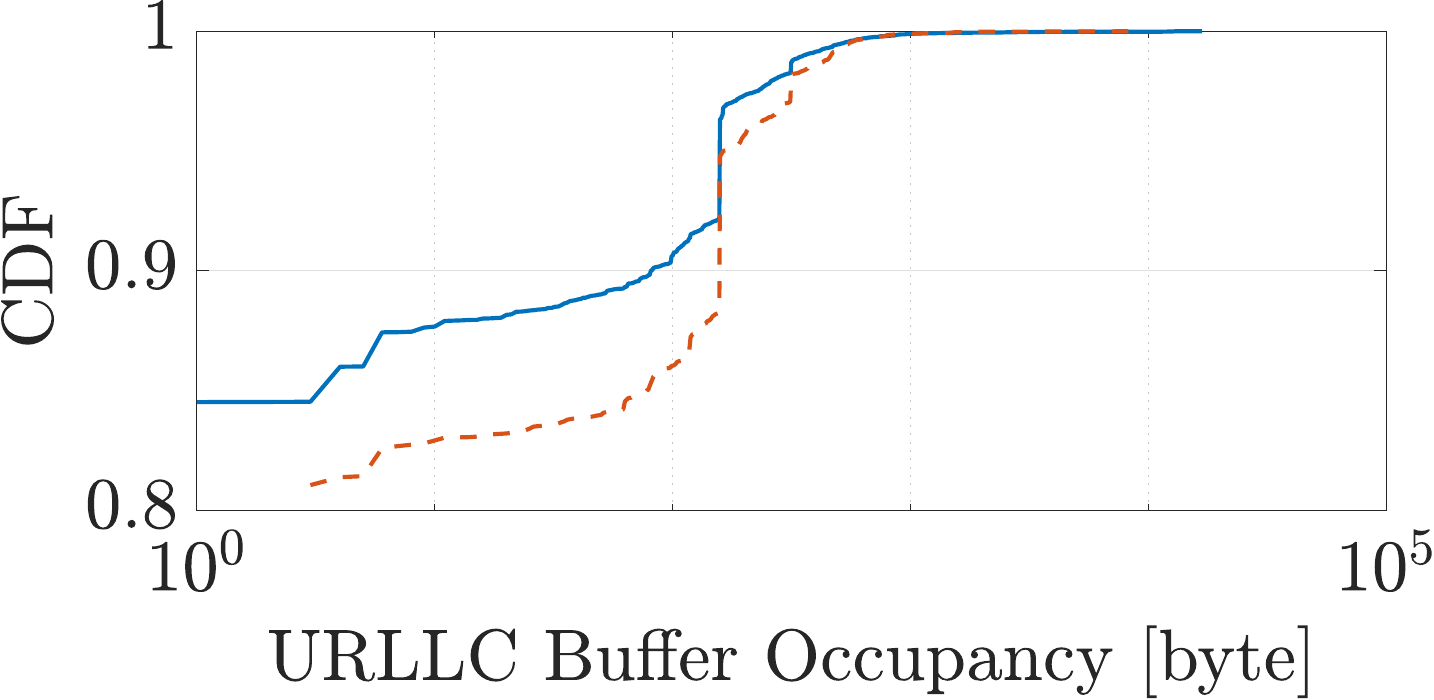}\label{fig:Figure8c}}
\caption{Performance evaluation with $4$ xApps and per-slice scheduling profile selection under PPO and DQN Architectures.}
\label{Figure8-1a}
\end{figure}

\begin{figure}[htbp]
\centering
\subfloat[\gls{embb} Throughput]{\includegraphics[width=2.85in]{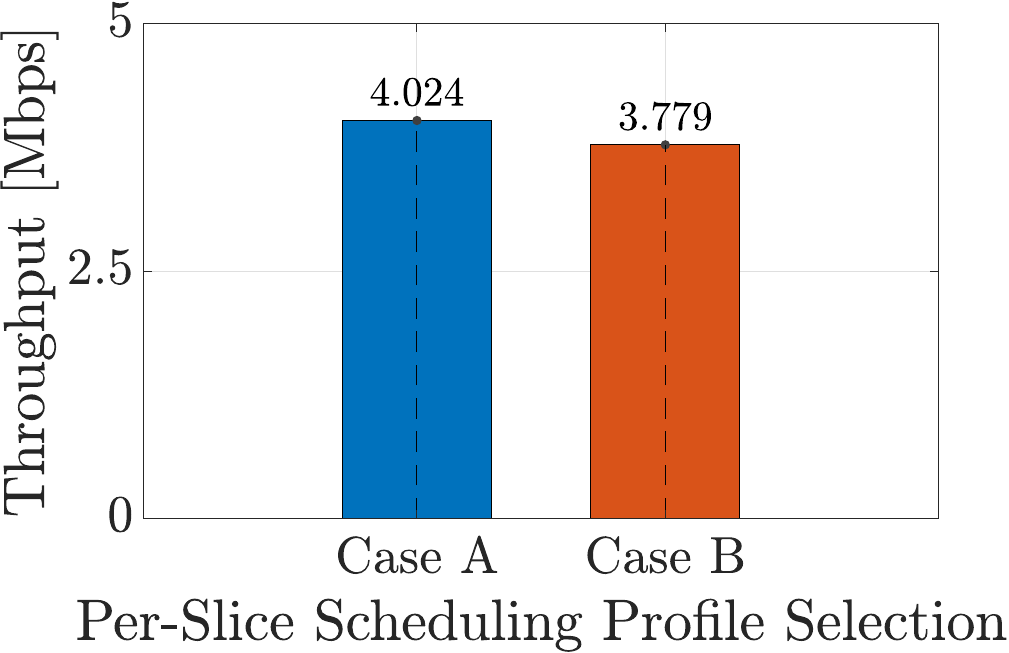}\label{fig:NewEntry1}}
\hfil
\subfloat[\gls{mmtc} Packets]{\includegraphics[width=2.85in]{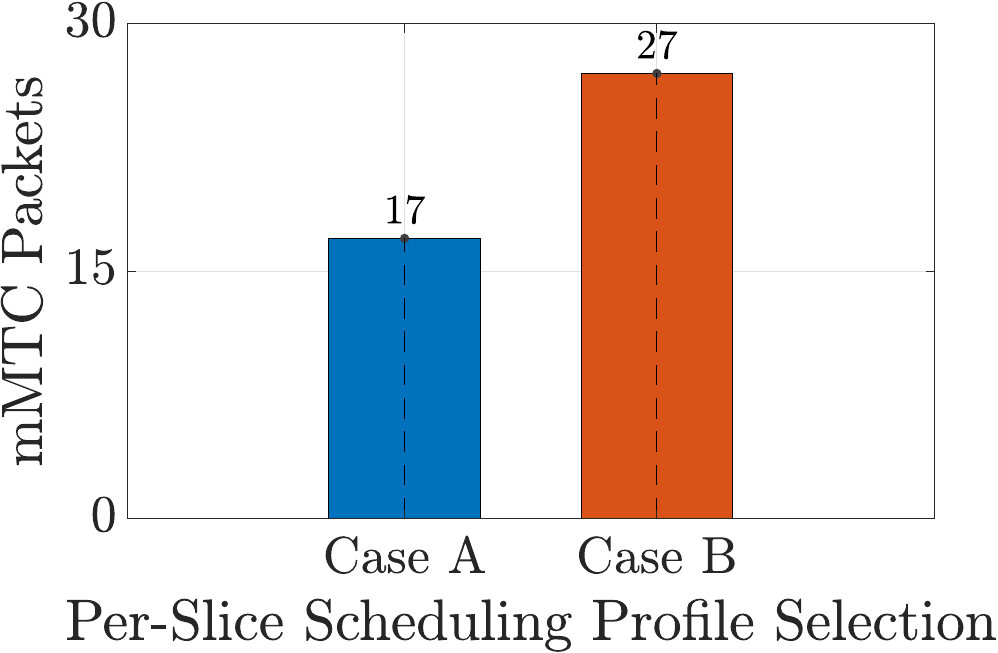}\label{fig:barplot-4xapps-mmtc}}
\caption{Median values obtained with a $4$-xApp setup and per-slice scheduling profile selection under PPO and DQN Architectures.}
\label{Figure9-1b}
\end{figure}

We begin the evaluation by focusing on the \gls{dqn} algorithm and the case of the joint-slice optimization. All results were produced with agents trained under the \texttt{Default} weight configuration of Table~\ref{table:weight-confs-list} and with a discount factor of $\gamma=0.95$. The findings presented in Figs.~\ref{Figure9-1a} and~\ref{fig:Figure9e} indicate that in the \gls{embb} slice, \gls{dqn} underperforms compared to \gls{ppo} for both $\gamma\in\{0.5,0.99\}$, with only \texttt{Slicing} delivering the highest reported throughput value at $2.733$\:Mbps, as reported in Fig.~\ref{fig:Figure9d}. With respect to \gls{mmtc}, Figs.~\ref{fig:Figure9b} and~\ref{fig:barplot_mmtc_dqn} depict a similar performance to the one shown in Figs.~\ref{fig:Figure4b} and~\ref{fig:Figure4e}, when examining the actions separately under their respective discount factor (i.e., $\gamma\in\{0.5,0.99\}$). In detail, we observe that in both cases, \texttt{Slicing} delivers the highest reported number of transmitted packets, followed by \texttt{Sched}, and \texttt{Sched \& Slicing}. It is also shown that the \texttt{Sched} xApp, in the case of \gls{dqn}, performs better in the \gls{mmtc} compared to the case with \gls{ppo}, delivering a median value of $32$ transmitted packets, which is $\sim28$\% more compared to \texttt{Sched 0.99} with \gls{ppo}, as shown in Fig.~\ref{fig:Figure4e}. Finally, we observe that slicing is the action that delivers the best performance in both \gls{embb} and \gls{mmtc}. Indeed, the competitive behavior observed with \gls{ppo} is mitigated in the case of \gls{dqn}, as shown in Fig.~\ref{fig:Figure9e}, since the actions that deliver the highest \gls{embb} throughput also deliver the highest number of \gls{mmtc} transmitted packets. On the \gls{urllc}, all configurations once again delivered optimal performance, reporting a median buffer occupancy of $0$~byte.

Both \gls{drl} architectures are model free, i.e., they do not rely on an explicit model of the environment. Instead, the \gls{drl} agent learns directly by interacting with the environment and makes decisions through trial and error. \gls{ppo} employs a trust region method~\cite{schulman2017proximal, kim2022adaptive} that helps in achieving more stable training, compared to \gls{dqn} which can suffer from divergence and slow convergence~\cite{zhang2021minibatch}. The trust region ensures that the old and current policy do not deviate significantly, which helps mitigate divergence issues. Actor and critic networks calculate the state-value, with the critic network reducing variance in action value estimates. This reduction in variance makes policy updates more reliable and exploration safer~\cite{lillicrap2015continuous}. On the contrary, \gls{dqn} does not implement trust regions and uses the target network to achieve stability during training~\cite{mnih2015human}. In environments affected by noisy observations and stochasticity, the aforementioned features, combined with the model's ability to explicitly model the policy's probability distribution over actions, make \gls{ppo} capable of adapting more efficiently to varying conditions. This ability to generalize better across tasks and environments can be achieved with \gls{dqn} but might require more fine-tuning and adjustments when transitioning to new tasks. Furthermore, \gls{dqn} relies on epsilon-greedy exploration, which can be less efficient, especially in complex environments~\cite{hessel2018rainbow,fortunato2017noisy}. \gls{ppo}, being an on-policy algorithm, manages to optimize the current policy directly based on the most recent experiences. On the other hand, \gls{dqn}, an off-policy algorithm, estimates the value of actions independently of the current policy. On-policy methods often work better when the policy is changing during training, especially in stochastic environments. Therefore, in the latter case where rapid policy changes are observed, off-policy methods may struggle to deal successfully with them, and they may require more extensive replay buffers or further fine-tuning to handle noisy environments effectively~\cite{schulman2015trust,fujimoto2019off,9068634}.

Lastly, \gls{ppo} has been shown to work more efficiently in problems with high-dimensional state and action spaces, as demonstrated in~\cite{kozlica2023deep}, owing to its policy-based and trust region approaches. In our work, the state and action spaces are high-dimensional. In detail, a total of $43$~actions are considered when resource allocation is performed jointly for all slices. For scheduling, the size of the action space is $27$, while for slicing, it is $16$. The former size encompasses all possible combinations of scheduling policies for all three slices (i.e., \gls{rr}, \gls{wf}, and \gls{pf}), while the combination for \gls{prb} allocation is given in Table~\ref{tab:feasible_prb_allocation} for a total of $50$~\glspl{prb}. Finally, the observation of the state, obtained through srsRAN, consists of $10$ independent measurements of $3$~\glspl{kpi}, i.e., \gls{dl} buffer occupancy, the \gls{dl} throughput and the number of transmitted packets on the \gls{dl}.

From the previous discussion, we have seen that \gls{dqn} performs poorly in the high-dimensional state and action spaces, such as the ones considered in this work. Therefore, we test its effectiveness in a reduced action space where the \gls{dqn} agent only controls scheduling decisions. Specifically, we consider the case of one \texttt{Slicing} xApp that serves all slices simultaneously, while the remaining $3$~xApps are dedicated to each slice and control the scheduling profile for the corresponding slice only. For instance, one xApp selects a scheduling profile for the \gls{embb} slice, another one focuses on the \gls{mmtc} slice, and the last one controls the \gls{urllc} slice only. This configuration allows for both testing and mixing different \gls{drl} architectures, such as \gls{ppo} and \gls{dqn}. It also enables experimentation with a setup in which the scheduling selection action follows the \gls{prb} allocation. Therefore, the preceding xApps will reconfigure a network where the action space is altered after the changes made by the \texttt{Slicing} xApp.

In Figs.~\ref{Figure8-1a} and~\ref{Figure9-1b}, we present the results obtained from testing a total of~$4$ xApps that perform resource allocation. For the purposes of the experiments we consider two cases, namely \emph{Case A} and \emph{Case B}. In both cases, the \texttt{Slicing} xApp embeds a \gls{drl} agent trained with \gls{ppo} under the discount factor, $\gamma=0.5$. In \emph{Case A}, the remaining \texttt{Sched} xApps also embed \gls{ppo} agents trained with $\gamma=0.5$. In \emph{Case B}, the xApps embed \gls{dqn} agents with a discount factor of $\gamma=0.95$, as described in Section~\ref{sec:oran_edge:pandora:architectures}. In Fig.~\ref{Figure8-1a}, we observe that \emph{Case A} performs better for the \gls{embb} slice. Indeed, the results reported in Fig.~\ref{Figure9-1b} indicate that \emph{Case A} achieves a median \gls{embb} throughput value of $4.024$\:Mbps, which is $\sim 6.5\%$ higher than that of \emph{Case B}. Regarding the \gls{mmtc} slice, the median for the corresponding \gls{kpm} metric was measured at $27$ transmitted packets, representing an increase of $\sim60\%$ compared to \emph{Case A}. For \gls{urllc}, both cases delivered optimal performance with a median buffer occupancy of $0$~byte. We notice that \emph{Case A} performs similarly to the case of jointly performing slicing and scheduling control with $\gamma=0.5$, as shown in Figs.~\ref{Figure4-1a} and~\ref{Figure4-1b}, by delivering one additional \gls{mmtc} packet, as depicted in Fig.~\ref{Figure9-1b}. With regard to \emph{Case B}, the obtained results closely match those reported with Hierarchical Control under Setup~$1$ in Fig.~\ref{Figure4-2b}. Precisely, as depicted in Fig.~\ref{fig:barplot-4xapps-mmtc}, the two configurations achieve identical performance on the \gls{mmtc} slice, by achieving the same median value of transmitted packets (i.e., $27$~packets). For the \gls{embb} slice, the setup with $4$~xApps (i.e., \emph{Case B}) exhibits a minimal drop of $\sim0.66\%$ in the median throughput value compared to the performance achieved with Setup~$1$. Therefore, \emph{Case B} provides a more flexible and heterogeneous setup where different \gls{drl} agents can coexist, without a significant performance degradation. Furthermore, the \gls{dqn} can still be leveraged in those cases where agents independently optimize the performance of each slice thanks to the reduced action space.

\subsubsection{Impact of Weight Configuration}\label{sec:oran_edge:pandora:weights}

\begin{table}[htbp]
\centering
\small
\caption{Weight Design}
\begin{tabular}{ccc}
\toprule
$w_{eMBB}$ & $w_{mMTC}$ & $w_{URLLC}$ \\
\midrule
$\alpha_{eMBB}\cdot\dfrac{1}{A}$ & $\beta_{mMTC}\cdot\dfrac{1}{B}$ & $\gamma_{URLLC}\cdot \left( -\dfrac{1}{C} \right)$ \\
\bottomrule
\end{tabular}
\label{table:weight-design}
\end{table}

\begin{figure}[H]
\centering
\subfloat[\gls{embb} Throughput]{\includegraphics[height=4cm]{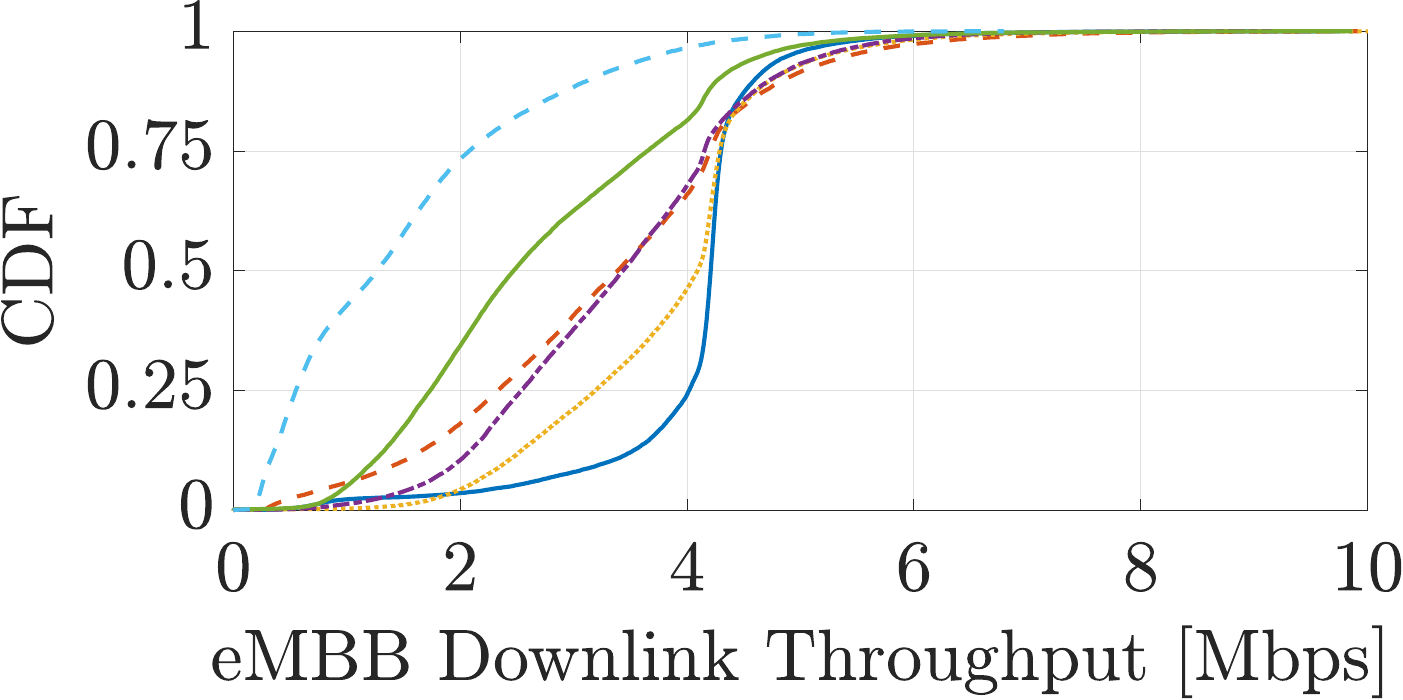}\label{Figure4a3}}
% \hfil
\subfloat[\gls{mmtc} Packets]{\includegraphics[height=4cm]{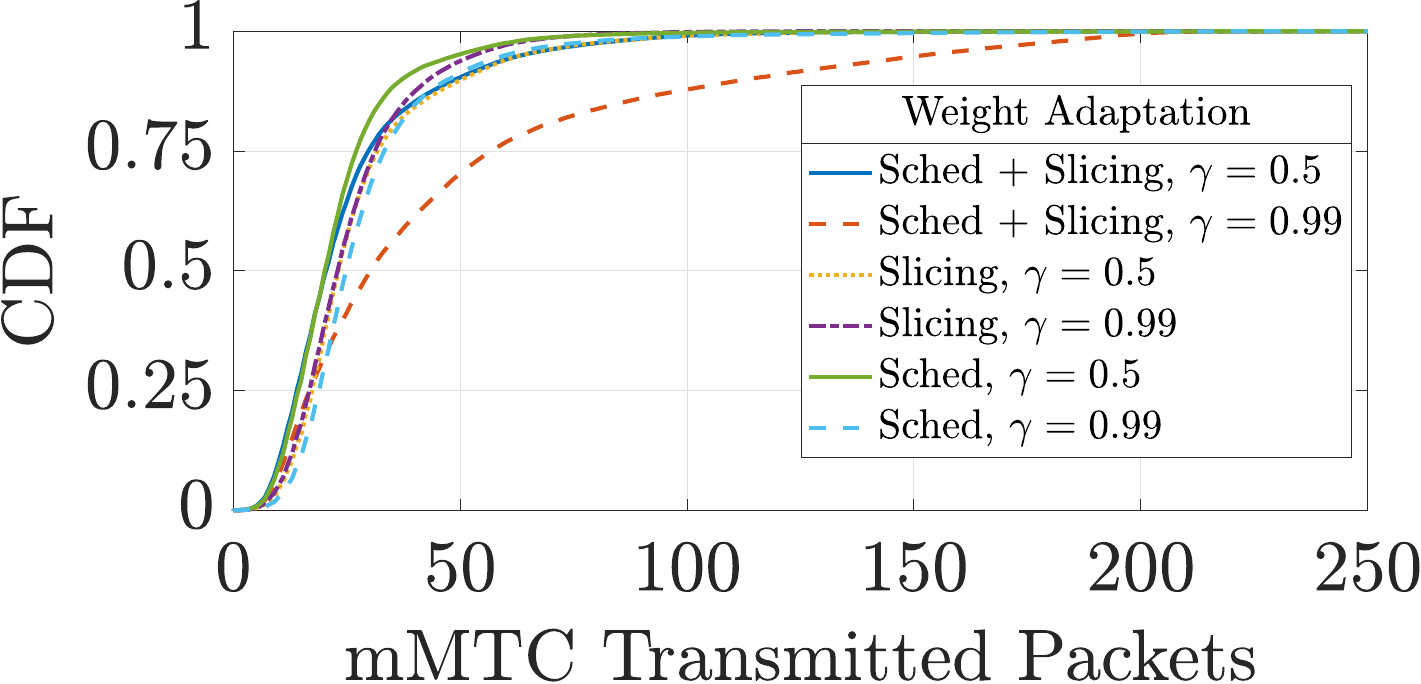}\label{Figure4b3}}
\hfil
\subfloat[\gls{urllc} Buffer Occupancy]{\includegraphics[height=4cm]{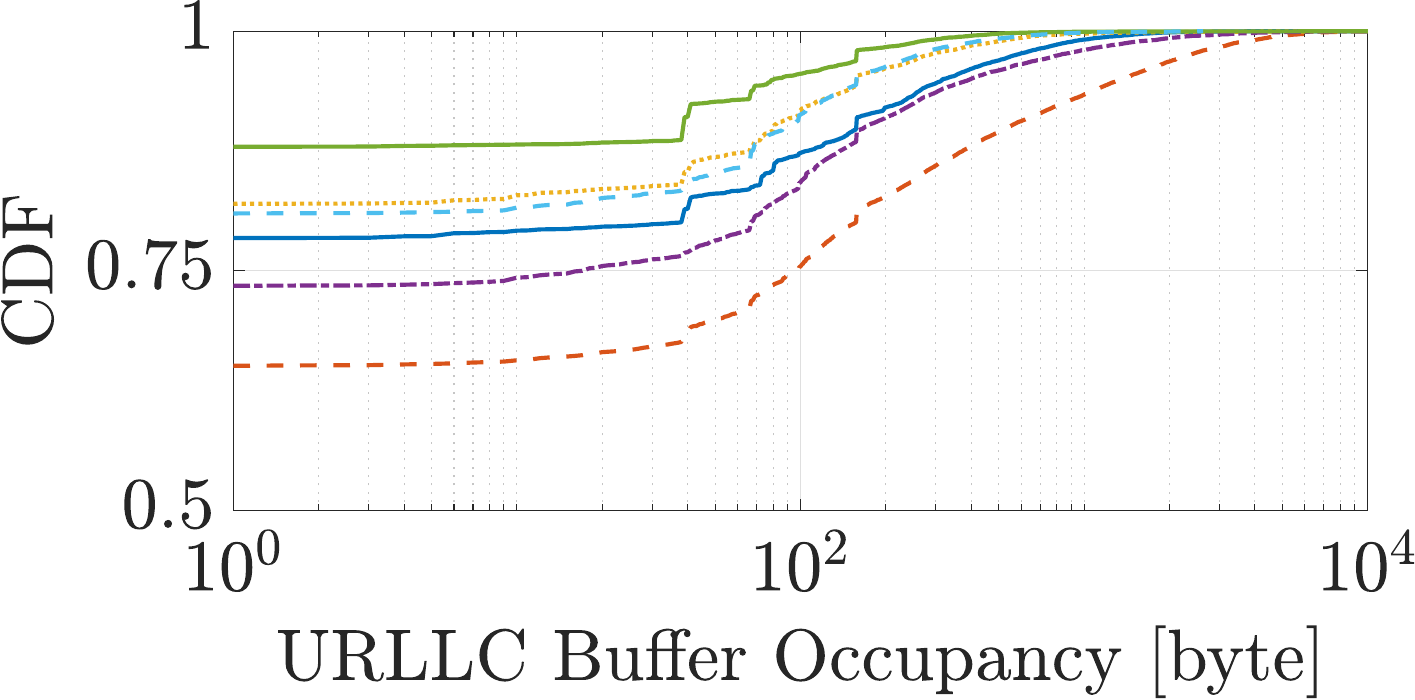}\label{Figure4c3}}
\caption{Performance evaluation under the \textit{Alternative} weight configuration for different action spaces and discount factors with the PPO DRL Architecture.}
\label{Figure4-3a}
\end{figure}

\begin{figure}[htbp]
\centering
\subfloat[\gls{embb} DL Throughput]{\includegraphics[width=2.85in]{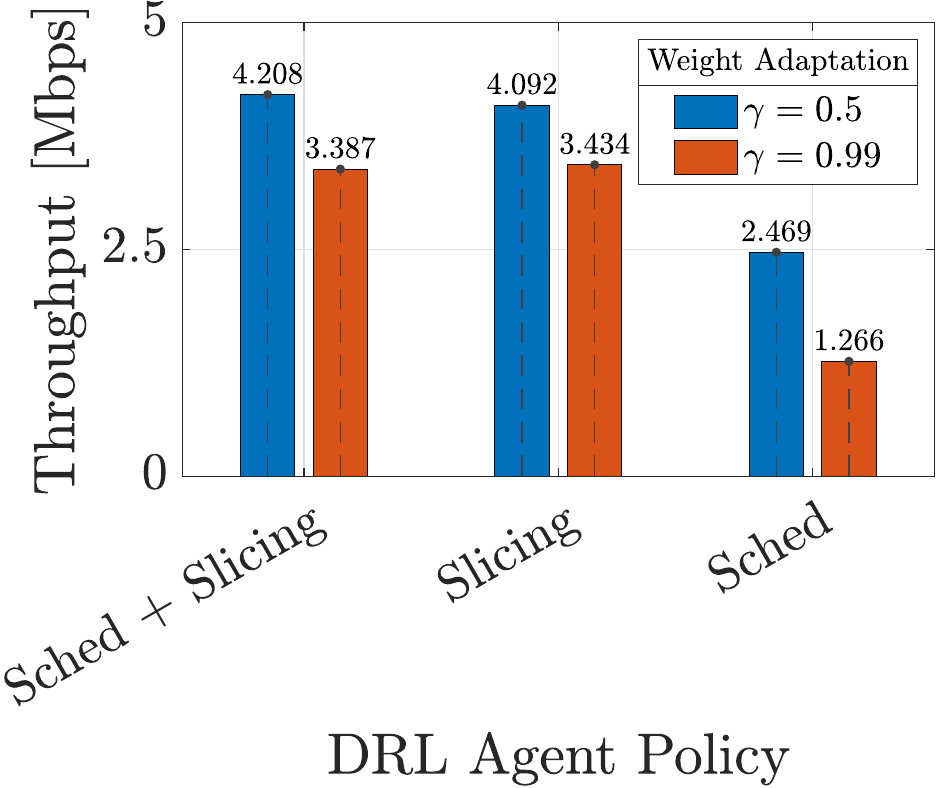}\label{Figure4d3}}
\hfil
\subfloat[\gls{mmtc} Packets]{\includegraphics[width=2.85in]{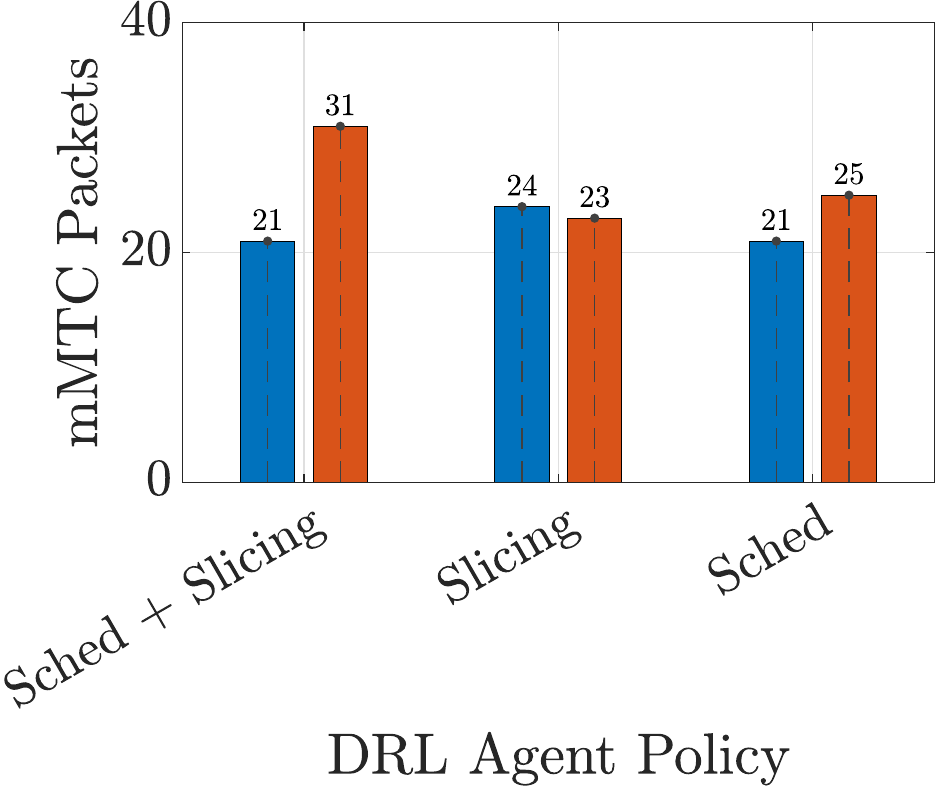}\label{Figure4e3}}
\caption{Median values under the \textit{Alternative} weight configuration for different action spaces and discount factors with the PPO DRL Architecture.}
\label{Figure4-3b}
\end{figure}

In this study, we consider different weight configurations to compute the cumulative average reward function in Eq.~\eqref{eq:weighted_reward}. The considered configurations are reported in Table~\ref{table:weight-confs-list}. The \textit{Alternative} weight configuration is computed by using the weights in Table~\ref{table:weight-design}, where $A, B, C, \alpha_{eMBB}, \beta_{mMTC}$, and $\gamma_{URLLC}$ are used to both scale and prioritize certain slices. Specifically, $A, B, C$ are used to scale the individual weights according to statistical information of corresponding \glspl{kpm}. For example, $A, B, C$ can represent either the average, minimum or maximum values reported \gls{kpm} per slice so as to scale the weight according to the dynamic range of the corresponding \gls{kpm}. Similarly, $\alpha_{eMBB}, \beta_{mMTC}$, and $\gamma_{URLLC}$ can be used to give priority to one slice or the other.

We set $\alpha_{eMBB}=1000$, $\beta_{mMTC}=456$ and $\gamma_{URLLC}=1$. As a reference for $A$, $B$ and $C$, we choose the historically maximum reported \gls{kpm} values for each slice, i.e., $A=13.88$\:Mbps, $B=304$, and $C=20186$~byte.

Based on these steps, we derive their respective weights $w_{eMBB}$, $w_{mMTC}$, $w_{URLLC}$. For example, the weight of \gls{mmtc} can be computed as $w_{mMTC} = \beta_{mMTC}\cdot\dfrac{1}{B} = 456/304 = 1.5$, as reported in the \textit{Alternative} configuration in Table~\ref{table:weight-confs-list}. The goal of comparing the two \textit{Default} and \textit{Alternative} weight configurations is to explore and understand the dynamics between \gls{mmtc} and \gls{embb} and the overall impact on the network performance. Specifically, since previous results have shown that the \gls{mmtc} can be penalized by the \gls{embb} slice, with the \textit{Alternative configuration} we aim at giving the former a weight that is $6\times$ larger than the \textit{Default} configuration.

Results for the \textit{Alternative} configuration are reported in Figs.~\ref{Figure4-3a} and~\ref{Figure4-3b}. In Fig.~\ref{Figure4a3}, \texttt{Sched \& Slicing 0.5} delivers the best \gls{embb} performance. Similarly to the results presented in Section~\ref{sec:oran_edge:pandora:discount}, \texttt{Sched \& Slicing 0.5} and \texttt{Slicing 0.5} are the best choices, with short-term reward design being ideal for \gls{embb}. In Fig.~\ref{Figure4b3}, the \textit{Alternative} weight configuration results in \texttt{Sched \& Slicing 0.99} being the best \gls{mmtc} choice and long-term rewards are better for \gls{mmtc} users. For \gls{urllc}, all policies perform well, with \texttt{Sched \& Slicing 0.5} performing slightly better compared to \texttt{Sched \& Slicing 0.99}.

Figs.~\ref{Figure4d3} and~\ref{Figure4e3} confirm that controlling scheduling alone does not improve performance in general. Similarly to our previous analysis, a high \gls{embb} performance (i.e., \texttt{Sched \& Slicing 0.5}) results in a degraded \gls{mmtc} performance. However, if compared with the \texttt{Default}, the \texttt{Alternative} weight configuration achieves a $31.25\%$ increase for \gls{mmtc}, with the same equally good \gls{urllc} performance and a $1\%$ throughput increase for \gls{embb} users.

\begin{table}[H]
\centering
\small
\caption{Design Options Catalog}
\begin{tabular}{ll}
\toprule
$\textbf{Option 1}$ & \texttt{Sched \& Slicing 0.5 - Alternative} \\
$\textbf{Option 2}$ & \texttt{Slicing 0.5 - Default} \\
$\textbf{Option 3}$ & \texttt{Hierarchical Control - Setup 1} \\
$\textbf{Option 4}$ & \texttt{Slicing 0.99 - Default} \\
\bottomrule
\end{tabular}
\label{table:design-catalogue}
\end{table}

In Table~\ref{table:design-catalogue} we summarize the design options that have delivered a good overall performance so far. Table~\ref{table:final-designs} indicates \gls{embb} and \gls{mmtc}'s dynamic and competitive relation. Option $2$ brings balance, in terms of throughput and transmitted packets, Option $1$ favors \gls{embb}, and Option $4$ boosts \gls{mmtc} but with a significant decrease in the \gls{qos} of the \gls{embb} slice.

\begin{table}[H]
\centering
\small
\caption{Design Options}
\begin{tabular}{lccc}
\toprule
 & \textbf{eMBB [Mbps]} & \textbf{mMTC [packet]} & \textbf{URLLC [byte]} \\
\midrule
$\textbf{Option 1}$ & $4.208$ & $21$ & $0$ \\
$\textbf{Option 2}$ & $4.114$ & $26$ & $0$ \\
$\textbf{Option 3}$ & $3.804$ & $27$ & $0$ \\
$\textbf{Option 4}$ & $3.636$ & $37$ & $0$ \\
\bottomrule
\end{tabular}
\label{table:final-designs}
\end{table}

\subsubsection{Impact of \texorpdfstring{\gls{ran}}{RAN} Control Timers}\label{sec:oran_edge:pandora:timers}

\begin{table}[H]
\centering
\small
\caption{RAN Control Timers}
\begin{tabular}{cccc}
\toprule
\textbf{Control Time} & \textbf{Set 1} & \textbf{Set 2} & \textbf{Set 3} \\
\midrule
DU Report & $1$\:s & $250$\:ms & $100$\:ms \\
\glspl{kpm} Log & $250$\:ms & $250$\:ms & $100$\:ms \\
Action Update & $250$\:ms & $250$\:ms & $100$\:ms \\
\bottomrule
\end{tabular}
\label{table:control-time}
\end{table}

\begin{figure}[H]
\centering
\subfloat[\gls{embb} Throughput]{\includegraphics[height=4cm]{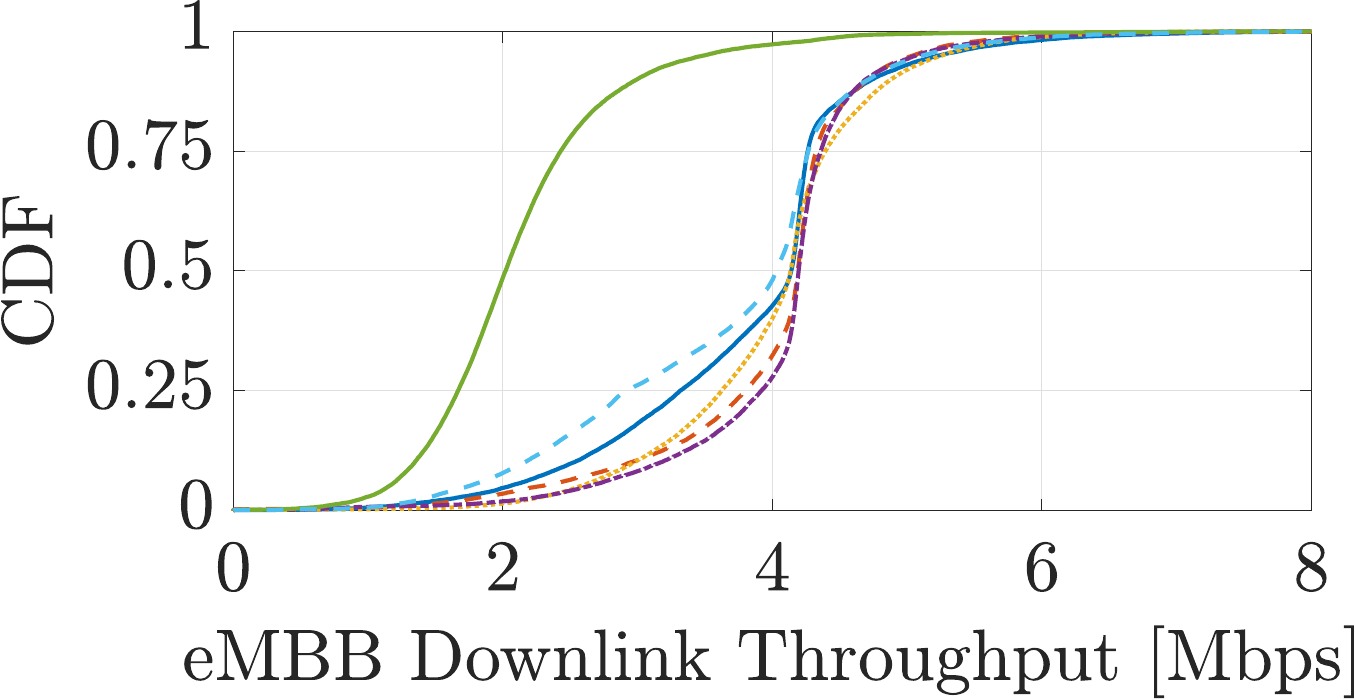}\label{Figure5a3}}
% \hfil
\subfloat[\gls{mmtc} Packets]{\includegraphics[height=4cm]{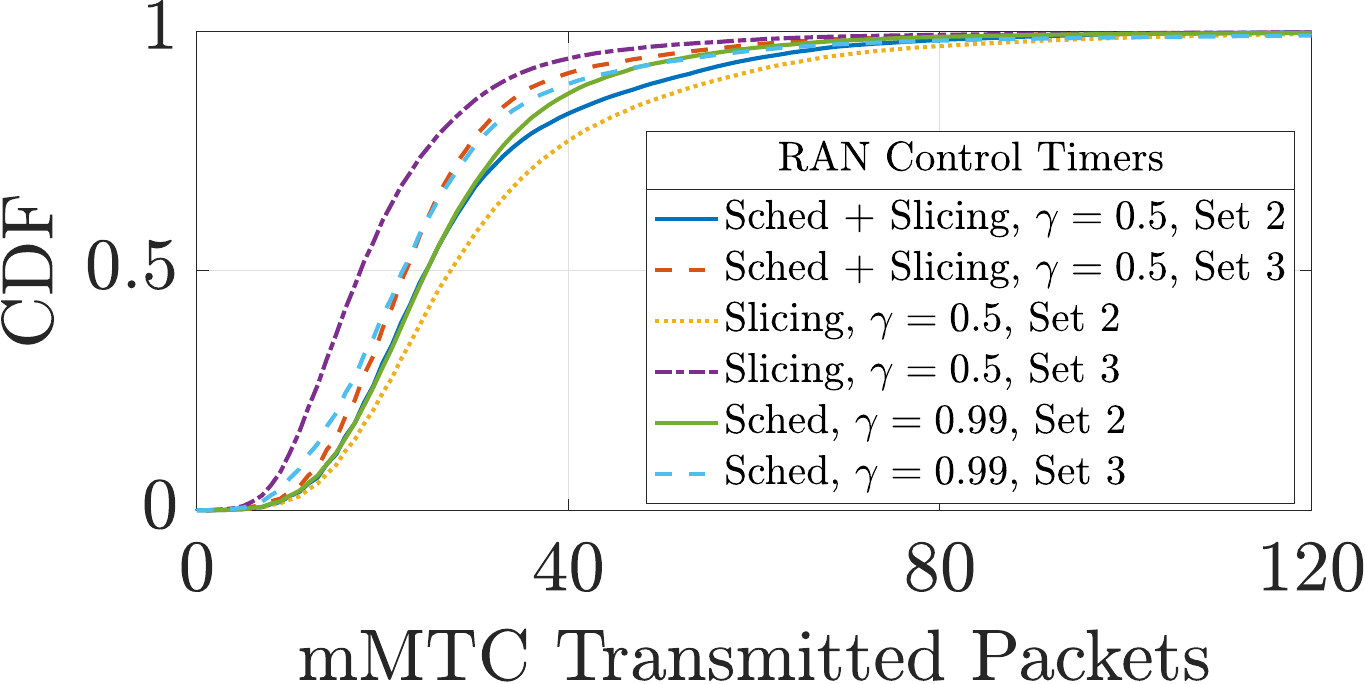}\label{Figure5b3}}
\hfil
\subfloat[\gls{urllc} Buffer Occupancy]{\includegraphics[height=4cm]{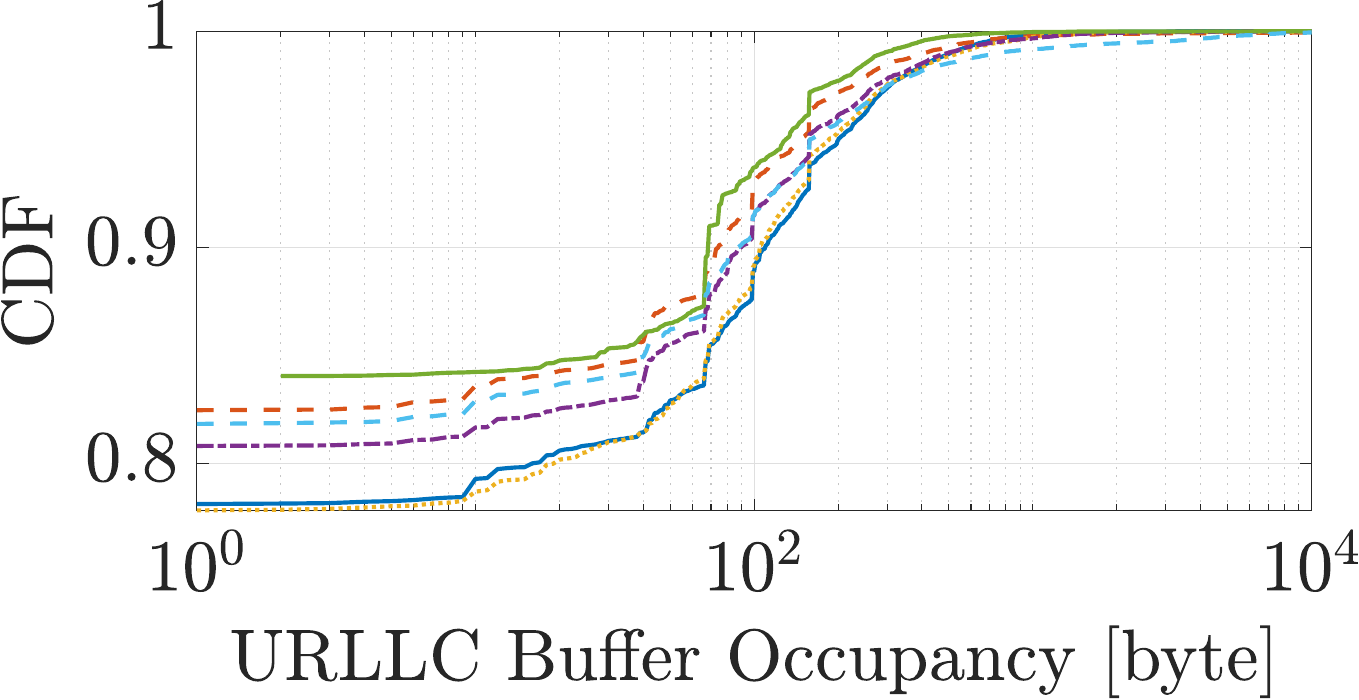}\label{Figure5c3}}
\caption{Performance evaluation under different action spaces, values of the $\gamma$ parameter and sets of \gls{ran} control timers with the PPO DRL Architecture.}
\label{Figure5-3a}
\end{figure}

\begin{figure}[htbp]
\centering
\subfloat[\gls{embb} DL Throughput]{\includegraphics[width=2.85in]{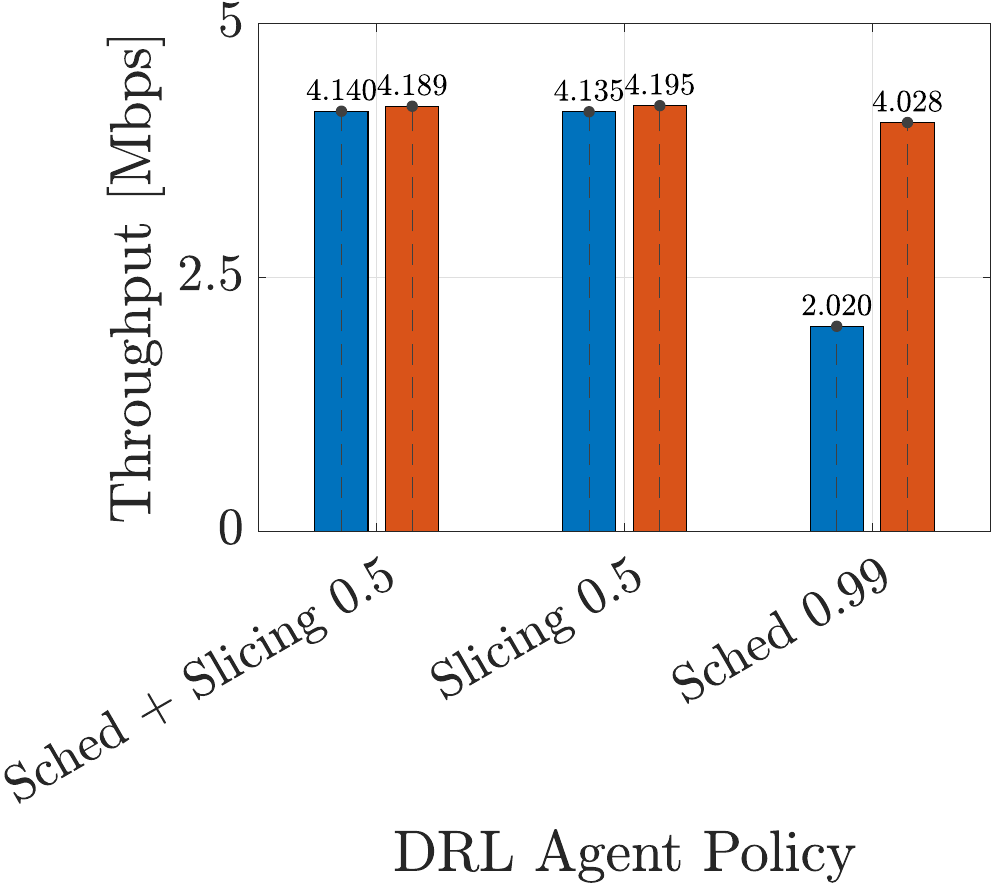}\label{ran-timers-bar-throughput}}
\hfil
\subfloat[\gls{mmtc} Packets]{\includegraphics[width=2.85in]{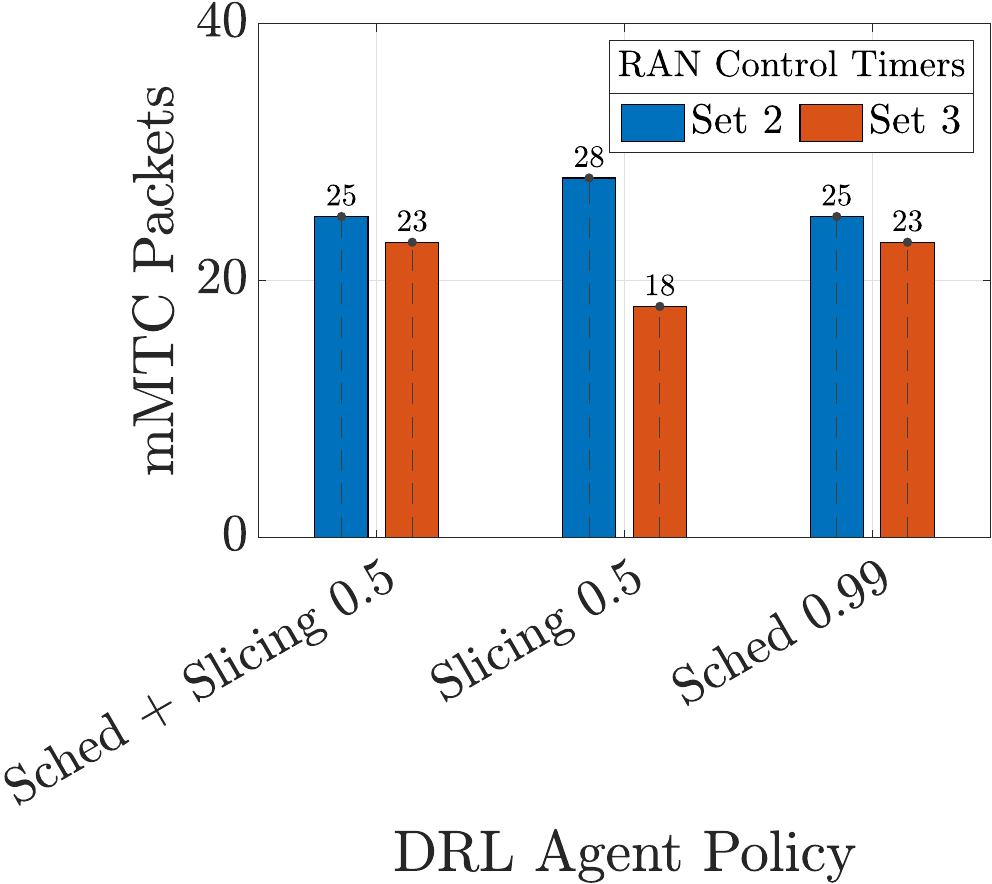}\label{ran-timers-bar-mmtc}}
\caption{Median values under different action spaces and \gls{ran} control timers with the PPO DRL Architecture.}
\label{Figure10-1b}
\end{figure}

We look into the case where we control three different \gls{ran} control timers, as reported in Table~\ref{table:control-time}, with the focus primarily steered onto Sets~$2$ and~$3$. In the latter cases, the \gls{kpm} collection granularity (i.e., \glspl{kpm} Log time) matches both the Action Update and the \gls{kpm} report granularities on the \gls{bs}'s \gls{du}. In detail, \glspl{kpm} are directly streamed into the \gls{ric}, upon collected, with an action being sent back through the E2 termination. This is in contrast with the case of Set $1$, where the \gls{du} report time is equal to $1$\:s, and hence some of the freshly captured data may not be reported. The findings of this evaluation are presented in Figs.~\ref{Figure5-3a} and~\ref{Figure10-1b}, and all configurations concern \gls{drl} agents trained under the \texttt{Default} weight configuration presented in Table~\ref{table:weight-confs-list}.

In Fig.~\ref{Figure5a3}, it is observed that all configurations result in a comparable high performance, delivering a median throughput value of $\sim 4$\:Mbps (Fig.~\ref{ran-timers-bar-throughput}) with the only exception of \texttt{Sched 0.99} under Set~$2$ which delivers poor throughput performance. The aforementioned configuration yields a performance similar to the one reported in Fig.~\ref{fig:Figure4d}, where \texttt{Sched 0.99} is evaluated under the time granularities of Set~$1$. Indeed, this combination achieves the same number of transmitted packets (i.e., $25$\:packets) with a negligible drop on the \gls{embb} throughput, which is now reported at $\sim 2.02$\:Mbps, a value approximately $5\%$ less than the $2.125$\:Mbps reported with Set~$1$. Additionally, the \texttt{Sched 0.99} xApp, when tested under Set~$3$, attains the maximum reported throughput value for controlling scheduling using a single xApp, (reaching $4.028$\:Mbps, a $\sim99\%$ increase compared to $2.02$\:Mbps achieved with Set~$2$). The latter is achieved while transmitting $23$ packets, which is two packets fewer than the maximum number of transmitted packets ever recorded when controlling scheduling alone, and with \gls{ppo}-implemented agents.

Based on these findings, scheduling should be controlled using small time granularities (i.e., $\sim 100$\:ms) to ensure high \gls{embb} throughput, while guaranteeing no performance degradation on the \gls{mmtc} slice. Furthermore, the results illustrated in Figs.~\ref{Figure5b3} and~\ref{ran-timers-bar-mmtc} indicate that both scheduling and slicing, as well as scheduling under Sets $2$ and $3$, can achieve the same performance on the \gls{mmtc}. However, \texttt{Sched \& Slicing 0.5} delivers higher \gls{embb} throughput, and hence, is preferred compared to \texttt{Sched 0.99}. In consistency with the reported findings so far, and when focusing on the three action profiles individually, we observe \gls{mmtc}'s and \gls{embb}'s competitive behavior (e.g., they heavily compete to gain radio resources). This trend is clearly illustrated in Figs.~\ref{Figure5a3},~\ref{Figure5b3},~\ref{ran-timers-bar-throughput}, and~\ref{ran-timers-bar-mmtc}, where \texttt{Slicing 0.5} under Set~$3$ achieves the highest throughput value at $4.195$\:Mbps, but the lowest number of transmitted packets (i.e., $18$ packets). The reported results in Figs.~\ref{ran-timers-bar-throughput} and~\ref{ran-timers-bar-mmtc} also indicate that Set~$3$ primarily boosts \gls{embb}'s performance, compared to Set~$2$ which is a better fit for \gls{mmtc} \glspl{ue}. Finally, in Fig.~\ref{Figure5c3}, we note that all configurations for the investigated set of granularities deliver identical performance on the \gls{urllc} slice and, although not shown in the figures, they consistently maintain a median buffer occupancy of $0$~byte to ensure low latency. It is also observed that both \texttt{Sched \& Slicing 0.5} and \texttt{Slicing 0.5} under Set~$2$ perform slightly better on the aforementioned slice, compared to \texttt{Sched 0.99} under Set~$2$, which performs slightly worse.

\subsection{Out-of-Sample Experimental Evaluation}\label{sec:oran_edge:pandora:outofsample}

\begin{table}[H]
\centering
\small
\caption{\gls{ml} Agent \& Network Condition Settings}
\begin{tabular}{ccccl}
\toprule
\textbf{Setting ID} & \textbf{\gls{ue} Speed [m/s]} & \textbf{Traffic Load} & \textbf{\gls{ran} Control Timers} & \textbf{Weight Configuration} \\
\midrule
\textbf{1} & 3 & Profile 1 & Set 1 & \texttt{Alternative/Default} \\
\textbf{2} & 3 & Profile 2 & Set 1 & \texttt{Alternative/Default} \\
\textbf{3} & 0 & Profile 2 & Set 1 & \texttt{Alternative} \\
\textbf{4} & 0 & Profile 2 & Set 2 & \texttt{Default} \\
\bottomrule
\end{tabular}
\label{table:net-settings}
\end{table}

\begin{figure}[H]
\centering
\subfloat[\gls{embb} Throughput]{\includegraphics[height=4cm]{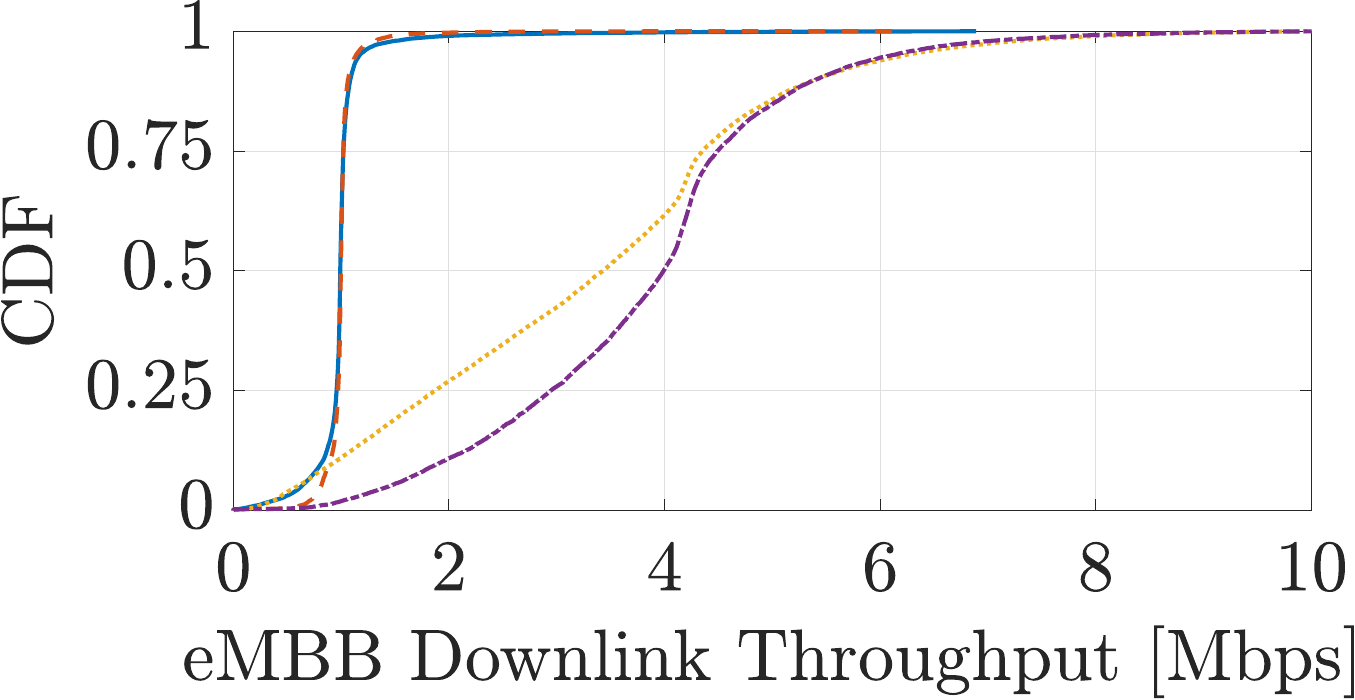}\label{Figure6a2}}
% \hfil
\subfloat[\gls{mmtc} Packets]{\includegraphics[height=4cm]{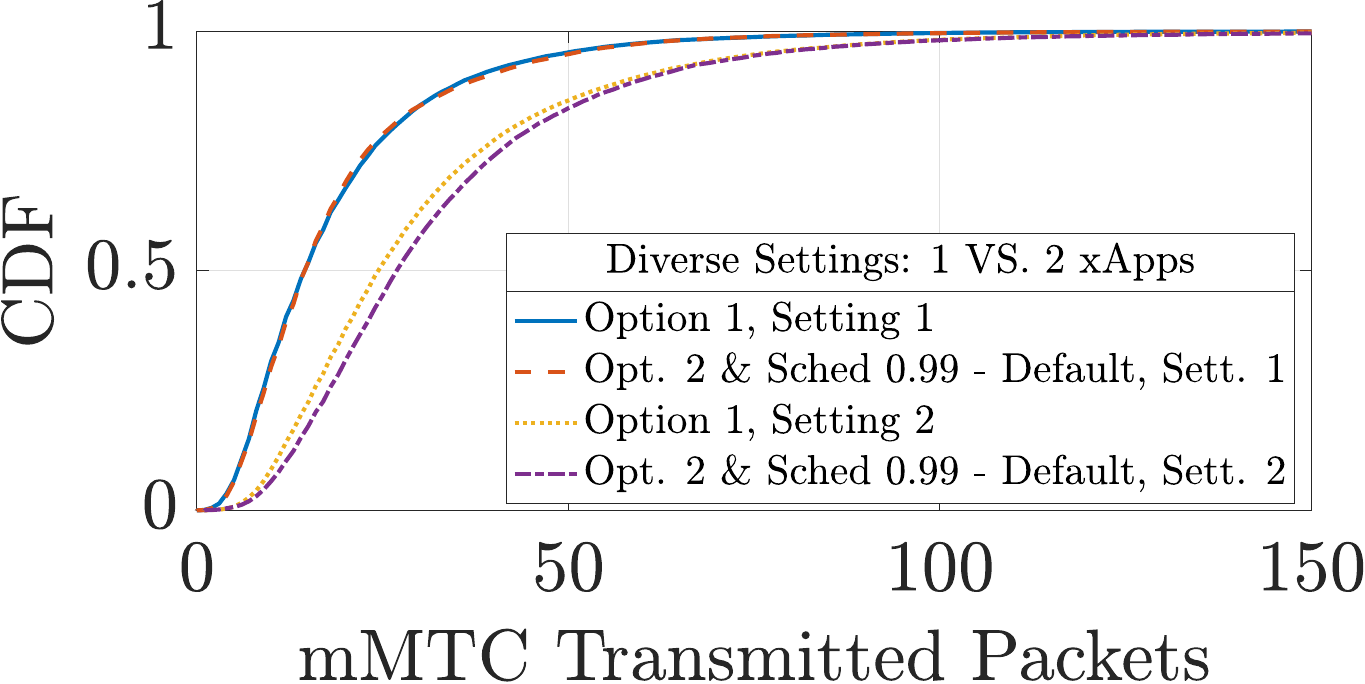}\label{Figure6b2}}
\hfil
\subfloat[\gls{urllc} Buffer Occupancy]{\includegraphics[height=4cm]{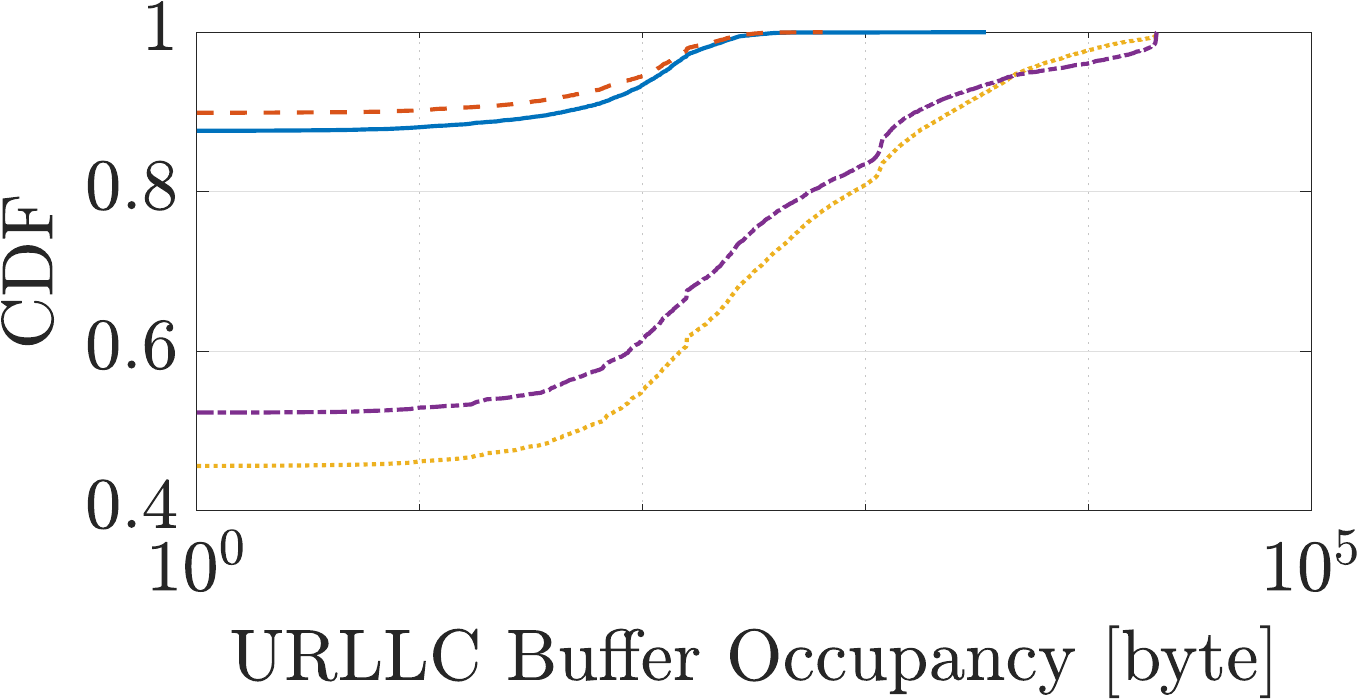}\label{Figure6c2}}
\caption{Performance evaluation under mobility with Settings~$1$ and~$2$ of Table~\ref{table:net-settings} and the PPO DRL Architecture.}
\label{Figure6-2a}
\end{figure}

\begin{figure}[H]
\centering
\subfloat[\gls{embb} Throughput]{\includegraphics[width=3in]{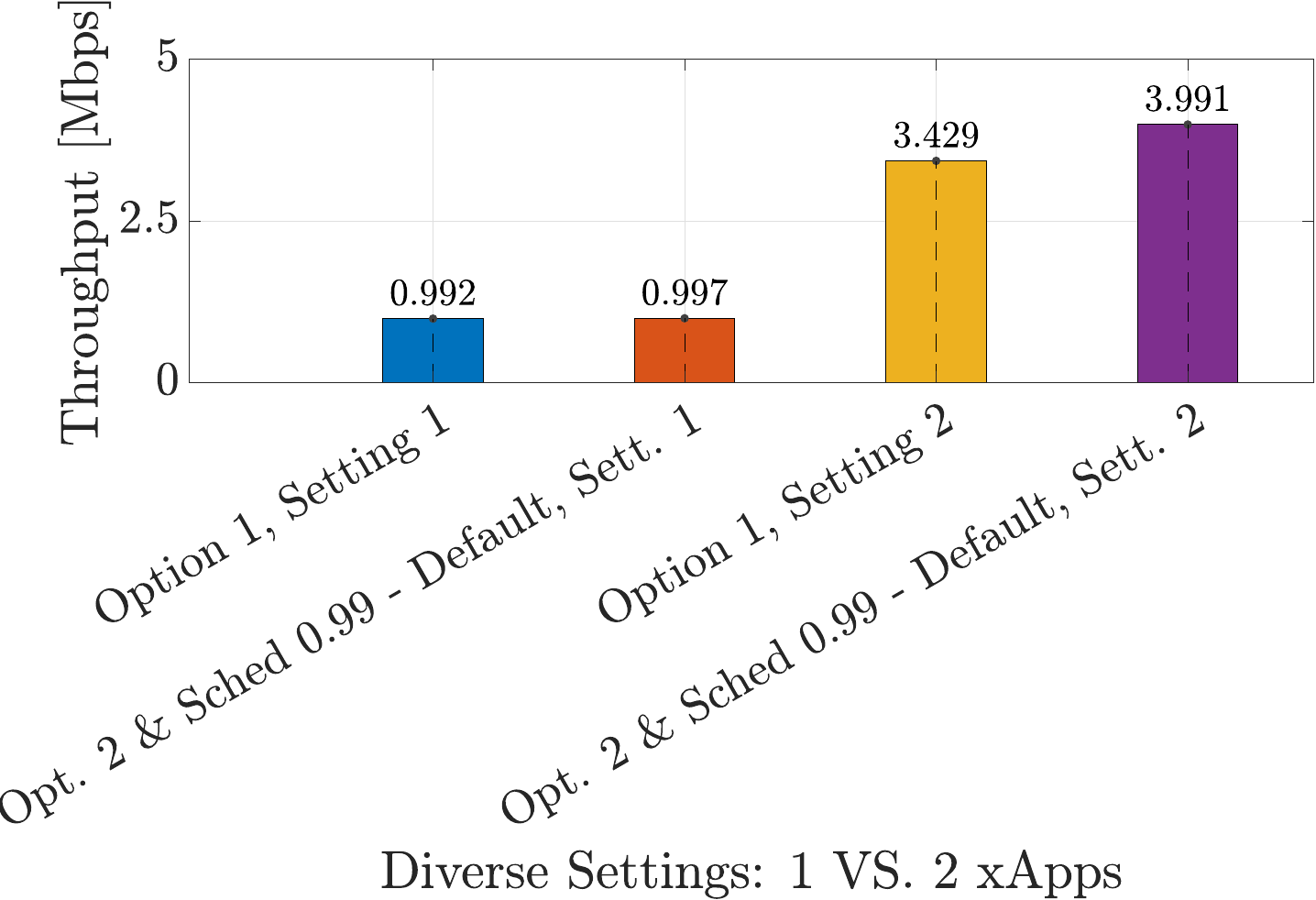}\label{mob_barplot_troughput}}
\hfil
\subfloat[\gls{mmtc} Packets]{\includegraphics[width=3in]{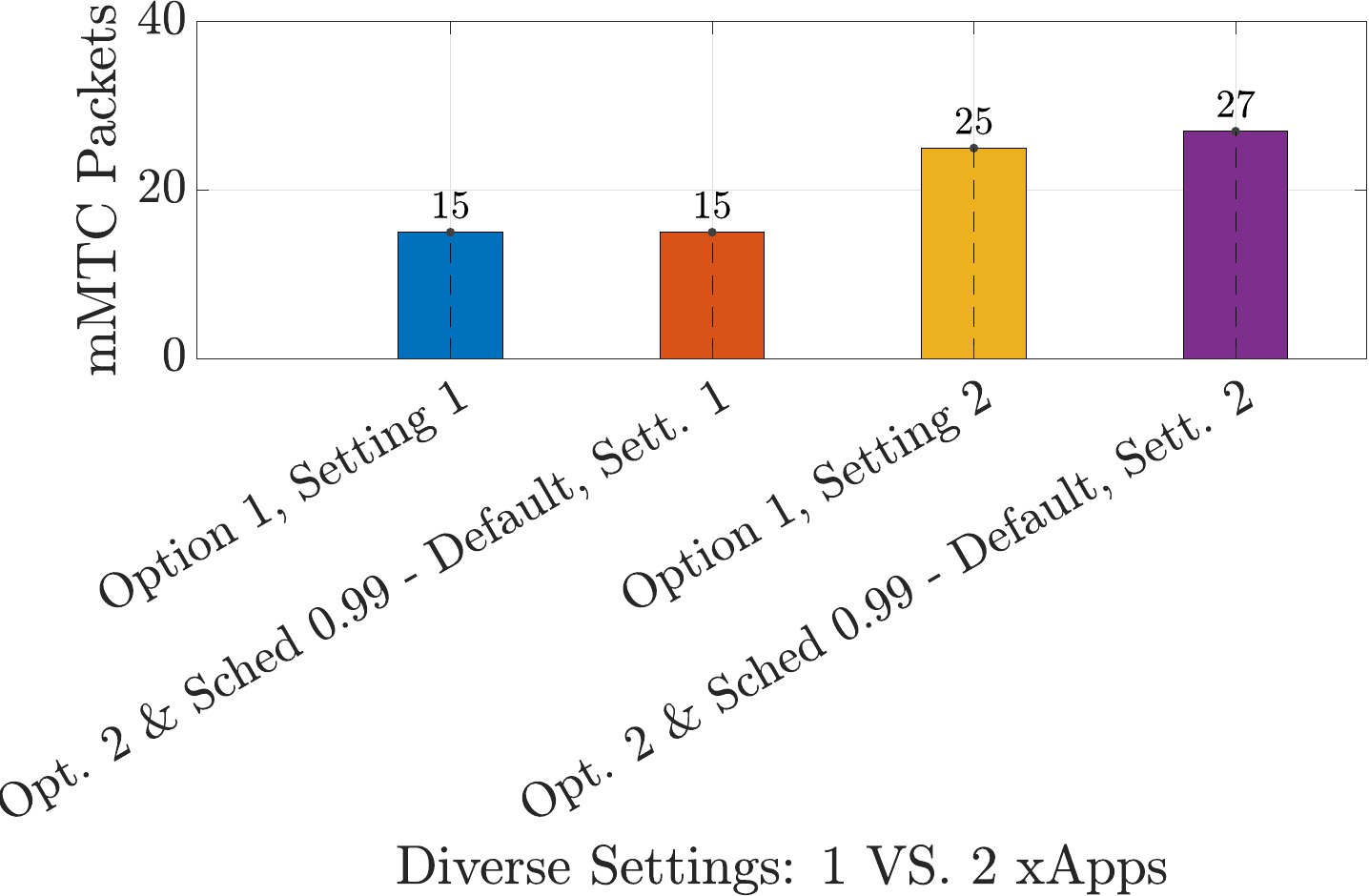}\label{mob_barplot_mmtc_packets}}
\caption{Median values obtained under mobility with Settings~$1$ and~$2$ of Table~\ref{table:net-settings} and the PPO DRL Architecture.}
\label{Figure12-1b}
\end{figure}

\begin{figure}[H]
\centering
\subfloat[\gls{embb} Throughput]{\includegraphics[height=4cm]{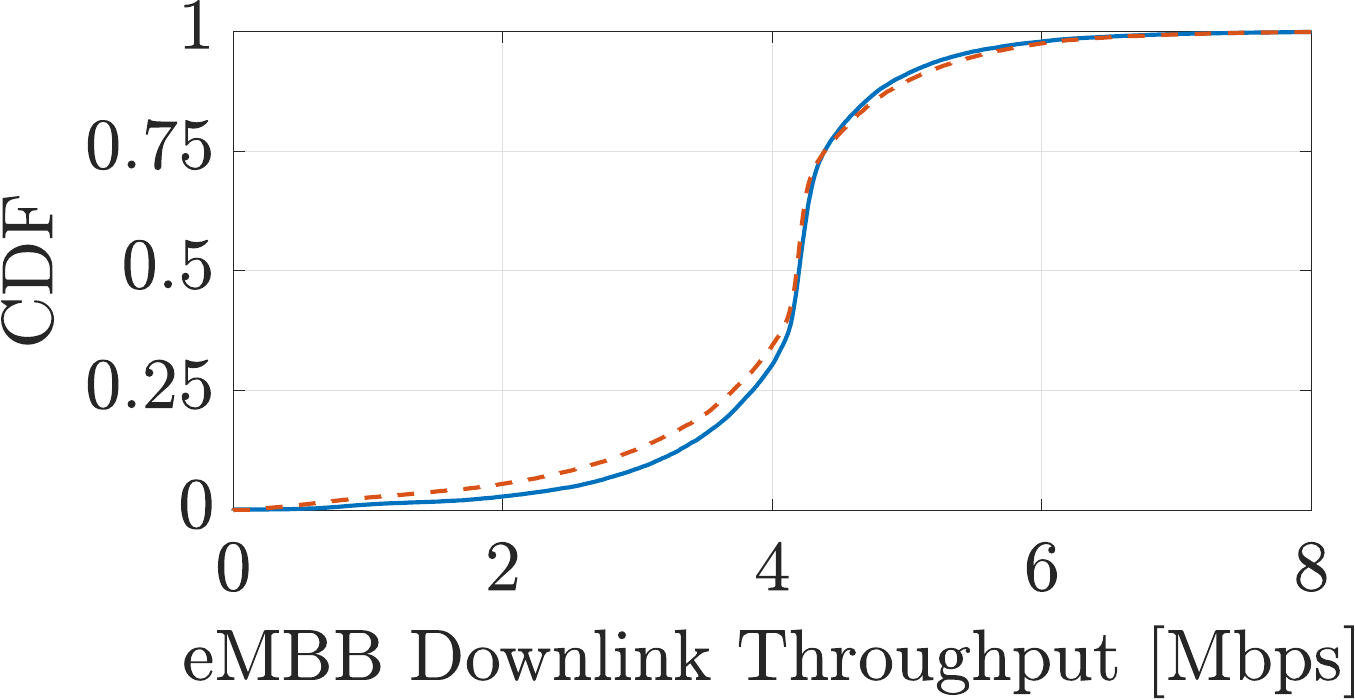}\label{comp2-embb-throughput}}
% \hfil
\subfloat[\gls{mmtc} Packets]{\includegraphics[height=4cm]{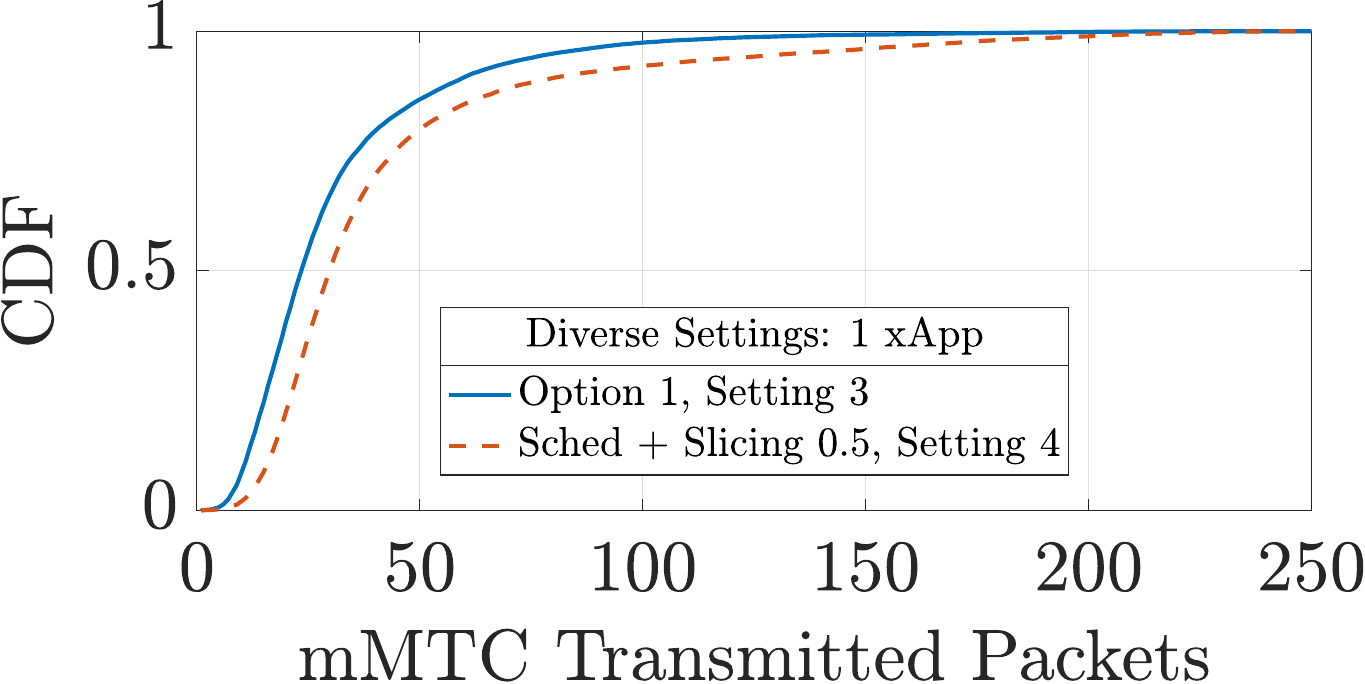}\label{comp2-tx-pkts-mmtc}}
\hfil
\subfloat[\gls{urllc} Buffer Occupancy]{\includegraphics[height=4cm]{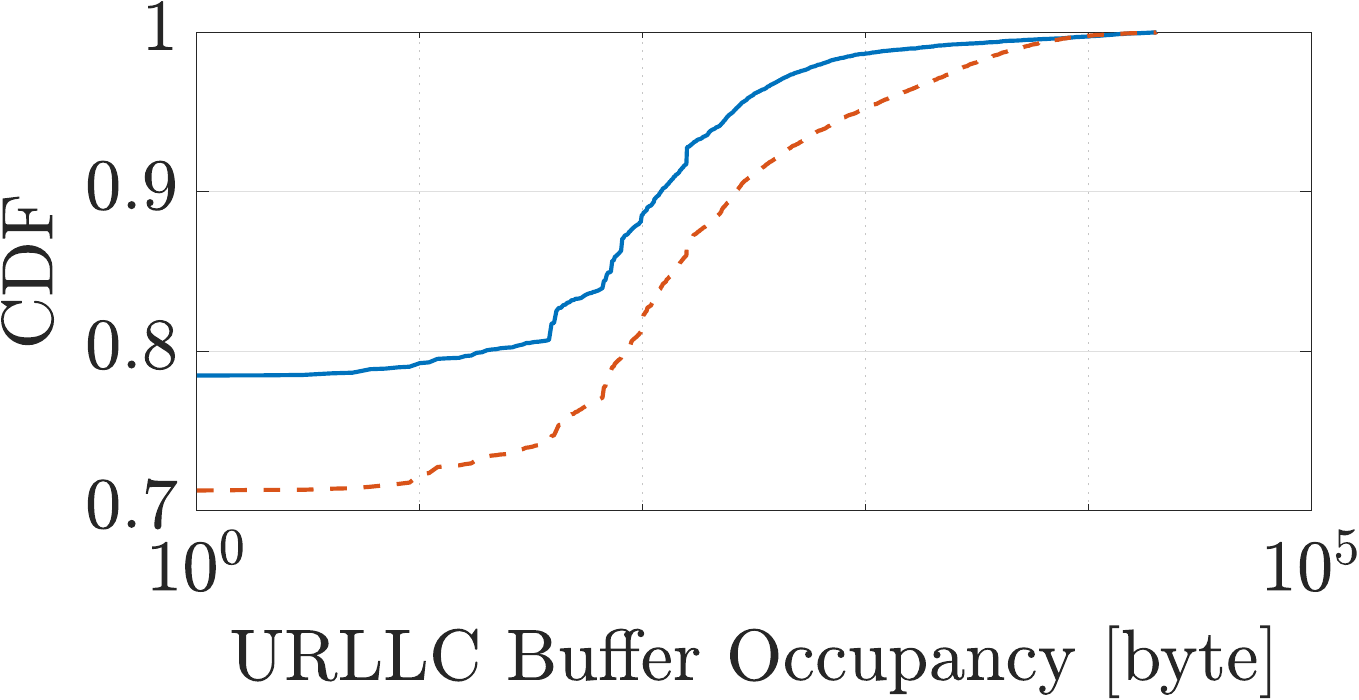}\label{Figure7c2}}
\caption{Performance evaluation focusing on the case of joint-slice optimization with a single xApp and the PPO DRL Architecture under Settings~$3$ and~$4$ of Table~\ref{table:net-settings}.}
\label{Figure7-2a}
\vspace{-0.5cm}
\end{figure}

\begin{figure}[htbp]
\centering
\subfloat[\gls{embb} Throughput]{\includegraphics[width=2.85in]{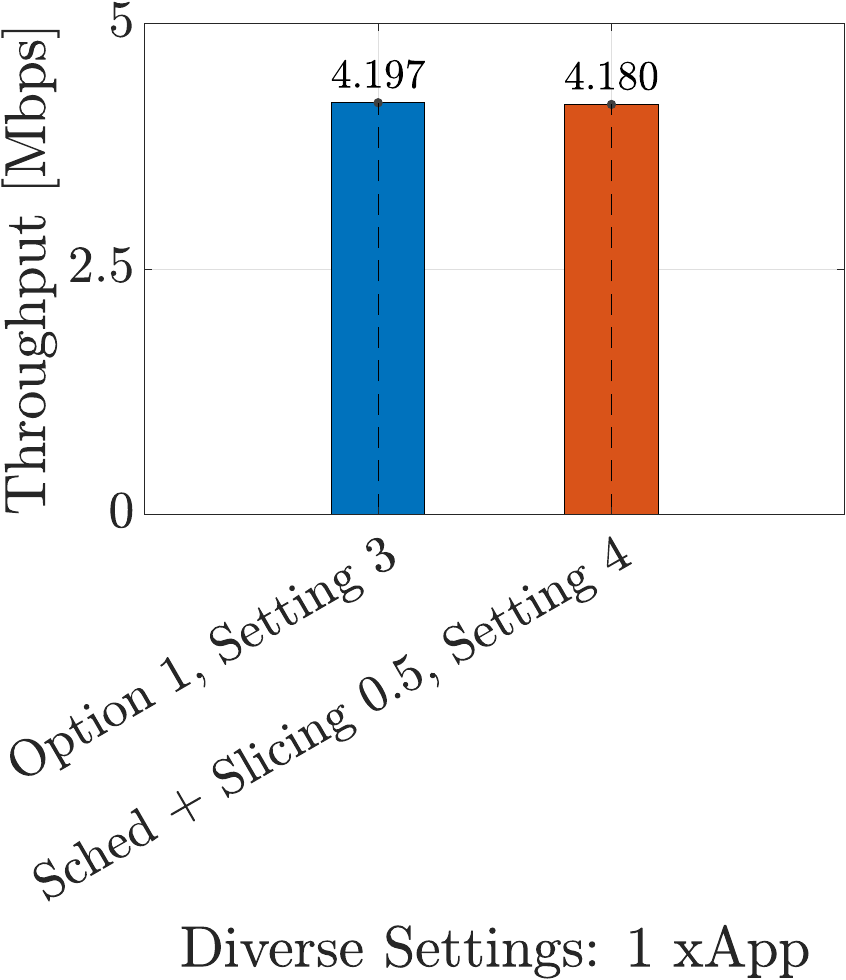}\label{comp2-embb-throughput_barplot}}
\hfil
\subfloat[\gls{mmtc} Packets]{\includegraphics[width=2.85in]{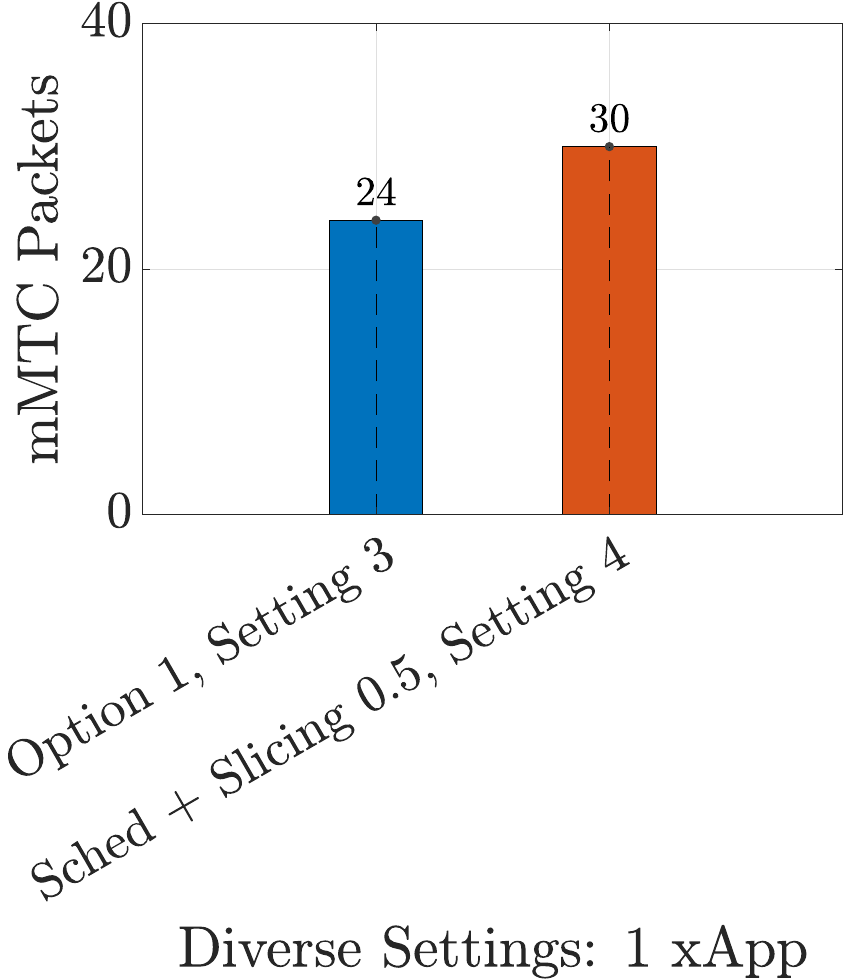}\label{comp2-urllc-bufsize_barplot}}
\caption{Median values obtained when focusing on the case of joint-slice optimization with a single xApp and the PPO DRL Architecture under Settings~$3$ and~$4$ of Table~\ref{table:net-settings}.}
\label{Figure11-1b}
\vspace{-0.55cm}
\end{figure}

In the Out-of-Sample experimental performance evaluation, we focus on the case of \textbf{Location 2}, and we test the framework's performance in static and mobile scenarios under two different traffic loads. Recall that our agents have been trained using data from \textbf{Location 1} only. Information regarding the traffic profiles and the \gls{ue} mobility can be found in Tables~\ref{table:traffic-profiles} and~\ref{table:net-settings}.

In Figs.~\ref{Figure6-2a} and~\ref{Figure12-1b}, we present the results collected when testing the framework under mobile \gls{ue} conditions and diverse traffic profiles for Settings~$1$ and $2$ as defined in Table~\ref{table:net-settings}. We juxtapose the configuration that delivered the highest \gls{embb} throughput (i.e., $4.208$\:Mbps), among those tested in this work, namely Option 1: \texttt{Sched \& Slicing 0.5 - Alternative} with the case of two xApps which jointly optimize the performance of all slices, namely Option 2: \texttt{Slicing 0.5}, and \texttt{Sched 0.99}. Finally, the set of granularities which are considered for this evaluation are given by Set~$1$ of Table~\ref{table:control-time}.

In Figs.~\ref{Figure6a2},~\ref{Figure6b2}, and~\ref{Figure6c2}, we observe that under the traffic conditions of Setting~$1$, both configurations deliver almost identical performance. This is clearly depicted in Figs.~\ref{mob_barplot_troughput} and~\ref{mob_barplot_mmtc_packets}, where the mentioned xApps achieve the same median values for \gls{embb} throughput and number of \gls{mmtc} transmitted packets. However, when the two configurations are tested under Setting~$2$ and specifically under Traffic Profile $2$, a slight performance degradation is observed in the case of joint-slice optimization with $1$~xApp. In detail, Figs.~\ref{Figure6a2} and~\ref{mob_barplot_troughput} indicate that the performance of \gls{embb} is optimized when using $2$~xApps, resulting in $\sim16$\% more throughput compared to the case with a single xApp. In the \gls{mmtc} slice, both configurations deliver similar performance, as shown in Fig.~\ref{Figure6b2}. However, when focusing on the former setup, it results in delivering two more packets as shown in Fig.~\ref{mob_barplot_mmtc_packets}. On the \gls{urllc}, it is noted that most of the configurations under both Settings resulted in a median buffer occupancy of $0$~byte. However, in the case of joint optimization with $1$~xApp, this value was reported at $52$~byte. Hence, in case of \gls{ue} mobility, \texttt{Sched \& Slicing 0.5 - Alternative} underperforms compared to joint optimization with \texttt{Slicing 0.5}, and \texttt{Sched 0.99}. Based on the reported findings, the simultaneous operation of two xApps can more successfully tackle possible \gls{ue} disconnections caused by mobile conditions, and therefore enhance network performance.

In Figs.~\ref{Figure7-2a} and~\ref{Figure11-1b} we include the results collected when testing the optimization framework under Settings~$3$ and $4$ in \textbf{Location 2}. In this evaluation, we once again compare Option 1: \texttt{Sched \& Slicing 0.5 - Alternative} with \texttt{Sched \& Slicing 0.5 - Default}. It is noted that the main difference between these two setups pertains to the fact that the former is evaluated under granularity Set~$1$, while the latter under Set~$2$. This comparison allows us to assess how collecting and reporting \glspl{kpm} as well as updating actions at smaller timescales (i.e., $250$\:ms) will impact the effectiveness of the optimization framework, under the traffic load defined in Profile~$2$. Additionally, the latter xApp is chosen due to the fact that, among all configurations evaluated in Section~\ref{sec:oran_edge:pandora:timers} in terms of \gls{ran} control timers, it has managed to provide the most significant performance boost in the network. In detail, when tested under Set~$2$, it has improved the performance of \gls{mmtc} by $\sim 56$\%, achieving a median value of $25$~packets, as shown in Fig.~\ref{ran-timers-bar-mmtc}. This is in contrast to when it was evaluated under Set~$1$, where it delivered a median value of $16$~packets (Fig.~\ref{fig:Figure4e}). With regard to the \gls{embb} slice, it still manages to maintain good overall performance, delivering a median throughput value of $4.140$\:Mbps (Fig.~\ref{ran-timers-bar-throughput}).

The results reported in Figs.~\ref{comp2-embb-throughput} and~\ref{comp2-embb-throughput_barplot} indicate that both configurations deliver similar performance in the \gls{embb} slice, with Option 1: \texttt{Sched \& Slicing 0.5 - Alternative} performing slightly better. However, on the \gls{mmtc} slice, \texttt{Sched \& Slicing 0.5 - Default} performs better, as depicted in Fig.~\ref{comp2-tx-pkts-mmtc}, by achieving a median value of $30$ transmitted packets (Fig.~\ref{comp2-urllc-bufsize_barplot}). \gls{urllc} \glspl{ue}, once again, achieve zero latency (i.e., buffers are emptied), and, as a result, the respective figures are omitted.

\subsection{Broader Evaluation of \texorpdfstring{\pandora}{PandORA}}\label{sec:oran_edge:pandora:broader}

\begin{figure}[H]
\centering
\subfloat[\gls{embb} \gls{prb} Ratio]{\includegraphics[height=4cm]{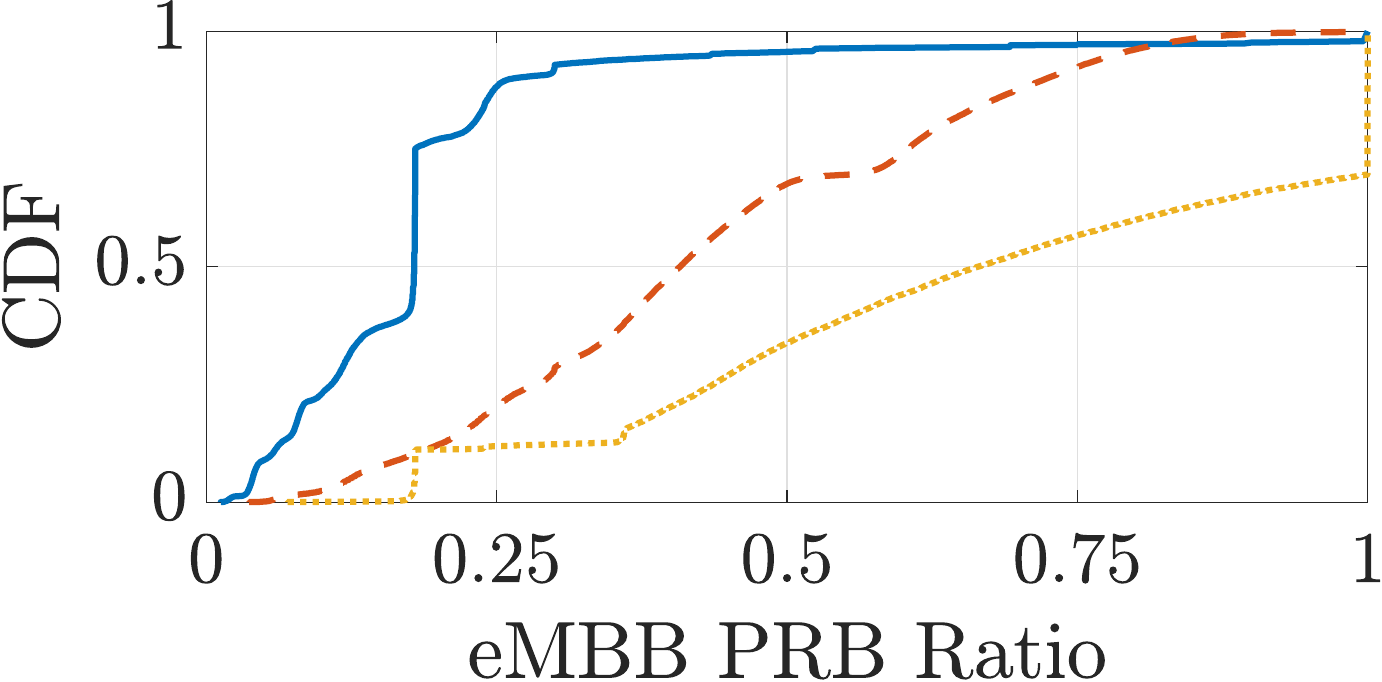}\label{fig:Figure-prbratio-1}}
% \hfil
\subfloat[\gls{mmtc} \gls{prb} Ratio]{\includegraphics[height=4cm]{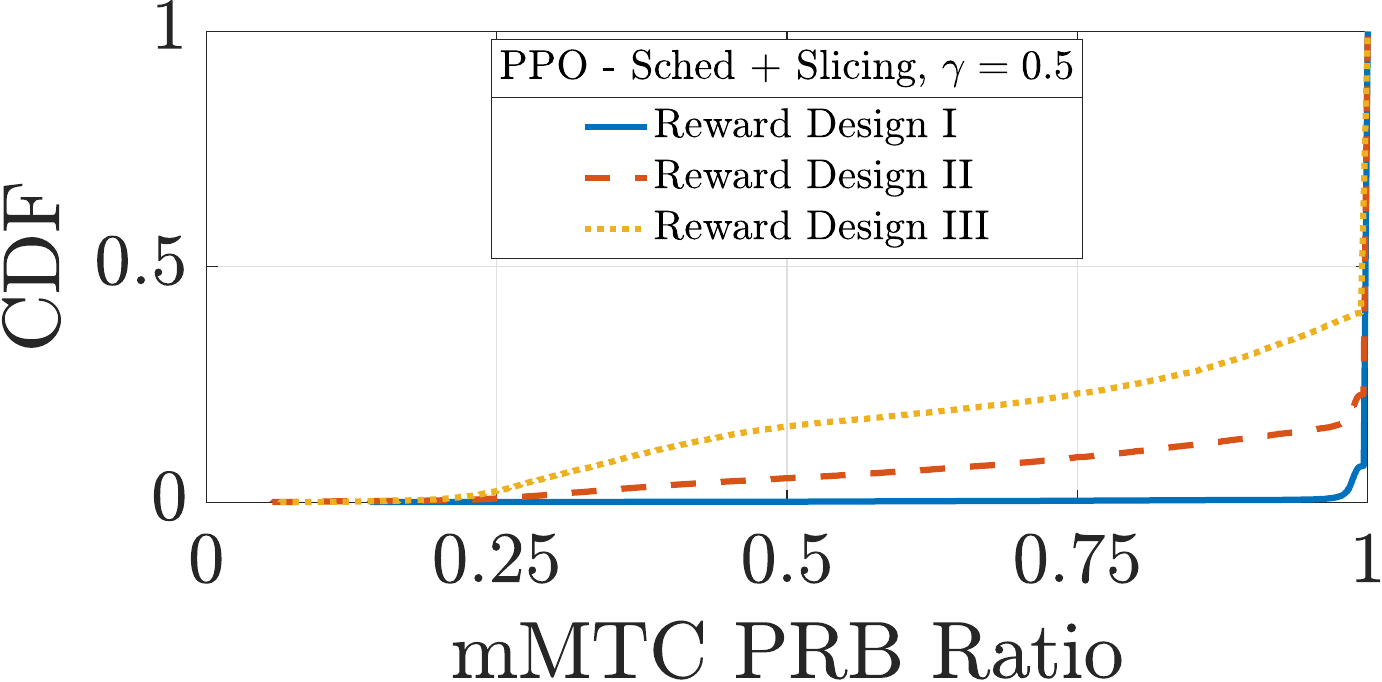}\label{fig:Figure-prbratio-2}}
\hfil
\subfloat[\gls{urllc} \gls{prb} Ratio]{\includegraphics[height=4cm]{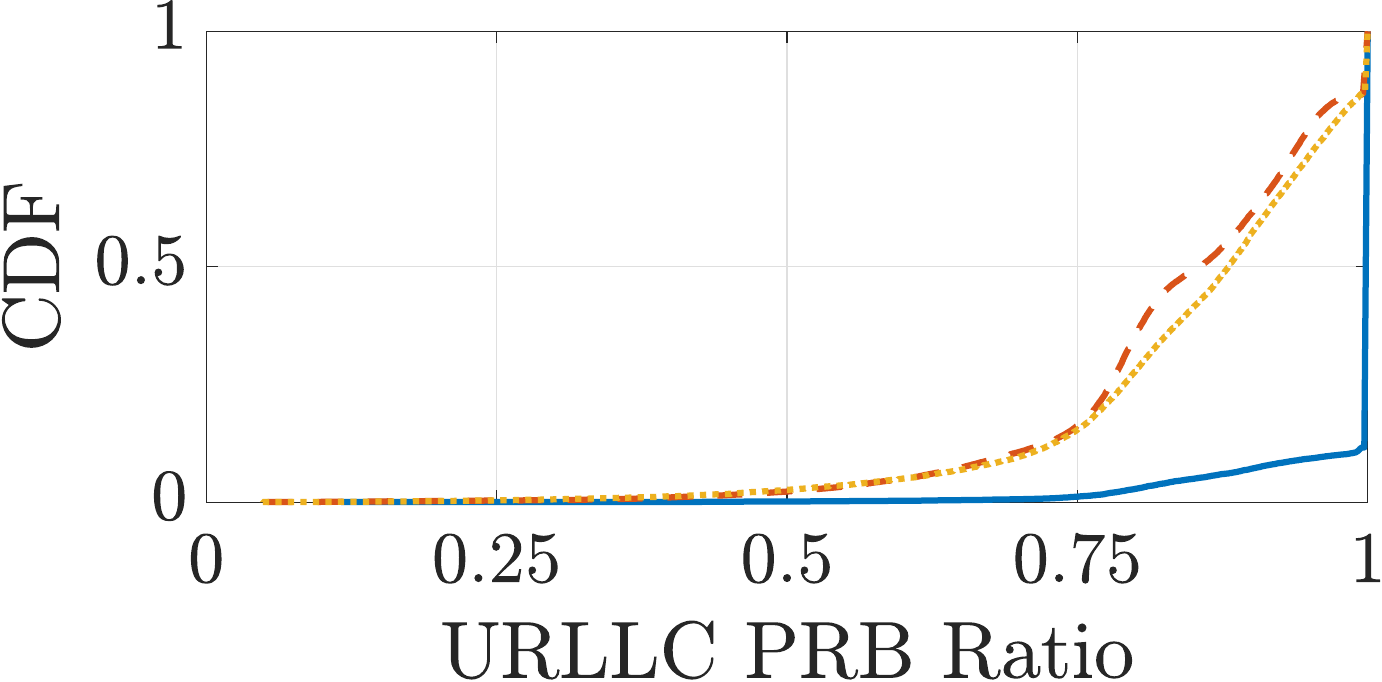}\label{fig:Figure-prbratio-3}}
\caption{\gls{ue} Satisfaction expressed in the form of PRB ratio for the Reward Designs of Table~\ref{table:reward-design}.}
\label{Figure-prbratio-reward}
\end{figure}

\begin{figure}[htbp]
\centering
\subfloat[\gls{embb} Throughput]{\includegraphics[height=4cm]{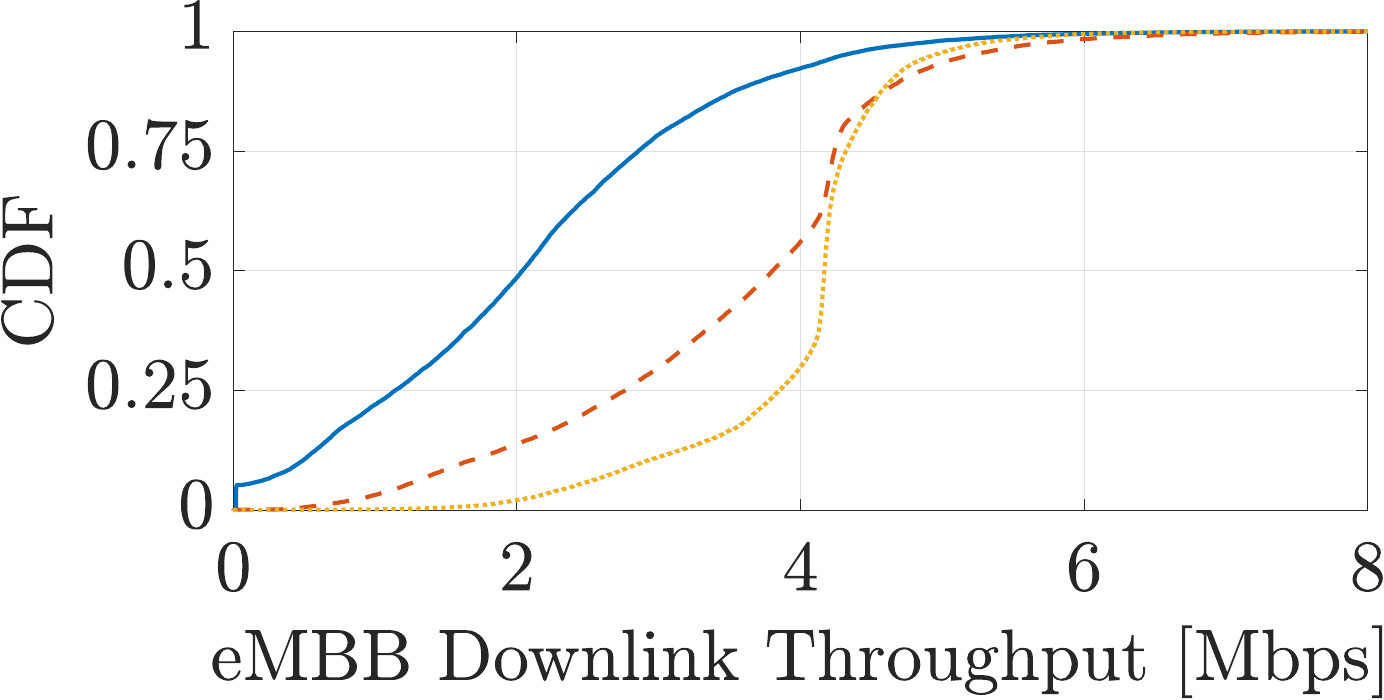}\label{fig:Figure-prbratio-1-embb}}
% \hfil
\subfloat[\gls{mmtc} Packets]{\includegraphics[height=4cm]{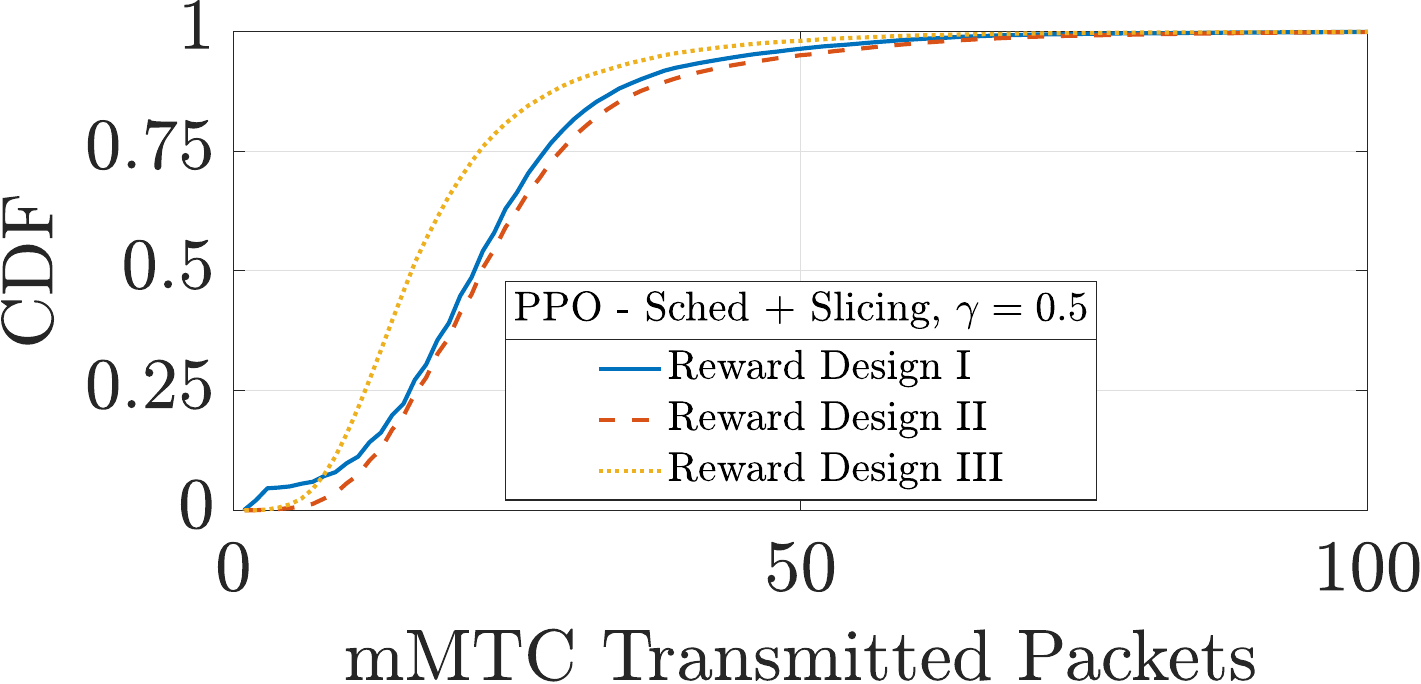}\label{fig:Figure-prbratio-2-mmtc}}
\hfil
\subfloat[\gls{urllc} Buffer Occupancy]{\includegraphics[height=4cm]{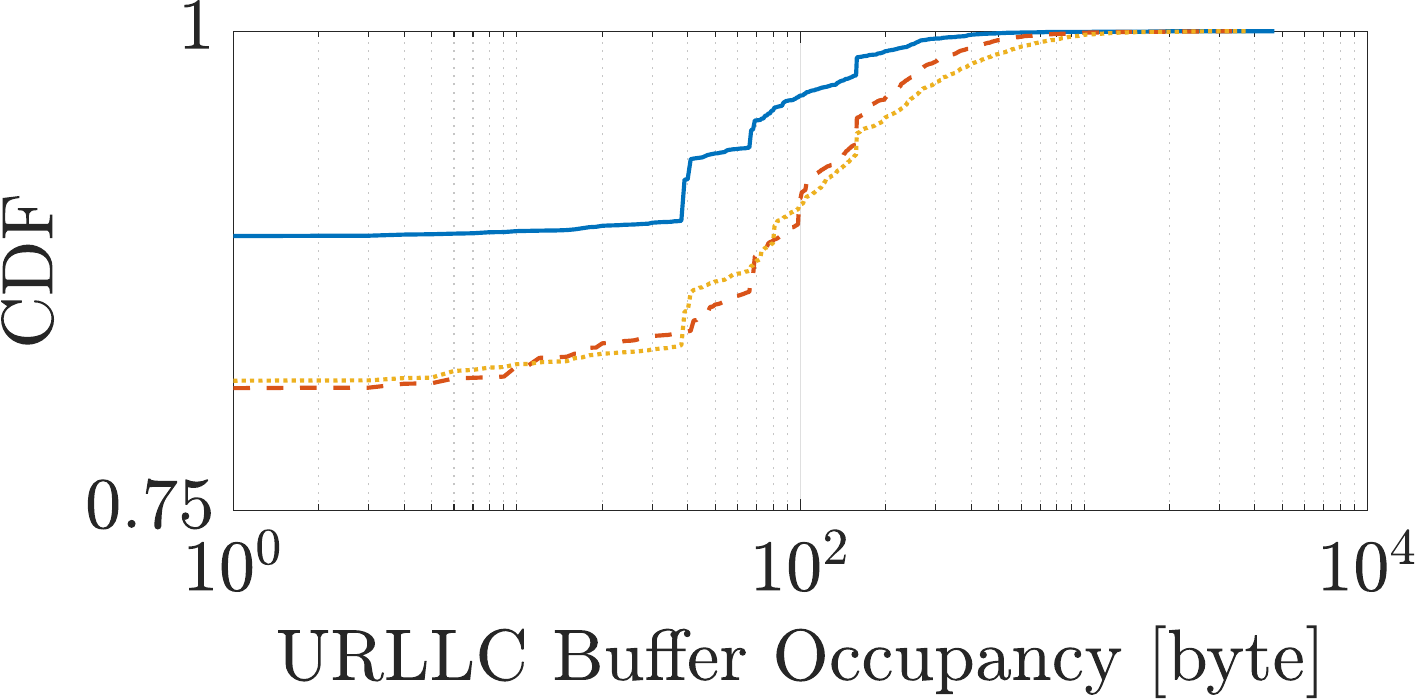}\label{fig:Figure-prbratio-3-urllc}}
\caption{Impact of the Reward Designs from Table~\ref{table:reward-design} on \gls{dl} \gls{embb} Throughput, \gls{dl} \gls{mmtc} and \gls{tx} Packets.}
\label{Figure-prbratio-reward-cdfs}
\end{figure}

\begin{figure}[htbp]
\centering
\subfloat[\gls{embb} DL Throughput]{\includegraphics[width=2.85in]{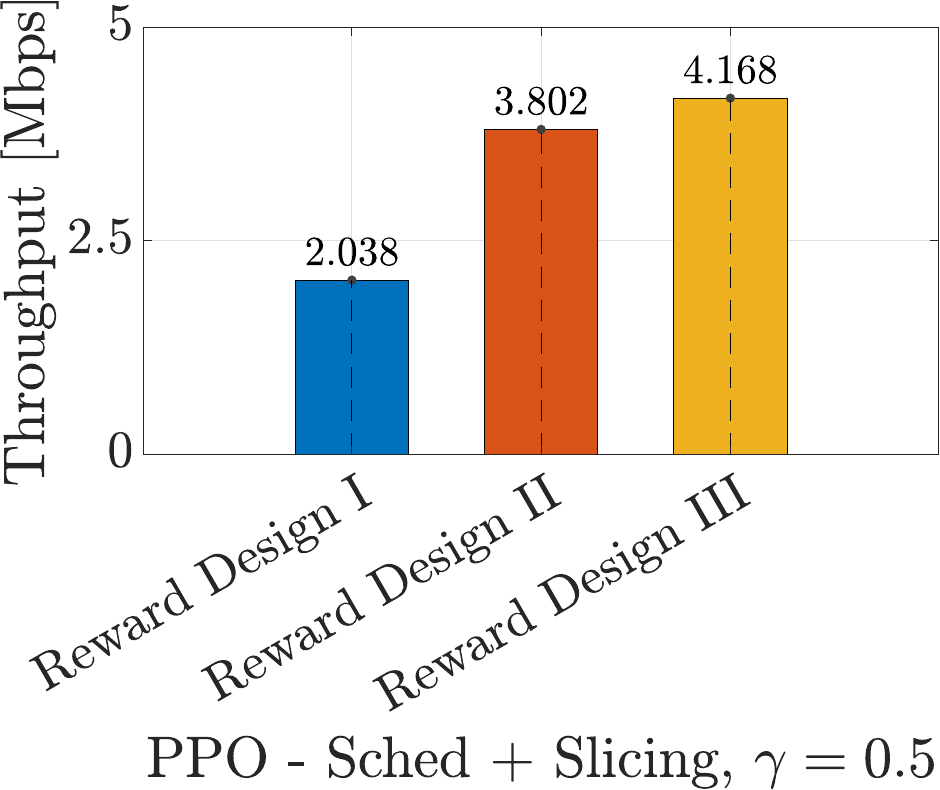}\label{fig:ebarplot_prbratiReward_embb}}
\hfil
\subfloat[\gls{mmtc} Packets]{\includegraphics[width=2.85in]{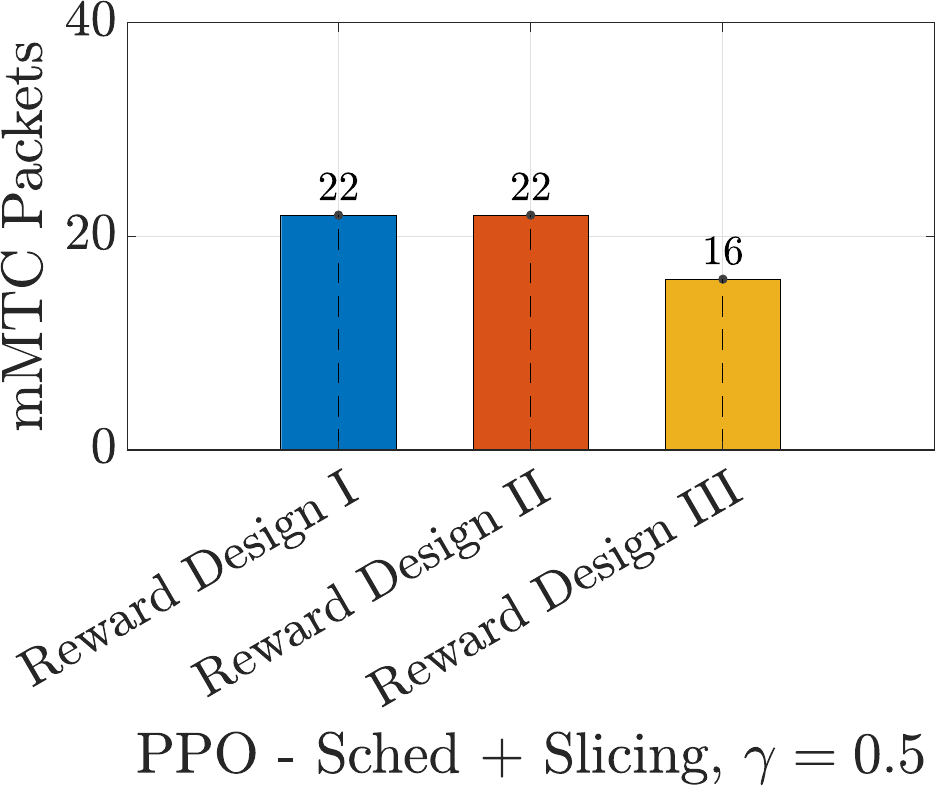}\label{fig:barplot_prbratiReward_mmtc}}
\caption{Impact of the Reward Designs from Table~\ref{table:reward-design} on the median values of \gls{dl} \gls{embb} throughput, \gls{dl} \gls{mmtc} \gls{tx} Packets, and \gls{dl} \gls{urllc} buffer occupancy.}
\label{barplot_prbratiReward}
\end{figure}

The goal of this section is to provide additional insights into \pandora by: (i) considering a broader set of evaluation metrics aimed at shedding light on resource utilization and \gls{ue} satisfaction; and (ii) testing across scenarios with a larger number of \glspl{ue}. Specifically, in the latter case, we examine the performance of our xApps in a scenario where we add additional \glspl{ue} to the \gls{mmtc} slice. In line with the previous sections, we will evaluate \pandora in setups previously seen during the training process (i.e., in-sample experimental evaluation in Location~$1$), as well as unseen conditions (i.e., out-of-sample experimental evaluation in Location~$2$). All subsequent results concern joint optimization with \texttt{Sched \& Slicing 0.5} and are tested using Sets $1$ and $2$ of \gls{ran} control timers from Table~\ref{table:control-time} and the Traffic Profiles listed in Table~\ref{table:traffic-profiles}, both in static and mobile scenarios (e.g., $3$~m/s).

It is noted that so far we have considered three main reward functions for the respective slices (i.e., maximization of \gls{dl} \gls{embb} throughput and \gls{mmtc} \gls{tx} Packets, and minimization of \gls{dl} buffer occupancy in the \gls{urllc}). In order to demonstrate \pandora's capabilities as well as to compare the reward functions evaluated so far, we additionally craft two new reward designs, which are defined in Table~\ref{table:reward-design}. In \emph{Reward Design I}, all slices are given the same priority (i.e., they are all assigned a weight of $1$). The reward for all slices is the maximization of the \gls{prb} Ratio, defined as $\text{\gls{prb} Ratio}=\frac{\text{Sum of Granted \glspl{prb}}}{\text{Sum of Requested \glspl{prb}}}, \text{where \gls{prb} Ratio} \in [0, 1]$, that represents the amount of the allocated \glspl{prb} per slice. For \emph{Reward Design II}, we consider the maximization of the \gls{embb} throughput and the minimization of the \gls{urllc} buffer occupancy, with their weights defined in Table~\ref{table:weight-confs-list} under the \texttt{Default} weight configuration. For the \gls{mmtc} slice, we consider the maximization of the \gls{prb} Ratio and the slice's weight is set to $0.5$. Lastly, \emph{Reward Design III} pertains to the joint scheduling and slicing optimization discussed and evaluated in Section~\ref{sec:oran_edge:pandora:discount} and shown in Table~\ref{table:reward-design}. The results of this experimental evaluation are illustrated in Figs.~\ref{Figure-prbratio-reward},~\ref{Figure-prbratio-reward-cdfs},~\ref{barplot_prbratiReward} and were collected under Traffic Profile~$2$ (i.e., $4$~Mbps \gls{embb} throughput, $44.6$~kbps \gls{mmtc} throughput and $89.3$~kbps \gls{urllc} throughput), in Location~$1$ (i.e., $50$~m radius from the \gls{bs}), and Set~$1$ of \gls{ran} control timers (see Table~\ref{table:control-time}), for a total of $6$ \glspl{ue}, uniformly distributed across the slices.

\begin{table}[htbp]
\centering
\small
\caption{\gls{drl} Reward Design Catalog for \texttt{Sched \& Slicing 0.5} with \gls{ppo} in Location~$1$ and Set~$1$ of \gls{ran} control timers.}
\begin{tabular}{c|ccc}
\toprule
\textbf{Reward Design ID} & \textbf{eMBB} & \textbf{mMTC} & \textbf{URLLC} \\
\midrule
\textbf{I} & \shortstack{$1$ \\ Max. \gls{prb} Ratio} & \shortstack{$1$ \\ Max. \gls{prb} Ratio} & \shortstack{$1$ \\ Max. \gls{prb} Ratio} \\
\midrule
\textbf{II} & \shortstack{$72.0440333$ \\ Max. \gls{dl} Throughput} & \shortstack{$0.5$ \\ Max. \gls{prb} Ratio} & \shortstack{$0.00005$ \\ Min. \gls{dl} Buffer Occupancy} \\
\midrule
\textbf{III} & \shortstack{$72.0440333$ \\ Max. \gls{dl} Throughput} & \shortstack{$0.229357798$ \\ Max. \gls{dl} \gls{tx} Packets} & \shortstack{$0.00005$ \\ Min. \gls{dl} Buffer Occupancy} \\
\bottomrule
\end{tabular}
\label{table:reward-design}
\end{table}

In Fig.~\ref{Figure-prbratio-reward}, we show the \gls{prb} Ratio for the three Reward Designs defined in Table~\ref{table:reward-design}. We observe that with \emph{Reward Design I}, \gls{embb} \glspl{ue} experience lower levels of \emph{satisfaction} (Fig.~\ref{fig:Figure-prbratio-1}), in terms of the number of allocated \glspl{prb}. On the contrary, both on the \gls{mmtc} and \gls{urllc} slices (Figs.~\ref{fig:Figure-prbratio-2} and~\ref{fig:Figure-prbratio-3}), \emph{Reward Design I} results in significant \gls{ue} satisfaction. The latter is due to the fact that both \gls{mmtc} and \gls{urllc} have lower traffic requirements to satisfy (i.e., $44.6$~kbps and $89.3$~kbps throughput, respectively), compared to \gls{embb} (i.e., $4$~Mbps throughput). With \emph{Reward Design II}, we observe an improvement in \gls{ue} satisfaction levels on the \gls{embb}, meaning that \glspl{ue} are granted the resources they request, along with equally good performance in \gls{mmtc} and \gls{urllc}. The results with \emph{Reward Design III} indicate that this combination of weight configuration (i.e., tailored to prioritize one slice over the other) and slice-specific rewards effectively result in good overall performance without penalizing one slice over the other. Notably, \emph{Reward Design III} is the configuration under which \gls{embb} \glspl{ue} experience more satisfaction. Finally, we can observe the competitive behavior among \gls{embb} and \gls{mmtc}. Indeed, the reward designs that deliver a higher \gls{ue} satisfaction for \gls{embb} (Fig.~\ref{fig:Figure-prbratio-1}) also deliver lower satisfaction for \gls{mmtc} (Fig.~\ref{fig:Figure-prbratio-2}).

In Figs.~\ref{Figure-prbratio-reward-cdfs} and~\ref{barplot_prbratiReward}, we observe the impact of the Reward Designs explored in Fig.~\ref{Figure-prbratio-reward} on the three \glspl{kpi} of interest, namely \gls{dl} \gls{embb} throughput, \gls{mmtc} \gls{tx} packets, and \gls{urllc} buffer occupancy. In detail, Figs.~\ref{fig:Figure-prbratio-1},~\ref{fig:Figure-prbratio-1-embb} and~\ref{fig:ebarplot_prbratiReward_embb} demonstrate a similar trend, since a higher value of \gls{prb} Ratio yields higher throughput. A similar trend is observed in Figs.~\ref{fig:Figure-prbratio-2},~\ref{fig:Figure-prbratio-2-mmtc} and~\ref{fig:barplot_prbratiReward_mmtc} for the \gls{mmtc} slice, with \emph{Reward Designs I} \& \emph{II} demonstrating similar performance (i.e., a median value of $22$ \gls{tx} packets as illustrated in Fig.~\ref{fig:barplot_prbratiReward_mmtc}). These results indicate that the maximization of the \gls{prb} Ratio serves as ideal reward for the aforementioned slice but with a penalty on the \gls{embb}. Specifically, even though \emph{Reward Designs I} \& \emph{II} result in $\sim37.5\%$ improvement on the performance of the \gls{mmtc}, they result in lower \gls{embb} throughput values as depicted in Fig.~\ref{fig:ebarplot_prbratiReward_embb}. On the \gls{urllc}, all \emph{Reward Designs} result in empty buffers, with \emph{Reward Design I} slightly resulting in higher performance, as seen in Fig.~\ref{fig:Figure-prbratio-3-urllc}.

\begin{figure}[htbp]
\centering
\subfloat[\gls{embb} \glspl{prb}]{\includegraphics[height=4cm]{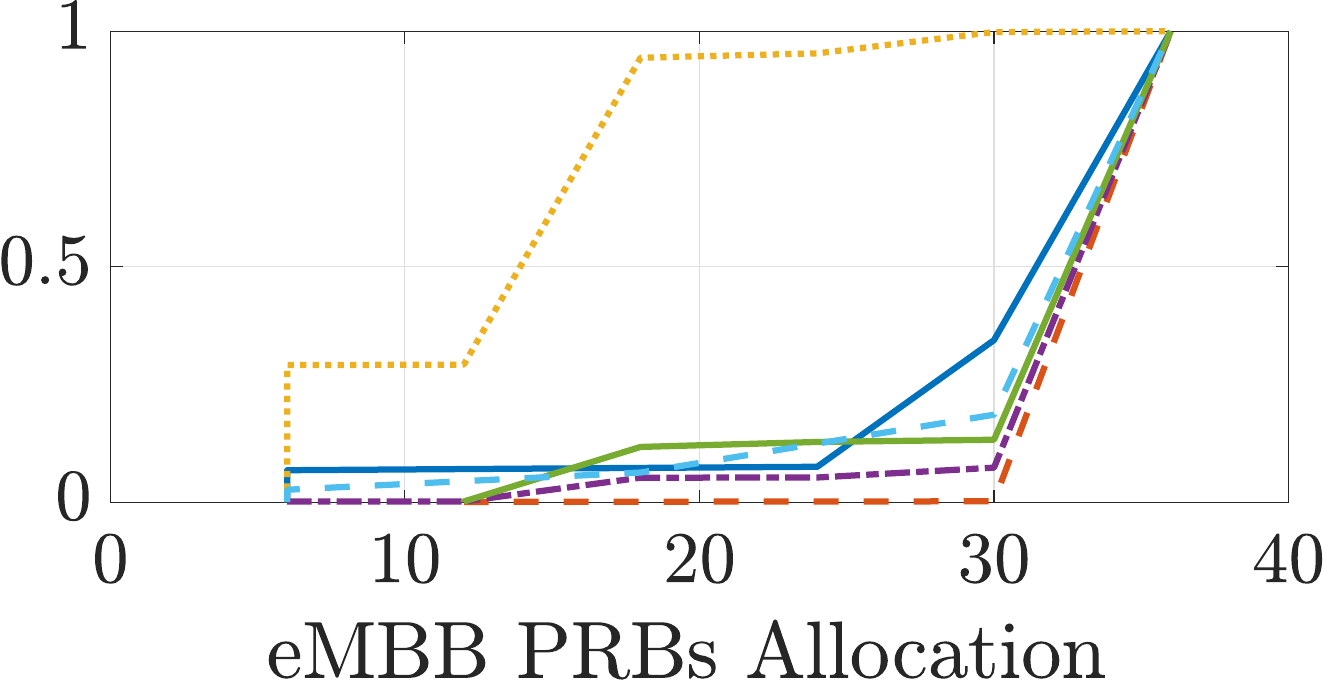}\label{fig:Figure-embb_prbs}}
% \hfil
\subfloat[\gls{mmtc} \glspl{prb}]{\includegraphics[height=4cm]{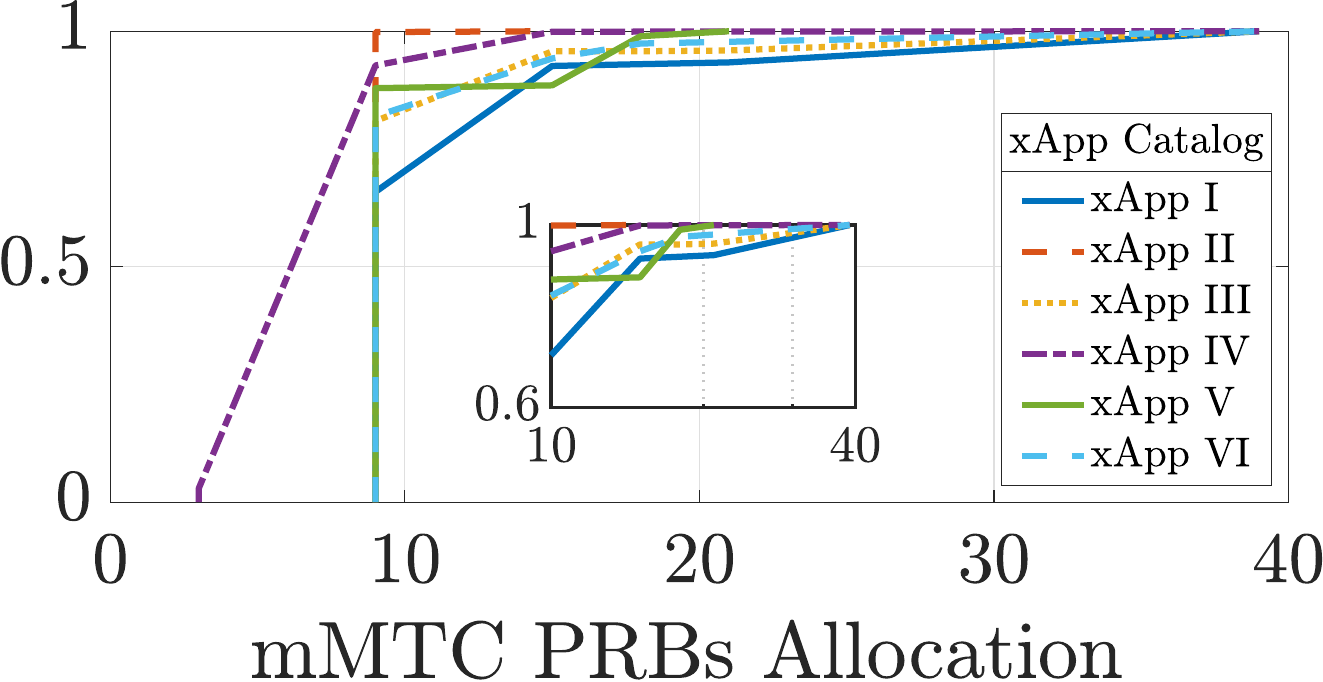}\label{fig:Figure-mmtc_prbs}}
\hfil
\subfloat[\gls{urllc} \glspl{prb}]{\includegraphics[height=4cm]{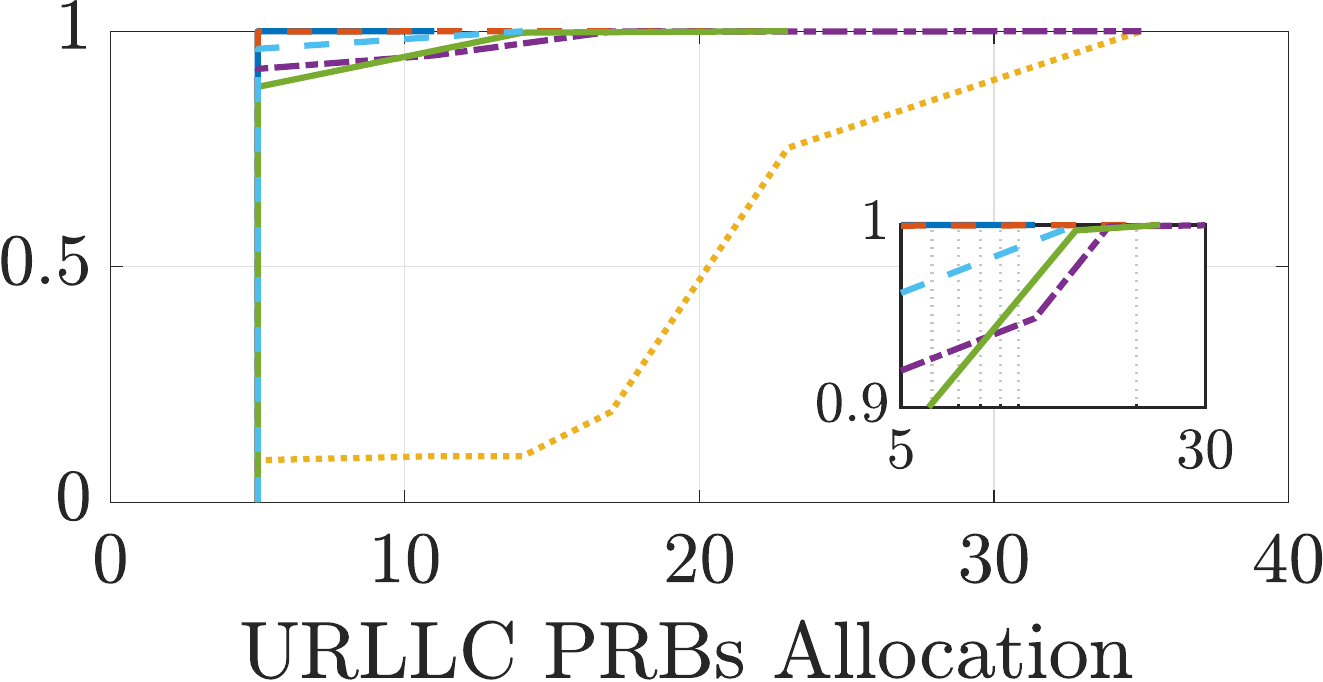}\label{fig:Figure-urllc_prbs}}
\caption{Selection of \gls{prb} Actions for the three slices from the xApp Catalog of Table~\ref{table:xapp-catalog-comp}.}
\label{Figure-cdfs_prbs}
\end{figure}

In Fig.~\ref{Figure-cdfs_prbs}, we include the distribution of actions in terms of selected \glspl{prb} for the xApp Catalog presented in Table~\ref{table:xapp-catalog-comp}. The performance evaluation results indicate that agents can make decisions resulting in diverse action distributions (summarized in Table~\ref{tab:feasible_prb_allocation}), which is due to varying rewards and design choices, impacting the performance achieved by the system.

We now consider the case where we increase the number of \glspl{ue} allocated to the \gls{mmtc} slice as illustrated by \emph{Use Cases II} \& \emph{III}. Specifically, \emph{Use Case II} involves a total of $7$ \glspl{ue}, while in \emph{Use Case III}, the setup comprises a total of $8$ \glspl{ue}, with $3$ and $4$ \glspl{ue} allocated to the \gls{mmtc} slice, respectively. \emph{Use Case I} corresponds to a setup of $6$ \glspl{ue} in total, and all deployments have been tested under Traffic Profile~$1$ of Table~\ref{table:traffic-profiles} and Set~$1$ of \gls{ran} control timers of Table~\ref{table:control-time} in Location~$2$ (i.e., $20$~m radius from the \gls{bs}). Finally, the xApp under evaluation corresponds to \emph{Reward Design III} of Table~\ref{table:reward-design}.

\begin{figure}[htbp]
\centering
\subfloat[\gls{embb} \gls{prb} Ratio]{\includegraphics[height=4cm]{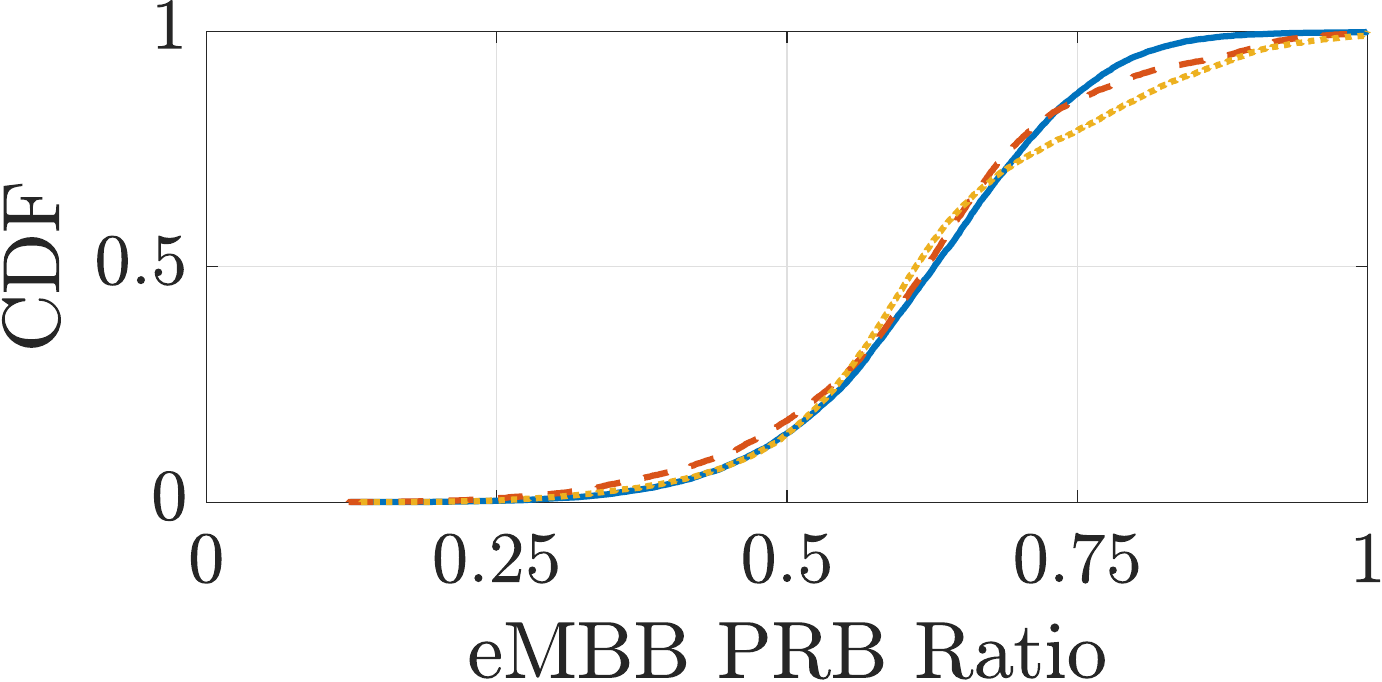}\label{fig:Figure-prbratio-1-embb_moreues}}
% \hfil
\subfloat[\gls{mmtc} \gls{prb} Ratio]{\includegraphics[height=4cm]{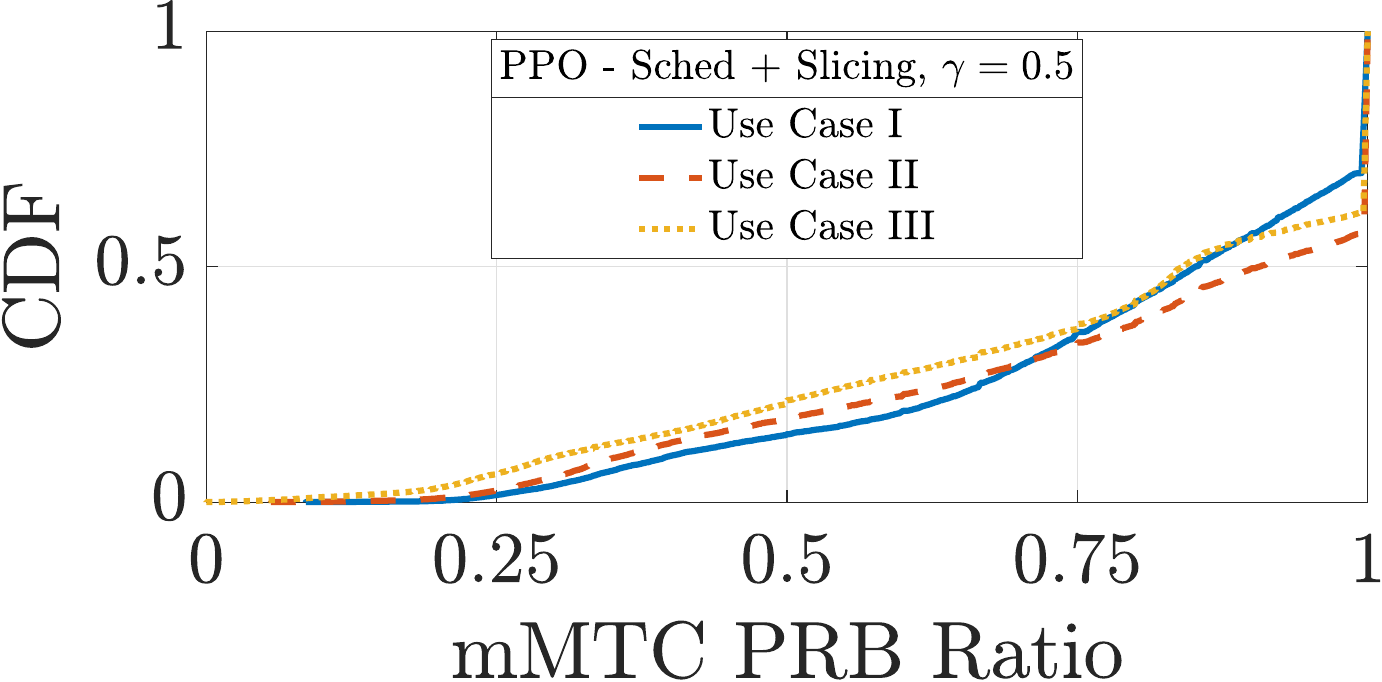}\label{fig:Figure-prbratio-2-mmtc_moreues}}
\hfil
\subfloat[\gls{urllc} \gls{prb} Ratio]{\includegraphics[height=4cm]{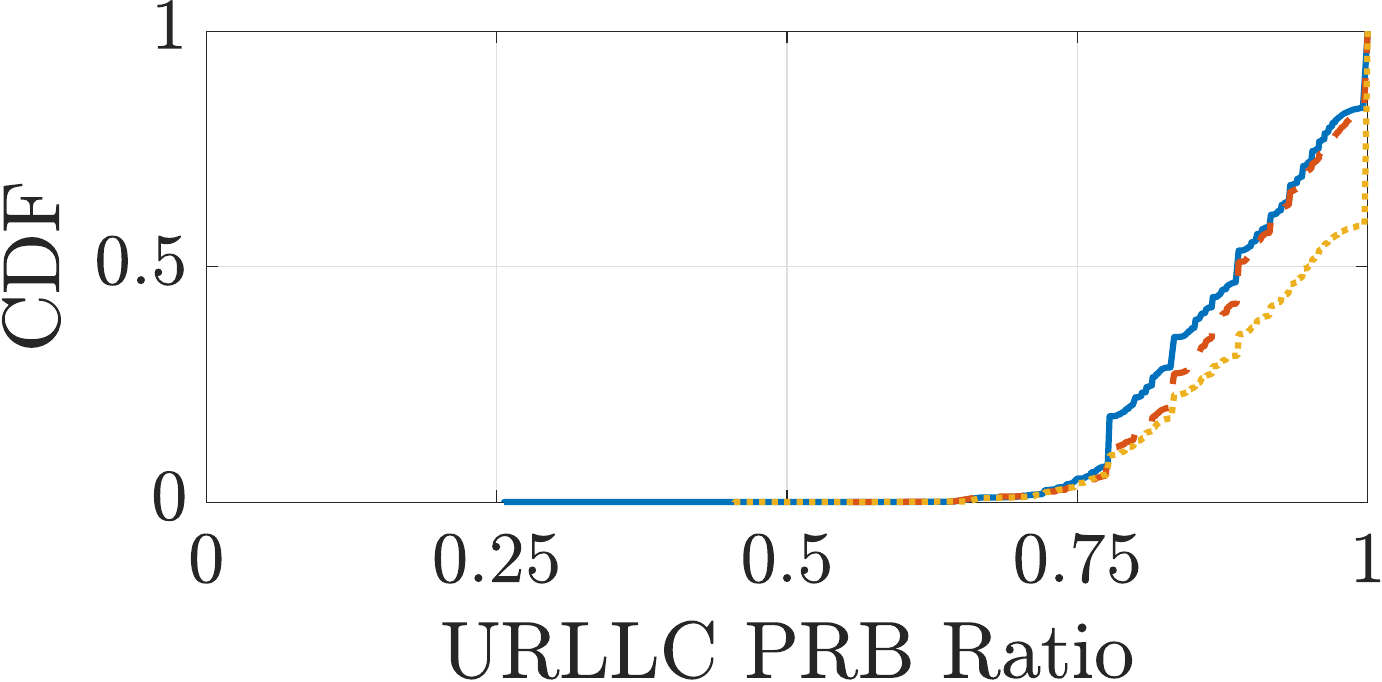}\label{fig:Figure-prbratio-3-urllc_moreues}}
\caption{Resource Utilization and \gls{ue} Satisfaction for a network deployment described by three different Use Cases.}
\label{Figure-prbratio-reward-cdfs_moreues}
\end{figure}

\begin{figure}[htbp]
\centering
\subfloat[\gls{embb} Throughput]{\includegraphics[height=4cm]{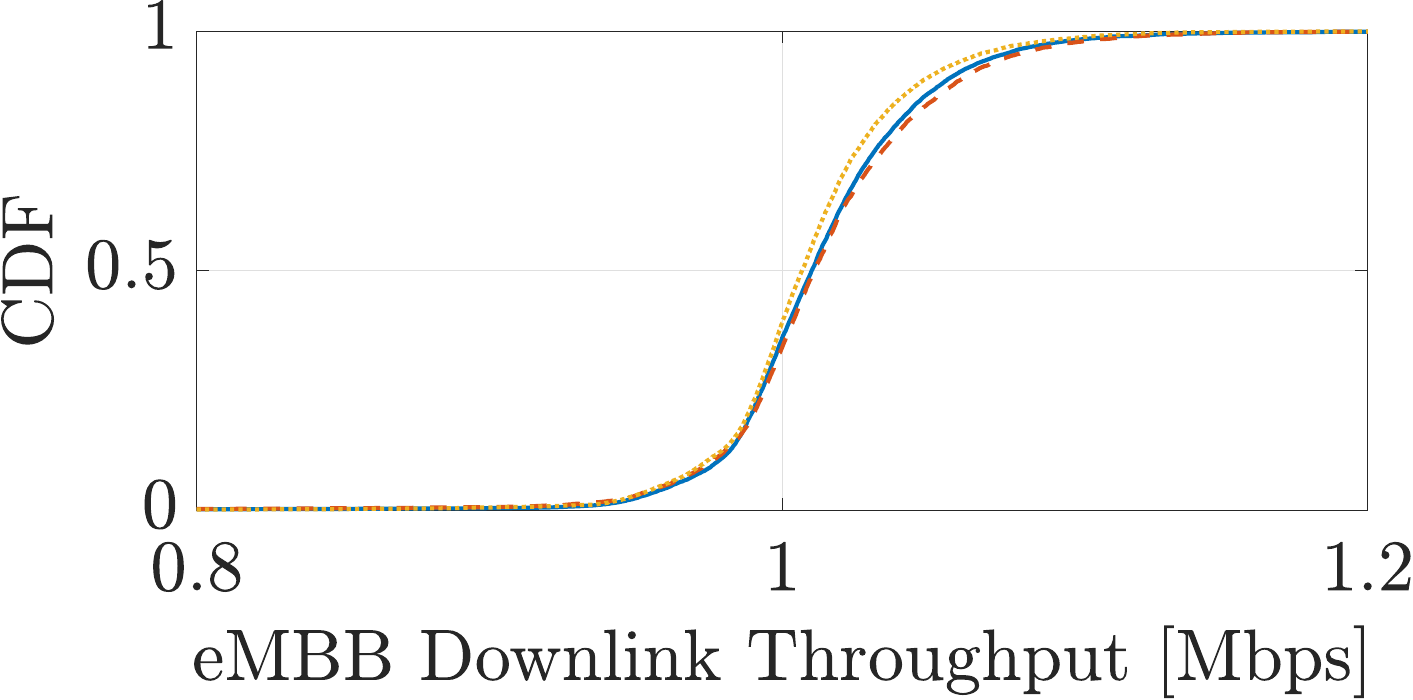}\label{fig:Figure-embb_moreues}}
% \hfil
\subfloat[\gls{mmtc} Packets]{\includegraphics[height=4cm]{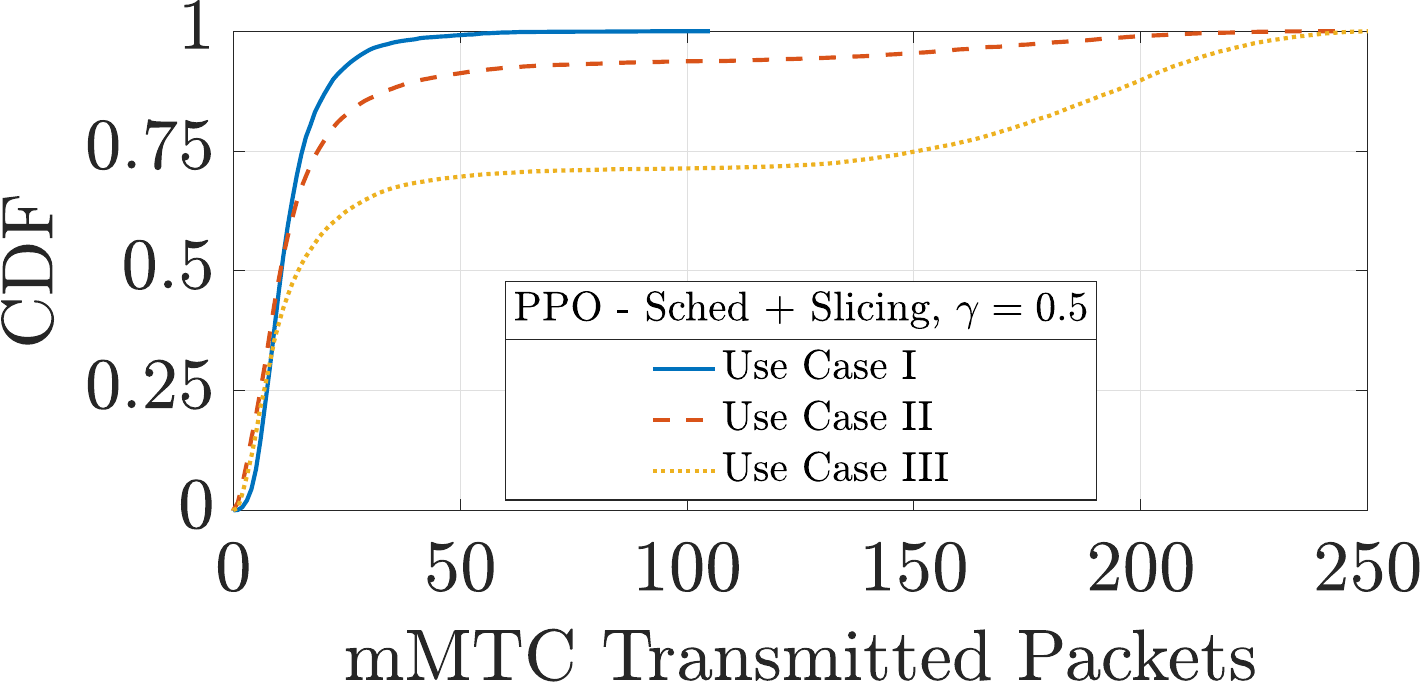}\label{fig:Figure-mmtc_moreues}}
\hfil
\subfloat[\gls{urllc} Buffer Occupancy]{\includegraphics[height=4cm]{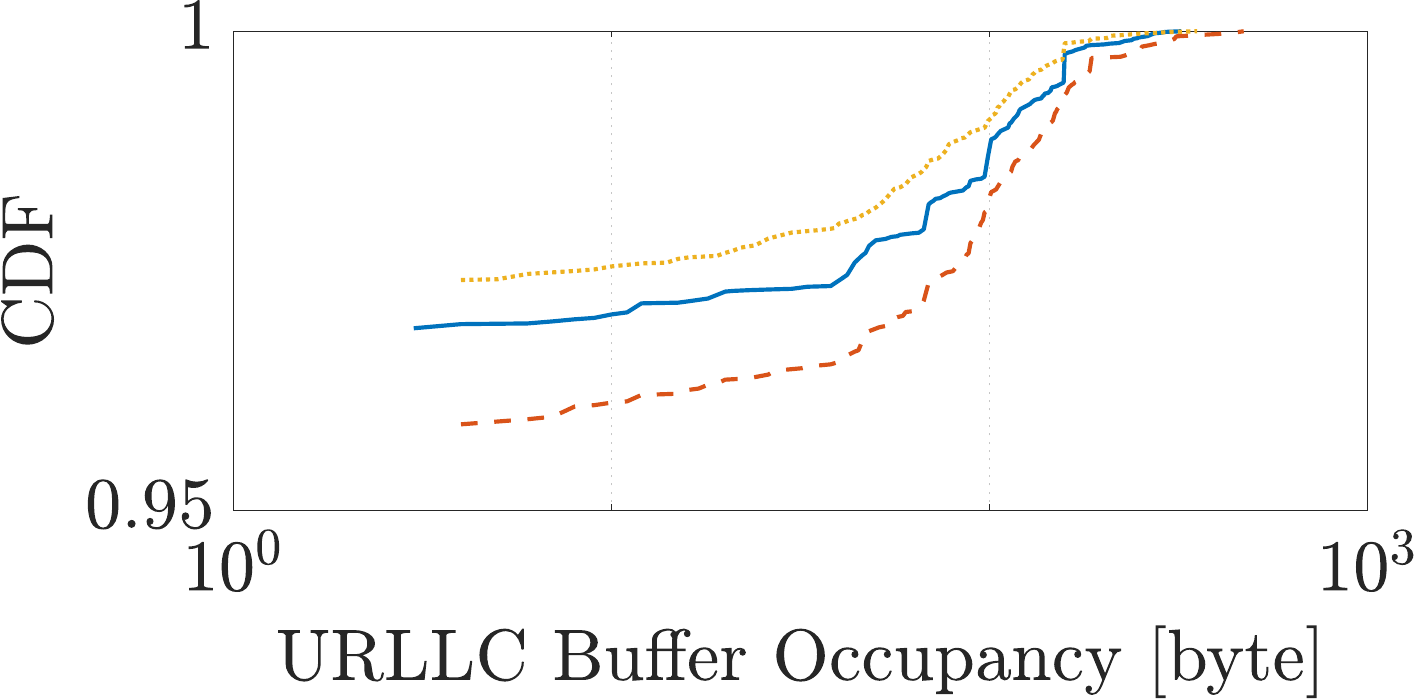}\label{fig:Figure-urllc_moreues}}
\caption{Impact of the network deployment described by three different Use Cases on \gls{dl} \gls{embb} Throughput, \gls{dl} \gls{mmtc} \gls{tx} Packets, and \gls{dl} \gls{urllc} Buffer Occupancy.}
\label{Figure-cdfs_moreues}
\end{figure}

\begin{figure}[htbp]
\centering
\subfloat[\gls{embb} DL Throughput]{\includegraphics[width=2.85in]{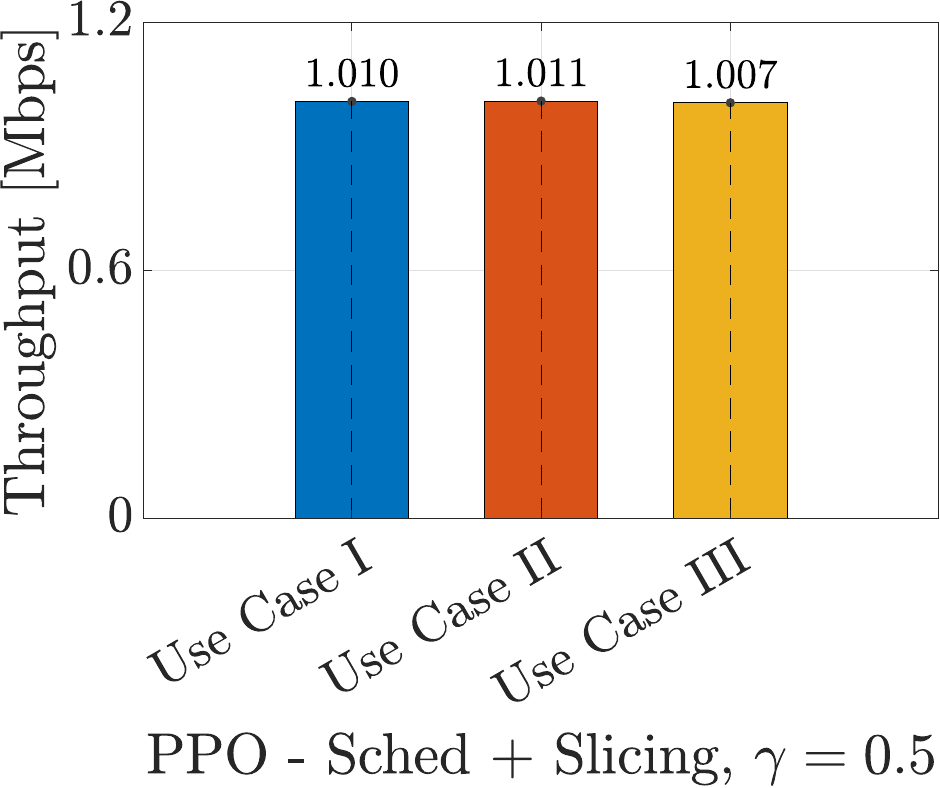}\label{fig:barplot_iot_embb}}
\hfil
\subfloat[\gls{mmtc} Packets]{\includegraphics[width=2.85in]{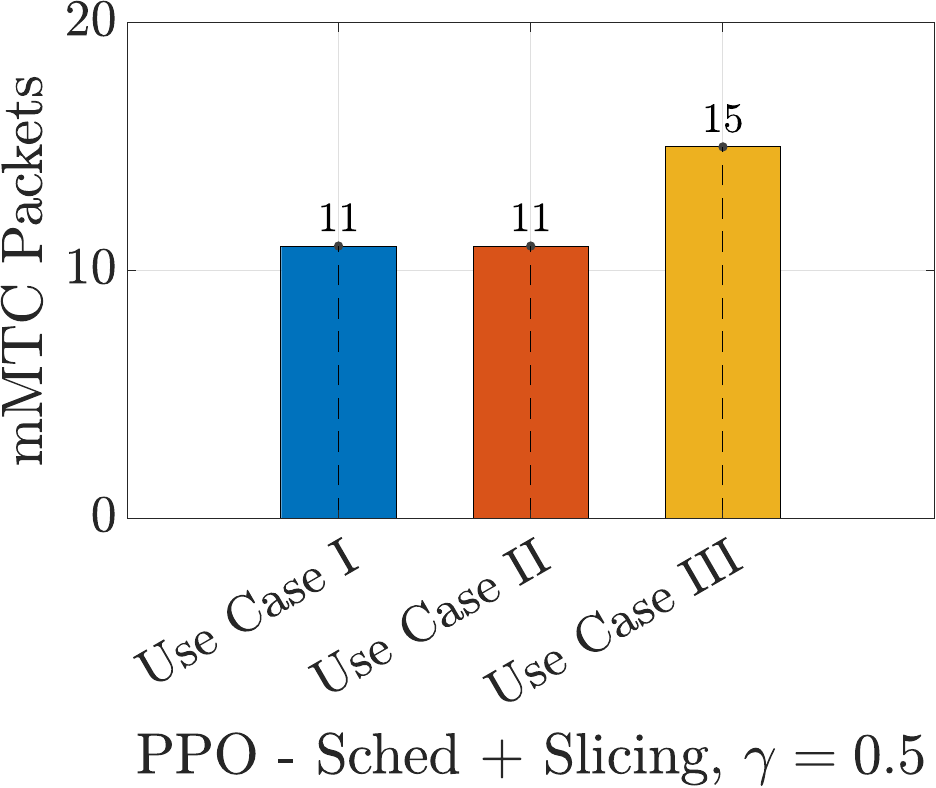}\label{fig:barplot_iot_mmtc}}
\caption{Impact of the network deployment described by three Use Cases on the median values of \gls{dl} \gls{embb} Throughput and \gls{dl} \gls{mmtc} \gls{tx} Packets.}
\label{barplot_iot}
\end{figure}

Figs.~\ref{Figure-prbratio-reward-cdfs_moreues},~\ref{Figure-cdfs_moreues} and~\ref{barplot_iot} demonstrate that even when the number of \glspl{ue} increases, xApps trained using \pandora still deliver good performance, with all \emph{Use Cases} resulting in good resource utilization (Fig.~\ref{Figure-prbratio-reward-cdfs_moreues}) for the three slices. Similarly, the improved \gls{ue} satisfaction is reflected on the \gls{embb} \gls{dl} throughput (Fig.~\ref{fig:Figure-embb_moreues}), with all \emph{Use Cases} enjoying a median throughput value of $\sim1$~Mbps (Fig.~\ref{fig:barplot_iot_embb}). For the \gls{mmtc} slice, and specifically in Fig.~\ref{fig:Figure-mmtc_moreues}, we observe that as the number of \glspl{ue} increases, the number of generated \gls{tx} packets also increases, with \emph{Use Case III} reporting the highest median value of $15$ \gls{dl} \gls{tx} packets (Fig.~\ref{fig:barplot_iot_mmtc}). In line with the results presented in the previous evaluation sections, the performance on the \gls{urllc} remains high (i.e., empty buffers), with \emph{Use Case III} performing slightly better as seen in Figs.~\ref{fig:Figure-prbratio-3-urllc_moreues} and~\ref{fig:Figure-urllc_moreues}.

\subsection{Conclusions}\label{sec:oran_edge:pandora:conclusions}

\newcommand{\MyHuge}{\fontsize{38}{44}\selectfont}
\begin{table}[htb]
\centering
\MyHuge
\setlength\abovecaptionskip{-.1cm}
\caption{xApp Catalog for \texttt{Sched \& Slicing 0.5} with the \gls{ppo} \gls{drl} Architecture.}
\begin{adjustbox}{width=1\linewidth}
\begin{tabular}{@{}c@{\hspace{1.5pt}}|cccc|l@{}}
\toprule
\thead{\MyHuge \textbf{xApp} \\ \MyHuge \textbf{ID}} & & \thead{\MyHuge \textbf{eMBB}} & \thead{\MyHuge \textbf{mMTC}} & \thead{\MyHuge \textbf{URLLC}} &  \thead{\MyHuge \textbf{Testing} \\ \MyHuge \textbf{Conditions}}   \\
\midrule
\textbf{I} & \shortstack{\emph{\textbf{Weight Config.}} \\ \emph{\textbf{Slice Reward}} \\ \emph{\textbf{Traffic}}} & \shortstack{\(\text{\MyHuge{72.0440333}}\) \\ Max. \gls{dl} Throughput \\ \(\text{\MyHuge{1}}\)~Mbps}  &  \shortstack{\(\text{\MyHuge{1.5}}\) \\ Max. \gls{dl} \gls{tx} Packets \\\ \(\text{\MyHuge{30}}\)~kbps}  & \shortstack{\(\text{\MyHuge{0.00005}}\) \\ Min. \gls{dl} Buffer Occupancy \\ \(\text{\MyHuge{10}}\)~kbps}  & \shortstack{\emph{mobility}, \emph{\gls{ran} Control} \\ \emph{Timer Set~\(\text{\MyHuge{1}}\)}, \emph{Location 2}} \\
\midrule
\textbf{II} & \shortstack{\emph{\textbf{Weight Config.}} \\ \emph{\textbf{Slice Reward}} \\ \emph{\textbf{Traffic}}} & \shortstack{\(\text{\MyHuge{72.0440333}}\) \\ Max.  \gls{dl} Throughput \\ \(\text{\MyHuge{4}}\)~Mbps}  & \shortstack{\(\text{\MyHuge{0.229357798}}\) \\ Max. \gls{dl} \gls{tx} Packets \\ \(\text{\MyHuge{44.6}}\)~kbps}  & \shortstack{\(\text{\MyHuge{0.00005}}\) \\ Min.  \gls{dl} Buffer Occupancy \\ \(\text{\MyHuge{89.3}}\)~kbps} & \shortstack{\emph{\gls{ran} Control} \\ \emph{Timer Set~\(\text{\MyHuge{2}}\)}, \emph{Location 2}}  \\
\midrule
\textbf{III} & \shortstack{\emph{\textbf{Weight Config.}} \\ \emph{\textbf{Slice Reward}} \\ \emph{\textbf{Traffic}}} & \shortstack{\(\text{\MyHuge{1}}\) \\ Max.  \gls{prb} Ratio \\ \(\text{\MyHuge{4}}\)~Mbps} & \shortstack{\(\text{\MyHuge{1}}\) \\ Max. \gls{prb} Ratio  \\ \(\text{\MyHuge{44.6}}\)~kbps}  &  \shortstack{\(\text{\MyHuge{1}}\) \\ Max. \gls{prb} Ratio  \\ \(\text{\MyHuge{89.3}}\)~kbps} & \shortstack{\emph{\gls{ran} Control} \\ \emph{Timer Set~\(\text{\MyHuge{1}}\)}, \emph{Location 1}} \\
\midrule
\textbf{IV} & \shortstack{\emph{\textbf{Weight Config.}} \\ \emph{\textbf{Slice Reward}} \\ \emph{\textbf{Traffic}}} & \shortstack{\(\text{\MyHuge{72.0440333}}\) \\ Max.  \gls{dl} Throughput \\ \(\text{\MyHuge{4}}\)~Mbps} & \shortstack{\(\text{\MyHuge{0.5}}\) \\ Max. \gls{prb} Ratio  \\ \(\text{\MyHuge{44.6}}\)~kbps}  &  \shortstack{\(\text{\MyHuge{0.00005}}\) \\ Min. \gls{dl} Buffer Occupancy \\ \(\text{\MyHuge{89.3}}\)~kbps} & \shortstack{ \emph{\gls{ran} Control} \\ \emph{Timer Set~\(\text{\MyHuge{1}}\)}, \emph{Location 1}} \\
\midrule
\textbf{V} & \shortstack{\emph{\textbf{Weight Config.}} \\ \emph{\textbf{Slice Reward}} \\ \emph{\textbf{Traffic}}} & \shortstack{\(\text{\MyHuge{72.0440333}}\) \\ Max.  \gls{dl} Throughput \\ \(\text{\MyHuge{4}}\)~Mbps} & \shortstack{\(\text{\MyHuge{0.229357798}}\) \\ Max. \gls{dl} \gls{tx} Packets \\ \(\text{\MyHuge{44.6}}\)~kbps}  &  \shortstack{\(\text{\MyHuge{0.00005}}\) \\ Min. \gls{dl} Buffer Occupancy \\ \(\text{\MyHuge{89.3}}\)~kbps} & \shortstack{\emph{\gls{ran} Control} \\ \emph{Timer Set~\(\text{\MyHuge{1}}\)}, \emph{Location 1}}  \\
\midrule
\textbf{VI} & \shortstack{\emph{\textbf{Weight Config.}} \\ \emph{\textbf{Slice Reward}} \\ \emph{\textbf{Traffic}}} & \shortstack{\(\text{\MyHuge{72.0440333}}\) \\ Max.  \gls{dl} Throughput \\ \(\text{\MyHuge{4}}\)~Mbps} & \shortstack{\(\text{\MyHuge{1.5}}\) \\ Max. \gls{dl} \gls{tx} Packets \\ \(\text{\MyHuge{44.6}}\)~kbps}  &  \shortstack{\(\text{\MyHuge{0.00005}}\) \\ Min. \gls{dl} Buffer Occupancy \\  \(\text{\MyHuge{89.3}}\)~kbps} & \shortstack{\ \emph{\gls{ran} Control} \\ \emph{Timer Set~\(\text{\MyHuge{1}}\)}, \emph{Location 1}}  \\
\bottomrule
\end{tabular}
\end{adjustbox}
\label{table:xapp-catalog-comp}
\vspace{-.55cm}
\end{table}

In this chapter, we presented \pandora, a comprehensive evaluation with insights on how to design \gls{drl} agents for \gls{ran} control. \pandora leverages a framework to streamline and automate \gls{drl} agent training and xApp on-boarding for extensive evaluation and testing of \gls{drl} agents with Open RAN in Colosseum. We investigated the impact of \gls{drl} design on the performance of an Open \gls{ran} system controlled by xApps embedding \gls{drl} agents that make decisions in near-real-time to compute efficient slicing and scheduling control policies. We benchmarked $23$~xApps trained using \gls{drl} agents with different action spaces, architectures, reward design and decision-making timescales. Additionally, we tested the \gls{drl} agents under various traffic and mobility conditions, considering different hierarchical setups and time granularities in locations both observed and unseen during the training phase. Our experimental results show that network slices with similar objectives (e.g., maximizing throughput and number of transmitted packets) might result in a competitive behavior that can be mitigated using proper weight and reward configurations, and possibly different architectures. Additionally, we have explored how the fine-tuning of \gls{ran} control timers can boost the performance of various xApps tasked with altering different action spaces. Based on the reported findings, the \pandora evaluation shows that \gls{drl} agents adapt well to network conditions encountered during the \gls{ai}/\gls{ml} training phase as well as unforeseen conditions, while ensuring high system resource utilization.

\section[Identifying Conflicts among DRL-based xApps]{Identifying Conflicts among \texorpdfstring{\gls{drl}}{DRL}-based xApps}
\label{sec:oran_edge:pacifista:conflicts}

\subsection{Introduction}
\label{sec:oran_edge:pacifista:intro}

The O-RAN ALLIANCE identifies several types of conflicts, including cell ON/OFF, beamforming, handover, antenna tilt, traffic steering, and others, and highlights the importance of conflict management frameworks~\cite{oran-wg3-con-mit}. Industry experience clearly shows that conflicts are a critical issue in \oran and require appropriate solutions to mitigate performance degradation and prevent service disruptions, outages, and potential financial losses. While these frameworks increase network complexity, they are necessary to ensure continuous and reliable operation.

Given its complexity and importance, conflict management has emerged as a key research area within the community and a critical enabler for the adoption of \oran. The O-RAN ALLIANCE classifies conflicts into direct, indirect, and implicit categories. Early efforts in \oran conflict management include real-time detection of conflicting control policies~\cite{adamczyk2023challenges,adamczyk2023conflict}, online deconfliction~\cite{wadud2024qacm}, coordination of \ai-based xApps and rApps through team learning to reduce conflict occurrence~\cite{iturria2022multi,zhang2022team}, and orchestration of xApp and rApp selection and deployment to avoid conflicts. While these works provide concrete solutions for mitigating conflicts in \oran, they do not fully capture the nuances of conflict severity or their impact on \glspl{kpm}. In addition, they generally do not account for conflicts that emerge only under specific operational conditions.

\begin{figure}[htbp]
    \centering
    \includegraphics[width=\columnwidth]{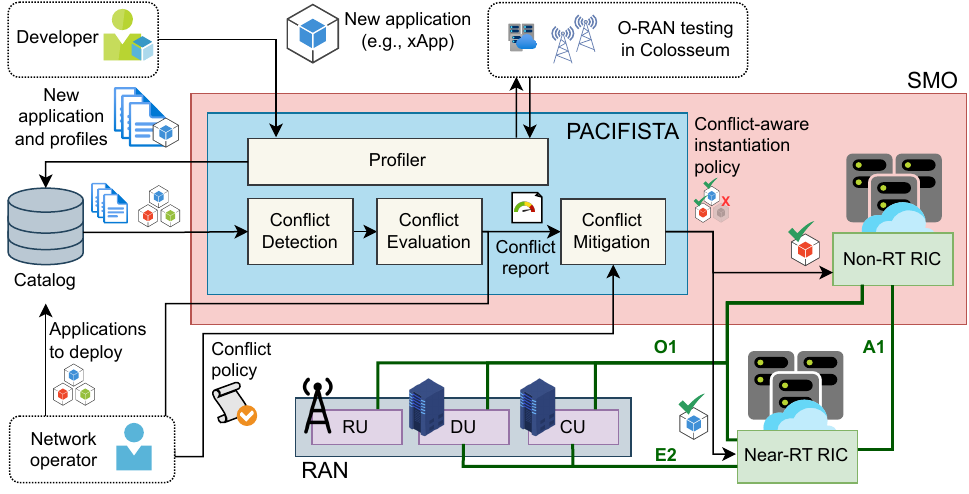}
    \caption{\name architecture and workflow~\cite{11007770}.}
    \label{fig:oran_edge:pacifista:architecture}
\end{figure}

In this section, we leverage \name~\cite{11007770}, an empirical and formal framework that uses statistical information from xApps to detect, characterize, and mitigate conflicts between applications. The architecture of \name is illustrated in Fig.~\ref{fig:oran_edge:pacifista:architecture}. \name operates through a profiling pipeline that tests \oran applications in a sandbox environment under predefined operational conditions, generating statistical profiles---in the form of \glspl{ecdf}---of control parameters and \kpms; these profiles are then analyzed through distance metrics (e.g., K-S distance for numerical parameters, $\chi$ distance for categorical ones) to detect and evaluate direct, indirect, and implicit conflicts. \name further combines a hierarchical graph representation with statistical models of \oran application behavior to identify relationships between control parameters and \glspl{kpm}. This enables \name to: (i)~determine whether two or more applications will generate conflicts; and (ii)~quantify the severity of such conflicts. Our contribution in this work is the set of \gls{drl}-based xApps for slicing and scheduling control introduced in~\cite{10437367}, whose conflict behavior under this framework we analyze below.

\name has a modular architecture (Fig.~\ref{fig:oran_edge:pacifista:architecture}) with four major logical blocks and a catalog of applications (e.g., xApps). \name executes four major tasks, namely application profiling, conflict detection, evaluation, and mitigation. The \textit{Profiler} runs on sandbox testing environments (e.g., \glspl{dt}, emulation environments) and generates profiles that describe the statistical behavior of \oran applications. The \textit{Conflict Detection Module} uses such profiles for detecting the occurrence of conflicts and identifying the affected parameters and \kpms. Upon detecting the existence of conflicts, the \textit{Conflict Evaluation Module}, which is executed on the production network, generates a report that summarizes how severe the conflict is, and how much it impacts the \kpms. Finally, the \textit{Conflict Mitigation Module} leverages the information included in the report to make informed decisions on the deployment of \oran applications. These decisions are made by using conflict management policies specified by the network operator, such as avoiding the deployment of an application that would generate too large of a conflict, or removing a subset of applications to reduce the severeness of conflicts below a certain threshold.

\subsection{Conflict Model and Distance Metrics}
\label{sec:oran_edge:pacifista:model}

To formalize how \name detects and evaluates conflicts among the \gls{drl}-based xApps considered in this work, we briefly summarize the conflict model and the statistical metrics that the framework adopts~\cite{11007770}. Let $\apps$ denote the set of \oran applications that can be deployed on the \rics, $\params$ the set of control parameters that they can act upon, and $\metrics$ the set of observable \kpms. For each application $a\in\apps$ and parameter $p\in\params$, we define an indicator $\alpha_{a,p}\in\{0,1\}$ such that $\alpha_{a,p}=1$ if $a$ controls $p$, and $\alpha_{a,p}=0$ otherwise. Accordingly, $\params_{a}=\{p\in\params : \alpha_{a,p}=1\}\subseteq\params$ identifies the subset of parameters controlled by application $a$. Analogously, an indicator $\epsilon_{p,k}\in\{0,1\}$ tracks whether parameter $p\in\params$ impacts \gls{kpm} $k\in\metrics$, with $\epsilon_{p,k}=1$ if such a dependency exists, and $\epsilon_{p,k}=0$ otherwise. For any 2-tuple $(p_1,p_2) \in \params \times \params$, let $\pi_{p_1,p_2}\in\{0,1\}$ be an indicator parameter such that $\pi_{p_1,p_2}=1$ if parameter $p_1$ impacts parameter $p_2$, $\pi_{p_1,p_2}=0$ otherwise. Formally, we have $E^{\mathrm{P}} = \{ (p_1,p_2) \in \params \times \params : \pi_{p_1,p_2} = 1 \}$. We consider the case where applications are profiled on a set $\ops$ of predefined operational conditions. Each operational condition $c\in\ops$ can specify, among others, the number of \glspl{ue}, cell load, and \gls{sinr}/\gls{cqi} conditions.

Building on this notation, \name formalizes the three classes of conflicts identified by the O-RAN ALLIANCE~\cite{oran-wg3-con-mit} as follows.
\begin{definition}[Direct Conflict]\label{def:oran_edge:pacifista:direct}
Let $p\in\params$ be any parameter. Let $a_1$ and $a_2$ be any two applications in $\apps$. We say that $a_1$ and $a_2$ are in direct conflict with respect to parameter $p$ if $p \in \params^{\mathrm{DC}}_{a_1,a_2} = \params_{a_1} \cap \params_{a_2}$.
\end{definition}
\begin{definition}[Parameter Conflict]\label{def:oran_edge:pacifista:param}
Let $a_1$ and $a_2$ be any two applications in $\apps$. We say that parameter $p_1\in\params_{a_1}$ controlled by application $a_1$ generates a Parameter conflict with parameter $p_2\in\params_{a_2}$ controlled by $a_2$ if $(p_1,p_2) \in E^{\mathrm{P}}$, i.e., $\pi_{p_1,p_2} = 1$.
\end{definition}
\begin{definition}[\gls{kpm} Conflict]\label{def:oran_edge:pacifista:kpm}
Let $a_1$ and $a_2$ be any two applications in $\apps$. We say that $a_1$ and $a_2$ generate a \gls{kpm} conflict with respect to \gls{kpm} $k\in\metrics$ if there exists at least one 2-tuple $(p_1,p_2) \in \params_{a_1} \times \params_{a_2}$, with $p_1 \neq p_2$, such that $\epsilon_{p_1,k} = \epsilon_{p_2,k} = 1$.
\end{definition}

In our analysis, the \gls{drl}-based xApps $a_6$, $a_7$, and $a_8$ from~\cite{10437367} share the parameter set $\params$ (slicing and scheduling policies) and the \gls{kpm} set $\metrics$ (throughput, buffer occupancy, transmitted packets). $a_6$ controls the scheduling policy only, $a_7$ controls the slicing policy only, while $a_8$ controls both the scheduling and the slicing policies. Since all three xApps influence overlapping \kpms (i.e., $\epsilon_{p,k}=1$ for the same $k\in\metrics$), \gls{kpm} conflicts are also expected.

To assess the severity of such conflicts, \name profiles each xApp under each operational condition $c\in\ops$ and extracts \glspl{ecdf} of the controlled parameters and observed \kpms. Pairs of one-dimensional generic \glspl{ecdf} $F_1(x)$ and $F_2(x)$, with $x\in\mathbb{R}$ being a random variable, are then compared via the distance functions summarized in Table~\ref{tab:oran_edge:pacifista:distances}. We write $D^{\mathrm{K\text{-}S}}_{i,j}$, $D^{\mathrm{INT}}_{i,j}$, and $D^{\chi}_{i,j}$ to denote the respective distances between the statistical profiles of xApps $a_i$ and $a_j$. For numerical variables (e.g., \gls{prb} allocation, throughput, buffer occupancy), \name uses the \gls{ks} and \gls{int} distances:
\begin{equation}
    D^{\mathrm{K\text{-}S}} = \max\lvert F_1(x)-F_2(x)\rvert, \qquad
    D^{\mathrm{INT}} = \sqrt{\tfrac{1}{L}\int \lvert F_1(x)-F_2(x)\rvert\,dx},
    \label{eq:oran_edge:pacifista:distances_num}
\end{equation}
with $L=\max(x)-\min(x)$. For categorical variables such as the scheduling policy (chosen among \gls{wf}, \gls{rr}, and \gls{pf}), \name relies on the $\chi$ distance $D^{\chi}=1-p\text{-value}$ from Pearson's Chi-Square test~\cite{hazra2016biostatistics}. By construction, all three distances take values in $[0,1]$, with larger values indicating more severe conflicts. We exploit this formulation in Section~\ref{sec:oran_edge:pacifista:setup} to quantify direct and \gls{kpm} conflicts among xApps $a_6$, $a_7$, and $a_8$ from~\cite{10437367}.

\begin{table}[htbp]
    \setlength\abovecaptionskip{5pt}
    \setlength\belowcaptionskip{0pt}
    \centering
    \caption{Distance functions used in \name to evaluate conflicts in \oran~\cite{11007770}.}
    \begin{tabular}{c c p{0.55\linewidth}}
        \toprule
        \textbf{ID} & \textbf{Equation}                                          & \textbf{Description}                                                                              \\
        \midrule
        \gls{ks}    & $\max \lvert F_1(x) - F_2(x)\rvert$                        & Maximum vertical distance between the two \glspl{ecdf}.                                           \\
        \midrule
        \gls{int}   & $\sqrt{\frac{1}{L}\int \lvert F_1(x) - F_2(x)\rvert\,dx}$  & Integral of the absolute distance between the two \glspl{ecdf}, with $L=\max(x)-\min(x)$.         \\
        \midrule
        $\chi$      & $1-\textrm{p-value}$                                       & Likelihood that data from two categorical distributions differ.                                   \\
        \bottomrule
    \end{tabular}
    \label{tab:oran_edge:pacifista:distances}
\end{table}

\subsection{Experimental Setup and Conflict Analysis}
\label{sec:oran_edge:pacifista:setup}

To fairly compare all xApps against the same repeatable operational conditions $c$, we benchmark them on the Rome Colosseum scenario~\cite{bonati2023openran}---which reproduces the real-world cellular deployment in a section of Rome, Italy.
We consider 6~\glspl{ue} uniformly distributed within $50$\:m from the \gls{bs}, which allocates them in three network slices (\gls{embb}, \gls{urllc}, and \gls{mmtc}) across $10$\:MHz of spectrum (50~\glspl{prb} grouped into 17~\glspl{rbg}).
We leveraged the \gls{mgen}~\cite{mgen} tool to serve \gls{dl} traffic to the slice \glspl{ue} as follows: (i)~\gls{embb} \glspl{ue} are served constant bit rate traffic at a rate of $4$\:Mbps; (ii)~\gls{urllc} are served Poisson-distributed traffic at an average rate of $89.29$\:kbps; and (iii)~\gls{mmtc} \glspl{ue} are served Poisson-distributed traffic at an average rate of $44.64$\:kbps.

Our xApps act on a combination of two control parameters: (i)~the \textit{network slicing policy}, by adjusting the number of \glspl{prb} allocated to each slice; and (ii)~the \textit{scheduling policy} used in downlink transmissions, chosen among \gls{wf}, \gls{rr}, and \gls{pf}.

These xApps are taken from \cite{10437367} and embed \gls{drl} agents controlling a combination of slicing (i.e., the portion of the available \glspl{prb} allocated to each slice) and scheduling (i.e., how the \glspl{prb} are internally allocated to users of each slice) policies to satisfy slice-specific intents. Specifically, they all aim at maximizing \gls{embb} throughput and number of \gls{mmtc} transmitted packets, while minimizing \gls{dl} buffer size for \gls{urllc} traffic (as a proxy of latency). They all target this goal via different action spaces. $a_6$ controls the scheduling policy only, $a_7$ controls the slicing policy only, while $a_8$ controls both the scheduling and the slicing policies. These are referred to as \texttt{\small Sched 0.5}, \texttt{\small Slicing 0.5}, \texttt{\small Sched \& Slicing 0.5} in \cite{10437367}, respectively.
These xApps embed agents trained using \gls{ppo}, a state-of-the-art \gls{rl} architecture~\cite{chen2023static}, and are deployed inside the Near-RT \gls{ric}. They receive real-time \glspl{kpm} via the E2 interface, make decisions based on network conditions such as downlink throughput, buffer occupancy, and the number of transmitted packets, and continuously adapt their control policies based on real-time feedback from the \gls{ran}. It is noted that the model-free architecture and trial-and-error approach of the \gls{ppo} algorithm are an ideal fit for enhancing the resource allocation process in stochastic environments, such as wireless channels. Ultimately, our goal is to demonstrate that xApps with similar intents,
even when controlling different parameters, are prone to generating minimal conflicts.

We consider xApps $a_6$-$a_8$ which use \gls{drl} agents with diverse action spaces to improve slice-specific \kpms. We present conflict analysis for \gls{embb} and \gls{urllc} slices. We report \gls{dl} throughput values for the former, and \gls{dl} buffer size for the latter.
Table~\ref{tab:oran_edge:pacifista:drl_distances} shows the K-S and INT distances for numerical parameters (i.e., slicing policies) and \kpms, and $\chi$ distance for scheduling policies, as well as the impact of conflicts on relevant \kpms. Recall that $a_8$ controls both slicing and scheduling policies, while $a_6$ and $a_7$ respectively control scheduling and slicing only. Therefore, there is no direct conflict on scheduling between $a_8$ and $a_7$, and no direct conflict on slicing between $a_8$ and $a_6$.
Since $a_7$ and $a_8$ both control slicing policies, they generate a direct conflict with K-S and INT distance equal to 0.11 and 0.23, respectively. The same holds for $a_6$ and $a_8$ which produce a direct conflict with respect to scheduling with a $\chi$ distance of 0.61, suggesting that the two xApps select different scheduling policies. In general, conflicts have low values due to the shared goal. However, we notice that controlling slicing policies (i.e., $a_7$ and $a_8$) results in lower \gls{kpm} conflicts for both throughput and buffer size (i.e., the largest distance in this case is $D^{\mathrm{K-S}}_{8,7}=0.14$), suggesting that controlling slicing policies under a shared goal makes it possible to achieve higher performance than scheduling control alone~\cite{10437367}, which allows the xApps to better satisfy the shared intent and produce less conflicts. We also notice that the larger action space (i.e., scheduling and slicing) allows $a_8$ to improve performance.
For example, Fig.~\ref{fig:oran_edge:pacifista:drl_throughput} shows how $a_8$ delivers higher \gls{embb} throughput than $a_7$.

\begin{table}[t]
    \addtolength{\tabcolsep}{-1pt}
    \setlength\abovecaptionskip{5pt}
    \setlength\belowcaptionskip{5pt}
    \centering
    \caption{Direct and \gls{kpm} conflict analysis taking $a_8$ as the reference xApp.}
    \begin{tabular}{ c c *{6}{c} }
        \toprule
        \textbf{Slice} & \textbf{Variable} & \multicolumn{3}{c}{\textbf{$a_8$-$a_6$}}    & \multicolumn{3}{c}{\textbf{$a_8$-$a_7$}}                                                                                                                                                                     \\
        \cmidrule(lr){3-5} \cmidrule(lr){6-8}
                       &                   & {$D^{\mathrm{K-S}}_{8,6}$}   \hspace{0.3cm} & {$D^{\mathrm{INT}}_{8,6}$} \hspace{0.3cm} & {$D^{\mathrm{\chi}}_{8,6}$} \hspace{0.3cm} & {$D^{\mathrm{K-S}}_{8,7}$} \hspace{0.3cm} & {$D^{\mathrm{INT}}_{8,7}$} \hspace{0.3cm} & {$D^{\mathrm{\chi}}_{8,7}$} \\ \midrule
        \gls{embb}     & {\glspl{prb}}     & 0                                           & 0                                         & $-$                                        & 0.11                                      & 0.23                                      & $-$                         \\
        \gls{embb}     & {Scheduling}      & $-$                                         & $-$                                       & 0.61                                       & $-$                                       & $-$                                       & 0                           \\
        \gls{embb}     & {Throughput}      & 0.46                                        & 0.27                                      & $-$                                        & 0.14                                      & 0.13                                      & $-$                         \\ \midrule
        \gls{urllc}    & {\glspl{prb}}     & 0                                           & 0                                         & $-$                                        & 0.10                                      & 0.16                                      & $-$                         \\
        \gls{urllc}    & {Scheduling}      & $-$                                         & $-$                                       & 0.19                                       & $-$                                       & $-$                                       & 0                           \\
        \gls{urllc}    & {Buffer Size}     & 0.06                                        & 0.01                                      & $-$                                        & 0.04                                      & 0.01                                      & $-$                         \\ \bottomrule
    \end{tabular}
    \label{tab:oran_edge:pacifista:drl_distances}
\end{table}

\begin{figure}[t]
    \setlength\abovecaptionskip{0pt}
    \setlength\belowcaptionskip{0pt}
    \centering
    \includegraphics[width=\columnwidth]{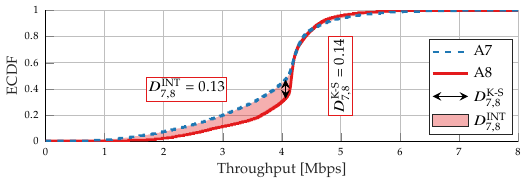}
    \caption{\glspl{ecdf} of throughput of the \gls{embb} slice for xApps $a_7$ and $a_8$, highlighting the expected \gls{kpm} conflict for that control parameter.}
    \label{fig:oran_edge:pacifista:drl_throughput}
\end{figure}

\subsection{Conclusions}
\label{sec:oran_edge:pacifista:conclusions}

In this section, we leveraged \name~\cite{11007770}, a framework to detect, characterize, and mitigate conflicts in the \oran ecosystem. \name leverages statistical information on \oran applications and a set of hierarchical graphs to determine the likelihood of conflict emergence and the corresponding severity. We focused our analysis on \gls{drl}-based xApps for slicing and scheduling control, showing how \name uses a formal, data-driven model to capture the impact of conflicts on control parameters and target \kpms, thus providing useful insights on how conflicts affect the network performance and the operator's intents. Specifically, we showed that the \gls{drl}-based xApps from~\cite{10437367}, despite operating on overlapping control parameters, produce low-severity conflicts due to their shared optimization intent---with \gls{kpm} distances remaining below $0.5$ across all slice--variable pairs considered, and direct conflict distances on slicing as low as $D^{\mathrm{K\text{-}S}}_{a_8,a_7}=0.11$. Future work will extend this analysis to larger xApp sets, richer action spaces, and more heterogeneous network deployments.
% --- EOF ---

%% file: tex/ris_propagation.tex
% !TEX root = ../thesis_dissertation template.tex
% ris_propagation.tex

\chapter[Optimizing Wireless Propagation with RIS]{Optimizing Wireless Propagation with \texorpdfstring{\gls{ris}}{RIS}}
\label{chap:ris_propagation}

\section[Cross-Layer System Level Experimental Evaluation of RIS]{Cross-Layer System Level Experimental Evaluation of \texorpdfstring{\gls{ris}}{RIS}}
\label{sec:ris_propagation:cross_layer}

\subsection{Introduction}
\label{sec:ris_propagation:cross_layer:intro}

\glspl{ris} are a promising technique for enhancing the performance of \gls{nextg} wireless communication systems in terms of both spectral and energy efficiency, as well as resource utilization. However, current \gls{ris} research has primarily focused on theoretical modeling and \gls{phy} layer considerations only. Full protocol stack emulation and accurate modeling of the propagation characteristics of the wireless channel are necessary for studying the benefits introduced by \gls{ris} technology across various spectrum bands %, enabling efficient resource allocation and utilization tailored to the specific needs of each band 
and use-cases.
In this chapter, we propose, for the first time:
(i) accurate \gls{phy} layer \gls{ris}-enabled channel modeling through \glspl{gbsm}, leveraging the \gls{quadriga} open-source statistical ray-tracer; (ii) optimized resource allocation with \glspl{ris} by comprehensively studying energy efficiency and power control on different portions of the spectrum through a single-leader multiple-followers Stackelberg game theoretical approach; (iii) full-stack emulation and performance evaluation of \gls{ris}-assisted channels with SCOPE/srsRAN for \gls{embb} and \gls{urllc} applications in the world's largest emulator of wireless systems with hardware-in-the-loop,
namely Colosseum~\cite{10973151}. Our findings indicate (i) the significant power savings in terms of energy efficiency achieved with \gls{ris}-assisted topologies, especially in the \gls{mmwave} band; and  (ii) the benefits introduced for Sub-6~GHz band \glspl{ue}, where the deployment of a relatively small \gls{ris} (e.g., in the order of $100$ \gls{ris} elements) can result in decreased levels of latency for \gls{urllc} services in resource-constrained environments. 

Undoubtedly, research in \gls{5g} wireless communication systems has been following an upward trend in recent years. \gls{5g} promises better connectivity anywhere, anytime, for everyone and everything, with increased coverage, better system capacity, extremely low latency~\cite{attar20225g}, improved performance in terms of both energy and spectral efficiency~\cite{shehab20215g,sobhi2020energy}, while also simultaneously satisfying the diverse \gls{qos} demands of its heterogeneous users~\cite{zhang2017network} across a variety of deployment scenarios and network topologies~\cite{wang2018survey}. Hence, \gls{5g} and beyond technologies are expected to become an indispensable component of future smart environments~\cite{lopes20235g,liu2021promoting,sharif2019compact,marabissi2018real,jiang2019smart,zhao2021nanogenerators,sanchez2021review}. Another prominent technology anticipated to accommodate the needs of future \gls{nextg} communication networks are \glspl{ris}. \glspl{ris} have recently attracted widespread interest in the research community as a part of a smart city environment for \gls{5g}, and beyond \gls{5g} wireless communication systems~\cite{rana2023review,zivuku2022maximizing,al2023emerging,kisseleff2020reconfigurable,liu2023integrated,salah2022paving,shehab20215g,makarfi2020reconfigurable,kamruzzaman2022key,bariah2023digital,dagiuklas2023journey,li2023liquid,mishra20236g,masouros2023guest,renzo2019smart}. Indeed, \glspl{ris} technologies have been proven to extend coverage~\cite{li2023liquid,liu2023integrated,salah2022paving,makarfi2020reconfigurable}, enhance both spectral and energy efficiency~\cite{shehab20215g,hou2020reconfigurable,9079457,guan2022irs}, while at the same time enabling the vision of smart and reconfigurable propagation environments~\cite{ji2022reconfigurable,renzo2019smart,dagiuklas2023journey}, which can be controlled through~\gls{sdr} techniques~\cite{liaskos2022software,banchs2021network}. Therefore, due to both the tremendous advantages that they introduce (e.g., decreased levels of latency~\cite{basharat2022exploring,liu2021reconfigurable}) and due to their presence in a multitude of use-cases scenarios (e.g., resource allocation in multi-\gls{ue} \gls{noma} systems~\cite{mu2021capacity}, localization and positioning~\cite{elzanaty2021reconfigurable}, assisting anti-jamming methods~\cite{yang2020intelligent}, and enhancing \gls{uav}-enabled communications~\cite{salih2023enhancing,diamanti2021energy,liu2021reconfigurable,li2020reconfigurable,basharat2021reconfigurable,ning2023intelligent,yu2022fair} among others), there is a clear need for further understanding of \gls{ris}-assisted channels through effective performance evaluation campaigns in different spectrum bands and use cases.

In this section, we aim to provide, for the first time, a holistic, system-level evaluation of \gls{ris}-assisted channels for \gls{nextg} communication systems in \gls{cv2x} environments across different portions of the spectrum, while also paving the way for their evaluation across various network slices on experimental platforms. Our key contributions are as follows:
\begin{enumerate}
    \item We leverage the capabilities of the fully reconfigurable, MATLAB-based~\cite{MATLAB:2021a} statistical ray-tracer \gls{quadriga}~\cite{quadrigacode}, which uses \gls{rf} data collected under various conditions to cascade the wireless links and include the effect of a \gls{ris} in the topology.
    \item We focus on a \gls{cv2x} setup comprising a \gls{uav} serving as a flying \gls{gnb} for multiple \glspl{ue}.
    \item Through a single-leader multiple-followers Stackelberg game we jointly optimize the overall received signal strength at the \gls{gnb}, and each \gls{ue}'s individual energy efficiency in a \gls{noma} system.
    \item Through this hierarchical game-theoretic approach, we determine both the phase shifts of the \gls{ris} reflecting elements, as well as the \gls{ul} transmission power of the \glspl{ue} for both the Sub-6~GHz and the \gls{mmwave} bands.
    \item We install the generated wireless topologies for the band that requires only a minimal number of \gls{ris} elements to achieve a significant path gain enhancement in the Colosseum Wireless Network and Channel Emulator~\cite{bonati2021colosseum}.
    \item We modify the \gls{ran} functionalities in terms of scheduling profiles (i.e., \gls{rr} and \gls{wf}) and slicing (i.e., allocation of the available \glspl{prb}) and we observe the performance gain provided through the \gls{ris} technology on the \gls{embb} and \gls{urllc} network slices.
\end{enumerate}

\subsubsection{Related Work}
\label{sec:ris_propagation:cross_layer:related}

The current literature hosts a plethora of \gls{ris} simulators for \gls{phy} layer channel modeling across various operating frequencies. Additionally, substantial efforts have been reported in the design, creation, and real-world deployment of such surfaces. In~\cite{basar2021reconfigurable,basar2020simris}, the authors present the \gls{mmwave} SimRIS simulator for indoor and microcell environments. The aforementioned works state that accurate \gls{phy} layer simulators that can operate in a variety of frequencies and applications, ideally using real data, are of utmost importance. The authors note that further research is needed to develop practical channel models for various propagation environments and to identify compelling use cases. Finally, the authors emphasize the need to explore the coexistence of \gls{embb} and \gls{urllc} \glspl{ue} in various frequency domains, as well as the importance of standardization and integration into existing networks. In~\cite{sihlbom2022reconfigurable}, a commercial \gls{ris} simulator is leveraged to perform system-level simulation in the mid and \gls{mmwave} bands within a real-world urban environment. The authors conclude by mentioning the importance of analytical frameworks for system-level analysis by using tools leveraging stochastic geometry. Regarding the sub-6~GHz band, in~\cite{sun20213d}, a 3D \gls{gbsm} relying on the statistical properties of the wireless channels is discussed, taking into account both multipath propagation and continuous-time evolution. A \gls{gbsm} channel model incorporating a \gls{ris} is also considered in~\cite{dang2021geometry}, highlighting the importance of leveraging small panels. Other \gls{nextg} \gls{ris}-assisted channel models, which discuss path-loss propagation and take into account geometry, are also presented in~\cite{dajer2022reconfigurable,liu2021reconfigurable}. Rayleigh and Rician channels are used to model the multipath effect, and their integration in \gls{ris}-assisted communication systems is discussed in detail in~\cite{zhou2023survey,kong2021channel}. In~\cite{tian2023reconfigurable}, a frequency-independent channel model for \gls{ris} is discussed, whereas a \gls{ris}-assisted channel model for \gls{v2x} environments that incorporates the Rician channel matrix is presented in~\cite{chen2021qos}. A \gls{ris}-assisted channel modeling approach for high frequencies is discussed in~\cite{boulogeorgos2021coverage}, while a signal processing perspective on multipath propagation with \gls{ris} is given in~\cite{bjornson2022reconfigurable}. Physics-based \gls{phy} models are presented in~\cite{yu2021smart,xu2023exploiting}, while other approaches include deploying \gls{ris} panels in the Sub-6~GHz~\cite{tang2021channel} and \gls{mmwave}~\cite{da2023varactor} bands~\cite{liu2022reconfigurable, liu2022simulation}, with a focus on energy efficiency~\cite{rossanese2022designing}.

Additionally, the contribution of \gls{ris} technology to energy efficiency through power control is a well-investigated topic. In~\cite{8741198}, a non-convex optimization framework for energy efficiency that jointly optimizes the \gls{tx} power of the \gls{bs} and the reflecting coefficients of the \gls{ris} on the \gls{dl} is considered. In~\cite{9950621}, the paper tackles energy efficiency maximization in an \gls{ul} \gls{ris}-aided \gls{mmwave} \gls{noma} network by jointly optimizing the \gls{ue} \gls{tx} power and the \gls{ris} phase shifts using iterative optimization techniques. In~\cite{9810453}, the usage of both \glspl{ris} and \gls{uav} is discussed in terms of providing energy-efficient communications on the \gls{dl} while ensuring the \gls{qos} demands of the \glspl{ue}. The proposed method uses \gls{sca} to iteratively determine a joint optimal solution for the \gls{uav} trajectory, \gls{ris} phase shifts, and the \gls{uav} \gls{tx} power. In~\cite{9628234}, the problem of energy efficiency maximization is modeled as a \gls{minlp} problem, with the objective of enhancing overall energy efficiency on the \gls{dl} by optimizing the \gls{bs}'s \gls{tx} power and \gls{ris} phase shifts. The work in~\cite{10096617} discusses the topic of energy efficiency maximization in a \gls{ris}-assisted network, where convex optimization and sequential approximation methods are employed to jointly optimize the \gls{tx} powers of the \glspl{ue} and the reflection coefficients of the \gls{ris} in the \gls{ul}. In~\cite{10107766}, a \gls{ml} approach based on the \gls{ppo} algorithm for the maximization of energy efficiency in \gls{ris}-aided networks is discussed. The optimization problem is formulated under constraints of the rate requirement of the \glspl{ue}, the power budget at the \gls{bs}, and the discrete phase shift coefficients of each reflecting element at the \gls{ris} on the \gls{dl}. In this approach, the \gls{bs}'s \gls{tx} power and the phase shift matrix of the \gls{ris} are jointly optimized. Similarly, in~\cite{9964251}, the joint energy efficiency maximization problem for a \gls{dl} \gls{noma} network is discussed, with the goal of jointly optimizing the beamforming vectors at the \gls{bs} and the coefficient matrices at the \gls{ris}, while also controlling the \gls{tx} power at the \gls{bs} using \gls{drl}. In~\cite{diamanti2021energy}, a two-tier Stackelberg game is used to jointly maximize the received signal strength at the \gls{enb} by calculating the \gls{ris} phase shifts and the \glspl{ue}' \gls{ul} transmission power in \gls{ris}-assisted, \gls{uav}-enabled networks. However, the aforementioned framework does not consider the frequency of operation, and therefore provides limited insights on energy-efficient resource allocation on different portions of the spectrum. In~\cite{diamanti2021prospect}, a Stackelberg game is also discussed for \gls{e2e} energy-efficient optimization through power control and \gls{ris} reconfiguration in \gls{ul} \gls{ris}-assisted, \gls{uav}-enabled \gls{iab} networks. However, the focus is directed towards the low Sub-6~GHz portion of the spectrum, disregarding emerging \gls{5g} frequency bands, such as the \gls{mmwave}.

Lastly, the generation of \gls{rf} scenarios for Wireless Network Channel Emulators has been a trending topic. In~\cite{tehrani2021creating}, a framework for creating \gls{rf} scenarios for large-scale, real-time wireless emulators such as Colosseum, leveraging a commercial ray-tracer is introduced. It leverages efficient clustering techniques and channel impulse response re-sampling to scale down the large input set of \gls{rf} data to fewer parameters. In~\cite{villa2022cast}, the CaST Toolchain is presented. It comprises a framework for creating \gls{rf} scenarios from ray-tracing models for \gls{fpga}-based emulation platforms and a \gls{sdn} channel sounder to characterize the emulated channels. In~\cite{rusca2023mobile}, a framework for creating mobile \gls{rf} scenarios, including the presence of \glspl{uav}, based on real collected data, is discussed.

The contributions discussed above clearly demonstrate the significance of \gls{ris}-assisted \gls{nextg} channel models, as well as the substantial gains achieved in terms of energy efficiency through the deployment of \glspl{ris} in various heterogeneous scenarios and use cases. At the same time, the interest in the validation and testing of channel models, leveraging the full cellular protocol stack, in high-fidelity and accurate network channel emulators is ever-growing.

However, despite the established contributions of hierarchical game-theoretic approaches to energy efficient power control in wireless networks~\cite{lasaulce2009introducing,he2011stackelberg}, there has been limited research on the benefits of energy-efficient power control in \gls{ris}-assisted and \gls{uav}-enabled \gls{cv2x} networks in the \gls{fr1} and \gls{fr2}. Moreover, despite the availability of frameworks for generating \gls{rf} scenarios in channel emulators, no prior work has focused on installing \gls{ris}-assisted scenarios in these emulators. Consequently, a research gap exists in cross-layer experimental evaluation of \gls{ris}-assisted channels utilizing the full protocol stack on experimental platforms. Therefore, the evaluation of \gls{ris}-assisted networks across different network slices (e.g., \gls{embb} and \gls{urllc}) is currently missing. This section aims to fill these gaps in the literature with the following original contributions: (i)~we propose accurate \gls{phy} layer \gls{ris}-enabled channel modeling through \glspl{gbsm} and \gls{quadriga}; (ii)~we cascade the generated channels to include the effect of a \gls{ris}; (iii)~we leverage hierarchical game theory to optimize the \gls{ris} elements' phase shifts; (iv)~we study energy-efficient power control across different spectrum bands; (v)~we implement the \gls{ris}-assisted channels on the Colosseum testbed; and (vi)~we perform full-protocol stack emulation using srsRAN~\cite{srsran} to evaluate the performance gain in \gls{embb} and \gls{urllc} network slicing by fine-tuning the \gls{bs}'s network functions.

The remainder of this section is organized as follows. Section~\ref{sec:ris_propagation:cross_layer:gbsm} provides an overview on \gls{gbsm}-based \gls{nextg} channel modeling with \gls{quadriga}. Section~\ref{sec:ris_propagation:cross_layer:energy} discusses the simulation of a \gls{ris}-assisted \gls{cv2x} setup with \gls{quadriga}, as well as its evaluation on energy efficiency in the Sub-6~GHz and the \gls{mmwave} bands. Section~\ref{sec:ris_propagation:cross_layer:colosseum} details the installation and evaluation of a \gls{ris}-assisted topology on the Colosseum wireless network and channel emulator, as well as its evaluation on the \gls{embb} and \gls{urllc} network slices. Finally, Section~\ref{sec:ris_propagation:cross_layer:summary} concludes the section and discusses possible future directions.

\subsection[GBSMs with QuaDRiGa]{\texorpdfstring{\glspl{gbsm}}{GBSMs} with \texorpdfstring{\gls{quadriga}}{QuaDRiGa}}
\label{sec:ris_propagation:cross_layer:gbsm}

\subsubsection[Channel Modeling with QuaDRiGa]{Channel Modeling with \texorpdfstring{\gls{quadriga}}{QuaDRiGa}}
\label{sec:ris_propagation:cross_layer:gbsm:modeling}

\begin{figure}[ht]
    \centering
    \subfloat[Location from Google Maps~\cite{googlemaps}.\label{fig:Penn_topologies}]{%
        \includegraphics[width=0.30\textwidth]{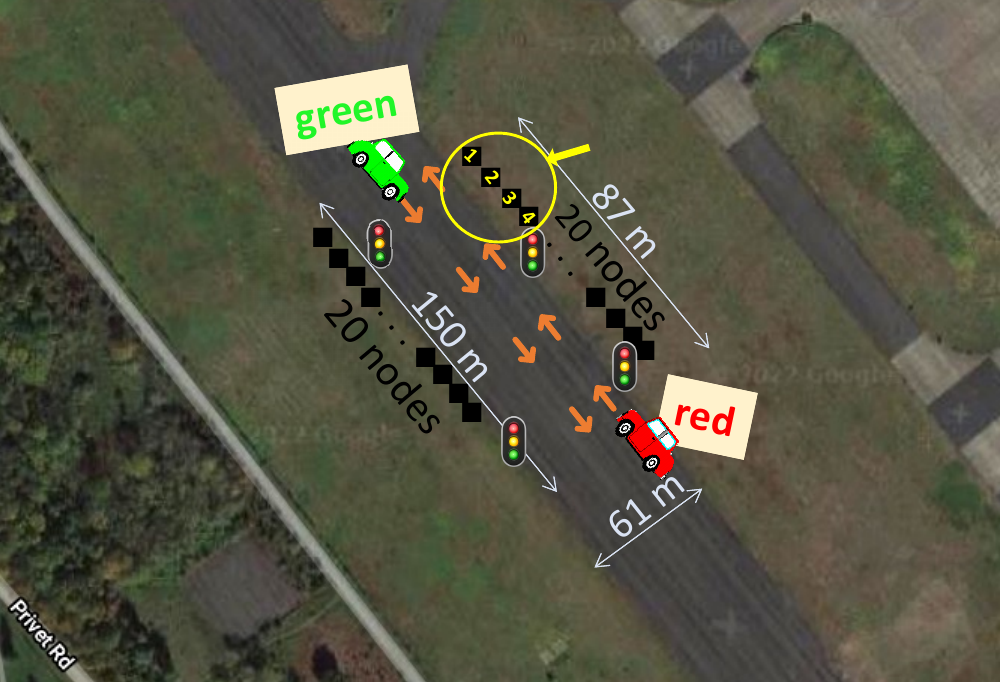}}
    \hfill
    \subfloat[Path Gain.\label{fig:v1_to_v2_100sims_0.1_updateRate}]{%
        \includegraphics[width=0.38\textwidth]{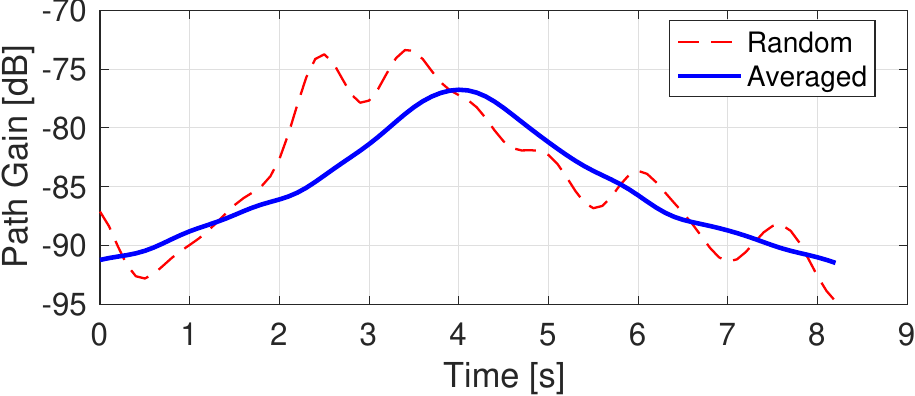}}
    \hfill
    \subfloat[Path Power vs.\ Delay vs.\ Distance.\label{fig:total}]{%
        \includegraphics[width=0.30\textwidth]{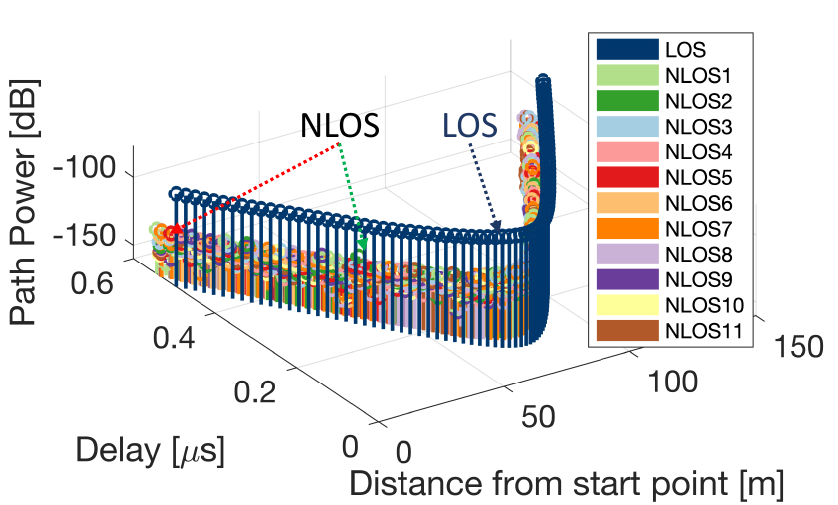}}

    \subfloat[Delay vs.\ Distance.\label{fig:total2}]{%
        \includegraphics[width=0.32\textwidth]{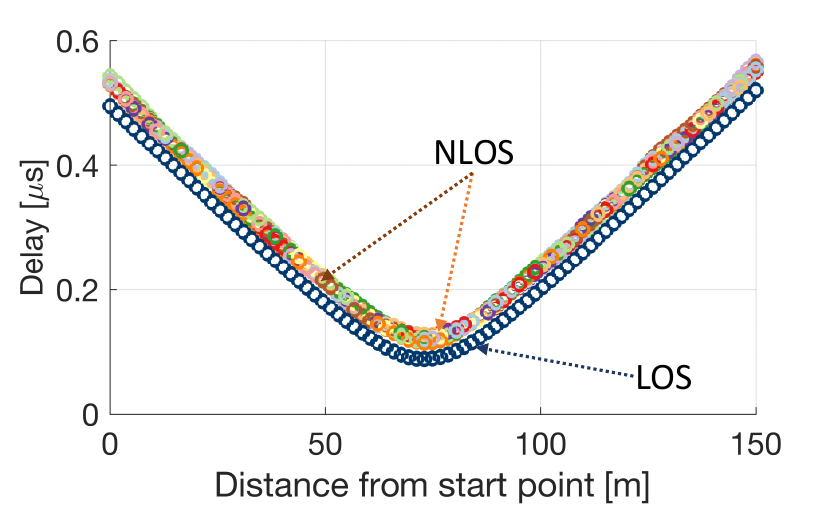}}
    % \hfill
    \subfloat[Path Power vs.\ Distance.\label{fig:total3}]{%
        \includegraphics[width=0.32\textwidth]{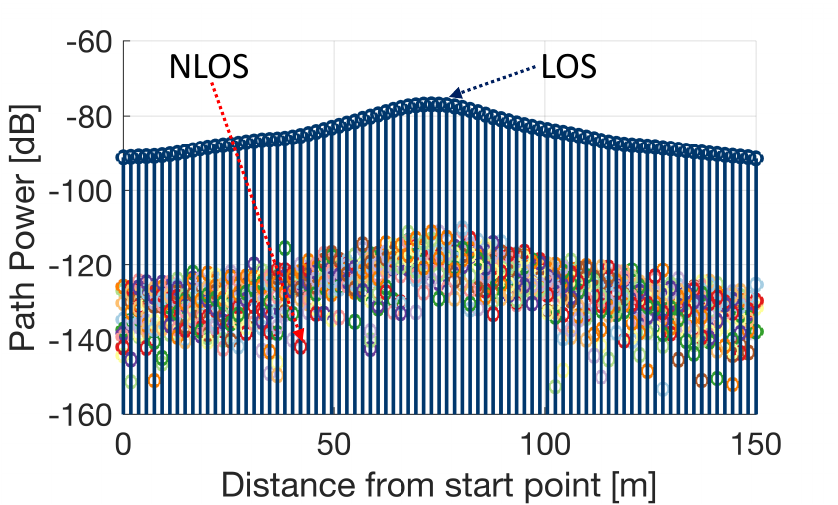}}

    \caption{Results averaged over $100$ channel realizations for the link describing the green vehicle's transmission to the red one in Willow Grove Naval Air Base in Horsham Township, Pennsylvania, USA.}
    \label{fig:V1toV2_pathpowers_LOS}
\end{figure}

The implementation of \glspl{gbsm} is fairly simple and computationally inexpensive, and
can be done through code-based simulators (e.g., NYUSIM~\cite{ju2019millimeter}, WINNER II/+~\cite{kyosti20074} and 3GPP-3D~\cite{ademaj20163gpp}), and Standardized Channel Models (e.g., the Spatial Channel Model (SCM), IMT-Advanced, METIS, mmMAGIC and the 5G-ALLSTAR~\cite{pang2022investigation}).
\gls{quadriga} extends the WINNER and 3GPP-3D models~\cite{mondal20153d}, with the advantage that it achieves a high level of precision~\cite{jaeckel2019f}.
It takes as input configuration files to describe a propagation scenario, and each file contains a parameter table with statistical information about the wireless channels (i.e., delay spread, Rician K-factor, shadow fading, \gls{xpr}, angular spread values and their properties such as correlation and distribution). The aforementioned files were created through the process of channel sounding either with measurements performed in urban cities~\cite{jaeckel2014quadriga, ranjkesh2015optimized}, or by the incorporation of statistical tables such as those included in~\cite{etsi19deliv}.
The modeling procedure can be divided into two parts: (i) a stochastic part that generates the \glspl{lsp}, e.g., path loss, shadowing and fading, and (ii) a geometry-based part that calculates the updated \glspl{ssp}, e.g., delay spreads~\cite{he2020investigation}.
In contrast to the classical ray-tracing approach, \gls{quadriga} does not use an exact geometric representation of the propagation environment.
The distribution of the scattering clusters (e.g., tree foliage and buildings) is random.
This adds simplicity into the implementation and decreases the computational cost without accuracy loss.
\gls{quadriga} simulates complicated \gls{nextg} heterogeneous environments~\cite{etsi5g},
by considering effects such as dual-mobility, 3D propagation and continuous time evolution. The generated parameters are spatially correlated,
while the wide-sense stationary (WSS) properties are kept both within a segment and during mobility. Finally, the simulation steps can be summarized as follows~\cite{arunachalaperumal2018enhanced, jaeckel2019f, jaeckel2014quadriga}.

\begin{enumerate}
\item \textbf{\textit{Definition of the \gls{ue} Input Variables.}}
The topology layout is defined  (i.e., \glspl{tx}/\glspl{rx}), location (i.e., coordinates) and antenna characteristics (i.e., carrier frequency and type). Second, the nodes' trajectories and their speed profile are determined. Last, a propagation scenario is set (\gls{los}/\gls{nlos}, or a combination). Examples of supported scenarios~\cite{jaeckel2019f,jaeckel2014quadriga,giordani2019path} are:

\begin{enumerate}
\item \textit{\textbf{Berlin/Dresden UMa}}
\begin{itemize}
\item Terrestrial \textit{macrocell} parameters extracted from measurements in Germany.
\end{itemize}

\item \textit{\textbf{WINNER model}}
\begin{itemize}
\item \textit{UMa C1 \& C2}: For \textit{macrocell} \glspl{bs} in urban \& sub-urban areas, respectively.
\item \textit{UMi B1}: For \textit{microcell} \glspl{bs} in urban areas.
\end{itemize}
\item \textit{\textbf{3GPP TR 37.885 Highway V2X }}
\begin{itemize}
    \item \textit{NLOS} due to dynamic blockages (e.g., vehicles).
\end{itemize}
\item \textit{\textbf{3GPP TR 38.901 NR model - \textit{UMa \& UMi}}}
\begin{itemize}
\item For center frequencies between 0.5 and 100 GHz.
\end{itemize}
\end{enumerate}

\item \textbf{\textit{Calculation of \glspl{lsp}.}} Out of the aforementioned parameter tables, the \glspl{lsp} are calculated. At this step, sources of scattering have been created, according to the multi-bounce approach, comprising a \gls{fbs} and a \gls{lbs}.

\item \textbf{\textit{Generation of the \gls{xpr}.}}
Antenna characteristics and polarization effects are taken into account.

\item \textbf{\textit{Calculation of the Channel Coefficients and Path Gain Generation.}}
The results of the previous steps along with \glspl{ssp} (e.g., angular characteristics of the paths) are combined to calculate the channel coefficients (i.e., the \glspl{mpc}). A last fine-tuning of the shadow fading and path loss parameters, along with the Rician factor are applied to produce the final path gains.

\item \textbf{\textit{Repetition of steps $1-4$ for successive channel traces and final Propagation Channels.}}
In case of mobility the process is repeated. All individual components produced by every step of the aforementioned procedure are combined to obtain the final channels.
\end{enumerate}

\subsubsection[Experimentation through Simulation with QuaDRiGa]{Experimentation through Simulation with \texorpdfstring{\gls{quadriga}}{QuaDRiGa}}
\label{sec:ris_propagation:cross_layer:gbsm:experimentation}

To test the software's effectiveness, we generate a \gls{cv2x} setup~\cite{mecklenbrauker2011vehicular} in which two vehicles approach each other under the 3GPP 37.885 Highway \gls{los} transmission conditions (Section~\ref{sec:ris_propagation:cross_layer:gbsm:modeling}). Both have a steady speed of $66$~km/h for a distance of $150$~m. The vehicles are equipped with omni-directional antennas mounted on their roof at $1.5$~m and operate at the center frequency of $5.9$~GHz. As it can be seen from the topology in Fig.~\ref{fig:Penn_topologies}, scattering objects such as $4$ \glspl{rsu} along with $40$ nodes on the highway's sides, placed $5$~m apart, representing other parked cars, are also present. The simulation's location corresponds to the Willow Grove Naval Air Base in Horsham Township, Pennsylvania, USA (\gls{gps} coordinates for latitude and longitude are $[40.202951, -75.149008]$), and the conversion from \gls{gps} to Cartesian coordinates has been done with MATLAB's~\cite{MATLAB:2021a} \texttt{latlon2local} function. Given the stochastic nature of the channels, and in order to obtain a good estimate, we observe the results every $0.1$~s and average them over $100$ channel realizations. In Fig.~\ref{fig:V1toV2_pathpowers_LOS}, we present the simulation's results, when the green transmits to the red vehicle, as depicted in Fig.~\ref{fig:Penn_topologies}. In Fig.~\ref{fig:v1_to_v2_100sims_0.1_updateRate} the continuous bell-shaped curve represents the averaged path gains, while the dotted represents the path gains for a random setup. The log-distance path loss~\cite{mecklenbrauker2011vehicular} model leveraged in the channel generation, explains the increase in path gains as the path loss decreases, particularly for a minimal distance between the vehicles. In the final path gain values in Fig.~\ref{fig:v1_to_v2_100sims_0.1_updateRate}, the main contribution comes from the direct \gls{los} component, while the contribution of the \gls{nlos} \glspl{mpc} is weighted to be insignificant, based on the input configuration file.
In Fig.~\ref{fig:total}, we include a 3D plot of the generated path powers versus delay and travelled distance, while in Fig.~\ref{fig:total2} and in Fig.~\ref{fig:total3}, we observe Fig.~\ref{fig:total} from two different perspectives. 12 paths are generated, one for the direct path, and 11 for the multiple copies of the signal, for each position of the mobile \gls{tx}/\gls{rx} (i.e., Fig.~\ref{fig:total3}). In Fig.~\ref{fig:total2}, the path power of the \gls{los} path attains a lower delay value compared to its \gls{nlos} counterparts for a given distance.
As the vehicles begin to approach each other, the delay values decrease, reaching their minimum around the midpoint of the track, where the two vehicles cross.

\subsubsection[GBSMs for Wireless Network Channel Emulators]{\texorpdfstring{\glspl{gbsm}}{GBSMs} for Wireless Network Channel Emulators}
\label{sec:ris_propagation:cross_layer:gbsm:emulators}

We install the aforementioned scenario in Colosseum, a publicly available testbed comprising $128$~\glspl{srn}. It consists of pairs of Dell PowerEdge R730 servers and NI \gls{usrp} X310 \glspl{sdr} that enable large-scale experimentation in diverse network deployments via the \gls{mchem} component. \gls{mchem} leverages \gls{fpga}-based \gls{fir} filters to replicate wireless environment conditions such as path loss, fading, attenuation, mobility, and interference pre-modeled through ray-tracing software, analytical models, or real-world measurements. The Colosseum \gls{tgen}, built on top of the \gls{mgen} \gls{tcp}/\gls{udp} traffic generator~\cite{mgen}, emulates different traffic profiles (e.g., multimedia content), demand, and distributions (e.g., Poisson, periodic), by emulating \gls{ip} traffic flows between the \glspl{srn}.

\begin{figure}[ht]
    \centering
    \subfloat[\gls{ul} Throughput {[Mbps]}.\label{fig:throughput}]{%
        \includegraphics[width=0.32\textwidth]{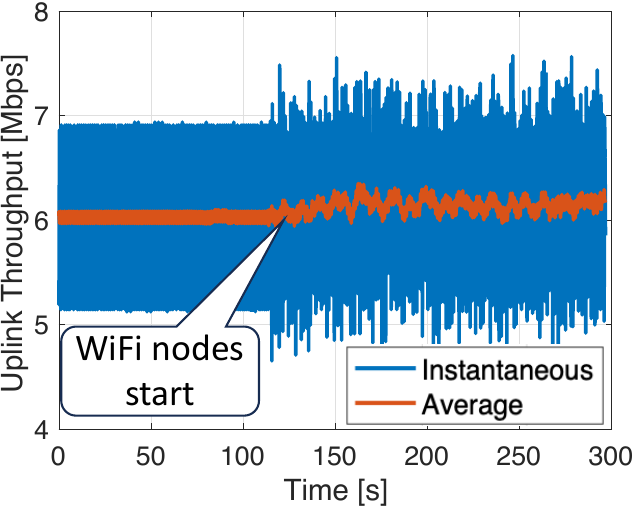}}
    \hspace{0.02\textwidth}
    \subfloat[\gls{ul} \gls{sinr}.\label{fig:snir}]{%
        \includegraphics[width=0.32\textwidth]{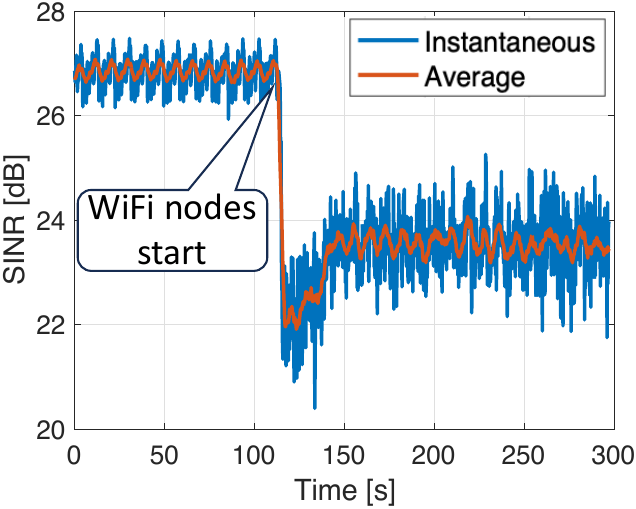}}
    \caption{\gls{ul} cellular metrics between \gls{bs}-node~$1$ and \gls{ue}-node~$2$.}
    \label{fig:43011cellularmetrics}
\end{figure}

We focus on nodes $1\rightarrow4$ as depicted in Fig.~\ref{fig:Penn_topologies}, which share a bandwidth of $10$~MHz. Nodes $1\rightarrow2$ are deployed through the srsRAN protocol stack~\cite{srsran}, while nodes $3\rightarrow4$ are deployed with the Wi-Fi Stack, which is based on the open-source GNU Radio implementation of the IEEE 802.11a/g/p standard~\cite{bloessl2017performance}. Cellular node $1$ and Wi-Fi node $3$ are \glspl{bs}, while nodes $2$ and $4$ serve as the respective cellular and Wi-Fi \glspl{ue}. Finally, \gls{udp} traffic is generated with \texttt{iPerf}.
In Figs.~\ref{fig:throughput} and~\ref{fig:snir}, we illustrate the \gls{ul} cellular \gls{sinr} and throughput between the cellular nodes.
Initially, an average \gls{sinr} of $27~\mathrm{dB}$ is observed, which drops to $22~\mathrm{dB}$, when the WiFi nodes start transmitting at $20$~dB (at approximately $t=120~\mathrm{s}$), before slightly recovering to $\sim24~\mathrm{dB}$. Similarly, the \gls{ul} throughput becomes unstable once WiFi transmissions commence.

Conclusively, with this simple yet richly detailed use-case, we are able to verify the suitability of \gls{quadriga} and \glspl{gbsm} for generating \gls{rf} scenarios for Colosseum, as well as the efficiency of the latter in generating and managing traffic among heterogeneous nodes. Notably, although the aforementioned scenario comprises multiple nodes, the focus is directed towards the aforementioned to demonstrate the software's capabilities, and hence an extensive evaluation is omitted. Readers interested in the creation of \gls{rf} scenarios for Colosseum can refer to~\cite{bonati2021colosseum,tehrani2021creating,villa2022cast}.

\subsection[Investigating Energy Efficiency with frequency-dependent GBSMs and RIS in C-V2X]{Investigating Energy Efficiency with frequency-dependent \texorpdfstring{\glspl{gbsm}}{GBSMs} and \texorpdfstring{\glspl{ris}}{RISs} in \texorpdfstring{\gls{cv2x}}{C-V2X}}
\label{sec:ris_propagation:cross_layer:energy}

Evidently, \gls{quadriga} is effective in providing accurate geometry-based stochastic channel modeling. Additionally, considering that the generated channels are frequency-dependent, \gls{quadriga} serves as an ideal candidate for investigating energy efficient power allocation across different spectrum bands and various use-cases. In this Section, we focus particularly on a \gls{cv2x} use-case~\cite{khan2022vehicle} involving a \gls{uav} acting as a flying \gls{gnb} serving multiple \glspl{ue}, to investigate energy efficiency in the Sub-6~GHz and \gls{mmwave} bands. Without loss of generality, we assume that the \gls{bs} simulated with \gls{quadriga} is a static \gls{uav} hovering at a certain height, while the generated channels are cascaded to simulate the inclusion of a \gls{ris} in the topology. We believe that the strong \gls{los} links provided by the \gls{uav} and the cascaded channels offered by \gls{ris} technology fit into the description of a \gls{cv2x} scenario for \gls{nextg} wireless environments and smart cities as described in~\cite{diamanti2021energy}.

\subsubsection[System Model: A RIS-assisted C-V2X setup with QuaDRiGa]{System Model: A \texorpdfstring{\gls{ris}}{RIS}-assisted \texorpdfstring{\gls{cv2x}}{C-V2X} setup with \texorpdfstring{\gls{quadriga}}{QuaDRiGa}}
\label{sec:ris_propagation:cross_layer:system_model}

\begin{figure}[ht]
    \centering
    \includegraphics[width=0.45\textwidth]{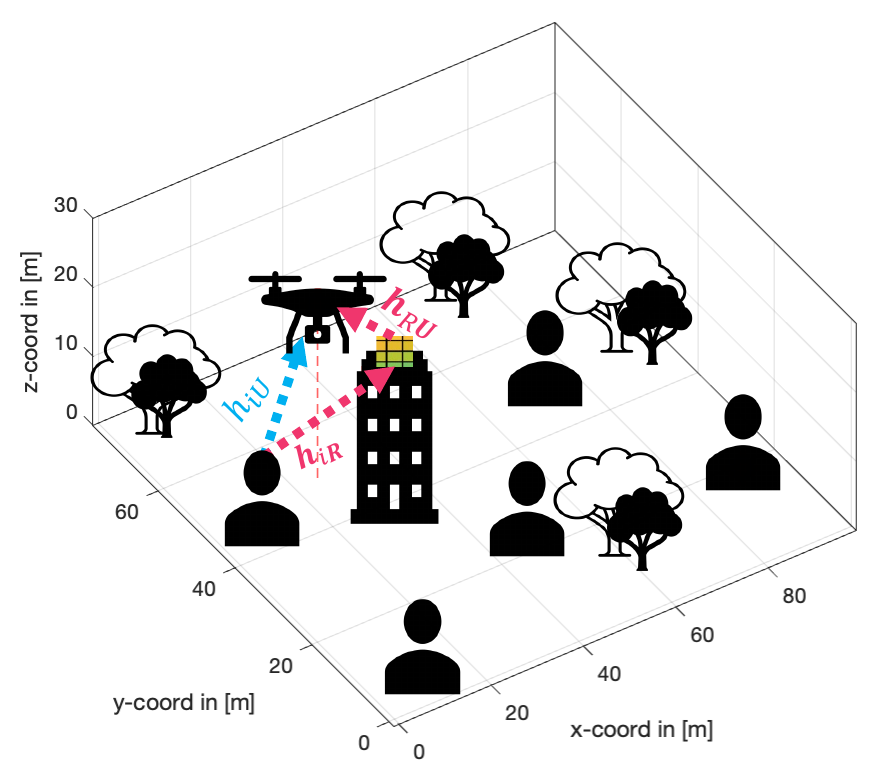}
    \caption{\gls{ris}-assisted and \gls{uav}-enabled communications system.}
    \label{fig:ris-uav}
\end{figure}

As illustrated in Fig.~\ref{fig:ris-uav}, a \gls{ris}-assisted and \gls{uav}-enabled wireless communications system is evaluated on the \gls{ul} direction. Specifically, the topology comprises a \gls{uav}, a building facade hosting a \gls{ris} and a set of mobile \glspl{ue}, denoted as $I=\{1,\dots,i,\dots, |I|\}$, which communicate directly with the \gls{uav}. The \gls{ris} consists of $|M|$ reflecting elements, the set of which is defined as $M=\{1, \dots , m, \dots , |M|\}$, and each \gls{ris} element's phase shift is $\theta_m \in [0,2\pi]$, $\forall~m~\in~M$. Finally, the corresponding diagonal phase-shift matrix is defined as $\boldsymbol{\Theta}=diag( e^{j\theta_1},\dots,e^{j\theta_m},\dots,e^{j\theta_{|M|}} )$.
The incoming signals at the \gls{uav} are combined, as in~\cite{diamanti2021energy}, so that each \gls{ue}'s $i$ signal is the outcome of the coherent addition of the direct and the reflected signal. The channel gain of the direct link between a user $i$ and the \gls{uav} is described as $h_{iU}$, the channel gain between a user $i$ and the \gls{ris} is defined as $\mathbf{h}_{iR}$, where $\mathbf{h}_{iR}$ can also be written as $\mathbf{h}_{iR}=[|h_{iR,1}| e^{j \omega_{1}}, \dots,|h_{iR,m}| e^{j \omega_{m}}, \dots,|h_{iR,|M|}| e^{j\omega_{|M|}}]^T$, and finally, the channel gain of the reflected link from the \gls{ris} to the \gls{uav} is denoted as $\mathbf{h}_{RU}$. Subsequently, the cascaded channel is given as $\mathbf{h}_{RU}^H\boldsymbol{\Theta}\mathbf{h}_{iR}$. Last, the steering vector for the \gls{ue} $i$ to the \gls{ris} \gls{los} link is defined as $\mathbf{h}_{iR}^{\gls{los}}=[1, e^{-j \frac{2 \pi}{\lambda} d \phi_{iR}}, \dots, e^{-j \frac{2 \pi}{\lambda}(|M|-1) d \phi_{iR}}]^{T}$, while the steering vector for the \gls{ris} to the \gls{uav} \gls{los} link is denoted as $\mathbf{h}_{RU}^{\gls{los}}=[1, e^{-j \frac{2 \pi}{\lambda} d \phi_{RU}}, \dots, e^{-j \frac{2 \pi}{\lambda}(|M|-1) d \phi_{RU}}]^T$. Moving forward, we keep using the same notation as in~\cite{diamanti2021energy}, which is also reported in Table~\ref{table:not-chmod}. The first \gls{ris} element, i.e., $m=1$, with its respective coordinates, i.e., $(x_R, y_R, z_R)$ [m], is used as a reference point in the following calculations, while for the \gls{uav}, its coordinates are given as follows $(x_U, y_U, z_U)$ [m]. It is noted that in~\cite{diamanti2021energy}, the featured channel model is frequency-independent.
In our study, we aim at observing the impact of the frequency of operation in terms of energy efficiency on realistic \gls{ris}-assisted wireless channels on two different 5G bands, namely the Sub-6~GHz and the \gls{mmwave} band. It is highlighted that \glspl{ris} are expected to work across multiple frequency ranges, including both the \gls{fr1} and \gls{fr2} bands~\cite{ETSI_RIS001}, to ensure interoperability across diverse deployment scenarios and applications. Given that \gls{quadriga} can reliably support the simulation of channels up to $100$~GHz, as mentioned in Section~\ref{sec:ris_propagation:cross_layer:gbsm}, the Sub-6~GHz and \gls{mmwave} bands are chosen. Therefore, we experiment with two different values for the center frequency, i.e., $5.9$~GHz and $28$~GHz, respectively. Finally, both the \glspl{ue} and the \gls{uav} are equipped with single-antenna omni-directional \glspl{tx}/\glspl{rx}. It is noted that although omni-directional antennas are commonly used in the Sub-6~GHz band~\cite{paul2022omni,azim2021multi,rani2022development,sharawi2017two,de2021radio,zaidi2020wide}, and they can be deployed on \gls{mmwave} communication systems~\cite{mao2018planar,ranvier2008low,fan2018wideband,maccartney2015millimeter,hasan2019dual}, this is not typical~\cite{sun2015synthesizing}. Instead, high-directional antennas are most commonly used in such scenarios~\cite{abirami2017review,niu2015survey,pan2011dual}, to combat severe propagation loss and achieve high antenna gains. Within the scope of this work, we focus on the use of omni-directional antennas for the \gls{mmwave} to provide insights into the performance enhancement introduced by \gls{ris} in cases where omni-directional antennas are convenient for deployment, such as in a \gls{cv2x} environment~\cite{mao2020compact}. While it is indeed true that in a dynamic environment such as the aforementioned, a directional antenna can improve \gls{rssi}, this increased directionality comes at the cost of precise alignment with the \gls{ris} and \glspl{ue}. Therefore, continuous beam tracking and adaptation are needed, which introduce additional overhead and latency that may outweigh the benefits of directive gains. Additionally, highly accurate and precise channel modeling with directional antennas in urban \gls{cv2x} environments, where the effects of multipath propagation are strong, can prove rather challenging, mainly due to the narrow \gls{mmwave} beams, which lead to position deviations, inaccuracies, and alignment errors, resulting in beam misalignment, signal loss, and significant performance drops.

\begin{table}[H]
\centering
\small
\caption{Channel Modeling Notation}
\begin{adjustbox}{width=0.85\linewidth}
\begin{tabular}{@{}ll@{}}
\toprule
\textbf{Symbol} & \textbf{Description} \\
\midrule
$|I|$ & Number of mobile \glspl{ue} \\
$I$ & Set of mobile \glspl{ue} \\
$|M|$ & Number of reflecting elements on the \gls{ris} \\
$M$ & Set of \gls{ris}'s reflecting elements \\
$\theta_m$, $\forall m \in M$ & Reflecting element's phase shift \\
$\mathbf{h}_{RU}$ & Channel gain for the \gls{ris} to \gls{uav} reflected link \\
$\boldsymbol{\Theta}$ & Diagonal phase-shift matrix \\
$\mathbf{h}_{iR}$ & Channel gain for the \gls{ue} $i$ to \gls{ris} link \\
$\mathbf{h}_{RU}^H\boldsymbol{\Theta}\mathbf{h}_{iR}$ & Cascaded channel gain \\
$h_{iU}$ & Channel gain for the \gls{ue} $i$ to \gls{uav} direct link \\
$d_{RU}$ [m] & Euclidean distance between the \gls{ris} and the \gls{uav} \\
$d_{iR}$ [m] & Distance between the \gls{ue} $i$ and the \gls{ris} \\
$d_{iU}$ [m] & Distance between the \gls{ue} $i$ and the \gls{uav} \\
$\phi_{RU}$ & Cosine of the signal's \gls{aod} for the \gls{ris} to \gls{uav} link \\
$\phi_{iR}$ & Cosine of the signal's \gls{aoa} for the \gls{ue} $i$ to \gls{ris} link \\
$\lambda$ [m] & Carrier wavelength \\
$d$ [m] & Antenna separation \\
$\mathbf{h}_{RU}^{\gls{los}}$ & Steering vector for the \gls{ris} to \gls{uav} \gls{los} link \\
$\mathbf{h}_{iR}^{\gls{los}}$ & Steering vector for the \gls{ue} $i$ to \gls{ris} \gls{los} link \\
$h^\prime_{xy}$ & \gls{mpc} gain, where $x \in \{i, R\}, y \in \{R, U\}$ \\
$|h^\prime_{xy}|, \varphi^\prime_{xy}$ & Magnitude and phase of the \gls{mpc} \\
$|N|$ & Number of generated \glspl{mpc} \\
\bottomrule
\end{tabular}
\end{adjustbox}
\label{table:not-chmod}
\end{table}

Last, the channels in~\cite{diamanti2021energy}, from which our use-case is inspired, rely on the Rice and Rayleigh mathematical models to emulate the wireless links. However, their simple yet effective implementation may not accurately represent real-world wireless environments~\cite{dang2021geometry}. For these reasons, we proceed by creating the channels with \gls{quadriga}. The channel generation process is described in Section~\ref{sec:ris_propagation:cross_layer:gbsm:modeling}, where for each generated link a number of \glspl{mpc} is produced based on the Channel Generator's input scenario, in our case the 3GPP-compliant scenario, 3GPP\textunderscore38.901\textunderscore UMa. Finally, we assume that the links between the \glspl{ue} and the \gls{uav} are \gls{nlos}, while those remaining, i.e., the channels between the \glspl{ue} and the \gls{ris}, and the \gls{ris} to \gls{uav}, are \gls{los}. The transmission conditions of the aforementioned links were selected to ensure that the cascaded channel created by the \gls{ris} is strong enough to redirect the \glspl{ue}' signals to the \gls{uav}, for both the Sub-6~GHz and, especially, the \gls{mmwave} band, which suffers from transmission losses. The wireless channels are observed every $0.1$~s and for their creation, the steps listed below are followed:

\begin{itemize}
\item Generate scenario-based \glspl{mpc} (i.e., $h_{xy}^\prime$) for every link.
\item Perform coherent summation of the \glspl{mpc}~\cite{tehrani2021creating} to obtain the link's channel gain, as reported in~\eqref{eq:mpc-def}:
\begin{equation}\label{eq:mpc-def}
    h^{\prime\prime}_{xy}=\sum_{j=1}^{|N|}\left|h_{xy}^\prime\right| \cdot e^{j \varphi_{xy}^\prime}.
\end{equation}
\item Average the channel gains (i.e., $h^{\prime\prime}_{xy}$) over $100$ different channel realizations to obtain the link's \textit{\textbf{final}} channel gain (e.g., $h_{iU}$).
\item Multiply the generated channels with the corresponding steering vectors, i.e., $\mathbf{h}_{RU}^{\gls{los}}$ and $\mathbf{h}_{iR}^{\gls{los}}$, to get the respective final channel gains (i.e., $\mathbf{h}_{RU}$, $\mathbf{h}_{iR}$).
\end{itemize}

The overall channel power gain for the \gls{ue} $i$ to \gls{uav} link is denoted as $G_{i}=|h_{iU}+\mathbf{h}_{RU}^H \boldsymbol{\Theta} \mathbf{h}_{iR}|^2$, and the \glspl{ue} are able to communicate concurrently with the \gls{uav} thanks to the \gls{noma} technique. The received channel gains (i.e., $G_{i}$) at the \gls{uav} are sorted and decoding starts from the \gls{ue} with the highest channel gain. Finally, the \gls{uav}'s \gls{rx} decodes the received superposed signal with the help of the \gls{sic} technique, which is implemented on its end. The interference sensed by each \gls{ue} is defined as follows in~\eqref{eq:eq1}:

\begin{equation} \label{eq:eq1}
    I_{i}=\sum_{j<i} G_{j} P_{j}+I_{0},
\end{equation}

\noindent where $P_{j}$ [W] is the \gls{ul} \gls{tx} power of the \gls{ue} $j$ and $I_0$ is the power of the zero-mean \gls{awgn}. Each \gls{ue}'s achieved \gls{sinr} is given as follows in~\eqref{eq:eq2}:
\begin{equation} \label{eq:eq2}
\gamma_{i}=\frac{P_{i} G_{i}}{I_{i}}.
\end{equation}

\subsubsection{Hierarchical Game-Theoretic Power Control}
\label{sec:ris_propagation:cross_layer:powercontrol}

In this Section, we aim to study how the frequency-dependent \gls{gbsm}-based system model presented in Section~\ref{sec:ris_propagation:cross_layer:system_model} will impact the energy efficiency in both the Sub-6~GHz and \gls{mmwave} bands. In brief, both the \glspl{ue} and the \gls{uav} indulge in a single-leader multiple-followers Stackelberg Game, where the target is to jointly maximize the \gls{uav}'s overall received signal strength and each \gls{ue}'s energy efficiency in a distributed manner. In detail, the \gls{uav} acting as a leader computes the \gls{ris} elements' effective phase shifts (and intelligently steers the signals reflected by the \gls{ris}) with a goal to enhance its received signal quality. In turn, the \glspl{ue}, acting as followers, observe the leader's action and through the formulation of a non-cooperative game aim at maximizing their energy efficiency, by determining their optimal \gls{ul} \gls{tx} power. Each \gls{ue}'s utility function is defined in~\eqref{eq:eq3}:
\begin{equation} \label{eq:eq3}
    U_{i}\left(P_{i}, \mathbf{P}_{-i}\right)=\frac{W \cdot\left(1-e^{-\alpha \gamma_{i}}\right)^{M}}{P_{i}},
\end{equation}

\noindent where $\mathbf{P}_{-i}$ is the \gls{tx} power vector of all the \glspl{ue} except for \gls{ue} $i$, $W$ [Hz] is the system's bandwidth, and $\alpha,M \in \mathbb{R}^{+}$ are parameters which control the utility function's slope. The adopted utility function for energy efficiency represents the trade-off between the achieved \gls{qos} satisfaction, i.e., \gls{sinr}, and the invested \gls{ul} \gls{tx} power~\cite{diamanti2021energy, tsiropoulou2015combined}. In detail, the numerator of the utility function, i.e., $(1 - e^{-\alpha \gamma_i})^M$, is a sigmoidal curve which expresses the \gls{ue} satisfaction, with respect to its \gls{qos} demands, while the denominator expresses the associated cost with reference to the transmission power levels. Because of the nature of the sigmoidal curve, small increases in $\gamma_i$ (by increasing the \gls{tx} power) will initially improve the utility. However, increasing $\gamma_i$ further requires significant power while providing diminishing returns in utility. As a result, the optimal strategy is to transmit at lower power levels in order to maintain higher energy efficiency.

Finally, the joint optimization problem is formulated as follows, with the respective problem solved by the \gls{uav} written as:
\begin{subequations}
\begin{equation}
\max_{\boldsymbol{\theta}} \sum_{\forall i \in I} P_{i} |h_{iU}+\mathbf{h}_{RU}^H \boldsymbol{\Theta} \mathbf{h}_{iR}|^2
\label{eq4a}
\end{equation}
\begin{equation}
\text{s.t.\quad}0\leq\theta_m \leq 2\pi, \forall m\in M.
\label{eq4b}
\end{equation}
\end{subequations}

With regard to the optimization problem solved by the \glspl{ue}, the latter ones participate in a non-cooperative game defined as $G=[I,\{A_i\}_{\forall i \in I}, \{U_i\}_{\forall i \in I}]$. In detail, $I$ is the set of \glspl{ue} (e.g., players), $A_{i}=[P_i^{min}, P_{i}^{max}]$ is each \gls{ue}'s strategy set, and $U_{i}$ is their utility function. $P_{i}^{min}$ [W] and $P_{i}^{max}$ [W] correspond to each \gls{ue}'s minimum and maximum \gls{ul} \gls{tx} power levels respectively, where $P_i^{min}$, is determined by the \gls{sic} prerequisite at the \gls{uav}'s \gls{rx} as follows:
\begin{equation} \label{eq5}
G_i P_i - I_i \geq P_{tol}, \forall i \in \{2,\dots,i,\dots, |I|\},
\end{equation}

\noindent such that each \gls{ue}'s signal is decoded successfully. According to~\eqref{eq5}, each \gls{ue}'s signal strength, i.e., $G_i P_i$, should be greater than or equal to the \gls{ue}'s $i$ sensed interference $I_i$, which is broadcasted by the \gls{uav} to all the \glspl{ue}, plus a minimum acceptable power level $P_{tol} \in \mathbb{R}^+$, which represents the \gls{sic} \gls{rx}'s tolerance/sensitivity. Finally, the maximum \gls{ul} \gls{tx} power level $P_i^{max}$ is imposed by the maximum power budget of the \gls{ue}'s device. The optimization problem is finally determined as follows:

\begin{subequations}\label{eq:7total}
\begin{equation}
\max_{P_{i} \in A_{i}} U_{i}\left(P_{i}, \mathbf{P}_{-i}\right)=\frac{W \cdot\left(1-e^{-\alpha \gamma_{i}}\right)^{M}}{P_{i}}, \forall i \in I
\label{eq7a}
\end{equation}
\begin{equation}
\text{s.t.\quad}G_i P_i - I_i \geq P_{tol}, \forall i \in \{2,\dots,i,\dots, |I|\}
\label{eq7b}
\end{equation}
\begin{equation}
P_i \leq P_i^{max}, \forall i \in I.
\label{eq7c}
\end{equation}
\end{subequations}

The non-cooperative game's, $G$, outcome is a Nash equilibrium, which determines the \glspl{ue}' optimal \gls{tx} power vector, i.e., $\mathbf{P}^{*}=[P_{1}^{*}, \dots, P_{i}^{*}, \dots, P_{|I|}^{*}]$.

The solution of the optimization problem taking place on the \gls{uav}'s side and defined in~\eqref{eq4a}-\eqref{eq4b} is the \gls{ris} elements' effective phase shifts' vector $\boldsymbol{\theta^{*}}$. It is noted that to maximize the overall received signal strength, the \gls{uav} only needs to calculate the effective \gls{ris} elements' phase shifts (i.e., $\boldsymbol{\theta^*}$) that maximize the overall channel power gain of the \glspl{ue}. The aforementioned optimization problem is non-convex, making it challenging to find a globally optimal solution.
As described in~\cite{diamanti2021energy,li2020reconfigurable}, when there is a single \gls{ue} in the system, the optimal phase shifts of the \gls{ris} elements are given by the following closed-form solution:
\begin{equation} \label{eq8}
\theta_{m}^{*}=\angle{h_{iU}}+\omega_{m}+\frac{2 \pi}{\lambda} d(m-1) \phi_{RU}, \forall m \in M.
\end{equation}
To derive the closed-form phase-shift solution in~\eqref{eq8} and maximize the received signal strength at the \gls{uav}, we aim to achieve phase alignment of the received signals from different transmission paths. Notably, this can be achieved through coherent signal construction of the incoming signals at the \gls{uav}. Indeed, the \gls{ue}'s channel power gain is maximized when the direct signal and the signal reflected by the \gls{ris} are perfectly aligned and coherently combined at the \gls{uav}'s \gls{rx}~\cite{li2020reconfigurable}. This generally holds true when the phase shifts of the direct and the cascaded signals are equal, as follows $\angle h_{iU} = \angle \left( \mathbf{h}_{RU}^{H} \mathbf{\Theta} \mathbf{h}_{iR} \right)$. Therefore, there's an optimal $1 \times |M|$ phase-shift vector $\boldsymbol{\theta^*} = \angle \mathbf{v}$ for the single-\gls{ue} case. Similarly, in the multi-\gls{ue} case, there exists a distinct reflection-coefficient vector $\mathbf{v}_i = [v_{i,1}, \ldots, v_{i,|M|}] \in \mathbb{C}^{|M| \times 1}$ for each \gls{ue} $i$ that maximizes its channel power gain. In the multi-\gls{ue} case, the solution to the leader's maximization problem is a linear combination of the phase shifts of the \gls{ris} elements that enhances each \gls{ue}'s signal strength, with the reflection-coefficient vectors $\mathbf{v}_i$ being distinct for each \gls{ue}.

On the other hand, the optimization problem solved on the \glspl{ue}' side will result in the determination of their optimal \gls{ul} \gls{tx} power vector $\mathbf{P^{*}}$, as mentioned, since given the \gls{ris} elements' effective phase shifts, each \gls{ue} determines its optimal \gls{ul} \gls{tx} power $P_{i}^{*}$ via non-cooperatively interacting with the rest of the \glspl{ue} in the action space. The Stackelberg equilibrium of this hierarchical game-theoretic approach can be given as $(\boldsymbol{\theta^{*}}, \mathbf{P^{*}})$.

\newtheorem{Definition}{\textbf{Definition}}
\begin{Definition}\label{def:nash}
\textbf{\textit{(Nash Equilibrium)}} The \gls{ul} \gls{tx} power vector $\mathbf{P^{*}}=[P_{1}^{*}, \dots, P_{i}^{*}, \dots, P_{I}^{*}]$ is a Nash equilibrium of the non-cooperative game $G=[I,\{A_i\}_{\forall i \in I}, \{U_i\}_{\forall i \in I}]$ if for every \gls{ue} $i\in I$, it holds that $U_i(P_i^*,\mathbf{P}_{-i})\geq U_i(P_i,\mathbf{P}_{-i}), \forall P_i\in A_i$.
\end{Definition}

At the Nash equilibrium, the \gls{ue} has no incentive to improve its achieved utility by altering its transmission power strategy, given the strategies of the rest of the players (i.e., the other \glspl{ue}).

\newtheorem{Theorem}{\textbf{Theorem}}
\begin{Theorem}\label{thm:nash}
\textbf{\textit{(Existence \& Uniqueness of a Nash Equilibrium)}} For the non-cooperative game given as $G=[I,\{A_i\}_{\forall i \in I}, \{U_i\}_{\forall i \in I}]$, there exists a unique Nash equilibrium point, which is defined as follows:
\begin{equation} \label{eqnash}
P_{i}^{*}= \max\left\{P_{i}^{min}, \min\left\{\frac{\gamma_{i}^{*} \cdot I_{i}}{W G_{i}}, P_{i}^{max}\right\}\right\}, \forall i \in I,
\end{equation}
where $\gamma_{i}^{*}$ represents the unique positive solution to the equation $\frac{\partial f(\gamma_i)}{\partial \gamma_{i}} \gamma_{i}-f(\gamma_{i})=0, f(\gamma_{i})=(1-e^{-\alpha \gamma_i})^{M}$.
\end{Theorem}

\begin{figure}[ht]
  \centering
  \includegraphics[width=0.55\textwidth]{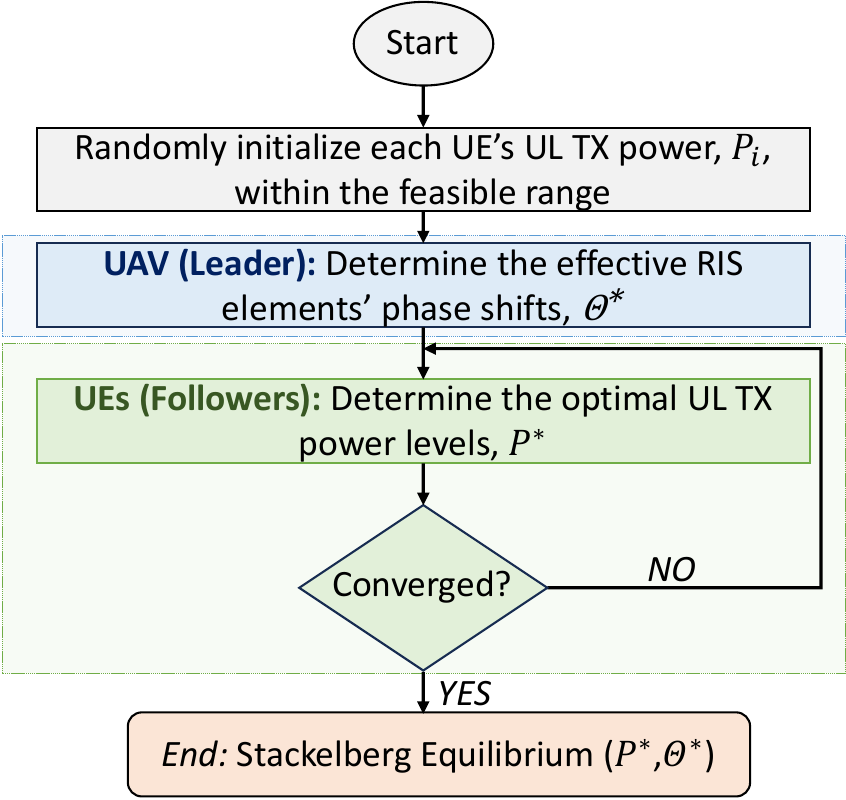}
  \caption{A hierarchical Stackelberg game-theoretic approach to energy-efficient power control.}
  \label{fig:game-flow}
\end{figure}

The Theorem's proof relies on the quasi-concavity property of the utility function $U_{i}(P_{i}, \mathbf{P}_{-i})$ with respect to the \gls{ul} \gls{tx} power $P_i$. The quasi-concavity property is proven and further explained in~\cite{tsiropoulou2015combined}, which also provides a reference guide with detailed steps of the proof. Therefore, given the \gls{ris} elements' effective phase shifts $\theta^{*}_{m}, \forall m \in M$, and the optimal transmission power vector $P_{i}^{*}, \forall i \in I$, the Stackelberg equilibrium is iteratively determined as presented in Fig.~\ref{fig:game-flow}. Notably, the convergence of the \glspl{ue}' strategies to the Nash equilibrium point is achieved through the implementation of a Best Response Dynamics algorithm~\cite{tsiropoulou2015combined}, as further detailed in Algorithm~\ref{alg:stackelberg_optimization}. Note that the superscript ($j$) indicates the iterations required for the non-cooperative game played among the \glspl{ue}. Additionally, it is highlighted that the maximum \gls{tx} power levels for the \glspl{ue} (as defined in Step $3$ of Algorithm~\ref{alg:stackelberg_optimization}) are determined by 3GPP standards and are set to $23$~dBm~\cite{etsi2022,etsi2022b,etsi2017}. The corresponding minimum value is set to $-20$~dBm.

\begin{algorithm}
\caption{Stackelberg Game-Theoretic Optimization}
\label{alg:stackelberg_optimization}
\begin{algorithmic}[1]
    \STATE Initialize the network topology, including the locations of \glspl{ue}, \gls{ris}, and the \gls{uav}.
    \STATE Generate the channels using \gls{quadriga}, as described in Section~\ref{sec:ris_propagation:cross_layer:system_model}.
    \STATE Initialize randomly $P_i \in [10^{-5}, P_i^{\max}]$.
    \STATE Determine \gls{ris} elements' phase-shift adaptation by solving \eqref{eq4a}--\eqref{eq4b} and calculate $G_i, \forall i \in I$.
    \STATE Sort users according to $G_i$, so decoding starts from the \gls{ue} with the highest channel gain.
    \STATE Set $j = 0$.
    \REPEAT
        \STATE Set $j = j + 1$.
        \FOR{$i \in I$}
            \STATE Determine the optimal \gls{ul} \gls{tx} power $P_i^*(j)$ by solving~\eqref{eq7a}-\eqref{eq7c}.
        \ENDFOR
    \UNTIL{$|P_i^*(j) - P_i^*(j-1)| \leq \epsilon, \forall i \in I$, where $\epsilon \approx 10^{-4}$.}
\end{algorithmic}
\end{algorithm}

We consider the following setup, where a \gls{uav} hovers at the point $(x_U=25,y_U=50,z_U=25)$ [m] of the three-dimensional space serving a \gls{noma} cluster of $|I|=5$ users. Their distance from the \gls{ris} is given as $d_{i,R}=[20, 27, 37, 58, 66]$~[m] and all are randomly distributed around the \gls{ris}'s reference point, which is given as $(x_R=30,y_R=40,z_R=20)$ [m]. The system's bandwidth is equal to $W=10$~MHz, and the \gls{awgn} power is $I_0=-140$~dBm for the Sub-6~GHz and $I_0=-160$~dBm for the \gls{mmwave} band, respectively. It is noted that \gls{awgn} noise is already added in the final path gain values through \gls{quadriga}~\cite{jaeckel2019f}. Specifically, in \gls{quadriga}, \gls{awgn} noise is used to generate spatially correlated \glspl{lsp}, which statistically characterize the propagation environment. These \glspl{lsp} indirectly shape the multipath environment and influence path gain.
Thus, the \gls{awgn} values (i.e., $I_0$), which are included in the calculation of the \gls{ue}'s interference, represent reference noise levels.
Indeed, these two distinct values are a correction applied to the \glspl{ue}' sensed interference, i.e., $I_{i}$, defined in~\eqref{eq:eq1}, so that the optimization framework can be evaluated for the respective spectrum bands under the same configuration of the utility's slope control parameters, defined as $\alpha=0.3$ and $M=3$. Finally, the frequencies are centered to $5.9$ and $28$~GHz for the respective bands, the mobile \glspl{ue}' maximum power budget is set to $P_i^{max}=23$~dBm, and the sensitivity of the \gls{uav}'s \gls{rx} is set to be equal to $P_{tol}=-150$~dBm.

It is finally noted that in the following, the performance evaluation of the \gls{ris}-assisted and \gls{uav}-enabled communication system depicted in Fig.~\ref{fig:ris-uav} will take place under $10$, $100$, and $1000$ \gls{ris} elements for \textit{both} the Sub-6~GHz and the \gls{mmwave} band. Although it generally holds true that for a given \gls{ris} dimension, a \gls{mmwave} \gls{ris} will host a higher number of \gls{ris} elements than a Sub-6~GHz \gls{ris}, given the same element spacing (e.g., $\lambda/2$), increasing the number of elements onboarded on the \gls{ris} also increases the complexity of its handling~\cite{rossanese2022designing}. Consequently, a trade-off emerges between hosting thousands of \gls{ris} elements and maintaining ease of operation, particularly when experimenting in the \gls{mmwave} band, where location inaccuracies are of utmost importance. For the aforementioned reasons, investigating the number of the elements onboarded on the \gls{ris} is deemed necessary.

\begin{figure}[ht]
    \centering
    \subfloat[Sub-6 GHz: allocated \gls{ul} \gls{tx} power per \gls{ue}.\label{fig:sub6ghz_p}]{%
        \includegraphics[width=0.46\columnwidth]{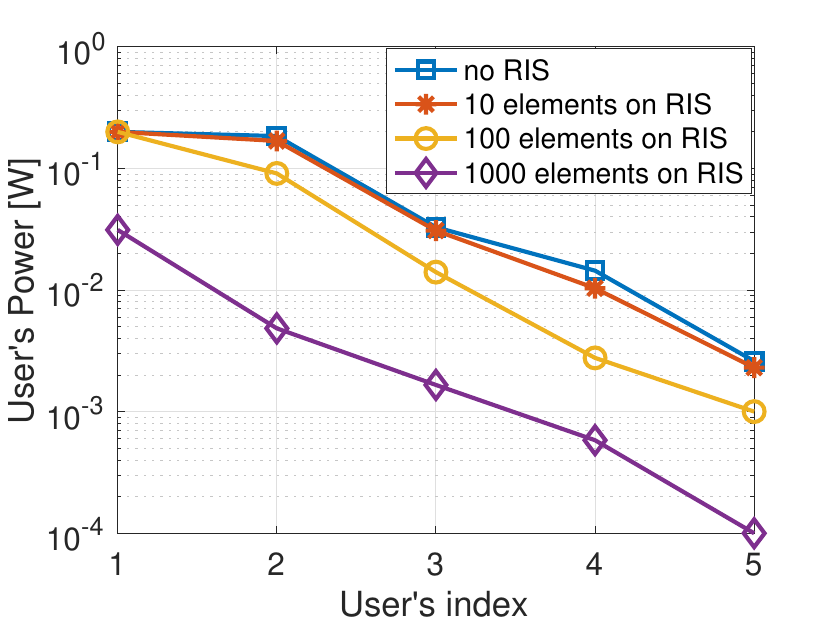}}
    \hfill
    \subfloat[Sub-6 GHz: achieved utility per \gls{ue}.\label{fig:sub6ghz_utility}]{%
        \includegraphics[width=0.46\columnwidth]{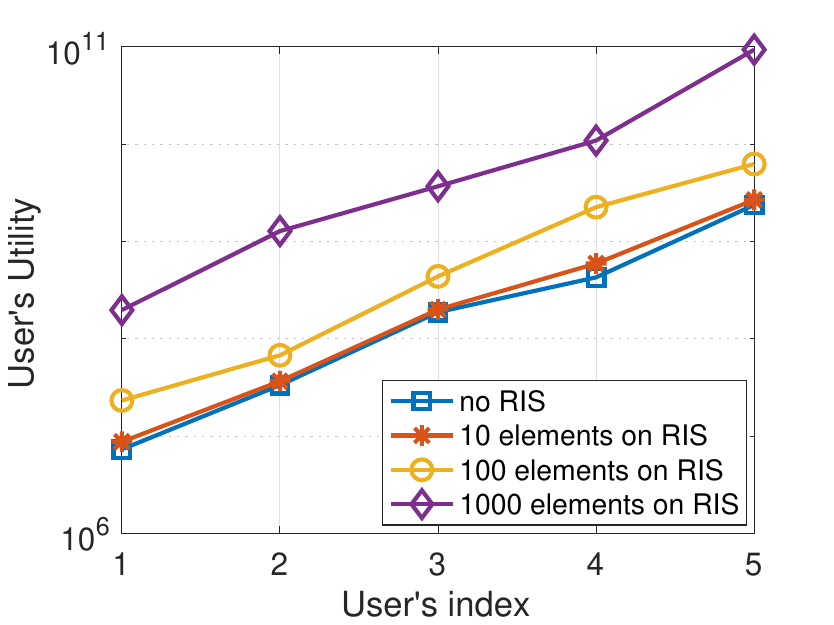}}

    \subfloat[\gls{mmwave}: allocated \gls{ul} \gls{tx} power per \gls{ue}.\label{fig:mmwave_p}]{%
        \includegraphics[width=0.46\columnwidth]{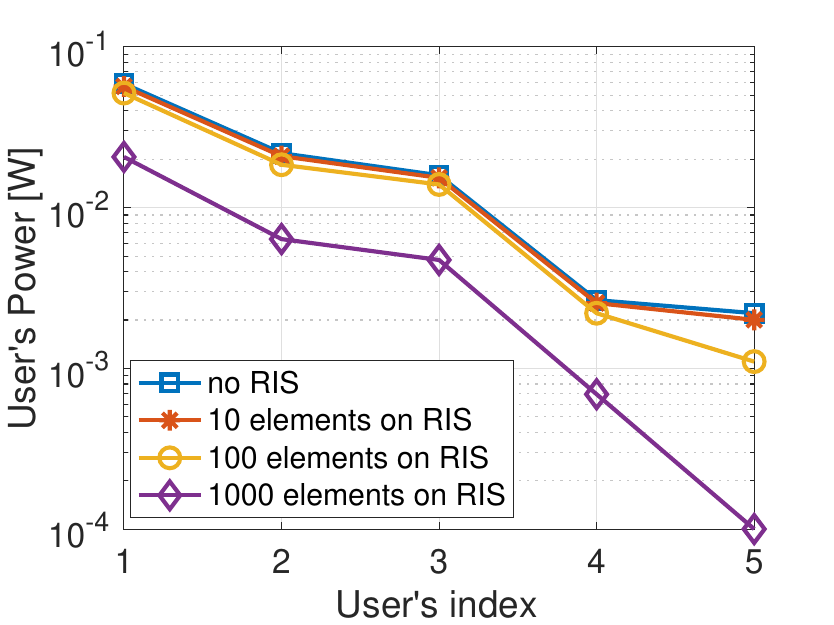}}
    \hfill
    \subfloat[\gls{mmwave}: achieved utility per \gls{ue}.\label{fig:mmwave_utility}]{%
        \includegraphics[width=0.46\columnwidth]{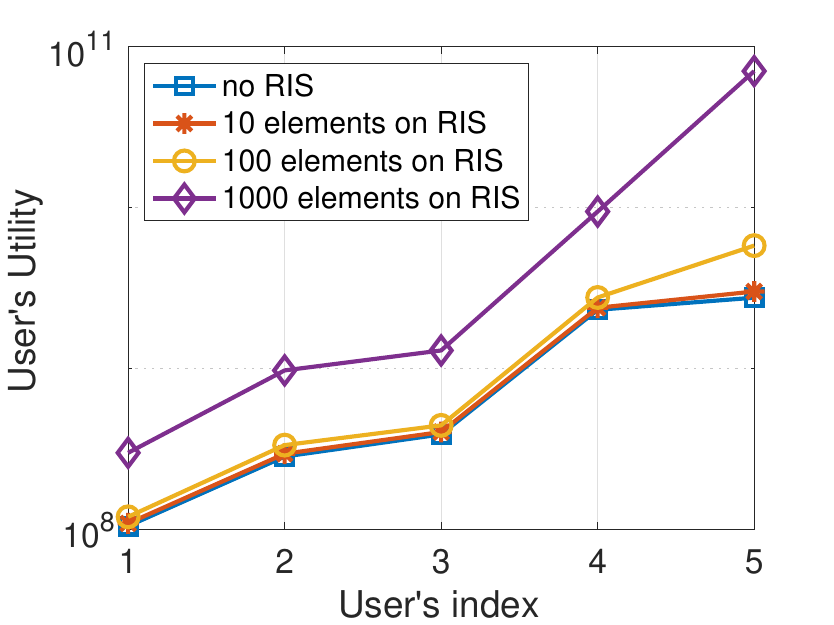}}
    \caption{Evaluation of the hierarchical game-theoretic power control framework per \gls{ue}, under different number of \gls{ris} elements and different bands.}
    \label{fig:powercontrol}
\end{figure}

The outcomes of the hierarchical game-theoretic power allocation in the Sub-6~GHz and \gls{mmwave} bands are observed by considering the allocated \gls{ul} \gls{tx} power level and achieved \gls{ue} utility both independently and collectively. In Fig.~\ref{fig:powercontrol}, we observe each \gls{ue}'s allocated \gls{ul} \gls{tx} power (Figs.~\ref{fig:sub6ghz_p},~\ref{fig:mmwave_p}) and achieved utility (Figs.~\ref{fig:sub6ghz_utility},~\ref{fig:mmwave_utility}), versus their respective index under different number of \gls{ris}'s reflecting elements both for the Sub-6~GHz and the \gls{mmwave} band. All \glspl{ue} are sorted in descending order with reference to their channel gains, where the lowest index represents the \gls{ue} with the highest channel gain, and the highest \gls{ue} index corresponds to the one with the lowest channel gain. Notably, the framework's power control~\cite{diamanti2021energy} enables low channel gain \glspl{ue} to transmit with lower power levels and achieve higher utilities. This behavior can be explained by the utility function which is affected by the \gls{ue}'s channel-gain-to-interference ratio, i.e., $\frac{G_i}{I_i}$, at the exponent of the utility's numerator (see~\eqref{eq:eq3}). In detail, low channel gain \glspl{ue} usually exhibit a low channel-gain-to-interference ratio which forces the maximum point of the utility function to take a low power value (i.e., $P_{i}$). This pattern is clearly illustrated in Fig.~\ref{fig:powercontrol} where we also observe that as the number of \gls{ris}'s reflecting elements increases, their achieved utilities also increase (Figs.~\ref{fig:sub6ghz_utility},~\ref{fig:mmwave_utility}), whereas the \glspl{ue}' allocated \gls{ul} \gls{tx} powers decrease (Figs.~\ref{fig:sub6ghz_p},~\ref{fig:mmwave_p}), as expected. With regard to the two bands under study, in Fig.~\ref{fig:sub6ghz_p} we observe that for the Sub-6~GHz band, $100$ reflecting elements on the \gls{ris} are enough for efficient power control, while for the \gls{mmwave} band, a minimum of $1000$ elements is needed as seen in Fig.~\ref{fig:mmwave_p}. This is due to the high propagation losses of the latter band~\cite{tapio2021survey}, which stress the need for bigger \gls{ris} surfaces~\cite{da2023varactor} to enable effective beamforming and therefore more efficient power allocation.

\begin{figure}[ht]
    \centering
    \subfloat[Sum allocated power across the \gls{noma} cluster.\label{fig:sub6ghz_p_sum}]{%
        \includegraphics[width=0.46\columnwidth]{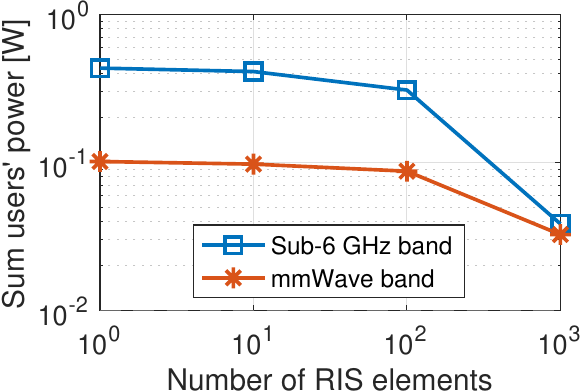}}
    \hfill
    \subfloat[Sum utility across the \gls{noma} cluster.\label{fig:sub6ghz_utility_sum}]{%
        \includegraphics[width=0.46\columnwidth]{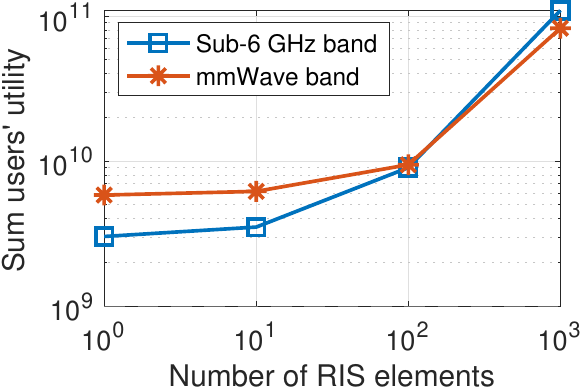}}
    \caption{Evaluation of the hierarchical game-theoretic power control framework on the entire system, under different number of \gls{ris} elements and different bands.}
    \label{fig:powercontrol_sum}
\end{figure}

The overall performance gain of the entire system (e.g., the \gls{noma} cluster) is illustrated in Fig.~\ref{fig:powercontrol_sum}, where we observe both the sum \glspl{ue}' allocated power levels and utilities as a function of a different number of \gls{ris} elements both for the Sub-6~GHz and the \gls{mmwave} band. We can deduce that for a small number \gls{ris} elements (e.g., $10^1$ \gls{ris} elements), the system's performance is comparable to the case without a \gls{ris} (i.e., $10^0$ \gls{ris} elements). When the number of \gls{ris} elements increases, the sum \glspl{ue}' powers decreases while their sum utility/satisfaction increases, as expected.
Notably, the increased number of \gls{ris} elements results in the reduced sum of \glspl{ue}' powers and thus in their increased sum utility/satisfaction.
We also observe that even if the system follows the same power allocation trend in the two bands, the sum of the allocated powers to the \gls{mmwave} \glspl{ue} is lower compared to the case of their Sub-6~GHz counterparts. It is reminded that the devised utility function for energy efficiency, captures the trade-off between increasing $\gamma_i$ (which improves the \gls{qos}) and minimizing $P_i$ (which reduces power consumption). Since the increase in $\gamma_i$ is limited by a low $\frac{G_i}{I_i}$, increasing $P_i$ will lead to diminishing returns in the numerator of the utility function. In detail, when the ratio $\frac{G_i}{I_i}$ is low, the \gls{sinr} i.e., $\gamma_i = \frac{P_i G_i}{I_i}$ becomes small. In this case, the exponential term $e^{-\alpha \gamma_i}$ approaches 1, meaning that even if the \gls{ue} increases $P_i$, the utility does not increase significantly because the \gls{sinr} is limited by the high interference. Thus, the utility function reaches its maximum at a lower power level $P_i$ because increasing power doesn't significantly improve the \gls{sinr}. Therefore, \glspl{ue} with low $\frac{G_i}{I_i}$ transmit at lower power levels to avoid unnecessary energy consumption, as increasing $P_i$ will not substantially improve their utility. Our power control framework favors \glspl{ue} with lower $\frac{G_i}{I_i}$ to transmit with lower power levels. In our work, \gls{mmwave} \glspl{ue} experience low $\frac{G_i}{I_i}$ levels, and hence have a lower \gls{cqi} compared to the \glspl{ue} of the Sub-6~GHz band they are compared to. As a result when ultimately compared, given the low $\frac{G_i}{I_i}$ levels of the former, they get to transmit with lower power levels compared to their Sub-6~GHz counterparts.

\begin{figure}[ht]
    \centering
    \subfloat[Sub-6 GHz path gains.\label{fig:sub6ghzpg}]{%
        \includegraphics[width=0.99\textwidth]{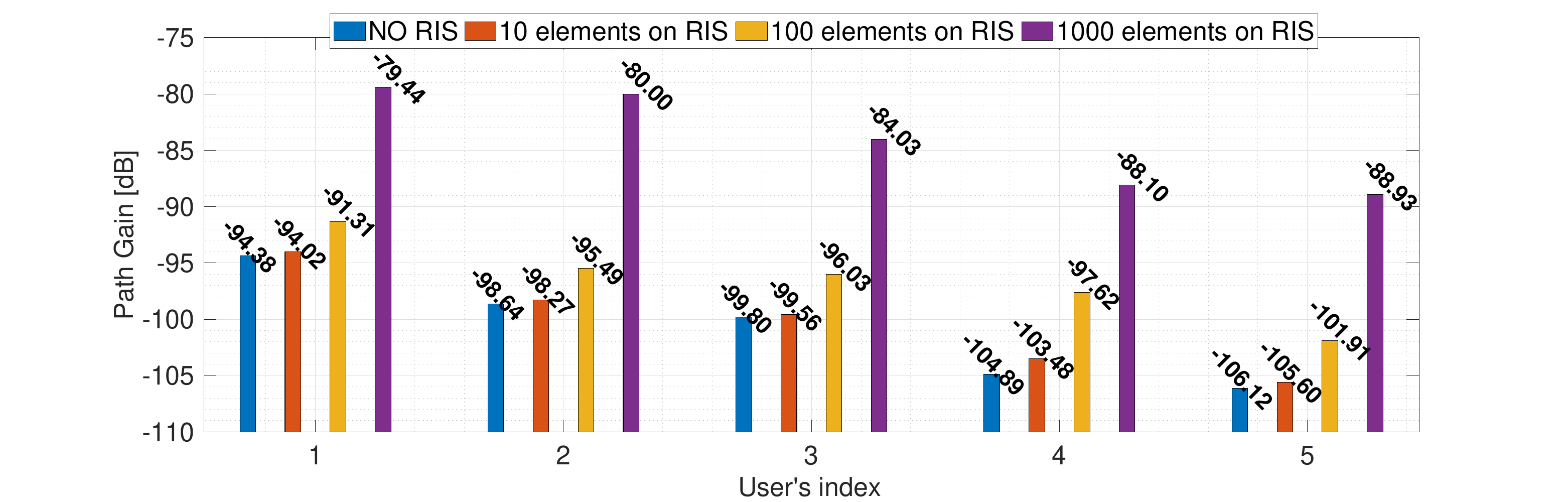}}
    \hfill
    \subfloat[\gls{mmwave} path gains.\label{fig:mmwavepg}]{%
        \includegraphics[width=0.99\textwidth]{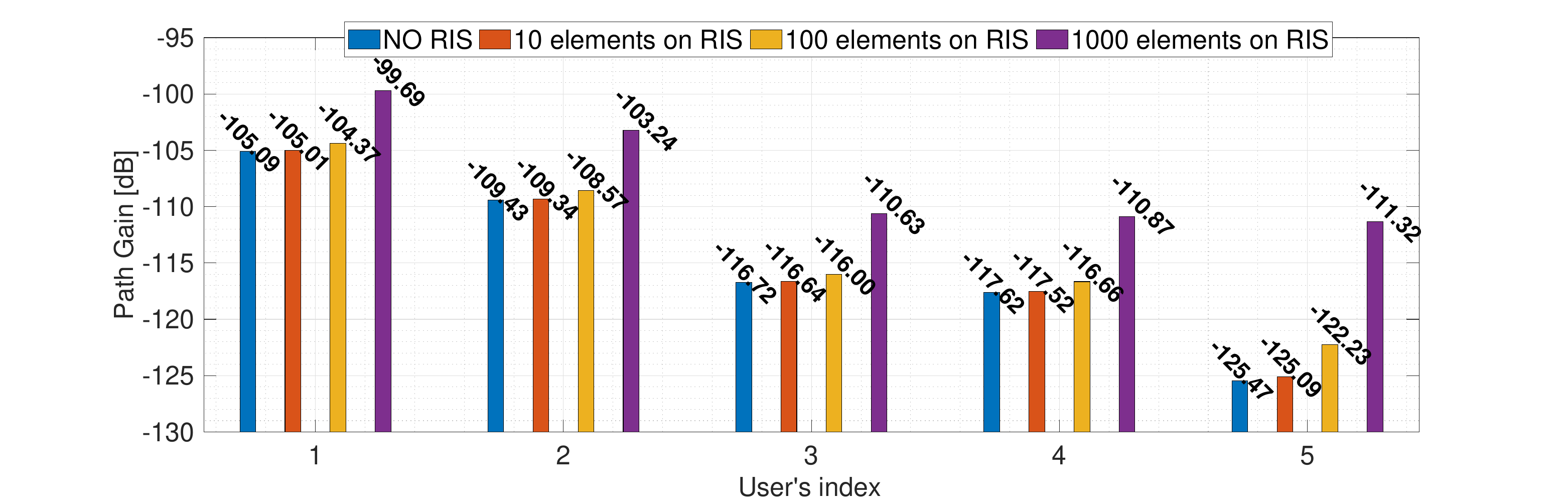}}
    \caption{Path gains for two different spectrum bands.}
    \label{fig:pathgainsris}
\end{figure}

Finally, in Fig.~\ref{fig:pathgainsris}, we present the path gains, $PG_i$, for every \gls{ue}, $i$, with and without the inclusion of the \gls{ris} both for the Sub-6~GHz and the \gls{mmwave} band. All the \glspl{ue} are sorted in descending order, and thus the \gls{ue} with the lowest index is the \gls{ue} with the best channel gain conditions, while the \gls{ue} with the biggest index corresponds to the \gls{ue} with the worst channel.
Specifically, Fig.~\ref{fig:sub6ghzpg} depicts the path gains for the Sub-6~GHz band with the carrier frequency centered at $5.9$~GHz. Based on the reported findings, the performance achieved with a minimum of~$10$ \gls{ris} elements closely resembles the performance without the \gls{ris}. A noticeable difference at $\sim|1.41|$~dB between the aforementioned cases is only observed for index~$4$, which, within the scope of this work, indicates a \gls{ue} with \emph{poor} \gls{cqi}.
Discernible differences can be observed for all the \glspl{ue} when focusing on the cases of $100$ or $1000$ \gls{ris} elements. In detail, an average difference of $\sim|4|$~dB is observed when collectively comparing all the \glspl{ue} without and with the existence of \gls{ris} in the topology under~$100$~\gls{ris} elements. Specifically, among all the \glspl{ue}, the \gls{ue} with index~$4$, achieves the most notable difference, by achieving a $|7.27|$~dB higher value at $-97.62$~dB in the presence of $100$~\gls{ris} elements. In the absence of \gls{ris}, the corresponding value is reported at $-104.89$~dB. Last, the benefits of including $1000$ elements on \gls{ris} are clearly illustrated in Fig.~\ref{fig:sub6ghzpg}, where the average path gain value for all the \glspl{ue} considered is marked at $-84.1$~dB, $\sim16.5\%$ higher compared to the average value of $-100.77$~dB achieved without the \gls{ris} inclusion.

Fig.~\ref{fig:mmwavepg} depicts the path gains obtained for the \gls{mmwave} band with the carrier frequency centered at~$28$~GHz. Contrary to the findings of the conducted evaluation in the Sub-6~GHz, in the \gls{mmwave} band, a minimum number of~$1000$ \gls{ris} elements is required to achieve distinguishable differences between the cases with and without a \gls{ris}. In this band, the \gls{ue} that significantly benefits from the \gls{ris} technology with $1000$ onboarded elements is the \gls{ue} that experiences the worst \gls{cqi} conditions, identified by \gls{ue} ID~$5$. Specifically, the latter \gls{ue} reports a $|14.15|$~dB higher value at~$-111.32$~dB compared to the case without the \gls{ris}, where the path gain is marked at $-125.47$~dB. On average, \glspl{ue} under the presence of $1000$~\gls{ris} elements in the topology report path gains $\sim|7.72|$~dB or $\sim7\%$ higher than in the absence of \gls{ris}. Precisely, the average path gain value of the \glspl{ue} in the case of~$1000$ \gls{ris} elements is reported at $-107.15$~dB, while in the case of \gls{ris} absence this value is indicated at~$-114.87$~dB.

Conclusively, in the Sub-6~GHz band, the average path gain value reported with $100$ \gls{ris} elements is marked at $-96.5$~dB, a value $\sim4\%$ higher than the average path gain of $-100.77$~dB reported in the absence of \gls{ris}. In the \gls{mmwave} band, with $1000$ \gls{ris} elements, the corresponding increase is found to be $\sim7\%$. Based on these findings, a minimum number of $100$ \gls{ris} elements is deemed necessary to support the deployment of \gls{ris} in the Sub-6~GHz band, while in the \gls{mmwave} this number is found to be $1000$.

\begin{figure}[ht]
    \centering
    \subfloat[Sub-6 GHz: power.\label{fig:sub6ghz_p_more_elems}]{%
        \includegraphics[width=0.46\columnwidth]{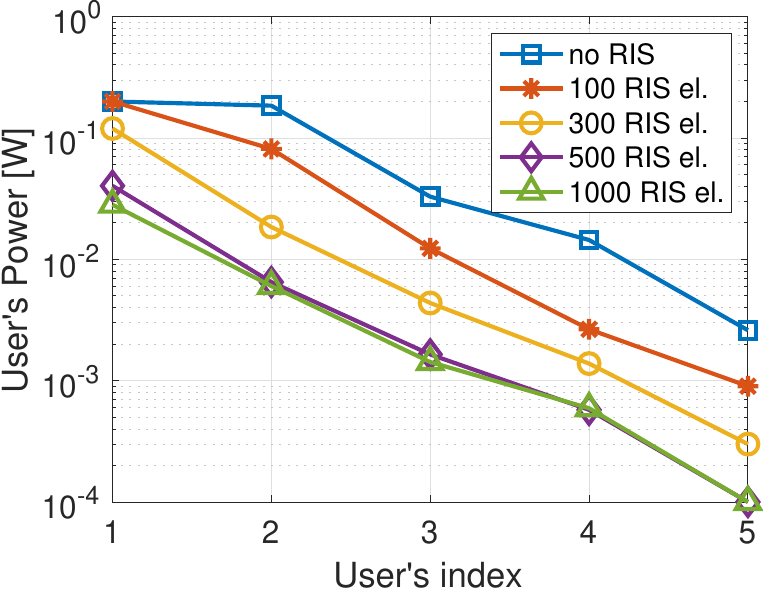}}
    \hfill
    \subfloat[Sub-6 GHz: utility.\label{fig:sub6ghz_utility_more_elems}]{%
        \includegraphics[width=0.46\columnwidth]{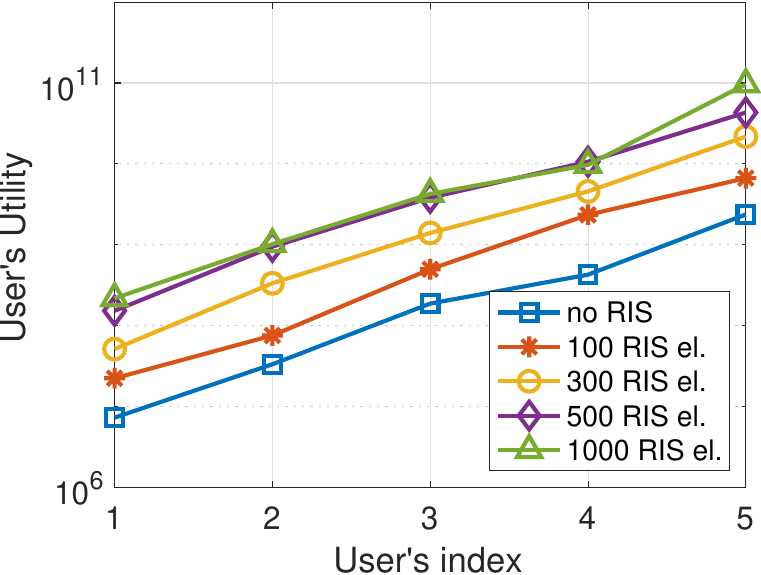}}

    \subfloat[\gls{mmwave}: power.\label{fig:mmwave_p_more_elems}]{%
        \includegraphics[width=0.46\columnwidth]{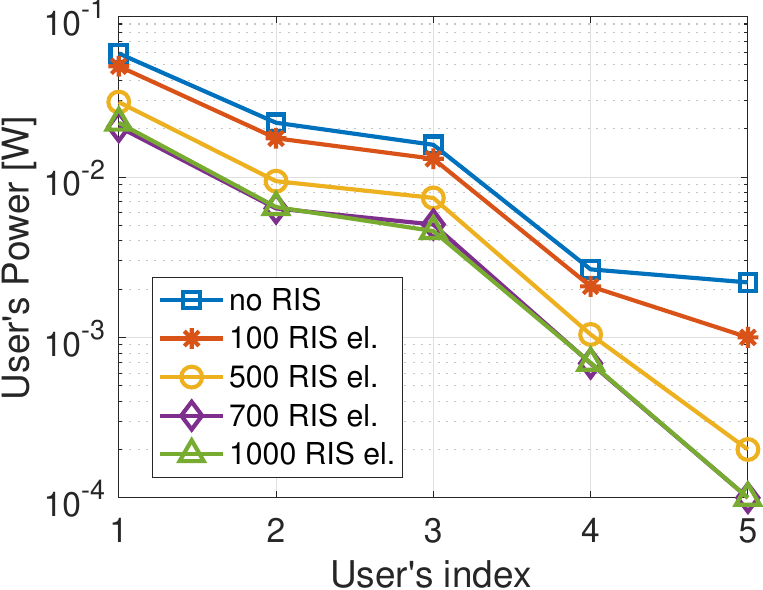}}
    \hfill
    \subfloat[\gls{mmwave}: utility.\label{fig:mmwave_utility_more_elems}]{%
        \includegraphics[width=0.46\columnwidth]{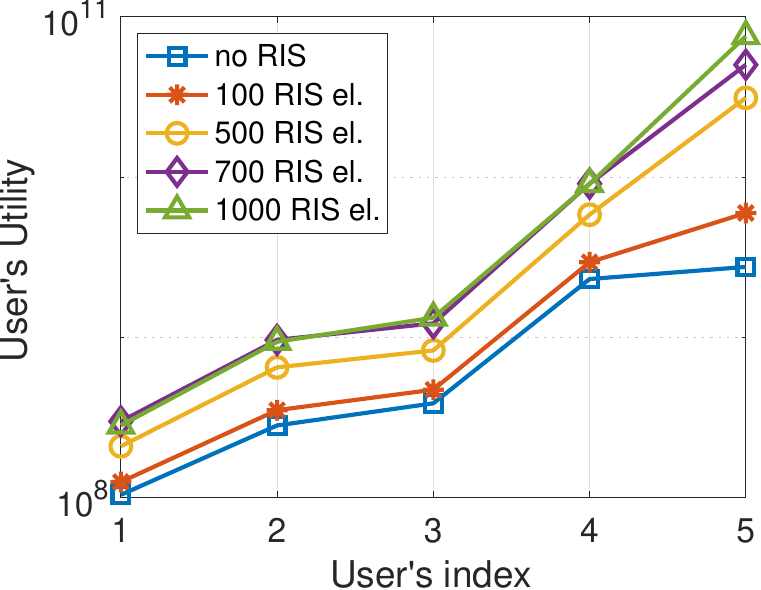}}
    \caption{Evaluation of the hierarchical game-theoretic power control framework per \gls{ue}, under different numbers of \gls{ris} elements per each band.}
    \label{fig:powercontrol_more_elems}
\end{figure}

It is finally noted that the performance evaluation of the Stackelberg game-theoretic optimization process involved elements in increasing orders of magnitude, representing successive powers of $10$. The aforementioned numbers not only represent typical values encountered in the literature~\cite{rossanese2022designing,bjornson2022reconfigurable}, but also meet the requirements outlined in~\cite{ETSI_RIS001}, which state that existing \gls{ris} elements can range from $10$ elements and above.
In the following, we also provide a performance evaluation under different sets of \gls{ris} elements for the respective bands, aiming to determine the minimal number of \gls{ris} elements each band requires to achieve higher levels of energy efficiency. Based on the experimental results in Fig.~\ref{fig:powercontrol_more_elems}, it is noted that for the Sub-6~GHz band, $500$ \gls{ris} elements result in similar performance to $1000$ (Figs.~\ref{fig:sub6ghz_p_more_elems} and~\ref{fig:sub6ghz_utility_more_elems}), while discernible performance improvements are observed from $100$ elements and above, with $300$ being an indicative value. For the \gls{mmwave} band, it is noted that the respective values are $700$ (resulting in similar performance to $1000$) and $500$ (indicative value for performance improvement), correspondingly (Figs.~\ref{fig:mmwave_p_more_elems} and~\ref{fig:mmwave_utility_more_elems}).

\subsubsection{Comparison with the state-of-the-art}
\label{sec:ris_propagation:cross_layer:soa}

In this subsection, we evaluate our optimization framework using two different utilities for energy efficiency, as shown in Table~\ref{table:utility-functions-comparison}. The utility function proposed in our work is labeled \textit{Proposed Utility} and is compared with the utility presented in~\cite{diamanti2021prospect}, referred to as \textit{Literature Utility}. To assess the performance of our proposed method in comparison with~\cite{diamanti2021prospect}, we also consider a \gls{noma} cluster of $4$ \glspl{ue}, a bandwidth of $5$~MHz, and the same noise levels as in~\cite{diamanti2021prospect}, and we evaluate the performance in the Sub-6~GHz band. Finally, we consider two cases: one without the inclusion of a \gls{ris} and the other with $100$ \gls{ris} elements onboard.

\begin{table}[H]
\centering
\small
\caption{Utility Functions Comparison}
\begin{adjustbox}{width=0.85\linewidth}
\begin{tabular}{@{}ll@{}}
\toprule
\textbf{Proposed Utility:} & $U_{i}\left(P_{i}, \mathbf{P}_{-i}\right)=\dfrac{W \cdot\left(1-e^{-\alpha \gamma_{i}}\right)^{M}}{P_{i}}$ \\
\textbf{Literature Utility:} & $EE_i({P}_i, \mathbf{P}_{-i}) = \dfrac{W \cdot \log_2(1 + \gamma_i)}{P_i}$ \\
\bottomrule
\end{tabular}
\end{adjustbox}
\label{table:utility-functions-comparison}
\end{table}

\begin{figure}[ht]
    \centering
    \subfloat[Allocated \gls{ul} \gls{tx} power per \gls{ue}.\label{fig:sub6ghz_propp}]{%
        \includegraphics[width=0.5\columnwidth]{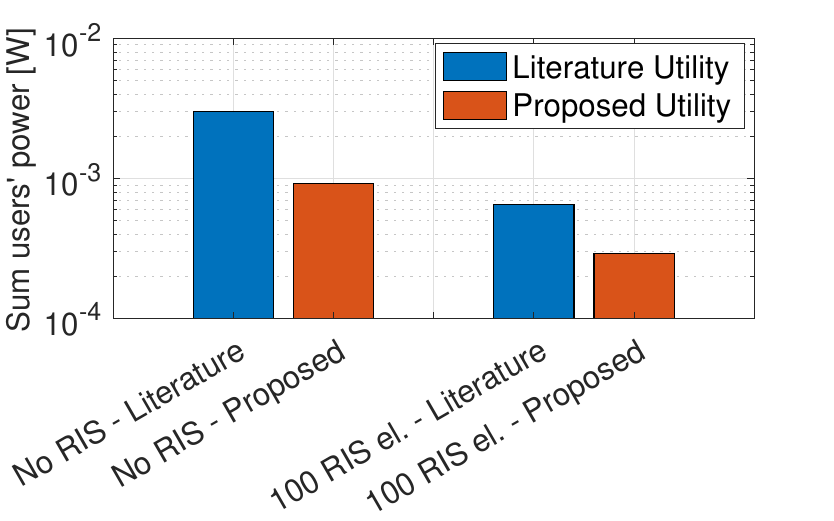}}
    \hfill
    \subfloat[Achieved utility per \gls{ue}.\label{fig:propo_uti}]{%
        \includegraphics[width=0.5\columnwidth]{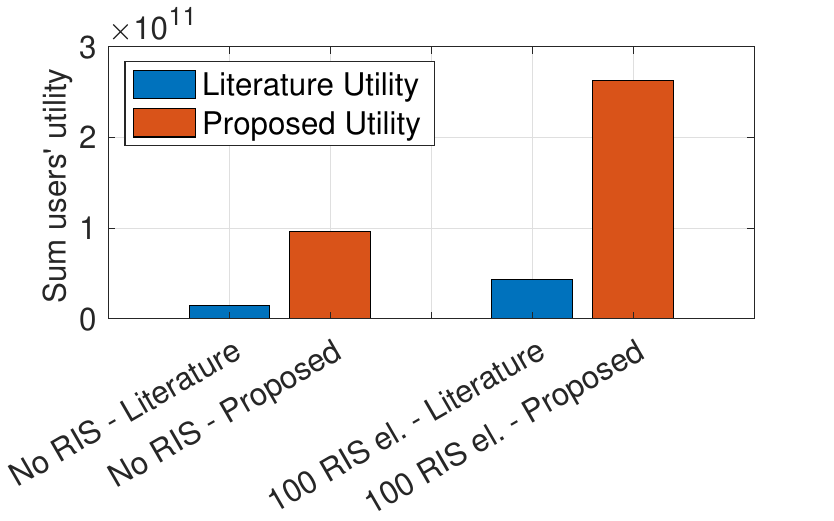}}
    \caption{Evaluation of the hierarchical game-theoretic power control framework under two different utilities given in Table~\ref{table:utility-functions-comparison}.}
    \label{fig:powercontrol_prop}
\end{figure}

The performance evaluation results reported in Fig.~\ref{fig:powercontrol_prop} clearly indicate the suitability of the proposed utility function in yielding lower transmission power levels (Fig.~\ref{fig:sub6ghz_propp}) and achieving higher energy efficiency (Fig.~\ref{fig:propo_uti}). This is due to the flexible design of our proposed utility through the control parameters, as well as its effective capture of the trade-off between achieving higher \gls{qos} and lower costs (e.g., reduced \gls{tx} power levels).

\subsection[Performance Evaluation of RIS-Assisted C-V2X Communication Systems on Colosseum]{Performance Evaluation of RIS-Assisted \texorpdfstring{\gls{cv2x}}{C-V2X} Communication Systems on Colosseum}
\label{sec:ris_propagation:cross_layer:colosseum}

As mentioned in Section~\ref{sec:ris_propagation:cross_layer:powercontrol}, the optimization problem solved by the \gls{uav} and defined in Eqs.~\eqref{eq4a}--\eqref{eq4b}, will result in the maximization of its total received signal strength through the individual maximization of each \gls{ue}'s incoming signal strength. Indeed, the latter is achieved through a linear combination of the \gls{ris} elements' effective phase shifts as given in~\eqref{eq8}, which will enable the creation of the cascaded channels denoted as $\mathbf{{h}}_{RU}^H\boldsymbol{\Theta}\mathbf{{h}}_{iR}$. In consistency with Section~\ref{sec:ris_propagation:cross_layer:system_model}, all the incoming \gls{ue} signals at the \gls{uav} are the result of the coherent summation of the direct, i.e., $h_{iU}$, and reflected signals. It is reminded that the path gains of the corresponding generated channels, without and with the inclusion of a different number of \gls{ris} elements are provided and illustrated in Fig.~\ref{fig:pathgainsris}.

As a next step, we proceed by installing the generated channels in Colosseum. In summary, the reference input scenario to be installed includes two key components: the path gain, represented as a complex coefficient value in the form of \gls{fir}, and the \gls{toa} value. The path gain is derived from the coherent summation of generated \glspl{mpc} across the wireless links and has been averaged across multiple experiments, as detailed in Section~\ref{sec:ris_propagation:cross_layer:system_model}. Meanwhile, the \gls{toa} values have been computed based on the Euclidean distance between the entities considered in the topology (i.e., \gls{uav}, \gls{ris}, and \glspl{ue}), given the speed of light (i.e., $3 \times 10^8$ [m/s]). These values collectively contribute to generating the time-variant \gls{cir} for each node pair within the topology. Finally, a tutorial on the installation of \gls{rf} scenarios in Colosseum can be found in~\cite{bonati2021colosseum,villa2022cast}, but as it is not the focus of this section, the detailed explanation is omitted.

\begin{figure}[ht]
    \centering
    \subfloat[eMBB DL throughput CDF — same scheduling policy across slices.\label{fig:ris_on_col1}]{%
        \includegraphics[width=0.5\textwidth]{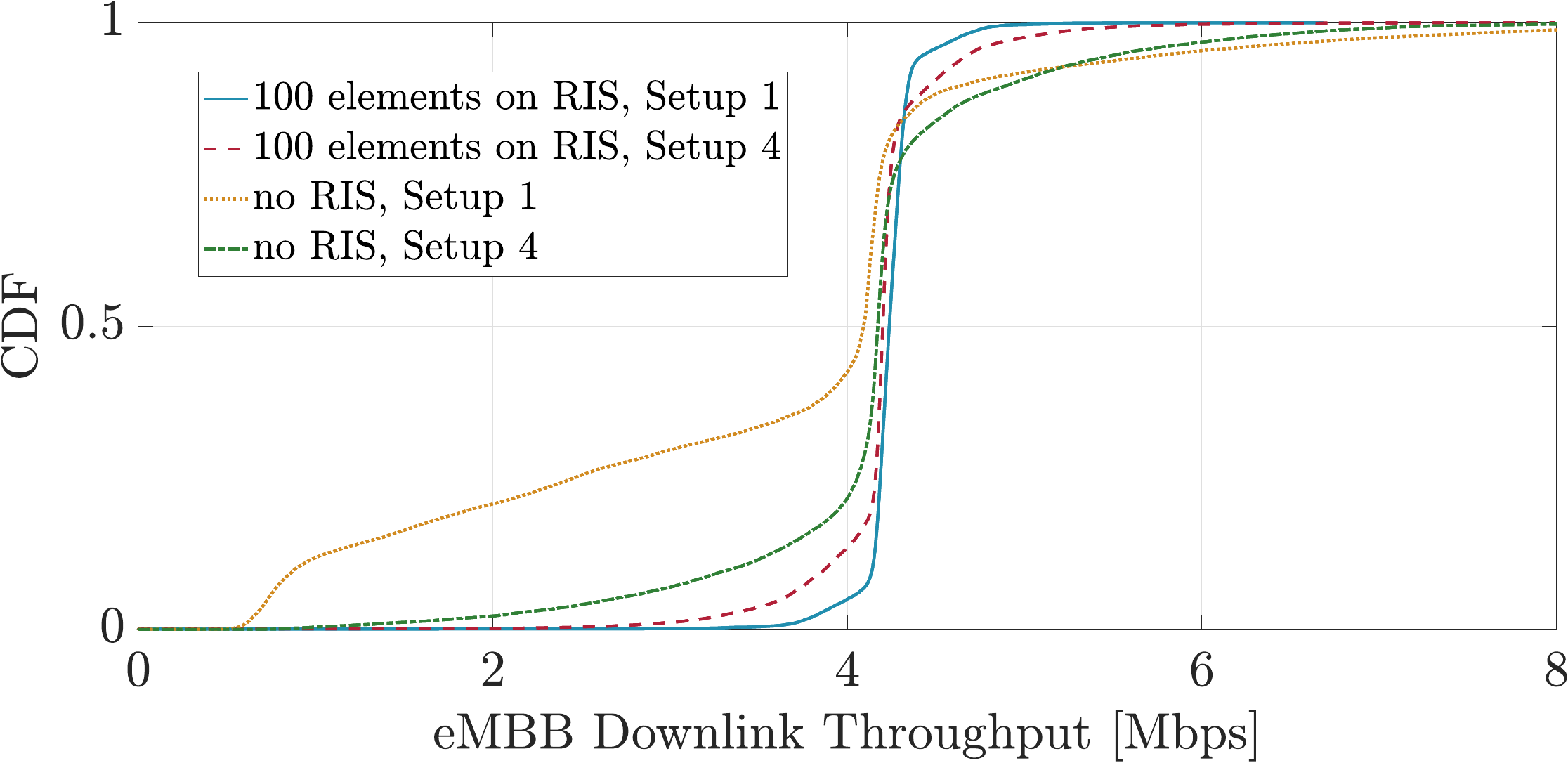}}
    \hfill
    \subfloat[URLLC buffer occupancy CDF — same scheduling policy across slices.\label{fig:ris_on_col2}]{%
        \includegraphics[width=0.5\textwidth]{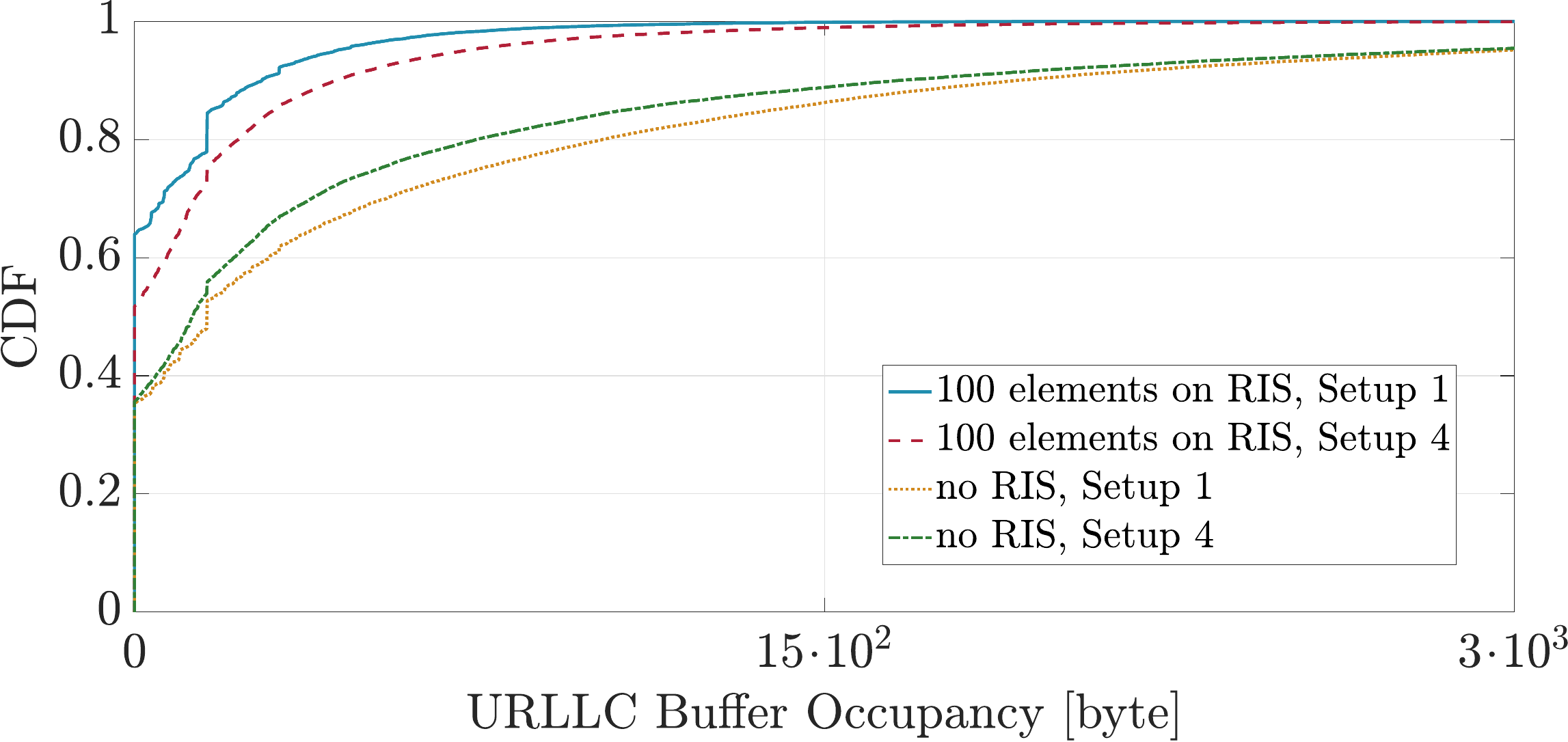}}

    \subfloat[eMBB DL throughput CDF — different scheduling policies across slices.\label{fig:ris_on_col1_caseb}]{%
        \includegraphics[width=0.5\textwidth]{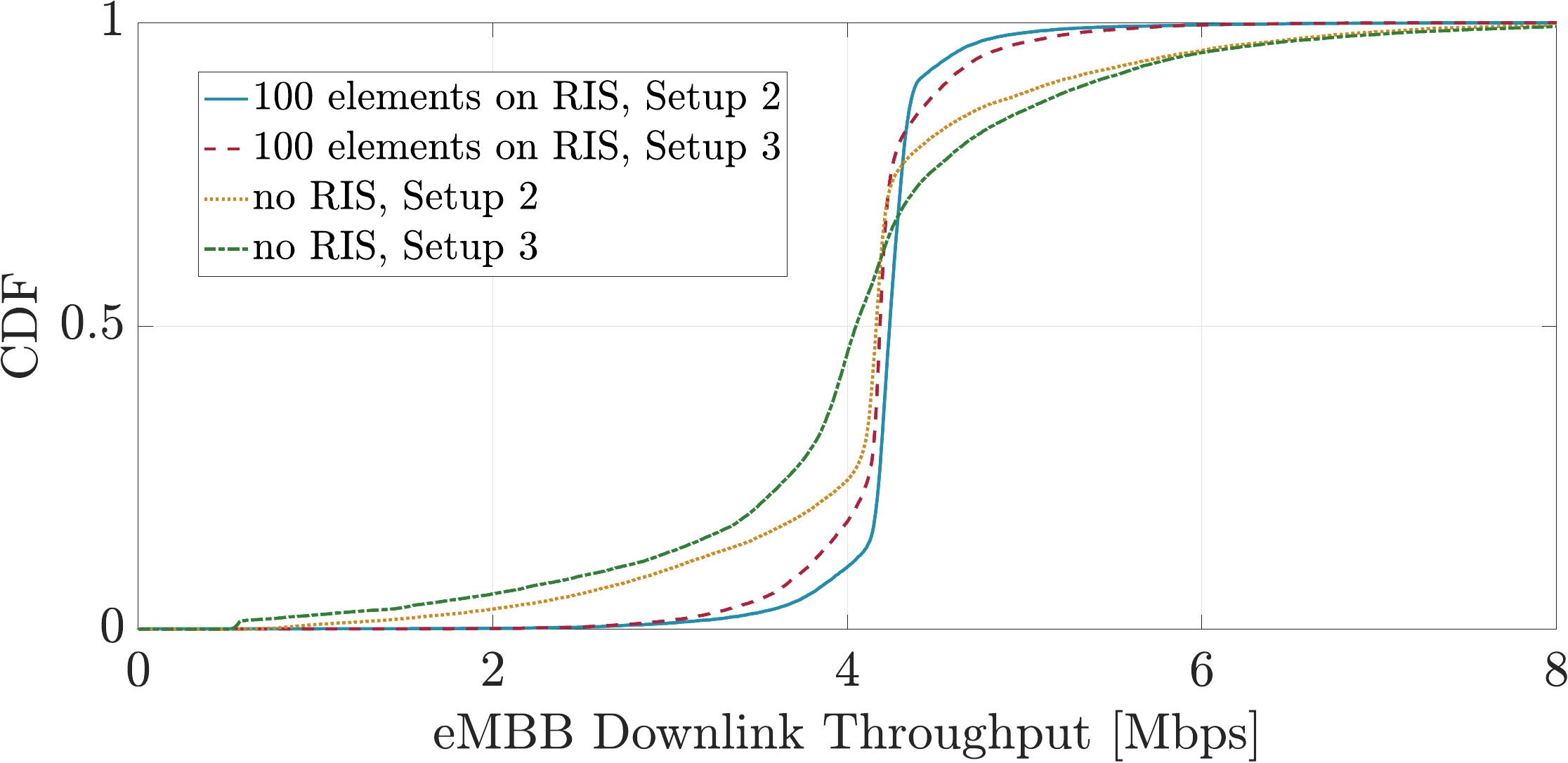}}
    \hfill
    \subfloat[URLLC buffer occupancy CDF — different scheduling policies across slices.\label{fig:ris_on_col2_caseb}]{%
        \includegraphics[width=0.5\textwidth]{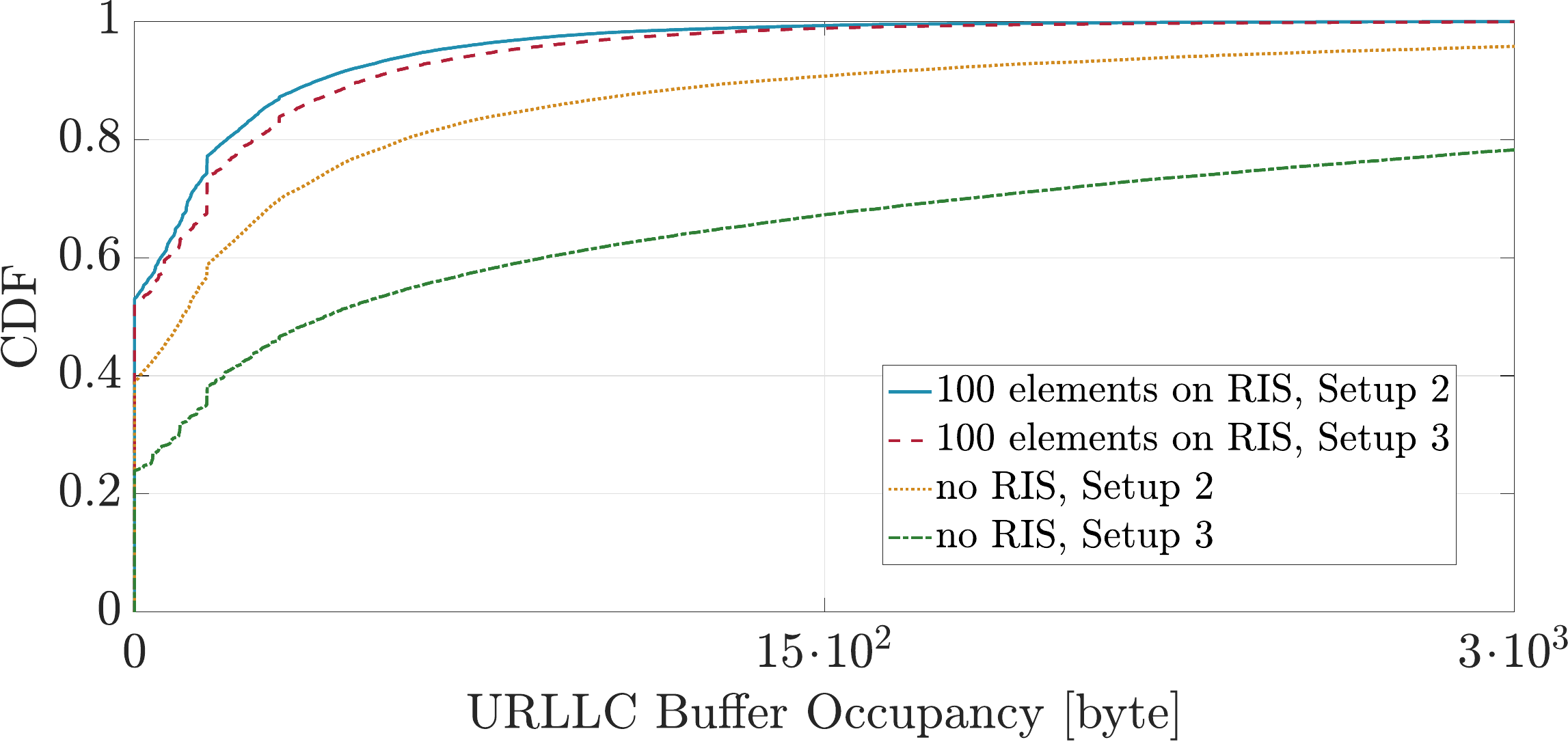}}

    \caption{Evaluation of a slice-based multi-\gls{ue} \gls{cv2x} communication system on Colosseum with and without the inclusion of \gls{ris} under different scheduling profiles.}
    \label{fig:ris_on_col_total}
\end{figure}

We focus on the case of the Sub-6~GHz band, where all the wireless channels were simulated with \gls{quadriga} at the center frequency of~$5.9$~GHz. We only consider the scenario with $100$~\gls{ris} elements for all the \gls{ris}-assisted generated channels. This approach allows us to explore how a relatively small number of \gls{ris} elements (typically large arrays are required~\cite{bjornson2020reconfigurable,siddiqi2022reconfigurable}), and an average numerical increase of $\sim4\%$ in the path gains between cases with and without \gls{ris} in the topology, can significantly improve network performance.
Additionally, since previous results (see Section~\ref{sec:ris_propagation:cross_layer:energy}) have shown that \gls{mmwave}~\glspl{ue} are mostly benefited in terms of energy efficiency (i.e., they get to transmit with lower power levels, especially for $1000$ \gls{ris} elements, while simultaneously satisfying their \gls{qos} demands), we aim at examining the advantages introduced by \gls{ris} technology in a realistic \gls{nextg} \gls{ran} scenario tested in the Sub-6~GHz band.
It is noted that although the channels were created at the frequency of~$5.9$~GHz using the \gls{quadriga} channel modeling simulator, the testing frequency for this scenario in Colosseum is set to $1$~GHz, which is the frequency that the \gls{mchem} is optimized to work~\cite{villa2023twinning}.

We leverage the capabilities of the SCOPE framework~\cite{bonati2021scope} to deploy an end-to-end \gls{ran} and core network through the srsRAN~\cite{gomez2016srslte} softwarized open-source protocol stack on Colosseum. Specifically, we deploy a 3GPP-compliant cellular network comprising a single \gls{bs} (i.e., the \gls{uav}) and $5$~\glspl{ue} distributed across $2$ different slices. These are: (i)~\gls{embb} that concerns high traffic modeling of high-quality multimedia content and streaming applications, and (ii)~\gls{urllc} for time-sensitive applications, such as autonomous driving in \gls{v2x} scenarios as detailed in~\cite{10437367}. Slice-based traffic is generated
by the \gls{mgen} \gls{tcp}/\gls{udp} traffic generator~\cite{mgen}, based on the following specifications: \gls{embb} \glspl{ue} request $4$\:Mbps constant bitrate, while \gls{urllc} \glspl{ue} generate $89.3$\:kbps Poisson traffic.
The bandwidth of the \gls{bs} is set equal to $10$~MHz (i.e., $50$ \glspl{prb}), and it is divided between the $2$ slices.
Specifically, we allocate
$45$ \glspl{prb} on the \gls{embb}, while the remaining $\sim5$ \glspl{prb} are assigned on the \gls{urllc}. In this way, we allocate $90$\% of the available resources to \gls{embb}, which has the greatest demand for higher throughput. Consequently, $10$\% of the available resources are allocated to support the requirements for low latency on the \gls{urllc}. Among the $5$ users considered, $3$ of them are \gls{urllc} \glspl{ue}, while the remaining $2$ are assigned on the \gls{embb} slice. It is also noted that the \glspl{ue} of the \gls{embb} are those who reported the highest path gain values, while the \glspl{ue} of the \gls{urllc} are the ones with the lowest reported \gls{cqi} values. With this allocation, we aim at exploring how the presence of the \gls{ris} in the topology will benefit the \gls{urllc} slice in a scenario where limited resources are available to serve $60\%$ of the available \glspl{ue}, especially when they experience poor transmission conditions (e.g., in the absence of \gls{ris} in the topology).
Finally, the \gls{kpm} observed for \gls{embb} is the \gls{dl} throughput, while the respective value for \gls{urllc} is the buffer occupancy as a proxy for latency~\cite{10437367}.

By using SCOPE's open \glspl{api}, we fine-tune the \gls{ran}'s functionalities, in terms of scheduling profile selection. Specifically, two \gls{mac}-layer scheduling algorithms are compared, namely the \gls{rr} and a fairer one, the \gls{wf}, in all possible combinations. In Table~\ref{table:sched-profiles}, we present all the configurations considered in terms of scheduling profiles (defined in Table~\ref{table:sched-profiles-catalog}) for the \gls{embb} and \gls{urllc} slices, which define how \glspl{prb} are internally allocated to \glspl{ue} belonging to
each slice~\cite{10766614}.

\begin{table}[H]
\centering
\small
\caption{Scheduling Profiles Catalog}
\begin{adjustbox}{width=0.35\linewidth}
\begin{tabular}{@{}ll@{}}
\toprule
\textbf{RR:} & Round Robin \\
\textbf{WF:} & Waterfilling \\
\bottomrule
\end{tabular}
\end{adjustbox}
\label{table:sched-profiles-catalog}
\end{table}

\begin{table}[H]
\centering
\small
\caption{Scheduling Profiles Configuration}
\begin{adjustbox}{width=0.45\linewidth}
\begin{tabular}{@{}ccc@{}}
\toprule
\textbf{Setup ID} & \textbf{eMBB} & \textbf{URLLC} \\ \midrule
\textbf{1} & RR & RR \\
\textbf{2} & RR & WF \\
\textbf{3} & WF & RR \\
\textbf{4} & WF & WF \\
\bottomrule
\end{tabular}
\end{adjustbox}
\label{table:sched-profiles}
\end{table}

The results of this comparative performance evaluation can be seen in Fig.~\ref{fig:ris_on_col_total}.
Finally, all subsequent results were produced by taking the median as the most representative statistical value of a dataset, and averaged over multiple repetitions of experiments in the \gls{dl} direction of the communication system\footnote{Note that all channels generated with \gls{quadriga} in the \gls{ris}-assisted and \gls{uav}-enabled scenarios are reciprocal. The focus of this Section is the evaluation of a \gls{ris}-assisted and \gls{uav}-enabled communication system on the \gls{dl} direction.}.

In Figs.~\ref{fig:ris_on_col1} and~\ref{fig:ris_on_col2}, we focus on the case where both slices are served under the same scheduling policy. Based on the reported findings, when focusing on the \gls{embb} slice, all configurations, both with and without the \gls{ris} inclusion, deliver a median throughput value of $\sim4$\:Mbps. However, Setup $1$ achieves one of the highest reported throughput values of $4.236$ Mbps with $100$ \gls{ris} elements. The corresponding value in the absence of \gls{ris} is found to be $4.091$\:Mbps. With Setup~$4$ and $100$ onboarded \gls{ris} elements, the median throughput value reaches $4.2$\:Mbps. The respective value in the absence of \gls{ris} is reported at $4.172$\:Mbps. On the \gls{urllc} slice, however, in the absence of \gls{ris}, both Setups~$1$ and $4$ underperform compared to the case where \gls{ris} is included in the topology. Specifically, with Setup~$1$, the median value of the buffer size is reported at $158$~byte, while for Setup~$4$, the corresponding value is marked at $127$~byte. In both setups with \gls{ris}, a median value of $0$~byte buffer occupancy is achieved, indicating zero latency.

In Figs.~\ref{fig:ris_on_col1_caseb} and~\ref{fig:ris_on_col2_caseb} we include the results collected when both slices are served under different scheduling policies. On the \gls{embb}, Setup~$2$ achieves the highest throughput value reported in the presence of $100$ elements on \gls{ris} at $4.238$\:Mbps, while Setup~$3$ achieves one of the lowest at $4.184$\:Mbps. The respective values in the absence of \gls{ris} are reported at $4.163$\:Mbps and $4.048$\:Mbps, correspondingly. On the \gls{urllc}, in the presence of \gls{ris}, all setups resulted in empty buffers, reporting a median value of~$0$~buffer occupancy. In the absence of \gls{ris}, with Setup~$2$ the median value for buffer occupancy is marked at~$105$~byte, while with Setup~$3$ the corresponding value is found to be~$412$ byte.

In Tables~\ref{table:ris-setups} and~\ref{table:no-ris-setups}, all the results obtained without and with the inclusion of $100$ \gls{ris} elements under different combinations of scheduling policies are collectively presented. Based on the reported findings in Table~\ref{table:ris-setups} we observe that in the presence of \gls{ris} in the topology, Setups~$1$ and~$2$ deliver almost identical performance on the \gls{embb}, reporting median throughput values of $4.236$\:Mbps and $4.238$\:Mbps, respectively. Setups~$3$~and~$4$ deliver slightly worse performance, at $4.184$\:Mbps and $4.2$\:Mbps, correspondingly. On the \gls{urllc}, all setups resulted in optimal performance by achieving a median value of $0$ byte buffer occupancy. The findings of Table~\ref{table:no-ris-setups} indicate that in the absence of \gls{ris} in the topology, Setups~$2$~and~$4$ yield similar performance, achieving median \gls{embb} throughput values of $4.163$\:Mbps and $4.172$\:Mbps. The aforementioned setups also lead to improved \gls{urllc} performance among those compared, achieving buffer occupancies of $105$~and~$127$ byte, respectively. Setups~$1$~and~$3$ achieve \gls{embb} throughputs of $4.091$\:Mbps and $4.048$\:Mbps \gls{embb}, respectively. The median values for buffer occupancy on the \gls{urllc} are $158$~and~$412$~byte, with the latter being the highest recorded in the absence of \gls{ris} in the topology. It is noted that all setups resulted in empty buffers on the \gls{urllc} when \gls{ris} was included in the topology. In the absence of \gls{ris}, the reported median values for buffer occupancy vary around the average value of $200$ byte and are included in Table~\ref{table:no-ris-setups}. When \gls{ris} is not included in the topology, all configurations result in increased buffer sizes, indicating higher levels of latency in the communication links.

\begin{table}[H]
\centering
\small
\caption{\glspl{kpm} for \gls{ris}-assisted and \gls{uav}-enabled \gls{cv2x} topologies.}
\begin{adjustbox}{width=0.65\linewidth}
\begin{tabular}{@{}ccc@{}}
\toprule
\textbf{Setup ID} & \textbf{eMBB} [Mbps] & \textbf{URLLC} [byte] \\ \midrule
\textbf{1} & $4.236$ & $0$ \\
\textbf{2} & $4.238$ & $0$ \\
\textbf{3} & $4.184$ & $0$ \\
\textbf{4} & $4.2$   & $0$ \\
\bottomrule
\end{tabular}
\end{adjustbox}
\label{table:ris-setups}
\end{table}

\begin{table}[H]
\centering
\small
\caption{\glspl{kpm} for \gls{uav}-enabled \gls{cv2x} topologies without \gls{ris}.}
\begin{adjustbox}{width=0.65\linewidth}
\begin{tabular}{@{}ccc@{}}
\toprule
\textbf{Setup ID} & \textbf{eMBB} [Mbps] & \textbf{URLLC} [byte] \\ \midrule
\textbf{1} & $4.091$ & $158$ \\
\textbf{2} & $4.163$ & $105$ \\
\textbf{3} & $4.048$ & $412$ \\
\textbf{4} & $4.172$ & $127$ \\
\bottomrule
\end{tabular}
\end{adjustbox}
\label{table:no-ris-setups}
\end{table}

Conclusively, in resource-constrained environments (e.g., limited availability of \glspl{prb}), where the \glspl{ue} may experience relatively poorer \gls{cqi} conditions, the presence of \gls{ris} enhances the \gls{qos} experience of the \gls{urllc} by resulting in a buffer occupancy of $0$ byte.

\subsection{Conclusions}
\label{sec:ris_propagation:cross_layer:summary}

In this work, we have leveraged \gls{quadriga}'s open-source code to simulate a \gls{ris}-assisted and \gls{uav}-enabled \gls{cv2x} topology which we evaluated under a different number of \gls{ris} elements both in the Sub-6~GHz and the \gls{mmwave} portion of the spectrum. Through the formulation of a hierarchical - one leader - multiple followers - Stackelberg-based game-theoretic optimization framework, we were able to study the achieved energy efficiency of the topology on the aforementioned bands. Our results on power control indicate significant power savings and \gls{ue} satisfaction, specifically for the \gls{mmwave} band. Finally, while focusing on the case of $100$ elements mounted on the \gls{ris} and the Sub-6~GHz band, we installed the generated channels on the Colosseum wireless network emulator and conducted an evaluation campaign under different setups of scheduling profiles, considering a fixed \gls{prb} allocation for both the \gls{embb} and \gls{urllc} network slices. The performance evaluation results from network emulation on Colosseum indicate that in resource-limited conditions (i.e., a low number of allocated \glspl{prb}), the presence of \gls{ris} will counteract potential network degradation by resulting in empty buffers, thus ensuring low latency levels on the \gls{urllc} slice. Some interesting extensions of the current work include focusing on the study of stochastic \gls{ue} mobility on the Colosseum Network Emulator and exploring the expected improvements introduced by the coexistence of \glspl{ris} and \glspl{uav}, particularly in eliminating potential network coverage holes. Additionally, comparing \gls{quadriga} \gls{ris}-assisted channels with other \gls{ris}-aided schemes and traditional beamforming represents an interesting direction for future research.

\section[Bridging Open RAN and RIS]{Bridging Open \texorpdfstring{\gls{ran}}{RAN} and \texorpdfstring{\gls{ris}}{RIS}}
\label{sec:ris_propagation:bridging}

\subsection{Introduction}
\label{sec:ris_propagation:bridging:intro}
\glspl{ris} serve as a transformative technology to revolutionize the cellular architecture of \gls{nextg} \glspl{ran}. Previous studies have demonstrated the capabilities of \glspl{ris} in optimizing wireless propagation, achieving high spectral efficiency, and improving resource utilization. At the same time, the transition to softwarized, disaggregated, and virtualized architectures, such as those being standardized by the \oran ALLIANCE, enables the vision of a reconfigurable Open \gls{ran}.
%FD: too long abstract, trying to shorten a bit
%, whose functionalities can be controlled through open, real-time closed loops.
%Through intelligent applications, namely xApps and rApps, the management and orchestration of the once-monolithic \gls{ran} is now enabled.
In this work~\cite{11152828}, we aim to integrate these technologies by studying how different resource allocation policies enhance the performance of \gls{ris}-assisted Open \glspl{ran}.
%In this work, we aim to integrate these technologies by studying how various resource allocation policies enhance the performance of \gls{ris}-assisted Open \glspl{ran}, taking a decisive step toward understanding the system-level integration of \glspl{ris} into \gls{nextg} cellular networks.
We perform a comparative analysis among various network configurations and show how proper network optimization can enhance the performance across the \gls{embb} and \gls{urllc} network slices, achieving up to $\sim34\%$ throughput improvement. Furthermore, leveraging the capabilities of OpenRAN Gym, we deploy an xApp on Colosseum, the world's largest wireless system emulator with hardware-in-the-loop, to control the \gls{bs}'s scheduling policy. Experimental results demonstrate that \gls{ris}-assisted topologies achieve high resource efficiency and low latency, regardless of the \gls{bs}'s scheduling policy.
%Furthermore, by leveraging the capabilities of OpenRAN Gym, we deploy an xApp controlling the \gls{enb}'s scheduling policy. Experimental results from a performance evaluation campaign in Colosseum, the world's largest emulator of wireless systems with hardware-in-the-loop, indicate that \gls{ris}-assisted topologies ensure high resource efficiency and zero latency, regardless of the \gls{enb}'s scheduling policy.

The Open \gls{ran} paradigm for \oran networks is pivotal in transforming cellular network design. Grounded in the principles of disaggregation, virtualization, and programmable radios, components are interconnected through open, standardized interfaces, ensuring multi-vendor interoperability. This approach enables the deployment of flexible architectures built on cloud-native principles. %while
The %resiliency and
reconfigurability of the \gls{ran} is further enhanced through \glspl{ric}, streaming \gls{ran} telemetry and implementing control policies via a closed-loop mechanism~\cite{polese2022understanding}. Both near- and non-real-time \glspl{ric}, deployed at the \gls{ran}'s edge, leverage xApps and rApps to optimize network management and resource allocation. This functionality further advances the integration of \gls{ai} for the \gls{ran} and \gls{ai} on the \gls{ran}, fostering the development of intelligent and adaptive \gls{ran} architectures~\cite{AIRANAlliance2024}.

At the same time, \glspl{ris} are envisioned to be an integral part of \gls{nextg} cellular architectures, enabling smart propagation environments that can be controlled through %\gls{sdr} techniques
programmable \gls{rf} components~\cite{liaskos2022software}. Thanks to their reconfigurability, \glspl{ris} can manipulate electromagnetic waves, reshaping wireless propagation to optimize spectrum utilization and efficiency, particularly in %This is also needed to overcome side-effects in
multi-operator networks~\cite{angjo2023side}. This reconfiguration capability aligns with \oran's objectives of developing adaptable and resource-efficient \glspl{ran}.

Intelligent control loops with xApps in \oran have attracted widespread attention from the research community~\cite{bonati2023openran}. In~\cite{10766614}, the authors explore the impact of xApps embedding \gls{drl} agents to manage the \gls{bs}'s slicing and scheduling control policies across various network deployments on the Colosseum testbed. In addition, several efforts have been conducted in the optimization of wireless propagation with \glspl{ris}. Some previous works~\cite{li2020reconfigurable,diamanti2021energy,diamanti2021prospect} have focused on the use of \glspl{ris} and constructive beams to achieve coherent signal construction at the \gls{bs}, for maximizing the achieved \gls{rssi}.  Finally, a system-level experimental evaluation of \gls{ris}-enabled deployments and their cross-layer optimization, leveraging the full protocol stack on the Colosseum testbed has been discussed in~\cite{10973151}.

\subsubsection{Contributions and Outline}
\label{sec:ris_propagation:bridging:contributions}

However, limited research has focused on bridging %the aforementioned %\gls{nextg}
%technologies.
\oran and \glspl{ris} at a system level.
In this study, we aim to address this research gap by evaluating network performance across various \gls{ris}-assisted topologies using the Colosseum Open \gls{ran} testbed~\cite{10643670}. We consider an Open \gls{ran} delivering services to the \gls{embb} and \gls{urllc} network slices. Using OpenRAN Gym~\cite{bonati2023openran}---an open-source framework for experimentation in \oran---we deploy this Open \gls{ran} on the Colosseum network emulator and manage it through an xApp that controls the \gls{bs}'s scheduling policy. By investigating various spectrum allocation policies and \gls{ris} deployments, we demonstrate how network performance and resource efficiency are impacted differently for each slice. %To the best of our knowledge, this is the first experimental study %to provide a comprehensive evaluation of design choices for
With this work, we aim to complement prior studies~\cite{10973151} by providing a comprehensive evaluation of \gls{ris}-assisted Open \glspl{ran} on an \oran-compliant testbed.
%This study complements prior research~\cite{10973151} by presenting a comprehensive evaluation of \gls{ris}-assisted Open \glspl{ran} implemented on an \oran-compliant testbed.
We believe that our findings and insights will contribute to a deeper understanding of the %potential
integration of \glspl{ris} in \gls{nextg} Open \glspl{ran}.

The remainder of this chapter is organized as follows. Section~\ref{sec:ris_propagation:bridging:system_model} describes the system model. Section~\ref{sec:ris_propagation:bridging:opt} presents the optimization framework for shaping wireless propagation with \glspl{ris}, while Section~\ref{sec:ris_propagation:bridging:strategies} discusses the resource allocation strategies. Section~\ref{sec:ris_propagation:bridging:expsetup} details the experimental setup, while Section~\ref{sec:ris_propagation:bridging:expevaluation} discusses the experimental results. Finally, Section~\ref{sec:ris_propagation:bridging:conclusion} draws our conclusions and outlines directions for future work.

\noindent

\subsection{System Model}
\label{sec:ris_propagation:bridging:system_model}

\subsubsection{Open RAN Framework}
\label{sec:ris_propagation:bridging:oranframe}
\begin{figure}[t!]
  \centering
  \includegraphics[width=\columnwidth]{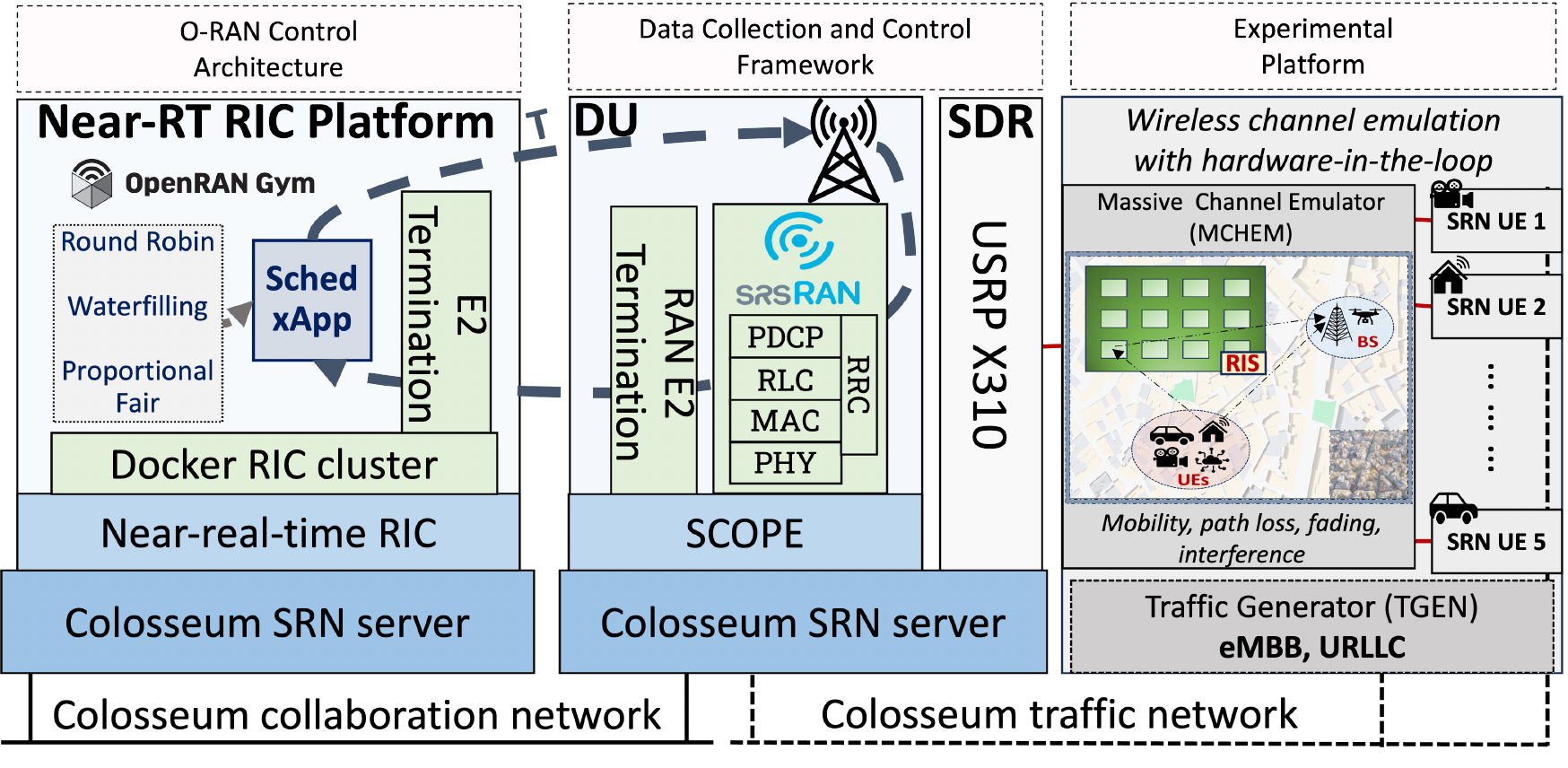} %pdf also, png orig
  \setlength\abovecaptionskip{-.2cm}
  \caption{Reference \oran testing architecture for \gls{ris}-enabled deployments, focusing on the \texttt{Sched} xApp Case Study, as described in Section~\ref{sec:ris_propagation:bridging:multiue-secc}.}
  \label{fig:bridging:ris-top}
  \vspace{-.9em}
\end{figure}

We investigate an Open \gls{ran} system where \glspl{ue} generate traffic with diverse profiles, which is classified into the \gls{embb} and \gls{urllc} network slices. We leverage an xApp tasked with reconfiguring the \gls{bs}'s \gls{mac}-layer scheduling policies. The xApp selects a dedicated scheduling profile among multiple resource allocation algorithms (i.e., \gls{rr}, \gls{wf}, and \gls{pf}), influencing how \glspl{prb} are internally allocated to the \glspl{ue} of each slice~\cite{10766614}. It is noted that for the \gls{rf} environment we consider various \gls{ris}-assisted network deployments. %, which will be described in the following sections.
In Fig.~\ref{fig:bridging:ris-top}, we outline the reference architecture considered in our work.

\subsubsection{Channel Model}
\label{sec:ris_propagation:bridging:channelmod}
We consider a topology consisting of a \gls{ris}, a \gls{bs}, and a set of \glspl{ue}, denoted by $I=\{1,\dots,i,\dots, |I|\}$, which establish direct communication with the \gls{bs}. The number of elements on the \gls{ris} is given as $|M|$ and their set is defined as $M=\{1, \dots , m, \dots , |M|\}$. The first \gls{ris} element, i.e., $m=1$, with its respective coordinates, i.e., $(x_R, y_R, z_R)$ [m], is used as a reference point in the following arithmetic calculations, while for the \gls{bs}, its coordinates are given as follows $(x_A, y_A, z_A)$ [m].
The channel gain corresponding to the direct link between a \gls{ue} $i$ and the \gls{bs} is given as $h_{iA}$, the channel gain between a \gls{ue} $i$ and the \gls{ris} is defined as $\mathbf{h}_{iR}$, where $\mathbf{h}_{iR}=[|h_{iR,1}| e^{j \omega_{1}}, \dots,|h_{iR,m}| e^{j \omega_{m}}, \dots,|h_{iR,|M|}| e^{j\omega_{|M|}}]^T$, and finally, the channel gain of the reflected link from the \gls{ris} to the \gls{bs} is denoted as $\mathbf{h}_{RA}$.
The diagonal phase-shift matrix is defined as $\boldsymbol{\Theta}=\mathrm{diag}( e^{j\theta_1},\dots,e^{j\theta_m},\dots,e^{j\theta_{|M|}} )$, while
each \gls{ris} element's phase shift is $\theta_m \in [0,2\pi]$, $\forall~m~\in~M$, and the cascaded channel is therefore given as $\mathbf{{h}}_{RA}^H\boldsymbol{\Theta}\mathbf{{h}}_{iR}$. Finally, the steering vector for the \gls{ue} $i$ to the \gls{ris} link is given as $\mathbf{h}_{iR}^{\prime\prime\prime}=[1, e^{-j \frac{2 \pi}{\lambda} d \phi_{iR}}, \dots, e^{-j \frac{2 \pi}{\lambda}(|M|-1) d \phi_{i R}}]^{T}$, while the steering vector for the \gls{ris} to the \gls{bs} link is denoted as $\mathbf{h}_{RA}^{\prime\prime\prime}=[1, e^{-j \frac{2 \pi}{\lambda} d \phi_{RA}}, \dots, e^{-j \frac{2 \pi}{\lambda}(|M|-1) d \phi_{RA}}]^T$. In the above calculations, $d$ is the Euclidean distance, $\lambda [m]$ is the carrier wavelength and $d [m]$ is the antenna separation. Additionally, $\phi_{RA}$ is the cosine of the signal's \gls{aod} for the \gls{ris} to the \gls{bs} link, and
$\phi_{iR}$ is the cosine of the signal's \gls{aoa} for the \gls{ue} $i$ to \gls{ris} link.
It is noted that we consider the links between the \glspl{ue} and the \gls{bs} to be \gls{nlos}, while the channels connecting the \glspl{ue} to the \gls{ris} and the \gls{ris} to the \gls{bs} are assumed to be \gls{los}. These transmission conditions are chosen to ensure that the cascaded channel formed by the \gls{ris} provides a sufficiently strong link for redirecting the \glspl{ue}' signals to the \gls{bs}.
%Such placement can represent a maneuverable flying \gls{bs} (e.g., a \gls{uav}) positioned near a \gls{ris} mounted on a building facade~\cite{diamanti2021energy,10973151}, to enhance capacity in emergency response scenarios, %but it is not limited to this use case. Another application could involve an \gls{enb} mounted on a building's roof, positioned close to a \gls{ris} %mounted on another building,
Such placement can represent a maneuverable flying \gls{ris} mounted on an \gls{uav}, capable of dynamically establishing \gls{los} to \glspl{ue} that would otherwise encounter \gls{nlos} relative to a fixed-position \gls{bs}.
%as well as traditional deployments where \glspl{ris} and \glspl{enb} are leveraged together to improve coverage.

Lastly, both the \gls{bs} and the \glspl{ue} are equipped with single-antenna omnidirectional \glspl{tx}/\glspl{rx}, with their center frequency set to $5.9$~GHz, where \glspl{ris} are considered for deployment\cite{ETSI_RIS001}. For the generation of the wireless links, we leverage \gls{quadriga}'s~\cite{jaeckel2014quadriga} statistical ray-tracing capabilities, which is based on channel sounding and geometry-based channel modeling. We observe the channels every $0.1$~s and select the 3GPP\textunderscore38.901\textunderscore UMa scenario for the channel generator. A number of \glspl{mpc} (denoted as $|N|$) are produced for each generated link, with the channel generation procedure based on the following flow adopted from~\cite{10973151}:
%\vspace{-0.15cm}
\noindent
\begin{enumerate}

\item Generate scenario-specific \glspl{mpc} (denoted as $h_{xy}^\prime$) for each link, leveraging \gls{quadriga}.

\item Combine the \glspl{mpc} coherently to compute the channel gain for every link, following Eq.~\eqref{bridging-eq:mpc-def}:
%{\small
\begin{equation}\label{bridging-eq:mpc-def}
\vspace{-0.12cm}
    \mathrm{h}^{\prime\prime}_{xy}=\sum_{j=1}^{|N|}\left|\textit{h}_{xy}^\prime\right| \cdot e^{j \varphi_{xy}^\prime},
\vspace{-0.2cm}
\end{equation}
%}
\noindent
The \gls{mpc} gain is denoted as $\textit{h}^\prime_{xy}$, where $\text{\(x \in \{i, R\}, \ y \in \{R, A\}\)}$, and
$|\textit{h}^\prime_{xy}|$, $\varphi^\prime_{xy}$ are the respective \gls{mpc} magnitude and phase.

\item Calculate the average of the channel gains (i.e., $\mathrm{h^{\prime\prime}_{xy}}$) over $100$ different channel realizations to determine the link's final channel gain (i.e., $h_{iA}$).

\item Multiply the computed channels by their corresponding steering vectors, denoted as $\mathbf{h}_{RA}^{\prime\prime\prime}$ and $\mathbf{h}_{iR}^{\prime\prime\prime}$, to obtain the respective final channel gains (i.e., $\mathbf{h}_{RA}$, $\mathbf{h}_{iR}$).

\item Calculate the overall channel power gain for the \gls{ue} $i$ to \gls{bs} link as follows $G_{i}=|h_{iA}+\mathbf{h}_{RA}^H \boldsymbol{\Theta} \mathbf{h}_{iR}|^2$.

\end{enumerate}

\subsection[Optimizing Wireless Propagation with RIS]{Optimizing Wireless Propagation with \texorpdfstring{\gls{ris}}{RIS}}
\label{sec:ris_propagation:bridging:opt}

In this work, the wireless environment is not treated as an uncontrollable entity but is instead subject to optimization. For this purpose, we leverage the capabilities of \glspl{ris} to reshape the wireless propagation by effectively adapting the phase shifts of the \gls{ris} elements. For these reasons, we formulate the following optimization problem defined in Eqs.~(\ref{bridging-eq4a})-(\ref{bridging-eq4b}), as follows:
\begin{subequations}
\begin{equation}
%\vspace{-3pt}
\max_{\boldsymbol{\theta}} \sum_{i=1}^{|I|} |h_{iA}+\mathbf{h}_{RA}^H \boldsymbol{\Theta} \mathbf{h}_{iR}|^2
\label{bridging-eq4a}
\end{equation}
\begin{equation}
\text{s.t.\quad}0\leq\theta_m \leq 2\pi, \forall m\in M.
\vspace{-5pt}
\label{bridging-eq4b}
\end{equation}
\end{subequations}

%\noindent
It is noted that the solution to the aforementioned optimization problem, which takes place on the \gls{bs}'s side, is the effective phase shift vector of the \gls{ris} elements, denoted as $\boldsymbol{\theta^{*}}$. In order to maximize the overall channel power gain of the \glspl{ue}, the \gls{bs} only needs to calculate the effective phase shifts of the \gls{ris} elements (i.e., \( \boldsymbol{\theta^*} \)). Given the non-convexity of the aforementioned optimization problem, computing a globally optimal solution is challenging. To derive a closed-form phase-shift solution, we aim to achieve coherent signal alignment for the incoming signals from different transmission paths at the \gls{bs}. Indeed, the \gls{ue}'s channel power gain is maximized when the direct signal and the signal reflected by the \gls{ris} are perfectly aligned and coherently combined at the \gls{bs}'s \gls{rx}~\cite{li2020reconfigurable,diamanti2021energy}. This condition is generally satisfied when the phase shifts of the direct and cascaded signals are aligned as $\angle h_{iA} = \angle \left( \mathbf{h}_{RA}^{H} \mathbf{\Theta} \mathbf{h}_{iR} \right)$. Therefore, in the single-\gls{ue} case there's an \(1 \times |M|\) phase-shift vector \(\boldsymbol{\theta^*} = \angle \mathbf{v}\), where the optimal phase shifts of the \gls{ris} elements are given as follows: %in Eq.~\eqref{eq8}:
\begin{equation} \label{bridging-eq8}
\theta_{m}^{*}=\angle{h_{iA}}+\omega_{m}+\frac{2 \pi}{\lambda} d(m-1) \phi_{RA}, \forall m \in \ M.
\end{equation}

%\noindent
In the multi-\gls{ue} case, the solution to the maximization problem is also a linear combination of the phase shifts of the \gls{ris} elements, with the reflection-coefficient vectors \( \mathbf{v}_i \) being distinct for each \gls{ue}. Therefore, there exists a distinct reflection-coefficient vector \( \mathbf{v}_i = [v_{i,1}, \ldots, v_{i,|M|}] \in \mathbb{C}^{|M| \times 1} \) for each \gls{ue} \( i \) that maximizes its channel power gain. To determine the set of suitable weights for each \gls{ue} that will maximize the aggregate channel power gain across all \glspl{ue} as defined in Eq.~\eqref{bridging-eq4a}, we adopt the research methodologies proposed in~\cite{8970580,diamanti2021prospect}. For each \gls{ue} $i$, we define an appropriate weight factor \( w_i \in [0,1] \) and compute the linear combination of the overall
\glspl{ue}' reflection-coefficients \( \mathbf{v}_i \) as follows:
\begin{equation}
\vspace{-0.18cm}
    \mathbf{v} = \sum_{i=1}^{|I|} w_i \mathbf{v}_i,
\label{bridging-ris-weigh-factor}
\vspace{-0.28cm}
\end{equation}
\noindent
such that the weight factors satisfy the condition \( \sum\limits_{i=1}^{|I|} w_i = 1\).
\noindent
In the multi-\gls{ue} case, the goal is to determine the optimal values of the weight factors (i.e., $w_i$) for the reflection-coefficient vector (i.e., $\mathbf{v}$) of Eq.~\ref{bridging-ris-weigh-factor}, which maximize the overall channel power gain of the \glspl{ue}, as defined in Eq.~\eqref{bridging-eq4a}. Therefore, in order to compute the effective \gls{ris} elements phase shifts, the optimization problem defined in Eqs.~\eqref{bridging-eq4a}-\eqref{bridging-eq4b} is formulated as follows in Eqs.~\eqref{bridging-eq4-weights}-\eqref{bridging-eq4-weights3}:
\begin{subequations}
\begin{equation}
\max_{\textbf{w}} \sum_{i=1}^{|I|} |h_{iA}+\mathbf{h}_{RA}^H \boldsymbol{\Theta} \mathbf{h}_{iR}|^2
\label{bridging-eq4-weights}
\end{equation}
\begin{equation}
\text{s.t.\quad}0\leq w_i \leq 1, \forall i\in I
%\vspace{-5pt}
\label{bridging-eq4-weights2}
\end{equation}
\begin{equation}
\sum_{i=1}^{|I|} w_i = 1
\vspace{-5pt}
\label{bridging-eq4-weights3}
\end{equation}
\end{subequations}
\noindent
Through backward induction~\cite{diamanti2021prospect}, the \gls{ris} reflection matrix \( \boldsymbol{\Theta} = \mathrm{diag}(e^{j \angle \mathbf{v}}) \) is computed to optimize the phase alignment. %across multiple \glspl{ue}.
This approach ensures that the final \gls{ris} configuration \( \boldsymbol{\Theta} \) maximizes both the channel gain of each \gls{ue} and the total channel power gain across all \glspl{ue}. Given that
\(\mathbf{w} = [w_1, \ldots, w_i, \ldots, w_{|I|}]\) is the vector of the \glspl{ue}' assigned weight factors, the optimization problem described in Eqs.~\eqref{bridging-eq4-weights}-\eqref{bridging-eq4-weights3} consists of a non-negative linear objective function and a set of constraints, which can be intuitively optimized to derive the optimal weights \( \mathbf{w}^* \). The derivation of the optimal weights \( \mathbf{w}^* \) leads to an effective \gls{ris} elements' reflection-coefficient vector
\( \mathbf{v}^* = [v_1, \ldots, v_m, \ldots, v_{|M|}] \), which subsequently determines the corresponding phase shifts of the \gls{ris} elements, expressed as
\( \boldsymbol{\theta}^* = [\theta_1, \ldots, \theta_m, \ldots, \theta_{|M|}] \).%\footnote{Extensions of the \oran control plane to carry \gls{ris} weights determined by the scheduler for real-time network adaptation are left for future work.}.

\subsection{Resource Allocation Strategies}
\label{sec:ris_propagation:bridging:strategies}

We consider three different case studies, encompassing both single and multi-\gls{ue} scenarios, which are evaluated under various resource management policies in terms of \gls{prb} allocation and scheduling profile selection. %to assess the contribution of \glspl{ris} to network performance and resource utilization.

\begin{comment}
%\subsection{Sliced-based evaluation}
\textbf{Case A}. Following the requirements outlined in~\cite{ETSI_RIS001}, which specify that existing \gls{ris} elements can range from $10$ and above, we evaluate three \gls{ris}-assisted topologies with $10$, $100$, and $1000$ elements onboarded on the \gls{ris}. %These topologies represent increasing orders of magnitude, corresponding to successive powers of $10$.
We consider a single \gls{ue} allocated to the \gls{embb} network slice, which is assigned $\sim!1!/!3$ of the available bandwidth resources.
\end{comment}
\textbf{Case Study A}. Following the requirements outlined in~\cite{ETSI_RIS001}, which specify that existing \gls{ris} panels can host a minimum of $10$ elements, we evaluate three network configurations, where the number of \gls{ris} elements increases by orders of magnitude. %, corresponding to successive powers of $10$.
We consider a single \gls{ue} allocated to the \gls{embb} network slice, which is assigned $\sim1/3$ of the available bandwidth resources. The \gls{csi} is varied to evaluate the contribution of \gls{ris} to network performance given the same resource availability, which is considered fixed. The results of this analysis are provided in Section~\ref{sec:ris_propagation:bridging:singleue-seca}.

\textbf{Case Study B}. Considering the trade-off between hosting thousands of \gls{ris} elements and maintaining operational simplicity, we focus on a \gls{ris} with $100$ elements, serving multiple \glspl{ue}, which are distributed across the \gls{embb} and \gls{urllc} network slices. By dividing the available spectrum between the two, %, allocating $40\%$ of the \glspl{ue} to the \gls{embb} slice and the remaining $60\%$ to the \gls{urllc}.
 we aim to demonstrate how a \gls{ris} can significantly enhance network performance, improve \gls{ue} satisfaction, and increase \gls{prb} utilization based on the capacity demands of each slice. The findings of this investigation are provided in Section~\ref{sec:ris_propagation:bridging:multiue-secb}.

%\subsection{xApp-based scheduling profile selection}
\textbf{Case Study C}. This case study focuses on an \oran infrastructure where an xApp controls the scheduling profile selection of the \gls{bs}'s \gls{du}. The xApp updates the scheduling profile of the \gls{bs}'s scheduler through a control loop by alternating its policy given the following options: \gls{rr}, \gls{wf}, and \gls{pf}. A multi-\gls{ue} use case is considered, where \gls{embb} \glspl{ue} are provisioned with $90\%$ of the available bandwidth, while \gls{urllc} \glspl{ue} are allocated the remaining $10\%$ of the spectrum resources. With this case study, we aim to study the impact of an xApp frequently changing the scheduling profile from fair scheduling (i.e., \gls{rr}) to spectral efficiency-driven (i.e., \gls{wf}), or balanced scheduling (i.e., \gls{pf}), analyzing its effect on resource allocation and overall network performance in both \gls{ris}-assisted and non-\gls{ris}-assisted topologies. The results of this exploration are provided in Section~\ref{sec:ris_propagation:bridging:multiue-secc}.

\subsection{Experimental Setup}
\label{sec:ris_propagation:bridging:expsetup}

To experimentally evaluate the \gls{ris}-assisted channels within an \oran ecosystem, we leverage the capabilities of OpenRAN Gym, an experimental toolbox for the end-to-end development, implementation, and testing of diverse solutions---including, but not limited to, \gls{ai}/\gls{ml} applications---within \oran. The software offers capabilities such as \gls{ran} and core network deployments using the srsRAN~\cite{gomez2016srslte} protocol stack. It supports large-scale data collection, testing, and fine-tuning of \gls{ran} functionalities by integrating open \glspl{api} %into srsRAN
for slicing and scheduling control, as well as \glspl{kpm} collection. Additionally, the software incorporates an \oran-compliant control architecture for executing xApps in the near-real-time \gls{ric}. Through the E2 interface, which bridges the \gls{ran} and \gls{ric}, along with its \glspl{sm}~\cite{polese2022understanding}, the streaming of \glspl{kpm} from the \gls{ran} and the execution of control actions by the xApps are enabled.
%We deploy OpenRAN Gym on the Colosseum Open RAN Digital Twin~\cite{10643670}, a publicly available testbed designed for testing \oran-based optimization frameworks in controlled environments

We deploy OpenRAN Gym on Colosseum, which features $128$~\glspl{srn} consisting of pairs of Dell PowerEdge R730 servers and NI \gls{usrp} X310 \glspl{sdr}, supporting large-scale experimentation in diverse network deployments. The testbed's \gls{mchem} component emulates the wireless environment by leveraging \gls{fpga}-based \gls{fir} filters. These filters simulate \gls{rf} conditions, including path loss, fading and attenuation based on models created through various frameworks (e.g., ray-tracing software, analytical models, or real-world measurements). Similarly, the Colosseum \gls{mgen} TCP/UDP traffic generator~\cite{mgen} emulates a variety of network traffic profiles (e.g., multimedia content) and demand distributions (e.g., Poisson, periodic).

\begin{figure}[b!]
  \vspace{-1em}
  \centering
  \includegraphics[width=\columnwidth]{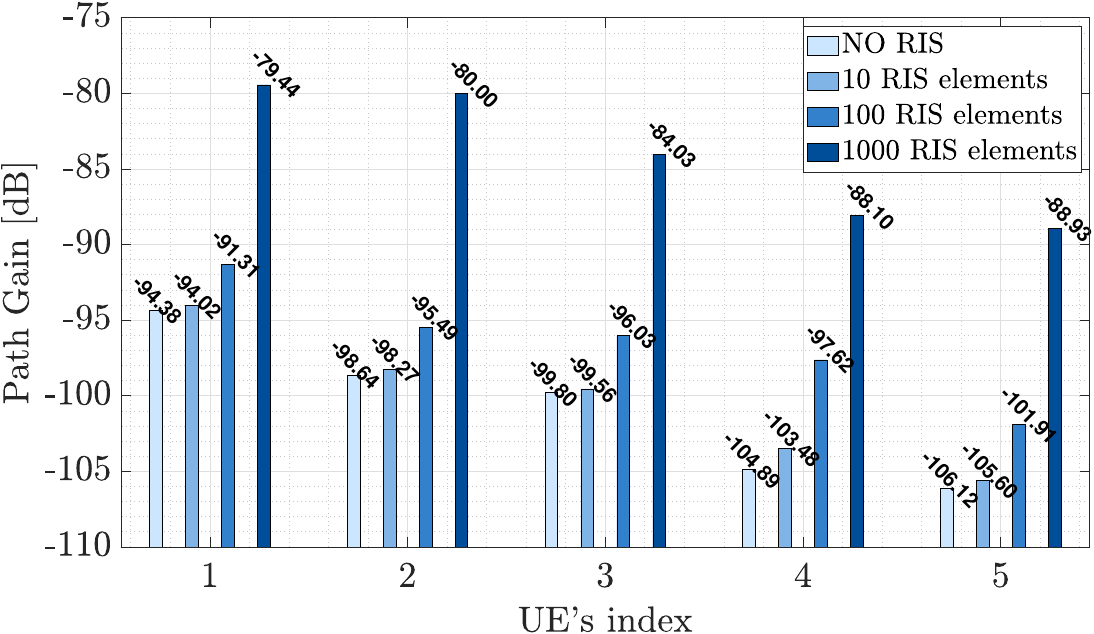}
  \setlength\abovecaptionskip{-.1cm}
  \vspace{-0.3cm}
  \caption{Path Gains under varying numbers of \gls{ris} elements.}
  \label{fig:bridging:pathgainsris}
\end{figure}

We consider a cellular network comprising one \gls{bs} and up to $5$~\glspl{ue} distributed across $2$ network slices. These are: (i)~\gls{embb}, for high-traffic scenarios involving multimedia content and streaming applications; and (ii)~\gls{urllc}, designed for time-critical applications such as vehicle coordination in \gls{cv2x} environments. The maximum bandwidth considered in this work is set to $10$\:MHz (i.e., $50$ \glspl{prb}) and is divided among the $2$ slices. Traffic is slice-based and generated according to the following specifications: \gls{embb} \glspl{ue} request a constant bitrate of $4$\:Mbps, while \gls{urllc} \glspl{ue} generate Poisson traffic at $89.3$\:kbps. In terms of physical deployment, the \gls{bs} is placed at the point $(x_A=25,y_A=50,z_A=25)$ [m] of the three-dimensional space.
Their distance from the \gls{ris} is given as $d_{i,R}=[20, 27, 37, 58, 66]$~[m] and all are uniformly distributed around the \gls{ris}'s reference point, denoted as $(x_R=30,y_R=40,z_R=20)$ [m]. %Both the \gls{enb} and the \gls{ris} are positioned in close proximity to ensure significantly strong \gls{los} links. Such placement can represent a maneuverable flying \gls{enb} (e.g., a \gls{uav}) positioned near a \gls{ris} mounted on a building facade~\cite{diamanti2021energy}, to enhance capacity in emergency response scenarios, %but it is not limited to this use case. Another application could involve an \gls{enb} mounted on a building's roof, positioned close to a \gls{ris} %mounted on another building,
%as well as traditional deployments where \glspl{ris} and \glspl{enb} are leveraged together to improve coverage.

%following the pipeline discussed in,
We install the generated links in Colosseum. The Colosseum scenario includes the path gain (i.e., a complex \gls{fir} coefficient) and the \gls{toa} values, %The path gain is derived as described in Section~\ref{channelmod},
%while the \gls{toa} values
which are calculated based on the Euclidean distance between the active entities (i.e., \gls{bs}, \gls{ris}, and \glspl{ue}), using the speed of light. % (i.e., $3 \times 10^8$ [m/s]).
These values are used to generate the time-variant \gls{cir} for each node pair in the network.

\begin{comment}
by using OpenRAN Gym and the Colosseum network emulator over

Each \gls{drl} model evaluated in the following sections takes as input \gls{ran} \glspl{kpm} such as throughput, buffer occupancy, number of \glspl{prb}, and outputs resource allocation policies (e.g., RAN slicing and/or scheduling) for \gls{ran} control.
\end{comment}

%\section{Experimental Evaluation\protect\footnote{It is noted that all the results presented in Section~\ref{expevaluation} have been averaged over multiple repetitions of the experiments.}}\label{expevaluation}

\subsection{Experimental Evaluation}
\label{sec:ris_propagation:bridging:expevaluation}

%\hspace{-15.5cm}
\begin{table}[H]
\vspace{-.55cm}
\centering
\renewcommand{\arraystretch}{1.05} % Increase row height by 20%
\Large
%\fontsize{28}{2}
% \setlength\abovecaptionskip{-.1cm}
\caption{Catalog of Network Configurations}
\begin{adjustbox}{width=1\linewidth}
%\begin{tabular}{@{}c@{\hskip 0cm}c@{}cccccc@{}}
\begin{tabular}{@{}ccccccc@{}}
%\begin{tabular}{@{\hskip 0.05cm}c@{\hskip 0.05cm}c@{\hskip 0.05cm}c@{\hskip 0.05cm}c@{\hskip 0.05cm}c@{\hskip 0.05cm}c@{\hskip 0.05cm}c@{\hskip 0.05cm}}
%\begin{tabular}{@{\hskip 0cm}c@{\hskip 0cm}c@{\hskip 1.5cm}c@{\hskip 0cm}c@{\hskip 0cm}c@{\hskip 0cm}c@{\hskip 0cm}c@{\hskip 0cm}}
\toprule
\multicolumn{1}{c}{\shortstack{\textbf{\centering Config.} \\ \textbf{ID}}} &
\multicolumn{1}{c}{ \shortstack{\textbf{Number} \\ \textbf{of \glspl{ue}}}} &
\multicolumn{1}{c}{ \shortstack{\textbf{\gls{ue}} \\ \textbf{ID(s)}}} &
\multicolumn{1}{c}{ \shortstack{\textbf{Slice} \\ \textbf{Type}}} &
\multicolumn{1}{c}{ \shortstack{\textbf{Bandwidth} \\ \textbf{[MHz]}}} &
\multicolumn{1}{c}{\shortstack{\textbf{\gls{ris}} \\ \textbf{Assisted}}} &
\multicolumn{1}{c}{\shortstack{\textbf{xApp} \\ \textbf{Enabled}}} \\
\midrule
\centering \textbf{I} & $1$ & $5$ & \text{\gls{embb}} & \text{$3.6$} & \text{No \gls{ris}} & \text{No}\\
\centering \textbf{II} & $1$ & $5$ & \text{\gls{embb}} & \text{$3.6$} & \text{$100$ \gls{ris} el.} & \text{No} \\
\centering \textbf{III} & $1$ & $1$ & \text{\gls{embb}} & \text{$3.6$} & \text{$10$ \gls{ris} el.}         & \text{No} \\
\centering \textbf{IV} & $1$ & $1$ & \text{\gls{embb}} & \text{$3.6$} & \text{$1000$ \gls{ris} el.}             & \text{No} \\
%\midrule
\centering \textbf{V} & $5$ & \text{$\{1, 2\}$ \& $\{3, 4, 5\}$} & \text{\gls{embb} \& \gls{urllc}} & \text{$5$ \& $5$} & \text{No \gls{ris}}         & \text{No}\\
\centering \textbf{VI} & $5$ & \text{$\{1, 2\}$ \& $\{3, 4, 5\}$}& \text{\gls{embb} \& \gls{urllc}} & \text{$5$ \& $5$} & \text{$100$ \gls{ris} el.}             & \text{No} \\
\centering \textbf{VII} & $5$ & \text{$\{1, 2\}$ \& $\{3, 4, 5\}$} & \text{\gls{embb} \& \gls{urllc}} & \text{$9$ \& $1$} & \text{No \gls{ris}} & \text{Yes}\\
\centering \textbf{VIII} & $5$ & \text{$\{1, 2\}$ \& $\{3, 4, 5\}$} & \text{\gls{embb} \& \gls{urllc}} & \text{$9$ \& $1$} & \text{$100$ \gls{ris} el.}             & \text{Yes}\\
\bottomrule
\end{tabular}
\end{adjustbox}
\label{table:bridging:net-topologies}
% \vspace{-.35cm}
\end{table}

% CASE STUDY A EXPERIMENTAL RESULTS
\begin{figure*}[t!]
\centering
\subfloat[\label{fig:bridging:singleue_thr}]{\includegraphics[height=4cm]{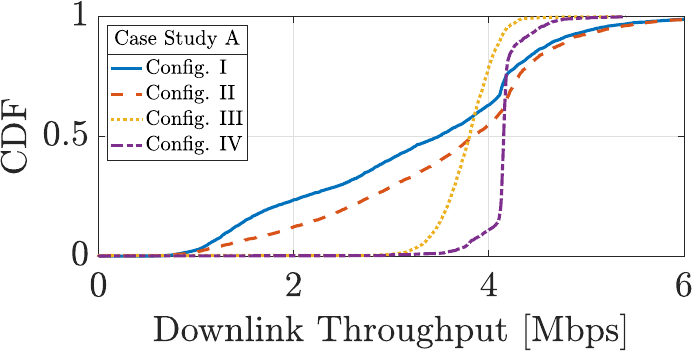}}
%
% \hfil
%
\centering
\subfloat[\label{fig:bridging:singleuecqi}]{\includegraphics[height=4cm]{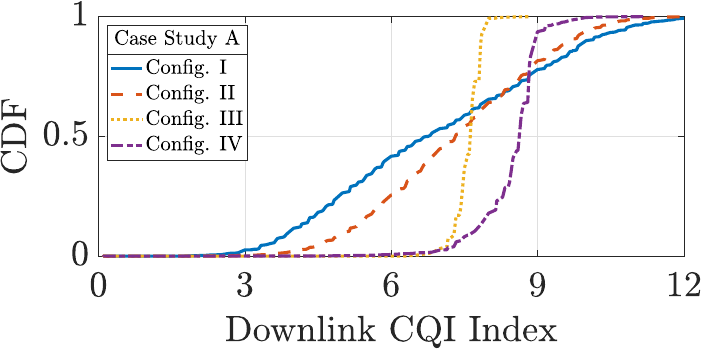}}
\hfil
\centering
\subfloat[\label{fig:bridging:singleuemcs}]{\includegraphics[height=4cm]{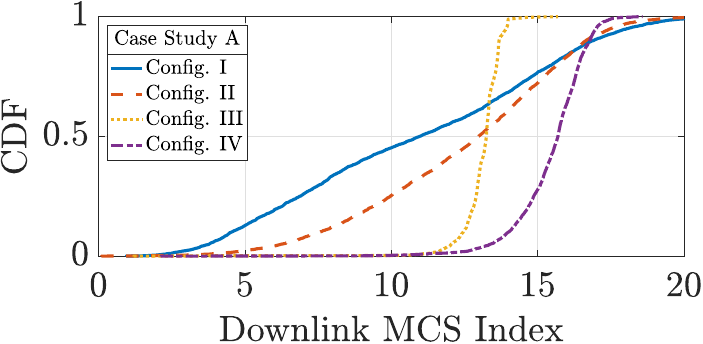}}
\setlength\abovecaptionskip{-.02cm}
\caption{Performance evaluation results for the single-\gls{ue} Case Study.}
\label{fig:bridging:singleue-total}
\vspace{-1em}
\end{figure*}

% CASE STUDY B EXPERIMENTAL RESULTS
\begin{figure*}[t!]
\vspace{-.5em}
\centering
\subfloat[\label{fig:bridging:multiue_thr}]{\includegraphics[height=4cm]{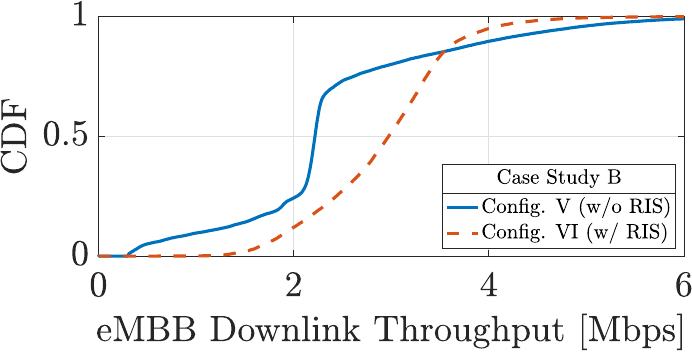}}
%
% \hfil
%
\centering
\subfloat[\label{fig:bridging:multiuebuf}]{\includegraphics[height=4cm]{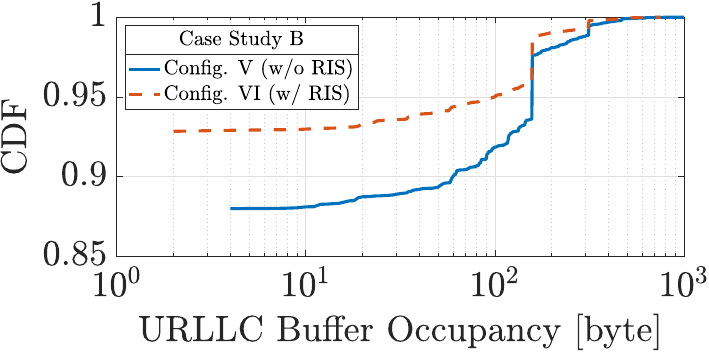}}
\hfil
\centering
\subfloat[\label{fig:bridging:multiueratio}]{\includegraphics[height=4cm]{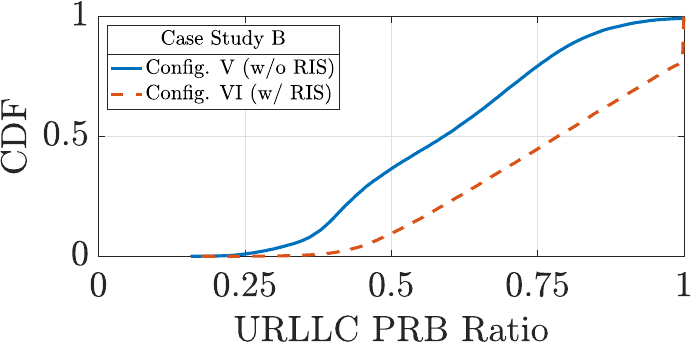}}
\setlength\abovecaptionskip{-.02cm}
\caption{Performance evaluation results for the multi-\gls{ue} Case Study.}
\label{fig:bridging:multiue-1}
\vspace{-1.5em}
\end{figure*}

% CASE STUDY C EXPERIMENTAL RESULTS
\begin{figure*}[t!]
\centering
% --- First Row ---
\subfloat[\label{fig:bridging:multiue_thr_c}]{\includegraphics[width=0.45\textwidth]{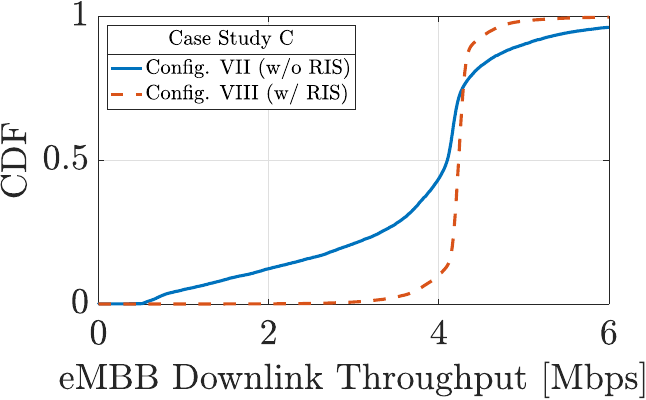}}
\hfill
\subfloat[\label{fig:bridging:multiuebuf_c}]{\includegraphics[width=0.45\textwidth]{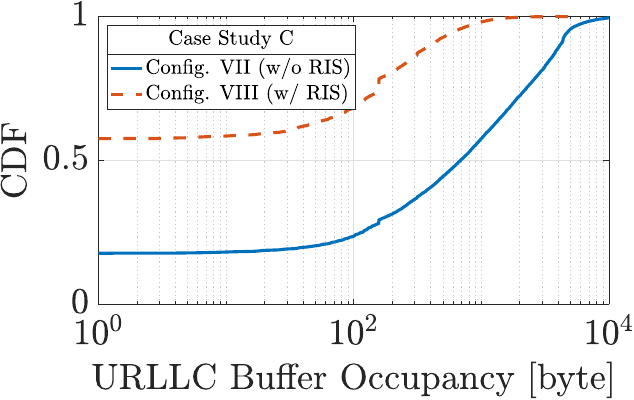}}

\vspace{1em} % Space between the two rows

% --- Second Row ---
\subfloat[\label{fig:bridging:multiueprb_c}]{\includegraphics[width=0.45\textwidth]{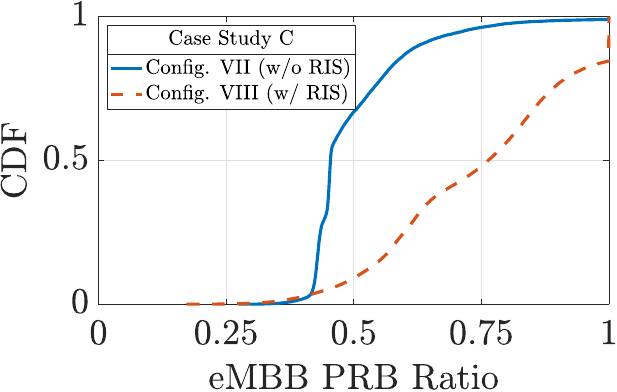}}
\hfill
\subfloat[\label{fig:bridging:multiue_case_c}]{\includegraphics[width=0.45\textwidth]{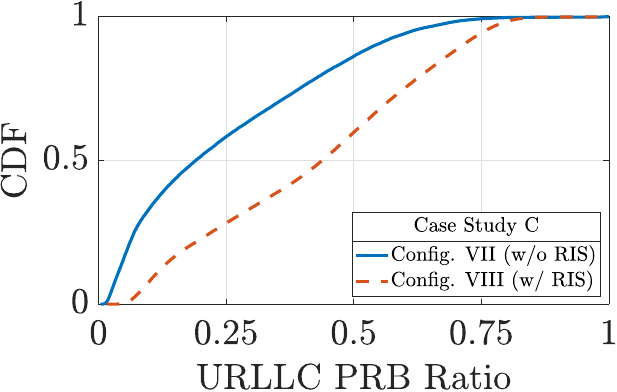}}

\caption{Performance evaluation results for the multi-\gls{ue} \oran\ Case Study.}
\label{fig:bridging:multiue-2}
\end{figure*}

\begin{figure}[t!]
\centering
% --- Figure (a) ---
\subfloat[\label{fig:bridging:baseline1}]{\includegraphics[width=0.45\textwidth]{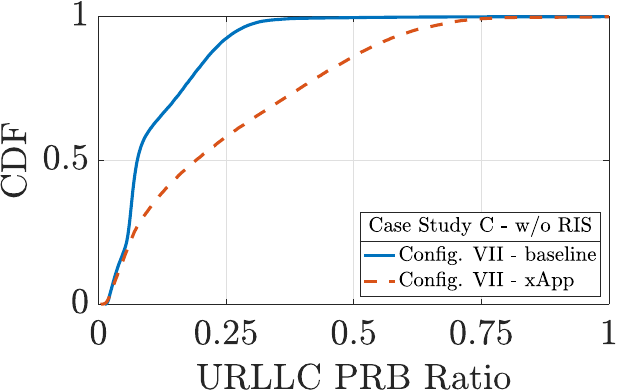}}
\hfill % Pushes the two figures to the margins to give them maximum space
%
% --- Figure (b) ---
\subfloat[\label{fig:bridging:baseline2}]{\includegraphics[width=0.45\textwidth]{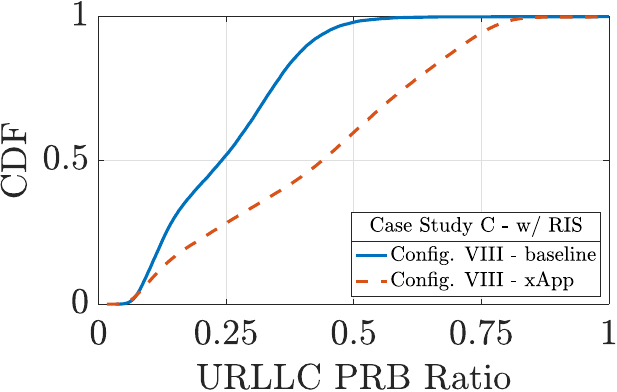}}
\vspace{5pt} % Adds a small gap between the plots and the caption
\caption{Performance evaluation results for the multi-\gls{ue} Case Study with and without the inclusion of the \texttt{Sched} xApp in \oran.}
\label{fig:bridging:baseline-total}
\end{figure}

In Fig.~\ref{fig:bridging:pathgainsris}, we present the path gains for each \gls{ue}, $i$, with and without a \gls{ris}, generated through the channel modeling process and the optimization procedure as described in Sections~\ref{sec:ris_propagation:bridging:channelmod} %. It is reminded that the generated channels result from %for deriving the effective phase shifts of the \gls{ris} elements,
and Section~\ref{sec:ris_propagation:bridging:opt}, respectively.
All the \glspl{ue} are sorted in descending order, such that the \gls{ue} with the lowest index corresponds to the \gls{ue} with the best channel, while the \gls{ue} with the highest index corresponds to the \gls{ue} with the worst. Based on the reported results, the performance with as few as~$10$ \gls{ris} elements is nearly identical to that observed without the \gls{ris}. Improvements in path gains are evident across all \glspl{ue} when examining scenarios with $100$ or $1000$ \gls{ris} elements.

In Table~\ref{table:bridging:net-topologies}, we present a catalog of the various \textit{network configurations} evaluated within the scope of this work. Configs. \textbf{I}-\textbf{IV} focus on a single \gls{ue}, whereas configs. \textbf{V}-\textbf{VIII} address the multi-\gls{ue} case study, where scheduling decisions influence intra-slice resource allocation. Regarding configs. \textbf{VII}-\textbf{VIII}, an xApp tasked with reconfiguring the scheduling profile selection is evaluated. The xApp updates the scheduling profile of the \gls{bs} at a granularity of $1$~s, assigning one of three scheduling profiles (i.e., \gls{rr}, \gls{wf}, and \gls{pf}) to the \glspl{ue} of the \gls{embb} and \gls{urllc} slices, %in all possible combinations and in a cyclic manner.
cycling through all possible combinations.

It is noted that all the results presented in this chapter have been averaged over multiple repetitions of the experiments.

\subsubsection{Single-UE Case Study A}
\label{sec:ris_propagation:bridging:singleue-seca}
In Fig.~\ref{fig:bridging:singleue-total}, %we present the experimental results for the single-\gls{ue} use case. Specifically,
we compare the \gls{ue} with the best channel conditions, as shown in Fig.~\ref{fig:bridging:pathgainsris} (i.e., \gls{ue} with ID $1$), to the \gls{ue} with the worst (i.e., \gls{ue} with ID $5$). %We assume that they generate \gls{embb} traffic based on the specifications outlined in Section~\ref{expsetup}.
%Finally, the topologies evaluated in this subsection are \textbf{I}-\textbf{IV}, with their descriptions provided in Table~\ref{table:net-topologies}.
In Fig.~\ref{fig:bridging:singleue_thr}, focusing on Configs. \textbf{I} and \textbf{II} (Table~\ref{table:bridging:net-topologies}), which correspond to the \gls{ue} with the worst channel, we observe that the median throughput reported with $100$ \gls{ris} elements is $3.834$\:Mbps, representing an increase of $\sim10\%$ compared to the $3.486$\:Mbps observed with no \gls{ris}. Similarly, when focusing on the \gls{ue} with the best channel (Configs. \textbf{III} and \textbf{IV}), we observe that the median throughput reported with $1000$ \gls{ris} elements is $4.154$~Mbps, representing an $\sim10\%$ increase compared to the $3.8$~Mbps achieved with only $10$ \gls{ris} elements.
It is reminded that according to the experimental results presented in Fig.~\ref{fig:bridging:pathgainsris}, the performance achieved with $10$ \gls{ris} elements is nearly identical to that observed without a \gls{ris} in the topology.
Based on the aforementioned experimental evaluation, we observe that increases in the order of $10^{2}$ in the size of the panels (e.g., from no \gls{ris} to $10^{2}$ \gls{ris}, or from $10^{1}$ \gls{ris} to $10^{3}$ \gls{ris}) can result in a $\sim10\%$ increase in the reported throughput under the same resource availability (e.g., $18$~\glspl{prb} or $3.6$~MHz bandwidth).

In Figs.~\ref{fig:bridging:singleuecqi} and~\ref{fig:bridging:singleuemcs}, we present the corresponding \gls{cqi} and \gls{mcs} values as reported by the protocol stack. The results in Fig.~\ref{fig:bridging:singleuecqi} demonstrate the channel improvement introduced by the \gls{ris} in the topology, particularly for the \gls{ue} with the best channel conditions (Configs. \textbf{III}-\textbf{IV}), which reports a $\sim13\%$
improvement in the median \gls{cqi} value (from $8$ to $9$). Similarly, the \gls{mcs} (Fig.~\ref{fig:bridging:singleuemcs}) increases from a median value of $13$ to $16$, enabling higher data rates given the same \gls{prb} availability. Regarding the \gls{ue} with the worst channel (Configs. \textbf{I}-\textbf{II}), the \gls{mcs} shows an increase of $\sim18\%$, rising from $11$ to $13$, while the \gls{cqi} remains relatively stable around $7$, with a modest increase of $\sim8\%$ for the \gls{ris} topology.

\subsubsection{Multi-UE Case Study B}
\label{sec:ris_propagation:bridging:multiue-secb}
We examine the multi-\gls{ue} case study, where \glspl{ue} are allocated to specific slices based on their traffic demands.  To express \gls{ue} satisfaction (which corresponds to high resource utilization), we define a new metric, i.e., the \gls{prb} Ratio. Its definition is given as $\text{\small \gls{prb} Ratio}=\frac{\text{Sum of Granted \glspl{prb}}}{\text{Sum of Requested \glspl{prb}}}, \text{where \gls{prb} Ratio} \in [0, 1]$, and it represents the amount of allocated \glspl{prb} to the slice according to the channel and traffic conditions. The analysis focuses on Configs. \textbf{V} and \textbf{VI} (Table~\ref{table:bridging:net-topologies}), where the \glspl{ue} with the best reported channel conditions (Fig.~\ref{fig:bridging:pathgainsris}) are grouped together on the \gls{embb} slice, while the remaining \glspl{ue} are assigned to the \gls{urllc}.
The \gls{wf} algorithm is selected for the scheduling profile of the \gls{embb} (which has the highest throughput requirements), while \gls{rr} is used for the \gls{urllc} (which has the lowest).
Fig.~\ref{fig:bridging:multiue-1} presents the complete performance evaluation results for the aforementioned case.
Fig.~\ref{fig:bridging:multiue_thr} illustrates the median \gls{embb} throughput for the \gls{ris} and non-\gls{ris} assisted topologies. The median throughput with a \gls{ris} is $2.973$\:Mbps, $\sim34\%$ higher than the $2.219$\:Mbps achieved without a \gls{ris}, under the same resource availability. Fig.~\ref{fig:bridging:multiuebuf} shows that both topologies (with and without a \gls{ris}) result in a median buffer occupancy of $0$ in the \gls{urllc}, indicating significantly reduced latency levels.\footnote{It is noted that since srsRAN does not directly measure latency, buffer occupancy is used as a proxy~\cite{10766614}.}
Fig.~\ref{fig:bridging:multiueratio} shows that although both topologies indicate near-zero latency, the inclusion of a \gls{ris} (Config. \textbf{VI}) leads to higher utilization efficiency of the available \glspl{prb} at the \gls{bs}'s scheduler.

\subsubsection[Multi-UE O-RAN Case Study C]{Multi-\texorpdfstring{\gls{ue}}{UE} \texorpdfstring{\oran}{O-RAN} Case Study C}
\label{sec:ris_propagation:bridging:multiue-secc}
In Fig.~\ref{fig:bridging:multiue-2}, focusing on Configs. \textbf{VII} and \textbf{VIII}, as detailed in Table~\ref{table:bridging:net-topologies}, we present the results of the experimental evaluation where the \texttt{Sched} xApp controls the \gls{bs}'s scheduling policy.
Fig.~\ref{fig:bridging:multiue_thr_c} shows that the median \gls{embb} throughput with a \gls{ris} in the topology is $4.234$\:Mbps, compared to $4.1$\:Mbps without a \gls{ris}. These high throughput values are reasonable, as the \gls{embb} \glspl{ue} are allocated $45$~\glspl{prb}, representing $90\%$ of the available resources. Fig.~\ref{fig:bridging:multiuebuf_c}, demonstrates that in the presence of a \gls{ris}, the median \gls{urllc} buffer occupancy is $0$, whereas in its absence, the corresponding value is $688$, indicating high latency levels. Therefore, the results show that the presence of $100$ \gls{ris} elements significantly reduces latency levels, regardless of the \gls{bs}'s scheduling policy. On the contrary, non-\gls{ris}-assisted topologies in resource-limited environments (i.e., with only $\sim10\%$ of the available bandwidth) experience performance degradation, even when the \gls{bs} frequently updates its scheduler from fair (e.g., \gls{rr}) to more efficient approaches (e.g., \gls{wf} or \gls{pf}) to accommodate the \glspl{ue} based on their distinct \gls{qos} demands.
Finally, Figs.~\ref{fig:bridging:multiueprb_c} and~\ref{fig:bridging:multiue_case_c} demonstrate that \gls{ris}-enabled topologies achieve higher resource utilization for both the \gls{embb} and \gls{urllc} slices. Figs.~\ref{fig:bridging:baseline1} and~\ref{fig:bridging:baseline2} compare the previously discussed topologies to the case where no xApp reconfigures the \gls{bs}'s scheduling policy. Notably, the keyword \texttt{baseline}, as shown in Fig.~\ref{fig:bridging:baseline-total}, refers to srsRAN's default scheduling policy, which is \gls{rr}. Fig.~\ref{fig:bridging:baseline1} shows that updating the \gls{bs}'s scheduling profile increases resource efficiency for \gls{urllc} in the absence of a \gls{ris} (Config. \textbf{VII}). However, in the presence of a \gls{ris} (Fig.~\ref{fig:bridging:baseline2}), resource utilization and hence \gls{ue} satisfaction improve significantly (Config. \textbf{VIII}). Therefore, both the presence of \gls{ris} in the topology and frequent updates of the \gls{bs}'s scheduling policy enhance the resource utilization efficiency.

\subsection{Conclusions}
\label{sec:ris_propagation:bridging:conclusion}
In this work, we focused on an Open \gls{ran} and investigated the impact of various network management policies on the performance of \gls{ris}-assisted topologies, leveraging the capabilities of OpenRAN Gym on Colosseum. Our experimental results demonstrate that a \gls{ris} deployment can enhance network throughput and resource utilization under the same bandwidth availability. In addition, our experimentation in an emulated \oran system, where the \gls{bs}'s scheduling policy is controlled by an xApp, demonstrates that \gls{ris}-assisted topologies achieve superior network performance and higher resource efficiency compared to their non-\gls{ris}-assisted counterparts, regardless of the scheduling policy employed. Part of our current and future work focuses on designing \gls{ai}/\gls{ml} solutions to dynamically control the selection of scheduling policies in such Open \glspl{ran}, leveraging \gls{csi} and traffic conditions in the decision-making processes. Extensions to the \oran control plane for carrying \gls{ris} weights, determined by the scheduler for real-time network adaptation, are also left for future work.
\vspace{-0.1cm}

% --- EOF ---

%% file: tex/ai_ran_link_adaptation.tex
% !TEX root = ../thesis_dissertation template.tex
% ai_ran_link_adaptation.tex

\chapter[AI-RAN Link Adaptation Leveraging Digital Twins]{\texorpdfstring{\gls{ai}}{AI}-\texorpdfstring{\gls{ran}}{RAN} Link Adaptation Leveraging \texorpdfstring{\glspl{dt}}{Digital Twins}}
\label{chap:ai_ran_link_adaptation}

% Local macro definitions for this chapter
\providecommand{\ariadne}{ARIADNE\xspace}
\providecommand{\sionnasys}{SIONNA-SYS\xspace}
\providecommand{\sionna}{SIONNA\xspace}
\providecommand{\nvidia}{NVIDIA\xspace}
\providecommand{\nokia}{Nokia\xspace}
\providecommand{\ack}{ACK\xspace}
\providecommand{\nack}{NACK\xspace}

\section{Introduction}\label{sec:ai:intro}

\gls{ai}-powered \gls{ran} networks have attracted 
significant attention from both industry and academia. Meanwhile, 
\glspl{dt} offer a safe playground for experimenting with 
\gls{ai}/\gls{ml}-based solutions for advanced \gls{ai}-\gls{ran} 
research.
By enabling the testing of online algorithms 
before deployment on the \gls{ran}, they reduce costs and safety risks 
associated with physical field testing. In this chapter, we propose \ariadne, an online \gls{rl}-based module that seamlessly 
integrates with \sionna and is tasked with performing link adaptation. 
We 
explore different design choices and demonstrate how \ariadne can 
surpass industry-standard and state-of-the-art methods by achieving up 
to $11\%$ and $20\%$ improvements in Spectral Efficiency, respectively.
Finally, we show that \gls{rl} learns a \gls{mcs} selection strategy that diverges from \gls{olla}, exhibiting either more conservative or more aggressive behavior depending on the configuration, a trend further corroborated by training offline on \gls{5g} \gls{ota} measurements.

Recent years have seen collaboration among leading industry entities such as \nokia and \nvidia to advance \gls{ai}-\gls{ran} research~\cite{nvidia_nokia2025}. \gls{ai}-native \gls{6g} is expected to emerge from simulation, with \glspl{dt} playing a key role in the train--simulate--deploy--optimize lifecycle~\cite{nokia_ran_dt2026}. Indeed, \glspl{dt} are envisioned to enable faster innovation by allowing the evaluation of ``what-if'' scenarios for future technologies that are not yet available in hardware~\cite{10643670}. By reducing the ``concept-to-live'' cycle, dense urban environments and complex \gls{6g} use cases can be simulated with greater accuracy, supporting the design and delivery of \gls{nextg} systems while accelerating the adoption and deployment of advanced \gls{ai} solutions~\cite{nokia_ran_dt2026}.

In this context, \nvidia's \sionna \gls{dt} environment~\cite{sionna2022} provides an all-in-one platform for wireless research, enabling the execution of system-level simulations over ray-traced channels. In particular, \sionnasys allows the on-the-fly integration of plugin \gls{ai}/\gls{ml} solutions, facilitating the testing of key features and capabilities essential to \gls{ai}-native \gls{6g} research. Simulated, controlled environments often provide a practical first step for developing \gls{ai}/\gls{ml} solutions before conducting \gls{ota} experiments, allowing for algorithm refinement and parameter optimization.

At the same time, a trending topic that has gained widespread attention in the research community is the optimization of the \gls{gnb}'s \gls{mac}-layer scheduler. This involves the optimal allocation of \glspl{prb} and selection of the \gls{mcs} index, as well as the tuning of power control mechanisms (e.g., 3GPP-compliant open and closed-loop power control) that adjust transmit power based on predefined \gls{snr} targets to meet \gls{qos} requirements for different slices and verticals. Importantly, by jointly optimizing power control and \gls{mcs} selection, Spectral Efficiency can be significantly improved, demonstrating the benefits of a coordinated approach for enhanced network management.

\subsection{Related Work}\label{sec:ai:relatedwork}

Numerous works have focused on \gls{rl} for \gls{mcs} selection~\cite{khedhri2025adaptive}, ranging from offline approaches~\cite{peri2025offline} to contextual \gls{mab} solutions~\cite{pulliyakode2017reinforcement,saxena2019contextual}. The authors in~\cite{zubow2021grgym} introduce GrGym, an \gls{rl}-based framework for \gls{mcs} selection in WiFi, which leverages a custom gym environment. In~\cite{leite2012flexible}, the authors formulate an \gls{rl} framework where a low discount factor enforces myopic \gls{mcs} selection based on the current channel state. The industry standard for link adaptation is \gls{olla}~\cite{pedersen2007frequency}, and modifications involving \gls{rl} have been introduced for intelligent on-the-fly adaptation~\cite{kela2022reinforcement}. More recently, the authors in~\cite{wiesmayr2025salad} propose a gradient descent approach with a learning rate that self-adapts online through knowledge distillation. However, none of the works mentioned above focuses on the integration of an \gls{ai} module on the fly within high-fidelity \glspl{dt}.

\subsection{Contributions}\label{sec:ai:contributions}

Motivated by the suitability of system-level simulations for \gls{dt} environments, we propose \ariadne~\cite{tsampazi2026ariadne}, a framework that leverages \nvidia's platform \sionna to integrate an \gls{rl}-module for \gls{ai}-\gls{ran} networks, enabling learning-based link adaptation through adaptive \gls{mcs} selection on channels simulated via \sionna ray tracing. Our module seamlessly integrates with the \sionnasys platform, enabling direct comparison with state-of-the-art link adaptation algorithms (such as the industry-standard \gls{olla}~\cite{pedersen2007frequency} and \nvidia's proposed link adaptation scheme, entitled \gls{salad}~\cite{wiesmayr2025salad}). In this chapter, we train and evaluate \ariadne on a variety of high-fidelity ray-traced channels generated with \sionna, and we demonstrate that online \gls{rl} outperforms both \gls{olla} and \gls{salad} in \gls{mcs} selection in terms of performance, as measured by the achieved Spectral Efficiency. Finally, we evaluate the suitability of \gls{rl} on offline \gls{5g} data collected from an \gls{ota} \gls{5g} testbed, demonstrating its effectiveness on real-world measurements.

\section{\texorpdfstring{\ariadne}{ARIADNE}: Learning Link Adaptation via Reinforcement Learning}\label{sec:ai:ariadne-intro}

Our \gls{rl} agent operates at the \textit{Fast Link Adaptation} level, where decisions (i.e., the selection of \gls{mcs} levels for all \glspl{ue}) are enforced on a per-slot basis. This requires precise knowledge of the channel quality. However, \gls{cqi} reports are subject to feedback delay~\cite{maggi2026sinr}. In addition, \gls{cqi} is typically reported over a wide band, while transmission may occur over a narrower band. As a result, the true \gls{sinr} is not directly observable and must be estimated online at each slot using only past feedback. Consequently, link adaptation requires jointly inferring the channel quality and selecting an appropriate \gls{mcs}~\cite{wiesmayr2025salad}. Therefore, for the current slot's channel quality, we consider two estimation modes: an \textit{Oracle} mode that establishes an upper bound using \sionna's simulated environment to calculate the \gls{sinr}, and a predictor mode for \gls{sinr} estimation that leverages previous \gls{cqi} reports.

\textbf{\textit{\gls{mdp} Formulation and Temporal Resolution.}} We formulate the \gls{mcs} selection problem as an \gls{mdp} defined by the tuple $(\mathcal{S}, \mathcal{A}, R, P, \beta)$. Critical to our design is a $1$~:~$1$ mapping between the \gls{rl} environment and the \gls{5g}~\gls{nr} \gls{phy}-layer, where each discrete environment step corresponds exactly to one \gls{5g}~\gls{nr} time slot.

\subsection{State Space}\label{sec:ai:state-space}

At each time slot $t$, the agent observes a state vector $\mathbf{s}_t \in \mathbb{R}^{(N_\mathrm{cqi} + N_\mathrm{harq} + 5) \cdot U}$, where $U$ is the number of \glspl{ue}. The state is composed of per-\gls{ue} feature vectors:
\begin{equation}
    \mathbf{s}_t = \left[\mathbf{s}_t^{(1)}, \ldots, \mathbf{s}_t^{(U)}\right],
\end{equation}
where each per-\gls{ue} component $\mathbf{s}_t^{(u)}$ is given by
\begin{equation}
    \mathbf{s}_t^{(u)} = \left[\tilde{\gamma}_{\mathrm{post},t}^{(u)}, \; \tilde{\gamma}_{\mathrm{eff},t-k}^{(u)}, \; h_{t-k}^{(u)},\; \tilde{m}_{t-1}^{(u)},\; \hat{B}_t^{(u)},\; \Delta_{\mathrm{offset},t}^{(u)},\; \bar{a}_t^{(u)}\right],
\end{equation}
and the respective components are defined as follows:
\begin{itemize}
    \item $\tilde{\gamma}_{\mathrm{post},t}^{(u)}$: An estimate of the current slot's normalized post-equalization \gls{sinr}, computed as the mean \gls{sinr} across allocated resource elements after \gls{lmmse} equalization. In \textit{Oracle} mode, this is derived from the current channel realization; in \textit{predictor} mode, it is estimated from past measurements.
    \item $\tilde{\gamma}_{\mathrm{eff},t-k}^{(u)}$, $k = 1, \ldots, N_{\mathrm{cqi}}$: The $N_{\mathrm{cqi}}$ most recent effective normalized \gls{sinr} values reported by the \gls{phy}-layer Abstraction.
    \item $h_{t-k}^{(u)}$, $k = 1, \ldots, N_{\mathrm{harq}}$: A window of $N_{\mathrm{harq}}$ prior \gls{harq} outcomes, where $h \in \{-1, 0, 1\}$ denotes unscheduled, \nack, and \ack, respectively.
    \item $\tilde{m}_{t-1}^{(u)}$: The normalized previously selected \gls{mcs} index.
    \item $\hat{B}_t^{(u)}$: The sliding-window \gls{bler} estimate over the last $N_{\mathrm{bler}}$ scheduled slots, calculated as $\hat{B}_t^{(u)} = \sum \mathrm{NACK} \,/\, \sum \mathrm{Scheduled}$.
    \item $\Delta_{\mathrm{offset},t}^{(u)}$: A normalized asymmetric feedback accumulator updated by $+0.1$ on \ack\ and $-0.9$ on \nack.\footnote{Both $\Delta_{\mathrm{offset},t}^{(u)}$ and $\bar{a}_t^{(u)}$ are derived from \gls{harq} feedback; the former weights \nack\ outcomes $9\times$ more heavily than \ack, making it more sensitive to link failures than the symmetric \ack\ rate $\bar{a}_t^{(u)}$.}
    \item $\bar{a}_t^{(u)}$: The running \ack\ rate over the full episode.
\end{itemize}

\subsection{Action Space}\label{sec:ai:action-space}

The agent performs link adaptation for the \gls{5g}~\gls{nr} \gls{pdsch}. The action $\mathbf{a}_t \in \{0, 1, \ldots, 28\}^U$ is a vector of \gls{mcs} indices, one per \gls{ue}, as defined in the \gls{5g}~\gls{nr} \gls{mcs} Table~$1$~\cite{3gpp38214}. The mapping is performed directly by the policy network as follows in~\eqref{eq:ai:mapping}:
\begin{equation}\label{eq:ai:mapping}
    \mathbf{a}_t = \pi(\mathbf{s}_t) = \left(m_t^{(1)}, \ldots, m_t^{(U)}\right).
\end{equation}

\subsection{Reward}\label{sec:ai:rewarddef}

We aim to capture the fundamental \gls{mcs} selection tradeoff. Higher \gls{mcs} indices result in higher Spectral Efficiency, but simultaneously increase the \gls{bler} and, consequently, the probability of decoding failure. Conversely, a conservative \gls{mcs} lowers the \gls{bler} but also reduces the Spectral Efficiency. To target the practical throughput--reliability tradeoff, we maximize a reward that captures the effective Spectral Efficiency, which is directly proportional to system goodput.

\textbf{\textit{Instantaneous Reward.}} At each environment step (i.e., each \gls{5g}~\gls{nr} time slot), the agent selects an \gls{mcs} index for each \gls{ue}, and the environment subsequently returns a scalar reward based on the successfully delivered Spectral Efficiency in that slot. The instantaneous reward $r_t$ at slot $t$ is defined as the total achieved Spectral Efficiency across all \glspl{ue}, as expressed in~\eqref{eq:ai:reward}:
\begin{equation}
    r_t = \sum_{u=1}^{U} Q(m_t^{(u)}) \cdot R_c(m_t^{(u)}) \cdot \mathbb{1} \left\{ \mathcal{H}_t^{(u)} = \mathrm{ACK} \right\},
    \label{eq:ai:reward}
\end{equation}
\noindent
where $Q(m_t^{(u)})$ and $R_c(m_t^{(u)})$ denote the modulation order and the effective code rate, respectively, associated with the \gls{mcs} index $m_t^{(u)}$ selected for \gls{ue} $u$ at slot $t$. The term $\mathbb{1}\{\cdot\}$ denotes the indicator function, which evaluates to $1$ if the \gls{harq} outcome $\mathcal{H}_t^{(u)}$ for the $u$-th \gls{ue} is a successful acknowledgment ($\mathrm{ACK}$), and $0$ otherwise.

\textbf{\textit{Expected Reward.}} In the \sionnasys simulator, the \gls{harq} feedback is generated by the \gls{phy} Abstraction module as a Bernoulli outcome with success probability $\Pr(\mathrm{ACK}_t^{(u)}=1\mid\mathbf{s}_t,a_t)=1-\mathrm{BLER}(m_t^{(u)},\gamma_t^{(u)})$, where the \gls{bler} depends on the selected \gls{mcs} index $m_t^{(u)}$ and the effective \gls{sinr} $\gamma_t^{(u)}$ computed from the post-equalization \gls{sinr} at slot $t$.

\noindent
Given that the nominal Spectral Efficiency is defined as $\mathrm{SE}^{(u)}_{\mathrm{nom},t} \triangleq Q(m_t^{(u)})\cdot R_c(m_t^{(u)})$ for a chosen \gls{mcs}, the expected per-slot reward is given in~\eqref{eq:ai:expected_reward_final}:
\begin{equation}
\mathbb{E}[r_t\mid\mathbf{s}_t, a_t]=\sum_{u=1}^{U}
      \underbrace{Q(m_t^{(u)}) \cdot R_c(m_t^{(u)})}_{
        \text{SE}_{\text{nom}}(m_{t})}
      \cdot \bigl(1-\text{\small{BLER}}(m_t^{(u)}, \gamma_t^{(u)})\bigr).
  \label{eq:ai:expected_reward_final}
\end{equation}

\textbf{\textit{Reward Interpretation.}} The expectation in~\eqref{eq:ai:expected_reward_final} evaluates the effective Spectral Efficiency, which is directly proportional to the system goodput. Our \gls{rl} agent is tasked with maximizing the immediate per-slot reward, which over an episode is equivalent to maximizing the time-average effective throughput. Therefore, our reward formulation naturally captures the fundamental link adaptation tradeoff, with the \gls{rl} agent learning a policy that selects the \gls{mcs} to maximize the product $\mathrm{SE}_{\mathrm{nom}}\cdot(1 - \mathrm{BLER})$, thereby learning the rate--reliability tradeoff through trial and error.

\textbf{\textit{Adaptive \gls{bler} Penalty.}} Although the reward in~\eqref{eq:ai:reward} implicitly penalizes \gls{bler} through zero Spectral Efficiency on \nack, this signal alone may not be sufficient to prevent persistently \textit{aggressive} \gls{mcs} selection. Link adaptation mechanisms such as \gls{olla} target a fixed \gls{bler} level (typically around $10\%$~\cite{3gpp38214,wiesmayr2025salad,pedersen2007frequency}), balancing throughput and reliability. To incorporate a similar notion of reliability control, we additionally augment the reward with a per-\gls{ue} penalty weight $\lambda_t^{(u)}$, governed by an integral controller that increases when the cumulative \gls{bler} exceeds a target $\tau$ and remains zero otherwise, as defined in~\eqref{eq:ai:penaltyreward}--\eqref{eq:ai:explicit-penalty-reward}:
\begin{equation}
r_t=\sum_{u}\Big[\mathrm{SE}_{\mathrm{nom},t}^{(u)}\mathbb{1}\{\mathcal{H}_t^{(u)}=\mathrm{ACK}\}
-\lambda_t^{(u)}\mathbb{1}\{\mathcal{H}_t^{(u)}=\mathrm{NACK}\}\Big],
\label{eq:ai:penaltyreward}
\end{equation}
\noindent
where
\begin{equation}
\lambda_t^{(u)}=\max\!\Big[0,\;k_E\!\!\sum_{\substack{i\le t\\ \mathcal{H}_i^{(u)}\neq\varnothing}}
\bigl(\mathbb{1}\{\mathcal{H}_i^{(u)}=\mathrm{NACK}\}-\tau\bigr)\Big].
\label{eq:ai:explicit-penalty-reward}
\end{equation}
\noindent
When $\lambda_t^{(u)} = 0$ for all $u \in \{1,\ldots,U\}$,~\eqref{eq:ai:penaltyreward} reduces to~\eqref{eq:ai:reward}.

\subsection{Transition Dynamics}\label{sec:ai:transition-dyn}

The environment transitions $P(\mathbf{s}_{t+1} | \mathbf{s}_t, \mathbf{a}_t)$ are determined by the \sionnasys simulator as detailed in Section~\ref{sec:ai:systemmodel}.

\section{System Model}\label{sec:ai:systemmodel}

\begin{figure}[H]
  \centering
  \includegraphics[width=0.99\textwidth, keepaspectratio]{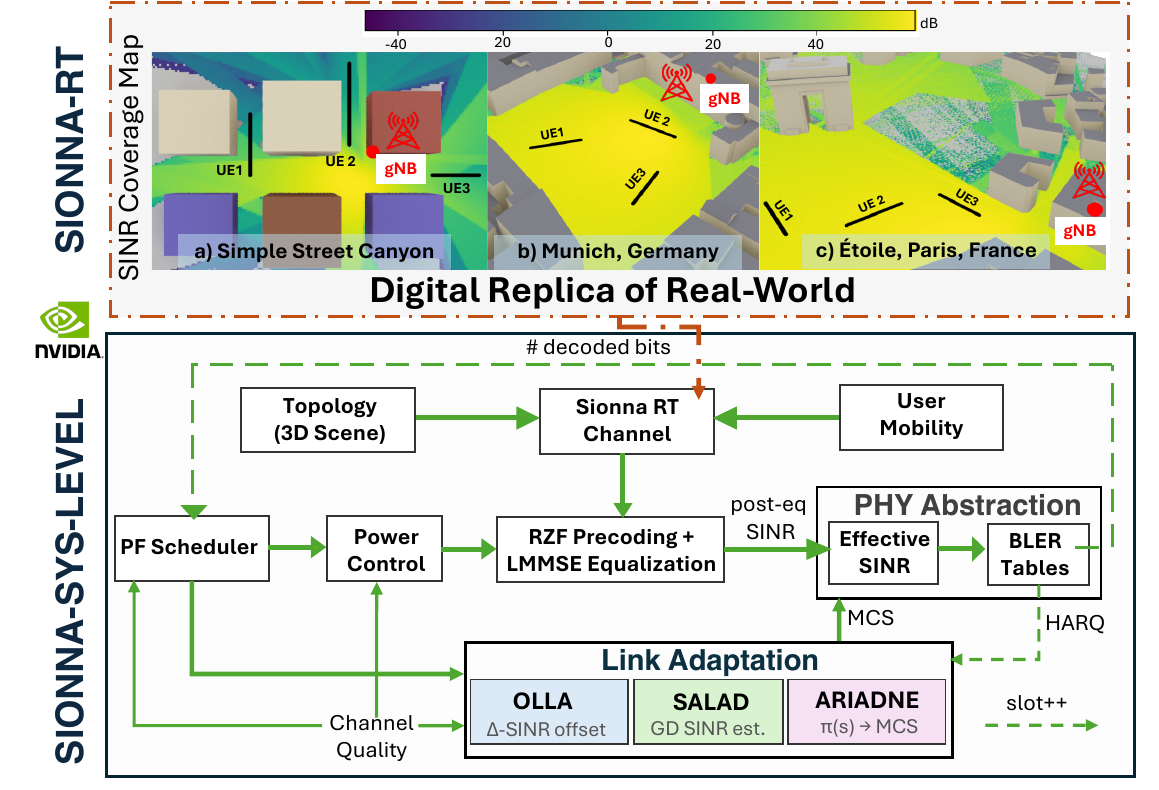}
  \caption{System-level simulations using \sionna over ray-traced channels for link adaptation with \gls{olla}, \ariadne, and \gls{salad}.}
  \label{fig:ai:syslevelsim}
\end{figure}

We compare three link adaptation strategies, namely \gls{olla}, \gls{salad}, and \ariadne. All operate within the same \sionnasys pipeline, and the decision-making component resides within the \gls{gnb}'s \gls{mac}-layer scheduler. Independent of the specific algorithm employed, this reflects a centralized architecture in which the \gls{gnb} collects \gls{cqi} and \gls{harq} feedback from all \glspl{ue}, executes the link adaptation algorithm internally, and assigns \gls{mcs} indices to scheduled \glspl{ue} for the upcoming transmission. The per-slot pipeline is illustrated in Fig.~\ref{fig:ai:syslevelsim}.

\subsection{System Architecture and Link Adaptation Workflow}\label{sec:ai:system-model-arch}

\begin{figure}[H]
\centering
\includegraphics[width=0.99\textwidth, keepaspectratio]{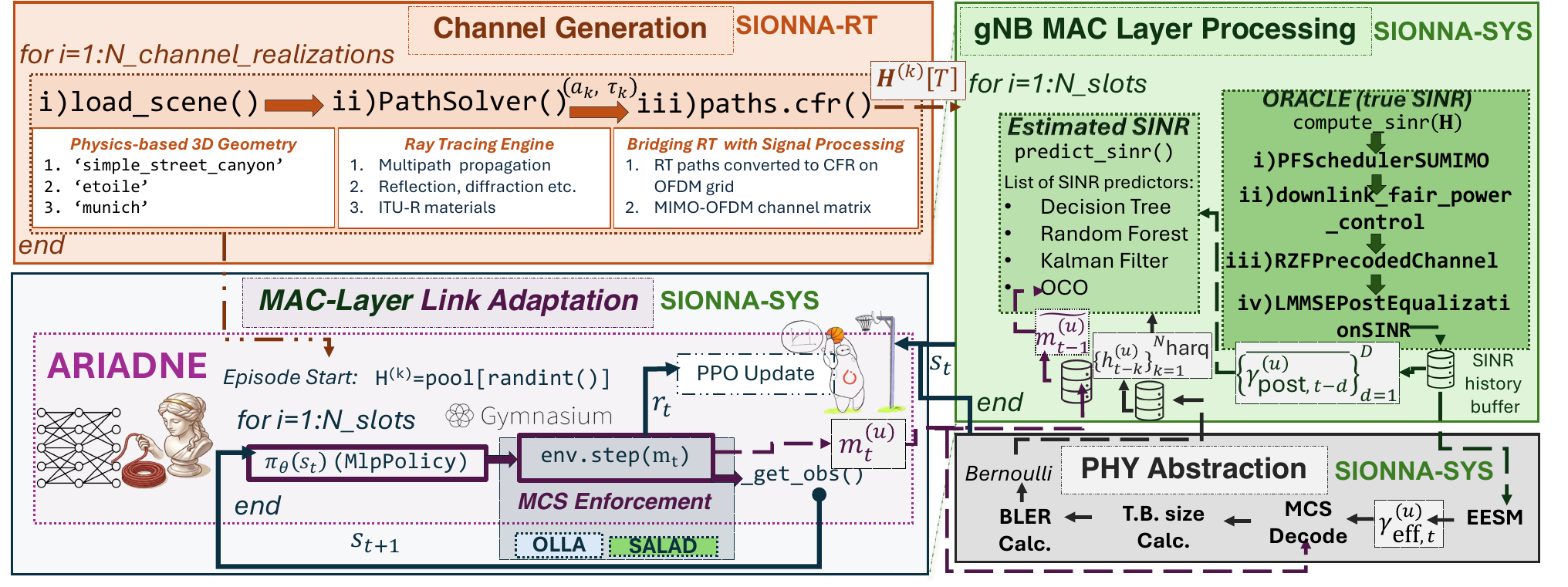}
\caption{End-to-end execution of \gls{mcs} selection in \sionna. The channel matrix is generated from a pool of ray-traced channel realizations to ensure diversity. The current \gls{sinr} is either directly used or estimated from previous observations. The post-equalization \gls{sinr} is then provided as input to the \gls{phy} Abstraction. The state $\mathbf{s}_t$ is observed by the \ariadne\ framework, which selects the corresponding \gls{mcs} index.}
\label{fig:ai:pipeline_total}
\end{figure}

The end-to-end pipeline is shown in Fig.~\ref{fig:ai:pipeline_total}. At the beginning of each simulation, a 3D scene (e.g., Urban Street Canyon, Munich, or \'{E}toile in Fig.~\ref{fig:ai:syslevelsim}) is loaded into \sionna's ray tracer to compute the channel frequency response $\mathbf{H}_t$ per slot. At each slot, the simulator executes proportional-fair scheduling, power control, and \gls{rzf} precoding with \gls{lmmse} equalization, yielding post-equalization \gls{sinr} values. Link adaptation then selects an \gls{mcs} index based on the \gls{sinr} and \gls{harq} feedback. This is the only component that differs across methods (Fig.~\ref{fig:ai:syslevelsim}). \gls{olla} maintains a per-\gls{ue} \gls{sinr} offset updated via \ack/\nack feedback and selects the highest \gls{mcs} under a target \gls{bler}, using the latest \gls{sinr} report and \gls{harq} outcome. \gls{salad} primarily relies on \ack/\nack feedback, estimating the \gls{sinr} online via recursive updates and adapting the \gls{bler} target through a feedback control loop. \ariadne observes a window of past \gls{cqi} and \gls{harq} feedback, together with a predicted \gls{sinr}, and directly maps this state to \gls{mcs} indices via a trained policy network. Finally, \gls{phy} Abstraction maps the selected \gls{mcs} and post-equalization \gls{sinr} to a \gls{bler}, from which the \gls{harq} outcome is sampled, yielding the achieved Spectral Efficiency. All steps except link adaptation are identical across methods, ensuring that any performance difference is solely due to the \gls{mcs} selection strategy.

\subsection{Implementation and Training}\label{sec:ai:ppo-design}

We train the agent using \gls{ppo}~\cite{schulman2017proximal} as implemented in Stable-Baselines3~\cite{stable-baselines3} within a custom Gymnasium environment~\cite{gymnasium2023} built on top of the \sionna system-level simulator, with \ariadne handling agent--environment interaction. The policy is parameterized by a two-layer \gls{mlp} with $64$ hidden units per layer, a learning rate of $3\times10^{-4}$, a clip range of $\varepsilon=0.2$, and an entropy coefficient of $0.05$ to encourage exploration. We set the discount factor $\beta=0$, treating the problem as a \textit{contextual bandit}. This reflects the observation that \gls{mcs} selection is primarily driven by instantaneous channel state, and temporal credit assignment for future slots does not improve performance. Finally, in our setup, we assume three mobile \glspl{ue} while the ray-traced channels correspond to \sionna's Simple Street Canyon scenario.

\section{Experimental Evaluation}\label{sec:ai:expevalresults}

\ariadne observes a configurable window of $N_{\mathrm{cqi}}$ past effective \gls{sinr} reports and $N_{\mathrm{harq}}$ prior \gls{harq} outcomes, as highlighted in Section~\ref{sec:ai:state-space}. We evaluate two configurations, as shown in Table~\ref{tab:ai:configurations}. In Setup~A, the observation window leverages $N_{\mathrm{cqi}}\!=\!3$ reports and $N_{\mathrm{harq}}\!=\!10$ outcomes, providing the agent with temporal context. In Setup~B, the observation is limited to a single past \gls{sinr} report and a single \gls{harq} outcome. Finally, all experimental results reported below have been collected over multiple experiments and channel realizations.

\begin{table}[H]
\centering
\caption{Observation Size configurations for \ariadne.}
\label{tab:ai:configurations}
\begin{tabular}{@{}lcc@{}}
\toprule
\textbf{Parameter} & \textbf{Setup~A} & \textbf{Setup~B} \\
\midrule
\gls{cqi} window ($N_{\mathrm{cqi}}$)   & 3  & 1 \\
\gls{harq} window ($N_{\mathrm{harq}}$) & 10 & 1 \\
\bottomrule
\end{tabular}
\end{table}

\subsection{Impact of Perfect and Imperfect Channel State Information}\label{sec:ai:perfectcsi}

We begin our experimental evaluation by quantifying the impact of various \gls{sinr} estimators on system performance, where the predicted current slot's normalized post-equalization \gls{sinr}, denoted as $\tilde{\gamma}_{\mathrm{post},t}^{(u)}$, serves as input to the \gls{rl} model, as described in Section~\ref{sec:ai:state-space}. Performance is evaluated in terms of the achieved mean \textit{\gls{se}} and mean \gls{bler}. In detail, when using the Oracle as input to \ariadne's \gls{rl} state, we rely on the ground-truth \gls{sinr} at the current slot, as reported by \sionnasys. When using one of the predictors, namely \gls{dt}~\cite{quinlan1986induction}, \gls{kf}~\cite{kalman1960}, \gls{rf}~\cite{breiman2001rf}, and \gls{oco}~\cite{maggi2026sinr}, we replace the Oracle with the predicted \gls{sinr} in order to assess how closely the resulting performance matches that of the Oracle, which serves as an upper bound. We then compare all the aforementioned \gls{rl}-based methods against state-of-the-art baselines, namely \gls{olla} and \gls{salad}. Finally, we also discard the current slot's predicted \gls{sinr} as input to the \gls{rl} model (denoted as \gls{dcqi} in the plots) and instead rely solely on past \gls{cqi} observations, evaluating the resulting performance across all methods.

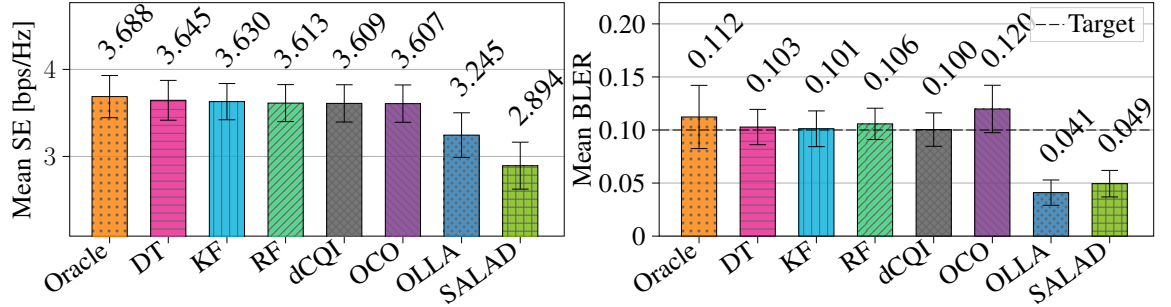
\begin{figure}[H]
    \centering
    \resizebox{\textwidth}{!}{\input{fig/predictors_summary_bar__2.tex}}
    \caption{Mean Spectral Efficiency and \gls{bler} for all \gls{rl} methods and baselines under Setup~A from Table~\ref{tab:ai:configurations}.}
    \label{fig:ai:predictors_summary}
\end{figure}

Regarding the inputs used by each \gls{sinr} predictor,\footnote{We do not aim to directly compare the predictors. Instead, we assess whether the performance of the \gls{rl} agent is affected by the predictor choice, focusing on the robustness of the approach. The \gls{kf} is a constant-velocity filter with \gls{harq}-driven bias correction, adaptive process noise, and innovation gating. The \gls{dt} and \gls{rf} are trained online on past \gls{sinr} observations, and retrained every $N_\text{slots}$. The \gls{oco} follows the \ack/\nack-only feedback setting of~\cite{maggi2026sinr}, using mirror descent with Fixed-Share expert mixing over $12$ experts defined on the grid $\eta \in \{0.5, 1, 2, 3\}$, $\beta \in \{0, 0.15, 0.3\}$, with $\alpha = 0$.} \gls{kf}, \gls{dt}, and \gls{rf} leverage past \gls{sinr} observations, \gls{kf} and \gls{oco} incorporate the most recent \gls{harq} feedback, while \gls{dt} and \gls{rf} rely on the \gls{harq} accumulator (i.e., $\Delta_{\mathrm{offset},t}^{(u)}$) to capture feedback trends. In contrast, in our implementation, \gls{oco} does not rely on explicit \gls{sinr} history and instead operates using only \gls{harq} outcomes and the previously selected \gls{mcs} index~\cite{maggi2026sinr}.

\begin{table}[H]
\centering
\setlength{\tabcolsep}{4pt}
\caption{Performance comparison with respect to the Oracle.}
\label{tab:ai:relative_performance}
\begin{tabular}{@{}lccc@{}}
\toprule
\textbf{Method} & \textbf{SE [bps/Hz]} & \textbf{$\Delta$SE vs. Oracle [\%]} & \textbf{BLER} \\
\midrule
Oracle \gls{rl}   & 3.688 & 0.0    & 0.112 \\
\gls{dt} \gls{rl} & 3.645 & -1.17  & 0.103 \\
\gls{kf} \gls{rl} & 3.630 & -1.57  & 0.101 \\
\gls{rf} \gls{rl} & 3.613 & -2.03  & 0.106 \\
dCQI \gls{rl}     & 3.609 & -2.14  & 0.100 \\
\gls{oco} \gls{rl}& 3.607 & -2.20  & 0.120 \\
\gls{olla}        & 3.245 & -12.01 & 0.041 \\
\gls{salad}       & 2.894 & -21.53 & 0.049 \\
\bottomrule
\end{tabular}
\end{table}

In Fig.~\ref{fig:ai:predictors_summary}, we observe that \ariadne, regardless of the predictor used, consistently outperforms \gls{olla} and \gls{salad}. Even without prediction of the current slot's \gls{sinr}, denoted as \gls{dcqi} in the plots, the \gls{rl} agent relying only on past \gls{cqi} reports outperforms \gls{olla} by $10\%$ in terms of spectral efficiency, achieving a mean spectral efficiency of $3.609$~bps/Hz compared to $3.245$~bps/Hz, at the cost of a higher mean \gls{bler} of $0.1$, whereas \gls{olla} achieves a \gls{bler} of $0.041$. Finally, \gls{olla} outperforms \gls{salad} by approximately $11\%$, while achieving similar \gls{bler} values. For the configuration of \gls{salad}, we use the default parameters as specified in~\cite{wiesmayr2025salad}. It is worth noting that \gls{salad} may require further parameter tuning for the specific scenario at hand, and therefore the parameterization used here may not be optimal for the current setup. This further highlights the advantage of \gls{rl} in adapting to changing channel conditions and dynamically adjusting its policy.

\begin{figure}[H]
    \centering
    \resizebox{\textwidth}{!}{\input{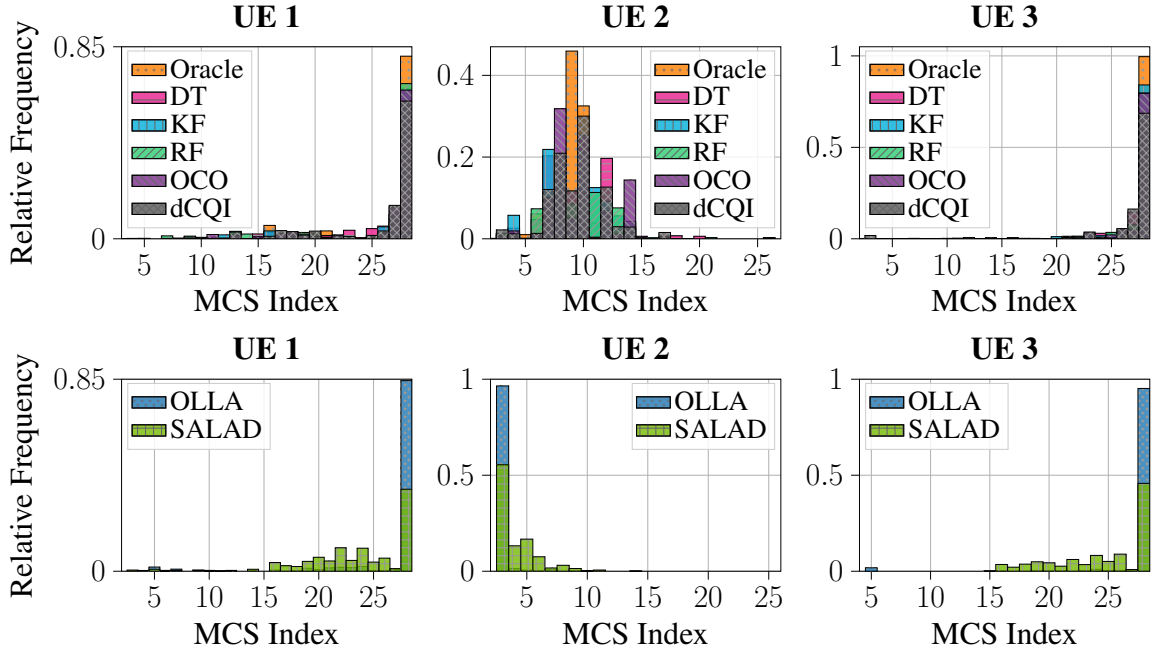}}
    \caption{Distribution of Selected \gls{mcs} Indices across Methods and \glspl{ue} under Setup~A from Table~\ref{tab:ai:configurations}.}
    \label{fig:ai:predictors_summary_2}
\end{figure}

In Table~\ref{tab:ai:relative_performance}, we provide a summary of how closely all methods perform relative to the Oracle when used as input to \ariadne. We observe that \gls{dt} achieves a Spectral Efficiency closest to the Oracle, with only $\sim1\%$ lower performance, reporting a mean Spectral Efficiency of $3.645$~bps/Hz compared to $3.688$~bps/Hz for the Oracle, followed by \gls{kf}, whose Spectral Efficiency is reported at $3.630$~bps/Hz. All \gls{rl} methods achieve a Spectral Efficiency close to the Oracle, within $\sim2\%$. Finally, \gls{olla} reports a mean Spectral Efficiency that is $\sim12\%$ lower compared to the Oracle, while \gls{salad} underperforms all methods by $\sim20\%$. However, all \gls{rl} methods report a mean \gls{bler} of $0.1$, indicating the tradeoff between high throughput and reliability. In contrast, \gls{olla} and \gls{salad} maintain the \gls{bler} well below the $10\%$ target, at the cost of lower Spectral Efficiency.

Fig.~\ref{fig:ai:predictors_summary_2} shows the relative frequency distribution of selected \gls{mcs} indices across different \gls{sinr} predictor methods and \glspl{ue}. Notably, the mean \gls{sinr} observed per \gls{ue} is $\sim26$~dB for \gls{ue}~$1$, $\sim6$~dB for \gls{ue}~$2$, and $\sim28$~dB for \gls{ue}~$3$. We observe that, for \gls{ue}~$1$ and \gls{ue}~$3$, all \gls{rl} methods select the highest \gls{mcs} indices, indicating agreement across methods in high-\gls{sinr} regimes. In contrast, for \gls{ue}~$2$, which experiences the lowest \gls{sinr}, the \gls{rl} agent adopts a more aggressive strategy, selecting \gls{mcs} indices between $5$ and $15$, with most selections concentrated around $8$--$12$. On the other hand, \gls{olla} and \gls{salad} predominantly select low \gls{mcs} indices, concentrated below $10$, indicating a more conservative approach.

\subsection{Impact of the Adaptive BLER Penalty}\label{sec:ai:adapt-eval}

\begin{figure}[H]
    \centering
    \resizebox{\textwidth}{!}{\input{fig/penalty_cdf_sinrpos.tex}}
    \caption{\glspl{cdf} of Spectral Efficiency, \gls{bler}, and \gls{mcs} under Setup~A from Table~\ref{tab:ai:configurations}, leveraging the \gls{dt} predictor.}
    \label{fig:ai:penalty_cdf_mcs}
\end{figure}
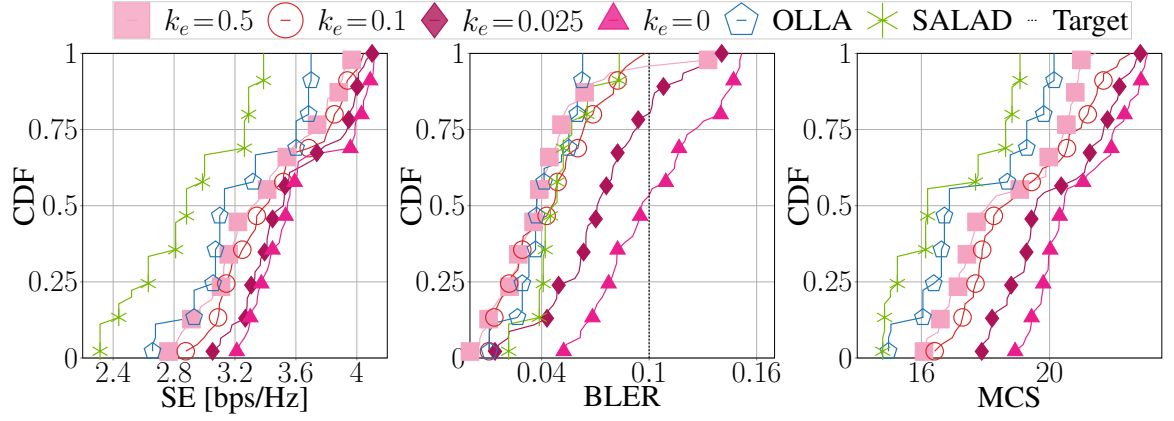

We evaluate the impact of different $k_E$ values, as defined in~\eqref{eq:ai:penaltyreward}--\eqref{eq:ai:explicit-penalty-reward}. Since the previous results do not explicitly constrain the \gls{bler} and yield values above $10\%$ for all predictor modes and the Oracle, in contrast to \gls{salad} and \gls{olla}, which maintain \gls{bler} below $0.1$, we vary $k_E \in \{0,\,0.025,\,0.1,\,0.5\}$. Here, $k_E=0$ corresponds to no penalty, $k_E=0.025$ provides a moderate penalty, and $k_E=0.1$ and $k_E=0.5$ impose increasingly strong penalties on the median \gls{bler}. The resulting performance is summarized in Table~\ref{tab:ai:ke_impact} and further illustrated in Fig.~\ref{fig:ai:penalty_cdf_mcs}. As expected, increasing $k_E$ leads to more conservative \gls{mcs} selection, reducing Spectral Efficiency while improving reliability. In particular, $k_E=0$ achieves the highest Spectral Efficiency but at the cost of a higher \gls{bler}, while larger values of $k_E$ progressively reduce the median \gls{bler} at the expense of Spectral Efficiency. Among all configurations, $k_E=0.1$ provides the best tradeoff, achieving high Spectral Efficiency while consistently maintaining the \gls{bler} below the $10\%$ target (Fig.~\ref{fig:ai:penalty_cdf_mcs}). Finally, \gls{olla} and \gls{salad} remain the most conservative.

\begin{table}[H]
\centering
\caption{Impact of $k_E$ on Median Spectral Efficiency, \gls{bler}, and \gls{mcs}.}
\label{tab:ai:ke_impact}
\begin{tabular}{@{}lccc@{}}
\toprule
$k_E$ & SE [bps/Hz] & BLER & MCS \\
\midrule
0     & 3.547 & 0.096 & 21 \\
0.025 & 3.520 & 0.072 & 20 \\
0.1   & 3.421 & 0.047 & 19 \\
0.5   & 3.342 & 0.037 & 18 \\
\midrule
OLLA  & 3.131 & 0.041 & 17 \\
SALAD & 2.882 & 0.049 & 16 \\
\bottomrule
\end{tabular}
\end{table}

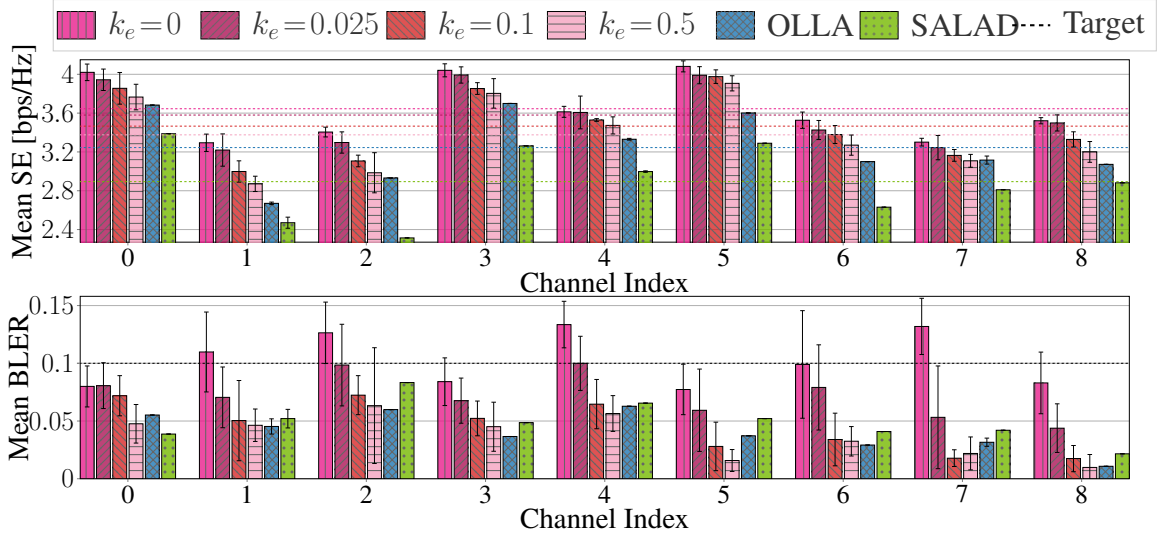
\begin{figure}[H]
    \centering
    \resizebox{\textwidth}{!}{\input{fig/penalty_perchannel_se_bler.tex}}
    \caption{Spectral Efficiency and \gls{bler} under Setup~A from Table~\ref{tab:ai:configurations}, using the \gls{dt} predictor across multiple channel realizations and experiments.}
    \label{fig:ai:penalty_se_all}
\end{figure}

\begin{figure}[H]
    \centering
    \resizebox{0.9\textwidth}{!}{\input{fig/obsSize_bar_sinrpos.tex}}
    \caption{Mean Spectral Efficiency and \gls{bler} for all \gls{rl} methods and baselines under Setup~A and~B from Table~\ref{tab:ai:configurations}, leveraging the \gls{dt} predictor.}
    \label{fig:ai:cqiSize_comparison}
\end{figure}
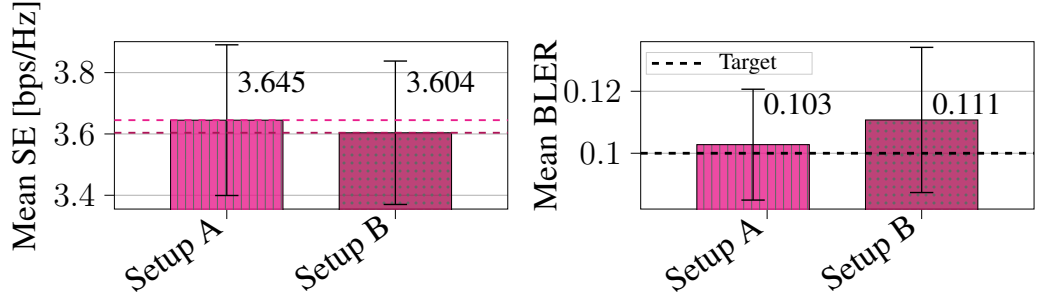

Finally, in Fig.~\ref{fig:ai:penalty_se_all}, we observe the mean Spectral Efficiency across each channel realization, averaged over multiple experiments. We observe that \gls{rl} with $k_E=0.1$ consistently meets the $10\%$ \gls{bler} target while achieving one of the highest Spectral Efficiency values among all considered methods, outperforming the industry-standard \gls{olla}. In contrast, \gls{rl} with $k_E=0$ and $k_E=0.025$ achieves higher Spectral Efficiency but often reaches or exceeds the \gls{bler} target.

\subsection{Impact of Observation Window Size}\label{sec:ai:obssize-eval}

We now proceed by evaluating how different observation window sizes impact the mean performance of the \gls{rl} agent. Once again, for the current slot's \gls{sinr} prediction, we leverage the \gls{dt}, as it achieved the closest performance to the Oracle in terms of Spectral Efficiency in the previous evaluation. The parameterization for this step is given in Table~\ref{tab:ai:configurations}. As shown in Fig.~\ref{fig:ai:cqiSize_comparison}, a smaller observation window (Setup~B) results in a slightly lower mean Spectral Efficiency of $3.604$~bps/Hz compared to $3.645$~bps/Hz achieved by Setup~A. Although both Spectral Efficiency values are comparable, Setup~A achieves a lower \gls{bler} ($0.103$) than Setup~B ($0.111$). Both values remain above the $10\%$ \gls{bler} target, which is expected since the adaptive penalty reward is not leveraged in these experiments. Overall, Setup~A is preferred, as it achieves slightly higher Spectral Efficiency while maintaining a lower \gls{bler}, enabling the \gls{rl} agent to select actions without significantly exceeding the \gls{bler} target.

\section{Site-Specific Training and Robustness Across Channel Scenarios}\label{sec:ai:site-specific}

\begin{figure}[H]
    \centering
    \resizebox{0.9\textwidth}{!}{\input{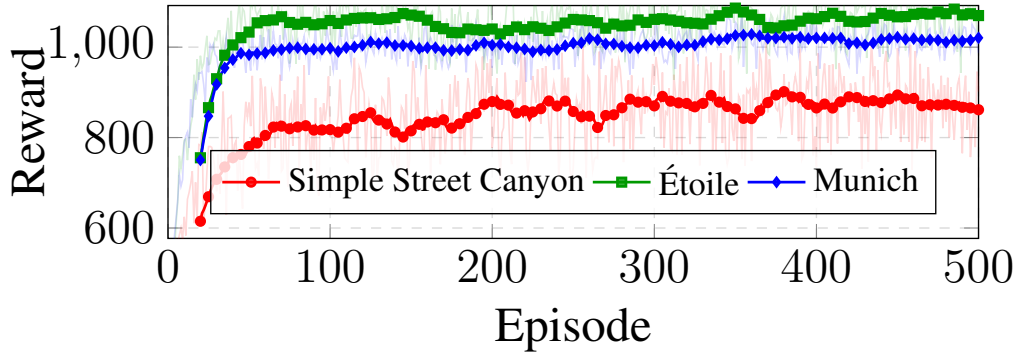}}
    \caption{Training reward over episodes.}
    \label{fig:ai:training_results}
\end{figure}

Fig.~\ref{fig:ai:training_results} shows the training reward evolution across three different channel scenarios in \sionna. The \gls{rl} agent consistently improves its performance and converges in all considered environments. This indicates that the learned policy adapts to varying channel characteristics.

\section[Future Integration of RL on OTA 5G Testbeds]{Future Integration of \texorpdfstring{\gls{rl}}{RL} on \texorpdfstring{\gls{ota}}{OTA} \texorpdfstring{\gls{5g}}{5G} Testbeds}\label{sec:ai:nextsteps}

\begin{figure}[H]
    \centering
    \resizebox{0.9\textwidth}{!}{\input{fig/fqi_oai_combined_apr15.tex}}

    \vspace{0.25cm}

    \resizebox{0.9\textwidth}{!}{\input{fig/fqi_csv_combined_apr15.tex}}
    \caption{Offline \gls{fqi} results on \gls{ota} data for \gls{ul} (top) and \gls{dl} (bottom). Left: Q-value convergence across Bellman iterations. Right: \gls{mcs} distribution of the learned policy vs.\ the \gls{oai} scheduler.}
    \label{fig:ai:fqi}
\end{figure}
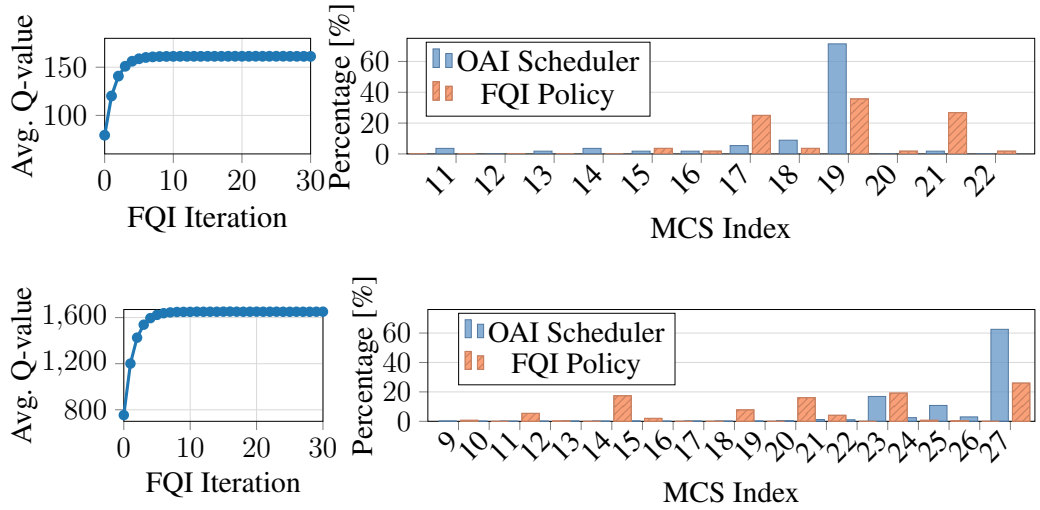

To examine whether reward-driven \gls{mcs} selection extends beyond simulation, we train an \gls{fqi} agent~\cite{ernst2005tree} on \gls{ota} measurements collected from a \gls{5g} testbed~\cite{villa2025x5g}, using both \gls{dl} and \gls{ul} data. The state consists of \gls{cqi}, \gls{rsrp}, and instantaneous \gls{bler}; the reward is a proxy for achieved throughput derived from logged measurements; and the Q-function is an \gls{rf} with $\gamma = 0.5$. As shown in Fig.~\ref{fig:ai:fqi}, Q-function targets stabilize across Bellman iterations, and the learned policy shifts \gls{mcs} selections away from the scheduler's preferred indices. Since these datasets reflect only the \gls{oai} \gls{olla} scheduler's choices, closed-loop evaluation is left for future work. The fact that both this offline agent and the online \gls{ppo} agent independently deviate from \gls{olla} suggests that reward-driven \gls{mcs} selection tends to identify operating points that rule-based schedulers do not.

\section{Conclusions}\label{sec:ai:conclusion}

We developed \ariadne, a framework that seamlessly integrates with \sionna, leveraging online \gls{rl} for \gls{mcs} selection. The use of system-level simulators, which enable the on-the-fly integration of \gls{ai} modules, has facilitated the exploration of various approaches, spanning reward formulation and observation window design, to study the impact of different design choices on link adaptation. Although the \sionna integration provides a practical first step toward the design of \gls{rl}-based link adaptation, future work will focus on the deployment of \ariadne on \gls{ota} \gls{5g} testbeds. Preliminary results on both \gls{dl} and \gls{ul} \gls{5g} data demonstrate the potential of learning-based approaches in real-world scenarios.

% --- EOF ---

%% file: fig/predictors_summary_bar__2.tex
% Add to preamble: \usetikzlibrary{patterns}
\begin{tikzpicture}

\definecolor{darkgray176}{RGB}{176,176,176}
\definecolor{lightgray204}{RGB}{204,204,204}
\definecolor{steelblue31119180}{RGB}{31,119,180}
\definecolor{barcolor2551270}{RGB}{255,127,0}
\definecolor{barcolor23330140}{RGB}{233,30,140}
\definecolor{barcolor0174219}{RGB}{0,174,219}
\definecolor{barcolor46204113}{RGB}{46,204,113}
\definecolor{barcolor858585}{RGB}{85,85,85}
\definecolor{barcolor12345139}{RGB}{123,45,139}
\definecolor{barcolor31119180}{RGB}{31,119,180}
\definecolor{barcolor1181850}{RGB}{118,185,0}

\begin{groupplot}[
  group style={
    group size=2 by 1,
    horizontal sep=2cm,
  },
  width=12cm,
  height=6.5cm,
  tick label style={font=\fontsize{16}{19}\selectfont},
  label style={font=\fontsize{16}{19}\selectfont},
  legend style={font=\fontsize{16}{19}\selectfont, fill opacity=0.8, draw opacity=1, text opacity=1, draw=lightgray204},
  title style={font=\fontsize{16}{19}\selectfont},
  xticklabel style={rotate=35.0, anchor=east, font=\fontsize{16}{19}\selectfont},
  scaled y ticks=false,
]

\nextgroupplot[
tick align=outside,
tick pos=left,
x grid style={darkgray176},
xmin=-0.68, xmax=7.75,
xtick style={color=black},
xtick={0,1,2,3,4,5,6,7},
xticklabels={Oracle,DT,KF,RF,dCQI,OCO,OLLA,SALAD},
y grid style={darkgray176},
ylabel={Mean SE [bps/Hz]},
ylabel style={yshift=-0.1cm},
ymajorgrids,
ymin=2.08423747303479, ymax=4.76702804101758,
ytick style={color=black},
]
% Oracle - dots
\draw[draw=none, fill=barcolor2551270, fill opacity=0.8] (axis cs:-0.3,0) rectangle (axis cs:0.3,3.68759320271841);
\draw[draw=black, fill=none, pattern=dots, pattern color=black!60] (axis cs:-0.3,0) rectangle (axis cs:0.3,3.68759320271841);
% DT - horizontal lines
\draw[draw=none, fill=barcolor23330140, fill opacity=0.8] (axis cs:0.7,0) rectangle (axis cs:1.3,3.64498262113612);
\draw[draw=black, fill=none, pattern=horizontal lines, pattern color=black!60] (axis cs:0.7,0) rectangle (axis cs:1.3,3.64498262113612);
% Kalman - vertical lines
\draw[draw=none, fill=barcolor0174219, fill opacity=0.8] (axis cs:1.7,0) rectangle (axis cs:2.3,3.62988119787029);
\draw[draw=black, fill=none, pattern=vertical lines, pattern color=black!60] (axis cs:1.7,0) rectangle (axis cs:2.3,3.62988119787029);
% RF - north east lines
\draw[draw=none, fill=barcolor46204113, fill opacity=0.8] (axis cs:2.7,0) rectangle (axis cs:3.3,3.61279679386807);
\draw[draw=black, fill=none, pattern=north east lines, pattern color=black!60] (axis cs:2.7,0) rectangle (axis cs:3.3,3.61279679386807);
% Raw SINR - crosshatch
\draw[draw=none, fill=barcolor858585, fill opacity=0.8] (axis cs:3.7,0) rectangle (axis cs:4.3,3.60947830041617);
\draw[draw=black, fill=none, pattern=crosshatch, pattern color=black!60] (axis cs:3.7,0) rectangle (axis cs:4.3,3.60947830041617);
% OCO - north west lines
\draw[draw=none, fill=barcolor12345139, fill opacity=0.8] (axis cs:4.7,0) rectangle (axis cs:5.3,3.60684388903407);
\draw[draw=black, fill=none, pattern=north west lines, pattern color=black!60] (axis cs:4.7,0) rectangle (axis cs:5.3,3.60684388903407);
% OLLA - crosshatch dots
\draw[draw=none, fill=barcolor31119180, fill opacity=0.8] (axis cs:5.7,0) rectangle (axis cs:6.3,3.24494255444711);
\draw[draw=black, fill=none, pattern=crosshatch dots, pattern color=black!60] (axis cs:5.7,0) rectangle (axis cs:6.3,3.24494255444711);
% SALAD - grid
\draw[draw=none, fill=barcolor1181850, fill opacity=0.8] (axis cs:6.7,0) rectangle (axis cs:7.3,2.89381360175917);
\draw[draw=black, fill=none, pattern=grid, pattern color=black!60] (axis cs:6.7,0) rectangle (axis cs:7.3,2.89381360175917);
\path [draw=black, line width=0.56pt] (axis cs:0,3.44539028104879) -- (axis cs:0,3.92979612438802);
\path [draw=black, line width=0.56pt] (axis cs:1,3.41628533686592) -- (axis cs:1,3.87367990540631);
\path [draw=black, line width=0.56pt] (axis cs:2,3.42164095780609) -- (axis cs:2,3.8381214379345);
\path [draw=black, line width=0.56pt] (axis cs:3,3.40034076162507) -- (axis cs:3,3.82525282611107);
\path [draw=black, line width=0.56pt] (axis cs:4,3.3953213461324) -- (axis cs:4,3.82363525469993);
\path [draw=black, line width=0.56pt] (axis cs:5,3.39238063571057) -- (axis cs:5,3.82130714235757);
\path [draw=black, line width=0.56pt] (axis cs:6,2.98836439019744) -- (axis cs:6,3.50152071869678);
\path [draw=black, line width=0.56pt] (axis cs:7,2.62395489218437) -- (axis cs:7,3.16367231133396);
\addplot [semithick, steelblue31119180, mark=-, mark size=5, mark options={solid,draw=black}, only marks]
table {%
0 3.44539028104879
1 3.41628533686592
2 3.42164095780609
3 3.40034076162507
4 3.3953213461324
5 3.39238063571057
6 2.98836439019744
7 2.62395489218437
};
\addplot [semithick, steelblue31119180, mark=-, mark size=5, mark options={solid,draw=black}, only marks]
table {%
0 3.92979612438802
1 3.87367990540631
2 3.8381214379345
3 3.82525282611107
4 3.82363525469993
5 3.82130714235757
6 3.50152071869678
7 3.16367231133396
};
\draw (axis cs:0,3.92979612438802) node[scale=1.6, anchor=south west, text=black, rotate=45.0]{3.688};
\draw (axis cs:1,3.87367990540631) node[scale=1.6, anchor=south west, text=black, rotate=45.0]{3.645};
\draw (axis cs:2,3.8381214379345) node[scale=1.6, anchor=south west, text=black, rotate=45.0]{3.630};
\draw (axis cs:3,3.82525282611107) node[scale=1.6, anchor=south west, text=black, rotate=45.0]{3.613};
\draw (axis cs:4,3.82363525469993) node[scale=1.6, anchor=south west, text=black, rotate=45.0]{3.609};
\draw (axis cs:5,3.82130714235757) node[scale=1.6, anchor=south west, text=black, rotate=45.0]{3.607};
\draw (axis cs:6,3.50152071869678) node[scale=1.6, anchor=south west, text=black, rotate=45.0]{3.245};
\draw (axis cs:7,3.16367231133396) node[scale=1.6, anchor=south west, text=black, rotate=45.0]{2.894};

\nextgroupplot[
legend cell align={left},
tick align=outside,
tick pos=left,
x grid style={darkgray176},
xmin=-0.68, xmax=7.75,
xtick style={color=black},
xtick={0,1,2,3,4,5,6,7},
xticklabels={Oracle,DT,KF,RF,dCQI,OCO,OLLA,SALAD},
y grid style={darkgray176},
ylabel={Mean BLER},
ylabel style={yshift=-0.1cm},
ymajorgrids,
ymin=0, ymax=0.22,
ytick style={color=black},
ytick={0, 0.05, 0.10, 0.15, 0.20},
yticklabels={0, 0.05, 0.10, 0.15, 0.20},
]
% Oracle - dots
\draw[draw=none, fill=barcolor2551270, fill opacity=0.8] (axis cs:-0.3,0) rectangle (axis cs:0.3,0.112281003957158);
\draw[draw=black, fill=none, pattern=dots, pattern color=black!60] (axis cs:-0.3,0) rectangle (axis cs:0.3,0.112281003957158);
% DT - horizontal lines
\draw[draw=none, fill=barcolor23330140, fill opacity=0.8] (axis cs:0.7,0) rectangle (axis cs:1.3,0.102777873499161);
\draw[draw=black, fill=none, pattern=horizontal lines, pattern color=black!60] (axis cs:0.7,0) rectangle (axis cs:1.3,0.102777873499161);
% Kalman - vertical lines
\draw[draw=none, fill=barcolor0174219, fill opacity=0.8] (axis cs:1.7,0) rectangle (axis cs:2.3,0.101156905464885);
\draw[draw=black, fill=none, pattern=vertical lines, pattern color=black!60] (axis cs:1.7,0) rectangle (axis cs:2.3,0.101156905464885);
% RF - north east lines
\draw[draw=none, fill=barcolor46204113, fill opacity=0.8] (axis cs:2.7,0) rectangle (axis cs:3.3,0.105762594092691);
\draw[draw=black, fill=none, pattern=north east lines, pattern color=black!60] (axis cs:2.7,0) rectangle (axis cs:3.3,0.105762594092691);
% Raw SINR - crosshatch
\draw[draw=none, fill=barcolor858585, fill opacity=0.8] (axis cs:3.7,0) rectangle (axis cs:4.3,0.100379247011796);
\draw[draw=black, fill=none, pattern=crosshatch, pattern color=black!60] (axis cs:3.7,0) rectangle (axis cs:4.3,0.100379247011796);
% OCO - north west lines
\draw[draw=none, fill=barcolor12345139, fill opacity=0.8] (axis cs:4.7,0) rectangle (axis cs:5.3,0.119812753072803);
\draw[draw=black, fill=none, pattern=north west lines, pattern color=black!60] (axis cs:4.7,0) rectangle (axis cs:5.3,0.119812753072803);
% OLLA - crosshatch dots
\draw[draw=none, fill=barcolor31119180, fill opacity=0.8] (axis cs:5.7,0) rectangle (axis cs:6.3,0.0409678559152416);
\draw[draw=black, fill=none, pattern=crosshatch dots, pattern color=black!60] (axis cs:5.7,0) rectangle (axis cs:6.3,0.0409678559152416);
% SALAD - grid
\draw[draw=none, fill=barcolor1181850, fill opacity=0.8] (axis cs:6.7,0) rectangle (axis cs:7.3,0.0494250684882088);
\draw[draw=black, fill=none, pattern=grid, pattern color=black!60] (axis cs:6.7,0) rectangle (axis cs:7.3,0.0494250684882088);
\path [draw=black, line width=0.56pt] (axis cs:0,0.0824889143509812) -- (axis cs:0,0.142073093563335);
\path [draw=black, line width=0.56pt] (axis cs:1,0.0861372308903222) -- (axis cs:1,0.119418516108);
\path [draw=black, line width=0.56pt] (axis cs:2,0.0843196106929769) -- (axis cs:2,0.117994200236794);
\path [draw=black, line width=0.56pt] (axis cs:3,0.0909994411197656) -- (axis cs:3,0.120525747065616);
\path [draw=black, line width=0.56pt] (axis cs:4,0.0846275524871097) -- (axis cs:4,0.116130941536482);
\path [draw=black, line width=0.56pt] (axis cs:5,0.0974772719684981) -- (axis cs:5,0.142148234177108);
\path [draw=black, line width=0.56pt] (axis cs:6,0.0290322780972475) -- (axis cs:6,0.0529034337332357);
\path [draw=black, line width=0.56pt] (axis cs:7,0.0369335526664347) -- (axis cs:7,0.0619165843099828);
\addplot [semithick, steelblue31119180, mark=-, mark size=5, mark options={solid,draw=black}, only marks, forget plot]
table {%
0 0.0824889143509812
1 0.0861372308903222
2 0.0843196106929769
3 0.0909994411197656
4 0.0846275524871097
5 0.0974772719684981
6 0.0290322780972475
7 0.0369335526664347
};
\addplot [semithick, steelblue31119180, mark=-, mark size=5, mark options={solid,draw=black}, only marks, forget plot]
table {%
0 0.142073093563335
1 0.119418516108
2 0.117994200236794
3 0.120525747065616
4 0.116130941536482
5 0.142148234177108
6 0.0529034337332357
7 0.0619165843099828
};
\addplot [line width=0.48pt, black, dash pattern=on 6pt off 2pt]
table {%
-0.68 0.1
7.68 0.1
};
\addlegendentry{Target}
\draw (axis cs:0,0.142073093563335) node[scale=1.6, anchor=south west, text=black, rotate=45.0]{0.112};
\draw (axis cs:1,0.119418516108) node[scale=1.6, anchor=south west, text=black, rotate=45.0]{0.103};
\draw (axis cs:2,0.117994200236794) node[scale=1.6, anchor=south west, text=black, rotate=45.0]{0.101};
\draw (axis cs:3,0.120525747065616) node[scale=1.6, anchor=south west, text=black, rotate=45.0]{0.106};
\draw (axis cs:4,0.116130941536482) node[scale=1.6, anchor=south west, text=black, rotate=45.0]{0.100};
\draw (axis cs:5,0.142148234177108) node[scale=1.6, anchor=south west, text=black, rotate=45.0]{0.120};
\draw (axis cs:6,0.0529034337332357) node[scale=1.6, anchor=south west, text=black, rotate=45.0]{0.041};
\draw (axis cs:7,0.0619165843099828) node[scale=1.6, anchor=south west, text=black, rotate=45.0]{0.049};
\end{groupplot}

\end{tikzpicture}

%% file: fig/penalty_cdf_sinrpos.tex
% Add to preamble:
% \usepgfplotslibrary{groupplots}
% \usetikzlibrary{calc}

\begin{tikzpicture}

\definecolor{darkgray176}{RGB}{176,176,176}
\definecolor{lightgray204}{RGB}{204,204,204}
\definecolor{keSoftRoseF4A3C0}{RGB}{244,163,192}
\definecolor{keRedD62828}{RGB}{214,40,40}
\definecolor{keBerryAD1457}{RGB}{173,20,87}
\definecolor{hotpinkE91E8C}{RGB}{233,30,140}
\definecolor{steelblue1F77B4}{RGB}{31,119,180}
\definecolor{limgreen76B900}{RGB}{118,185,0}

\begin{groupplot}[
  group style={
    group size=3 by 1,
    horizontal sep=5cm,
  },
  width=20cm,
  height=20cm,
  ylabel style={yshift=-10pt, xshift=15pt},
  % tick label style={font=\Huge},
  % label style={font=\Huge},
tick label style={font=\fontsize{52}{62}\selectfont},  %% <-- was \Huge
  label style={font=\fontsize{52}{62}\selectfont},       %% <-- was \Huge
  tick align=outside,
  tick pos=left,
  x grid style={darkgray176},
  y grid style={darkgray176},
  xmajorgrids,
  ymajorgrids,
  every axis plot/.append style={
    mark options={scale=7.5.5}
  },
]

%% ---- LEFT PLOT: SE CDF ----
\nextgroupplot[
  legend to name=CommonLegend,
  legend columns=7,
  legend style={
    font=\fontsize{120}{144}\selectfont,
    draw=lightgray204,
    fill opacity=0.8,
    draw opacity=1,
    text opacity=1,
    column sep=0.8cm,
    /tikz/every even column/.append style={column sep=0.8cm},
    /tikz/every mark/.append style={scale=7.5},
    legend image post style={mark size=4pt}, 
  },
  xlabel={SE [bps/Hz]},
  ylabel={CDF},
  ymin=0, ymax=1,
  ytick={0, 0.25, 0.5, 0.75, 1.0},
  xmin=2.2, xmax=4.2,
  xtick distance=0.4,
  ylabel style={yshift=-8pt},
  every axis plot/.append style={
    mark options={scale=7.5}
  },
]
\addplot [line width=2pt, keSoftRoseF4A3C0,  mark=square*, mark repeat=5]
table {%
2.7634 0.0213
2.8028 0.0426
2.8294 0.0638
2.8504 0.0851
2.8903 0.1064
2.9137 0.1277
2.9578 0.1489
2.9792 0.1702
3.0460 0.1915
3.0652 0.2128
3.1083 0.2340
3.1150 0.2553
3.1253 0.2766
3.1264 0.2979
3.1352 0.3191
3.1588 0.3404
3.1791 0.3617
3.1963 0.3830
3.2124 0.4043
3.2144 0.4255
3.2193 0.4468
3.2703 0.4681
3.2950 0.4894
3.3417 0.5106
3.3787 0.5319
3.4111 0.5532
3.4247 0.5745
3.4307 0.5957
3.4693 0.6170
3.5012 0.6383
3.5400 0.6596
3.5609 0.6809
3.5986 0.7021
3.6564 0.7234
3.7328 0.7447
3.7391 0.7660
3.7792 0.7872
3.8033 0.8085
3.8053 0.8298
3.8093 0.8511
3.8815 0.8723
3.8826 0.8936
3.9040 0.9149
3.9111 0.9362
3.9265 0.9574
3.9660 0.9787
3.9836 1.0000
};
\addlegendentry{{\fontsize{52}{62}\selectfont $k_e\!=\!0.5$}}

\addplot [line width=2pt, keRedD62828, solid, mark=o, mark repeat=5]
table {%
2.8764 0.0222
2.9528 0.0444
2.9970 0.0667
3.0301 0.0889
3.0768 0.1111
3.0889 0.1333
3.0961 0.1556
3.0979 0.1778
3.1142 0.2000
3.1333 0.2222
3.1429 0.2444
3.1582 0.2667
3.1733 0.2889
3.1817 0.3111
3.2277 0.3333
3.2482 0.3556
3.2762 0.3778
3.3019 0.4000
3.3331 0.4222
3.3413 0.4444
3.3433 0.4667
3.3775 0.4889
3.4210 0.5111
3.4387 0.5333
3.4610 0.5556
3.5170 0.5778
3.5188 0.6000
3.5310 0.6222
3.5352 0.6444
3.5438 0.6667
3.6844 0.6889
3.7866 0.7111
3.7995 0.7333
3.8189 0.7556
3.8435 0.7778
3.8536 0.8000
3.8877 0.8222
3.9059 0.8444
3.9175 0.8667
3.9242 0.8889
3.9384 0.9111
3.9748 0.9333
4.0158 0.9556
4.0297 0.9778
4.0449 1.0000
};
\addlegendentry{{\fontsize{52}{62}\selectfont $k_e\!=\!0.1$}}

\addplot [line width=2pt, keBerryAD1457, solid, mark=diamond*, mark repeat=5]
table {%
3.0531 0.0217
3.1013 0.0435
3.1063 0.0652
3.1695 0.0870
3.1803 0.1087
3.2677 0.1304
3.2757 0.1522
3.2794 0.1739
3.2803 0.1957
3.3005 0.2174
3.3066 0.2391
3.3302 0.2609
3.3314 0.2826
3.3659 0.3043
3.3691 0.3261
3.3943 0.3478
3.4100 0.3696
3.4111 0.3913
3.4195 0.4130
3.4275 0.4348
3.4467 0.4565
3.4613 0.4783
3.5173 0.5000
3.5229 0.5217
3.5248 0.5435
3.5289 0.5652
3.5539 0.5870
3.5584 0.6087
3.6417 0.6304
3.7188 0.6522
3.7376 0.6739
3.7890 0.6957
3.8923 0.7174
3.9266 0.7391
3.9441 0.7609
3.9470 0.7826
3.9508 0.8043
3.9772 0.8261
3.9852 0.8478
3.9940 0.8696
4.0025 0.8913
4.0101 0.9130
4.0110 0.9348
4.0450 0.9565
4.0612 0.9783
4.1003 1.0000
};
\addlegendentry{{\fontsize{54}{66}\selectfont $k_e\!=\!0.025$}}

\addplot [line width=2pt, hotpinkE91E8C, solid, mark=triangle*, mark repeat=5]
table {%
3.2117 0.0222
3.2313 0.0444
3.2621 0.0667
3.2856 0.0889
3.2963 0.1111
3.3006 0.1333
3.3134 0.1556
3.3437 0.1778
3.3474 0.2000
3.3554 0.2222
3.3729 0.2444
3.3887 0.2667
3.4125 0.2889
3.4253 0.3111
3.4332 0.3333
3.4463 0.3556
3.4902 0.3778
3.4987 0.4000
3.5079 0.4222
3.5291 0.4444
3.5324 0.4667
3.5424 0.4889
3.5468 0.5111
3.5605 0.5333
3.5624 0.5556
3.5917 0.5778
3.6076 0.6000
3.6112 0.6222
3.6158 0.6444
3.6855 0.6667
3.9579 0.6889
3.9702 0.7111
3.9877 0.7333
3.9877 0.7556
3.9934 0.7778
4.0304 0.8000
4.0434 0.8222
4.0441 0.8444
4.0632 0.8667
4.0733 0.8889
4.0878 0.9111
4.1063 0.9333
4.1101 0.9556
4.1129 0.9778
4.1471 1.0000
};
\addlegendentry{{\fontsize{52}{62}\selectfont $k_e\!=\!0$}}

\addplot [line width=2pt, steelblue1F77B4, solid, mark=pentagon, mark repeat=5]
table {%
2.6588 0.0222
2.6588 0.0444
2.6777 0.0667
2.6777 0.0889
2.6777 0.1111
2.9316 0.1333
2.9316 0.1556
2.9316 0.1778
2.9316 0.2000
2.9316 0.2222
3.0561 0.2444
3.0719 0.2667
3.0719 0.2889
3.0719 0.3111
3.0719 0.3333
3.0719 0.3556
3.1000 0.3778
3.1000 0.4000
3.1000 0.4222
3.1000 0.4444
3.1000 0.4667
3.1310 0.4889
3.1310 0.5111
3.1310 0.5333
3.1310 0.5556
3.3179 0.5778
3.3340 0.6000
3.3340 0.6222
3.3340 0.6444
3.3340 0.6667
3.6013 0.6889
3.6013 0.7111
3.6013 0.7333
3.6013 0.7556
3.6013 0.7778
3.6830 0.8000
3.6830 0.8222
3.6830 0.8444
3.6830 0.8667
3.6830 0.8889
3.6997 0.9111
3.6997 0.9333
3.6997 0.9556
3.6997 0.9778
3.6997 1.0000
};
\addlegendentry{{\fontsize{52}{62}\selectfont OLLA}}

\addplot [line width=2pt, limgreen76B900, solid, mark=asterisk, mark repeat=5]
table {%
2.3140 0.0222
2.3140 0.0444
2.3140 0.0667
2.3140 0.0889
2.3140 0.1111
2.4370 0.1333
2.4370 0.1556
2.4370 0.1778
2.5209 0.2000
2.5209 0.2222
2.6307 0.2444
2.6307 0.2667
2.6307 0.2889
2.6307 0.3111
2.6307 0.3333
2.8103 0.3556
2.8103 0.3778
2.8103 0.4000
2.8103 0.4222
2.8103 0.4444
2.8818 0.4667
2.8818 0.4889
2.8818 0.5111
2.8818 0.5333
2.8818 0.5556
2.9866 0.5778
3.0003 0.6000
3.0003 0.6222
3.0003 0.6444
3.0003 0.6667
3.2620 0.6889
3.2620 0.7111
3.2620 0.7333
3.2620 0.7556
3.2620 0.7778
3.2896 0.8000
3.2896 0.8222
3.2896 0.8444
3.2896 0.8667
3.2896 0.8889
3.3877 0.9111
3.3877 0.9333
3.3877 0.9556
3.3877 0.9778
3.3877 1.0000
};
\addlegendentry{{\fontsize{52}{62}\selectfont SALAD}}

% Dummy target line — legend entry only, invisible in this plot
\addplot [line width=2pt, black, dashed]
  coordinates {(2.2, -1) (4.2, -1)};
  
\addlegendentry{{\fontsize{52}{62}\selectfont Target}}

%% ---- MIDDLE PLOT: BLER CDF ----
\nextgroupplot[
  every axis plot/.append style={
    mark options={scale=7.5}
  },
  xlabel={BLER},
  ylabel={CDF},
  ylabel style={yshift=-8pt},
  ymin=0, ymax=1,
  ytick={0, 0.25, 0.5, 0.75, 1.0},
  xmin=0, xmax=0.17,
  xtick={0.04, 0.1, 0.16},
  scaled x ticks=false,
  x tick label style={
    /pgf/number format/fixed,
    /pgf/number format/precision=2,
  },
]
\addplot [line width=2pt, keSoftRoseF4A3C0, mark=square*, mark repeat=5]
table {%
0.0000 0.0213
0.0054 0.0426
0.0055 0.0638
0.0069 0.0851
0.0103 0.1064
0.0104 0.1277
0.0105 0.1489
0.0136 0.1702
0.0160 0.1915
0.0209 0.2128
0.0221 0.2340
0.0241 0.2553
0.0242 0.2766
0.0267 0.2979
0.0267 0.3191
0.0270 0.3404
0.0275 0.3617
0.0298 0.3830
0.0319 0.4043
0.0353 0.4255
0.0355 0.4468
0.0357 0.4681
0.0368 0.4894
0.0374 0.5106
0.0380 0.5319
0.0387 0.5532
0.0398 0.5745
0.0409 0.5957
0.0430 0.6170
0.0440 0.6383
0.0443 0.6596
0.0468 0.6809
0.0489 0.7021
0.0492 0.7234
0.0511 0.7447
0.0511 0.7660
0.0517 0.7872
0.0532 0.8085
0.0552 0.8298
0.0636 0.8511
0.0640 0.8723
0.0642 0.8936
0.0749 0.9149
0.0782 0.9362
0.0983 0.9574
0.1329 0.9787
0.1420 1.0000
};

\addplot [line width=2pt, keRedD62828, solid, mark=o, mark repeat=5]
table {%
0.0106 0.0222
0.0107 0.0444
0.0108 0.0667
0.0109 0.0889
0.0136 0.1111
0.0136 0.1333
0.0143 0.1556
0.0159 0.1778
0.0178 0.2000
0.0210 0.2222
0.0219 0.2444
0.0231 0.2667
0.0267 0.2889
0.0273 0.3111
0.0276 0.3333
0.0292 0.3556
0.0292 0.3778
0.0298 0.4000
0.0351 0.4222
0.0361 0.4444
0.0428 0.4667
0.0447 0.4889
0.0471 0.5111
0.0479 0.5333
0.0479 0.5556
0.0489 0.5778
0.0511 0.6000
0.0532 0.6222
0.0535 0.6444
0.0543 0.6667
0.0602 0.6889
0.0610 0.7111
0.0625 0.7333
0.0636 0.7556
0.0647 0.7778
0.0686 0.8000
0.0695 0.8222
0.0745 0.8444
0.0800 0.8667
0.0802 0.8889
0.0824 0.9111
0.0838 0.9333
0.0870 0.9556
0.0920 0.9778
0.0978 1.0000
};

\addplot [line width=2pt, keBerryAD1457, solid, mark=diamond*, mark repeat=5]
table {%
0.0140 0.0217
0.0160 0.0435
0.0213 0.0652
0.0219 0.0870
0.0357 0.1087
0.0430 0.1304
0.0432 0.1522
0.0465 0.1739
0.0484 0.1957
0.0486 0.2174
0.0495 0.2391
0.0585 0.2609
0.0608 0.2826
0.0615 0.3043
0.0618 0.3261
0.0634 0.3478
0.0647 0.3696
0.0647 0.3913
0.0649 0.4130
0.0659 0.4348
0.0702 0.4565
0.0707 0.4783
0.0710 0.5000
0.0723 0.5217
0.0743 0.5435
0.0761 0.5652
0.0789 0.5870
0.0794 0.6087
0.0800 0.6304
0.0824 0.6522
0.0824 0.6739
0.0867 0.6957
0.0894 0.7174
0.0909 0.7391
0.0939 0.7609
0.0939 0.7826
0.1005 0.8043
0.1016 0.8261
0.1020 0.8478
0.1030 0.8696
0.1080 0.8913
0.1183 0.9130
0.1190 0.9348
0.1285 0.9565
0.1353 0.9783
0.1405 1.0000
};

\addplot [line width=2pt, hotpinkE91E8C, solid, mark=triangle*, mark repeat=5]
table {%
0.0523 0.0222
0.0538 0.0444
0.0615 0.0667
0.0618 0.0889
0.0645 0.1111
0.0684 0.1333
0.0695 0.1556
0.0710 0.1778
0.0739 0.2000
0.0757 0.2222
0.0774 0.2444
0.0794 0.2667
0.0806 0.2889
0.0819 0.3111
0.0820 0.3333
0.0824 0.3556
0.0862 0.3778
0.0909 0.4000
0.0924 0.4222
0.0939 0.4444
0.0950 0.4667
0.0952 0.4889
0.0964 0.5111
0.1000 0.5333
0.1016 0.5556
0.1093 0.5778
0.1105 0.6000
0.1141 0.6222
0.1145 0.6444
0.1160 0.6667
0.1167 0.6889
0.1198 0.7111
0.1212 0.7333
0.1268 0.7556
0.1304 0.7778
0.1399 0.8000
0.1402 0.8222
0.1404 0.8444
0.1429 0.8667
0.1460 0.8889
0.1471 0.9111
0.1488 0.9333
0.1506 0.9556
0.1506 0.9778
0.1517 1.0000
};

\addplot [line width=2pt, steelblue1F77B4, solid, mark=pentagon, mark repeat=5]
table {%
0.0108 0.0222
0.0108 0.0444
0.0108 0.0667
0.0108 0.0889
0.0108 0.1111
0.0265 0.1333
0.0292 0.1556
0.0292 0.1778
0.0292 0.2000
0.0292 0.2222
0.0292 0.2444
0.0329 0.2667
0.0329 0.2889
0.0329 0.3111
0.0329 0.3333
0.0366 0.3556
0.0366 0.3778
0.0366 0.4000
0.0366 0.4222
0.0366 0.4444
0.0372 0.4667
0.0372 0.4889
0.0372 0.5111
0.0372 0.5333
0.0372 0.5556
0.0414 0.5778
0.0414 0.6000
0.0414 0.6222
0.0511 0.6444
0.0511 0.6667
0.0552 0.6889
0.0552 0.7111
0.0552 0.7333
0.0552 0.7556
0.0552 0.7778
0.0599 0.8000
0.0599 0.8222
0.0599 0.8444
0.0599 0.8667
0.0599 0.8889
0.0625 0.9111
0.0629 0.9333
0.0629 0.9556
0.0629 0.9778
0.0629 1.0000
};

\addplot [line width=2pt, limgreen76B900, solid, mark=asterisk, mark repeat=5]
table {%
0.0216 0.0222
0.0216 0.0444
0.0216 0.0667
0.0216 0.0889
0.0216 0.1111
0.0387 0.1333
0.0387 0.1556
0.0387 0.1778
0.0387 0.2000
0.0387 0.2222
0.0409 0.2444
0.0409 0.2667
0.0409 0.2889
0.0409 0.3111
0.0409 0.3333
0.0420 0.3556
0.0420 0.3778
0.0420 0.4000
0.0420 0.4222
0.0420 0.4444
0.0449 0.4667
0.0449 0.4889
0.0486 0.5111
0.0486 0.5333
0.0486 0.5556
0.0486 0.5778
0.0486 0.6000
0.0521 0.6222
0.0521 0.6444
0.0521 0.6667
0.0521 0.6889
0.0521 0.7111
0.0568 0.7333
0.0568 0.7556
0.0568 0.7778
0.0652 0.8000
0.0656 0.8222
0.0656 0.8444
0.0656 0.8667
0.0656 0.8889
0.0833 0.9111
0.0833 0.9333
0.0833 0.9556
0.0833 0.9778
0.0833 1.0000
};

% Vertical target line at BLER = 0.1
\addplot [line width=2pt, black, dashed]
table {%
0.1000 0
0.1000 1
};

%% ---- RIGHT PLOT: MCS CDF ----
\nextgroupplot[
  xlabel={MCS},
  ylabel={CDF},
  ylabel style={yshift=-8pt},
  ymin=0, ymax=1,
  ytick={0, 0.25, 0.5, 0.75, 1.0},
  xmin=14, xmax=23.5,
  xtick={0, 4, 8, 12, 16, 20, 24, 28},
  every axis plot/.append style={
    mark options={scale=7.5}
  },
]
\addplot [line width=2pt, keSoftRoseF4A3C0, mark=square*, mark repeat=5]
table {%
16.0819 0.0213
16.1250 0.0426
16.1779 0.0638
16.5260 0.0851
16.5655 0.1064
16.5934 0.1277
16.6621 0.1489
16.8581 0.1702
16.8851 0.1915
16.9251 0.2128
17.1429 0.2340
17.1733 0.2553
17.1828 0.2766
17.3516 0.2979
17.3953 0.3191
17.4199 0.3404
17.5097 0.3617
17.5696 0.3830
17.6294 0.4043
17.6879 0.4255
17.7202 0.4468
17.7977 0.4681
18.0109 0.4894
18.1576 0.5106
18.3373 0.5319
19.0702 0.5532
19.2614 0.5745
19.8807 0.5957
19.8840 0.6170
19.9128 0.6383
19.9721 0.6596
20.1957 0.6809
20.3529 0.7021
20.3906 0.7234
20.4526 0.7447
20.5227 0.7660
20.5780 0.7872
20.6429 0.8085
20.6593 0.8298
20.7968 0.8511
20.8011 0.8723
20.8503 0.8936
20.8936 0.9149
20.9096 0.9362
20.9781 0.9574
20.9895 0.9787
21.4332 1.0000
};

\addplot [line width=2pt, keRedD62828, solid, mark=o, mark repeat=5]
table {%
16.4152 0.0222
16.6250 0.0444
16.9667 0.0667
17.0699 0.0889
17.2313 0.1111
17.2834 0.1333
17.3370 0.1556
17.3643 0.1778
17.6216 0.2000
17.6287 0.2222
17.6882 0.2444
17.7891 0.2667
17.7956 0.2889
17.8232 0.3111
17.8889 0.3333
17.8966 0.3556
17.9518 0.3778
17.9882 0.4000
18.2114 0.4222
18.2514 0.4444
18.2609 0.4667
18.4136 0.4889
18.7225 0.5111
18.9706 0.5333
19.0409 0.5556
19.4413 0.5778
19.6705 0.6000
19.7803 0.6222
20.2857 0.6444
20.4709 0.6667
20.5419 0.6889
20.5928 0.7111
20.6138 0.7333
20.9840 0.7556
21.0266 0.7778
21.1902 0.8000
21.2350 0.8222
21.3797 0.8444
21.4728 0.8667
21.5000 0.8889
21.6791 0.9111
21.6845 0.9333
22.0934 0.9556
22.1862 0.9778
22.5312 1.0000
};

\addplot [line width=2pt, keBerryAD1457, solid, mark=diamond*, mark repeat=5]
table {%
17.8837 0.0217
17.8980 0.0435
18.0979 0.0652
18.1338 0.0870
18.2044 0.1087
18.2061 0.1304
18.3476 0.1522
18.6703 0.1739
18.6765 0.1957
18.7770 0.2174
18.8012 0.2391
18.8772 0.2609
19.1964 0.2826
19.2324 0.3043
19.2471 0.3261
19.2771 0.3478
19.2840 0.3696
19.3876 0.3913
19.4118 0.4130
19.4242 0.4348
19.4343 0.4565
19.5220 0.4783
19.5893 0.5000
19.6185 0.5217
19.8412 0.5435
20.3631 0.5652
20.8245 0.5870
21.0865 0.6087
21.1229 0.6304
21.2210 0.6522
21.2527 0.6739
21.3371 0.6957
21.6103 0.7174
21.6667 0.7391
21.7043 0.7609
21.8054 0.7826
21.8864 0.8043
21.9471 0.8261
22.1148 0.8478
22.1522 0.8696
22.1793 0.8913
22.2053 0.9130
22.4974 0.9348
22.6409 0.9565
22.7005 0.9783
22.8287 1.0000
};

\addplot [line width=2pt, hotpinkE91E8C, solid, mark=triangle*, mark repeat=5]
table {%
18.9240 0.0222
19.1761 0.0444
19.2643 0.0667
19.3897 0.0889
19.4266 0.1111
19.4382 0.1333
19.5349 0.1556
19.7299 0.1778
19.7518 0.2000
19.7907 0.2222
19.8012 0.2444
19.8251 0.2667
19.9518 0.2889
19.9624 0.3111
19.9674 0.3333
20.0180 0.3556
20.0482 0.3778
20.1905 0.4000
20.2857 0.4222
20.2970 0.4444
20.3133 0.4667
20.4293 0.4889
20.4573 0.5111
20.4762 0.5333
20.5595 0.5556
21.0444 0.5778
21.1657 0.6000
21.4213 0.6222
21.6229 0.6444
21.9034 0.6667
21.9270 0.6889
21.9365 0.7111
21.9516 0.7333
22.0378 0.7556
22.3102 0.7778
22.3368 0.8000
22.4652 0.8222
22.5137 0.8444
22.7598 0.8667
22.7912 0.8889
22.8492 0.9111
22.8508 0.9333
22.9617 0.9556
22.9947 0.9778
23.0591 1.0000
};

\addplot [line width=2pt, steelblue1F77B4, solid, mark=pentagon, mark repeat=5]
table {%
14.9763 0.0222
14.9763 0.0444
14.9763 0.0667
15.0511 0.0889
15.0511 0.1111
16.0484 0.1333
16.0484 0.1556
16.0484 0.1778
16.0484 0.2000
16.0484 0.2222
16.3841 0.2444
16.6257 0.2667
16.6257 0.2889
16.6257 0.3111
16.6257 0.3333
16.6257 0.3556
16.7186 0.3778
16.7186 0.4000
16.7186 0.4222
16.7186 0.4444
16.7186 0.4667
16.8750 0.4889
16.8750 0.5111
16.8750 0.5333
16.8750 0.5556
18.6818 0.5778
18.7714 0.6000
18.7714 0.6222
18.7714 0.6444
18.7714 0.6667
19.2872 0.6889
19.2872 0.7111
19.2872 0.7333
19.2872 0.7556
19.2872 0.7778
19.8168 0.8000
19.8168 0.8222
19.8168 0.8444
19.8168 0.8667
19.8168 0.8889
20.1381 0.9111
20.1381 0.9333
20.1381 0.9556
20.1381 0.9778
20.1381 1.0000
};

\addplot [line width=2pt, limgreen76B900, solid, mark=asterisk, mark repeat=5]
table {%
14.7955 0.0222
14.7955 0.0444
14.7955 0.0667
14.8512 0.0889
14.8512 0.1111
14.8512 0.1333
14.8512 0.1556
14.8512 0.1778
15.0337 0.2000
15.0337 0.2222
15.2515 0.2444
15.2515 0.2667
15.2515 0.2889
15.2515 0.3111
15.2515 0.3333
16.1329 0.3556
16.1329 0.3778
16.1329 0.4000
16.1329 0.4222
16.1329 0.4444
16.1946 0.4667
16.1946 0.4889
16.1946 0.5111
16.1946 0.5333
16.1946 0.5556
17.6902 0.5778
17.7705 0.6000
17.7705 0.6222
17.7705 0.6444
17.7705 0.6667
18.6270 0.6889
18.6270 0.7111
18.6270 0.7333
18.6270 0.7556
18.6270 0.7778
18.8281 0.8000
18.8281 0.8222
18.8281 0.8444
18.8281 0.8667
18.8281 0.8889
19.0773 0.9111
19.0773 0.9333
19.0773 0.9556
19.0773 0.9778
19.0773 1.0000
};

\end{groupplot}

% Common horizontal legend centred above all plots
\node at ($(group c1r1.north)!0.5!(group c3r1.north)+(0,2cm)$)
  {\pgfplotslegendfromname{CommonLegend}};

% Enlarge bounding box so legend is not clipped
\path
  (current bounding box.south west)
  rectangle
  ($(current bounding box.north east)+(0,2.0cm)$);

\end{tikzpicture}

%% file: fig/penalty_perchannel_se_bler.tex
% Preamble: \usetikzlibrary{patterns, calc}
%           \usepgfplotslibrary{groupplots}

\begin{tikzpicture}

\definecolor{darkgray176}{RGB}{176,176,176}
\definecolor{lightgray204}{RGB}{204,204,204}
\definecolor{steelblue31119180}{RGB}{31,119,180}
\definecolor{hotpinkE91E8C}{RGB}{233,30,140}
\definecolor{keBerryAD1457}{RGB}{173,20,87}
\definecolor{keRedD62828}{RGB}{214,40,40}
\definecolor{keSoftRoseF4A3C0}{RGB}{244,163,192}
\definecolor{steelblue1F77B4}{RGB}{31,119,180}
\definecolor{limgreen76B900}{RGB}{118,185,0}

\begin{groupplot}[
  group style={group size=1 by 2, vertical sep=2.5cm},
  width=50cm,
  height=10cm,
  tick label style={font=\fontsize{36}{40}\selectfont},
  label style={font=\fontsize{38}{45}\selectfont},
  title style={font=\fontsize{38}{45}\selectfont},
  xticklabel style={font=\fontsize{34}{40}\selectfont},
  ylabel style={yshift=-0.1cm},
  scaled y ticks=true,
]

%% ---- TOP: SE ----
\nextgroupplot[
  tick align=outside,
  tick pos=left,
  ymajorgrids,
  y grid style={darkgray176},
  ymin=2.27, ymax=4.15,
  ytick distance=0.4,
  xlabel={Channel Index},
  ylabel={Mean SE [bps/Hz]},
  xtick={0,1,2,3,4,5,6,7,8},
  xticklabels={0,1,2,3,4,5,6,7,8},
  xmin=0, xmax=8,
  enlarge x limits=0.05,
  scaled y ticks=false,
]

% Direct RL
\draw[draw=none, fill=hotpinkE91E8C, fill opacity=0.8] (axis cs:-0.4000,2.000) rectangle (axis cs:-0.2833,4.0204);
\draw[draw=black, fill=none, pattern=vertical lines, pattern color=black!60] (axis cs:-0.4000,2.000) rectangle (axis cs:-0.2833,4.0204);
\draw[draw=none, fill=hotpinkE91E8C, fill opacity=0.8] (axis cs:0.6000,2.000) rectangle (axis cs:0.7167,3.2944);
\draw[draw=black, fill=none, pattern=vertical lines, pattern color=black!60] (axis cs:0.6000,2.000) rectangle (axis cs:0.7167,3.2944);
\draw[draw=none, fill=hotpinkE91E8C, fill opacity=0.8] (axis cs:1.6000,2.000) rectangle (axis cs:1.7167,3.4049);
\draw[draw=black, fill=none, pattern=vertical lines, pattern color=black!60] (axis cs:1.6000,2.000) rectangle (axis cs:1.7167,3.4049);
\draw[draw=none, fill=hotpinkE91E8C, fill opacity=0.8] (axis cs:2.6000,2.000) rectangle (axis cs:2.7167,4.0411);
\draw[draw=black, fill=none, pattern=vertical lines, pattern color=black!60] (axis cs:2.6000,2.000) rectangle (axis cs:2.7167,4.0411);
\draw[draw=none, fill=hotpinkE91E8C, fill opacity=0.8] (axis cs:3.6000,2.000) rectangle (axis cs:3.7167,3.6133);
\draw[draw=black, fill=none, pattern=vertical lines, pattern color=black!60] (axis cs:3.6000,2.000) rectangle (axis cs:3.7167,3.6133);
\draw[draw=none, fill=hotpinkE91E8C, fill opacity=0.8] (axis cs:4.6000,2.000) rectangle (axis cs:4.7167,4.0816);
\draw[draw=black, fill=none, pattern=vertical lines, pattern color=black!60] (axis cs:4.6000,2.000) rectangle (axis cs:4.7167,4.0816);
\draw[draw=none, fill=hotpinkE91E8C, fill opacity=0.8] (axis cs:5.6000,2.000) rectangle (axis cs:5.7167,3.5267);
\draw[draw=black, fill=none, pattern=vertical lines, pattern color=black!60] (axis cs:5.6000,2.000) rectangle (axis cs:5.7167,3.5267);
\draw[draw=none, fill=hotpinkE91E8C, fill opacity=0.8] (axis cs:6.6000,2.000) rectangle (axis cs:6.7167,3.3010);
\draw[draw=black, fill=none, pattern=vertical lines, pattern color=black!60] (axis cs:6.6000,2.000) rectangle (axis cs:6.7167,3.3010);
\draw[draw=none, fill=hotpinkE91E8C, fill opacity=0.8] (axis cs:7.6000,2.000) rectangle (axis cs:7.7167,3.5214);
\draw[draw=black, fill=none, pattern=vertical lines, pattern color=black!60] (axis cs:7.6000,2.000) rectangle (axis cs:7.7167,3.5214);

% k_e=0.025
\draw[draw=none, fill=keBerryAD1457, fill opacity=0.8] (axis cs:-0.2633,2.000) rectangle (axis cs:-0.1467,3.9436);
\draw[draw=black, fill=none, pattern=north east lines, pattern color=black!60] (axis cs:-0.2633,2.000) rectangle (axis cs:-0.1467,3.9436);
\draw[draw=none, fill=keBerryAD1457, fill opacity=0.8] (axis cs:0.7367,2.000) rectangle (axis cs:0.8533,3.2193);
\draw[draw=black, fill=none, pattern=north east lines, pattern color=black!60] (axis cs:0.7367,2.000) rectangle (axis cs:0.8533,3.2193);
\draw[draw=none, fill=keBerryAD1457, fill opacity=0.8] (axis cs:1.7367,2.000) rectangle (axis cs:1.8533,3.2971);
\draw[draw=black, fill=none, pattern=north east lines, pattern color=black!60] (axis cs:1.7367,2.000) rectangle (axis cs:1.8533,3.2971);
\draw[draw=none, fill=keBerryAD1457, fill opacity=0.8] (axis cs:2.7367,2.000) rectangle (axis cs:2.8533,3.9930);
\draw[draw=black, fill=none, pattern=north east lines, pattern color=black!60] (axis cs:2.7367,2.000) rectangle (axis cs:2.8533,3.9930);
\draw[draw=none, fill=keBerryAD1457, fill opacity=0.8] (axis cs:3.7367,2.000) rectangle (axis cs:3.8533,3.6070);
\draw[draw=black, fill=none, pattern=north east lines, pattern color=black!60] (axis cs:3.7367,2.000) rectangle (axis cs:3.8533,3.6070);
\draw[draw=none, fill=keBerryAD1457, fill opacity=0.8] (axis cs:4.7367,2.000) rectangle (axis cs:4.8533,3.9907);
\draw[draw=black, fill=none, pattern=north east lines, pattern color=black!60] (axis cs:4.7367,2.000) rectangle (axis cs:4.8533,3.9907);
\draw[draw=none, fill=keBerryAD1457, fill opacity=0.8] (axis cs:5.7367,2.000) rectangle (axis cs:5.8533,3.4259);
\draw[draw=black, fill=none, pattern=north east lines, pattern color=black!60] (axis cs:5.7367,2.000) rectangle (axis cs:5.8533,3.4259);
\draw[draw=none, fill=keBerryAD1457, fill opacity=0.8] (axis cs:6.7367,2.000) rectangle (axis cs:6.8533,3.2445);
\draw[draw=black, fill=none, pattern=north east lines, pattern color=black!60] (axis cs:6.7367,2.000) rectangle (axis cs:6.8533,3.2445);
\draw[draw=none, fill=keBerryAD1457, fill opacity=0.8] (axis cs:7.7367,2.000) rectangle (axis cs:7.8533,3.4987);
\draw[draw=black, fill=none, pattern=north east lines, pattern color=black!60] (axis cs:7.7367,2.000) rectangle (axis cs:7.8533,3.4987);

% k_e=0.1
\draw[draw=none, fill=keRedD62828, fill opacity=0.8] (axis cs:-0.1267,2.000) rectangle (axis cs:-0.0100,3.8557);
\draw[draw=black, fill=none, pattern=north west lines, pattern color=black!60] (axis cs:-0.1267,2.000) rectangle (axis cs:-0.0100,3.8557);
\draw[draw=none, fill=keRedD62828, fill opacity=0.8] (axis cs:0.8733,2.000) rectangle (axis cs:0.9900,2.9984);
\draw[draw=black, fill=none, pattern=north west lines, pattern color=black!60] (axis cs:0.8733,2.000) rectangle (axis cs:0.9900,2.9984);
\draw[draw=none, fill=keRedD62828, fill opacity=0.8] (axis cs:1.8733,2.000) rectangle (axis cs:1.9900,3.1067);
\draw[draw=black, fill=none, pattern=north west lines, pattern color=black!60] (axis cs:1.8733,2.000) rectangle (axis cs:1.9900,3.1067);
\draw[draw=none, fill=keRedD62828, fill opacity=0.8] (axis cs:2.8733,2.000) rectangle (axis cs:2.9900,3.8534);
\draw[draw=black, fill=none, pattern=north west lines, pattern color=black!60] (axis cs:2.8733,2.000) rectangle (axis cs:2.9900,3.8534);
\draw[draw=none, fill=keRedD62828, fill opacity=0.8] (axis cs:3.8733,2.000) rectangle (axis cs:3.9900,3.5292);
\draw[draw=black, fill=none, pattern=north west lines, pattern color=black!60] (axis cs:3.8733,2.000) rectangle (axis cs:3.9900,3.5292);
\draw[draw=none, fill=keRedD62828, fill opacity=0.8] (axis cs:4.8733,2.000) rectangle (axis cs:4.9900,3.9760);
\draw[draw=black, fill=none, pattern=north west lines, pattern color=black!60] (axis cs:4.8733,2.000) rectangle (axis cs:4.9900,3.9760);
\draw[draw=none, fill=keRedD62828, fill opacity=0.8] (axis cs:5.8733,2.000) rectangle (axis cs:5.9900,3.3793);
\draw[draw=black, fill=none, pattern=north west lines, pattern color=black!60] (axis cs:5.8733,2.000) rectangle (axis cs:5.9900,3.3793);
\draw[draw=none, fill=keRedD62828, fill opacity=0.8] (axis cs:6.8733,2.000) rectangle (axis cs:6.9900,3.1643);
\draw[draw=black, fill=none, pattern=north west lines, pattern color=black!60] (axis cs:6.8733,2.000) rectangle (axis cs:6.9900,3.1643);
\draw[draw=none, fill=keRedD62828, fill opacity=0.8] (axis cs:7.8733,2.000) rectangle (axis cs:7.9900,3.3291);
\draw[draw=black, fill=none, pattern=north west lines, pattern color=black!60] (axis cs:7.8733,2.000) rectangle (axis cs:7.9900,3.3291);

% k_e=0.5
\draw[draw=none, fill=keSoftRoseF4A3C0, fill opacity=0.8] (axis cs:0.0100,2.000) rectangle (axis cs:0.1267,3.7656);
\draw[draw=black, fill=none, pattern=horizontal lines, pattern color=black!60] (axis cs:0.0100,2.000) rectangle (axis cs:0.1267,3.7656);
\draw[draw=none, fill=keSoftRoseF4A3C0, fill opacity=0.8] (axis cs:1.0100,2.000) rectangle (axis cs:1.1267,2.8708);
\draw[draw=black, fill=none, pattern=horizontal lines, pattern color=black!60] (axis cs:1.0100,2.000) rectangle (axis cs:1.1267,2.8708);
\draw[draw=none, fill=keSoftRoseF4A3C0, fill opacity=0.8] (axis cs:2.0100,2.000) rectangle (axis cs:2.1267,2.9854);
\draw[draw=black, fill=none, pattern=horizontal lines, pattern color=black!60] (axis cs:2.0100,2.000) rectangle (axis cs:2.1267,2.9854);
\draw[draw=none, fill=keSoftRoseF4A3C0, fill opacity=0.8] (axis cs:3.0100,2.000) rectangle (axis cs:3.1267,3.8034);
\draw[draw=black, fill=none, pattern=horizontal lines, pattern color=black!60] (axis cs:3.0100,2.000) rectangle (axis cs:3.1267,3.8034);
\draw[draw=none, fill=keSoftRoseF4A3C0, fill opacity=0.8] (axis cs:4.0100,2.000) rectangle (axis cs:4.1267,3.4735);
\draw[draw=black, fill=none, pattern=horizontal lines, pattern color=black!60] (axis cs:4.0100,2.000) rectangle (axis cs:4.1267,3.4735);
\draw[draw=none, fill=keSoftRoseF4A3C0, fill opacity=0.8] (axis cs:5.0100,2.000) rectangle (axis cs:5.1267,3.9069);
\draw[draw=black, fill=none, pattern=horizontal lines, pattern color=black!60] (axis cs:5.0100,2.000) rectangle (axis cs:5.1267,3.9069);
\draw[draw=none, fill=keSoftRoseF4A3C0, fill opacity=0.8] (axis cs:6.0100,2.000) rectangle (axis cs:6.1267,3.2701);
\draw[draw=black, fill=none, pattern=horizontal lines, pattern color=black!60] (axis cs:6.0100,2.000) rectangle (axis cs:6.1267,3.2701);
\draw[draw=none, fill=keSoftRoseF4A3C0, fill opacity=0.8] (axis cs:7.0100,2.000) rectangle (axis cs:7.1267,3.1084);
\draw[draw=black, fill=none, pattern=horizontal lines, pattern color=black!60] (axis cs:7.0100,2.000) rectangle (axis cs:7.1267,3.1084);
\draw[draw=none, fill=keSoftRoseF4A3C0, fill opacity=0.8] (axis cs:8.0100,2.000) rectangle (axis cs:8.1267,3.2002);
\draw[draw=black, fill=none, pattern=horizontal lines, pattern color=black!60] (axis cs:8.0100,2.000) rectangle (axis cs:8.1267,3.2002);

% OLLA
\draw[draw=none, fill=steelblue1F77B4, fill opacity=0.8] (axis cs:0.1467,2.000) rectangle (axis cs:0.2633,3.6830);
\draw[draw=black, fill=none, pattern=crosshatch, pattern color=black!60] (axis cs:0.1467,2.000) rectangle (axis cs:0.2633,3.6830);
\draw[draw=none, fill=steelblue1F77B4, fill opacity=0.8] (axis cs:1.1467,2.000) rectangle (axis cs:1.2633,2.6701);
\draw[draw=black, fill=none, pattern=crosshatch, pattern color=black!60] (axis cs:1.1467,2.000) rectangle (axis cs:1.2633,2.6701);
\draw[draw=none, fill=steelblue1F77B4, fill opacity=0.8] (axis cs:2.1467,2.000) rectangle (axis cs:2.2633,2.9316);
\draw[draw=black, fill=none, pattern=crosshatch, pattern color=black!60] (axis cs:2.1467,2.000) rectangle (axis cs:2.2633,2.9316);
\draw[draw=none, fill=steelblue1F77B4, fill opacity=0.8] (axis cs:3.1467,2.000) rectangle (axis cs:3.2633,3.6997);
\draw[draw=black, fill=none, pattern=crosshatch, pattern color=black!60] (axis cs:3.1467,2.000) rectangle (axis cs:3.2633,3.6997);
\draw[draw=none, fill=steelblue1F77B4, fill opacity=0.8] (axis cs:4.1467,2.000) rectangle (axis cs:4.2633,3.3308);
\draw[draw=black, fill=none, pattern=crosshatch, pattern color=black!60] (axis cs:4.1467,2.000) rectangle (axis cs:4.2633,3.3308);
\draw[draw=none, fill=steelblue1F77B4, fill opacity=0.8] (axis cs:5.1467,2.000) rectangle (axis cs:5.2633,3.6013);
\draw[draw=black, fill=none, pattern=crosshatch, pattern color=black!60] (axis cs:5.1467,2.000) rectangle (axis cs:5.2633,3.6013);
\draw[draw=none, fill=steelblue1F77B4, fill opacity=0.8] (axis cs:6.1467,2.000) rectangle (axis cs:6.2633,3.1000);
\draw[draw=black, fill=none, pattern=crosshatch, pattern color=black!60] (axis cs:6.1467,2.000) rectangle (axis cs:6.2633,3.1000);
\draw[draw=none, fill=steelblue1F77B4, fill opacity=0.8] (axis cs:7.1467,2.000) rectangle (axis cs:7.2633,3.1160);
\draw[draw=black, fill=none, pattern=crosshatch, pattern color=black!60] (axis cs:7.1467,2.000) rectangle (axis cs:7.2633,3.1160);
\draw[draw=none, fill=steelblue1F77B4, fill opacity=0.8] (axis cs:8.1467,2.000) rectangle (axis cs:8.2633,3.0719);
\draw[draw=black, fill=none, pattern=crosshatch, pattern color=black!60] (axis cs:8.1467,2.000) rectangle (axis cs:8.2633,3.0719);

% SALAD
\draw[draw=none, fill=limgreen76B900, fill opacity=0.8] (axis cs:0.2833,2.000) rectangle (axis cs:0.4000,3.3877);
\draw[draw=black, fill=none, pattern=dots, pattern color=black!60] (axis cs:0.2833,2.000) rectangle (axis cs:0.4000,3.3877);
\draw[draw=none, fill=limgreen76B900, fill opacity=0.8] (axis cs:1.2833,2.000) rectangle (axis cs:1.4000,2.4706);
\draw[draw=black, fill=none, pattern=dots, pattern color=black!60] (axis cs:1.2833,2.000) rectangle (axis cs:1.4000,2.4706);
\draw[draw=none, fill=limgreen76B900, fill opacity=0.8] (axis cs:2.2833,2.000) rectangle (axis cs:2.4000,2.3140);
\draw[draw=black, fill=none, pattern=dots, pattern color=black!60] (axis cs:2.2833,2.000) rectangle (axis cs:2.4000,2.3140);
\draw[draw=none, fill=limgreen76B900, fill opacity=0.8] (axis cs:3.2833,2.000) rectangle (axis cs:3.4000,3.2620);
\draw[draw=black, fill=none, pattern=dots, pattern color=black!60] (axis cs:3.2833,2.000) rectangle (axis cs:3.4000,3.2620);
\draw[draw=none, fill=limgreen76B900, fill opacity=0.8] (axis cs:4.2833,2.000) rectangle (axis cs:4.4000,2.9975);
\draw[draw=black, fill=none, pattern=dots, pattern color=black!60] (axis cs:4.2833,2.000) rectangle (axis cs:4.4000,2.9975);
\draw[draw=none, fill=limgreen76B900, fill opacity=0.8] (axis cs:5.2833,2.000) rectangle (axis cs:5.4000,3.2896);
\draw[draw=black, fill=none, pattern=dots, pattern color=black!60] (axis cs:5.2833,2.000) rectangle (axis cs:5.4000,3.2896);
\draw[draw=none, fill=limgreen76B900, fill opacity=0.8] (axis cs:6.2833,2.000) rectangle (axis cs:6.4000,2.6307);
\draw[draw=black, fill=none, pattern=dots, pattern color=black!60] (axis cs:6.2833,2.000) rectangle (axis cs:6.4000,2.6307);
\draw[draw=none, fill=limgreen76B900, fill opacity=0.8] (axis cs:7.2833,2.000) rectangle (axis cs:7.4000,2.8103);
\draw[draw=black, fill=none, pattern=dots, pattern color=black!60] (axis cs:7.2833,2.000) rectangle (axis cs:7.4000,2.8103);
\draw[draw=none, fill=limgreen76B900, fill opacity=0.8] (axis cs:8.2833,2.000) rectangle (axis cs:8.4000,2.8818);
\draw[draw=black, fill=none, pattern=dots, pattern color=black!60] (axis cs:8.2833,2.000) rectangle (axis cs:8.4000,2.8818);

% Whisker error bars — SE plot
\path [draw=black, line width=0.56pt] (axis cs:-0.3417,3.9352) -- (axis cs:-0.3417,4.1056);
\path [draw=black, line width=0.56pt] (axis cs:0.6583,3.2050) -- (axis cs:0.6583,3.3838);
\path [draw=black, line width=0.56pt] (axis cs:1.6583,3.3545) -- (axis cs:1.6583,3.4552);
\path [draw=black, line width=0.56pt] (axis cs:2.6583,3.9740) -- (axis cs:2.6583,4.1082);
\path [draw=black, line width=0.56pt] (axis cs:3.6583,3.5568) -- (axis cs:3.6583,3.6698);
\path [draw=black, line width=0.56pt] (axis cs:4.6583,4.0250) -- (axis cs:4.6583,4.1382);
\path [draw=black, line width=0.56pt] (axis cs:5.6583,3.4426) -- (axis cs:5.6583,3.6109);
\path [draw=black, line width=0.56pt] (axis cs:6.6583,3.2613) -- (axis cs:6.6583,3.3406);
\path [draw=black, line width=0.56pt] (axis cs:7.6583,3.4896) -- (axis cs:7.6583,3.5533);
\addplot [forget plot, semithick, steelblue31119180, mark=-, mark size=3, mark options={solid,draw=black}, only marks] table {%
-0.3417 3.9352
-0.3417 4.1056
0.6583 3.2050
0.6583 3.3838
1.6583 3.3545
1.6583 3.4552
2.6583 3.9740
2.6583 4.1082
3.6583 3.5568
3.6583 3.6698
4.6583 4.0250
4.6583 4.1382
5.6583 3.4426
5.6583 3.6109
6.6583 3.2613
6.6583 3.3406
7.6583 3.4896
7.6583 3.5533
};
\path [draw=black, line width=0.56pt] (axis cs:-0.2050,3.8331) -- (axis cs:-0.2050,4.0540);
\path [draw=black, line width=0.56pt] (axis cs:0.7950,3.0530) -- (axis cs:0.7950,3.3857);
\path [draw=black, line width=0.56pt] (axis cs:1.7950,3.1875) -- (axis cs:1.7950,3.4068);
\path [draw=black, line width=0.56pt] (axis cs:2.7950,3.9092) -- (axis cs:2.7950,4.0769);
\path [draw=black, line width=0.56pt] (axis cs:3.7950,3.4381) -- (axis cs:3.7950,3.7758);
\path [draw=black, line width=0.56pt] (axis cs:4.7950,3.9025) -- (axis cs:4.7950,4.0789);
\path [draw=black, line width=0.56pt] (axis cs:5.7950,3.3275) -- (axis cs:5.7950,3.5243);
\path [draw=black, line width=0.56pt] (axis cs:6.7950,3.1192) -- (axis cs:6.7950,3.3699);
\path [draw=black, line width=0.56pt] (axis cs:7.7950,3.4158) -- (axis cs:7.7950,3.5817);
\addplot [forget plot, semithick, steelblue31119180, mark=-, mark size=3, mark options={solid,draw=black}, only marks] table {%
-0.2050 3.8331
-0.2050 4.0540
0.7950 3.0530
0.7950 3.3857
1.7950 3.1875
1.7950 3.4068
2.7950 3.9092
2.7950 4.0769
3.7950 3.4381
3.7950 3.7758
4.7950 3.9025
4.7950 4.0789
5.7950 3.3275
5.7950 3.5243
6.7950 3.1192
6.7950 3.3699
7.7950 3.4158
7.7950 3.5817
};
\path [draw=black, line width=0.56pt] (axis cs:-0.0683,3.6925) -- (axis cs:-0.0683,4.0189);
\path [draw=black, line width=0.56pt] (axis cs:0.9317,2.8885) -- (axis cs:0.9317,3.1082);
\path [draw=black, line width=0.56pt] (axis cs:1.9317,3.0467) -- (axis cs:1.9317,3.1668);
\path [draw=black, line width=0.56pt] (axis cs:2.9317,3.7930) -- (axis cs:2.9317,3.9139);
\path [draw=black, line width=0.56pt] (axis cs:3.9317,3.5151) -- (axis cs:3.9317,3.5432);
\path [draw=black, line width=0.56pt] (axis cs:4.9317,3.9061) -- (axis cs:4.9317,4.0458);
\path [draw=black, line width=0.56pt] (axis cs:5.9317,3.2870) -- (axis cs:5.9317,3.4717);
\path [draw=black, line width=0.56pt] (axis cs:6.9317,3.1038) -- (axis cs:6.9317,3.2249);
\path [draw=black, line width=0.56pt] (axis cs:7.9317,3.2508) -- (axis cs:7.9317,3.4074);
\addplot [forget plot, semithick, steelblue31119180, mark=-, mark size=3, mark options={solid,draw=black}, only marks] table {%
-0.0683 3.6925
-0.0683 4.0189
0.9317 2.8885
0.9317 3.1082
1.9317 3.0467
1.9317 3.1668
2.9317 3.7930
2.9317 3.9139
3.9317 3.5151
3.9317 3.5432
4.9317 3.9061
4.9317 4.0458
5.9317 3.2870
5.9317 3.4717
6.9317 3.1038
6.9317 3.2249
7.9317 3.2508
7.9317 3.4074
};
\path [draw=black, line width=0.56pt] (axis cs:0.0683,3.6339) -- (axis cs:0.0683,3.8973);
\path [draw=black, line width=0.56pt] (axis cs:1.0683,2.7919) -- (axis cs:1.0683,2.9498);
\path [draw=black, line width=0.56pt] (axis cs:2.0683,2.7786) -- (axis cs:2.0683,3.1921);
\path [draw=black, line width=0.56pt] (axis cs:3.0683,3.6514) -- (axis cs:3.0683,3.9553);
\path [draw=black, line width=0.56pt] (axis cs:4.0683,3.3853) -- (axis cs:4.0683,3.5616);
\path [draw=black, line width=0.56pt] (axis cs:5.0683,3.8289) -- (axis cs:5.0683,3.9850);
\path [draw=black, line width=0.56pt] (axis cs:6.0683,3.1661) -- (axis cs:6.0683,3.3741);
\path [draw=black, line width=0.56pt] (axis cs:7.0683,3.0422) -- (axis cs:7.0683,3.1746);
\path [draw=black, line width=0.56pt] (axis cs:8.0683,3.0931) -- (axis cs:8.0683,3.3073);
\addplot [forget plot, semithick, steelblue31119180, mark=-, mark size=3, mark options={solid,draw=black}, only marks] table {%
0.0683 3.6339
0.0683 3.8973
1.0683 2.7919
1.0683 2.9498
2.0683 2.7786
2.0683 3.1921
3.0683 3.6514
3.0683 3.9553
4.0683 3.3853
4.0683 3.5616
5.0683 3.8289
5.0683 3.9850
6.0683 3.1661
6.0683 3.3741
7.0683 3.0422
7.0683 3.1746
8.0683 3.0931
8.0683 3.3073
};
\path [draw=black, line width=0.56pt] (axis cs:0.2050,3.6830) -- (axis cs:0.2050,3.6830);
\path [draw=black, line width=0.56pt] (axis cs:1.2050,2.6573) -- (axis cs:1.2050,2.6830);
\path [draw=black, line width=0.56pt] (axis cs:2.2050,2.9316) -- (axis cs:2.2050,2.9316);
\path [draw=black, line width=0.56pt] (axis cs:3.2050,3.6997) -- (axis cs:3.2050,3.6997);
\path [draw=black, line width=0.56pt] (axis cs:4.2050,3.3218) -- (axis cs:4.2050,3.3398);
\path [draw=black, line width=0.56pt] (axis cs:5.2050,3.6013) -- (axis cs:5.2050,3.6013);
\path [draw=black, line width=0.56pt] (axis cs:6.2050,3.1000) -- (axis cs:6.2050,3.1000);
\path [draw=black, line width=0.56pt] (axis cs:7.2050,3.0745) -- (axis cs:7.2050,3.1576);
\path [draw=black, line width=0.56pt] (axis cs:8.2050,3.0719) -- (axis cs:8.2050,3.0719);
\addplot [forget plot, semithick, steelblue31119180, mark=-, mark size=3, mark options={solid,draw=black}, only marks] table {%
0.2050 3.6830
0.2050 3.6830
1.2050 2.6573
1.2050 2.6830
2.2050 2.9316
2.2050 2.9316
3.2050 3.6997
3.2050 3.6997
4.2050 3.3218
4.2050 3.3398
5.2050 3.6013
5.2050 3.6013
6.2050 3.1000
6.2050 3.1000
7.2050 3.0745
7.2050 3.1576
8.2050 3.0719
8.2050 3.0719
};
\path [draw=black, line width=0.56pt] (axis cs:0.3417,3.3877) -- (axis cs:0.3417,3.3877);
\path [draw=black, line width=0.56pt] (axis cs:1.3417,2.4135) -- (axis cs:1.3417,2.5277);
\path [draw=black, line width=0.56pt] (axis cs:2.3417,2.3140) -- (axis cs:2.3417,2.3140);
\path [draw=black, line width=0.56pt] (axis cs:3.3417,3.2620) -- (axis cs:3.3417,3.2620);
\path [draw=black, line width=0.56pt] (axis cs:4.3417,2.9900) -- (axis cs:4.3417,3.0051);
\path [draw=black, line width=0.56pt] (axis cs:5.3417,3.2896) -- (axis cs:5.3417,3.2896);
\path [draw=black, line width=0.56pt] (axis cs:6.3417,2.6307) -- (axis cs:6.3417,2.6307);
\path [draw=black, line width=0.56pt] (axis cs:7.3417,2.8103) -- (axis cs:7.3417,2.8103);
\path [draw=black, line width=0.56pt] (axis cs:8.3417,2.8818) -- (axis cs:8.3417,2.8818);
\addplot [forget plot, semithick, steelblue31119180, mark=-, mark size=3, mark options={solid,draw=black}, only marks] table {%
0.3417 3.3877
0.3417 3.3877
1.3417 2.4135
1.3417 2.5277
2.3417 2.3140
2.3417 2.3140
3.3417 3.2620
3.3417 3.2620
4.3417 2.9900
4.3417 3.0051
5.3417 3.2896
5.3417 3.2896
6.3417 2.6307
6.3417 2.6307
7.3417 2.8103
7.3417 2.8103
8.3417 2.8818
8.3417 2.8818
};

% Dashed mean lines
\addplot[hotpinkE91E8C, dashed, line width=0.8pt, forget plot] coordinates {(-0.6000,3.6450) (8.6000,3.6450)};
\addplot[keBerryAD1457, dashed, line width=0.8pt, forget plot] coordinates {(-0.6000,3.5800) (8.6000,3.5800)};
\addplot[keRedD62828, dashed, line width=0.8pt, forget plot] coordinates {(-0.6000,3.4658) (8.6000,3.4658)};
\addplot[keSoftRoseF4A3C0, dashed, line width=0.8pt, forget plot] coordinates {(-0.6000,3.3760) (8.6000,3.3760)};
\addplot[steelblue1F77B4, dashed, line width=0.8pt, forget plot] coordinates {(-0.6000,3.2449) (8.6000,3.2449)};
\addplot[limgreen76B900, dashed, line width=0.8pt, forget plot] coordinates {(-0.6000,2.8938) (8.6000,2.8938)};

%% ---- BOTTOM: BLER ----
\nextgroupplot[
  tick align=outside,
  tick pos=left,
  ymajorgrids,
  y grid style={darkgray176},
  ymin=0.000, ymax=0.158,
  ytick={0, 0.05, 0.1, 0.15},
  scaled y ticks=false,
  y tick label style={/pgf/number format/fixed, /pgf/number format/precision=2},
  xlabel={Channel Index},
  ylabel={Mean BLER},
  ylabel style={yshift=-0.45cm},
  xtick={0,1,2,3,4,5,6,7,8},
  xticklabels={0,1,2,3,4,5,6,7,8},
  xmin=0, xmax=8,
  enlarge x limits=0.05,
  scaled y ticks=false,
  extra y ticks={0.1},
  extra y tick labels={},
  extra y tick style={grid=major, grid style={black, dashed, line width=0.8pt}},
]

% Direct RL
\draw[draw=none, fill=hotpinkE91E8C, fill opacity=0.8] (axis cs:-0.4000,0.000) rectangle (axis cs:-0.2833,0.0800);
\draw[draw=black, fill=none, pattern=vertical lines, pattern color=black!60] (axis cs:-0.4000,0.000) rectangle (axis cs:-0.2833,0.0800);
\draw[draw=none, fill=hotpinkE91E8C, fill opacity=0.8] (axis cs:0.6000,0.000) rectangle (axis cs:0.7167,0.1098);
\draw[draw=black, fill=none, pattern=vertical lines, pattern color=black!60] (axis cs:0.6000,0.000) rectangle (axis cs:0.7167,0.1098);
\draw[draw=none, fill=hotpinkE91E8C, fill opacity=0.8] (axis cs:1.6000,0.000) rectangle (axis cs:1.7167,0.1264);
\draw[draw=black, fill=none, pattern=vertical lines, pattern color=black!60] (axis cs:1.6000,0.000) rectangle (axis cs:1.7167,0.1264);
\draw[draw=none, fill=hotpinkE91E8C, fill opacity=0.8] (axis cs:2.6000,0.000) rectangle (axis cs:2.7167,0.0841);
\draw[draw=black, fill=none, pattern=vertical lines, pattern color=black!60] (axis cs:2.6000,0.000) rectangle (axis cs:2.7167,0.0841);
\draw[draw=none, fill=hotpinkE91E8C, fill opacity=0.8] (axis cs:3.6000,0.000) rectangle (axis cs:3.7167,0.1335);
\draw[draw=black, fill=none, pattern=vertical lines, pattern color=black!60] (axis cs:3.6000,0.000) rectangle (axis cs:3.7167,0.1335);
\draw[draw=none, fill=hotpinkE91E8C, fill opacity=0.8] (axis cs:4.6000,0.000) rectangle (axis cs:4.7167,0.0773);
\draw[draw=black, fill=none, pattern=vertical lines, pattern color=black!60] (axis cs:4.6000,0.000) rectangle (axis cs:4.7167,0.0773);
\draw[draw=none, fill=hotpinkE91E8C, fill opacity=0.8] (axis cs:5.6000,0.000) rectangle (axis cs:5.7167,0.0990);
\draw[draw=black, fill=none, pattern=vertical lines, pattern color=black!60] (axis cs:5.6000,0.000) rectangle (axis cs:5.7167,0.0990);
\draw[draw=none, fill=hotpinkE91E8C, fill opacity=0.8] (axis cs:6.6000,0.000) rectangle (axis cs:6.7167,0.1319);
\draw[draw=black, fill=none, pattern=vertical lines, pattern color=black!60] (axis cs:6.6000,0.000) rectangle (axis cs:6.7167,0.1319);
\draw[draw=none, fill=hotpinkE91E8C, fill opacity=0.8] (axis cs:7.6000,0.000) rectangle (axis cs:7.7167,0.0830);
\draw[draw=black, fill=none, pattern=vertical lines, pattern color=black!60] (axis cs:7.6000,0.000) rectangle (axis cs:7.7167,0.0830);

% k_e=0.025
\draw[draw=none, fill=keBerryAD1457, fill opacity=0.8] (axis cs:-0.2633,0.000) rectangle (axis cs:-0.1467,0.0806);
\draw[draw=black, fill=none, pattern=north east lines, pattern color=black!60] (axis cs:-0.2633,0.000) rectangle (axis cs:-0.1467,0.0806);
\draw[draw=none, fill=keBerryAD1457, fill opacity=0.8] (axis cs:0.7367,0.000) rectangle (axis cs:0.8533,0.0705);
\draw[draw=black, fill=none, pattern=north east lines, pattern color=black!60] (axis cs:0.7367,0.000) rectangle (axis cs:0.8533,0.0705);
\draw[draw=none, fill=keBerryAD1457, fill opacity=0.8] (axis cs:1.7367,0.000) rectangle (axis cs:1.8533,0.0985);
\draw[draw=black, fill=none, pattern=north east lines, pattern color=black!60] (axis cs:1.7367,0.000) rectangle (axis cs:1.8533,0.0985);
\draw[draw=none, fill=keBerryAD1457, fill opacity=0.8] (axis cs:2.7367,0.000) rectangle (axis cs:2.8533,0.0676);
\draw[draw=black, fill=none, pattern=north east lines, pattern color=black!60] (axis cs:2.7367,0.000) rectangle (axis cs:2.8533,0.0676);
\draw[draw=none, fill=keBerryAD1457, fill opacity=0.8] (axis cs:3.7367,0.000) rectangle (axis cs:3.8533,0.1000);
\draw[draw=black, fill=none, pattern=north east lines, pattern color=black!60] (axis cs:3.7367,0.000) rectangle (axis cs:3.8533,0.1000);
\draw[draw=none, fill=keBerryAD1457, fill opacity=0.8] (axis cs:4.7367,0.000) rectangle (axis cs:4.8533,0.0593);
\draw[draw=black, fill=none, pattern=north east lines, pattern color=black!60] (axis cs:4.7367,0.000) rectangle (axis cs:4.8533,0.0593);
\draw[draw=none, fill=keBerryAD1457, fill opacity=0.8] (axis cs:5.7367,0.000) rectangle (axis cs:5.8533,0.0791);
\draw[draw=black, fill=none, pattern=north east lines, pattern color=black!60] (axis cs:5.7367,0.000) rectangle (axis cs:5.8533,0.0791);
\draw[draw=none, fill=keBerryAD1457, fill opacity=0.8] (axis cs:6.7367,0.000) rectangle (axis cs:6.8533,0.0532);
\draw[draw=black, fill=none, pattern=north east lines, pattern color=black!60] (axis cs:6.7367,0.000) rectangle (axis cs:6.8533,0.0532);
\draw[draw=none, fill=keBerryAD1457, fill opacity=0.8] (axis cs:7.7367,0.000) rectangle (axis cs:7.8533,0.0438);
\draw[draw=black, fill=none, pattern=north east lines, pattern color=black!60] (axis cs:7.7367,0.000) rectangle (axis cs:7.8533,0.0438);

% k_e=0.1
\draw[draw=none, fill=keRedD62828, fill opacity=0.8] (axis cs:-0.1267,0.000) rectangle (axis cs:-0.0100,0.0719);
\draw[draw=black, fill=none, pattern=north west lines, pattern color=black!60] (axis cs:-0.1267,0.000) rectangle (axis cs:-0.0100,0.0719);
\draw[draw=none, fill=keRedD62828, fill opacity=0.8] (axis cs:0.8733,0.000) rectangle (axis cs:0.9900,0.0504);
\draw[draw=black, fill=none, pattern=north west lines, pattern color=black!60] (axis cs:0.8733,0.000) rectangle (axis cs:0.9900,0.0504);
\draw[draw=none, fill=keRedD62828, fill opacity=0.8] (axis cs:1.8733,0.000) rectangle (axis cs:1.9900,0.0724);
\draw[draw=black, fill=none, pattern=north west lines, pattern color=black!60] (axis cs:1.8733,0.000) rectangle (axis cs:1.9900,0.0724);
\draw[draw=none, fill=keRedD62828, fill opacity=0.8] (axis cs:2.8733,0.000) rectangle (axis cs:2.9900,0.0523);
\draw[draw=black, fill=none, pattern=north west lines, pattern color=black!60] (axis cs:2.8733,0.000) rectangle (axis cs:2.9900,0.0523);
\draw[draw=none, fill=keRedD62828, fill opacity=0.8] (axis cs:3.8733,0.000) rectangle (axis cs:3.9900,0.0646);
\draw[draw=black, fill=none, pattern=north west lines, pattern color=black!60] (axis cs:3.8733,0.000) rectangle (axis cs:3.9900,0.0646);
\draw[draw=none, fill=keRedD62828, fill opacity=0.8] (axis cs:4.8733,0.000) rectangle (axis cs:4.9900,0.0280);
\draw[draw=black, fill=none, pattern=north west lines, pattern color=black!60] (axis cs:4.8733,0.000) rectangle (axis cs:4.9900,0.0280);
\draw[draw=none, fill=keRedD62828, fill opacity=0.8] (axis cs:5.8733,0.000) rectangle (axis cs:5.9900,0.0340);
\draw[draw=black, fill=none, pattern=north west lines, pattern color=black!60] (axis cs:5.8733,0.000) rectangle (axis cs:5.9900,0.0340);
\draw[draw=none, fill=keRedD62828, fill opacity=0.8] (axis cs:6.8733,0.000) rectangle (axis cs:6.9900,0.0178);
\draw[draw=black, fill=none, pattern=north west lines, pattern color=black!60] (axis cs:6.8733,0.000) rectangle (axis cs:6.9900,0.0178);
\draw[draw=none, fill=keRedD62828, fill opacity=0.8] (axis cs:7.8733,0.000) rectangle (axis cs:7.9900,0.0175);
\draw[draw=black, fill=none, pattern=north west lines, pattern color=black!60] (axis cs:7.8733,0.000) rectangle (axis cs:7.9900,0.0175);

% k_e=0.5
\draw[draw=none, fill=keSoftRoseF4A3C0, fill opacity=0.8] (axis cs:0.0100,0.000) rectangle (axis cs:0.1267,0.0476);
\draw[draw=black, fill=none, pattern=horizontal lines, pattern color=black!60] (axis cs:0.0100,0.000) rectangle (axis cs:0.1267,0.0476);
\draw[draw=none, fill=keSoftRoseF4A3C0, fill opacity=0.8] (axis cs:1.0100,0.000) rectangle (axis cs:1.1267,0.0463);
\draw[draw=black, fill=none, pattern=horizontal lines, pattern color=black!60] (axis cs:1.0100,0.000) rectangle (axis cs:1.1267,0.0463);
\draw[draw=none, fill=keSoftRoseF4A3C0, fill opacity=0.8] (axis cs:2.0100,0.000) rectangle (axis cs:2.1267,0.0633);
\draw[draw=black, fill=none, pattern=horizontal lines, pattern color=black!60] (axis cs:2.0100,0.000) rectangle (axis cs:2.1267,0.0633);
\draw[draw=none, fill=keSoftRoseF4A3C0, fill opacity=0.8] (axis cs:3.0100,0.000) rectangle (axis cs:3.1267,0.0451);
\draw[draw=black, fill=none, pattern=horizontal lines, pattern color=black!60] (axis cs:3.0100,0.000) rectangle (axis cs:3.1267,0.0451);
\draw[draw=none, fill=keSoftRoseF4A3C0, fill opacity=0.8] (axis cs:4.0100,0.000) rectangle (axis cs:4.1267,0.0565);
\draw[draw=black, fill=none, pattern=horizontal lines, pattern color=black!60] (axis cs:4.0100,0.000) rectangle (axis cs:4.1267,0.0565);
\draw[draw=none, fill=keSoftRoseF4A3C0, fill opacity=0.8] (axis cs:5.0100,0.000) rectangle (axis cs:5.1267,0.0158);
\draw[draw=black, fill=none, pattern=horizontal lines, pattern color=black!60] (axis cs:5.0100,0.000) rectangle (axis cs:5.1267,0.0158);
\draw[draw=none, fill=keSoftRoseF4A3C0, fill opacity=0.8] (axis cs:6.0100,0.000) rectangle (axis cs:6.1267,0.0325);
\draw[draw=black, fill=none, pattern=horizontal lines, pattern color=black!60] (axis cs:6.0100,0.000) rectangle (axis cs:6.1267,0.0325);
\draw[draw=none, fill=keSoftRoseF4A3C0, fill opacity=0.8] (axis cs:7.0100,0.000) rectangle (axis cs:7.1267,0.0219);
\draw[draw=black, fill=none, pattern=horizontal lines, pattern color=black!60] (axis cs:7.0100,0.000) rectangle (axis cs:7.1267,0.0219);
\draw[draw=none, fill=keSoftRoseF4A3C0, fill opacity=0.8] (axis cs:8.0100,0.000) rectangle (axis cs:8.1267,0.0098);
\draw[draw=black, fill=none, pattern=horizontal lines, pattern color=black!60] (axis cs:8.0100,0.000) rectangle (axis cs:8.1267,0.0098);

% OLLA
\draw[draw=none, fill=steelblue1F77B4, fill opacity=0.8] (axis cs:0.1467,0.000) rectangle (axis cs:0.2633,0.0552);
\draw[draw=black, fill=none, pattern=crosshatch, pattern color=black!60] (axis cs:0.1467,0.000) rectangle (axis cs:0.2633,0.0552);
\draw[draw=none, fill=steelblue1F77B4, fill opacity=0.8] (axis cs:1.1467,0.000) rectangle (axis cs:1.2633,0.0453);
\draw[draw=black, fill=none, pattern=crosshatch, pattern color=black!60] (axis cs:1.1467,0.000) rectangle (axis cs:1.2633,0.0453);
\draw[draw=none, fill=steelblue1F77B4, fill opacity=0.8] (axis cs:2.1467,0.000) rectangle (axis cs:2.2633,0.0599);
\draw[draw=black, fill=none, pattern=crosshatch, pattern color=black!60] (axis cs:2.1467,0.000) rectangle (axis cs:2.2633,0.0599);
\draw[draw=none, fill=steelblue1F77B4, fill opacity=0.8] (axis cs:3.1467,0.000) rectangle (axis cs:3.2633,0.0366);
\draw[draw=black, fill=none, pattern=crosshatch, pattern color=black!60] (axis cs:3.1467,0.000) rectangle (axis cs:3.2633,0.0366);
\draw[draw=none, fill=steelblue1F77B4, fill opacity=0.8] (axis cs:4.1467,0.000) rectangle (axis cs:4.2633,0.0628);
\draw[draw=black, fill=none, pattern=crosshatch, pattern color=black!60] (axis cs:4.1467,0.000) rectangle (axis cs:4.2633,0.0628);
\draw[draw=none, fill=steelblue1F77B4, fill opacity=0.8] (axis cs:5.1467,0.000) rectangle (axis cs:5.2633,0.0372);
\draw[draw=black, fill=none, pattern=crosshatch, pattern color=black!60] (axis cs:5.1467,0.000) rectangle (axis cs:5.2633,0.0372);
\draw[draw=none, fill=steelblue1F77B4, fill opacity=0.8] (axis cs:6.1467,0.000) rectangle (axis cs:6.2633,0.0292);
\draw[draw=black, fill=none, pattern=crosshatch, pattern color=black!60] (axis cs:6.1467,0.000) rectangle (axis cs:6.2633,0.0292);
\draw[draw=none, fill=steelblue1F77B4, fill opacity=0.8] (axis cs:7.1467,0.000) rectangle (axis cs:7.2633,0.0316);
\draw[draw=black, fill=none, pattern=crosshatch, pattern color=black!60] (axis cs:7.1467,0.000) rectangle (axis cs:7.2633,0.0316);
\draw[draw=none, fill=steelblue1F77B4, fill opacity=0.8] (axis cs:8.1467,0.000) rectangle (axis cs:8.2633,0.0108);
\draw[draw=black, fill=none, pattern=crosshatch, pattern color=black!60] (axis cs:8.1467,0.000) rectangle (axis cs:8.2633,0.0108);

% SALAD
\draw[draw=none, fill=limgreen76B900, fill opacity=0.8] (axis cs:0.2833,0.000) rectangle (axis cs:0.4000,0.0387);
\draw[draw=black, fill=none, pattern=dots, pattern color=black!60] (axis cs:0.2833,0.000) rectangle (axis cs:0.4000,0.0387);
\draw[draw=none, fill=limgreen76B900, fill opacity=0.8] (axis cs:1.2833,0.000) rectangle (axis cs:1.4000,0.0521);
\draw[draw=black, fill=none, pattern=dots, pattern color=black!60] (axis cs:1.2833,0.000) rectangle (axis cs:1.4000,0.0521);
\draw[draw=none, fill=limgreen76B900, fill opacity=0.8] (axis cs:2.2833,0.000) rectangle (axis cs:2.4000,0.0833);
\draw[draw=black, fill=none, pattern=dots, pattern color=black!60] (axis cs:2.2833,0.000) rectangle (axis cs:2.4000,0.0833);
\draw[draw=none, fill=limgreen76B900, fill opacity=0.8] (axis cs:3.2833,0.000) rectangle (axis cs:3.4000,0.0486);
\draw[draw=black, fill=none, pattern=dots, pattern color=black!60] (axis cs:3.2833,0.000) rectangle (axis cs:3.4000,0.0486);
\draw[draw=none, fill=limgreen76B900, fill opacity=0.8] (axis cs:4.2833,0.000) rectangle (axis cs:4.4000,0.0655);
\draw[draw=black, fill=none, pattern=dots, pattern color=black!60] (axis cs:4.2833,0.000) rectangle (axis cs:4.4000,0.0655);
\draw[draw=none, fill=limgreen76B900, fill opacity=0.8] (axis cs:5.2833,0.000) rectangle (axis cs:5.4000,0.0521);
\draw[draw=black, fill=none, pattern=dots, pattern color=black!60] (axis cs:5.2833,0.000) rectangle (axis cs:5.4000,0.0521);
\draw[draw=none, fill=limgreen76B900, fill opacity=0.8] (axis cs:6.2833,0.000) rectangle (axis cs:6.4000,0.0409);
\draw[draw=black, fill=none, pattern=dots, pattern color=black!60] (axis cs:6.2833,0.000) rectangle (axis cs:6.4000,0.0409);
\draw[draw=none, fill=limgreen76B900, fill opacity=0.8] (axis cs:7.2833,0.000) rectangle (axis cs:7.4000,0.0420);
\draw[draw=black, fill=none, pattern=dots, pattern color=black!60] (axis cs:7.2833,0.000) rectangle (axis cs:7.4000,0.0420);
\draw[draw=none, fill=limgreen76B900, fill opacity=0.8] (axis cs:8.2833,0.000) rectangle (axis cs:8.4000,0.0216);
\draw[draw=black, fill=none, pattern=dots, pattern color=black!60] (axis cs:8.2833,0.000) rectangle (axis cs:8.4000,0.0216);

% Whisker error bars — BLER plot
\path [draw=black, line width=0.56pt] (axis cs:-0.3417,0.0622) -- (axis cs:-0.3417,0.0977);
\path [draw=black, line width=0.56pt] (axis cs:0.6583,0.0752) -- (axis cs:0.6583,0.1444);
\path [draw=black, line width=0.56pt] (axis cs:1.6583,0.0998) -- (axis cs:1.6583,0.1530);
\path [draw=black, line width=0.56pt] (axis cs:2.6583,0.0634) -- (axis cs:2.6583,0.1047);
\path [draw=black, line width=0.56pt] (axis cs:3.6583,0.1134) -- (axis cs:3.6583,0.1537);
\path [draw=black, line width=0.56pt] (axis cs:4.6583,0.0555) -- (axis cs:4.6583,0.0992);
\path [draw=black, line width=0.56pt] (axis cs:5.6583,0.0524) -- (axis cs:5.6583,0.1456);
\path [draw=black, line width=0.56pt] (axis cs:6.6583,0.1076) -- (axis cs:6.6583,0.1563);
\path [draw=black, line width=0.56pt] (axis cs:7.6583,0.0563) -- (axis cs:7.6583,0.1097);
\addplot [forget plot, semithick, steelblue31119180, mark=-, mark size=3, mark options={solid,draw=black}, only marks] table {%
-0.3417 0.0622
-0.3417 0.0977
0.6583 0.0752
0.6583 0.1444
1.6583 0.0998
1.6583 0.1530
2.6583 0.0634
2.6583 0.1047
3.6583 0.1134
3.6583 0.1537
4.6583 0.0555
4.6583 0.0992
5.6583 0.0524
5.6583 0.1456
6.6583 0.1076
6.6583 0.1563
7.6583 0.0563
7.6583 0.1097
};
\path [draw=black, line width=0.56pt] (axis cs:-0.2050,0.0608) -- (axis cs:-0.2050,0.1005);
\path [draw=black, line width=0.56pt] (axis cs:0.7950,0.0442) -- (axis cs:0.7950,0.0968);
\path [draw=black, line width=0.56pt] (axis cs:1.7950,0.0631) -- (axis cs:1.7950,0.1338);
\path [draw=black, line width=0.56pt] (axis cs:2.7950,0.0481) -- (axis cs:2.7950,0.0872);
\path [draw=black, line width=0.56pt] (axis cs:3.7950,0.0765) -- (axis cs:3.7950,0.1234);
\path [draw=black, line width=0.56pt] (axis cs:4.7950,0.0237) -- (axis cs:4.7950,0.0950);
\path [draw=black, line width=0.56pt] (axis cs:5.7950,0.0422) -- (axis cs:5.7950,0.1160);
\path [draw=black, line width=0.56pt] (axis cs:6.7950,0.0087) -- (axis cs:6.7950,0.0977);
\path [draw=black, line width=0.56pt] (axis cs:7.7950,0.0228) -- (axis cs:7.7950,0.0649);
\addplot [forget plot, semithick, steelblue31119180, mark=-, mark size=3, mark options={solid,draw=black}, only marks] table {%
-0.2050 0.0608
-0.2050 0.1005
0.7950 0.0442
0.7950 0.0968
1.7950 0.0631
1.7950 0.1338
2.7950 0.0481
2.7950 0.0872
3.7950 0.0765
3.7950 0.1234
4.7950 0.0237
4.7950 0.0950
5.7950 0.0422
5.7950 0.1160
6.7950 0.0087
6.7950 0.0977
7.7950 0.0228
7.7950 0.0649
};
\path [draw=black, line width=0.56pt] (axis cs:-0.0683,0.0545) -- (axis cs:-0.0683,0.0893);
\path [draw=black, line width=0.56pt] (axis cs:0.9317,0.0156) -- (axis cs:0.9317,0.0851);
\path [draw=black, line width=0.56pt] (axis cs:1.9317,0.0555) -- (axis cs:1.9317,0.0893);
\path [draw=black, line width=0.56pt] (axis cs:2.9317,0.0372) -- (axis cs:2.9317,0.0673);
\path [draw=black, line width=0.56pt] (axis cs:3.9317,0.0433) -- (axis cs:3.9317,0.0860);
\path [draw=black, line width=0.56pt] (axis cs:4.9317,0.0070) -- (axis cs:4.9317,0.0490);
\path [draw=black, line width=0.56pt] (axis cs:5.9317,0.0112) -- (axis cs:5.9317,0.0568);
\path [draw=black, line width=0.56pt] (axis cs:6.9317,0.0106) -- (axis cs:6.9317,0.0251);
\path [draw=black, line width=0.56pt] (axis cs:7.9317,0.0061) -- (axis cs:7.9317,0.0288);
\addplot [forget plot, semithick, steelblue31119180, mark=-, mark size=3, mark options={solid,draw=black}, only marks] table {%
-0.0683 0.0545
-0.0683 0.0893
0.9317 0.0156
0.9317 0.0851
1.9317 0.0555
1.9317 0.0893
2.9317 0.0372
2.9317 0.0673
3.9317 0.0433
3.9317 0.0860
4.9317 0.0070
4.9317 0.0490
5.9317 0.0112
5.9317 0.0568
6.9317 0.0106
6.9317 0.0251
7.9317 0.0061
7.9317 0.0288
};
\path [draw=black, line width=0.56pt] (axis cs:0.0683,0.0309) -- (axis cs:0.0683,0.0643);
\path [draw=black, line width=0.56pt] (axis cs:1.0683,0.0323) -- (axis cs:1.0683,0.0604);
\path [draw=black, line width=0.56pt] (axis cs:2.0683,0.0132) -- (axis cs:2.0683,0.1135);
\path [draw=black, line width=0.56pt] (axis cs:3.0683,0.0238) -- (axis cs:3.0683,0.0663);
\path [draw=black, line width=0.56pt] (axis cs:4.0683,0.0410) -- (axis cs:4.0683,0.0720);
\path [draw=black, line width=0.56pt] (axis cs:5.0683,0.0063) -- (axis cs:5.0683,0.0253);
\path [draw=black, line width=0.56pt] (axis cs:6.0683,0.0197) -- (axis cs:6.0683,0.0452);
\path [draw=black, line width=0.56pt] (axis cs:7.0683,0.0077) -- (axis cs:7.0683,0.0362);
\path [draw=black, line width=0.56pt] (axis cs:8.0683,-0.0014) -- (axis cs:8.0683,0.0210);
\addplot [forget plot, semithick, steelblue31119180, mark=-, mark size=3, mark options={solid,draw=black}, only marks] table {%
0.0683 0.0309
0.0683 0.0643
1.0683 0.0323
1.0683 0.0604
2.0683 0.0132
2.0683 0.1135
3.0683 0.0238
3.0683 0.0663
4.0683 0.0410
4.0683 0.0720
5.0683 0.0063
5.0683 0.0253
6.0683 0.0197
6.0683 0.0452
7.0683 0.0077
7.0683 0.0362
8.0683 -0.0014
8.0683 0.0210
};
\path [draw=black, line width=0.56pt] (axis cs:0.2050,0.0552) -- (axis cs:0.2050,0.0552);
\path [draw=black, line width=0.56pt] (axis cs:1.2050,0.0387) -- (axis cs:1.2050,0.0519);
\path [draw=black, line width=0.56pt] (axis cs:2.2050,0.0599) -- (axis cs:2.2050,0.0599);
\path [draw=black, line width=0.56pt] (axis cs:3.2050,0.0366) -- (axis cs:3.2050,0.0366);
\path [draw=black, line width=0.56pt] (axis cs:4.2050,0.0626) -- (axis cs:4.2050,0.0630);
\path [draw=black, line width=0.56pt] (axis cs:5.2050,0.0372) -- (axis cs:5.2050,0.0372);
\path [draw=black, line width=0.56pt] (axis cs:6.2050,0.0292) -- (axis cs:6.2050,0.0292);
\path [draw=black, line width=0.56pt] (axis cs:7.2050,0.0281) -- (axis cs:7.2050,0.0352);
\path [draw=black, line width=0.56pt] (axis cs:8.2050,0.0108) -- (axis cs:8.2050,0.0108);
\addplot [forget plot, semithick, steelblue31119180, mark=-, mark size=3, mark options={solid,draw=black}, only marks] table {%
0.2050 0.0552
0.2050 0.0552
1.2050 0.0387
1.2050 0.0519
2.2050 0.0599
2.2050 0.0599
3.2050 0.0366
3.2050 0.0366
4.2050 0.0626
4.2050 0.0630
5.2050 0.0372
5.2050 0.0372
6.2050 0.0292
6.2050 0.0292
7.2050 0.0281
7.2050 0.0352
8.2050 0.0108
8.2050 0.0108
};
\path [draw=black, line width=0.56pt] (axis cs:0.3417,0.0387) -- (axis cs:0.3417,0.0387);
\path [draw=black, line width=0.56pt] (axis cs:1.3417,0.0440) -- (axis cs:1.3417,0.0601);
\path [draw=black, line width=0.56pt] (axis cs:2.3417,0.0833) -- (axis cs:2.3417,0.0833);
\path [draw=black, line width=0.56pt] (axis cs:3.3417,0.0486) -- (axis cs:3.3417,0.0486);
\path [draw=black, line width=0.56pt] (axis cs:4.3417,0.0653) -- (axis cs:4.3417,0.0657);
\path [draw=black, line width=0.56pt] (axis cs:5.3417,0.0521) -- (axis cs:5.3417,0.0521);
\path [draw=black, line width=0.56pt] (axis cs:6.3417,0.0409) -- (axis cs:6.3417,0.0409);
\path [draw=black, line width=0.56pt] (axis cs:7.3417,0.0420) -- (axis cs:7.3417,0.0420);
\path [draw=black, line width=0.56pt] (axis cs:8.3417,0.0216) -- (axis cs:8.3417,0.0216);
\addplot [forget plot, semithick, steelblue31119180, mark=-, mark size=3, mark options={solid,draw=black}, only marks] table {%
0.3417 0.0387
0.3417 0.0387
1.3417 0.0440
1.3417 0.0601
2.3417 0.0833
2.3417 0.0833
3.3417 0.0486
3.3417 0.0486
4.3417 0.0653
4.3417 0.0657
5.3417 0.0521
5.3417 0.0521
6.3417 0.0409
6.3417 0.0409
7.3417 0.0420
7.3417 0.0420
8.3417 0.0216
8.3417 0.0216
};

\addplot[black, dashed, line width=0.8pt, forget plot]
  coordinates {(-0.6, 0.1) (8.6, 0.1)};

\end{groupplot}

% Manual legend centred above top panel
\node[anchor=south, draw=lightgray204, fill=white, fill opacity=0.8,
  inner sep=5pt] at
  ($(group c1r1.north)+(0,0.4cm)$) {%
  \scalebox{1.95}{%
  \begin{tikzpicture}[every node/.style={font=\fontsize{22}{27}\selectfont}]
    \draw[fill=hotpinkE91E8C, fill opacity=0.8, draw=black,
      postaction={pattern=vertical lines, pattern color=black!60}]
      (0.00,0.00) rectangle (0.90,0.65);
    \node[anchor=west] at (1.02,0.325) {$k_e\!=\!0$};
    \draw[fill=keBerryAD1457, fill opacity=0.8, draw=black,
      postaction={pattern=north east lines, pattern color=black!60}]
      (3.40,0.00) rectangle (4.30,0.65);
    \node[anchor=west] at (4.42,0.325) {$k_e\!=\!0.025$};
    \draw[fill=keRedD62828, fill opacity=0.8, draw=black,
      postaction={pattern=north west lines, pattern color=black!60}]
      (7.80,0.00) rectangle (8.70,0.65);
    \node[anchor=west] at (8.82,0.325) {$k_e\!=\!0.1$};
    \draw[fill=keSoftRoseF4A3C0, fill opacity=0.8, draw=black,
      postaction={pattern=horizontal lines, pattern color=black!60}]
      (11.60,0.00) rectangle (12.50,0.65);
    \node[anchor=west] at (12.62,0.325) {$k_e\!=\!0.5$};
    \draw[fill=steelblue1F77B4, fill opacity=0.8, draw=black,
      postaction={pattern=crosshatch, pattern color=black!60}]
      (15.60,0.00) rectangle (16.50,0.65);
    \node[anchor=west] at (16.62,0.325) {OLLA};
    \draw[fill=limgreen76B900, fill opacity=0.8, draw=black,
      postaction={pattern=dots, pattern color=black!60}]
      (19.00,0.00) rectangle (19.90,0.65);
    \node[anchor=west] at (20.02,0.325) {SALAD};
    \draw[black, dashed, line width=1.4pt]
      (22.60,0.325) -- (23.50,0.325);
    \node[anchor=west] at (23.62,0.325) {Target};
  \end{tikzpicture}%
  }%
};

\end{tikzpicture}

%% file: fig/obsSize_bar_sinrpos.tex
% Preamble: \usetikzlibrary{patterns, calc}
%           \usepgfplotslibrary{groupplots}

\begin{tikzpicture}

\definecolor{darkgray176}{RGB}{176,176,176}
\definecolor{lightgray204}{RGB}{204,204,204}
\definecolor{steelblue31119180}{RGB}{31,119,180}
\definecolor{hotpinkE91E8C}{RGB}{233,30,140}
\definecolor{keBerryAD1457}{RGB}{173,20,87}

\begin{groupplot}[
  group style={group size=2 by 1, horizontal sep=2cm},
  width=7.5cm,
  height=4.1cm,
  tick label style={font=\Large},
  label style={font=\Large},
  title style={font=\Large},
  xticklabel style={rotate=35, anchor=east, font=\Large},
  ylabel style={yshift=-0.1cm},
  scaled y ticks=false,
]

%% ---- SE ----
\nextgroupplot[
  tick align=outside, tick pos=left,
  x grid style={darkgray176}, y grid style={darkgray176},
  xmin=-0.4, xmax=1.0,
  xtick={0, 0.6},
  xticklabels={Setup A, Setup B},
  xtick style={color=black}, ytick style={color=black},
  ymajorgrids,
  ymin=3.355, ymax=3.901,
  ylabel={Mean SE [bps/Hz]},
]
\draw[draw=none, fill=hotpinkE91E8C, fill opacity=0.8] (axis cs:-0.2,2.000) rectangle (axis cs:0.2,3.64498);
\draw[draw=black, fill=none, pattern=vertical lines, pattern color=black!60] (axis cs:-0.2,2.000) rectangle (axis cs:0.2,3.64498);
\draw[draw=none, fill=keBerryAD1457, fill opacity=0.8] (axis cs:0.4,2.000) rectangle (axis cs:0.8,3.60388);
\draw[draw=black, fill=none, pattern=dots, pattern color=black!60] (axis cs:0.4,2.000) rectangle (axis cs:0.8,3.60388);
\path [draw=black, line width=0.56pt] (axis cs:0,3.39924) -- (axis cs:0,3.89072);
\path [draw=black, line width=0.56pt] (axis cs:0.6,3.37002) -- (axis cs:0.6,3.83775);
\addplot [forget plot, semithick, steelblue31119180, mark=-, mark size=5, mark options={solid,draw=black}, only marks] table {%
0   3.39924
0.6 3.37002
};
\addplot [forget plot, semithick, steelblue31119180, mark=-, mark size=5, mark options={solid,draw=black}, only marks] table {%
0   3.89072
0.6 3.83775
};
\addplot[hotpinkE91E8C, dashed, line width=0.8pt, forget plot] coordinates {(-0.4,3.64498) (1.0,3.64498)};
\addplot[keBerryAD1457, dashed, line width=0.8pt, forget plot] coordinates {(-0.4,3.60388) (1.0,3.60388)};
\draw (axis cs:0,3.7) node[scale=1.2, anchor=south west, text=black]{3.645};
\draw (axis cs:0.6,3.7) node[scale=1.2, anchor=south west, text=black]{3.604};

%% ---- BLER ----
\nextgroupplot[
  tick align=outside, tick pos=left,
  x grid style={darkgray176}, y grid style={darkgray176},
  xmin=-0.4, xmax=1.0,
  xtick={0.05, 0.6},
  xticklabels={Setup A, Setup B},
  xtick style={color=black}, ytick style={color=black},
  ymajorgrids,
  ymin=0.082, ymax=0.136,
  ylabel={Mean BLER},
  ylabel style={yshift=-2pt},
]
\draw[draw=none, fill=hotpinkE91E8C, fill opacity=0.8] (axis cs:-0.2,0.000) rectangle (axis cs:0.2,0.102778);
\draw[draw=black, fill=none, pattern=vertical lines, pattern color=black!60] (axis cs:-0.2,0.000) rectangle (axis cs:0.2,0.102778);
\draw[draw=none, fill=keBerryAD1457, fill opacity=0.8] (axis cs:0.4,0.000) rectangle (axis cs:0.8,0.110746);
\draw[draw=black, fill=none, pattern=dots, pattern color=black!60] (axis cs:0.4,0.000) rectangle (axis cs:0.8,0.110746);
\path [draw=black, line width=0.56pt] (axis cs:0,0.0848971) -- (axis cs:0,0.120659);
\path [draw=black, line width=0.56pt] (axis cs:0.6,0.0873295) -- (axis cs:0.6,0.134162);
\addplot [forget plot, semithick, steelblue31119180, mark=-, mark size=5, mark options={solid,draw=black}, only marks] table {%
0   0.0848971
0.6 0.0873295
};
\addplot [forget plot, semithick, steelblue31119180, mark=-, mark size=5, mark options={solid,draw=black}, only marks] table {%
0   0.120659
0.6 0.134162
};
\addplot [forget plot, black, dashed, line width=1.2pt] coordinates {(-0.4,0.1) (1.0,0.1)};
\draw (axis cs:0,0.109) node[scale=1.2, anchor=south west, text=black]{0.103};
\draw (axis cs:0.6,0.109) node[scale=1.2, anchor=south west, text=black]{0.111};
% Inline Target legend inside BLER plot (left, avoids Setup B bar)
\draw[draw=lightgray204, fill=white, fill opacity=0.8]
  (axis cs:-0.38,0.1255) rectangle (axis cs:0.21,0.1325);
\draw[black, dashed, line width=1.2pt]
  (axis cs:-0.36,0.129) -- (axis cs:-0.18,0.129);
\draw (axis cs:-0.15,0.129) node[scale=0.9, anchor=west, text=black]{\normalsize Target};

\end{groupplot}

\end{tikzpicture}

%% file: fig/fqi_oai_combined_apr15.tex
% Combined FQI figure: Q-value convergence (left) + MCS distribution (right)
% Preamble:
%   \usepackage{pgfplots}
%   \pgfplotsset{compat=1.18}
%   \usepgfplotslibrary{groupplots}
%   \usetikzlibrary{patterns}

\begin{tikzpicture}

\definecolor{bellmanblue}{RGB}{31,119,180}
\definecolor{schedblue}{RGB}{100,149,200}
\definecolor{fqicoral}{RGB}{250,160,120}

\begin{groupplot}[
  group style={
    group size=2 by 1,
    horizontal sep=1.6cm,
  },
  height=3.5cm,
  tick align=outside,
  tick pos=left,
  xmajorgrids,
  ymajorgrids,
  x grid style={black!15},
  y grid style={black!15},
  tick label style={font=\Large},
  label style={font=\Large},
]

%% ---- LEFT: Q-value convergence ----
\nextgroupplot[
  width=5cm,
  ylabel style={yshift=-0.2cm},
  xlabel={FQI Iteration},
  ylabel={Avg.\ Q-value},
  xmin=0, xmax=30,
  ymin=60, ymax=180,
  legend style={at={(0.97,0.35)}, anchor=east, font=\Large},
]

\addplot [line width=1.5pt, bellmanblue, mark=*, mark size=1.8pt]
table {%
0  79.41
1  120.15
2  140.77
3  150.99
4  156.21
5  158.79
6  160.13
7  160.74
8  161.09
9  161.21
10 161.28
11 161.33
12 161.31
13 161.35
14 161.31
15 161.35
16 161.34
17 161.34
18 161.36
19 161.38
20 161.37
21 161.34
22 161.40
23 161.37
24 161.38
25 161.38
26 161.41
27 161.33
28 161.31
29 161.30
30 161.25
};

%% ---- RIGHT: MCS distribution ----
\nextgroupplot[
  ybar,
  bar width=8.5pt,
  width=12cm,
  ylabel style={yshift=-0.2cm},
  xlabel={MCS Index},
  ylabel={Percentage [\%]},
  xtick={11,12,13,14,15,16,17,18,19,20,21,22},
  xticklabels={11,12,13,14,15,16,17,18,19,20,21,22},
  x tick label style={rotate=45, anchor=east, font=\Large},
  xmin=10, xmax=22.8,
  ymin=0, ymax=75,
  legend style={at={(0.03,0.97)}, anchor=north west, font=\Large},
  xmajorgrids=false,
]

\addplot [fill=schedblue, fill opacity=0.75, draw=schedblue!80!black]
coordinates {
  (0,  0.0)
  (1,  0.0)
  (2,  0.0)
  (3,  0.0)
  (4,  0.0)
  (5,  0.0)
  (6,  0.0)
  (7,  0.0)
  (8,  0.0)
  (9,  0.0)
  (10,  0.0)
  (11,  3.6)
  (12,  0.0)
  (13,  1.8)
  (14,  3.6)
  (15,  1.8)
  (16,  1.8)
  (17,  5.4)
  (18,  8.9)
  (19,  71.4)
  (20,  0.0)
  (21,  1.8)
  (22,  0.0)
};
\addlegendentry{OAI Scheduler}

\addplot [
  fill=fqicoral,
  fill opacity=1,
  draw=fqicoral!80!black,
  postaction={pattern=north east lines, pattern color=fqicoral!80!black}
]
coordinates {
  (0,  0.0)
  (1,  0.0)
  (2,  0.0)
  (3,  0.0)
  (4,  0.0)
  (5,  0.0)
  (6,  0.0)
  (7,  0.0)
  (8,  0.0)
  (9,  0.0)
  (10,  0.0)
  (11,  0.0)
  (12,  0.0)
  (13,  0.0)
  (14,  0.0)
  (15,  3.6)
  (16,  1.8)
  (17,  25.0)
  (18,  3.6)
  (19,  35.7)
  (20,  1.8)
  (21,  26.8)
  (22,  1.8)
};
\addlegendentry{FQI Policy}

\end{groupplot}
\end{tikzpicture}

%% file: fig/fqi_csv_combined_apr15.tex
% Combined FQI figure — metrics.csv (DL data, apr15)
% LEFT: Q-value convergence  RIGHT: MCS distribution
% Preamble:
% \usepackage{pgfplots}
% \pgfplotsset{compat=1.18}
% \usepgfplotslibrary{groupplots}
% \usetikzlibrary{patterns}

\begin{tikzpicture}

\definecolor{bellmanblue}{RGB}{31,119,180}
\definecolor{schedblue}{RGB}{100,149,200}
\definecolor{fqicoral}{RGB}{250,160,120}

\begin{groupplot}[
  group style={
    group size=2 by 1,
    horizontal sep=1.8cm,
  },
  height=3.5cm,
  tick align=outside,
  tick pos=left,
  ymajorgrids,
  y grid style={black!15},
  tick label style={font=\Large},
  label style={font=\Large},
]

%% ---- LEFT: Q-value convergence ----
\nextgroupplot[
  width=5cm,
  ylabel style={yshift=-0.2cm},
  xlabel={FQI Iteration},
  ylabel={Avg.\ Q-value},
  xmajorgrids,
  x grid style={black!15},
  xmin=0, xmax=30,
  ymin=700, ymax=1670,
  ytick={800,1200,1600},
  legend style={at={(0.97,0.25)}, anchor=east, font=\Large},
]

\addplot [line width=1.5pt, bellmanblue, mark=*, mark size=1.8pt]
table {%
0  754.10
1  1200.96
2  1425.94
3  1538.28
4  1595.28
5  1623.44
6  1637.97
7  1644.87
8  1648.54
9  1649.37
10  1649.71
11  1651.14
12  1650.57
13  1650.88
14  1650.74
15  1651.49
16  1652.05
17  1651.20
18  1650.65
19  1650.89
20  1651.31
21  1650.78
22  1650.95
23  1651.25
24  1650.52
25  1650.35
26  1650.87
27  1650.55
28  1650.35
29  1651.01
30  1651.13
};

%% ---- RIGHT: MCS distribution ----
\nextgroupplot[
  ybar,
  bar width=8.5pt,
  width=12cm,
  ylabel style={yshift=-0.2cm},
  xlabel={MCS Index},
  ylabel={Percentage [\%]},
  xtick={9,10,11,12,13,14,15,16,17,18,19,20,21,22,23,24,25,26,27},
  xticklabels={9,10,11,12,13,14,15,16,17,18,19,20,21,22,23,24,25,26,27},
  x tick label style={rotate=45, anchor=east, font=\Large},
  xmin=8, xmax=27.8,
  ymin=0, ymax=76,
  legend style={at={(0.05,0.97)}, anchor=north west, font=\Large},
  xmajorgrids=false,
]

\addplot [fill=schedblue, fill opacity=0.75, draw=schedblue!80!black]
coordinates {
  (9,  0.17)
  (10,  0.11)
  (11,  0.10)
  (12,  0.09)
  (13,  0.20)
  (14,  0.12)
  (15,  0.24)
  (16,  0.10)
  (17,  0.19)
  (18,  0.14)
  (19,  0.19)
  (20,  0.37)
  (21,  1.21)
  (22,  1.05)
  (23,  16.91)
  (24,  2.49)
  (25,  10.79)
  (26,  3.00)
  (27,  62.54)
};
\addlegendentry{OAI Scheduler}

\addplot [
  fill=fqicoral,
  fill opacity=1,
  draw=fqicoral!80!black,
  postaction={pattern=north east lines, pattern color=fqicoral!80!black}
]
coordinates {
  (9,  0.70)
  (10,  0.00)
  (11,  5.40)
  (12,  0.10)
  (13,  0.00)
  (14,  17.40)
  (15,  2.00)
  (16,  0.00)
  (17,  0.00)
  (18,  7.80)
  (19,  0.00)
  (20,  16.10)
  (21,  4.10)
  (22,  0.00)
  (23,  19.20)
  (24,  0.70)
  (25,  0.50)
  (26,  0.00)
  (27,  26.00)
};
\addlegendentry{FQI Policy}

\end{groupplot}
\end{tikzpicture}

%% file: tex/spectrum_sharing.tex
% !TEX root = ../thesis_dissertation template.tex
% spectrum_sharing.tex

\chapter[Enabling Spectrum Sharing in FR3 for Cellular and NTN Systems]{Enabling Spectrum Sharing in \texorpdfstring{\gls{fr3}}{FR3} for Cellular and \texorpdfstring{\gls{ntn}}{NTN} Systems}
\label{chap:spectrum_sharing}

% Local macro definitions for this chapter
\newcommand{\R}{\Delta\textrm{RSS}}
\newcommand{\quadriga}{QuaDRiGa\xspace}

% Listings style for MATLAB code in this chapter
\lstset{
  language=Matlab,
  basicstyle=\ttfamily\small,
  keywordstyle=\color{blue},
  commentstyle=\color{green!50!black},
  numbers=left,
  numberstyle=\tiny,
  frame=single,
  breaklines=true,
  captionpos=b
}

\section{Introduction}\label{sec:ss:intro}

\gls{fr3}, encompassing frequencies between $7.125$ and $24.25$~GHz, is an emerging frequency band for \gls{6g} applications. The upper mid-band, as it is frequently referred to, represents the sweet spot between coverage and capacity, providing better range than mmWaves and higher bandwidth than the sub-6~GHz band. Despite these advantages, the spectrum is already occupied by incumbent systems such as satellites (e.g., Starlink), and sharing it with terrestrial cellular applications results in spectrum conflicts, only exacerbating the existing spectrum scarcity. This chapter~\cite{tsampazi2026satellite} investigates the impact of two state-of-the-art methods, namely \gls{qos}-Aware Power Control and Interference Nulling, as well as their joint application, on interference mitigation toward non-terrestrial links while maintaining acceptable \gls{qos} on terrestrial networks.
Our simulation results demonstrate the advantages and disadvantages of each method, pinpointing how interference nulling can maintain high average performance and how power control is more appropriate for risk-averse scenarios to enhance fairness in terrestrial \gls{qos}. Finally, we showcase how the two can complement each other to enhance fairness in terrestrial \gls{qos} and increase the \gls{gnb}'s energy efficiency, while suppressing interference toward incumbents.

Indeed, \gls{fr3} has been identified as the \textit{golden band} for \gls{6g}, as it satisfies both capacity and coverage requirements~\cite{bazzi2025upper}. However, the rapid deployment of dense satellite communication systems introduces new challenges that necessitate robust interference management strategies. In particular, coexistence between terrestrial and non-terrestrial networks remains a critical issue for the successful commercial deployment of \gls{6g} systems. Therefore, the implementation of advanced interference management techniques is essential.

The challenge of interference mitigation in coexistence scenarios has been extensively studied in the literature~\cite{heydarishahreza2024spectrum}. Proposed strategies range from the use of highly directional antennas to adaptive beamforming and null steering designed to suppress radiation in undesired directions. Furthermore, spectrum masking, power control, and diversity techniques are commonly employed to minimize interference effects~\cite{agarwal2023coexistence}. While these approaches are effective for managing interference in the \gls{fr3}, they often face practical challenges related to hardware limitations and limited spectrum availability~\cite{agarwal2023coexistence}.

Specifically, in~\cite{niloy2024ascent, niloy2023interference}, the authors underscore the importance of site-specific deployment strategies and context-aware policies in ensuring efficient coexistence between terrestrial and satellite systems. In~\cite{niloy2024ascent}, a spectrum-sharing scenario is examined in which a terrestrial network operating in the upper mid-band interferes with incumbent satellite communications. Interference mitigation is achieved via a closed-loop feedback mechanism that enables real-time adaptation of spectrum-sharing policies. This approach balances satellite protection with efficient spectrum utilization by dynamically adjusting the \gls{ez} radius whenever interference levels exceed predefined thresholds. In~\cite{niloy2023interference}, spectral overlap is discussed with a focus on \gls{los} links and angular alignment between the \gls{gnb} and satellite \glspl{rx}, which are identified as critical factors affecting interference. In~\cite{wu2023space}, the authors propose an optimization framework to maximize the \gls{wsr} in multi-\gls{ue} multibeam \gls{leo} satellite networks by optimizing the beamforming weights to balance signal quality and interference mitigation. The work in~\cite{ma2023resource} optimizes beamforming and power in dense \gls{leo} satellite networks to maximize system fairness. Beamforming vectors are calculated using \gls{sca} to maximize the minimum \gls{ue} rate, while satisfying power and interference constraints. If power limits are exceeded, power scaling is applied to adjust the beamforming vectors. In~\cite{kang2024terrestrial}, the authors focus on the trade-off between terrestrial \gls{snr} maximization and interference suppression, represented by the tunable parameter~$\lambda$, and solve it as a joint maximization--minimization problem. Finally, in~\cite{wadaskar2025satellite}, the authors address satellite-terrestrial coexistence in \gls{fr3} by designing hybrid \gls{ttd} array precoders at the \gls{gnb} that steer wideband nulls toward satellites while preserving main-lobe gain for terrestrial \glspl{ue}.

\subsection{Contributions and Outline} \label{sec:ss:contributions}
\begin{figure}[H]
  \centering
  \includegraphics[width=0.7\textwidth, keepaspectratio]{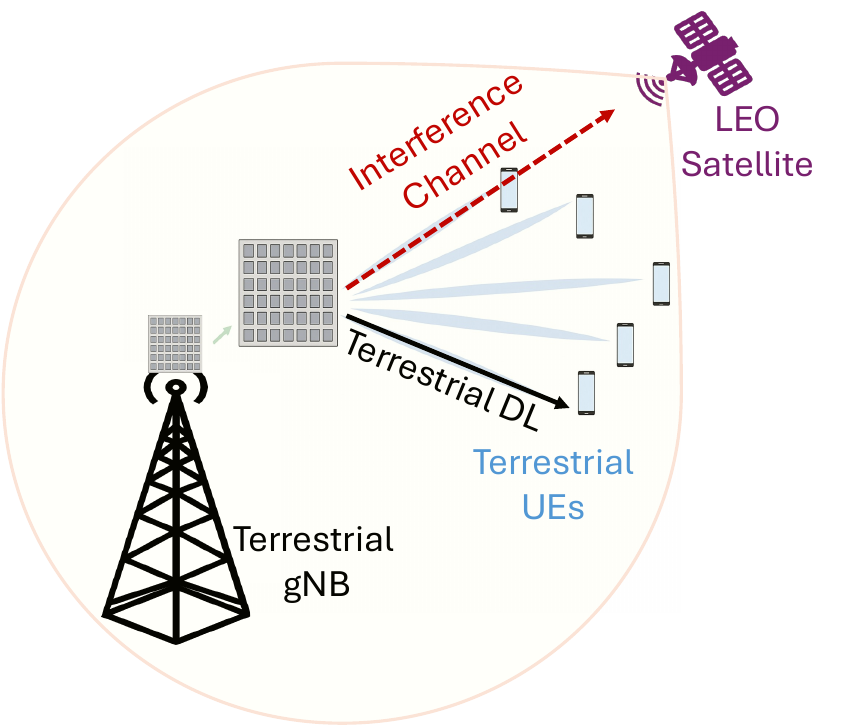}
  \caption{Interference from terrestrial cellular \gls{dl} transmissions to satellite \gls{ul} in terrestrial-\gls{leo} satellite coexistence scenario.}
  \label{fig:coexistence-top}
\end{figure}

However, despite the aforementioned contributions, prior work on spectrum sharing and coexistence has not thoroughly compared widely adopted interference suppression techniques, such as power control and spatial nulling. In this chapter, we aim to address this gap by considering the worst-case interference caused to satellites, which results from \gls{dl} transmissions in the terrestrial network leaking into the satellite \gls{ul}~\cite{kang2024terrestrial}, as shown in Fig.~\ref{fig:coexistence-top}. Additionally, contrary to prior works~\cite{wadaskar2025satellite, kang2024terrestrial}, which focus on a limited number of satellites (e.g., up to $10$ satellites within a constellation) sharing spectrum with terrestrial cellular networks, we consider deployments involving denser constellations. This allows us to evaluate the impact of interference suppression strategies on scalable coexistence scenarios, thereby pinpointing their strengths and limitations.
Our contributions can be summarized as follows:
\begin{itemize}
    \item We systematically investigate the impact of Interference Nulling on the trade-off between terrestrial \gls{qos} and interference suppression at \gls{leo} satellites.
    \item We formulate an optimization problem to suppress interference at the incumbents by minimizing the \gls{gnb}'s \gls{dl} transmit power, subject to throughput and \gls{inr} constraints.
    \item We provide a comparative analysis between power control strategies and interference nulling techniques, pinpointing the strengths and limitations of each method while identifying suggested use-cases.
    \item We propose a joint interference nulling and power control strategy and benchmark it against standalone baselines.
    \item We conduct our analysis in the context of dense satellite constellations, considering $40$ \gls{leo} satellites to ensure the practical applicability of the proposed solutions under challenging scenarios.
    \item We also evaluate the system's performance using the energy efficiency \gls{kpm}, a critical metric for designing \gls{6g} terrestrial networks.
\end{itemize}

The remainder of this chapter is organized as follows. Section~\ref{sec:ss:sysmodel} describes the system model, and Section~\ref{sec:ss:strategies} presents the spectrum sharing strategies. Section~\ref{sec:ss:setup} details the simulation setup, Section~\ref{sec:ss:kpms} introduces the \glspl{kpm} used for performance evaluation, and Section~\ref{sec:ss:results} discusses the simulation results. Finally, Section~\ref{sec:ss:conclusions} concludes the chapter and outlines directions for future work.

\section{System Model}\label{sec:ss:sysmodel}
We focus on a topology comprising a terrestrial \gls{gnb} serving $K$ \glspl{ue} in the \gls{dl}. The \gls{gnb} is equipped with an $N_{\text{az}} \times N_{\text{el}}$ uniform planar array with $N_t = N_{\text{az}} \cdot N_{\text{el}}$ transmit antennas, while each \gls{ue} employs an $N_{\text{az}}^{(R)} \times N_{\text{el}}^{(R)}$ array with $N_r = N_{\text{az}}^{(R)} \cdot N_{\text{el}}^{(R)}$ receive antennas. Both the \gls{gnb} and the \glspl{ue} employ beamforming, selecting transmit and receive beamforming vectors $\boldsymbol{w}_t$ and $\boldsymbol{w}_r$, respectively. It is noted that we employ time-division scheduling, where the \gls{gnb} serves one \gls{ue} per time resource and computes beamforming vectors $\boldsymbol{w}_t$ and $\boldsymbol{w}_r$ for the scheduled \gls{ue}. We additionally consider $N_{\text{sat}}$ \gls{leo} satellites visible from the \gls{gnb} location, sharing the same spectrum. Each satellite is assumed to have a beam focused on the same area covered by the terrestrial \gls{gnb}, making the satellite susceptible to interference from terrestrial transmissions. Finally, the \gls{mimo} channel matrix between the \gls{gnb} and \gls{ue} $k$ is denoted as $\boldsymbol{H}_{\text{ter}}^{k} \in \mathbb{C}^{N_r \times N_t}$, for each \gls{ue} $k \in \{1, \ldots, K\}$, while $\boldsymbol{h}_j \in \mathbb{C}^{N_t \times 1}$ denotes the \gls{mimo} channel vector from the \gls{gnb} to the $j$-th satellite, where $j = 1, \ldots, N_{\text{sat}}$. The time interval during which satellite trajectories are visible to the \gls{gnb} is given as $N_{\text{slots}}$. Therefore, at time slot $t_i$ and for user $k$, the terrestrial channel matrix is denoted as $\mathbf{H}_{\text{ter}}^{k}(t) \in \mathbb{C}^{N_r \times N_t}$.

\section{Terrestrial-Satellite Spectrum Sharing Strategies}\label{sec:ss:strategies}

We begin by comparing two state-of-the-art methods, namely Interference Nulling and \gls{qos}-Aware Power Control, to investigate their impact on terrestrial network \gls{qos}, while ensuring that the interference at the satellite remains below the \gls{itu} reference levels in satellite--terrestrial spectrum sharing scenarios. Finally, we propose and evaluate a new coexistence method that combines Interference Nulling and \gls{qos}-Aware Power Control. The performance of the proposed algorithm is systematically evaluated and compared against the standalone methods.

\subsection{Interference Nulling}\label{sec:ss:interf-null}

The first algorithm we consider is Interference Nulling~\cite{kang2024terrestrial}, where the beamformed gain at the terrestrial \gls{ue} is traded off against the interference leakage toward the satellites via a regularization parameter, $\lambda \geq 0$. The solutions to this optimization problem are the terrestrial transmit and receive beamforming vectors $\boldsymbol{w}_t^{\text{opt}}$ and $\boldsymbol{w}_r^{\text{opt}}$, respectively, as detailed in~\eqref{eq:nulling-opt}:

\begin{equation}\label{eq:nulling-opt}
\begin{aligned}
\boldsymbol{w}_r^{\text{opt}}, \boldsymbol{w}_t^{\text{opt}} &= \arg\max_{\boldsymbol{w}_r, \boldsymbol{w}_t} \left[ |\boldsymbol{w}_r^{\text{H}} \tilde{\boldsymbol{H}}_{\text{ter}} \boldsymbol{w}_t|^2 - \lambda \sum_{j=1}^{N_{\text{sat}}} \|\tilde{\boldsymbol{h}}_j^{\text{H}} \boldsymbol{w}_t\|^2 \right] \\
&\text{subject to\quad} \|\boldsymbol{w}_t\|^2 = 1 \quad \text{and} \quad \|\boldsymbol{w}_r\|^2 = 1.
\end{aligned}
\end{equation}

\noindent
$\tilde{\boldsymbol{H}}_{\text{ter}}$ and $\tilde{\boldsymbol{h}}_j$ denote the normalized versions of the \gls{mimo} multi-path channels $\boldsymbol{H}_{\text{ter}}$ and $\boldsymbol{h}_j$, respectively, as normalizing the channel matrices compensates for the substantial path loss difference between terrestrial and non-terrestrial (i.e., satellite) links, thereby maintaining $\lambda$ within a manageable range. To solve the optimization problem in~\eqref{eq:nulling-opt}, we follow the approach in~\cite{kang2024terrestrial}, where $\boldsymbol{w}_r^{\text{opt}}$ is selected as the maximum left singular vector of $\tilde{\boldsymbol{H}}_{\text{ter}}$, and $\boldsymbol{w}_t^{\text{opt}}$ is chosen as the maximum eigenvector of $\boldsymbol{M} = \tilde{\boldsymbol{H}}_{\text{ter}}^{\text{H}} \boldsymbol{w}_r \boldsymbol{w}_r^{\text{H}} \tilde{\boldsymbol{H}}_{\text{ter}} - \lambda \sum_{j=1}^{N_{\text{sat}}} \tilde{\boldsymbol{h}}_j \tilde{\boldsymbol{h}}_j^{\text{H}}$. It is noted that performing \gls{svd} on $\boldsymbol{H}_{\text{ter}}$ is equivalent to performing \gls{svd} on $\boldsymbol{M}$ for $\lambda = 0$, which is also equivalent to applying eigenvalue decomposition to $\boldsymbol{M}$ for $\lambda = 0$.
Therefore, all the aforementioned techniques yield the same solution for the \gls{gnb}'s transmit beamforming vector (i.e., $\boldsymbol{w}_t$) when only beamforming is applied, without accounting for signal leakage toward the satellites. This equivalence serves as a numerical verification of our implementation, confirming that the framework from~\cite{kang2024terrestrial} correctly reduces to standard \gls{svd}-based beamforming when $\lambda = 0$.

\subsection{\texorpdfstring{\gls{qos}}{QoS}-Aware Power Control}\label{sec:ss:qos-pc}
In this section, we propose and formulate a power control optimization problem that minimizes the \gls{gnb}'s transmit power while maintaining terrestrial \gls{qos}, in order to ensure satellite protection. We focus on the optimization of the \gls{dl} transmit power of a single \gls{gnb}, and we assume that power allocation decisions are made in a time-division manner, where, at each time slot, a single \gls{ue} is scheduled. The interference-plus-noise levels experienced in the terrestrial network due to nearby operating \glspl{gnb} are given in~\eqref{eq:terr-snir}:
\begin{equation}\label{eq:terr-snir}
{I_{i}^{\text{ter}}=\sum_{n \neq i} G_n P_n + \sigma^2},
\end{equation}

\noindent
where $\sigma^2$ is the \gls{awgn} power and $\sum_{n \neq i} G_n P_n$ is the sum of inter-cell interference power from all neighboring cell sites. The \gls{sinr} for each \gls{gnb}-\gls{ue} link is given in~\eqref{eq:sinr-ter}:

\begin{equation}\label{eq:sinr-ter}
\text{\gls{sinr}}_i^{\text{ter}} = \frac{P_i \left| \boldsymbol{w}_r\boldsymbol{H}_{\text{ter}} \boldsymbol{w}_t \right|^2}{I_{i}^{\text{ter}}},
\end{equation}

\noindent
while the achievable rate for bandwidth $W$ is given by the Shannon Capacity formula as follows in~\eqref{eq:rate}:

\begin{equation}\label{eq:rate}
R_i(P_i) = W \log_2\left(1 + \text{SINR}_i^{\text{ter}}\right).
\end{equation}

\textbf{\textit{\gls{qos}-Aware Utility Function.}} The utility function we consider in our optimization problem is given in~\eqref{eq:PC_obj} as:
\begin{equation}\label{eq:PC_obj}
U_i(P_i) = \frac{W \cdot \left(1 - e^{-\alpha \, \text{SINR}_i^{\text{ter}}}\right)^{M}}{P_i},
\end{equation}

\noindent
and is defined based on a network economics approach~\cite{tsiropoulou2015combined,10973151}, capturing the trade-off between the terrestrial network's \gls{qos} satisfaction and the transmission power invested by the \gls{gnb}. This formulation ensures that the \gls{gnb} selects transmission power levels that strike a balance between meeting \gls{qos} demands in terms of achieved terrestrial \gls{sinr} and power savings. In detail, due to the sigmoidal nature of the utility function numerator, small increases in transmission power initially improve the utility by enhancing the \gls{sinr}. However, as the \gls{sinr} increases, the utility function begins to saturate, meaning that further increases in transmission power yield diminishing returns in \gls{qos} satisfaction.
Beyond this saturation point, increasing power levels further provides little improvement in utility, effectively preventing unnecessary power expenditure. Consequently, the sigmoidal function drives power allocation toward an optimum where the desired \gls{qos} is achieved with less power consumption. Finally, $\alpha$ and $M$ are control parameters governing the slope of the utility function. To ensure that the utility function remains quasi-concave with an interior optimum over the feasible power range, these parameters are tuned based on the interference-plus-noise levels experienced in the terrestrial network~\cite{tsiropoulou2015combined}, as given in~\eqref{eq:terr-snir}.

For satellite $j$ and a \gls{gnb} employing transmit beamforming vector $\boldsymbol{w}_t$, the \gls{inr} is calculated as follows in~\eqref{eq:inr}:
\begin{equation}\label{eq:inr}
\text{INR}_j = P_{i} + 10 \log_{10}|\boldsymbol{w}_t^H \boldsymbol{h}_{j}|^2 + G/T - L_a - 10 \log_{10}(W\cdot k_B),
\end{equation}
\noindent
where $G/T$ is the satellite antenna gain-to-noise-temperature ratio, $L_a$ accounts for atmospheric and scintillation losses~\cite{3GPP38821}, and $k_B$ is the Boltzmann constant.

Finally, the optimization problem is ultimately formulated in~\eqref{eq:power_control_obj}--\eqref{eq:power_control_power}:
\begin{subequations}
\label{eq:power_control_optimization}
\begin{align}
\underset{P_i}{\text{maximize}} \quad & U_i(P_i)\label{eq:power_control_obj}  \\
\text{subject to} \quad & R_i(P_i) \geq \epsilon R_i^{\text{max}}, \label{eq:power_control_rate} \\
& \text{INR}_j(P_i) \leq \text{INR}_{\max}, \label{eq:power_control_inr} \\
& P_i^{\min} \leq P_i \leq P_i^{\max}, \quad \forall i \in I, \label{eq:power_control_power}
\end{align}
\end{subequations}

\noindent
and pertains to the selection of the power budget at the \gls{gnb} side that satisfies the operator's throughput constraints, while providing maximum interference reduction to the satellites, according to \gls{itu}'s requirements. Therefore, for each link $i \in I$, the \gls{gnb}-\gls{ue} link is allocated the full transmit power budget of the \gls{gnb}, denoted as $P_i$, and which satisfies the constraints defined in~\eqref{eq:power_control_rate}-\eqref{eq:power_control_power}. $U_i(P_i)$ is the \gls{gnb}'s utility function and $R_i(P_i)$ is the achievable rate for power level $P_i$.
\noindent
Finally, $P_{\min}$ and $P_{\max}$ are the minimum and maximum transmit power level of the \gls{gnb} respectively, $\text{INR}_{\max}$ is the maximum interference level leaked toward the satellite as permitted by the \gls{itu} (e.g., $-6$~dB~\cite{ITU-R-RS.2017-0}), $\epsilon \in (0,1]$ is the throughput threshold ratio (e.g., $\epsilon = 0.9$ for maintaining $90\%$ of the maximum theoretical throughput $R_i(P_i)$ attained at $P_{i}=P_{\max}$).

\subsection{Joint Interference Nulling and \texorpdfstring{\gls{qos}}{QoS}-Aware Power Control}\label{sec:ss:joint-opt}

\begin{figure}[H]
  \centering
  \includegraphics[width=0.7\textwidth, keepaspectratio]{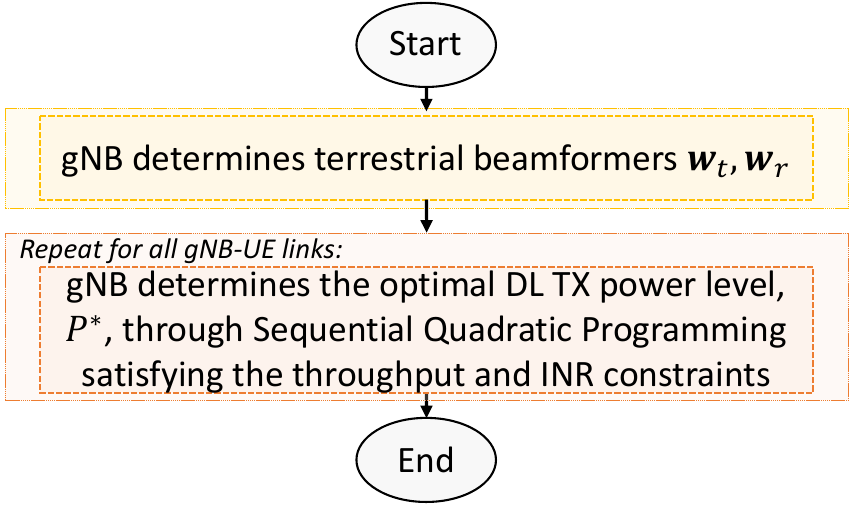}
  \caption{End-to-end resource management framework overview.}
  \label{fig:flowchart}
\end{figure}

The algorithm formulated in~\eqref{eq:nulling-opt} captures the trade-off between ensuring \gls{qos} for the terrestrial network and minimizing interference leakage toward the satellites through the regularization parameter $\lambda$, while the formulation in~\eqref{eq:power_control_obj}-\eqref{eq:power_control_power} ensures that only power budgets satisfying the throughput constraint are considered, and among those, the one providing maximum interference reduction to the satellite is ultimately selected. The formulation for the joint selection of the power level at the \gls{gnb} and the beamforming vectors is given in Algorithm~\ref{alg:joint-beamforming-power-control} and summarized in Fig.~\ref{fig:flowchart}.

\begin{algorithm}[H]
\caption{Joint Interference Nulling and \gls{qos}-Aware Power Control}
\label{alg:joint-beamforming-power-control}
\begin{algorithmic}[1]
\REQUIRE Channel matrix $\boldsymbol{H}_{\text{ter}}$ for scheduled \gls{ue} $k$, satellite channels $\{\boldsymbol{h}_j\}_{j=1}^{N_{\text{sat}}}$, regularization parameter $\lambda \geq 0$, utility parameters $\alpha$, $M$, throughput threshold $\epsilon$, INR cap $\text{INR}_{\max}$, power bounds $[P_{\min}, P_{\max}]$, interference-plus-noise $I_i^{\text{ter}}$, bandwidth $W$
\ENSURE Beamforming vectors $\boldsymbol{w}_t^{\text{opt}}$, $\boldsymbol{w}_r^{\text{opt}}$, and power $P_i^{\text{opt}}$
\STATE \textbf{Phase 1: Beamforming Vector Selection}
\STATE Normalize: $\tilde{\boldsymbol{H}}_{\text{ter}} \leftarrow \boldsymbol{H}_{\text{ter}} / \|\boldsymbol{H}_{\text{ter}}\|_F$, \\ $\tilde{\boldsymbol{h}}_j \leftarrow \boldsymbol{h}_j / \|\boldsymbol{h}_j\|_2$, $\forall j \in J$
\STATE Compute \gls{svd}: $\tilde{\boldsymbol{H}}_{\text{ter}} = \boldsymbol{U} \boldsymbol{\Sigma} \boldsymbol{V}^H$
\STATE Assign the optimal receive beamformer as the dominant left singular vector: $\boldsymbol{w}_r^{\text{opt}} \leftarrow \boldsymbol{u}_1$
\STATE Form $\boldsymbol{M} \leftarrow \tilde{\boldsymbol{H}}_{\text{ter}}^H \boldsymbol{w}_r^{\text{opt}} (\boldsymbol{w}_r^{\text{opt}})^H \tilde{\boldsymbol{H}}_{\text{ter}} - \lambda \sum_{j=1}^{N_{\text{sat}}} \tilde{\boldsymbol{h}}_j \tilde{\boldsymbol{h}}_j^H$
\STATE Compute eigendecomposition: $\boldsymbol{M} = \boldsymbol{Q} \boldsymbol{\Lambda} \boldsymbol{Q}^H$
\STATE Set the transmit beamformer to the principal eigenvector: $\boldsymbol{w}_t^{\text{opt}} \leftarrow \boldsymbol{q}_1$
\STATE \textbf{Phase 2: Power Control Optimization}
\STATE Compute the terrestrial channel gain: \\ $G_i^{\text{ter}} \leftarrow |(\boldsymbol{w}_r^{\text{opt}})^H \boldsymbol{H}_{\text{ter}} \boldsymbol{w}_t^{\text{opt}}|^2$
\STATE Compute $R_i^{\text{max}} \leftarrow W \log_2(1 + P_{\max} G_i^{\text{ter}} / I_i^{\text{ter}})$, \\ $R_i^{\min} \leftarrow \epsilon R_i^{\text{max}}$
\STATE Compute the satellite channel gain:\\ $G_j^{\text{sat}} \leftarrow |(\boldsymbol{w}_t^{\text{opt}})^H \boldsymbol{h}_j|^2$, $\forall j \in J$
\STATE Define the Utility function: \\ $f(P_i) \leftarrow -W (1-e^{-\alpha P_i G_i^{\text{ter}} / I_i^{\text{ter}}})^{M} / P_i$
\STATE Set Throughput constraint: \\ $c_1(P_i) \leftarrow R_i^{\min} - W \log_2(1 + P_i G_i^{\text{ter}} / I_i^{\text{ter}})$
\STATE Set INR constraint: \\ $c_2(P_i) \leftarrow (P_i G_j^{\text{sat}} / N_{\text{sat}}) - 10^{\text{INR}_{\max}/10}$, $\forall j \in J$
\STATE Solve $\min_{P_i} f(P_i)$ subject to $c_1(P_i) \leq 0$, $c_2(P_i) \leq 0$ and $P_{\min} \leq P_i \leq P_{\max}$ via \gls{sqp}
\RETURN $\boldsymbol{w}_t^{\text{opt}}$, $\boldsymbol{w}_r^{\text{opt}}$, $P_i^{\text{opt}}$
\end{algorithmic}
\end{algorithm}

\section{Simulation Setup}\label{sec:ss:setup}

\subsection{Terrestrial Channel Model}\label{sec:ss:ter-channelmod}

\begin{figure}[H]
\centering
\subfloat[Terrestrial \gls{los} Path Gain as given by QuaDRiGa's \texttt{power\_map} with the 3GPP\_38.901\_RMa\_LOS channel.]{\includegraphics[width=0.48\textwidth]{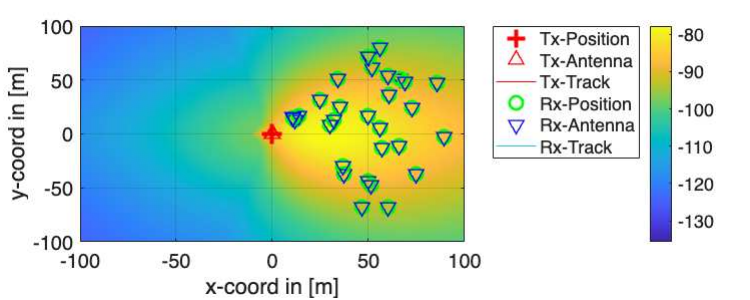}
\label{fig:ter-topology}}
\hfill
\subfloat[Complete topology overview of the terrestrial and \gls{ntn} network from \gls{gnb}'s perspective in \quadriga.]
{\includegraphics[width=0.48\textwidth]{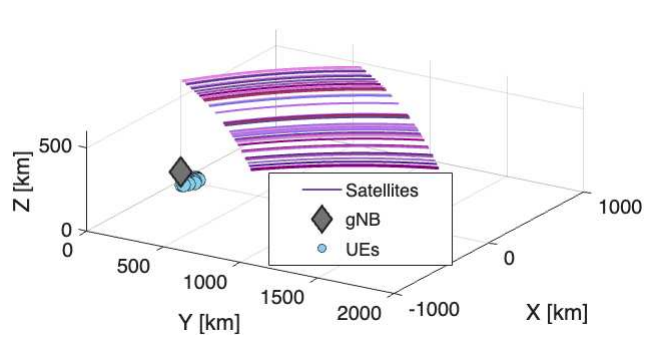}
\label{fig:full-topology}}
\caption{Simulation topology for terrestrial-satellite spectrum sharing.}
\label{fig:topology-combined}
\end{figure}

Without loss of generality, we assume that all \glspl{ue} are within the coverage of a single \gls{gnb} sector, as shown in Fig.~\ref{fig:ter-topology}, and that all satellites are \textit{visible} from the \gls{gnb}'s sector for most of their trajectory segments.
In order to create the terrestrial links, we leverage \gls{quadriga}~\cite{jaeckel2014quadriga}, a quasi-deterministic geometry-based stochastic channel model simulator compliant with 3GPP TR~38.901 specifications~\cite{3gpp38901}. A detailed tutorial on channel modeling with \quadriga can be found in~\cite{jaeckel2019f,10973151}, but full details are omitted in this chapter. For the creation of the terrestrial links, we use the parameters defined in Table~\ref{tab:sim_params}. It is noted that the geographical area of the Earth station considered in our simulation is located in central Spain, with geographical coordinates given as $38^\circ51'00.0''\text{N}, 5^\circ00'00.0''\text{W}$.

\subsection{\texorpdfstring{\gls{ntn}}{NTN} Channel Model}\label{sec:ss:ntn-channelmod}
\begin{figure}[H]
\centering
\subfloat[Horsham Township, PA ($C_{in}=95.4$, $D_{in}=-70.57$, $E_{in}=0$, $F_{in}=40.42$)]{\includegraphics[width=0.5\textwidth]{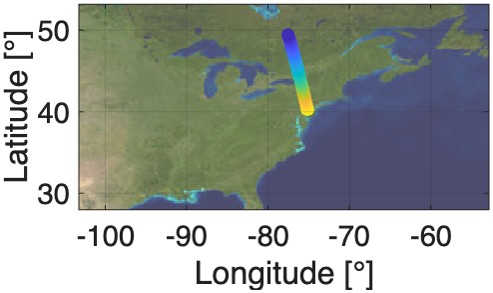}
\label{fig:penns}}
\hfill
\subfloat[Central Spain, EU ($C_{in}=63.4$, $D_{in}=-28.8$, $E_{in}=44.55$, $F_{in}=0$)]{\includegraphics[width=0.5\textwidth]{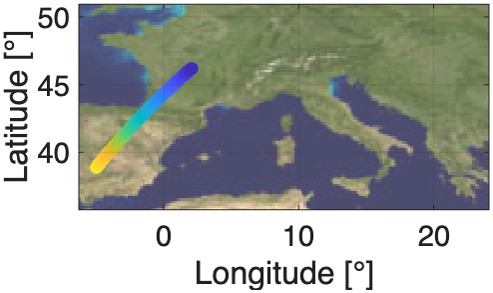}
\label{fig:spain}}
\caption{Satellite overview with \quadriga's \texttt{visualize\_lotlan} in various geographic regions. The trajectories shown correspond to different parameterizations of the orbital elements defined in Listing~\ref{lst:satellite}.}
\label{fig:locations}
\end{figure}

For the creation of the non-terrestrial links, we leverage \quadriga's satellite channel modeling approach as it has been discussed in~\cite{jaeckel20225g}. While a descriptive tutorial on satellite channel modeling is omitted, we provide an example of the configuration used in our work to define the satellite's trajectories as follows in Listing~\ref{lst:satellite}, while the configuration of the \gls{ntn} network is provided in Table~\ref{tab:sim_params}. In Fig.~\ref{fig:locations}, we show example satellite orbits created using
\quadriga's \texttt{qd\_satellite} constructor, initialized with different
Keplerian orbital elements as defined in Listing~\ref{lst:satellite}. In our work, we consider the satellite constellation corresponding to Fig.~\ref{fig:spain}. For each of the $N_{\text{sat}}$ satellites, and in order to ensure distinct trajectories, we vary their inclination (i.e., Cin of Listing~\ref{lst:satellite}) by an offset parameter~$\Delta$, as depicted in Fig.~\ref{fig:full-topology}. Finally, the satellite trajectory, given the initialized parameters, is determined using \quadriga's \texttt{orbit\_predictor}.

\begin{lstlisting}[float=htbp, floatplacement=htbp, language=Matlab,
basicstyle=\ttfamily\scriptsize,
linewidth=\textwidth,
numbers=left,
numbersep=5pt,
xleftmargin=7pt,
caption={LEO satellite constellation definition in QuaDRiGa.},
label={lst:satellite}]
h_qd_sat = qd_satellite(constellation, Ain, Bin, Cin, Din, Ein, Fin);
% Ain: Semimajor axis [km]
% Bin: Orbital eccentricity [0-1]
% Cin: Orbital inclination [deg]
% Din: Ascending node longitude [deg]
% Ein: Argument of periapsis [deg]
% Fin: True anomaly [deg]
\end{lstlisting}

\subsection{Considered Constellation}\label{sec:ss:constellation}
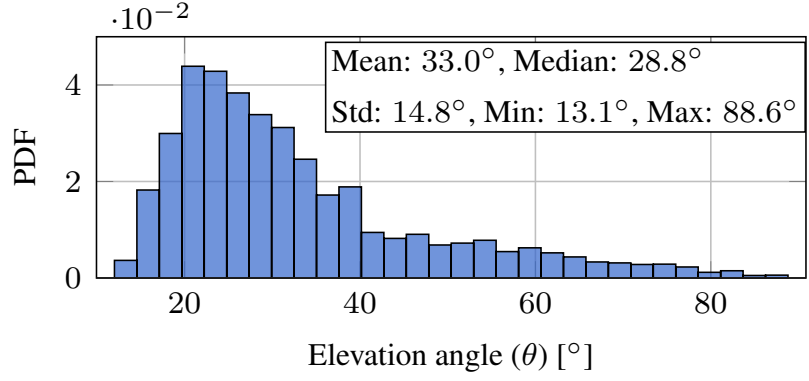
\begin{figure}[H]
    \centering
    \resizebox{0.7\textwidth}{!}{\input{fig/Elevation_Angle_PDF_gNB1_NoSats40.tex}}
    \caption{\gls{pdf} of the satellite elevation angles for the constellation of $N_{\text{sat}} = 40$ satellites considered in this work.}
    \label{fig:elevation-angle-pdf}
\end{figure}
To study the interference leakage from the \gls{gnb} toward the satellite \gls{ul}, we focus on lower elevation angles, as these correspond to scenarios with considerable terrestrial-to-satellite interference. Previous studies~\cite{li2002analytical,kang2024terrestrial} have shown that such angles may lead to increased interference due to the higher antenna gain of the \gls{gnb} in the direction of the satellite \gls{ul}. In addition, the probability of encountering satellites at lower elevation angles is higher, which represents a realistic use-case scenario.
The \gls{pdf} of the satellite elevation angles observed from the considered constellation is shown in Fig.~\ref{fig:elevation-angle-pdf}.

\begin{table}[H]
\caption{Simulation Parameters}
\centering
\label{tab:sim_params}
\small
\setlength{\tabcolsep}{4pt}
\begin{tabular}{|c|c|}
\hline
\multicolumn{2}{|c|}{\textbf{Terrestrial Network}} \\
\hline
Carrier frequency  & $7.125$ GHz   \\
\gls{ue} noise figure & $7$~dB \\
Cell radius \& Number of \glspl{ue} & $100$~m \& $30$ \\
Antenna array size & $8 \times 8$ (\gls{gnb}) \& $2 \times 1$ (\gls{ue}) \\
Antenna Configuration & \texttt{3gpp-3d}, $12^\circ$ downtilt (\gls{gnb})\\
Antenna height & $50$~m (\gls{gnb}) \& $1.6$~m (\gls{ue}) \\
Channel model & 3GPP 38.901 Rural Macro LOS \\
\hline
\multicolumn{2}{c}{} \\
\hline
\multicolumn{2}{|c|}{\textbf{Satellite Network}} \\
\hline
Number of satellites $(N_{\text{sat}})$  & $40$  \\
 $G/T$ & $13$~dB/K\\
Satellite Antenna Gain \& Altitude & $32$~dBi \& $600$~km\\
Satellite Antenna Configuration & \texttt{parabolic}, $0.25$~m aperture radius\\
Trajectory duration $(N_\text{slots})$ & $150$ (satellite visible to \gls{gnb} throughout)\\
Channel model & QuaDRiGa NTN Rural LOS \\
\hline
\multicolumn{2}{c}{} \\
\hline
\multicolumn{2}{|c|}{\textbf{Power Control Parameters}} \\
\hline
Bandwidth (W) & $30$ MHz \\
 $\text{INR}_\text{max}$ (target \gls{inr}) & $-6$~dB \\
Inter-cell Interference $(I_{i}^{ter})$ & $-73$~dBm \\
$\alpha$ \& $M$ (sigmoid utility parameters) & $10^{-3}$ \& $3$ \\
\gls{gnb} min \& max TX power  & $10$ \& $33$ dBm \\
Throughput thresholds ($\epsilon$) & $\epsilon \in \{0.85, 0.98\}$ \\
\hline
\end{tabular}
\end{table}

\section{Key Performance Evaluation Metrics}\label{sec:ss:kpms}
We introduce the \glspl{kpm} used
to evaluate the performance of the algorithms. Recall that the goal of this work is to suppress interference toward the satellites, while maintaining acceptable \gls{qos} in the terrestrial network.

We evaluate the \gls{inr} at the satellite, as defined in~\eqref{eq:inr}, to measure the effectiveness of the interference mitigation methods.
For the terrestrial network, we consider the following \glspl{kpm}.

\noindent\textbf{\textit{Terrestrial \gls{rss} degradation.}}
We define the terrestrial \gls{rss} loss at \gls{ue} $k$ as:
\begin{equation}
\label{eq:rss_degradation}
\Delta\textrm{RSS}_k
= 10\log_{10}\left(\frac{P_{\text{max}} \cdot |\boldsymbol{w}_k^{\mathrm{opt}H}\boldsymbol{H}_{\mathrm{ter}}\boldsymbol{w}_{t,\lambda=0}|^2}{\hat{P}_i \cdot |\boldsymbol{w}_k^{\mathrm{opt}H}\boldsymbol{H}_{\mathrm{ter}}\widehat{\boldsymbol{w}}_{t,\lambda}|^2}\right),
\end{equation}
\noindent
where $\boldsymbol{w}_k^{\mathrm{opt}}$ and $\widehat{\boldsymbol{w}}_{t,\lambda}$ are the \gls{ue} $k$'s and the \gls{gnb} beamforming vectors, respectively, obtained as outlined in Section~\ref{sec:ss:interf-null} with the corresponding $\lambda$ value.
$P_{\text{max}}$ and $\hat{P}_i<P_{\max}$ denote the maximum available and the selected transmit power of the \gls{gnb}, respectively.
The numerator of~\eqref{eq:rss_degradation} corresponds to the best performance case for the terrestrial network, i.e., when no interference mitigation measure is adopted, while the denominator quantifies the impact of the proposed methods (interference nulling and power control) on the terrestrial network \gls{qos}.
Specifically,
\begin{itemize}
\item when evaluating the interference nulling algorithm, without power control, we set $\hat{P}_i=P_{\text{max}}$
\item when evaluating the power control optimization, we consider $\lambda=0$, so $\hat{w}_{t,\lambda}=w_{t,\lambda=0}$ and~\eqref{eq:rss_degradation} reduces to $\Delta\textrm{RSS}_k
= 10\log_{10}\left(P_{\text{max}}/\hat{P}_i\right)$
\item when power control and interference nulling are jointly applied, $\hat{\boldsymbol{w}}_{t,\lambda}$ (for $\lambda>0$) is determined by interference nulling and $\hat{P}_i$ by the power control algorithm
\end{itemize}
$\tilde{\R_k}$ represents the mean \gls{rss} degradation for \gls{ue} $k$, averaged across all $N_\text{slots}$ time slots of the satellites' trajectories.

\textbf{\textit{\gls{jfi}.}} To quantify the fairness of \gls{rss} degradation across the terrestrial \glspl{ue}, we employ the \acrfull{jfi}~\cite{jain1984quantitative}.
For a set of $K$ \glspl{ue} with average \gls{rss} degradation $\{\tilde{\R_1}, \tilde{\R_2}, \ldots, \tilde{\R_K}\}$, the index is defined as:
\begin{equation}
J = \frac{\left(\sum_{k=1}^{K} \tilde{\R_k}\right)^2}{K \cdot \sum_{k=1}^{K} \tilde{\R_k}^2}.
\label{eq:jain_index}
\end{equation}
\noindent
The index ranges from $1/K$ (worst-case fairness, where degradation is concentrated on a single \gls{ue}) to $1$ (perfect fairness, where all \glspl{ue} experience identical degradation).
A value closer to unity indicates a more equitable distribution of \gls{rss} degradation.
This allows us to compare the fairness characteristics of different interference management methods, including interference nulling with various regularization parameters $\lambda$ and power control strategies.

\section{Performance Evaluation}\label{sec:ss:results}
In this section, we proceed by evaluating and comparing the optimization methods presented in Section~\ref{sec:ss:strategies}, identifying their key advantages and limitations, and investigating how they can complement each other.
It is noted that our simulation findings have been averaged over multiple repetitions. The parameters considered in this chapter for the system-level experimental analysis are given in Table~\ref{tab:sim_params}, while the optimization considers the scheduling decisions of a single \gls{gnb}, given the interference from nearby cells. Multi-cell scheduling interactions are not considered within the scope of this work. Finally, the constrained optimization problem defined in~\eqref{eq:power_control_obj}--\eqref{eq:power_control_power} was solved by leveraging MATLAB's~\cite{MATLAB:2025} \texttt{fmincon} with \gls{sqp}.

\subsection{Impact of Interference Nulling on Beamforming Gain}\label{sec:ss:eval1}

\begin{figure}[H]
\centering
\subfloat[]{\includegraphics[height=4.55cm, width=0.4\textwidth]{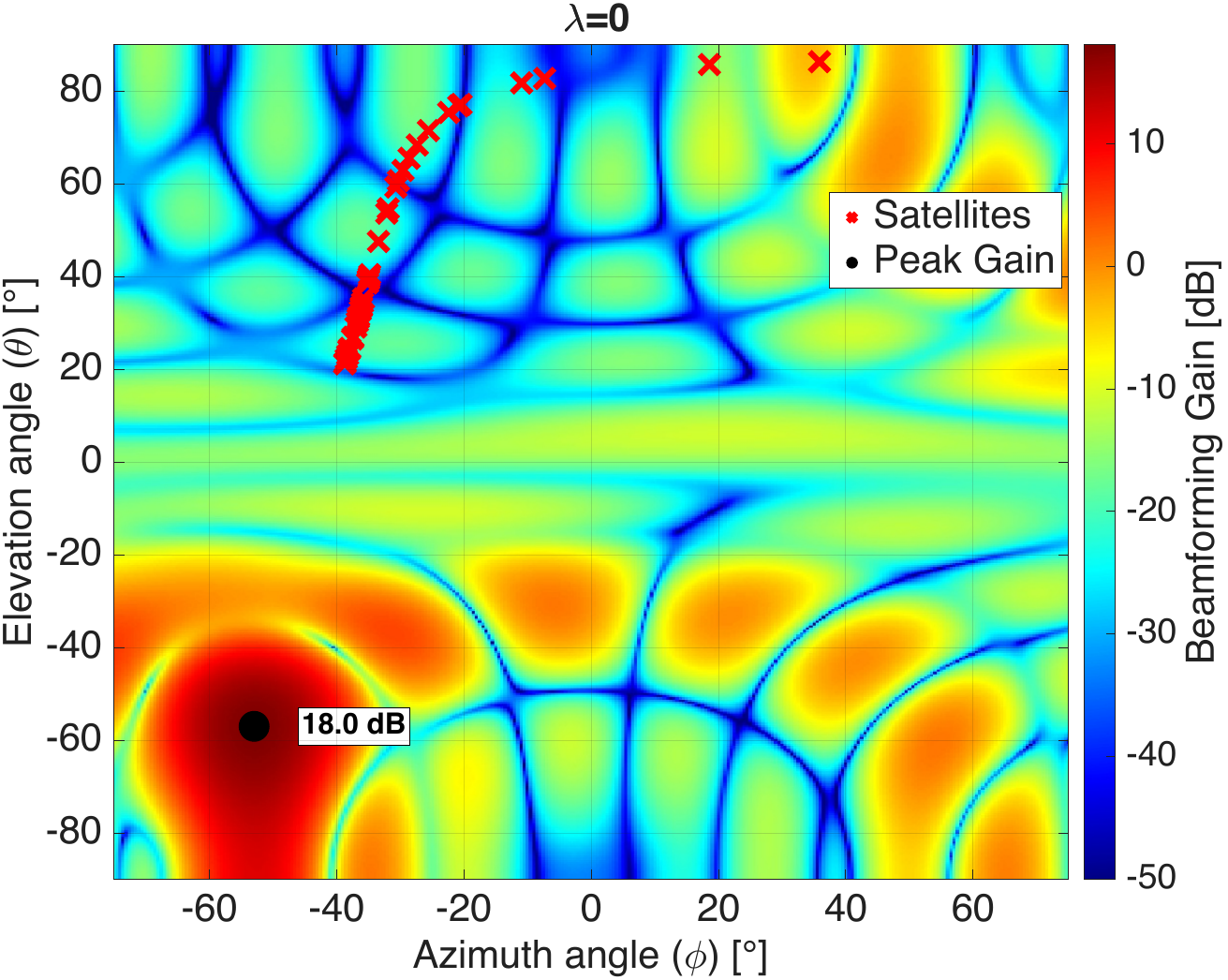}
\label{fig:bf-sim-start-0}}
% \hfill
\subfloat[]{\includegraphics[height=4.55cm, width=0.4\textwidth]{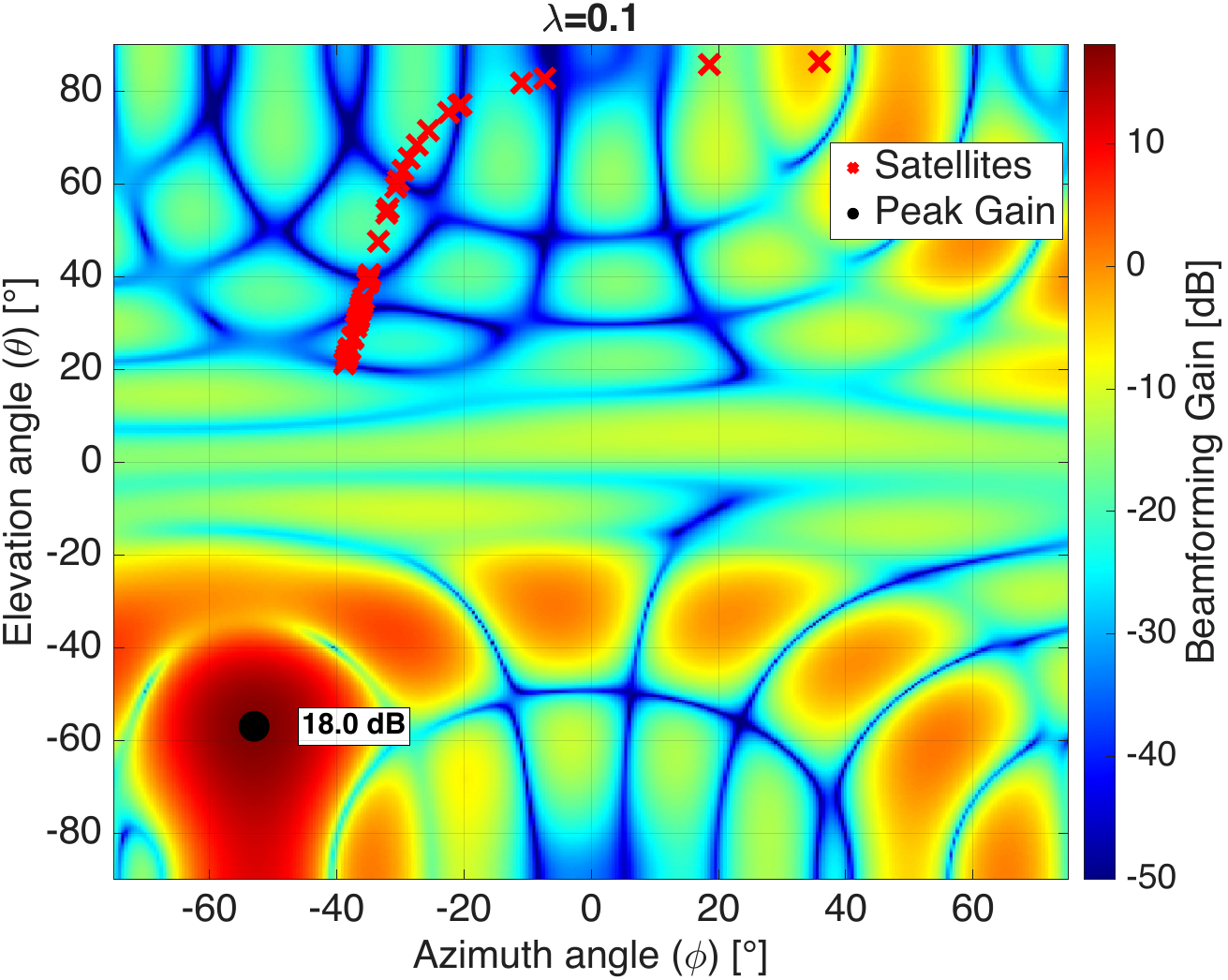}
\label{fig:bf-sim-start-01}}
\hfill
\subfloat[]{\includegraphics[height=4.55cm, width=0.4\textwidth]{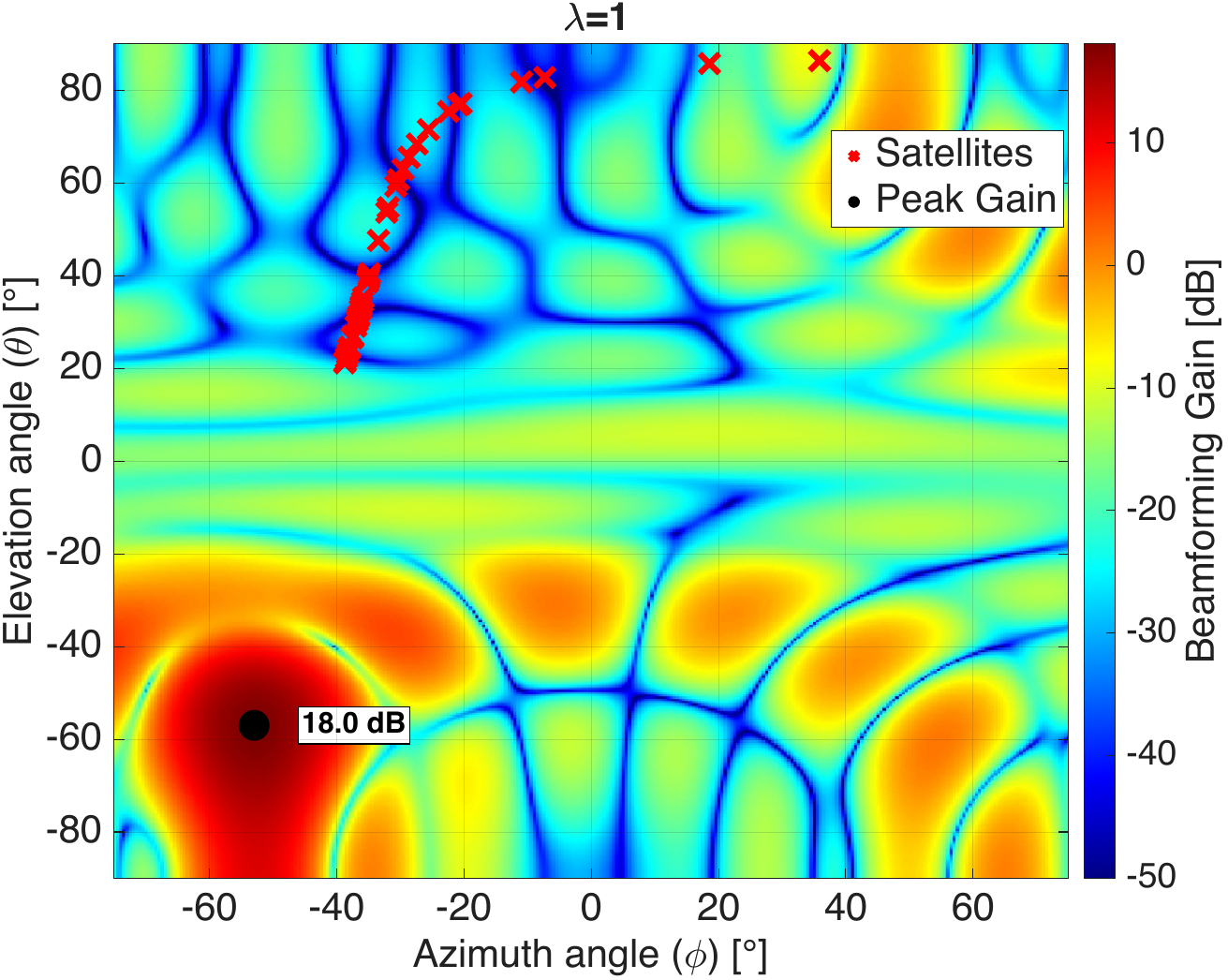}
\label{fig:bf-sim-start-1}}
% \hfill
\subfloat[]{\includegraphics[height=4.55cm, width=0.4\textwidth]{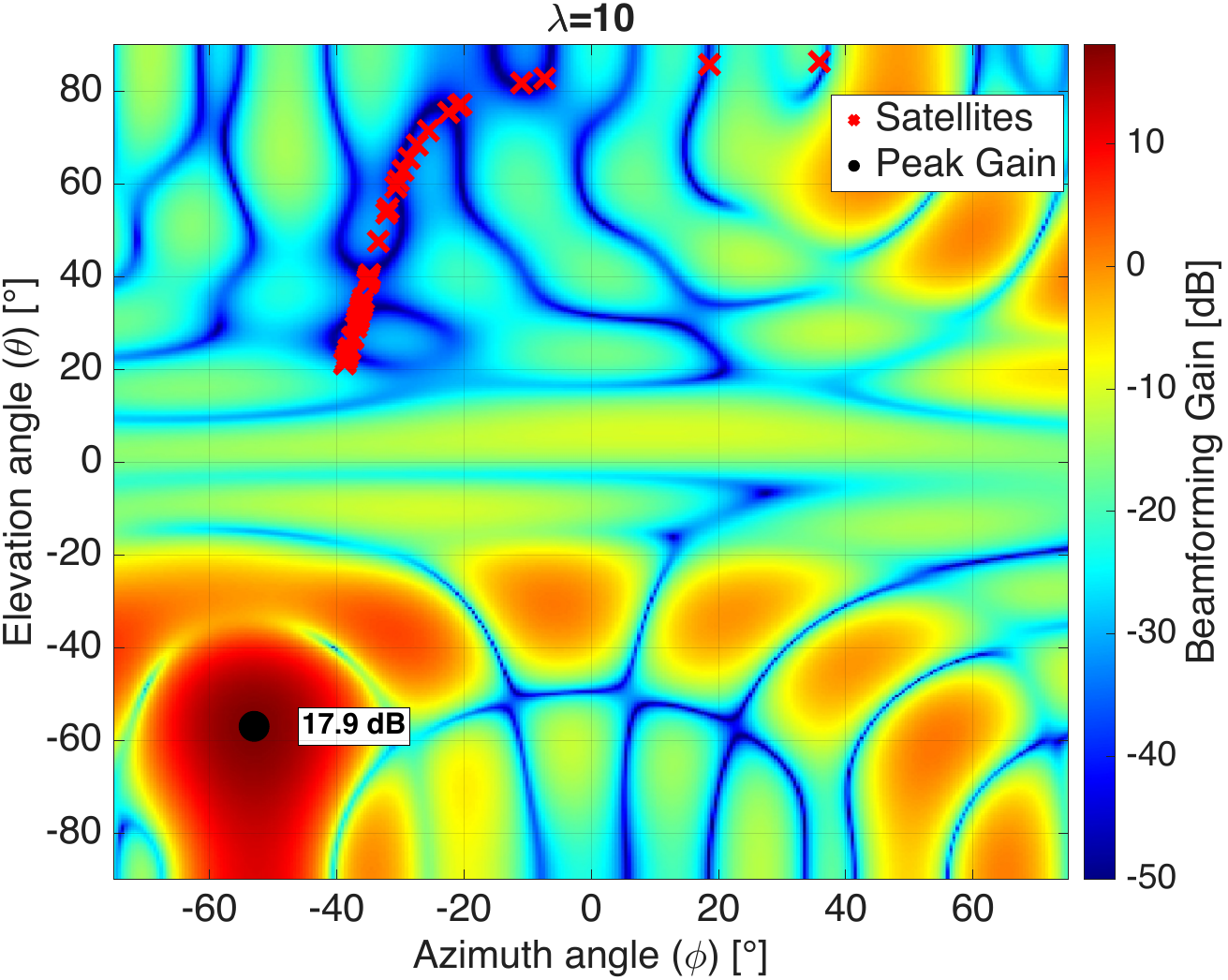}
\label{fig:bf-sim-start-10}}
\caption{Beamforming gain of the considered \gls{gnb} with a $8\times 8$ \gls{upa} and $\lambda\in\{0,0.1,1,10\}$ for time instance ($t_{l}$) along the satellites' trajectories.
As $\lambda$ increases, a greater priority is given to the interference nulling than to the terrestrial network \gls{qos}, leading to deeper nulls toward the satellites and lower gain in the \gls{ue} direction.
However, the degradation is minimal: $18$~dB at $\lambda = 0$ (Fig.~\ref{fig:bf-sim-start-0}), to $17.9$~dB at $\lambda = 10$ (Fig.~\ref{fig:bf-sim-start-10}), close to the theoretical maximum of $18.06$~dB.}
\label{bf-sim-start}
\end{figure}

We begin by investigating and visually interpreting the Interference Nulling algorithm, examining the beamforming gain patterns. To show a more intuitive correspondence between the nulls and lobes and the physical locations of the satellites and the \glspl{ue}, respectively, in this subsection we run the algorithm considering a simplified \gls{los}-only channel instead of the \quadriga channel matrix by replacing $\tilde{\boldsymbol{h}}_j$ with $\boldsymbol{e}(\theta_j, \phi_j)$, where $(\theta_j, \phi_j)$ are the elevation and azimuth angles to the $j$-th victim satellite, and $\boldsymbol{e}(\theta_j, \phi_j)$ is the spatial signature of the \gls{gnb} antenna array in that direction. In addition, since satellite trajectories can be derived from publicly available ephemeris data, this approach can reduce the real-time complexity associated with multi-path channel estimation, while still achieving interference suppression close to that of full multi-path nulling, as reported in~\cite{kang2024terrestrial}.

Fig.~\ref{bf-sim-start} illustrates the resulting transmit beamforming gains $|\boldsymbol{e}^H(\theta, \phi)\boldsymbol{w}_t|^2$ across the angular space of a \gls{gnb} equipped with an $8\times 8$ \gls{upa}, as obtained by solving the interference nulling algorithm in~\eqref{eq:nulling-opt} for different values of $\lambda$ and a given satellite configuration. Specifically, the $N_{\text{sat}} = 40$ satellite locations are marked with red crosses, while the main beam direction (i.e., the peak gain) toward the terrestrial \gls{ue} is indicated by a black circle. The results qualitatively show that nulls can be effectively placed at all satellite locations while maintaining a high beamforming gain toward the desired terrestrial \gls{ue}. As $\lambda$ increases from $0$ to $10$, interference nulling is prioritized over terrestrial \gls{qos}, and deeper nulls appear at the satellite locations, providing stronger interference suppression. This comes at a minimal cost to the terrestrial \gls{ue} gain, which decreases from $18$~dB at $\lambda = 0$ (Fig.~\ref{fig:bf-sim-start-0}) to $17.9$~dB at $\lambda = 10$ (Fig.~\ref{fig:bf-sim-start-10}). Finally, the beamforming gain of $17.9$~dB remains close to the theoretical maximum of $10 \log_{10}(64) = 18.06$~dB, indicating minimal impact on terrestrial link quality.

\subsection{\texorpdfstring{\gls{qos}}{QoS}-Aware Power Control for Risk-Averse Scenarios}\label{sec:ss:riskaverse}
As explained in Section~\ref{sec:ss:qos-pc}, the goal of the power control algorithm is to strike a balance between the interference toward the satellites and the terrestrial network \gls{qos}.

\begin{figure}[H]
\centering
\includegraphics[width=0.75\textwidth]{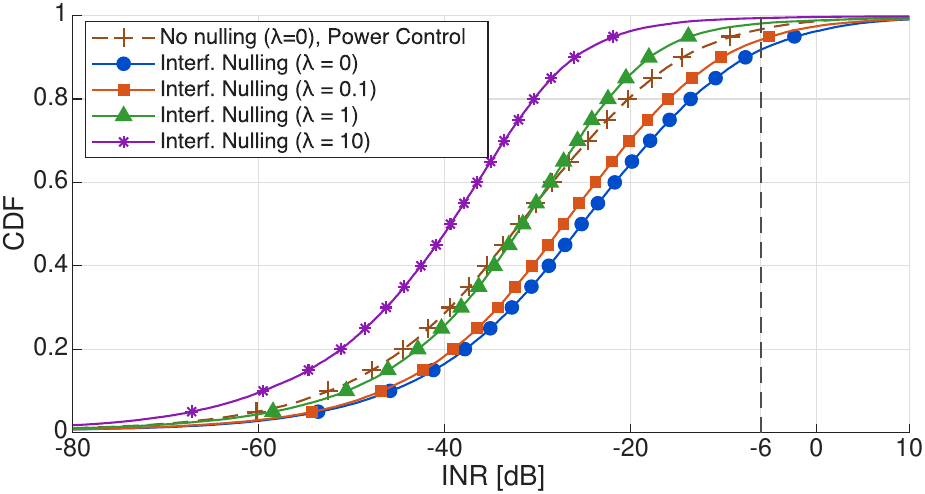}
\caption{\gls{cdf} of the \gls{inr} as defined in~\eqref{eq:inr}: Power Control \& Standard Beamforming vs. Interference Nulling.}
\label{fig:inr-cdf-pc-null}
\end{figure}

\begin{figure}[H]
    \centering
    \begin{minipage}{0.31\textwidth} % Reduced to 0.31
        \centering
        \input{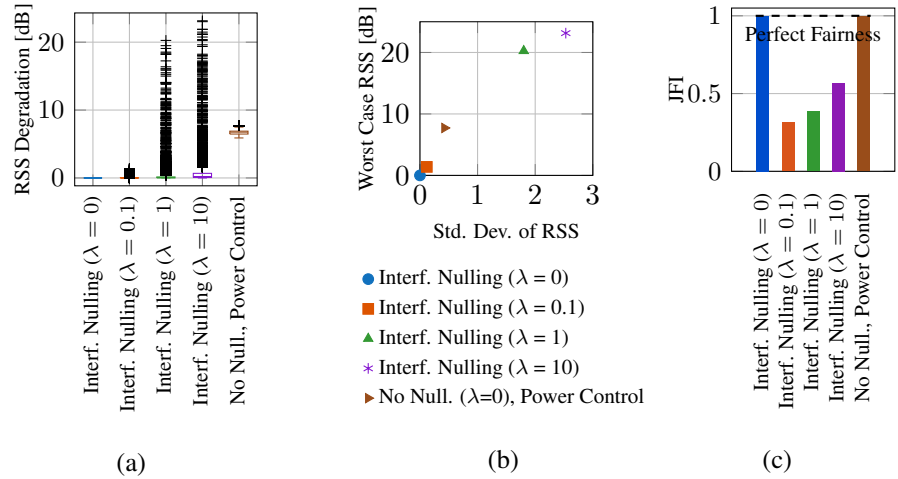}
        \centerline{\small (a)}
    \end{minipage}%
    \hspace{0.1cm} 
    \begin{minipage}{0.31\textwidth} % Reduced to 0.31
        \centering
        \input{fig/Fairness_Scatter_WorstCaseVsVariability_AllUEs30_NoSats40_gNB1.tex}
        \centerline{\small (b)}
    \end{minipage}%
    \hspace{-1.2cm}
    \begin{minipage}{0.31\textwidth} % Reduced to 0.31
        \centering
        \input{fig/Fairness_BarChart_JainsIndex_AllUEs30_NoSats40_gNB1.tex}
        \centerline{\small (c)}
    \end{minipage}
    \caption{Worst-case evaluation and fairness analysis for the terrestrial network topology described in Table~\ref{tab:sim_params}.}
    \label{fig:fairness}
\end{figure}

\noindent\textbf{\gls{inr} Reduction.}
In Figs.~\ref{fig:inr-cdf-pc-null} and~\ref{fig:fairness}, we compare the simulation results obtained using the algorithm defined in~\eqref{eq:nulling-opt} with those obtained through the power control formulation defined in~\eqref{eq:power_control_obj}--\eqref{eq:power_control_power}.
It is noted that the terms \textit{Interference Nulling ($\lambda=0$)} and \textit{No Nulling} can be used interchangeably.
They both refer to the case where the trade-off between terrestrial \gls{qos} maximization and interference suppression—represented by the tunable parameter $\lambda$—is not considered, as explained in Section~\ref{sec:ss:interf-null}. In Fig.~\ref{fig:inr-cdf-pc-null}, we observe the \gls{inr} at the satellites for the simulation setup described in Table~\ref{tab:sim_params} and a throughput threshold of $\epsilon=0.85$.
We observe that Interference Nulling with $\lambda=10$ achieves the most interference reduction toward the satellites with a median \gls{inr} of $\sim40$~dB, while Interference Nulling with $\lambda=0.1$ performs almost identically to the case with Interference Nulling with $\lambda=0$ (equivalent to No Nulling). Similarly, Interference Nulling with $\lambda=1$ yields similar performance to Power Control ($\lambda=0$).
Finally, all methods successfully meet the \gls{itu} constraint by keeping the \gls{inr} at the satellites below $-6$~dB on average, with Interference Nulling for $\lambda \in \{1, 10\}$ and the combination of No Nulling and Power Control performing best.

\noindent\textbf{\gls{rss} Degradation.} In Fig.~\ref{fig:fairness}, we analyze the \gls{ue} \gls{rss} degradation.
In Fig.~\ref{fig:fairness}(a), the simulation results indicate that although all Interference Nulling methods—regardless of the $\lambda$ parameter—achieve near-zero median \gls{rss} degradation on the terrestrial links, they exhibit extreme outliers, particularly as $\lambda$ increases. Conversely, Power Control combined with No Nulling results in a higher median \gls{rss} degradation, due to $P_i < P_{\rm max}$, but avoids the extreme outliers observed with $\lambda=1$ and $\lambda=10$. This approach maintains the worst-case degradation for all terrestrial links below $7$~dB, in contrast to the Interference Nulling methods, where the degradation can reach up to $\sim 20$~dB (Fig.~\ref{fig:fairness}(b)). This degradation indicates that \glspl{ue} at certain locations cannot be effectively served by the \gls{gnb} when using the interference nulling algorithm.
To quantify this effect, we resort to the \gls{jfi}, reported in Fig.~\ref{fig:fairness}(c). We observe that \textit{Perfect Fairness} is achieved only for Interference Nulling with $\lambda=0$, whether performed standalone or combined with Power Control. In the former case, No Nulling represents the scenario where no \gls{qos} compromise is performed on the terrestrial network, resulting in a \gls{jfi} equal to unity. Similarly, the \gls{jfi} for the combined method is also equal to unity, as all terrestrial links are degraded equally. This can be attributed to the fact that Power Control reduces the transmit power across all links. In contrast, Interference Nulling compromises the channel quality of each \gls{gnb}--\gls{ue} link individually, depending on its specific channel state. Consequently, certain \glspl{ue} experience more severe \gls{qos} degradation, particularly as $\lambda$ increases to values of $1$ and $10$. Also, as $\lambda$ increases, a greater number of links are affected and experience similar levels of \gls{rss} degradation. Consequently, the \gls{jfi} also increases with $\lambda$; for instance, $\lambda=10$ results in more links with comparable and high \gls{qos} degradation compared to the degradation observed when $\lambda=0.1$ is used in the optimization. Therefore, except for the case when a simple \gls{svd} is performed toward the \gls{ue} ($\lambda=0$), the \gls{jfi} of the interference nulling method remains below $0.6$.

The aforementioned simulation results indicate that Interference Nulling with $\lambda=\{1, 10\}$, as well as No Nulling combined with Power Control, can keep the \gls{inr} at the satellites well under the \gls{itu} thresholds on average, thereby successfully minimizing interference at the satellites. However, Power Control is more suitable for risk-averse scenarios, since it manages to maintain an equal degradation for all links while avoiding extreme outliers for specific \glspl{ue}. Such an approach is well-suited for tactical or public safety applications, where the goal is to guarantee a reliable link for all \glspl{ue}, whereas Interference Nulling maintains high overall performance at the cost of compromising certain links to achieve a high average.

\begin{figure}[H]
\centering
\subfloat[]{
    \includegraphics[width=0.3\textwidth]{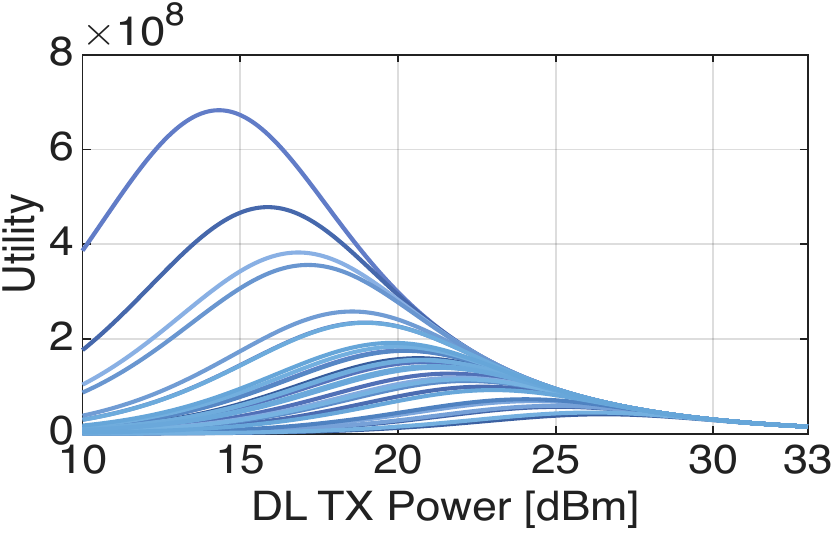}
    \label{fig:utility_all}
}
\hfill
\subfloat[]{
    \includegraphics[width=0.3\textwidth]{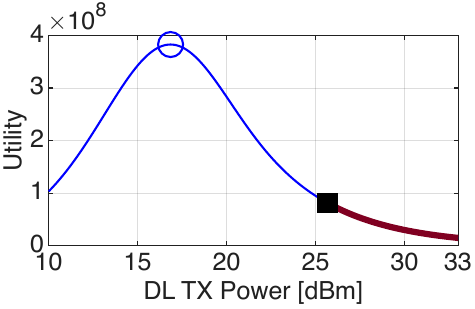}
    \label{fig:utility_ue11}
}
\hfill
\subfloat[]{
    \includegraphics[width=0.3\textwidth]{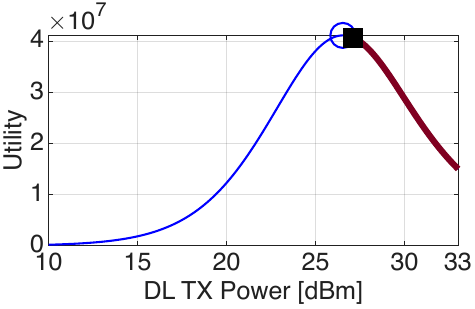}
    \label{fig:utility_ue9}
}
\caption{Utility function analysis for the terrestrial network topology described in Table~\ref{tab:sim_params}.}
\label{fig:utility}
\end{figure}

In Fig.~\ref{fig:utility}, the plots show the quasi-concavity, across the power levels considered, of the bell-shaped utility function formulated in~\eqref{eq:power_control_obj}. Fig.~\ref{fig:utility_all} showcases the optimal \gls{dl} transmit power chosen by the \gls{gnb} for each \gls{gnb}--\gls{ue} link, given the time-division-based scheduling decisions made by the \gls{gnb} scheduler for a power budget $[P_{\min}, P_{\max}] = [10, 33]$~dBm. Additionally, Fig.~\ref{fig:utility_all} displays the achieved utility per \gls{ue}, depicting the various individual \gls{qos} satisfaction levels. Indeed, the varying utility curves indicate that different power levels are selected for each active link, resulting in distinct utility peaks.
Figs.~\ref{fig:utility_ue11} and~\ref{fig:utility_ue9} demonstrate the selected power level versus the achieved utility at two distinct time instances. Both figures showcase, with a blue circle, the optimal power level that achieves the highest utility and \gls{qos} satisfaction (wrt. terrestrial \gls{sinr}) without unnecessary energy expenditures, as explained in detail in Section~\ref{sec:ss:qos-pc}. The shaded areas represent the power levels that satisfy the \gls{inr} and throughput constraints defined in~\eqref{eq:power_control_rate}--\eqref{eq:power_control_inr}, while the solid black square indicates the minimum power level that satisfies the throughput requirements while also meeting the \gls{itu} interference allowance constraints as selected by the power control formulation defined in~\eqref{eq:power_control_obj}--\eqref{eq:power_control_power}. It is noted that for the time instance represented by Fig.~\ref{fig:utility_ue9}, the optimal \gls{dl} transmit power level chosen by the \gls{gnb} closely matches the peak of the utility function. Conversely, for the time instance shown in Fig.~\ref{fig:utility_ue11}, the power level selected by the optimization is pushed to higher values to satisfy both the throughput and \gls{inr} constraints. Therefore, the utility function in this work inherently captures the trade-off between \gls{qos} satisfaction and unnecessary power expenditure. This allows the system to shift from energy-efficient operations to more \gls{qos}-driven regimes while still maintaining a balance in total energy consumption.

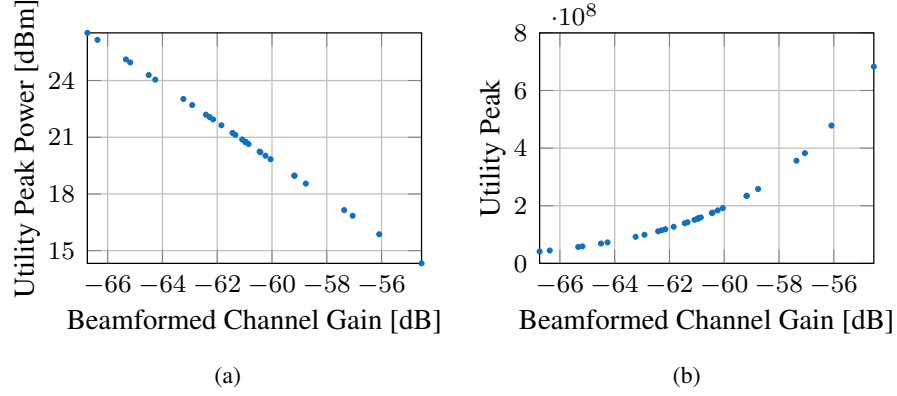
\begin{figure}[H]
\centering
\subfloat[]{
    \scalebox{1.2}{\input{fig/BellPeakPower_vs_BeamformedChannelGain.tex}}
    \label{fig:utility-peak}}
% \hfill
\subfloat[]{
    \scalebox{1.2}{\input{fig/UtilityPeak_vs_BeamformedChannelGain.tex}}
    \label{fig:bell-peak}}
\caption{Achieved utility and transmit power selection for the \gls{gnb} \gls{dl} under varying terrestrial channel conditions.}
\label{fig:utility-analysis}
\end{figure}

Fig.~\ref{fig:utility-analysis} shows the \gls{gnb}'s \gls{dl} transmit power and the associated achieved utility relative to the beamformed channel gain (i.e., $|\mathbf{w}_r^{\mathsf{H}} \tilde{\mathbf{H}}_{\text{ter}} \mathbf{w}_t|^2$) of the \gls{gnb}--\gls{ue} links.
These results are obtained for the case where only \gls{svd} is employed on $\mathbf{H}_{\text{ter}}$, which is equivalent to interference nulling with $\lambda=0$. We observe that as the beamformed channel gain improves, the selected transmit power for the \gls{gnb}--\gls{ue} link decreases (Fig.~\ref{fig:utility-peak}), while the achieved utility—and consequently the \gls{qos} satisfaction of the link—increases (Fig.~\ref{fig:bell-peak}). Therefore, the power control framework proposed in this work allows the \gls{gnb} to select lower transmit power levels for terrestrial links with higher channel gains on the \gls{dl}, thereby avoiding excess interference toward the incumbents.

\subsection{Joint \texorpdfstring{\gls{qos}}{QoS}-Aware Power Control and Interference Nulling}\label{sec:ss:combined-spectrum}
In this section, we present the simulation results when both Interference Nulling and Power Control are jointly considered and optimized to select the terrestrial beamforming weights and the \gls{gnb}'s \gls{dl} transmit power as summarized in Fig.~\ref{fig:flowchart}. We aim to explore how Interference Nulling with a less aggressive parameterization (i.e., lower $\lambda$ values) can be combined with Power Control to effectively minimize the \gls{inr} at the satellites while maintaining an acceptable terrestrial \gls{qos}.

\begin{figure}[H]
  \centering
  \includegraphics[width=0.75\textwidth]{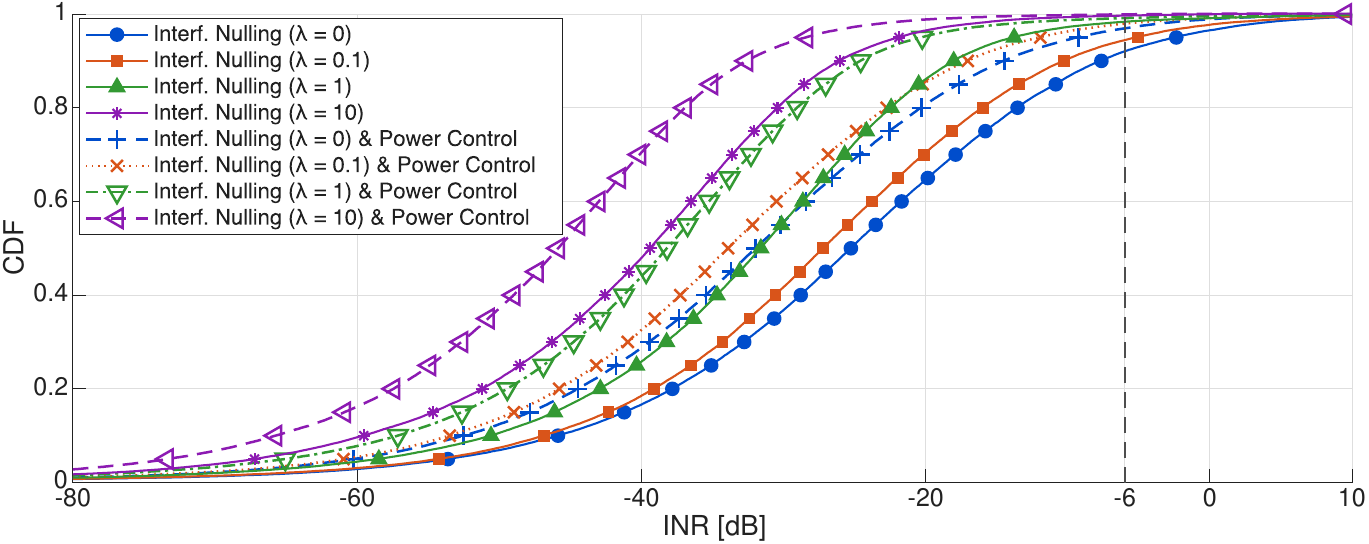}
  \caption{\gls{cdf} of the \gls{inr} as defined in~\eqref{eq:inr} for all optimization strategies.}
  \label{fig:inr-cdf-all}
\end{figure}

\noindent\textbf{\gls{inr} Reduction.} In Fig.~\ref{fig:inr-cdf-all}, we present the simulation results and the effect on the \gls{inr} at the satellites for all the methods considered in this work, namely Interference Nulling, No Nulling with Power Control, and the case of joint optimization through Interference Nulling and \gls{qos}-Aware Power Control. We observe that the most significant interference reduction towards the satellites is achieved through Interference Nulling with $\lambda=10$ combined with Power Control at a median \gls{inr} of $\sim46$~dB, while the least interference suppression is observed with Interference Nulling for $\lambda \in \{0, 0.1\}$. The most aggressive nulling technique observed with Interference Nulling ($\lambda=10$) yields similar performance to Interference Nulling with $\lambda=1$ combined with Power Control at a median \gls{inr} of $\sim38$~dB. Similarly, No Nulling combined with Power Control results in identical performance to Interference Nulling with $\lambda=1$. Therefore, when both power and the terrestrial beamformers are optimized jointly, the interference reduction at the satellites can be achieved with less aggressive nulling techniques (i.e., lower $\lambda$ values) given the proper adjustment of the \gls{gnb}'s \gls{dl} transmit power. Finally, \textit{Interf. Nulling ($\lambda=0.1$)} combined with Power Control provides a balanced trade-off, striking a middle ground between the less and more aggressive interference nulling techniques, by successfully keeping the interference leakage towards the satellites below the \gls{itu} interference levels.

\begin{figure}[H]
  \centering
  \includegraphics[width=0.5\textwidth]{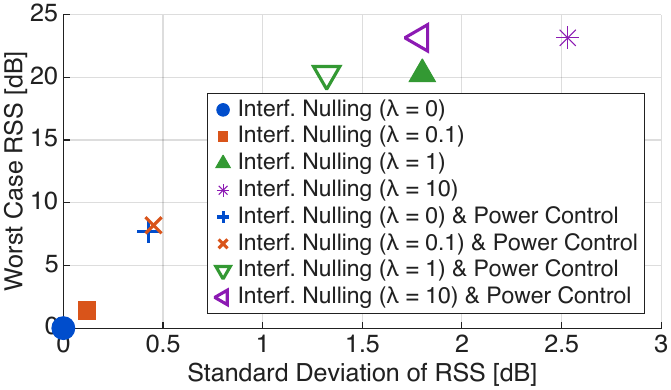}
  \caption{Worst-case \gls{rss} degradation as defined in~\eqref{eq:rss_degradation} for all optimization strategies.}
  \label{fig:worst-case-all}
\end{figure}

\noindent\textbf{\gls{rss} Degradation.} In Fig.~\ref{fig:worst-case-all}, we present the worst-case analysis results for all the methods. It is observed that Interference Nulling with $\lambda \in \{1, 10\}$, both standalone and when combined with Power Control, always results in the worst \gls{rss} values reported in our work; however, the joint optimization of the beamformers and the power results in more uniform \gls{rss} degradation across the terrestrial links (e.g., lower standard deviation). Additionally, Interference Nulling with $\lambda \in \{0, 0.1\}$ results in minimal, near-zero \gls{rss} degradation. Finally, No Nulling combined with Power Control achieves the same worst-case \gls{rss} as Interference Nulling with $\lambda=0.1$ also combined with Power Control, with the latter, however, achieving the best interference reduction among the two (Fig.~\ref{fig:inr-cdf-all}).

\subsection{Combined \texorpdfstring{\gls{qos}}{QoS}-Aware Power Control and Interference Nulling for Energy Efficiency Optimization}\label{sec:ss:combined-ee}

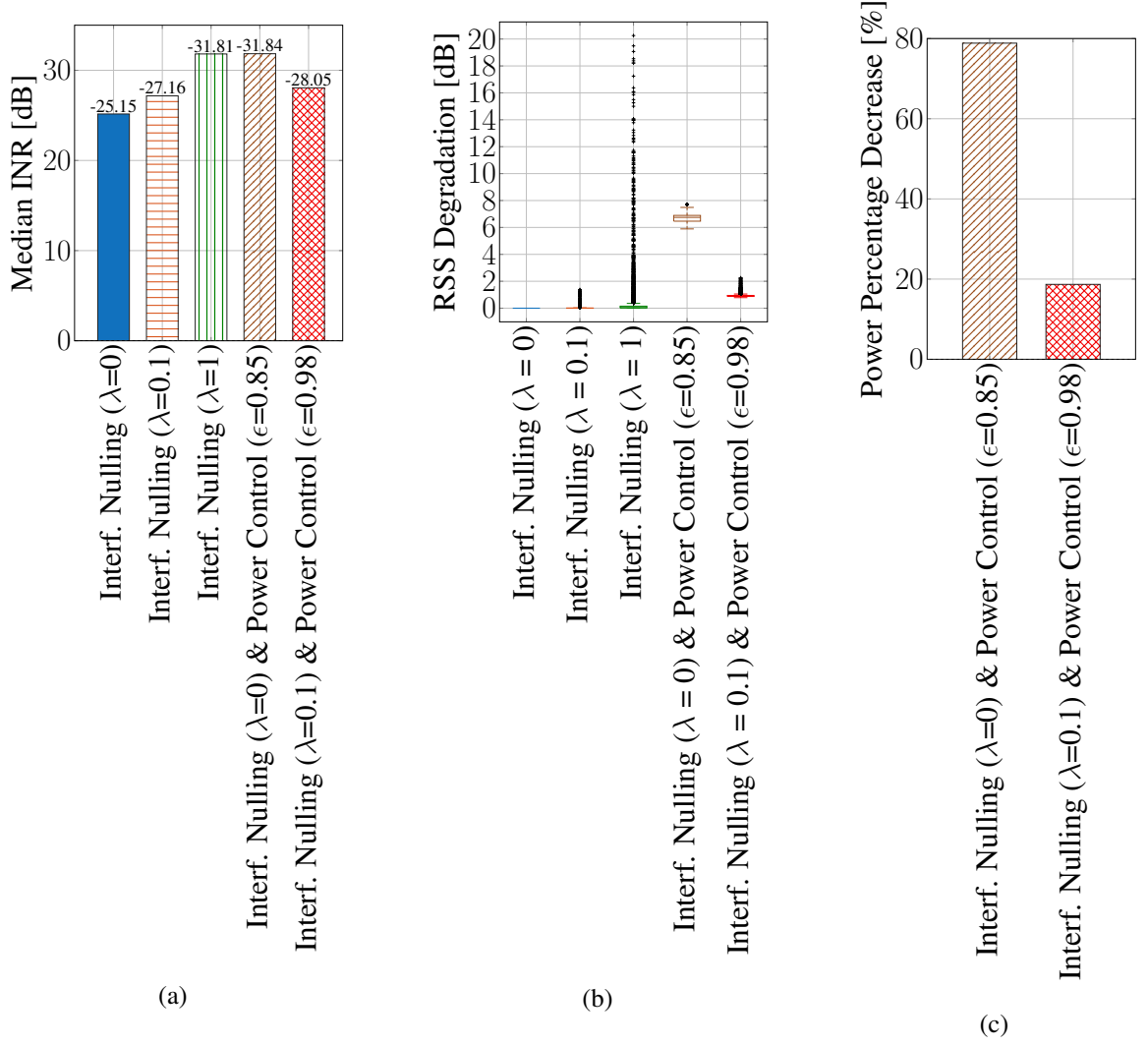
\begin{figure}[H]
    \centering
    \begin{minipage}{0.3\textwidth}
        \centering
        \resizebox{\linewidth}{!}{\input{fig/Combined_INR_Median_BarPlot.tex}}
        \centerline{\small (a)}
    \end{minipage}%
    \hfill
    \begin{minipage}{0.3\textwidth}
        \centering
        \resizebox{\linewidth}{!}{\input{fig/Combined_RSS_Degradation_BoxPlot.tex}}
        \centerline{\small (b)}
    \end{minipage}%
    \hfill
    \begin{minipage}{0.25\textwidth}
        \centering
        \resizebox{\linewidth}{!}{\input{fig/Combined_Power_PercentDecrease_BarPlot.tex}}
        \centerline{\small (c)}
    \end{minipage}
    \caption{Performance evaluation of joint interference nulling and power control: (a) \gls{inr}, (b) \gls{rss} degradation, (c) power savings.}
    \label{fig:combined-plots}
\end{figure}

\begin{figure}[ht]
    \centering
    % --- First Row: Two larger figures ---
    \begin{minipage}{0.5\textwidth}
        \centering
        \resizebox{\linewidth}{!}{\input{fig/lambda1_rss_degradation_cdf_NoSats2.tex}}
        \centerline{\small (a) $N_{\text{sat}} = 2$}
    \end{minipage}%
    \hfill
    \begin{minipage}{0.5\textwidth}
        \centering
        \resizebox{\linewidth}{!}{\input{fig/lambda1_rss_degradation_cdf_NoSats10.tex}}
        \centerline{\small (b) $N_{\text{sat}} = 10$}
    \end{minipage}

    \vspace{15pt} % Vertical gap between the rows

    % --- Second Row: Third figure centered below ---
    \begin{minipage}{0.5\textwidth}
        \centering
        \resizebox{\linewidth}{!}{\input{fig/lambda1_rss_degradation_cdf_NoSats40.tex}}
        \centerline{\small (c) $N_{\text{sat}} = 40$}
    \end{minipage}

    \caption{\gls{cdf} of \gls{rss} degradation for $\lambda = 1$ with varying number of satellites.}
    \label{fig:rss-degradation-satellites}
\end{figure}
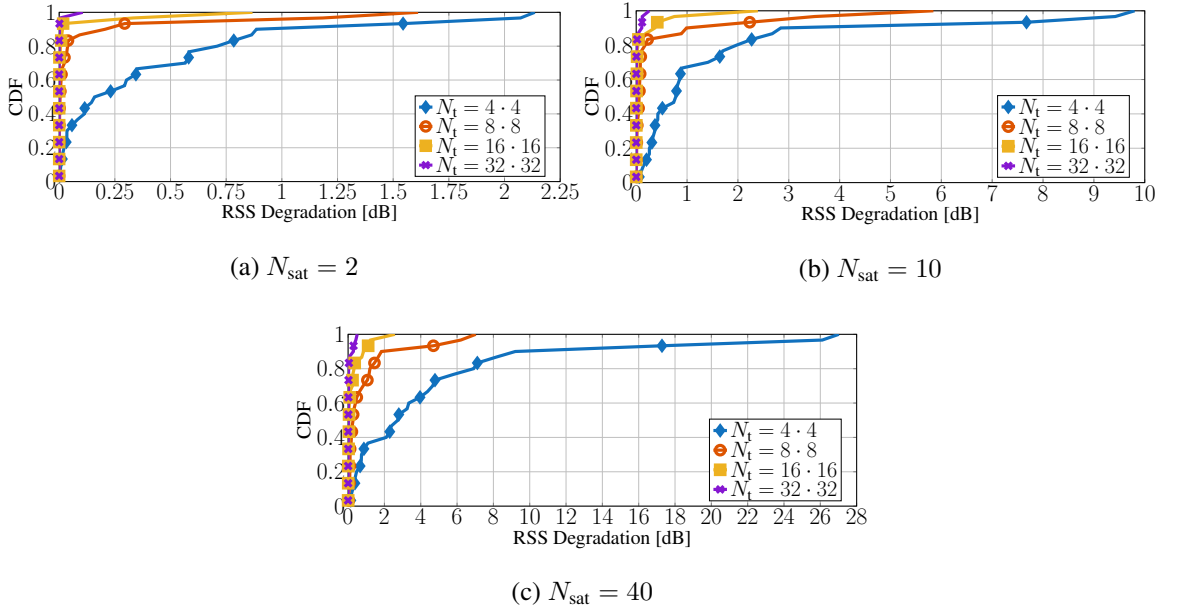

\begin{figure}[H]
    \centering
    % --- First Row ---
    \begin{minipage}{0.48\textwidth}
        \centering
        \resizebox{\linewidth}{!}{\input{fig/Fig7_AngularPositions_gNB1_NoSats40_Nt4x4_lambda1_snap1.tex}}
        \vspace{2pt}
        \centerline{\small (a) $4 \times 4$ array}
    \end{minipage}%
    \hfill
    \begin{minipage}{0.48\textwidth}
        \centering
        \resizebox{\linewidth}{!}{\input{fig/Fig7_AngularPositions_gNB1_NoSats40_Nt8x8_lambda1_snap1.tex}}
        \vspace{2pt}
        \centerline{\small (b) $8 \times 8$ array}
    \end{minipage}

    \vspace{10pt} % Adds vertical space between the two rows

    % --- Second Row ---
    \begin{minipage}{0.48\textwidth}
        \centering
        \resizebox{\linewidth}{!}{\input{fig/Fig7_AngularPositions_gNB1_NoSats40_Nt16x16_lambda1_snap1}}
        \vspace{2pt}
        \centerline{\small (c) $16 \times 16$ array}
    \end{minipage}%
    \hfill
    \begin{minipage}{0.48\textwidth}
        \centering
        \resizebox{\linewidth}{!}{\input{fig/Fig7_AngularPositions_gNB1_NoSats40_Nt32x32_lambda1_snap1}}
        \vspace{2pt}
        \centerline{\small (d) $32 \times 32$ array}
    \end{minipage}

    \caption{Angular positions for the complete topology for different antenna sizes with $N_{\text{sat}} = 40$ satellites and Interference Nulling with $\lambda=1$.}
    \label{fig:angular-total_plot}
\end{figure}
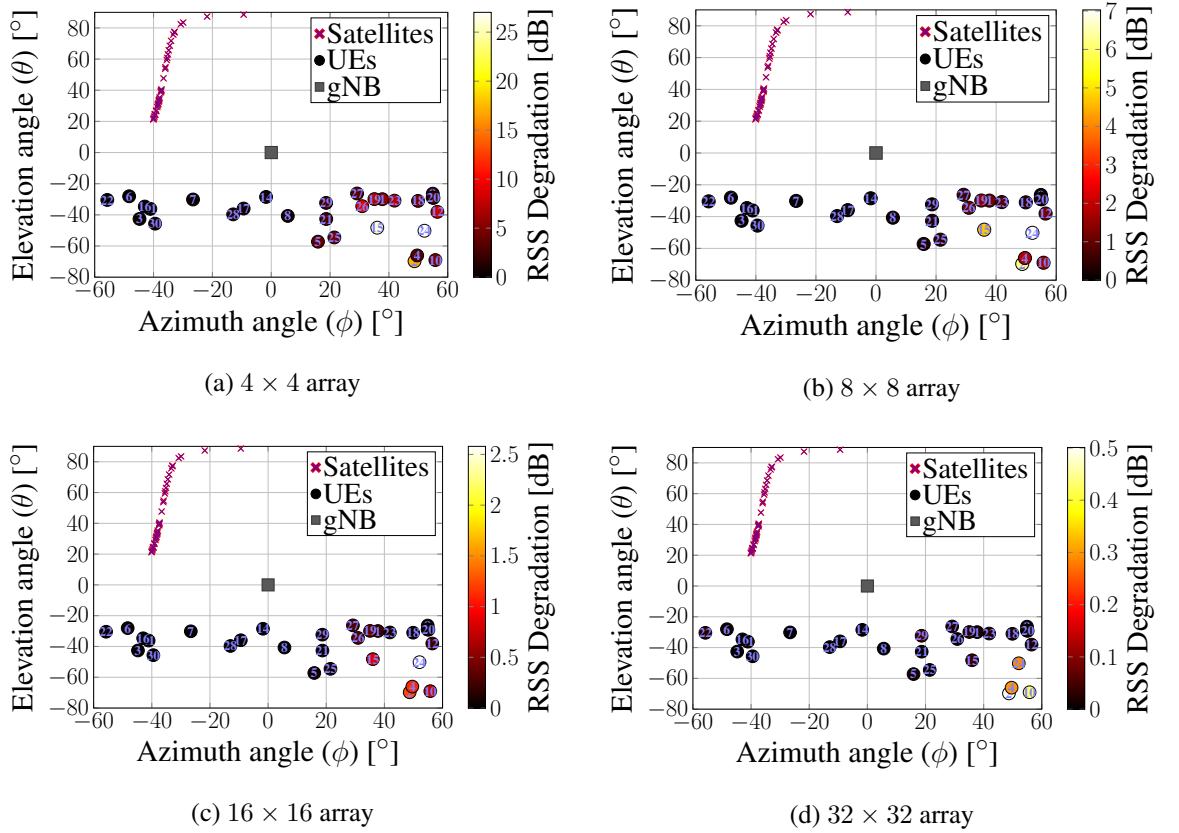

\begin{figure}[H]
  \centering
  \subfloat[Avg. Satellite Channel Gain]{
    \includegraphics[width=0.99\textwidth]{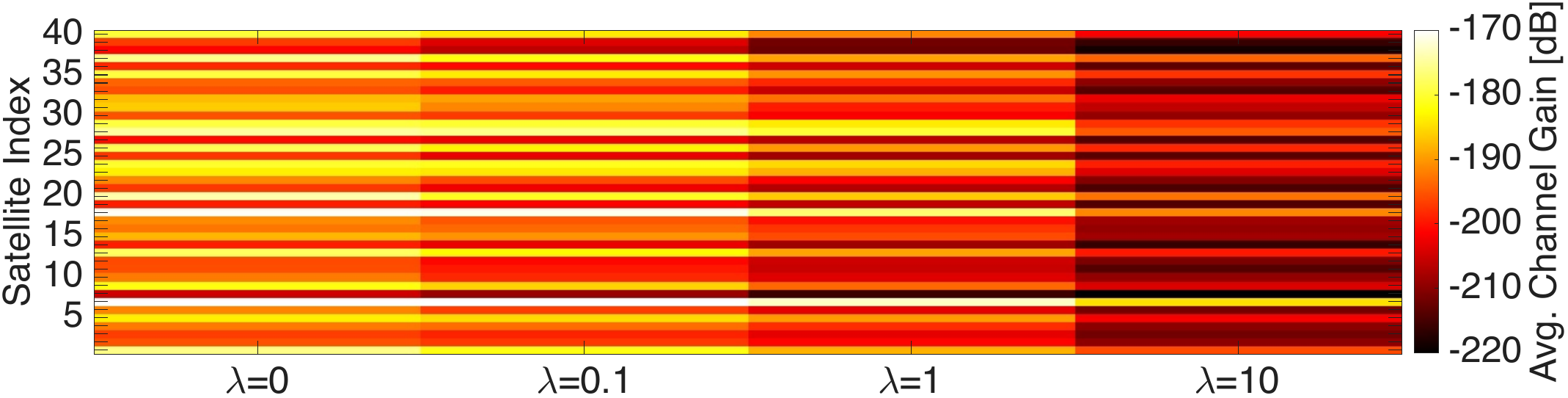}
    \label{fig:channel_strength}
  }\\
  \subfloat[Terrestrial \gls{ue} \gls{rss} Degradation]{
    \includegraphics[width=0.99\textwidth]{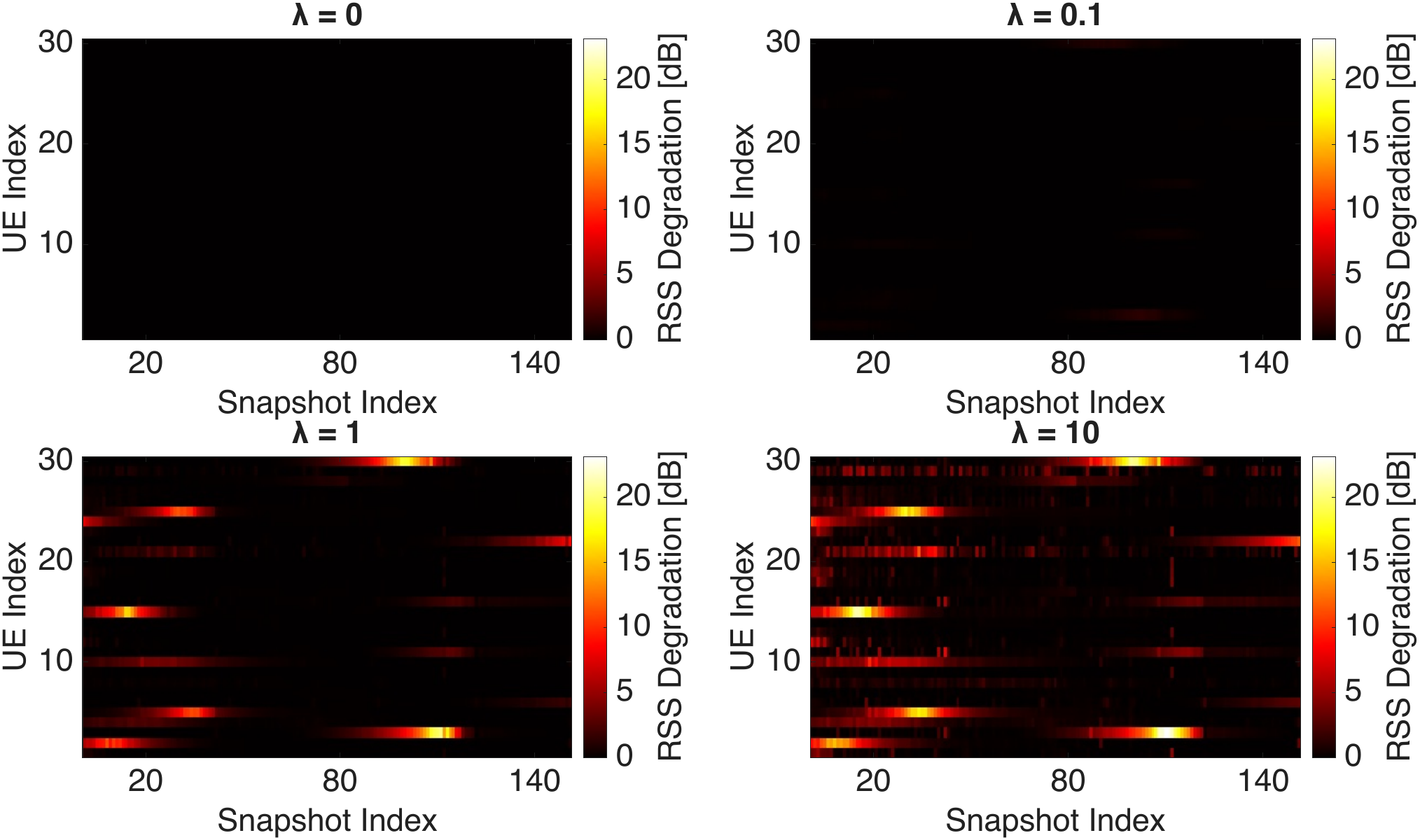}
    \label{fig:heatmap_deg}
  }
 \caption{Channel analysis: (a) Average signal leakage towards all satellites, and (b) \gls{ue} \gls{rss} degradation over satellite movement as defined in~\eqref{eq:rss_degradation} for all Interference Nulling techniques.}
  \label{fig:channel_analysis}
\end{figure}

\noindent\textbf{\gls{inr} Reduction, \gls{rss} Degradation and Power Savings.} In Fig.~\ref{fig:combined-plots}, we present the simulation results obtained when both Interference Nulling and Power Control are applied. We focus on the case with $\lambda = 1$, which was previously identified as providing the best trade-off between interference mitigation and acceptable \gls{qos} in the terrestrial network when only \textit{Interference Nulling} was considered. We additionally consider two different throughput thresholds (i.e., $\epsilon=\{0.85, 0.98\}$), as given in Table~\ref{tab:sim_params}, which represent different thresholds for the \gls{qos} requirements in the terrestrial network, as explained in Section~\ref{sec:ss:qos-pc}. In Fig.~\ref{fig:combined-plots}(a), we observe that Interference Nulling with $\lambda=1$ yields identical performance to Interference Nulling with $\lambda=0$ when power control is applied with $\epsilon=0.85$, providing an $\sim 7$~dB decrease in the median interference leakage at the satellite compared to the baseline (i.e., only Interference Nulling with $\lambda = 0$). Additionally, Interference Nulling with $\lambda = 0.1$ combined with power control with $\epsilon = 0.98$ (i.e., only a $2\%$ loss in the \gls{dl} throughput) manages to reduce the \gls{inr} at the satellites by approximately $3$~dB, compared to Interference Nulling alone (i.e., $\lambda = 0.1$),
for which the corresponding interference reduction compared to the baseline (i.e., $\lambda = 0$) is about $2$~dB.
Likewise, as Fig.~\ref{fig:combined-plots}(b) demonstrates, Interference Nulling combined with power control manages to balance the degradation in the \gls{qos} across the \glspl{ue}. In particular, the worst \gls{rss} degradation is reported to be less than $8$~dB when power control with $\epsilon=0.85$ is considered, compared to worst-case values of about $20$~dB obtained with Interference Nulling with $\lambda=1$. In contrast, power control with $\epsilon = 0.98$ combined with Interference Nulling with $\lambda=0.1$ manages to maintain an almost uniform degradation of approximately $1$~dB across all \glspl{ue}, with the worst \gls{rss} degradation reported at $\sim2$~dB. Therefore, power control with a dedicated throughput threshold can be combined with lower $\lambda$ values to balance the \gls{qos} degradation among all \glspl{ue}, while suppressing the interference leakage toward the satellites and additionally providing energy savings (Fig.~\ref{fig:combined-plots}(c)). Indeed, Fig.~\ref{fig:combined-plots}(c) demonstrates the energy savings achieved with both of the aforementioned power control techniques. Intuitively, more aggressive power control (i.e., $\epsilon = 0.85$) manages to save up to $80\%$ of the energy at the \gls{gnb}, at the cost of a $6.6$~dB median degradation across all terrestrial \glspl{ue}. Interference Nulling with $\lambda = 0.1$ combined with less aggressive power control ($\epsilon = 0.98$), on the other hand, also manages to save up to $20\%$ of the energy at the terrestrial \gls{gnb}, providing the best trade-off among all considered techniques in terms of terrestrial \gls{qos} and interference suppression.

\subsection{Case for Power Control: Limitations of Interference Nulling}\label{sec:ss:nulling-limitations}

In Fig.~\ref{fig:rss-degradation-satellites}, we present the \gls{cdf} of the \gls{rss} degradation for various antenna array sizes and number of satellites, focusing on the Interference Nulling technique with $\lambda = 1$. We observe that for a small constellation size (e.g., $N_{\text{sat}} = 2$ in Fig.~\ref{fig:rss-degradation-satellites}(a)), Interference Nulling can effectively keep the terrestrial \gls{qos} degradation low for all considered antenna sizes. Specifically, when larger antenna arrays are employed (i.e., $N_t \in \{256, 1024\}$), the degradation is maintained below $1$ and $0.25$ dB, respectively. However, as the number of satellites increases to $10$ and $40$ (Figs.~\ref{fig:rss-degradation-satellites}(b) and \ref{fig:rss-degradation-satellites}(c)), larger antenna arrays are required to keep the \gls{rss} degradation at low levels. For instance, in the case of $40$ satellites, an array size of $N_t = 1024$ elements is necessary to maintain the degradation below $1$ dB (Fig.~\ref{fig:rss-degradation-satellites}(c)). The simulation results demonstrate that antenna arrays with a higher number of elements yield more numerous and precise beams. This improved terrestrial performance aligns with the fundamental relationship \gls{dof} $= N_t - 1$, wherein increased element count provides additional \gls{dof} for null steering without compromising main lobe gain~\cite{balanis2016antenna,van2002optimum}.

Our analysis in Fig.~\ref{fig:angular-total_plot} further reveals that the degradation in terrestrial \gls{ue} performance due to Interference Nulling is primarily driven by the \gls{ue} angular position with respect to the \gls{gnb} \gls{upa} and the satellite position. Specifically, \glspl{ue} at
positive azimuth angles (e.g., $\phi>0^\circ$), symmetric to the satellite positions, experience the highest degradation. Indeed, most affected \glspl{ue} are located at positive azimuth angles (e.g., $\phi=20^\circ$ to $60^\circ$) and negative elevations (e.g., $\theta<-20^\circ$), while satellites are located at negative azimuth angles (e.g., $\phi=-20^\circ$ to $-60^\circ$) and positive elevations (e.g., $\theta>20^\circ$). This represents a fundamental geometric trade-off inherent to null steering: constraining the beamformer to create nulls toward satellites at positive elevations inherently reduces the achievable gain for \glspl{ue} at extreme negative elevations, with the effect most pronounced for \glspl{ue} in the opposite azimuth quadrant. Importantly, this effect persists even with larger antenna arrays; while increasing the array size from $4\times4$ to $32\times32$ significantly reduces the degradation magnitude (from $25$~dB to $0.5$~dB), the same \glspl{ue} consistently exhibit the worst relative performance (e.g., $\theta<-40^\circ$ and $40^\circ<\phi<60^\circ$). As further detailed in \cite{wadaskar2025satellite}, the array response satisfies $\mathbf{a}(\theta_{az}, \theta_{el})=\mathbf{a}(-\theta_{az},-\theta_{el})$ since $\sin(-\theta_{az})\sin(-\theta_{el})=\sin(\theta_{az})\sin(\theta_{el})$ and $\cos(-\theta_{el})=\cos(\theta_{el})$. This symmetry implies that steering nulls toward satellites at $(\theta_{az}^{\text{sat}}, \theta_{el}^{\text{sat}})$ simultaneously affects beamforming gain at $(-\theta_{az}^{\text{sat}}, -\theta_{el}^{\text{sat}})$, limiting the \gls{dof} for terrestrial beamforming. These findings demonstrate that null steering introduces an unavoidable geometric coupling between the nulled and served directions, which can only be partially mitigated through increased array dimensions.

In Fig.~\ref{fig:channel_analysis}, we present simulation results showing the effect of the Interference Nulling strategies (formulated in Section~\ref{sec:ss:interf-null}) on the channels for different $\lambda$ values. Fig.~\ref{fig:channel_analysis}(a) illustrates how the signal leakage towards the satellites (averaged across all snapshots of the satellite movement and, denoted as $|\tilde{\boldsymbol{h}}_j^{\text{H}} \boldsymbol{w}_t|$), is mitigated—particularly as $\lambda$ increases—demonstrating the effectiveness of the interference nulling techniques, especially for $\lambda$ values of $1$ and $10$. Fig.~\ref{fig:channel_analysis}(b) shows the effects of the interference nulling techniques on all \glspl{ue} throughout the satellite movement. Notably, $\lambda=0.1$ incurs minimal degradation on the terrestrial network across all satellite snapshot indices, whereas $\lambda=1$, and particularly $\lambda=10$, demonstrate a stronger impact on specific \glspl{ue}, which can experience up to $\sim 20$~dB \gls{rss} degradation. Therefore, Fig.~\ref{fig:channel_analysis} indicates that, although the average \gls{rss} degradation remains near zero, specific \glspl{ue} can experience severe \gls{qos} loss at certain points along the satellite trajectories, highlighting the limitations and significant trade-offs of Interference Nulling for satellite interference suppression while maintaining acceptable terrestrial \gls{qos}.

\section{Conclusions}\label{sec:ss:conclusions}
We have performed a comparative analysis of two state-of-the-art techniques used for interference suppression to satellites, namely \gls{qos}-aware Power Control and Interference Nulling, in spectrum sharing scenarios for the \gls{fr3} band. Our simulation results demonstrated the suitability of power control techniques for risk-averse scenarios, while pinpointing the specific geometric and array-size limitations of interference nulling. Specifically, we demonstrated that interference nulling can severely degrade \glspl{ue} at certain angular positions; however, this degradation can be significantly mitigated by employing larger antenna array sizes. Finally, we proposed the joint application of interference nulling and power control techniques to reduce interference towards the incumbents while ensuring a more energy-efficient operation and fair \gls{qos} distribution in the terrestrial network. Our future work will focus on the on-demand and adaptive per-\gls{ue} optimization of power and beamforming. Specifically, we aim to apply power control selectively only for \glspl{ue} experiencing significant \gls{qos} degradation, thereby maintaining the \gls{rss} degradation at manageable levels while preserving system-wide performance. Another research focus will be the joint optimization of transmit power levels and beamforming for \gls{mu}-\gls{mimo} systems, where power allocation decisions can be made proportionally fair to the links' channel quality.

% --- EOF ---

%% file: fig/Elevation_Angle_PDF_gNB1_NoSats40.tex
\definecolor{mycolor1}{rgb}{0.20000,0.40000,0.80000}%
\definecolor{mycolor2}{rgb}{0.12941,0.12941,0.12941}%
\begin{tikzpicture}

\begin{axis}[%
width=2.5in,
height=0.85in,
at={(1.083in,0.785in)},
scale only axis,
bar shift auto,
clip=false,
xmin=9.95200000000002,
xmax=90.848,
xlabel style={font=\scriptsize},
xlabel={Elevation angle ($\theta$)  [$^\circ$]},
ymin=0,
ymax=0.05,
ylabel style={font=\scriptsize},
ylabel={PDF},
axis background/.style={fill=white},
xmajorgrids,
ymajorgrids,
tick label style={font=\scriptsize},
legend style={legend cell align=left, align=left, font=\scriptsize}
]
\addplot[ybar, bar width=2.56, fill=mycolor1, fill opacity=0.7, draw=black, area legend] table[row sep=crcr] {%
13.28	0.00364583333333333\\
15.84	0.0182291666666667\\
18.4	0.0299479166666667\\
20.96	0.0438802083333333\\
23.52	0.0428385416666667\\
26.08	0.0383463541666667\\
28.64	0.0338541666666666\\
31.2	0.0311848958333333\\
33.76	0.024609375\\
36.32	0.0171875\\
38.88	0.0188802083333334\\
41.44	0.00944010416666666\\
44	0.00820312499999999\\
46.56	0.00904947916666666\\
49.12	0.00683593750000001\\
51.68	0.00722656249999999\\
54.24	0.00781249999999999\\
56.8	0.00546875000000001\\
59.36	0.00624999999999999\\
61.92	0.00520833333333333\\
64.48	0.00436197916666669\\
67.04	0.0033203125\\
69.6	0.003125\\
72.16	0.00279947916666666\\
74.72	0.00286458333333333\\
77.28	0.00227864583333333\\
79.84	0.001171875\\
82.4	0.00149739583333333\\
84.96	0.000520833333333336\\
87.52	0.0005859375\\
};
\addplot[forget plot, color=mycolor2] table[row sep=crcr] {%
9.95200000000002	0\\
90.848	0\\
};
% \addlegendentry{Histogram}

\node[fill=white, below left, align=left, inner sep=1.5, font=\scriptsize, draw=black]
at (rel axis cs:1.0,0.98) {
Mean: $33.0^\circ$, Median: $28.8^\circ$\\
Std: $14.8^\circ$, Min: $13.1^\circ$, Max: $88.6^\circ$
};
\end{axis}
\end{tikzpicture}%

%% file: fig/Fairness_Scatter_WorstCaseVsVariability_AllUEs30_NoSats40_gNB1.tex
\definecolor{mycolor1}{rgb}{0.06600,0.44300,0.74500}%
\definecolor{mycolor2}{rgb}{0.86600,0.32900,0.00000}%
\definecolor{mycolor3}{rgb}{0.52100,0.08600,0.81900}%
\definecolor{mycolor4}{rgb}{0.60000,0.30000,0.10000}%
\definecolor{mycolor5}{rgb}{0.12941,0.12941,0.12941}%
\begin{tikzpicture}
\begin{axis}[%
width=0.9in,
height=0.8in,
at={(0in,0in)},
scale only axis,
xmin=0,
xmax=3,
xlabel style={font=\scriptsize},
xlabel={Std. Dev. of RSS},
ymin=0,
ymax=25,
ylabel style={font=\scriptsize,yshift=-6pt},
ylabel={Worst Case RSS [dB]},
axis background/.style={fill=white},
xmajorgrids,
ymajorgrids,
legend style={at={(0.5,-0.54)}, anchor=north, legend columns=1, font=\scriptsize, legend cell align=left, draw=none}
]
\addplot[only marks, mark=*, mark options={}, mark size=2pt, draw=mycolor1, fill=mycolor1] table[row sep=crcr]{%
x	y\\
0	0\\
};
\addlegendentry{Interf. Nulling (λ = 0)}
\addplot[only marks, mark=square*, mark options={}, mark size=2pt, draw=mycolor2, fill=mycolor2] table[row sep=crcr]{%
x	y\\
0.119136109520964	1.38457900093323\\
};
\addlegendentry{Interf. Nulling (λ = 0.1)}
\addplot[only marks, mark=triangle*, mark options={}, mark size=2pt, draw=green!60!violet, fill=green!60!violet] table[row sep=crcr]{%
x	y\\
1.80113549433475	20.2563255543226\\
};
\addlegendentry{Interf. Nulling (λ = 1)}
\addplot[only marks, mark=asterisk, mark options={}, mark size=2pt, draw=mycolor3, fill=mycolor3] table[row sep=crcr]{%
x	y\\
2.53068651206873	23.1469268862214\\
};
\addlegendentry{Interf. Nulling (λ = 10)}
\addplot[only marks, mark=triangle*, mark options={rotate=270}, mark size=2pt, draw=mycolor4, fill=mycolor4] table[row sep=crcr]{%
x	y\\
0.425536896631504	7.72084358179715\\
};
\addlegendentry{No Null. (λ=0), Power Control}
\end{axis}
\end{tikzpicture}%

%% file: fig/Fairness_BarChart_JainsIndex_AllUEs30_NoSats40_gNB1.tex
\definecolor{mycolor1}{rgb}{0.12941,0.12941,0.12941}%
\definecolor{bar1}{rgb}{0.0,0.30,0.80}% Deep Blue
\definecolor{bar2}{rgb}{0.85,0.33,0.10}% Strong Orange
\definecolor{bar3}{rgb}{0.20,0.60,0.20}% Rich Green
\definecolor{bar4}{rgb}{0.55,0.10,0.70}% Purple
\definecolor{bar5}{rgb}{0.6,0.3,0.1}% Brown
\begin{tikzpicture}[font=\small]
\begin{axis}[%
xtick style={draw=none},
width=0.85in,
height=0.85in,
ybar=0pt,
bar width=0.5,
scale only axis,
xmin=0.5,
xmax=5.5,
xtick={1,2,3,4,5},
xticklabels={{Interf. Nulling ($\lambda=0$)},{Interf. Nulling ($\lambda=0.1$)},{Interf. Nulling ($\lambda=1$)},{Interf. Nulling ($\lambda=10$)},{No Null., Power Control}},
xticklabel style={rotate=90, anchor=east, font=\scriptsize},
ymin=0,
ymax=1.05,
ylabel={JFI},
ylabel style={font=\scriptsize,yshift=-6pt},
axis background/.style={fill=white},
ymajorgrids,
clip=false,
enlarge x limits=0.15
]
\addplot[ybar, fill=bar1, draw=none, bar shift=0pt] coordinates {(1,1)};
\addplot[ybar, fill=bar2, draw=none, bar shift=0pt] coordinates {(2,0.313931050330797)};
\addplot[ybar, fill=bar3, draw=none, bar shift=0pt] coordinates {(3,0.384930621221513)};
\addplot[ybar, fill=bar4, draw=none, bar shift=0pt] coordinates {(4,0.568601454422722)};
\addplot[ybar, fill=bar5, draw=none, bar shift=0pt] coordinates {(5,0.995996633216235)};
\draw[black, dashed, thick, shorten >=2pt, shorten <=2pt] (axis cs:0.5,1) -- (axis cs:5.5,1) node[pos=0.5, below, font=\scriptsize] {Perfect Fairness};
\end{axis}
\end{tikzpicture}%

%% file: fig/BellPeakPower_vs_BeamformedChannelGain.tex
\definecolor{mycolor1}{rgb}{0.06600,0.44300,0.74500}%
\definecolor{mycolor2}{rgb}{0.12941,0.12941,0.12941}%
\begin{tikzpicture}

\begin{axis}[%
width=1.45in,
height=1in,
at={(0in,0in)},
scale only axis,
xmin=-66.7414407112592,
xmax=-54.5390548267721,
xlabel style={font=\footnotesize, yshift=3pt},
xlabel={Beamformed Channel Gain [dB]},
ymin=14.3261630815408,
ymax=26.5222611305653,
ylabel style={font=\footnotesize, yshift=-5pt},
ylabel={Utility Peak Power [dBm]},
axis background/.style={fill=white},
xmajorgrids,
ymajorgrids,
xtick distance=2,
ytick distance=3,
tick label style={font=\scriptsize},
]
\addplot[only marks, mark=*, mark options={}, mark size=0.75pt, color=mycolor1, fill=mycolor1, forget plot] table[row sep=crcr]{%
x	y\\
-54.5390548267721	14.3261630815408\\
-56.0853117830236	15.8679339669835\\
-57.0549047640633	16.8459229614807\\
-57.3642684368731	17.1450725362681\\
-58.7656648480108	18.5487743871936\\
-59.1773579986775	18.9629814907454\\
-59.1870662917992	18.9744872436218\\
-60.0527322308536	19.8374187093547\\
-60.2373958154455	20.0215107553777\\
-60.4246773111609	20.2056028014007\\
-60.4512180735895	20.24012006003\\
-60.8547619030881	20.6428214107054\\
-60.9480832308826	20.7348674337169\\
-60.9703486010627	20.7578789394697\\
-60.975643892087	20.7578789394697\\
-61.0879726891535	20.8729364682341\\
-61.3406787638822	21.1260630315158\\
-61.4417166629414	21.2296148074037\\
-61.8460283089614	21.632316158079\\
-62.1514299447453	21.9429714857429\\
-62.2807620095314	22.0695347673837\\
-62.4124910199673	22.1960980490245\\
-62.9119693359557	22.7023511755878\\
-63.2352739257249	23.0245122561281\\
-64.2614520537951	24.0485242621311\\
-64.5010760438395	24.2901450725363\\
-65.1763161067657	24.9574787393697\\
-65.331879238669	25.1185592796398\\
-66.3718723611911	26.1540770385193\\
-66.7414407112592	26.5222611305653\\
};
\end{axis}
\end{tikzpicture}%

%% file: fig/UtilityPeak_vs_BeamformedChannelGain.tex
\definecolor{mycolor1}{rgb}{0.06600,0.44300,0.74500}%
\definecolor{mycolor2}{rgb}{0.12941,0.12941,0.12941}%
\begin{tikzpicture}

\begin{axis}[%
width=1.45in,
height=1in,
at={(0in,0in)},
scale only axis,
xmin=-66.7414407112592,
xmax=-54.5390548267721,
xlabel style={font=\footnotesize, yshift=3pt},
xlabel={Beamformed Channel Gain [dB]},
ymin=0,
ymax=800000000,
ylabel style={font=\footnotesize, yshift=-5pt},
ylabel={Utility Peak},
axis background/.style={fill=white},
xmajorgrids,
ymajorgrids,
xtick distance=2,
tick label style={font=\scriptsize},
]
\addplot[only marks, mark=*, mark options={}, mark size=0.75pt, color=mycolor1, fill=mycolor1, forget plot] table[row sep=crcr]{%
x	y\\
-54.5390548267721	682984832.72717\\
-56.0853117830236	478393450.997838\\
-57.0549047640633	382671147.056436\\
-57.3642684368731	356360320.76784\\
-58.7656648480108	258077392.387138\\
-59.1773579986775	234736573.010031\\
-59.1870662917992	234212408.970654\\
-60.0527322308536	191886011.414123\\
-60.2373958154455	183897968.647588\\
-60.4246773111609	176136138.666808\\
-60.4512180735895	175063105.905671\\
-60.8547619030881	159529244.039796\\
-60.9480832308826	156137861.941875\\
-60.9703486010627	155339415.005414\\
-60.975643892087	155150070.061073\\
-61.0879726891535	151188681.152079\\
-61.3406787638822	142642414.127655\\
-61.4417166629414	139362150.902709\\
-61.8460283089614	126973686.430852\\
-62.1514299447453	118351323.948957\\
-62.2807620095314	114878894.162343\\
-62.4124910199673	111446743.140309\\
-62.9119693359557	99338900.1232685\\
-63.2352739257249	92212333.8895675\\
-64.2614520537951	72806698.4671001\\
-64.5010760438395	68898349.4275156\\
-65.1763161067657	58977264.4658116\\
-65.331879238669	56902143.9077569\\
-66.3718723611911	44784643.9598756\\
-66.7414407112592	41131269.6978792\\
};
\end{axis}
\end{tikzpicture}%

%% file: fig/Combined_INR_Median_BarPlot.tex
% \documentclass[tikz]{standalone}
% \usepackage[T1]{fontenc}
% \usepackage[utf8]{inputenc}
% \usepackage{pgfplots}
% \usepackage{grffile}
% \pgfplotsset{compat=newest}
% \usetikzlibrary{plotmarks}
% \usetikzlibrary{arrows.meta}
% \usepgfplotslibrary{patchplots}
\usetikzlibrary{patterns}
\definecolor{mycolor1}{rgb}{0.06600,0.44300,0.74500}%
\definecolor{mycolor2}{rgb}{0.8500,0.3250,0.0980}%Orange
\definecolor{mycolor3}{rgb}{0.00,0.50,0.00}% 
\definecolor{mycolor4}{rgb}{0.60000,0.30000,0.10000}% Brown
\definecolor{mycolor5}{rgb}{0.95,0.00,0.00}% Very dark red

\definecolor{mycolorOutline}{rgb}{0.12941,0.12941,0.12941}%  % Dark gray for outline
\renewcommand{\Huge}{\fontsize{35}{42}\selectfont}
\begin{tikzpicture}
\begin{axis}[%
width=4.25in,
height=5in,
ytick distance=10,
at={(1.62in,2.45in)},
scale only axis,
xmin=0.2,
xmax=5.7,
xtick={1,2,3,4,5},
xticklabels={{Interf. Nulling ($\lambda$=0)},{Interf. Nulling ($\lambda$=0.1)},{Interf. Nulling ($\lambda$=1)},{Interf. Nulling ($\lambda$=0) \& Power Control ($\epsilon$=0.85)},{Interf. Nulling ($\lambda$=0.1) \& Power Control ($\epsilon$=0.98)}},
xticklabel style={rotate=90, anchor=east, font=\Huge},
yticklabel style={font=\Huge},
xlabel style={font=\Huge},
ymin=0,
ymax=35,
ylabel style={font=\Huge},
ylabel={Median INR [dB]},
axis background/.style={fill=white},
xmajorgrids,
ymajorgrids,
]
% Bar 1 - Blue with solid fill (no pattern)
\draw[fill=mycolor1, draw=mycolorOutline] (axis cs:0.675,0) rectangle (axis cs:1.325,25.1537084627834);
% Bar 2 - Orange with horizontal lines
\draw[pattern=horizontal lines, pattern color=mycolor2, draw=mycolorOutline] (axis cs:1.675,0) rectangle (axis cs:2.325,27.1641126299116);
% Bar 3 - Green-violet with vertical lines
\draw[pattern=vertical lines, pattern color=mycolor3, draw=mycolorOutline] (axis cs:2.675,0) rectangle (axis cs:3.325,31.811866293842);
% Bar 4 - Brown with north east lines (diagonal)
\draw[pattern=north east lines, pattern color=mycolor4, draw=mycolorOutline] (axis cs:3.675,0) rectangle (axis cs:4.325,31.8377824954678);
% Bar 5 - Sienna with crosshatch
\draw[pattern=crosshatch, pattern color=mycolor5, draw=mycolorOutline] (axis cs:4.675,0) rectangle (axis cs:5.325,28.0534984510458);
\node[above, align=center, inner sep=2pt, font=\huge]
at (axis cs:1,25.1537084627834) {-25.15};
\node[above, align=center, inner sep=2pt, font=\huge]
at (axis cs:2,27.1641126299116) {-27.16};
\node[above, align=center, inner sep=2pt, font=\huge]
at (axis cs:3,31.811866293842) {-31.81};
\node[above, align=center, inner sep=2pt, font=\huge]
at (axis cs:4,31.8377824954678) {-31.84};
\node[above, align=center, inner sep=2pt, font=\huge]
at (axis cs:5,28.0534984510458) {-28.05};
\addplot [color=black, dashed, line width=1.0pt, forget plot]
  table[row sep=crcr]{%
-0.225	0\\
6.225	0\\
};
\end{axis}
\end{tikzpicture}%

%% file: fig/Combined_RSS_Degradation_BoxPlot.tex
% \documentclass[tikz]{standalone}
% \usepackage[T1]{fontenc}
% \usepackage[utf8]{inputenc}
% \usepackage{pgfplots}
% \usepackage{grffile}
% \pgfplotsset{compat=newest}
% \usetikzlibrary{plotmarks}
% \usetikzlibrary{arrows.meta}
% \usepgfplotslibrary{patchplots}
% \usepackage{amsmath}

% \begin{document}
\definecolor{mycolor1}{rgb}{0.05600,0.44300,0.74500}%Blue (slightly different value)
\definecolor{mycolor2}{rgb}{0.8500,0.3250,0.0980}%Orange (slightly different value)
\definecolor{mycolor3}{rgb}{0.00,0.50,0.00}% Green (currently it's purple 0.521,0.086,0.819)
\definecolor{mycolor4}{rgb}{0.60000,0.30000,0.10000}%
\definecolor{mycolor5}{rgb}{0.95,0.00,0.00}% Very dark red

% \definecolor{mycolor1}{rgb}{0.06600,0.44300,0.74500}%
% \definecolor{mycolor2}{rgb}{0.86600,0.32900,0.00000}%
% \definecolor{mycolor3}{rgb}{0.52100,0.08600,0.81900}%

\renewcommand{\Huge}{\fontsize{35}{42}\selectfont}

% \definecolor{mycolor1}{rgb}{0.06600,0.44300,0.74500}%
% % \definecolor{mycolor1}{rgb}{0.12941,0.12941,0.12941}%
% \definecolor{mycolor2}{rgb}{0.06667,0.44314,0.74510}%
% \definecolor{mycolor3}{rgb}{0.86667,0.32941,0.00000}%
\begin{tikzpicture}

\begin{axis}[%
width=4.2in,
height=4.7in,
at={(1.625in,2.452in)},
scale only axis,
unbounded coords=jump,
xmin=0.5,
xmax=5.5,
xtick={1,2,3,4,5},
xticklabels={{Interf. Nulling (λ = 0)},{Interf. Nulling (λ = 0.1)},{Interf. Nulling (λ = 1)},{Interf. Nulling (λ = 0) \& Power Control ($\epsilon$=0.85)},{Interf. Nulling (λ = 0.1) \& Power Control ($\epsilon$=0.98)}},
xticklabel style={rotate=90, anchor= east, font=\Huge},
yticklabel style={font=\Huge},
xlabel style={font=\Huge},
ymin=-1.01281627771613,
ymax=21.2691418320387,
ylabel style={font=\Huge},
ylabel={RSS Degradation [dB]},
axis background/.style={fill=white},
xmajorgrids,
ymajorgrids
]
\addplot [color=mycolor1, dashed, forget plot]
  table[row sep=crcr]{%
1	0\\
1	0\\
};
\addplot [color=mycolor2, dashed, forget plot]
  table[row sep=crcr]{%
2	0.0120706950301062\\
2	0.0285342531283967\\
};
\addplot [color=mycolor3, dashed, forget plot]
  table[row sep=crcr]{%
3	0.15147149953962\\
3	0.357040278456689\\
};
\addplot [color=mycolor4, dashed, forget plot]
  table[row sep=crcr]{%
4	6.89408029975809\\
4	7.48895850666693\\
};
\addplot [color=mycolor5, dashed, forget plot]
  table[row sep=crcr]{%
5	0.944843272276778\\
5	1.05332432665653\\
};
% Lower whiskers
\addplot [color=mycolor1, dashed, forget plot]
  table[row sep=crcr]{%
1	0\\
1	0\\
};
\addplot [color=mycolor2, dashed, forget plot]
  table[row sep=crcr]{%
2	2.60698031351785e-06\\
2	0.000943787758792978\\
};
\addplot [color=mycolor3, dashed, forget plot]
  table[row sep=crcr]{%
3	0.000167233660447924\\
3	0.0143603675333529\\
};
\addplot [color=mycolor4, dashed, forget plot]
  table[row sep=crcr]{%
4	5.89188728196996\\
4	6.46551160350306\\
};
\addplot [color=mycolor5, dashed, forget plot]
  table[row sep=crcr]{%
5	0.785532744606824\\
5	0.872175638756635\\
};
% Upper caps
\addplot [color=mycolor1, forget plot]
  table[row sep=crcr]{%
0.875	0\\
1.125	0\\
};
\addplot [color=mycolor2, forget plot]
  table[row sep=crcr]{%
1.875	0.0285342531283967\\
2.125	0.0285342531283967\\
};
\addplot [color=mycolor3, forget plot]
  table[row sep=crcr]{%
2.875	0.357040278456689\\
3.125	0.357040278456689\\
};
\addplot [color=mycolor4, forget plot]
  table[row sep=crcr]{%
3.875	7.48895850666693\\
4.125	7.48895850666693\\
};
\addplot [color=mycolor5, forget plot]
  table[row sep=crcr]{%
4.875	1.05332432665653\\
5.125	1.05332432665653\\
};
% Lower caps
\addplot [color=mycolor1, forget plot]
  table[row sep=crcr]{%
0.875	0\\
1.125	0\\
};
\addplot [color=mycolor2, forget plot]
  table[row sep=crcr]{%
1.875	2.60698031351785e-06\\
2.125	2.60698031351785e-06\\
};
\addplot [color=mycolor3, forget plot]
  table[row sep=crcr]{%
2.875	0.000167233660447924\\
3.125	0.000167233660447924\\
};
\addplot [color=mycolor4, forget plot]
  table[row sep=crcr]{%
3.875	5.89188728196996\\
4.125	5.89188728196996\\
};
\addplot [color=mycolor5, forget plot]
  table[row sep=crcr]{%
4.875	0.785532744606824\\
5.125	0.785532744606824\\
};
% Box outlines
\addplot [color=mycolor1, forget plot]
  table[row sep=crcr]{%
0.75	0\\
0.75	0\\
1.25	0\\
1.25	0\\
0.75	0\\
};
\addplot [color=mycolor2, forget plot]
  table[row sep=crcr]{%
1.75	0.000943787758792978\\
1.75	0.0120706950301062\\
2.25	0.0120706950301062\\
2.25	0.000943787758792978\\
1.75	0.000943787758792978\\
};
\addplot [color=mycolor3, forget plot]
  table[row sep=crcr]{%
2.75	0.0143603675333529\\
2.75	0.15147149953962\\
3.25	0.15147149953962\\
3.25	0.0143603675333529\\
2.75	0.0143603675333529\\
};
\addplot [color=mycolor4, forget plot]
  table[row sep=crcr]{%
3.75	6.46551160350306\\
3.75	6.89408029975809\\
4.25	6.89408029975809\\
4.25	6.46551160350306\\
3.75	6.46551160350306\\
};
\addplot [color=mycolor5, forget plot]
  table[row sep=crcr]{%
4.75	0.872175638756635\\
4.75	0.944843272276778\\
5.25	0.944843272276778\\
5.25	0.872175638756635\\
4.75	0.872175638756635\\
};
% Median lines
\addplot [color=mycolor1, forget plot]
  table[row sep=crcr]{%
0.75	0\\
1.25	0\\
};
\addplot [color=mycolor2, forget plot]
  table[row sep=crcr]{%
1.75	0.0034467713007926\\
2.25	0.0034467713007926\\
};
\addplot [color=mycolor3, forget plot]
  table[row sep=crcr]{%
2.75	0.048260436650288\\
3.25	0.048260436650288\\
};
\addplot [color=mycolor4, forget plot]
  table[row sep=crcr]{%
3.75	6.7473090457329\\
4.25	6.7473090457329\\
};
\addplot [color=mycolor5, forget plot]
  table[row sep=crcr]{%
4.75	0.908026505808365\\
5.25	0.908026505808365\\
};
% Outliers position 1 (none)
% \addplot [color=mycolor1, only marks, mark=+, mark options={solid, fill=mycolor1, draw=mycolor1}, forget plot]
%   table[row sep=crcr]{%
% nan	nan\\
% };
\addplot [color=mycolor5, only marks, mark=+, mark options={solid, fill=black, draw=black}, forget plot]
  table[row sep=crcr]{%
2	0.0288594609782575\\
2	0.0291884764381823\\
2	0.0292063882403262\\
2	0.0292403925949861\\
2	0.0292630226644481\\
2	0.0293074060116396\\
2	0.0293449615355665\\
2	0.0293649819750424\\
2	0.0295163132402567\\
2	0.0295794482532595\\
2	0.0296757613268952\\
2	0.029913797472945\\
2	0.0299538773256947\\
2	0.0300718399864453\\
2	0.0301560995057363\\
2	0.0301608716856441\\
2	0.0304168149472368\\
2	0.0305600177534329\\
2	0.0305777664949283\\
2	0.0306816315951428\\
2	0.0306944498545661\\
2	0.0307233635561984\\
2	0.0307365887367364\\
2	0.0307827703818687\\
2	0.0308814583606106\\
2	0.030923687644331\\
2	0.0310522139903529\\
2	0.0310881978683614\\
2	0.0311310147240544\\
2	0.0311527508058205\\
2	0.0312789480412024\\
2	0.0313250508876179\\
2	0.0313655570618582\\
2	0.0314609397524651\\
2	0.0314671960265216\\
2	0.0315112963228152\\
2	0.0315216993748351\\
2	0.0315936932280554\\
2	0.0316199545934195\\
2	0.0318621243626222\\
2	0.0321903869693164\\
2	0.0323629548959228\\
2	0.0324031108846695\\
2	0.0324066699665304\\
2	0.0324787574461975\\
2	0.0325004616055848\\
2	0.0325555481349368\\
2	0.0326042685281596\\
2	0.0326050303939928\\
2	0.0326059330888639\\
2	0.032708092780149\\
2	0.0327358808832412\\
2	0.0330962053300786\\
2	0.0332627992044962\\
2	0.0334188552140129\\
2	0.0337996635090157\\
2	0.0338098326634592\\
2	0.0338541641177155\\
2	0.0339223992784341\\
2	0.0340983191444485\\
2	0.034103506357311\\
2	0.0341566973866176\\
2	0.034225007216379\\
2	0.0345556387186074\\
2	0.034650512716493\\
2	0.0347200111483384\\
2	0.0350698071614998\\
2	0.0352540054247376\\
2	0.0352658206600181\\
2	0.0352757115794025\\
2	0.0353006237795913\\
2	0.0353140231087596\\
2	0.03538587643154\\
2	0.0353920120089944\\
2	0.0356036455813237\\
2	0.0356110162808801\\
2	0.0356655117682569\\
2	0.0358246680201292\\
2	0.0359676773547593\\
2	0.0360845961371831\\
2	0.0361627090609562\\
2	0.0362003274904223\\
2	0.0363753827204547\\
2	0.036382905586518\\
2	0.0364919161662866\\
2	0.0369060715549061\\
2	0.0371179367106817\\
2	0.0371245274435088\\
2	0.0373601908237685\\
2	0.0375765965621201\\
2	0.0377661829464734\\
2	0.0378627977517435\\
2	0.0381274942946137\\
2	0.0382395237752539\\
2	0.0385054839515076\\
2	0.0385644980119351\\
2	0.0387025417432878\\
2	0.0388018755902012\\
2	0.0388421226734261\\
2	0.0389606451949977\\
2	0.0393208605659262\\
2	0.0394217461225818\\
2	0.0395962162646843\\
2	0.0402465154304763\\
2	0.0402934087333915\\
2	0.0405365278436194\\
2	0.040588120938495\\
2	0.040939481735422\\
2	0.0410133264879415\\
2	0.0411290627039375\\
2	0.0413685185857315\\
2	0.0415187554931124\\
2	0.041586089799142\\
2	0.0416342479413046\\
2	0.0417758459247753\\
2	0.0418846323981334\\
2	0.0424573885726688\\
2	0.0425099566281203\\
2	0.0425154188585355\\
2	0.0425181139850705\\
2	0.0425346766788901\\
2	0.0426566097860169\\
2	0.04270044175453\\
2	0.0428914234892195\\
2	0.0431419385572978\\
2	0.0431952071520673\\
2	0.044289312013757\\
2	0.0446008433534934\\
2	0.0449890795454577\\
2	0.0451077779405916\\
2	0.0452948560454708\\
2	0.045390262092217\\
2	0.0459209147747285\\
2	0.0459780210811629\\
2	0.0461784603284221\\
2	0.0462484717934855\\
2	0.04633791840302\\
2	0.0466238235150299\\
2	0.046727432882794\\
2	0.0467492523442994\\
2	0.0470967453586151\\
2	0.0473002273605404\\
2	0.0474576345647702\\
2	0.0477622638625409\\
2	0.0477976911173358\\
2	0.0480029841601015\\
2	0.0481400858826373\\
2	0.0482275147658167\\
2	0.0485054810325291\\
2	0.0489204196596616\\
2	0.0491186938188848\\
2	0.0491209238735821\\
2	0.0493016561343986\\
2	0.0493269497847663\\
2	0.0493619007381042\\
2	0.0495644849745917\\
2	0.0498134153934086\\
2	0.0506233735793515\\
2	0.0507344047220025\\
2	0.050921344196366\\
2	0.0509273478868272\\
2	0.0510337770724309\\
2	0.0516473272684426\\
2	0.0517929314109139\\
2	0.0518082912991137\\
2	0.0519649507372516\\
2	0.0519654933120066\\
2	0.0521905297030812\\
2	0.0523177545948617\\
2	0.0524672962601769\\
2	0.0526225809989025\\
2	0.0526567451282615\\
2	0.0530140438892695\\
2	0.0536543974335959\\
2	0.0544720124699774\\
2	0.0549206495156953\\
2	0.0551533644100436\\
2	0.0553075620448666\\
2	0.0553516711616139\\
2	0.0557056221033764\\
2	0.0557698721976179\\
2	0.055935339698734\\
2	0.0559384782593954\\
2	0.0562528026324369\\
2	0.0563989688392075\\
2	0.0565851328090028\\
2	0.0566650871257606\\
2	0.056796589312797\\
2	0.05688508863037\\
2	0.0570573874126932\\
2	0.0574666034416757\\
2	0.057972179477753\\
2	0.0582957910240219\\
2	0.0583906199081891\\
2	0.0587869163315109\\
2	0.0588052849601104\\
2	0.0589057362570626\\
2	0.0591957081872675\\
2	0.0592083615379045\\
2	0.0604214998819845\\
2	0.0607763693047033\\
2	0.0608276073261455\\
2	0.0617406989977158\\
2	0.0624374906843238\\
2	0.0625363459892303\\
2	0.0628151941122121\\
2	0.0631198568845551\\
2	0.0631797239526805\\
2	0.0634214552618726\\
2	0.0634538479963235\\
2	0.0634898915662582\\
2	0.0638164384024596\\
2	0.0638957495898104\\
2	0.0644276780815098\\
2	0.0645188283882453\\
2	0.0646496428517875\\
2	0.0647134364827251\\
2	0.0648347289558315\\
2	0.0650535454809182\\
2	0.065322081693856\\
2	0.0655107569644203\\
2	0.0655153721617014\\
2	0.0655508585169507\\
2	0.0656924901741938\\
2	0.0657210377485668\\
2	0.0662359705453794\\
2	0.0671126695651619\\
2	0.0677137149188557\\
2	0.067810772847497\\
2	0.0678276658448353\\
2	0.0682296959996917\\
2	0.068478421297978\\
2	0.0686980416759884\\
2	0.0688284465182118\\
2	0.0695803223912452\\
2	0.0700668743249523\\
2	0.070406517615902\\
2	0.0706272729323558\\
2	0.070895529918715\\
2	0.0720959104703113\\
2	0.0727077398236605\\
2	0.0727464516914146\\
2	0.0729729010251981\\
2	0.0738172152504137\\
2	0.0744828171337586\\
2	0.0745538330982284\\
2	0.0746134756769445\\
2	0.0747309008219306\\
2	0.075097662092953\\
2	0.0760077750713792\\
2	0.076188427718424\\
2	0.0775784502517164\\
2	0.0782737449639079\\
2	0.0792065554654944\\
2	0.0796828449409028\\
2	0.0807503033478042\\
2	0.0809002120598729\\
2	0.0813566397022187\\
2	0.0819454509529647\\
2	0.0827088732881242\\
2	0.0828315018306708\\
2	0.083188716659947\\
2	0.0835607304978765\\
2	0.0839531034755783\\
2	0.0847883737756287\\
2	0.0860792479816069\\
2	0.0867542490698847\\
2	0.0870043177806885\\
2	0.0872903042073889\\
2	0.0881580663228856\\
2	0.0885340985575009\\
2	0.0893355341053961\\
2	0.0894070639965377\\
2	0.0894796883973348\\
2	0.0909878496037812\\
2	0.0914763932041376\\
2	0.0917491282329868\\
2	0.0923169803705392\\
2	0.0927666693330062\\
2	0.0934630625155149\\
2	0.0936230350943899\\
2	0.0943133129453612\\
2	0.0944193676412546\\
2	0.0946148613526027\\
2	0.0946313627298664\\
2	0.0946886464598868\\
2	0.0947553103095499\\
2	0.0955651690657171\\
2	0.0956918407263783\\
2	0.096331823921788\\
2	0.0970484743184074\\
2	0.0979160559624931\\
2	0.09810574893421\\
2	0.0997155765140399\\
2	0.0998758044421254\\
2	0.100273573803737\\
2	0.101550782064978\\
2	0.101721319152216\\
2	0.102002587568108\\
2	0.102714769889444\\
2	0.103663314835906\\
2	0.104450722906352\\
2	0.105959864018208\\
2	0.106115228292834\\
2	0.106203304060382\\
2	0.106470184047327\\
2	0.10726444531798\\
2	0.108706290515382\\
2	0.109461998397221\\
2	0.1100841699874\\
2	0.110358004426122\\
2	0.110642989444968\\
2	0.110822999348412\\
2	0.110931190239055\\
2	0.112521259203945\\
2	0.113379332726175\\
2	0.113714469198943\\
2	0.113736072770768\\
2	0.114284713063516\\
2	0.114475663427086\\
2	0.11458191147743\\
2	0.11514551125112\\
2	0.115156650412136\\
2	0.116265422037064\\
2	0.116441811701489\\
2	0.116586433230773\\
2	0.116969941575534\\
2	0.117402327844264\\
2	0.117488151180642\\
2	0.117767033293501\\
2	0.117962452238996\\
2	0.118255844854793\\
2	0.118471380013021\\
2	0.119170533478624\\
2	0.119437653000723\\
2	0.119600575403427\\
2	0.119995058711496\\
2	0.120053022305675\\
2	0.120196921422555\\
2	0.120513364247064\\
2	0.120787819028452\\
2	0.121405664279373\\
2	0.121444976691214\\
2	0.122115265819371\\
2	0.122326992914538\\
2	0.123105222096817\\
2	0.12313773421412\\
2	0.123769533654157\\
2	0.124302153276824\\
2	0.125724162356894\\
2	0.126054252220002\\
2	0.126120598975432\\
2	0.127902498493225\\
2	0.127994111856273\\
2	0.129100740899788\\
2	0.129184752163511\\
2	0.129630507005444\\
2	0.129694615500364\\
2	0.130202703118338\\
2	0.130716267070975\\
2	0.132407568668296\\
2	0.132487646213623\\
2	0.132752723303442\\
2	0.135184585330361\\
2	0.135874218330119\\
2	0.136471732646065\\
2	0.136753702860058\\
2	0.137098312099657\\
2	0.137716511688926\\
2	0.137977058044674\\
2	0.138453457379895\\
2	0.139078387803192\\
2	0.139270870444438\\
2	0.139682741940623\\
2	0.139814006514755\\
2	0.140434197563482\\
2	0.140454531359486\\
2	0.140604649119076\\
2	0.140865798873596\\
2	0.140868975982769\\
2	0.141298742446785\\
2	0.141304599819684\\
2	0.141676882282476\\
2	0.142653213513243\\
2	0.14428793546276\\
2	0.144393284048942\\
2	0.144855977656692\\
2	0.145531150938714\\
2	0.145788907736907\\
2	0.145970842728906\\
2	0.146249660323066\\
2	0.146559739678504\\
2	0.147289392121253\\
2	0.148138440405934\\
2	0.149065781274463\\
2	0.151536576398923\\
2	0.151763772062457\\
2	0.152354982615434\\
2	0.152877930504419\\
2	0.153025951416925\\
2	0.154055899675486\\
2	0.154112529822613\\
2	0.15602934266625\\
2	0.15659727609443\\
2	0.156615174820076\\
2	0.156801944876131\\
2	0.156815288340412\\
2	0.157223864145322\\
2	0.157786491023536\\
2	0.158079119528688\\
2	0.159939949439471\\
2	0.160884262604904\\
2	0.161933200741233\\
2	0.162621057832591\\
2	0.164178899965428\\
2	0.165162736482444\\
2	0.165702473379199\\
2	0.166142222815659\\
2	0.169633954741721\\
2	0.169725408836758\\
2	0.170025404487818\\
2	0.170230974369284\\
2	0.170474933442928\\
2	0.171724395134884\\
2	0.172331564998921\\
2	0.172485396614906\\
2	0.173291395493319\\
2	0.173874260469203\\
2	0.174940328488473\\
2	0.175129166717236\\
2	0.17527221876493\\
2	0.175730124787565\\
2	0.17887164498864\\
2	0.180103027688199\\
2	0.180776325822411\\
2	0.180941226875479\\
2	0.183042200615789\\
2	0.185441095295031\\
2	0.185657235341387\\
2	0.186242485513056\\
2	0.186870395931583\\
2	0.188905648434935\\
2	0.189249763770296\\
2	0.190181118209106\\
2	0.191729281446807\\
2	0.19355476236286\\
2	0.193995306882021\\
2	0.195701285377205\\
2	0.196756716011665\\
2	0.196935727960224\\
2	0.197968823120316\\
2	0.198068039218498\\
2	0.200072593308321\\
2	0.202634143892781\\
2	0.203154372253105\\
2	0.20396596343697\\
2	0.203983700017531\\
2	0.204052924492823\\
2	0.20424154036246\\
2	0.206882793304495\\
2	0.207889581673565\\
2	0.208233671165447\\
2	0.20828903349056\\
2	0.208389937329553\\
2	0.209710461838295\\
2	0.210634492596248\\
2	0.213280551800198\\
2	0.215249575093297\\
2	0.217344239854171\\
2	0.217376595018731\\
2	0.218239204706049\\
2	0.219724260852418\\
2	0.21976158316533\\
2	0.220008761244155\\
2	0.220838464307598\\
2	0.221119989181138\\
2	0.222318503171842\\
2	0.222392399666316\\
2	0.223970225505744\\
2	0.226409637114764\\
2	0.226916082312893\\
2	0.227471403646691\\
2	0.228772509722367\\
2	0.229721477436181\\
2	0.229916997377294\\
2	0.230246302099618\\
2	0.230739290252165\\
2	0.233939263356561\\
2	0.234287071159366\\
2	0.234575771649549\\
2	0.234637706820523\\
2	0.235979992084258\\
2	0.236308202763142\\
2	0.236434751351691\\
2	0.23655620569384\\
2	0.236603693778502\\
2	0.238446232508501\\
2	0.238942116073668\\
2	0.23901947807955\\
2	0.240233426984089\\
2	0.240545768099264\\
2	0.240927141500916\\
2	0.241105997807261\\
2	0.242214933101715\\
2	0.242249823141971\\
2	0.243840401902996\\
2	0.243955184513812\\
2	0.24401307411266\\
2	0.244596350614156\\
2	0.247541967523035\\
2	0.249135321664767\\
2	0.249143409277864\\
2	0.250428621115651\\
2	0.251453730669013\\
2	0.252580995007672\\
2	0.252794894160073\\
2	0.253258222402478\\
2	0.253594776367761\\
2	0.253654712164762\\
2	0.254725183321364\\
2	0.255800004628585\\
2	0.256274350387472\\
2	0.256288247449123\\
2	0.258645101409439\\
2	0.258832246347117\\
2	0.259270777278913\\
2	0.260132343661974\\
2	0.26121340450371\\
2	0.261216234464907\\
2	0.264395471631443\\
2	0.268121023607989\\
2	0.268231802040247\\
2	0.269625585087797\\
2	0.269777127618912\\
2	0.273001253479426\\
2	0.275382095857327\\
2	0.277101714184234\\
2	0.277888307470403\\
2	0.278764595211679\\
2	0.279879684658581\\
2	0.280128198448331\\
2	0.281337865323406\\
2	0.281367794801913\\
2	0.28158103135507\\
2	0.281612604584136\\
2	0.282816116167078\\
2	0.284023741998702\\
2	0.284587608851658\\
2	0.285389340339571\\
2	0.28605022947464\\
2	0.28975722144654\\
2	0.292778586502808\\
2	0.294398557898439\\
2	0.295262204148397\\
2	0.295934806753131\\
2	0.296188142134921\\
2	0.296604082378294\\
2	0.299563976610515\\
2	0.299899691583584\\
2	0.301463240546255\\
2	0.302547331098696\\
2	0.303446904415804\\
2	0.30729425506121\\
2	0.308469317996613\\
2	0.310120414273423\\
2	0.310993344417222\\
2	0.31265713608558\\
2	0.313120031000696\\
2	0.31409323923273\\
2	0.314375159095477\\
2	0.318843735930858\\
2	0.319966502259181\\
2	0.323657707898827\\
2	0.324538262620854\\
2	0.32508969024077\\
2	0.326570635484105\\
2	0.332665976761522\\
2	0.333349439767745\\
2	0.333423172964151\\
2	0.335183770011513\\
2	0.337429491099575\\
2	0.340510386002321\\
2	0.341540894234735\\
2	0.343375664623281\\
2	0.346927188361145\\
2	0.347337157908042\\
2	0.347611075789964\\
2	0.349929618213842\\
2	0.355416160421384\\
2	0.356427288294931\\
2	0.363188332192586\\
2	0.363363281467031\\
2	0.36446070734842\\
2	0.367709602883698\\
2	0.370518827464499\\
2	0.370678538317028\\
2	0.37262129497087\\
2	0.378338726415139\\
2	0.379883020921297\\
2	0.386374582078884\\
2	0.389568169885329\\
2	0.39403336860712\\
2	0.397194777824164\\
2	0.401310608814765\\
2	0.402864060475681\\
2	0.407368015507307\\
2	0.408928536861793\\
2	0.413536575473268\\
2	0.414735721721394\\
2	0.4197578864507\\
2	0.422015709796355\\
2	0.425700769623443\\
2	0.4279443619277\\
2	0.4285546609743\\
2	0.431385182345098\\
2	0.434871157100076\\
2	0.440005032714475\\
2	0.441809213231095\\
2	0.445144205612614\\
2	0.44846050779609\\
2	0.450260384926448\\
2	0.450358705123734\\
2	0.450443825426829\\
2	0.464515099960994\\
2	0.46460129889959\\
2	0.467589132465049\\
2	0.470152915329185\\
2	0.476322595470655\\
2	0.477152106878046\\
2	0.483092975164807\\
2	0.492834750473862\\
2	0.496058830300795\\
2	0.498677132395324\\
2	0.517174391910229\\
2	0.51792853246331\\
2	0.525116885123191\\
2	0.527517968961933\\
2	0.530718592588699\\
2	0.540824554667757\\
2	0.54301105508921\\
2	0.547390217813724\\
2	0.550666607132446\\
2	0.551419183273415\\
2	0.556204529910912\\
2	0.563996545430523\\
2	0.572712907977055\\
2	0.575791183825576\\
2	0.578085089865433\\
2	0.58378579204857\\
2	0.590259897966866\\
2	0.626047798937418\\
2	0.633624393527729\\
2	0.648395495964343\\
2	0.651181138440727\\
2	0.654233508074055\\
2	0.678489670215285\\
2	0.707436773953713\\
2	0.710783231238951\\
2	0.732219600664756\\
2	0.734312005085724\\
2	0.751846792906436\\
2	0.757040288106408\\
2	0.806072878190598\\
2	0.869354569516985\\
2	0.870401882366134\\
2	0.88446956910272\\
2	0.886500937697985\\
2	0.896017226830691\\
2	0.899986043101359\\
2	0.908520712080986\\
2	0.941247721298867\\
2	0.956809918143205\\
2	0.982211344961005\\
2	0.99222575187597\\
2	1.02037757875232\\
2	1.05637460467743\\
2	1.0575620022219\\
2	1.05894759864294\\
2	1.07376448837766\\
2	1.08751348782769\\
2	1.11474182004327\\
2	1.1154710999609\\
2	1.17066309469618\\
2	1.19519456108872\\
2	1.20015339303912\\
2	1.22933078655613\\
2	1.24743881530771\\
2	1.25124861879485\\
2	1.28762738014058\\
2	1.29414446416066\\
2	1.36207562779133\\
2	1.3628710461299\\
2	1.38457900093323\\
};
\addplot [color=mycolor1, only marks, mark=+, mark options={solid, fill=black, draw=black}, forget plot]
  table[row sep=crcr]{%
3	0.36143818317275\\
3	0.362813035738776\\
3	0.364807443730622\\
3	0.364896113500318\\
3	0.365200204770598\\
3	0.366628483951652\\
3	0.367052968963711\\
3	0.36903443283554\\
3	0.370485908930324\\
3	0.371463076895701\\
3	0.374080320928592\\
3	0.375864867864623\\
3	0.376632489643893\\
3	0.377009957924656\\
3	0.385150624287988\\
3	0.385251054818201\\
3	0.386728122692501\\
3	0.387584792750824\\
3	0.388125547324884\\
3	0.388621225290699\\
3	0.390039312328981\\
3	0.390777879521253\\
3	0.391135218276284\\
3	0.391351149807488\\
3	0.391477457528366\\
3	0.391554213265605\\
3	0.393430241953808\\
3	0.393502630067356\\
3	0.39404898038051\\
3	0.394474282797551\\
3	0.394966949096003\\
3	0.396977890163312\\
3	0.397266611032717\\
3	0.398651226339837\\
3	0.399306259976146\\
3	0.399881156387462\\
3	0.400921871111301\\
3	0.401194681961982\\
3	0.402848870820769\\
3	0.403383872884932\\
3	0.404048045376308\\
3	0.404351997425092\\
3	0.404443026145794\\
3	0.404556579301116\\
3	0.409805678233131\\
3	0.410392964284044\\
3	0.412217966875532\\
3	0.413694443051875\\
3	0.414588512339977\\
3	0.416712321741071\\
3	0.418468693063777\\
3	0.418651570691355\\
3	0.419318266872568\\
3	0.419332459058419\\
3	0.41954363970303\\
3	0.419923834366552\\
3	0.423377818081732\\
3	0.423668480918631\\
3	0.424399788098127\\
3	0.428231996975047\\
3	0.429189636457147\\
3	0.430860573503298\\
3	0.430915018230181\\
3	0.4309550051505\\
3	0.431450436783795\\
3	0.431970597267689\\
3	0.432668266860519\\
3	0.432823188012042\\
3	0.433047756740208\\
3	0.433442373290517\\
3	0.436022099401029\\
3	0.436789049848696\\
3	0.4408927279979\\
3	0.440962414769057\\
3	0.442537308320138\\
3	0.443427760018534\\
3	0.445080781468984\\
3	0.447507198069712\\
3	0.450381013029133\\
3	0.452129162878257\\
3	0.45485137646534\\
3	0.455071974269854\\
3	0.461769851171553\\
3	0.462414302876764\\
3	0.463439848454253\\
3	0.46485334826979\\
3	0.46495961328194\\
3	0.466823423533189\\
3	0.467211625596841\\
3	0.469523678522061\\
3	0.469896592840795\\
3	0.473192053360008\\
3	0.476678000677277\\
3	0.476951085022368\\
3	0.479372113884939\\
3	0.479384601372188\\
3	0.485531482261612\\
3	0.485583686194339\\
3	0.485819144377306\\
3	0.48870238508585\\
3	0.489341244205157\\
3	0.489771074649722\\
3	0.490726934962878\\
3	0.494643246576281\\
3	0.494998936423776\\
3	0.497967895481426\\
3	0.501856781519496\\
3	0.502576331007027\\
3	0.503644364878473\\
3	0.504740299409093\\
3	0.50546713937452\\
3	0.505480010366598\\
3	0.505857635719816\\
3	0.506338072346776\\
3	0.515258697410923\\
3	0.516018217564699\\
3	0.518898356964701\\
3	0.519024645654151\\
3	0.519074437135613\\
3	0.519860144314426\\
3	0.520553503721571\\
3	0.521037543416604\\
3	0.521131046416374\\
3	0.523634488955842\\
3	0.523728705850425\\
3	0.526402668834943\\
3	0.527119225436494\\
3	0.52786444774222\\
3	0.528212304669938\\
3	0.528605231065071\\
3	0.529831016879571\\
3	0.53239031063897\\
3	0.532632359085571\\
3	0.540569281803177\\
3	0.541472515276052\\
3	0.54572538454412\\
3	0.546895943225714\\
3	0.547709557975099\\
3	0.548163097386141\\
3	0.550752475464858\\
3	0.550931422825456\\
3	0.5521656723847\\
3	0.552572543743728\\
3	0.553498614549424\\
3	0.553890998751175\\
3	0.553930721842048\\
3	0.556366020916886\\
3	0.559580513788886\\
3	0.560816874175715\\
3	0.561579926855857\\
3	0.562348823916369\\
3	0.563259344636736\\
3	0.563659264321379\\
3	0.567442367740645\\
3	0.567691861325084\\
3	0.567902139787986\\
3	0.569190699459848\\
3	0.56979281001683\\
3	0.575721725928649\\
3	0.576174600570723\\
3	0.589754420572117\\
3	0.597856820130381\\
3	0.599660065938826\\
3	0.605477875670442\\
3	0.606036283667264\\
3	0.60733280411258\\
3	0.60835779715608\\
3	0.608421447980612\\
3	0.612533850620872\\
3	0.612970583910221\\
3	0.615114746973123\\
3	0.616095263320841\\
3	0.616244048109798\\
3	0.633865871532682\\
3	0.636358893157839\\
3	0.636564436757318\\
3	0.638513079476601\\
3	0.645076627014999\\
3	0.651419132910344\\
3	0.651687348942583\\
3	0.652110013145524\\
3	0.653888989661286\\
3	0.653995734291545\\
3	0.656149042061224\\
3	0.661540090564009\\
3	0.665338587775404\\
3	0.669371555702525\\
3	0.676574200315478\\
3	0.677010001785951\\
3	0.692169814721077\\
3	0.69418740023044\\
3	0.694310453192842\\
3	0.694346681802834\\
3	0.694818926052822\\
3	0.695855680006741\\
3	0.69595558079414\\
3	0.701681198340305\\
3	0.708005121722763\\
3	0.710369572885123\\
3	0.713074383038601\\
3	0.713811983230476\\
3	0.714660668946484\\
3	0.715905549422814\\
3	0.717639009762395\\
3	0.718448106076727\\
3	0.719136548912399\\
3	0.719442079966936\\
3	0.724632163820642\\
3	0.725073460123996\\
3	0.726418747253271\\
3	0.72812511103367\\
3	0.729432977090993\\
3	0.731463859428239\\
3	0.731936383949775\\
3	0.73252794848395\\
3	0.732607810614641\\
3	0.73400796276116\\
3	0.735572851624689\\
3	0.737060440754905\\
3	0.738351932677863\\
3	0.742250313202194\\
3	0.743902350949525\\
3	0.744494273083093\\
3	0.747565748815721\\
3	0.750300246797745\\
3	0.752797126063904\\
3	0.757079429463527\\
3	0.757402906648088\\
3	0.760878418138714\\
3	0.767543285371128\\
3	0.768474620620899\\
3	0.772151002621063\\
3	0.778990333583125\\
3	0.781915907280384\\
3	0.782982799414148\\
3	0.785525038780177\\
3	0.786826480365805\\
3	0.788950838235942\\
3	0.791234309784847\\
3	0.791360553250135\\
3	0.793236656232435\\
3	0.796124662028729\\
3	0.798449811961627\\
3	0.800180657304407\\
3	0.806230984261206\\
3	0.806281668325758\\
3	0.808490077576202\\
3	0.808701286188711\\
3	0.809852433688188\\
3	0.816079451303166\\
3	0.816393019115296\\
3	0.820307086408386\\
3	0.821112893898992\\
3	0.823383694891307\\
3	0.823394999281549\\
3	0.82382154093275\\
3	0.826376839172855\\
3	0.834386260772566\\
3	0.839303408802713\\
3	0.839470560426354\\
3	0.842353419165579\\
3	0.843593223412893\\
3	0.84782698888832\\
3	0.847975464976543\\
3	0.850615807261184\\
3	0.851699504957114\\
3	0.870404712117617\\
3	0.873249799916303\\
3	0.874757601601312\\
3	0.878199350863125\\
3	0.881370955384064\\
3	0.894705373103677\\
3	0.900555093748585\\
3	0.903345144652029\\
3	0.903346671595248\\
3	0.904209679278781\\
3	0.907107641203106\\
3	0.912902993374278\\
3	0.916438107787474\\
3	0.917367283503718\\
3	0.920334292749495\\
3	0.922742666670688\\
3	0.926384332586627\\
3	0.927775555336636\\
3	0.932689767482691\\
3	0.935072746230659\\
3	0.93601519540693\\
3	0.941165515160082\\
3	0.954954587841144\\
3	0.955634700803175\\
3	0.956287510582417\\
3	0.956998257055599\\
3	0.957343037769204\\
3	0.958233371571101\\
3	0.962295295997754\\
3	0.971189452860255\\
3	0.981369419404516\\
3	0.982661624543653\\
3	0.98843925551414\\
3	0.988488805294013\\
3	0.992299419880853\\
3	0.993206700444948\\
3	0.994895510731161\\
3	0.996676025397699\\
3	1.00310723714239\\
3	1.00710141806989\\
3	1.01591924504596\\
3	1.01960410580772\\
3	1.02645545238477\\
3	1.03898445198522\\
3	1.04177499502041\\
3	1.0444499554553\\
3	1.04604725715857\\
3	1.05432357207953\\
3	1.06403909511708\\
3	1.06523724782224\\
3	1.07126445965924\\
3	1.07718590148277\\
3	1.08815749265729\\
3	1.09111429064504\\
3	1.110671228931\\
3	1.1110651702301\\
3	1.11240909471762\\
3	1.13781225023444\\
3	1.14416317902325\\
3	1.14586718091612\\
3	1.1544052925561\\
3	1.16231721749188\\
3	1.16509070983407\\
3	1.16830570567967\\
3	1.17118808854861\\
3	1.17153025889052\\
3	1.17276548910849\\
3	1.17941879450602\\
3	1.18236745200756\\
3	1.18498463562458\\
3	1.20982660279936\\
3	1.21611945715874\\
3	1.22880397039976\\
3	1.2335735348363\\
3	1.23845279814777\\
3	1.24934528977903\\
3	1.25820081212279\\
3	1.2715830201159\\
3	1.2730060977056\\
3	1.27808370569395\\
3	1.28585371989875\\
3	1.2890352998115\\
3	1.30088426635504\\
3	1.30455230442847\\
3	1.31917288845567\\
3	1.32411625145363\\
3	1.33003411497104\\
3	1.33331806351799\\
3	1.34545252854527\\
3	1.35135374174838\\
3	1.35183442841825\\
3	1.35448710918009\\
3	1.35564822719789\\
3	1.36332080055587\\
3	1.36917643895891\\
3	1.3750006636479\\
3	1.37943941577212\\
3	1.41136797458408\\
3	1.41423992528071\\
3	1.41747149718399\\
3	1.42376437294476\\
3	1.42644432816835\\
3	1.43568655678029\\
3	1.43783170897076\\
3	1.44045202754089\\
3	1.44522711095664\\
3	1.44718327499726\\
3	1.44889898551002\\
3	1.45346477347871\\
3	1.45473292708377\\
3	1.46081856384639\\
3	1.46860710906135\\
3	1.47022633647283\\
3	1.48421157420266\\
3	1.48630685778535\\
3	1.49207455048512\\
3	1.49542510842811\\
3	1.50834105893885\\
3	1.51532258946908\\
3	1.52952378794041\\
3	1.53598918380603\\
3	1.54306221009649\\
3	1.55509181029486\\
3	1.55578978451092\\
3	1.55784737366796\\
3	1.55813075799232\\
3	1.56136564714541\\
3	1.56195062221512\\
3	1.56610311561734\\
3	1.57031909586246\\
3	1.58024722591079\\
3	1.58685860882469\\
3	1.59286670922919\\
3	1.59296212769296\\
3	1.5953225593954\\
3	1.61253021667572\\
3	1.62049085885243\\
3	1.62533588386016\\
3	1.62600549101693\\
3	1.63490470342732\\
3	1.64964310517015\\
3	1.65115291373724\\
3	1.65828353846288\\
3	1.67891977792743\\
3	1.70385628052516\\
3	1.70586060780181\\
3	1.70968958183269\\
3	1.72745975694189\\
3	1.72790533360949\\
3	1.73258151754323\\
3	1.74299709590239\\
3	1.74879119371814\\
3	1.74996670966478\\
3	1.76257270191918\\
3	1.77199794336454\\
3	1.77631438643272\\
3	1.77901654321707\\
3	1.78212948456507\\
3	1.7860498642281\\
3	1.80392926694254\\
3	1.81041322429364\\
3	1.81173685276207\\
3	1.81572863735833\\
3	1.816427952544\\
3	1.84137473706646\\
3	1.84635283064711\\
3	1.85014985465105\\
3	1.85580364516147\\
3	1.86163363059037\\
3	1.86552483183781\\
3	1.89450548627953\\
3	1.90339342308851\\
3	1.904655965406\\
3	1.9165091582763\\
3	1.9309476076864\\
3	1.93401130336724\\
3	1.93859438927716\\
3	1.95598642842026\\
3	1.95686827314079\\
3	1.95701651078272\\
3	1.96326434299287\\
3	1.9674850797379\\
3	1.97508456861392\\
3	1.98971497075089\\
3	2.01071549961044\\
3	2.01511263689928\\
3	2.01996142871007\\
3	2.02521843636044\\
3	2.02653532159168\\
3	2.02810141998802\\
3	2.03873082709032\\
3	2.05659782053071\\
3	2.08103944350338\\
3	2.08354364859213\\
3	2.10927609545393\\
3	2.11853336701888\\
3	2.12654602971295\\
3	2.13778442022992\\
3	2.13984776874227\\
3	2.13991534532973\\
3	2.17548480567012\\
3	2.17766863157898\\
3	2.18471384277706\\
3	2.19404483261823\\
3	2.19668321252788\\
3	2.2068991383949\\
3	2.21073573179611\\
3	2.21171663328422\\
3	2.22148420813995\\
3	2.24840598345615\\
3	2.26603360854096\\
3	2.28138119213474\\
3	2.28502805035535\\
3	2.29521983740702\\
3	2.29667952605331\\
3	2.29855289071585\\
3	2.3034524030982\\
3	2.30634453842702\\
3	2.30942368618964\\
3	2.30959452599448\\
3	2.3159685076717\\
3	2.31721537505987\\
3	2.3284749286634\\
3	2.34330559983705\\
3	2.34764325810685\\
3	2.38895939817595\\
3	2.39125542204318\\
3	2.39243872414688\\
3	2.40417001636631\\
3	2.40812152104302\\
3	2.42393862339475\\
3	2.45367883506446\\
3	2.45614385915961\\
3	2.45842063614358\\
3	2.46030307779741\\
3	2.47014061215046\\
3	2.4828959537843\\
3	2.48428241439256\\
3	2.48964680906385\\
3	2.53749326984238\\
3	2.54449802335302\\
3	2.55848773141554\\
3	2.56661564382953\\
3	2.57868645802542\\
3	2.61882627321017\\
3	2.62167304639112\\
3	2.64670527442347\\
3	2.65894029035108\\
3	2.68518669383067\\
3	2.68612711221464\\
3	2.68691801625002\\
3	2.6975744797564\\
3	2.70617043874819\\
3	2.72354937405416\\
3	2.72450054532624\\
3	2.72603851068064\\
3	2.73496908446982\\
3	2.76646735728328\\
3	2.78167348234018\\
3	2.80665612966505\\
3	2.82334694954037\\
3	2.82434469036839\\
3	2.83768291174759\\
3	2.87279542184519\\
3	2.87723472532586\\
3	2.88678170332265\\
3	2.89161239740509\\
3	2.8958751485395\\
3	2.90086750799593\\
3	2.92052931790552\\
3	2.93289944245122\\
3	2.93742139237351\\
3	2.9605825879335\\
3	2.97609995120132\\
3	3.01716061583913\\
3	3.02051039907504\\
3	3.03452917408131\\
3	3.03929992025683\\
3	3.06714691632053\\
3	3.07825417719469\\
3	3.0893992900054\\
3	3.12331262475613\\
3	3.13501981805186\\
3	3.15559897877648\\
3	3.15888862380511\\
3	3.20657092871004\\
3	3.22143047962147\\
3	3.2251316005368\\
3	3.23126646005095\\
3	3.30331857184658\\
3	3.30848022143112\\
3	3.31855693774441\\
3	3.3309331297951\\
3	3.33331586496567\\
3	3.33443833601924\\
3	3.37222228220287\\
3	3.38457049154867\\
3	3.40588648093932\\
3	3.4289871972637\\
3	3.50966746428425\\
3	3.55835090570666\\
3	3.60345627492858\\
3	3.64294851177873\\
3	3.65877459318436\\
3	3.70712141727394\\
3	3.74442415443631\\
3	3.75220567959255\\
3	3.76914981928422\\
3	3.771048899743\\
3	3.78383469305058\\
3	3.79222755799649\\
3	3.79908528745444\\
3	3.82333153403008\\
3	3.84338250681232\\
3	3.87887202369784\\
3	3.88438766556582\\
3	3.90601130886098\\
3	3.90842044122605\\
3	3.92314892025415\\
3	3.92663065439336\\
3	3.9766260214813\\
3	4.09685859728969\\
3	4.12307320225104\\
3	4.27806695642014\\
3	4.36057504039169\\
3	4.39716252309543\\
3	4.40387981185406\\
3	4.53856341333728\\
3	4.54546787387588\\
3	4.55266288306834\\
3	4.58005386079221\\
3	4.60544601523522\\
3	4.63274837955202\\
3	4.64895818317226\\
3	4.65601867954272\\
3	4.6627183523656\\
3	4.74509474155412\\
3	4.99029378789894\\
3	5.01392124032212\\
3	5.01965594376816\\
3	5.02466172487338\\
3	5.04777803667334\\
3	5.06014868493198\\
3	5.06025771006288\\
3	5.08442580609675\\
3	5.10627078054395\\
3	5.11166877509619\\
3	5.13230091503886\\
3	5.20982944072318\\
3	5.41387545512164\\
3	5.41581727021335\\
3	5.42112777256538\\
3	5.45583282734889\\
3	5.496916900688\\
3	5.51651588236574\\
3	5.53755914095189\\
3	5.65179770586067\\
3	5.72094499010419\\
3	5.72445619608725\\
3	5.74271139419971\\
3	5.78426647121502\\
3	5.88453264805805\\
3	5.94145878060744\\
3	5.94446150363778\\
3	6.06372929680271\\
3	6.16357637687066\\
3	6.18870115316351\\
3	6.22496755605127\\
3	6.39939016968274\\
3	6.47638140119018\\
3	6.50626259138389\\
3	6.50884840619765\\
3	6.55945889475895\\
3	6.63080451854071\\
3	6.6580351660736\\
3	6.67938367166258\\
3	6.68677906847739\\
3	6.72235161526103\\
3	6.73561722604616\\
3	6.86245196467597\\
3	6.91446579758066\\
3	6.91798410338717\\
3	6.953052522513\\
3	7.01469797167091\\
3	7.11389204157933\\
3	7.2932863407762\\
3	7.32467645903896\\
3	7.36038953252933\\
3	7.3939264366484\\
3	7.45502947452606\\
3	7.68623974226725\\
3	7.76387466383292\\
3	7.79805641414598\\
3	7.83688876991698\\
3	7.92329561392296\\
3	7.92719261912433\\
3	7.93164257875926\\
3	7.95208240994172\\
3	8.00440475824552\\
3	8.10504878256903\\
3	8.16327826632266\\
3	8.16612386990266\\
3	8.19272823882309\\
3	8.42188117723003\\
3	8.45528973845345\\
3	8.65386859905654\\
3	8.70856983510833\\
3	8.74794060148864\\
3	8.77101179440452\\
3	8.80960928679303\\
3	8.84743884276927\\
3	8.86002180705619\\
3	8.92194903934748\\
3	9.0415506811819\\
3	9.10075750510023\\
3	9.14958432805449\\
3	9.33125618541799\\
3	9.35243755509489\\
3	9.41404817165332\\
3	9.4662929474576\\
3	9.51560985610249\\
3	9.56945054541745\\
3	9.65822877945276\\
3	10.0185947237988\\
3	10.0725574945165\\
3	10.1031338710951\\
3	10.207678172127\\
3	10.2618227847401\\
3	10.4022468931047\\
3	10.422984740122\\
3	10.4725812582752\\
3	10.6662443373353\\
3	10.6885888372419\\
3	10.7671609338938\\
3	10.8261031569829\\
3	10.9504792438571\\
3	11.0182350577668\\
3	11.0910207524314\\
3	11.1819541187613\\
3	11.3063608393617\\
3	11.5518880033519\\
3	11.5633268304358\\
3	11.5860979001028\\
3	11.6040543126213\\
3	11.6337317563751\\
3	11.7209958799855\\
3	12.355414938545\\
3	12.4432354717935\\
3	12.5609047765926\\
3	12.641612728269\\
3	12.8342714952765\\
3	13.0521661118246\\
3	13.2339836189496\\
3	13.7852447182785\\
3	13.9589436050297\\
3	13.9811048289065\\
3	13.986371121545\\
3	14.0440849809202\\
3	14.0755711799266\\
3	15.0035848753097\\
3	15.2606833057913\\
3	15.4967693100527\\
3	15.8655406950492\\
3	16.0402380040474\\
3	16.0860181637742\\
3	16.3733277772431\\
3	17.2429118370184\\
3	18.2117903305029\\
3	18.2236611517041\\
3	18.3824581798352\\
3	18.5343120481947\\
3	19.0725807745999\\
3	19.4882514824557\\
3	20.2563255543226\\
};
\addplot [color=mycolor1, only marks, mark=+, mark options={solid, fill=black, draw=black}, forget plot]
  table[row sep=crcr]{%
4	7.72084358179715\\
4	7.72084358179715\\
4	7.72084358179715\\
4	7.72084358179715\\
4	7.72084358179715\\
4	7.72084358179715\\
4	7.72084358179715\\
4	7.72084358179715\\
4	7.72084358179715\\
4	7.72084358179715\\
4	7.72084358179715\\
4	7.72084358179715\\
4	7.72084358179715\\
4	7.72084358179715\\
4	7.72084358179715\\
4	7.72084358179715\\
4	7.72084358179715\\
4	7.72084358179715\\
4	7.72084358179715\\
4	7.72084358179715\\
4	7.72084358179715\\
4	7.72084358179715\\
4	7.72084358179715\\
4	7.72084358179715\\
4	7.72084358179715\\
4	7.72084358179715\\
4	7.72084358179715\\
4	7.72084358179715\\
4	7.72084358179715\\
4	7.72084358179715\\
4	7.72084358179715\\
4	7.72084358179715\\
4	7.72084358179715\\
4	7.72084358179715\\
4	7.72084358179715\\
4	7.72084358179715\\
4	7.72084358179715\\
4	7.72084358179715\\
4	7.72084358179715\\
4	7.72084358179715\\
4	7.72084358179715\\
4	7.72084358179715\\
4	7.72084358179715\\
4	7.72084358179715\\
4	7.72084358179715\\
4	7.72084358179715\\
4	7.72084358179715\\
4	7.72084358179715\\
4	7.72084358179715\\
4	7.72084358179715\\
4	7.72084358179715\\
4	7.72084358179715\\
4	7.72084358179715\\
4	7.72084358179715\\
4	7.72084358179715\\
4	7.72084358179715\\
4	7.72084358179715\\
4	7.72084358179715\\
4	7.72084358179715\\
4	7.72084358179715\\
4	7.72084358179715\\
4	7.72084358179715\\
4	7.72084358179715\\
4	7.72084358179715\\
4	7.72084358179715\\
4	7.72084358179715\\
4	7.72084358179715\\
4	7.72084358179715\\
4	7.72084358179715\\
4	7.72084358179715\\
4	7.72084358179715\\
4	7.72084358179715\\
4	7.72084358179715\\
4	7.72084358179715\\
4	7.72084358179715\\
4	7.72084358179715\\
4	7.72084358179715\\
4	7.72084358179715\\
4	7.72084358179715\\
4	7.72084358179715\\
4	7.72084358179715\\
4	7.72084358179715\\
4	7.72084358179715\\
4	7.72084358179715\\
4	7.72084358179715\\
4	7.72084358179715\\
4	7.72084358179715\\
4	7.72084358179715\\
4	7.72084358179715\\
4	7.72084358179715\\
4	7.72084358179715\\
4	7.72084358179715\\
4	7.72084358179715\\
4	7.72084358179715\\
4	7.72084358179715\\
4	7.72084358179715\\
4	7.72084358179715\\
4	7.72084358179715\\
4	7.72084358179715\\
4	7.72084358179715\\
4	7.72084358179715\\
4	7.72084358179715\\
4	7.72084358179715\\
4	7.72084358179715\\
4	7.72084358179715\\
4	7.72084358179715\\
4	7.72084358179715\\
4	7.72084358179715\\
4	7.72084358179715\\
4	7.72084358179715\\
4	7.72084358179715\\
4	7.72084358179715\\
4	7.72084358179715\\
4	7.72084358179715\\
4	7.72084358179715\\
4	7.72084358179715\\
4	7.72084358179715\\
4	7.72084358179715\\
4	7.72084358179715\\
4	7.72084358179715\\
4	7.72084358179715\\
4	7.72084358179715\\
4	7.72084358179715\\
4	7.72084358179715\\
4	7.72084358179715\\
4	7.72084358179715\\
4	7.72084358179715\\
4	7.72084358179715\\
4	7.72084358179715\\
4	7.72084358179715\\
4	7.72084358179715\\
4	7.72084358179715\\
4	7.72084358179715\\
4	7.72084358179715\\
4	7.72084358179715\\
4	7.72084358179715\\
4	7.72084358179715\\
4	7.72084358179715\\
4	7.72084358179715\\
4	7.72084358179715\\
4	7.72084358179715\\
4	7.72084358179715\\
4	7.72084358179715\\
4	7.72084358179715\\
4	7.72084358179715\\
4	7.72084358179715\\
4	7.72084358179715\\
4	7.72084358179715\\
4	7.72084358179715\\
4	7.72084358179715\\
4	7.72084358179715\\
};
\addplot [color=mycolor5, only marks, mark=+, mark options={solid, fill=black, draw=black}, forget plot]
  table[row sep=crcr]{%
5	1.05434598272101\\
5	1.05491856071683\\
5	1.05498860133936\\
5	1.05568997221872\\
5	1.05569210348009\\
5	1.0559006714554\\
5	1.05604067001545\\
5	1.05628976306448\\
5	1.05823952292405\\
5	1.06022642680484\\
5	1.06084750016778\\
5	1.06236266582448\\
5	1.06321949199173\\
5	1.06351365415265\\
5	1.06368232389521\\
5	1.06492113278786\\
5	1.06527286600249\\
5	1.06583217495165\\
5	1.06660217152305\\
5	1.07000590924387\\
5	1.0704536568854\\
5	1.07159242730246\\
5	1.0729222778054\\
5	1.07371094055792\\
5	1.07381067485643\\
5	1.07594148934305\\
5	1.07715171316474\\
5	1.0772998963006\\
5	1.07767627132575\\
5	1.07801293257155\\
5	1.07916962454631\\
5	1.07990858992333\\
5	1.0804864642634\\
5	1.08076449325435\\
5	1.08077182891614\\
5	1.08335808908542\\
5	1.08413272589397\\
5	1.08464394742803\\
5	1.08535770924168\\
5	1.08565010349031\\
5	1.08630894245723\\
5	1.08654592884627\\
5	1.08771622869114\\
5	1.08782940030697\\
5	1.08908130349473\\
5	1.08934787570361\\
5	1.0897036226651\\
5	1.09008763042965\\
5	1.09012242889369\\
5	1.09220121733289\\
5	1.09248415022924\\
5	1.0929506206001\\
5	1.09355299259792\\
5	1.09440446519822\\
5	1.09450525330388\\
5	1.09702776125151\\
5	1.09734579338449\\
5	1.09737157587141\\
5	1.09737975524783\\
5	1.09749888334601\\
5	1.09751207939987\\
5	1.09901402965345\\
5	1.0993020534607\\
5	1.09968021649053\\
5	1.1003895990996\\
5	1.10082049995996\\
5	1.10139670272107\\
5	1.10296512290318\\
5	1.10425623185932\\
5	1.10645339836055\\
5	1.10705704980066\\
5	1.11363900396896\\
5	1.11375244629246\\
5	1.11411511115506\\
5	1.11414793800966\\
5	1.11456002079768\\
5	1.11458530368099\\
5	1.11499489196142\\
5	1.11507890273017\\
5	1.11513111178531\\
5	1.11520774919371\\
5	1.11893623127504\\
5	1.11919557035335\\
5	1.11920642244416\\
5	1.11990061056741\\
5	1.12003738580328\\
5	1.12206419646869\\
5	1.12224825096743\\
5	1.12378019055074\\
5	1.12636256425005\\
5	1.12731400593143\\
5	1.12850413421151\\
5	1.12850802615712\\
5	1.13024976146791\\
5	1.1303197147034\\
5	1.13245793173268\\
5	1.13257320435391\\
5	1.13277229250589\\
5	1.1342537244483\\
5	1.13562260888641\\
5	1.13610467757642\\
5	1.13782155846462\\
5	1.13827829643715\\
5	1.13929408420358\\
5	1.14053489430838\\
5	1.14054218678268\\
5	1.14090042514197\\
5	1.14093412554456\\
5	1.14096905543383\\
5	1.14122836827585\\
5	1.14172366885815\\
5	1.1421687755653\\
5	1.14356321975254\\
5	1.1440061099213\\
5	1.14450668115083\\
5	1.14611519652929\\
5	1.14646193940294\\
5	1.14807367039319\\
5	1.14828496979523\\
5	1.14930471790299\\
5	1.14981267166733\\
5	1.15002695361999\\
5	1.15029386880314\\
5	1.15100304563681\\
5	1.1511088261366\\
5	1.15292938120356\\
5	1.15478332629316\\
5	1.15490696721023\\
5	1.15538305703694\\
5	1.15579563658518\\
5	1.15596136718305\\
5	1.15653075159945\\
5	1.15691664364611\\
5	1.15818055722794\\
5	1.15936855082113\\
5	1.15964810202601\\
5	1.15967069869614\\
5	1.16091045890421\\
5	1.16324186833671\\
5	1.16369467130673\\
5	1.16403436282401\\
5	1.16557598332235\\
5	1.16573001934358\\
5	1.16612351217633\\
5	1.16985921602351\\
5	1.17032034883449\\
5	1.17043488247652\\
5	1.17235143746384\\
5	1.1725719622447\\
5	1.17313607926401\\
5	1.17448016707552\\
5	1.17685613514718\\
5	1.17734627192756\\
5	1.17834711285313\\
5	1.17875730977379\\
5	1.18436956761269\\
5	1.18453954436955\\
5	1.18607462124532\\
5	1.18637679186216\\
5	1.18903266862401\\
5	1.19053509514569\\
5	1.19094042592397\\
5	1.19126501306955\\
5	1.19205828390061\\
5	1.19599794089888\\
5	1.19603254321462\\
5	1.19605195295297\\
5	1.19684318389873\\
5	1.19775765979894\\
5	1.19957284265931\\
5	1.20205538113415\\
5	1.20718416705858\\
5	1.21033085025244\\
5	1.21831438167861\\
5	1.22436468154959\\
5	1.22601528556079\\
5	1.23054258454879\\
5	1.23106357489578\\
5	1.23177959273262\\
5	1.23192773030105\\
5	1.23562957071865\\
5	1.23710719659915\\
5	1.23852750346715\\
5	1.24171239642985\\
5	1.24285230990333\\
5	1.25086815027533\\
5	1.25431599174658\\
5	1.25945741079212\\
5	1.25955905607161\\
5	1.25963100418415\\
5	1.26132895621214\\
5	1.2624636770324\\
5	1.27553085411399\\
5	1.29076705813637\\
5	1.29246709710138\\
5	1.2992865271752\\
5	1.30282520583073\\
5	1.3045421460028\\
5	1.30719153902024\\
5	1.3089334387252\\
5	1.31456387773642\\
5	1.31485580007347\\
5	1.31710293176273\\
5	1.31775683160775\\
5	1.31891217128042\\
5	1.32224896114407\\
5	1.32539883805059\\
5	1.33036152596493\\
5	1.33298822679705\\
5	1.34607603186262\\
5	1.34616753660016\\
5	1.34692277169047\\
5	1.34919615474658\\
5	1.35591827896518\\
5	1.35722801786655\\
5	1.35791959021983\\
5	1.35840526492834\\
5	1.36031789654142\\
5	1.36142054512848\\
5	1.36433682072138\\
5	1.36784523712514\\
5	1.37456902332672\\
5	1.37695108622402\\
5	1.37787489732763\\
5	1.38104568558649\\
5	1.38473495066715\\
5	1.38480397563892\\
5	1.38675965758896\\
5	1.38853260729542\\
5	1.39111815069152\\
5	1.39125091492734\\
5	1.39296271364252\\
5	1.39395939920977\\
5	1.39776162833665\\
5	1.40735489122411\\
5	1.41144704452424\\
5	1.41743364392081\\
5	1.41920466902407\\
5	1.41961149770963\\
5	1.42060483872642\\
5	1.42865044318484\\
5	1.43068069430561\\
5	1.43530950752732\\
5	1.43817119108525\\
5	1.43947812427051\\
5	1.4413900175228\\
5	1.44208956434735\\
5	1.4454401460566\\
5	1.44586951590269\\
5	1.44697484256194\\
5	1.4510770984491\\
5	1.45148901952166\\
5	1.45943995830864\\
5	1.46117285450718\\
5	1.46748087753037\\
5	1.47465754293892\\
5	1.48270815624867\\
5	1.48591104051358\\
5	1.48833017567469\\
5	1.49068276054604\\
5	1.49361826330758\\
5	1.49854336104327\\
5	1.49975344723958\\
5	1.49992975499195\\
5	1.51507826205326\\
5	1.53063457812075\\
5	1.54010453700446\\
5	1.55251092084168\\
5	1.55260356652526\\
5	1.55738641726428\\
5	1.55847110191084\\
5	1.55898578321467\\
5	1.56115361566466\\
5	1.56282557170339\\
5	1.61346196648861\\
5	1.64063376090285\\
5	1.64678942382513\\
5	1.65566767452589\\
5	1.66245433192396\\
5	1.70879707121443\\
5	1.71421139229287\\
5	1.7230822003487\\
5	1.72395391254918\\
5	1.72509501859571\\
5	1.7915703774323\\
5	1.80336198964273\\
5	1.81852916013945\\
5	1.82976367048707\\
5	1.83703199604727\\
5	1.8431483809782\\
5	1.86278961576939\\
5	1.89672659817175\\
5	1.91299650078031\\
5	1.92653237758686\\
5	1.95030821845631\\
5	1.98848163914857\\
5	1.99798250626391\\
5	2.00573103988968\\
5	2.00690586315177\\
5	2.03227707357649\\
5	2.05591443473498\\
5	2.09088820681756\\
5	2.10157588045896\\
5	2.1299485677763\\
5	2.17099944328227\\
5	2.19806554093426\\
5	2.20501279773997\\
5	2.22340815841459\\
5	2.23354552152835\\
5	2.24361979095836\\
5	2.26105470439201\\
};
\end{axis}
\end{tikzpicture}%
% \end{document}

%% file: fig/Combined_Power_PercentDecrease_BarPlot.tex
\definecolor{mycolor2}{rgb}{0.12941,0.12941,0.12941}%
\definecolor{mycolor4}{rgb}{0.60000,0.30000,0.10000}% Brown
\definecolor{mycolor5}{rgb}{0.95,0.00,0.00}% Very dark red
\usetikzlibrary{patterns}
\renewcommand{\Huge}{\fontsize{35}{42}\selectfont}
\begin{tikzpicture}
\begin{axis}[%
width=3.25in,
height=5in,
ytick distance=20,
at={(1.625in,2.452in)},
scale only axis,
bar shift auto,
xmin=0.25,
xmax=2.75,
xtick={1,2},
xticklabels={{Interf. Nulling ($\lambda$=0) \& Power Control ($\epsilon$=0.85)},{Interf. Nulling ($\lambda$=0.1) \& Power Control ($\epsilon$=0.98)}},
xticklabel style={rotate=90, anchor=east, font=\Huge},
yticklabel style={font=\Huge},
xlabel style={font=\Huge},
ymin=0,
ymax=80,
ylabel style={font=\Huge},
ylabel={Power Percentage Decrease [\%]},
axis background/.style={fill=white},
xmajorgrids,
ymajorgrids
]
% Bar 1 - Brown with north east lines (diagonal)
\draw[pattern=north east lines, pattern color=mycolor4, draw=mycolor2] (axis cs:0.675,0) rectangle (axis cs:1.325,78.8519703093871);
% Bar 2 - Red with crosshatch
\draw[pattern=crosshatch, pattern color=mycolor5, draw=mycolor2] (axis cs:1.675,0) rectangle (axis cs:2.325,18.6858407834501);
\addplot[forget plot, color=mycolor2] table[row sep=crcr] {%
-0.225    0\\
3.225    0\\
};
\addplot [color=black, dashed, forget plot]
  table[row sep=crcr]{%
-0.225    0\\
3.225    0\\
};
\end{axis}
\end{tikzpicture}%
% \end{document}

%% file: fig/lambda1_rss_degradation_cdf_NoSats2.tex
\definecolor{mycolor1}{rgb}{0.06600,0.44300,0.74500}%
\definecolor{mycolor2}{rgb}{0.86600,0.32900,0.00000}%
\definecolor{mycolor3}{rgb}{0.92900,0.69400,0.12500}%
\definecolor{mycolor4}{rgb}{0.52100,0.08600,0.81900}%
\definecolor{mycolor5}{rgb}{0.12941,0.12941,0.12941}%
\begin{tikzpicture}

\begin{axis}[%
width=8.88in,
height=3in,
at={(1.49in,0.738in)},
scale only axis,
xmin=0,
xmax=2.25,
xlabel={RSS Degradation [dB]},
ymin=0,
ymax=1,
ylabel={CDF},
axis background/.style={fill=white},
axis lines=box,
xmajorgrids,
ymajorgrids,
xtick distance=0.25,
ytick distance=0.2,
% Font sizes
% tick label style={font=\LARGE},
tick label style={font=\fontsize{30pt}{36pt}\selectfont},
xlabel style={font=\Huge},
ylabel style={font=\Huge},
legend style={font=\fontsize{28pt}{33pt}\selectfont, at={(0.97,0.03)}, anchor=south east, legend cell align=left, align=left}
]
\addplot [color=mycolor1, line width=4.0pt, mark=diamond*, mark size=6pt, mark repeat=3]
  table[row sep=crcr]{%
0.00185246207490491	0.0333333333333333\\
0.00509833290503491	0.0666666666666667\\
0.00809021092469777	0.1\\
0.0146075921803634	0.133333333333333\\
0.0160225575306208	0.166666666666667\\
0.0221507919148604	0.2\\
0.0319431163924386	0.233333333333333\\
0.0345364472297768	0.266666666666667\\
0.0347671416410916	0.3\\
0.0573645340633523	0.333333333333333\\
0.0825293959738525	0.366666666666667\\
0.113088431759504	0.4\\
0.114547951820576	0.433333333333333\\
0.144701775987775	0.466666666666667\\
0.158769357969512	0.5\\
0.230977863586902	0.533333333333333\\
0.293532237685226	0.566666666666667\\
0.302134751317143	0.6\\
0.344620881876663	0.633333333333333\\
0.350458822258672	0.666666666666667\\
0.565976394958641	0.7\\
0.581580907905105	0.733333333333333\\
0.582749627679227	0.766666666666667\\
0.709188839354131	0.8\\
0.784143570896785	0.833333333333333\\
0.864392213439205	0.866666666666667\\
0.885804064515884	0.9\\
1.54448563416248	0.933333333333333\\
2.07196611415236	0.966666666666667\\
2.13603830802803	1\\
};
\addlegendentry{$N_\text{t}=4\cdot4$}

\addplot [color=mycolor2, line width=4.0pt, mark=o, mark size=6pt, mark repeat=3]
  table[row sep=crcr]{%
6.40404585158867e-05	0.0333333333333333\\
0.000288495681511648	0.0666666666666667\\
0.000818528114426156	0.1\\
0.000862901919099855	0.133333333333333\\
0.000989036175149204	0.166666666666667\\
0.00101421100033435	0.2\\
0.00108147914424792	0.233333333333333\\
0.00111143922880703	0.266666666666667\\
0.0011833240804884	0.3\\
0.00134089297166161	0.333333333333333\\
0.00175659525129856	0.366666666666667\\
0.00223266171781281	0.4\\
0.0022655636365451	0.433333333333333\\
0.00331225459686888	0.466666666666667\\
0.00420602750156023	0.5\\
0.00566927972439358	0.533333333333333\\
0.00688299639290808	0.566666666666667\\
0.00865094193644288	0.6\\
0.0108205636699158	0.633333333333333\\
0.0123896889310632	0.666666666666667\\
0.023579116514832	0.7\\
0.0240075444170943	0.733333333333333\\
0.0344349230054297	0.766666666666667\\
0.0378774399232419	0.8\\
0.03968767678035	0.833333333333333\\
0.0918590485979865	0.866666666666667\\
0.206378893222341	0.9\\
0.293936616270972	0.933333333333333\\
1.18413294276332	0.966666666666667\\
1.61189597845284	1\\
};
\addlegendentry{$N_\text{t}=8\cdot8$}

\addplot [color=mycolor3, line width=4.0pt, mark=square*, mark size=6pt, mark repeat=3]
  table[row sep=crcr]{%
4.83186715542002e-06	0.0333333333333333\\
7.15451905509577e-06	0.0666666666666667\\
8.07984918448124e-06	0.1\\
8.45377660705411e-06	0.133333333333333\\
1.80876026339194e-05	0.166666666666667\\
0.000112193983191309	0.2\\
0.000126550352795987	0.233333333333333\\
0.000127111515119988	0.266666666666667\\
0.000141837284801798	0.3\\
0.000169173352690975	0.333333333333333\\
0.000178873454215494	0.366666666666667\\
0.000179176420447109	0.4\\
0.00018232685232262	0.433333333333333\\
0.000204637029501553	0.466666666666667\\
0.000413461381619056	0.5\\
0.00045873874214206	0.533333333333333\\
0.000538547244257568	0.566666666666667\\
0.000578965181708905	0.6\\
0.000748809661357579	0.633333333333333\\
0.000767112114162681	0.666666666666667\\
0.000801422919700636	0.7\\
0.000810577861862333	0.733333333333333\\
0.00464953053821828	0.766666666666667\\
0.00625083938742894	0.8\\
0.0089135723692456	0.833333333333333\\
0.0115159027884242	0.866666666666667\\
0.0149595151533708	0.9\\
0.0177762685689154	0.933333333333333\\
0.331838926262096	0.966666666666667\\
0.868238749945629	1\\
};
\addlegendentry{$N_\text{t}=16\cdot16$}

\addplot [color=mycolor4, line width=4.0pt, mark=x, mark size=6pt, mark repeat=3]
  table[row sep=crcr]{%
2.0708128085789e-06	0.0333333333333333\\
2.88999527562197e-06	0.0666666666666667\\
3.84238212270627e-06	0.1\\
7.29092175551612e-06	0.133333333333333\\
7.62676942353363e-06	0.166666666666667\\
8.29067612927337e-06	0.2\\
8.58479951520024e-06	0.233333333333333\\
9.50307013301094e-06	0.266666666666667\\
1.02855259547355e-05	0.3\\
2.05461723721586e-05	0.333333333333333\\
2.07409576262786e-05	0.366666666666667\\
2.10356916191024e-05	0.4\\
2.60770661809399e-05	0.433333333333333\\
2.76275140205225e-05	0.466666666666667\\
2.82479329229413e-05	0.5\\
2.8749855124729e-05	0.533333333333333\\
3.06723190623979e-05	0.566666666666667\\
5.9056737232768e-05	0.6\\
7.53634905244553e-05	0.633333333333333\\
0.000111811807117766	0.666666666666667\\
0.000115298117781328	0.7\\
0.000117724686314078	0.733333333333333\\
0.000223232690936574	0.766666666666667\\
0.00026503858686534	0.8\\
0.000560141635132469	0.833333333333333\\
0.000971968296514627	0.866666666666667\\
0.0009827151714882	0.9\\
0.00105036974027681	0.933333333333333\\
0.00387157926862293	0.966666666666667\\
0.10575336679986	1\\
};
\addlegendentry{$N_\text{t}=32\cdot32$}

\end{axis}
\end{tikzpicture}%

%% file: fig/lambda1_rss_degradation_cdf_NoSats10.tex
\definecolor{mycolor1}{rgb}{0.06600,0.44300,0.74500}%
\definecolor{mycolor2}{rgb}{0.86600,0.32900,0.00000}%
\definecolor{mycolor3}{rgb}{0.92900,0.69400,0.12500}%
\definecolor{mycolor4}{rgb}{0.52100,0.08600,0.81900}%
\definecolor{mycolor5}{rgb}{0.12941,0.12941,0.12941}%
\begin{tikzpicture}

\begin{axis}[%
width=8.88in,
height=3in,
at={(1.49in,0.738in)},
scale only axis,
xmin=0,
xmax=10,
xlabel={RSS Degradation [dB]},
ymin=0,
ymax=1,
ylabel={CDF},
axis background/.style={fill=white},
axis lines=box,
xmajorgrids,
ymajorgrids,
xtick distance=1,
ytick distance=0.2,
% Font sizes
% tick label style={font=\LARGE},
tick label style={font=\fontsize{30pt}{36pt}\selectfont},
xlabel style={font=\Huge},
ylabel style={font=\Huge},
legend style={font=\fontsize{28pt}{33pt}\selectfont, at={(0.97,0.03)}, anchor=south east, legend cell align=left, align=left}
]
\addplot [color=mycolor1, line width=4.0pt, mark=diamond*, mark size=6pt, mark repeat=3]
  table[row sep=crcr]{%
0.0611440028466047	0.0333333333333333\\
0.100743603567605	0.0666666666666667\\
0.14534769003968	0.1\\
0.201912315325306	0.133333333333333\\
0.258920696658482	0.166666666666667\\
0.259483663777716	0.2\\
0.307993632588901	0.233333333333333\\
0.331420025696819	0.266666666666667\\
0.365219562586999	0.3\\
0.36900418818323	0.333333333333333\\
0.423691322314855	0.366666666666667\\
0.425662052392153	0.4\\
0.514900022029413	0.433333333333333\\
0.737490770618864	0.466666666666667\\
0.778060699456951	0.5\\
0.792184639770753	0.533333333333333\\
0.812882512734538	0.566666666666667\\
0.851820888814258	0.6\\
0.870269136902936	0.633333333333333\\
0.897718754205092	0.666666666666667\\
1.41036138623223	0.7\\
1.64317377297736	0.733333333333333\\
1.70150220128834	0.766666666666667\\
1.97316351038906	0.8\\
2.27037694445967	0.833333333333333\\
2.67528871416514	0.866666666666667\\
2.84343559614091	0.9\\
7.67488843263015	0.933333333333333\\
9.42089859911793	0.966666666666667\\
9.80084229387366	1\\
};
\addlegendentry{$N_\text{t}=4\cdot4$}

\addplot [color=mycolor2, line width=4.0pt, mark=o, mark size=6pt, mark repeat=3]
  table[row sep=crcr]{%
0.00199443846603832	0.0333333333333333\\
0.00275707072813541	0.0666666666666667\\
0.00643708539460161	0.1\\
0.00667511643362909	0.133333333333333\\
0.00861901708103901	0.166666666666667\\
0.0095643898564886	0.2\\
0.0119733279778033	0.233333333333333\\
0.013962290220192	0.266666666666667\\
0.0168266379426772	0.3\\
0.021640685517219	0.333333333333333\\
0.0286467396706449	0.366666666666667\\
0.0438161843529442	0.4\\
0.0472933171248899	0.433333333333333\\
0.0537858878712069	0.466666666666667\\
0.056543150222863	0.5\\
0.0638151494935463	0.533333333333333\\
0.0660087405787577	0.566666666666667\\
0.0704007964711532	0.6\\
0.0781736128380582	0.633333333333333\\
0.0797576951823083	0.666666666666667\\
0.0812009973875893	0.7\\
0.0900530781377855	0.733333333333333\\
0.103750521935142	0.766666666666667\\
0.175852416937464	0.8\\
0.22483002423008	0.833333333333333\\
0.887701368052659	0.866666666666667\\
0.989062753346291	0.9\\
2.23242487879653	0.933333333333333\\
3.50902428301262	0.966666666666667\\
5.84368538391629	1\\
};
\addlegendentry{$N_\text{t}=8\cdot8$}

\addplot [color=mycolor3, line width=4.0pt, mark=square*, mark size=6pt, mark repeat=3]
  table[row sep=crcr]{%
0.000378498824302296	0.0333333333333333\\
0.00037925857025216	0.0666666666666667\\
0.00043884349511449	0.1\\
0.000546481070850689	0.133333333333333\\
0.000878565368150282	0.166666666666667\\
0.000918662965218198	0.2\\
0.001072092466995	0.233333333333333\\
0.00131667862896512	0.266666666666667\\
0.0016585418203277	0.3\\
0.00292325662994008	0.333333333333333\\
0.00366619646133799	0.366666666666667\\
0.00372669288522728	0.4\\
0.0037949628634118	0.433333333333333\\
0.00616204000413708	0.466666666666667\\
0.00709423047048688	0.5\\
0.00713094518483374	0.533333333333333\\
0.0100348228124021	0.566666666666667\\
0.0101234311900763	0.6\\
0.0145516793360326	0.633333333333333\\
0.0146485610069104	0.666666666666667\\
0.0175285610731069	0.7\\
0.0180778173849741	0.733333333333333\\
0.020346534853278	0.766666666666667\\
0.0344155902313644	0.8\\
0.0436970081105299	0.833333333333333\\
0.182408646368562	0.866666666666667\\
0.368193715397182	0.9\\
0.41465524712271	0.933333333333333\\
0.750227826678967	0.966666666666667\\
2.39346790144014	1\\
};
\addlegendentry{$N_\text{t}=16\cdot16$}

\addplot [color=mycolor4, line width=4.0pt, mark=x, mark size=6pt, mark repeat=3]
  table[row sep=crcr]{%
5.69285969134685e-05	0.0333333333333333\\
6.90112046332523e-05	0.0666666666666667\\
8.38726928855117e-05	0.1\\
0.000103824500398719	0.133333333333333\\
0.000146583067445744	0.166666666666667\\
0.000190573115701482	0.2\\
0.000205073693601723	0.233333333333333\\
0.000223158472966334	0.266666666666667\\
0.000330771322748681	0.3\\
0.000392273520265973	0.333333333333333\\
0.000452217337699372	0.366666666666667\\
0.000473505833207249	0.4\\
0.00048646097524304	0.433333333333333\\
0.00077992477197946	0.466666666666667\\
0.00111082648766845	0.5\\
0.00161379041376248	0.533333333333333\\
0.00165250589060079	0.566666666666667\\
0.00165533771967061	0.6\\
0.00175420251123978	0.633333333333333\\
0.0027297170880593	0.666666666666667\\
0.00290520014237938	0.7\\
0.00436187296864286	0.733333333333333\\
0.0043799408905278	0.766666666666667\\
0.0138754402275154	0.8\\
0.0180755445517012	0.833333333333333\\
0.0339796439679897	0.866666666666667\\
0.109407136919732	0.9\\
0.115922096705749	0.933333333333333\\
0.119481553428364	0.966666666666667\\
0.256789911032274	1\\
};
\addlegendentry{$N_\text{t}=32\cdot32$}

\end{axis}
\end{tikzpicture}%

%% file: fig/lambda1_rss_degradation_cdf_NoSats40.tex
\definecolor{mycolor1}{rgb}{0.06600,0.44300,0.74500}%
\definecolor{mycolor2}{rgb}{0.86600,0.32900,0.00000}%
\definecolor{mycolor3}{rgb}{0.92900,0.69400,0.12500}%
\definecolor{mycolor4}{rgb}{0.52100,0.08600,0.81900}%
\definecolor{mycolor5}{rgb}{0.12941,0.12941,0.12941}%
\begin{tikzpicture}

\begin{axis}[%
width=8.88in,
height=3in,
at={(1.49in,0.738in)},
scale only axis,
xmin=0,
xmax=28,
xlabel={RSS Degradation [dB]},
ymin=0,
ymax=1,
ylabel={CDF},
axis background/.style={fill=white},
axis lines=box,
xmajorgrids,
ymajorgrids,
xtick distance=2,
ytick distance=0.2,
% Font sizes
% tick label style={font=\LARGE},
tick label style={font=\fontsize{30pt}{36pt}\selectfont},
legend style={font=\fontsize{28pt}{33pt}\selectfont, at={(0.97,0.03)}, anchor=south east, legend cell align=left, align=left},
xlabel style={font=\Huge},
ylabel style={font=\Huge},
]
\addplot [color=mycolor1, line width=4.0pt, mark=diamond*, mark size=6pt, mark repeat=3]
  table[row sep=crcr]{%
0.150784343088984	0.0333333333333333\\
0.176914073082185	0.0666666666666667\\
0.256672393079011	0.1\\
0.374520791798248	0.133333333333333\\
0.390850283009693	0.166666666666667\\
0.463928630917365	0.2\\
0.660972834173043	0.233333333333333\\
0.727498811943934	0.266666666666667\\
0.738859283319612	0.3\\
0.870364445131341	0.333333333333333\\
1.09073328647381	0.366666666666667\\
2.09841827396183	0.4\\
2.29449110556897	0.433333333333333\\
2.38004350254743	0.466666666666667\\
2.73820113813495	0.5\\
2.79227469357596	0.533333333333333\\
3.24902225239094	0.566666666666667\\
3.34243593878675	0.6\\
3.95889267995385	0.633333333333333\\
4.37178303680653	0.666666666666667\\
4.65550714980012	0.7\\
4.78421102063783	0.733333333333333\\
5.8023756506039	0.766666666666667\\
6.91029166425237	0.8\\
7.11520769985376	0.833333333333333\\
8.13239961598975	0.866666666666667\\
9.20437524880089	0.9\\
17.2825473626598	0.933333333333333\\
26.1080597363247	0.966666666666667\\
27.0195885482789	1\\
};
\addlegendentry{$N_\text{t}=4\cdot4$}

\addplot [color=mycolor2, line width=4.0pt, mark=o, mark size=6pt, mark repeat=3]
  table[row sep=crcr]{%
0.0353844029983243	0.0333333333333333\\
0.0359226627796493	0.0666666666666667\\
0.0735084649833161	0.1\\
0.0800493418378649	0.133333333333333\\
0.0900529690103972	0.166666666666667\\
0.0954740351454275	0.2\\
0.101035443987375	0.233333333333333\\
0.107031334285951	0.266666666666667\\
0.10846999536044	0.3\\
0.11223016205084	0.333333333333333\\
0.164194498084884	0.366666666666667\\
0.181789830779762	0.4\\
0.209124603070517	0.433333333333333\\
0.225588519906638	0.466666666666667\\
0.27284707711987	0.5\\
0.278875658928573	0.533333333333333\\
0.327672488149953	0.566666666666667\\
0.40546088276772	0.6\\
0.478903960388396	0.633333333333333\\
0.627838662487565	0.666666666666667\\
0.827276346293179	0.7\\
1.06208144768094	0.733333333333333\\
1.15299337946193	0.766666666666667\\
1.20098611676122	0.8\\
1.45237885182073	0.833333333333333\\
1.65177592229646	0.866666666666667\\
1.83524733582869	0.9\\
4.69713182714115	0.933333333333333\\
6.20347511361891	0.966666666666667\\
7.02786845283056	1\\
};
\addlegendentry{$N_\text{t}=8\cdot8$}

\addplot [color=mycolor3, line width=4.0pt, mark=square*, mark size=6pt, mark repeat=3]
  table[row sep=crcr]{%
0.00363246326951044	0.0333333333333333\\
0.00600739494783401	0.0666666666666667\\
0.00824580286201115	0.1\\
0.00852528028816245	0.133333333333333\\
0.0109564843783053	0.166666666666667\\
0.0148508753473851	0.2\\
0.0156075045102645	0.233333333333333\\
0.0164385381009032	0.266666666666667\\
0.019345562985881	0.3\\
0.0265246330259201	0.333333333333333\\
0.042757607495839	0.366666666666667\\
0.0430453267660803	0.4\\
0.0464891499235645	0.433333333333333\\
0.0486133723411174	0.466666666666667\\
0.0522213624640864	0.5\\
0.0545972765655792	0.533333333333333\\
0.0568562800707586	0.566666666666667\\
0.0703160428447785	0.6\\
0.0895500797894316	0.633333333333333\\
0.102088187939036	0.666666666666667\\
0.237054937535303	0.7\\
0.243200362593343	0.733333333333333\\
0.260910637355988	0.766666666666667\\
0.328477666441404	0.8\\
0.363070990237147	0.833333333333333\\
0.751462851012756	0.866666666666667\\
0.848642473196218	0.9\\
1.10487999618769	0.933333333333333\\
1.229093606269	0.966666666666667\\
2.57941551638496	1\\
};
\addlegendentry{$N_\text{t}=16\cdot16$}

\addplot [color=mycolor4, line width=4.0pt, mark=x, mark size=6pt, mark repeat=3]
  table[row sep=crcr]{%
0.00053604606142364	0.0333333333333333\\
0.00215064372876698	0.0666666666666667\\
0.00227824012899382	0.1\\
0.00230512435903118	0.133333333333333\\
0.00254223303965103	0.166666666666667\\
0.00269770036058914	0.2\\
0.00422339864310101	0.233333333333333\\
0.00448547062847651	0.266666666666667\\
0.00469700475910989	0.3\\
0.00797671810265893	0.333333333333333\\
0.00832471853217746	0.366666666666667\\
0.00923760990169438	0.4\\
0.0111108826964904	0.433333333333333\\
0.0117700120953546	0.466666666666667\\
0.0130320635061294	0.5\\
0.0137457542058332	0.533333333333333\\
0.0175182167026937	0.566666666666667\\
0.0185532357503033	0.6\\
0.0191846790322462	0.633333333333333\\
0.0213266323616321	0.666666666666667\\
0.0213282133482906	0.7\\
0.0234641454562958	0.733333333333333\\
0.0252851676630107	0.766666666666667\\
0.0311061385115086	0.8\\
0.0440242160841996	0.833333333333333\\
0.0498812081392195	0.866666666666667\\
0.290989445818881	0.9\\
0.299169440655581	0.933333333333333\\
0.426713421045725	0.966666666666667\\
0.500959309984685	1\\
};
\addlegendentry{$N_\text{t}=32\cdot32$}

\end{axis}
\end{tikzpicture}%

%% file: fig/Fig7_AngularPositions_gNB1_NoSats40_Nt4x4_lambda1_snap1.tex
\definecolor{mycolor1}{rgb}{0.06600,0.44300,0.74500}%
\definecolor{mycolor2}{rgb}{0.86600,0.32900,0.00000}%
\definecolor{mycolor3}{rgb}{0.20000,0.55000,1.00000}%
\definecolor{mycolor4}{rgb}{0.81900,0.01500,0.54500}%
\definecolor{mycolor5}{rgb}{0.12941,0.12941,0.12941}%
\begin{tikzpicture}

\tikzset{centered/.style={anchor=center}}

\begin{axis}[%
width=4in,
height=3in,
at={(1.254in,0.802in)},
tick label style={font=\fontsize{20pt}{26pt}\selectfont},
ylabel style={font=\Huge},
% label style={font=\Huge},
% label style={font=\Huge},
scale only axis,
point meta min=0,
point meta max=27.019588548163,
colormap={mymap}{[1pt] rgb(0pt)=(0.0104,0.0000,0.0000); rgb(1pt)=(0.0208,0.0000,0.0000); rgb(2pt)=(0.0312,0.0000,0.0000); rgb(3pt)=(0.0417,0.0000,0.0000); rgb(4pt)=(0.0521,0.0000,0.0000); rgb(5pt)=(0.0625,0.0000,0.0000); rgb(6pt)=(0.0729,0.0000,0.0000); rgb(7pt)=(0.0833,0.0000,0.0000); rgb(8pt)=(0.0938,0.0000,0.0000); rgb(9pt)=(0.1042,0.0000,0.0000); rgb(10pt)=(0.1146,0.0000,0.0000); rgb(11pt)=(0.1250,0.0000,0.0000); rgb(12pt)=(0.1354,0.0000,0.0000); rgb(13pt)=(0.1458,0.0000,0.0000); rgb(14pt)=(0.1562,0.0000,0.0000); rgb(15pt)=(0.1667,0.0000,0.0000); rgb(16pt)=(0.1771,0.0000,0.0000); rgb(17pt)=(0.1875,0.0000,0.0000); rgb(18pt)=(0.1979,0.0000,0.0000); rgb(19pt)=(0.2083,0.0000,0.0000); rgb(20pt)=(0.2188,0.0000,0.0000); rgb(21pt)=(0.2292,0.0000,0.0000); rgb(22pt)=(0.2396,0.0000,0.0000); rgb(23pt)=(0.2500,0.0000,0.0000); rgb(24pt)=(0.2604,0.0000,0.0000); rgb(25pt)=(0.2708,0.0000,0.0000); rgb(26pt)=(0.2812,0.0000,0.0000); rgb(27pt)=(0.2917,0.0000,0.0000); rgb(28pt)=(0.3021,0.0000,0.0000); rgb(29pt)=(0.3125,0.0000,0.0000); rgb(30pt)=(0.3229,0.0000,0.0000); rgb(31pt)=(0.3333,0.0000,0.0000); rgb(32pt)=(0.3438,0.0000,0.0000); rgb(33pt)=(0.3542,0.0000,0.0000); rgb(34pt)=(0.3646,0.0000,0.0000); rgb(35pt)=(0.3750,0.0000,0.0000); rgb(36pt)=(0.3854,0.0000,0.0000); rgb(37pt)=(0.3958,0.0000,0.0000); rgb(38pt)=(0.4062,0.0000,0.0000); rgb(39pt)=(0.4167,0.0000,0.0000); rgb(40pt)=(0.4271,0.0000,0.0000); rgb(41pt)=(0.4375,0.0000,0.0000); rgb(42pt)=(0.4479,0.0000,0.0000); rgb(43pt)=(0.4583,0.0000,0.0000); rgb(44pt)=(0.4688,0.0000,0.0000); rgb(45pt)=(0.4792,0.0000,0.0000); rgb(46pt)=(0.4896,0.0000,0.0000); rgb(47pt)=(0.5000,0.0000,0.0000); rgb(48pt)=(0.5104,0.0000,0.0000); rgb(49pt)=(0.5208,0.0000,0.0000); rgb(50pt)=(0.5312,0.0000,0.0000); rgb(51pt)=(0.5417,0.0000,0.0000); rgb(52pt)=(0.5521,0.0000,0.0000); rgb(53pt)=(0.5625,0.0000,0.0000); rgb(54pt)=(0.5729,0.0000,0.0000); rgb(55pt)=(0.5833,0.0000,0.0000); rgb(56pt)=(0.5938,0.0000,0.0000); rgb(57pt)=(0.6042,0.0000,0.0000); rgb(58pt)=(0.6146,0.0000,0.0000); rgb(59pt)=(0.6250,0.0000,0.0000); rgb(60pt)=(0.6354,0.0000,0.0000); rgb(61pt)=(0.6458,0.0000,0.0000); rgb(62pt)=(0.6562,0.0000,0.0000); rgb(63pt)=(0.6667,0.0000,0.0000); rgb(64pt)=(0.6771,0.0000,0.0000); rgb(65pt)=(0.6875,0.0000,0.0000); rgb(66pt)=(0.6979,0.0000,0.0000); rgb(67pt)=(0.7083,0.0000,0.0000); rgb(68pt)=(0.7188,0.0000,0.0000); rgb(69pt)=(0.7292,0.0000,0.0000); rgb(70pt)=(0.7396,0.0000,0.0000); rgb(71pt)=(0.7500,0.0000,0.0000); rgb(72pt)=(0.7604,0.0000,0.0000); rgb(73pt)=(0.7708,0.0000,0.0000); rgb(74pt)=(0.7812,0.0000,0.0000); rgb(75pt)=(0.7917,0.0000,0.0000); rgb(76pt)=(0.8021,0.0000,0.0000); rgb(77pt)=(0.8125,0.0000,0.0000); rgb(78pt)=(0.8229,0.0000,0.0000); rgb(79pt)=(0.8333,0.0000,0.0000); rgb(80pt)=(0.8438,0.0000,0.0000); rgb(81pt)=(0.8542,0.0000,0.0000); rgb(82pt)=(0.8646,0.0000,0.0000); rgb(83pt)=(0.8750,0.0000,0.0000); rgb(84pt)=(0.8854,0.0000,0.0000); rgb(85pt)=(0.8958,0.0000,0.0000); rgb(86pt)=(0.9062,0.0000,0.0000); rgb(87pt)=(0.9167,0.0000,0.0000); rgb(88pt)=(0.9271,0.0000,0.0000); rgb(89pt)=(0.9375,0.0000,0.0000); rgb(90pt)=(0.9479,0.0000,0.0000); rgb(91pt)=(0.9583,0.0000,0.0000); rgb(92pt)=(0.9688,0.0000,0.0000); rgb(93pt)=(0.9792,0.0000,0.0000); rgb(94pt)=(0.9896,0.0000,0.0000); rgb(95pt)=(1.0000,0.0000,0.0000); rgb(96pt)=(1.0000,0.0104,0.0000); rgb(97pt)=(1.0000,0.0208,0.0000); rgb(98pt)=(1.0000,0.0312,0.0000); rgb(99pt)=(1.0000,0.0417,0.0000); rgb(100pt)=(1.0000,0.0521,0.0000); rgb(101pt)=(1.0000,0.0625,0.0000); rgb(102pt)=(1.0000,0.0729,0.0000); rgb(103pt)=(1.0000,0.0833,0.0000); rgb(104pt)=(1.0000,0.0938,0.0000); rgb(105pt)=(1.0000,0.1042,0.0000); rgb(106pt)=(1.0000,0.1146,0.0000); rgb(107pt)=(1.0000,0.1250,0.0000); rgb(108pt)=(1.0000,0.1354,0.0000); rgb(109pt)=(1.0000,0.1458,0.0000); rgb(110pt)=(1.0000,0.1562,0.0000); rgb(111pt)=(1.0000,0.1667,0.0000); rgb(112pt)=(1.0000,0.1771,0.0000); rgb(113pt)=(1.0000,0.1875,0.0000); rgb(114pt)=(1.0000,0.1979,0.0000); rgb(115pt)=(1.0000,0.2083,0.0000); rgb(116pt)=(1.0000,0.2188,0.0000); rgb(117pt)=(1.0000,0.2292,0.0000); rgb(118pt)=(1.0000,0.2396,0.0000); rgb(119pt)=(1.0000,0.2500,0.0000); rgb(120pt)=(1.0000,0.2604,0.0000); rgb(121pt)=(1.0000,0.2708,0.0000); rgb(122pt)=(1.0000,0.2812,0.0000); rgb(123pt)=(1.0000,0.2917,0.0000); rgb(124pt)=(1.0000,0.3021,0.0000); rgb(125pt)=(1.0000,0.3125,0.0000); rgb(126pt)=(1.0000,0.3229,0.0000); rgb(127pt)=(1.0000,0.3333,0.0000); rgb(128pt)=(1.0000,0.3438,0.0000); rgb(129pt)=(1.0000,0.3542,0.0000); rgb(130pt)=(1.0000,0.3646,0.0000); rgb(131pt)=(1.0000,0.3750,0.0000); rgb(132pt)=(1.0000,0.3854,0.0000); rgb(133pt)=(1.0000,0.3958,0.0000); rgb(134pt)=(1.0000,0.4062,0.0000); rgb(135pt)=(1.0000,0.4167,0.0000); rgb(136pt)=(1.0000,0.4271,0.0000); rgb(137pt)=(1.0000,0.4375,0.0000); rgb(138pt)=(1.0000,0.4479,0.0000); rgb(139pt)=(1.0000,0.4583,0.0000); rgb(140pt)=(1.0000,0.4688,0.0000); rgb(141pt)=(1.0000,0.4792,0.0000); rgb(142pt)=(1.0000,0.4896,0.0000); rgb(143pt)=(1.0000,0.5000,0.0000); rgb(144pt)=(1.0000,0.5104,0.0000); rgb(145pt)=(1.0000,0.5208,0.0000); rgb(146pt)=(1.0000,0.5312,0.0000); rgb(147pt)=(1.0000,0.5417,0.0000); rgb(148pt)=(1.0000,0.5521,0.0000); rgb(149pt)=(1.0000,0.5625,0.0000); rgb(150pt)=(1.0000,0.5729,0.0000); rgb(151pt)=(1.0000,0.5833,0.0000); rgb(152pt)=(1.0000,0.5938,0.0000); rgb(153pt)=(1.0000,0.6042,0.0000); rgb(154pt)=(1.0000,0.6146,0.0000); rgb(155pt)=(1.0000,0.6250,0.0000); rgb(156pt)=(1.0000,0.6354,0.0000); rgb(157pt)=(1.0000,0.6458,0.0000); rgb(158pt)=(1.0000,0.6562,0.0000); rgb(159pt)=(1.0000,0.6667,0.0000); rgb(160pt)=(1.0000,0.6771,0.0000); rgb(161pt)=(1.0000,0.6875,0.0000); rgb(162pt)=(1.0000,0.6979,0.0000); rgb(163pt)=(1.0000,0.7083,0.0000); rgb(164pt)=(1.0000,0.7188,0.0000); rgb(165pt)=(1.0000,0.7292,0.0000); rgb(166pt)=(1.0000,0.7396,0.0000); rgb(167pt)=(1.0000,0.7500,0.0000); rgb(168pt)=(1.0000,0.7604,0.0000); rgb(169pt)=(1.0000,0.7708,0.0000); rgb(170pt)=(1.0000,0.7812,0.0000); rgb(171pt)=(1.0000,0.7917,0.0000); rgb(172pt)=(1.0000,0.8021,0.0000); rgb(173pt)=(1.0000,0.8125,0.0000); rgb(174pt)=(1.0000,0.8229,0.0000); rgb(175pt)=(1.0000,0.8333,0.0000); rgb(176pt)=(1.0000,0.8438,0.0000); rgb(177pt)=(1.0000,0.8542,0.0000); rgb(178pt)=(1.0000,0.8646,0.0000); rgb(179pt)=(1.0000,0.8750,0.0000); rgb(180pt)=(1.0000,0.8854,0.0000); rgb(181pt)=(1.0000,0.8958,0.0000); rgb(182pt)=(1.0000,0.9062,0.0000); rgb(183pt)=(1.0000,0.9167,0.0000); rgb(184pt)=(1.0000,0.9271,0.0000); rgb(185pt)=(1.0000,0.9375,0.0000); rgb(186pt)=(1.0000,0.9479,0.0000); rgb(187pt)=(1.0000,0.9583,0.0000); rgb(188pt)=(1.0000,0.9688,0.0000); rgb(189pt)=(1.0000,0.9792,0.0000); rgb(190pt)=(1.0000,0.9896,0.0000); rgb(191pt)=(1.0000,1.0000,0.0000); rgb(192pt)=(1.0000,1.0000,0.0156); rgb(193pt)=(1.0000,1.0000,0.0312); rgb(194pt)=(1.0000,1.0000,0.0469); rgb(195pt)=(1.0000,1.0000,0.0625); rgb(196pt)=(1.0000,1.0000,0.0781); rgb(197pt)=(1.0000,1.0000,0.0938); rgb(198pt)=(1.0000,1.0000,0.1094); rgb(199pt)=(1.0000,1.0000,0.1250); rgb(200pt)=(1.0000,1.0000,0.1406); rgb(201pt)=(1.0000,1.0000,0.1562); rgb(202pt)=(1.0000,1.0000,0.1719); rgb(203pt)=(1.0000,1.0000,0.1875); rgb(204pt)=(1.0000,1.0000,0.2031); rgb(205pt)=(1.0000,1.0000,0.2188); rgb(206pt)=(1.0000,1.0000,0.2344); rgb(207pt)=(1.0000,1.0000,0.2500); rgb(208pt)=(1.0000,1.0000,0.2656); rgb(209pt)=(1.0000,1.0000,0.2812); rgb(210pt)=(1.0000,1.0000,0.2969); rgb(211pt)=(1.0000,1.0000,0.3125); rgb(212pt)=(1.0000,1.0000,0.3281); rgb(213pt)=(1.0000,1.0000,0.3438); rgb(214pt)=(1.0000,1.0000,0.3594); rgb(215pt)=(1.0000,1.0000,0.3750); rgb(216pt)=(1.0000,1.0000,0.3906); rgb(217pt)=(1.0000,1.0000,0.4062); rgb(218pt)=(1.0000,1.0000,0.4219); rgb(219pt)=(1.0000,1.0000,0.4375); rgb(220pt)=(1.0000,1.0000,0.4531); rgb(221pt)=(1.0000,1.0000,0.4688); rgb(222pt)=(1.0000,1.0000,0.4844); rgb(223pt)=(1.0000,1.0000,0.5000); rgb(224pt)=(1.0000,1.0000,0.5156); rgb(225pt)=(1.0000,1.0000,0.5312); rgb(226pt)=(1.0000,1.0000,0.5469); rgb(227pt)=(1.0000,1.0000,0.5625); rgb(228pt)=(1.0000,1.0000,0.5781); rgb(229pt)=(1.0000,1.0000,0.5938); rgb(230pt)=(1.0000,1.0000,0.6094); rgb(231pt)=(1.0000,1.0000,0.6250); rgb(232pt)=(1.0000,1.0000,0.6406); rgb(233pt)=(1.0000,1.0000,0.6562); rgb(234pt)=(1.0000,1.0000,0.6719); rgb(235pt)=(1.0000,1.0000,0.6875); rgb(236pt)=(1.0000,1.0000,0.7031); rgb(237pt)=(1.0000,1.0000,0.7188); rgb(238pt)=(1.0000,1.0000,0.7344); rgb(239pt)=(1.0000,1.0000,0.7500); rgb(240pt)=(1.0000,1.0000,0.7656); rgb(241pt)=(1.0000,1.0000,0.7812); rgb(242pt)=(1.0000,1.0000,0.7969); rgb(243pt)=(1.0000,1.0000,0.8125); rgb(244pt)=(1.0000,1.0000,0.8281); rgb(245pt)=(1.0000,1.0000,0.8438); rgb(246pt)=(1.0000,1.0000,0.8594); rgb(247pt)=(1.0000,1.0000,0.8750); rgb(248pt)=(1.0000,1.0000,0.8906); rgb(249pt)=(1.0000,1.0000,0.9062); rgb(250pt)=(1.0000,1.0000,0.9219); rgb(251pt)=(1.0000,1.0000,0.9375); rgb(252pt)=(1.0000,1.0000,0.9531); rgb(253pt)=(1.0000,1.0000,0.9688); rgb(254pt)=(1.0000,1.0000,0.9844); rgb(255pt)=(1.0000,1.0000,1.0000)},
unbounded coords=jump,
xmin=-60,
xmax=60,
xlabel style={font=\fontsize{48}{50}\selectfont},
xlabel={Azimuth angle ($\phi$) [$^\circ$]},
ymin=-80,
ymax=90,
ylabel={Elevation angle ($\theta$)  [$^\circ$]},
axis background/.style={fill=white},
axis x line*=bottom,
axis y line*=left,
axis lines=box,
xmajorgrids,
ymajorgrids,
legend style={legend cell align=left, align=left, font=\fontsize{24pt}{30pt}\selectfont},
colorbar,
colorbar style={ylabel style={font=\Huge}, ylabel={RSS Degradation [dB]}},
ylabel style={font=\Huge},
xlabel style={font=\Huge}
]
\addplot[only marks, mark=x, mark options={}, mark size=3.7081pt, draw=red, forget plot] table[row sep=crcr]{%
x	y\\
-29.913045582444	83.467765732328\\
-37.331931452409	40.4928270911073\\
-37.3857422966607	39.9916557051464\\
-35.9439569233537	54.7399846405927\\
-37.5269899201166	38.701041726755\\
-34.482955942135	68.571425380883\\
-21.7903138594012	87.3036461855286\\
-35.3590521980698	60.8044080841969\\
-37.4622811396456	39.2877865860583\\
-39.8770623483742	22.3084666776452\\
-36.0344904942684	53.7743108355357\\
-40.0258095303973	21.5407390551593\\
-32.9240737430791	77.1674587065132\\
-39.4837572938755	24.4660524470951\\
-39.7182234865812	23.1567969959555\\
-9.3842841700594	88.5587373359836\\
-40.0811069804453	21.2616663862124\\
-38.4744854472882	30.9966619233646\\
-33.2810070244342	75.7123388212849\\
-38.2464753540496	32.7009762780514\\
-34.0322772560431	71.7109984649821\\
-36.6104860955641	47.6621228806058\\
-39.0673229358711	26.9736348692644\\
-38.572719949421	30.2902485963232\\
-39.5095811731233	24.3183914878462\\
-39.5360311186428	24.1680555591401\\
-38.0295560008468	34.4089627280394\\
-37.4386976793572	39.5035329639839\\
-35.4971620049957	59.4140569750991\\
-38.4119146725443	31.4552687770997\\
-38.1253631600541	33.6440065917086\\
-38.7325162382167	29.1758023523534\\
-38.1612483762703	33.3618143942112\\
-39.4617982029872	24.5923064862472\\
-32.8391518954244	77.4788441887845\\
-35.1281877548394	63.0432220259435\\
-34.8388622868693	65.6664667507763\\
-39.1234937378509	26.6209722730844\\
-30.7436550927979	82.3942579890017\\
-37.921571474812	35.2914649946493\\
};
\addplot[only marks, mark=x, mark options={}, mark size=2.9580pt, draw=violet, forget plot] table[row sep=crcr]{%
x	y\\
-29.913045582444	83.467765732328\\
-37.331931452409	40.4928270911073\\
-37.3857422966607	39.9916557051464\\
-35.9439569233537	54.7399846405927\\
-37.5269899201166	38.701041726755\\
-34.482955942135	68.571425380883\\
-21.7903138594012	87.3036461855286\\
-35.3590521980698	60.8044080841969\\
-37.4622811396456	39.2877865860583\\
-39.8770623483742	22.3084666776452\\
-36.0344904942684	53.7743108355357\\
-40.0258095303973	21.5407390551593\\
-32.9240737430791	77.1674587065132\\
-39.4837572938755	24.4660524470951\\
-39.7182234865812	23.1567969959555\\
-9.3842841700594	88.5587373359836\\
-40.0811069804453	21.2616663862124\\
-38.4744854472882	30.9966619233646\\
-33.2810070244342	75.7123388212849\\
-38.2464753540496	32.7009762780514\\
-34.0322772560431	71.7109984649821\\
-36.6104860955641	47.6621228806058\\
-39.0673229358711	26.9736348692644\\
-38.572719949421	30.2902485963232\\
-39.5095811731233	24.3183914878462\\
-39.5360311186428	24.1680555591401\\
-38.0295560008468	34.4089627280394\\
-37.4386976793572	39.5035329639839\\
-35.4971620049957	59.4140569750991\\
-38.4119146725443	31.4552687770997\\
-38.1253631600541	33.6440065917086\\
-38.7325162382167	29.1758023523534\\
-38.1612483762703	33.3618143942112\\
-39.4617982029872	24.5923064862472\\
-32.8391518954244	77.4788441887845\\
-35.1281877548394	63.0432220259435\\
-34.8388622868693	65.6664667507763\\
-39.1234937378509	26.6209722730844\\
-30.7436550927979	82.3942579890017\\
-37.921571474812	35.2914649946493\\
};
\addplot[scatter, only marks, mark=*, mark size=5.25pt, scatter src=explicit, 
scatter/use mapped color={mark options={}, draw=black, fill=mapped color},
nodes near coords*={\lbl},
visualization depends on={value \thisrow{label} \as \lbl},
every node near coord/.style={anchor=center, font=\normalsize\bfseries, text=blue!50!white},
forget plot] table[row sep=crcr, meta=color]{%
x	y	color	label\\
37.7668423671815	-29.9423050138247	5.8023756506032	1\\
48.6950324490743	-69.7645275237754	17.2825473626577	2\\
-44.7615820447793	-42.6060384406171	0.463928630917197	3\\
49.6051027366823	-66.060826815162	3.24902225238944	4\\
15.8831095470491	-57.2215006161576	3.9588926799532	5\\
-48.2951514000708	-28.0739086317017	0.15078434308861	6\\
-26.5802137359542	-30.1411392153783	0.390850283009616	7\\
5.62578230459806	-40.6791707652455	0.738859283318925	8\\
54.9008202521157	-26.4051109118662	2.0984182739604	9\\
55.7866242239132	-69.0199817599323	4.78421102063689	10\\
-41.0864301986942	-36.1556298212871	0.374520791798151	11\\
56.4711338112739	-38.0803530296104	8.13239961598806	12\\
54.8600337891535	-28.9505767671532	2.29449110556811	13\\
-1.75492215325905	-28.49130456038	0.660972834172395	14\\
36.033656266656	-48.2301649180406	26.1080597362645	15\\
-42.9736393647342	-34.6675349149033	0.25667239307891	16\\
-9.388646084847	-35.9447130169263	0.727498811943655	17\\
49.888263022688	-31.0495392279534	4.65550714979922	18\\
35.0648795471465	-29.8842348248309	7.11520769985283	19\\
55.1390911671484	-29.1237983691689	2.38004350254664	20\\
18.6888838987904	-42.6523993143263	2.79227469357544	21\\
-55.7145985710973	-30.4156854858531	0.176914073081323	22\\
41.8955167042533	-30.8789029755063	6.91029166425149	23\\
52.0791897309061	-50.2001497515931	27.019588548163	24\\
21.4482185829328	-54.5214598118171	4.37178303680595	25\\
30.9288156694	-34.4346089565159	9.20437524879967	26\\
29.1758961749899	-26.2915051902991	3.34243593878636	27\\
-12.9327576558998	-39.6785136795441	1.09073328647354	28\\
18.6573468213068	-32.3197625786399	2.7382011381343	29\\
-39.4575974626126	-45.6532354648834	0.870364445131145	30\\
};
\addplot[only marks, mark=square*, mark options={}, mark size=5pt, draw=white!20!black, fill=white!35!black, forget plot] table[row sep=crcr]{%
x	y\\
0	0\\
};

\addlegendimage{legend image code/.code={%
  \draw[red, line width=3.0pt] plot[mark=x, mark size=5.0pt] coordinates {(0.3cm,0cm)};
  \draw[violet, line width=2.0pt] plot[mark=x, mark size=4.0pt] coordinates {(0.3cm,0cm)};
}}
\addlegendentry{Satellites}

\addlegendimage{only marks, mark=*, mark size=4.0pt, draw=black, fill=black}
\addlegendentry{UEs}

\addlegendimage{only marks, mark=square*, mark size=3.5pt, draw=white!20!black, fill=white!35!black}
\addlegendentry{gNB}

\end{axis}
\end{tikzpicture}%

%% file: fig/Fig7_AngularPositions_gNB1_NoSats40_Nt8x8_lambda1_snap1.tex
\definecolor{mycolor1}{rgb}{0.06600,0.44300,0.74500}%
\definecolor{mycolor2}{rgb}{0.86600,0.32900,0.00000}%
\definecolor{mycolor3}{rgb}{0.20000,0.55000,1.00000}%
\definecolor{mycolor4}{rgb}{0.81900,0.01500,0.54500}%
\definecolor{mycolor5}{rgb}{0.12941,0.12941,0.12941}%
\begin{tikzpicture}

\tikzset{centered/.style={anchor=center}}

\begin{axis}[%
width=4in,
height=3in,
tick label style={font=\fontsize{20pt}{26pt}\selectfont},
at={(1.254in,0.802in)},
scale only axis,
point meta min=0,
point meta max=7.02786845283029,
colormap={mymap}{[1pt] rgb(0pt)=(0.0104,0.0000,0.0000); rgb(1pt)=(0.0208,0.0000,0.0000); rgb(2pt)=(0.0312,0.0000,0.0000); rgb(3pt)=(0.0417,0.0000,0.0000); rgb(4pt)=(0.0521,0.0000,0.0000); rgb(5pt)=(0.0625,0.0000,0.0000); rgb(6pt)=(0.0729,0.0000,0.0000); rgb(7pt)=(0.0833,0.0000,0.0000); rgb(8pt)=(0.0938,0.0000,0.0000); rgb(9pt)=(0.1042,0.0000,0.0000); rgb(10pt)=(0.1146,0.0000,0.0000); rgb(11pt)=(0.1250,0.0000,0.0000); rgb(12pt)=(0.1354,0.0000,0.0000); rgb(13pt)=(0.1458,0.0000,0.0000); rgb(14pt)=(0.1562,0.0000,0.0000); rgb(15pt)=(0.1667,0.0000,0.0000); rgb(16pt)=(0.1771,0.0000,0.0000); rgb(17pt)=(0.1875,0.0000,0.0000); rgb(18pt)=(0.1979,0.0000,0.0000); rgb(19pt)=(0.2083,0.0000,0.0000); rgb(20pt)=(0.2188,0.0000,0.0000); rgb(21pt)=(0.2292,0.0000,0.0000); rgb(22pt)=(0.2396,0.0000,0.0000); rgb(23pt)=(0.2500,0.0000,0.0000); rgb(24pt)=(0.2604,0.0000,0.0000); rgb(25pt)=(0.2708,0.0000,0.0000); rgb(26pt)=(0.2812,0.0000,0.0000); rgb(27pt)=(0.2917,0.0000,0.0000); rgb(28pt)=(0.3021,0.0000,0.0000); rgb(29pt)=(0.3125,0.0000,0.0000); rgb(30pt)=(0.3229,0.0000,0.0000); rgb(31pt)=(0.3333,0.0000,0.0000); rgb(32pt)=(0.3438,0.0000,0.0000); rgb(33pt)=(0.3542,0.0000,0.0000); rgb(34pt)=(0.3646,0.0000,0.0000); rgb(35pt)=(0.3750,0.0000,0.0000); rgb(36pt)=(0.3854,0.0000,0.0000); rgb(37pt)=(0.3958,0.0000,0.0000); rgb(38pt)=(0.4062,0.0000,0.0000); rgb(39pt)=(0.4167,0.0000,0.0000); rgb(40pt)=(0.4271,0.0000,0.0000); rgb(41pt)=(0.4375,0.0000,0.0000); rgb(42pt)=(0.4479,0.0000,0.0000); rgb(43pt)=(0.4583,0.0000,0.0000); rgb(44pt)=(0.4688,0.0000,0.0000); rgb(45pt)=(0.4792,0.0000,0.0000); rgb(46pt)=(0.4896,0.0000,0.0000); rgb(47pt)=(0.5000,0.0000,0.0000); rgb(48pt)=(0.5104,0.0000,0.0000); rgb(49pt)=(0.5208,0.0000,0.0000); rgb(50pt)=(0.5312,0.0000,0.0000); rgb(51pt)=(0.5417,0.0000,0.0000); rgb(52pt)=(0.5521,0.0000,0.0000); rgb(53pt)=(0.5625,0.0000,0.0000); rgb(54pt)=(0.5729,0.0000,0.0000); rgb(55pt)=(0.5833,0.0000,0.0000); rgb(56pt)=(0.5938,0.0000,0.0000); rgb(57pt)=(0.6042,0.0000,0.0000); rgb(58pt)=(0.6146,0.0000,0.0000); rgb(59pt)=(0.6250,0.0000,0.0000); rgb(60pt)=(0.6354,0.0000,0.0000); rgb(61pt)=(0.6458,0.0000,0.0000); rgb(62pt)=(0.6562,0.0000,0.0000); rgb(63pt)=(0.6667,0.0000,0.0000); rgb(64pt)=(0.6771,0.0000,0.0000); rgb(65pt)=(0.6875,0.0000,0.0000); rgb(66pt)=(0.6979,0.0000,0.0000); rgb(67pt)=(0.7083,0.0000,0.0000); rgb(68pt)=(0.7188,0.0000,0.0000); rgb(69pt)=(0.7292,0.0000,0.0000); rgb(70pt)=(0.7396,0.0000,0.0000); rgb(71pt)=(0.7500,0.0000,0.0000); rgb(72pt)=(0.7604,0.0000,0.0000); rgb(73pt)=(0.7708,0.0000,0.0000); rgb(74pt)=(0.7812,0.0000,0.0000); rgb(75pt)=(0.7917,0.0000,0.0000); rgb(76pt)=(0.8021,0.0000,0.0000); rgb(77pt)=(0.8125,0.0000,0.0000); rgb(78pt)=(0.8229,0.0000,0.0000); rgb(79pt)=(0.8333,0.0000,0.0000); rgb(80pt)=(0.8438,0.0000,0.0000); rgb(81pt)=(0.8542,0.0000,0.0000); rgb(82pt)=(0.8646,0.0000,0.0000); rgb(83pt)=(0.8750,0.0000,0.0000); rgb(84pt)=(0.8854,0.0000,0.0000); rgb(85pt)=(0.8958,0.0000,0.0000); rgb(86pt)=(0.9062,0.0000,0.0000); rgb(87pt)=(0.9167,0.0000,0.0000); rgb(88pt)=(0.9271,0.0000,0.0000); rgb(89pt)=(0.9375,0.0000,0.0000); rgb(90pt)=(0.9479,0.0000,0.0000); rgb(91pt)=(0.9583,0.0000,0.0000); rgb(92pt)=(0.9688,0.0000,0.0000); rgb(93pt)=(0.9792,0.0000,0.0000); rgb(94pt)=(0.9896,0.0000,0.0000); rgb(95pt)=(1.0000,0.0000,0.0000); rgb(96pt)=(1.0000,0.0104,0.0000); rgb(97pt)=(1.0000,0.0208,0.0000); rgb(98pt)=(1.0000,0.0312,0.0000); rgb(99pt)=(1.0000,0.0417,0.0000); rgb(100pt)=(1.0000,0.0521,0.0000); rgb(101pt)=(1.0000,0.0625,0.0000); rgb(102pt)=(1.0000,0.0729,0.0000); rgb(103pt)=(1.0000,0.0833,0.0000); rgb(104pt)=(1.0000,0.0938,0.0000); rgb(105pt)=(1.0000,0.1042,0.0000); rgb(106pt)=(1.0000,0.1146,0.0000); rgb(107pt)=(1.0000,0.1250,0.0000); rgb(108pt)=(1.0000,0.1354,0.0000); rgb(109pt)=(1.0000,0.1458,0.0000); rgb(110pt)=(1.0000,0.1562,0.0000); rgb(111pt)=(1.0000,0.1667,0.0000); rgb(112pt)=(1.0000,0.1771,0.0000); rgb(113pt)=(1.0000,0.1875,0.0000); rgb(114pt)=(1.0000,0.1979,0.0000); rgb(115pt)=(1.0000,0.2083,0.0000); rgb(116pt)=(1.0000,0.2188,0.0000); rgb(117pt)=(1.0000,0.2292,0.0000); rgb(118pt)=(1.0000,0.2396,0.0000); rgb(119pt)=(1.0000,0.2500,0.0000); rgb(120pt)=(1.0000,0.2604,0.0000); rgb(121pt)=(1.0000,0.2708,0.0000); rgb(122pt)=(1.0000,0.2812,0.0000); rgb(123pt)=(1.0000,0.2917,0.0000); rgb(124pt)=(1.0000,0.3021,0.0000); rgb(125pt)=(1.0000,0.3125,0.0000); rgb(126pt)=(1.0000,0.3229,0.0000); rgb(127pt)=(1.0000,0.3333,0.0000); rgb(128pt)=(1.0000,0.3438,0.0000); rgb(129pt)=(1.0000,0.3542,0.0000); rgb(130pt)=(1.0000,0.3646,0.0000); rgb(131pt)=(1.0000,0.3750,0.0000); rgb(132pt)=(1.0000,0.3854,0.0000); rgb(133pt)=(1.0000,0.3958,0.0000); rgb(134pt)=(1.0000,0.4062,0.0000); rgb(135pt)=(1.0000,0.4167,0.0000); rgb(136pt)=(1.0000,0.4271,0.0000); rgb(137pt)=(1.0000,0.4375,0.0000); rgb(138pt)=(1.0000,0.4479,0.0000); rgb(139pt)=(1.0000,0.4583,0.0000); rgb(140pt)=(1.0000,0.4688,0.0000); rgb(141pt)=(1.0000,0.4792,0.0000); rgb(142pt)=(1.0000,0.4896,0.0000); rgb(143pt)=(1.0000,0.5000,0.0000); rgb(144pt)=(1.0000,0.5104,0.0000); rgb(145pt)=(1.0000,0.5208,0.0000); rgb(146pt)=(1.0000,0.5312,0.0000); rgb(147pt)=(1.0000,0.5417,0.0000); rgb(148pt)=(1.0000,0.5521,0.0000); rgb(149pt)=(1.0000,0.5625,0.0000); rgb(150pt)=(1.0000,0.5729,0.0000); rgb(151pt)=(1.0000,0.5833,0.0000); rgb(152pt)=(1.0000,0.5938,0.0000); rgb(153pt)=(1.0000,0.6042,0.0000); rgb(154pt)=(1.0000,0.6146,0.0000); rgb(155pt)=(1.0000,0.6250,0.0000); rgb(156pt)=(1.0000,0.6354,0.0000); rgb(157pt)=(1.0000,0.6458,0.0000); rgb(158pt)=(1.0000,0.6562,0.0000); rgb(159pt)=(1.0000,0.6667,0.0000); rgb(160pt)=(1.0000,0.6771,0.0000); rgb(161pt)=(1.0000,0.6875,0.0000); rgb(162pt)=(1.0000,0.6979,0.0000); rgb(163pt)=(1.0000,0.7083,0.0000); rgb(164pt)=(1.0000,0.7188,0.0000); rgb(165pt)=(1.0000,0.7292,0.0000); rgb(166pt)=(1.0000,0.7396,0.0000); rgb(167pt)=(1.0000,0.7500,0.0000); rgb(168pt)=(1.0000,0.7604,0.0000); rgb(169pt)=(1.0000,0.7708,0.0000); rgb(170pt)=(1.0000,0.7812,0.0000); rgb(171pt)=(1.0000,0.7917,0.0000); rgb(172pt)=(1.0000,0.8021,0.0000); rgb(173pt)=(1.0000,0.8125,0.0000); rgb(174pt)=(1.0000,0.8229,0.0000); rgb(175pt)=(1.0000,0.8333,0.0000); rgb(176pt)=(1.0000,0.8438,0.0000); rgb(177pt)=(1.0000,0.8542,0.0000); rgb(178pt)=(1.0000,0.8646,0.0000); rgb(179pt)=(1.0000,0.8750,0.0000); rgb(180pt)=(1.0000,0.8854,0.0000); rgb(181pt)=(1.0000,0.8958,0.0000); rgb(182pt)=(1.0000,0.9062,0.0000); rgb(183pt)=(1.0000,0.9167,0.0000); rgb(184pt)=(1.0000,0.9271,0.0000); rgb(185pt)=(1.0000,0.9375,0.0000); rgb(186pt)=(1.0000,0.9479,0.0000); rgb(187pt)=(1.0000,0.9583,0.0000); rgb(188pt)=(1.0000,0.9688,0.0000); rgb(189pt)=(1.0000,0.9792,0.0000); rgb(190pt)=(1.0000,0.9896,0.0000); rgb(191pt)=(1.0000,1.0000,0.0000); rgb(192pt)=(1.0000,1.0000,0.0156); rgb(193pt)=(1.0000,1.0000,0.0312); rgb(194pt)=(1.0000,1.0000,0.0469); rgb(195pt)=(1.0000,1.0000,0.0625); rgb(196pt)=(1.0000,1.0000,0.0781); rgb(197pt)=(1.0000,1.0000,0.0938); rgb(198pt)=(1.0000,1.0000,0.1094); rgb(199pt)=(1.0000,1.0000,0.1250); rgb(200pt)=(1.0000,1.0000,0.1406); rgb(201pt)=(1.0000,1.0000,0.1562); rgb(202pt)=(1.0000,1.0000,0.1719); rgb(203pt)=(1.0000,1.0000,0.1875); rgb(204pt)=(1.0000,1.0000,0.2031); rgb(205pt)=(1.0000,1.0000,0.2188); rgb(206pt)=(1.0000,1.0000,0.2344); rgb(207pt)=(1.0000,1.0000,0.2500); rgb(208pt)=(1.0000,1.0000,0.2656); rgb(209pt)=(1.0000,1.0000,0.2812); rgb(210pt)=(1.0000,1.0000,0.2969); rgb(211pt)=(1.0000,1.0000,0.3125); rgb(212pt)=(1.0000,1.0000,0.3281); rgb(213pt)=(1.0000,1.0000,0.3438); rgb(214pt)=(1.0000,1.0000,0.3594); rgb(215pt)=(1.0000,1.0000,0.3750); rgb(216pt)=(1.0000,1.0000,0.3906); rgb(217pt)=(1.0000,1.0000,0.4062); rgb(218pt)=(1.0000,1.0000,0.4219); rgb(219pt)=(1.0000,1.0000,0.4375); rgb(220pt)=(1.0000,1.0000,0.4531); rgb(221pt)=(1.0000,1.0000,0.4688); rgb(222pt)=(1.0000,1.0000,0.4844); rgb(223pt)=(1.0000,1.0000,0.5000); rgb(224pt)=(1.0000,1.0000,0.5156); rgb(225pt)=(1.0000,1.0000,0.5312); rgb(226pt)=(1.0000,1.0000,0.5469); rgb(227pt)=(1.0000,1.0000,0.5625); rgb(228pt)=(1.0000,1.0000,0.5781); rgb(229pt)=(1.0000,1.0000,0.5938); rgb(230pt)=(1.0000,1.0000,0.6094); rgb(231pt)=(1.0000,1.0000,0.6250); rgb(232pt)=(1.0000,1.0000,0.6406); rgb(233pt)=(1.0000,1.0000,0.6562); rgb(234pt)=(1.0000,1.0000,0.6719); rgb(235pt)=(1.0000,1.0000,0.6875); rgb(236pt)=(1.0000,1.0000,0.7031); rgb(237pt)=(1.0000,1.0000,0.7188); rgb(238pt)=(1.0000,1.0000,0.7344); rgb(239pt)=(1.0000,1.0000,0.7500); rgb(240pt)=(1.0000,1.0000,0.7656); rgb(241pt)=(1.0000,1.0000,0.7812); rgb(242pt)=(1.0000,1.0000,0.7969); rgb(243pt)=(1.0000,1.0000,0.8125); rgb(244pt)=(1.0000,1.0000,0.8281); rgb(245pt)=(1.0000,1.0000,0.8438); rgb(246pt)=(1.0000,1.0000,0.8594); rgb(247pt)=(1.0000,1.0000,0.8750); rgb(248pt)=(1.0000,1.0000,0.8906); rgb(249pt)=(1.0000,1.0000,0.9062); rgb(250pt)=(1.0000,1.0000,0.9219); rgb(251pt)=(1.0000,1.0000,0.9375); rgb(252pt)=(1.0000,1.0000,0.9531); rgb(253pt)=(1.0000,1.0000,0.9688); rgb(254pt)=(1.0000,1.0000,0.9844); rgb(255pt)=(1.0000,1.0000,1.0000)},
unbounded coords=jump,
xmin=-60,
xmax=60,
xlabel style={font=\fontsize{48}{50}\selectfont},
xlabel={Azimuth angle ($\phi$) [$^\circ$]},
ymin=-80,
ymax=90,
ylabel style={font=\fontsize{48}{50}\selectfont},
ylabel={Elevation angle ($\theta$) [$^\circ$]},
axis background/.style={fill=white},
axis x line*=bottom,
axis y line*=left,
axis lines=box,
xmajorgrids,
ymajorgrids,
legend style={legend cell align=left, align=left, font=\fontsize{24pt}{30pt}\selectfont},
colorbar,
colorbar style={ylabel style={font=\Huge}, ylabel={RSS Degradation [dB]}},
ylabel style={font=\Huge},
xlabel style={font=\Huge}
]
\addplot[only marks, mark=x, mark options={}, mark size=3.7081pt, draw=red, forget plot] table[row sep=crcr]{%
x	y\\
-29.913045582444	83.467765732328\\
-37.331931452409	40.4928270911073\\
-37.3857422966607	39.9916557051464\\
-35.9439569233537	54.7399846405927\\
-37.5269899201166	38.701041726755\\
-34.482955942135	68.571425380883\\
-21.7903138594012	87.3036461855286\\
-35.3590521980698	60.8044080841969\\
-37.4622811396456	39.2877865860583\\
-39.8770623483742	22.3084666776452\\
-36.0344904942684	53.7743108355357\\
-40.0258095303973	21.5407390551593\\
-32.9240737430791	77.1674587065132\\
-39.4837572938755	24.4660524470951\\
-39.7182234865812	23.1567969959555\\
-9.3842841700594	88.5587373359836\\
-40.0811069804453	21.2616663862124\\
-38.4744854472882	30.9966619233646\\
-33.2810070244342	75.7123388212849\\
-38.2464753540496	32.7009762780514\\
-34.0322772560431	71.7109984649821\\
-36.6104860955641	47.6621228806058\\
-39.0673229358711	26.9736348692644\\
-38.572719949421	30.2902485963232\\
-39.5095811731233	24.3183914878462\\
-39.5360311186428	24.1680555591401\\
-38.0295560008468	34.4089627280394\\
-37.4386976793572	39.5035329639839\\
-35.4971620049957	59.4140569750991\\
-38.4119146725443	31.4552687770997\\
-38.1253631600541	33.6440065917086\\
-38.7325162382167	29.1758023523534\\
-38.1612483762703	33.3618143942112\\
-39.4617982029872	24.5923064862472\\
-32.8391518954244	77.4788441887845\\
-35.1281877548394	63.0432220259435\\
-34.8388622868693	65.6664667507763\\
-39.1234937378509	26.6209722730844\\
-30.7436550927979	82.3942579890017\\
-37.921571474812	35.2914649946493\\
};
\addplot[only marks, mark=x, mark options={}, mark size=2.9580pt, draw=violet, forget plot] table[row sep=crcr]{%
x	y\\
-29.913045582444	83.467765732328\\
-37.331931452409	40.4928270911073\\
-37.3857422966607	39.9916557051464\\
-35.9439569233537	54.7399846405927\\
-37.5269899201166	38.701041726755\\
-34.482955942135	68.571425380883\\
-21.7903138594012	87.3036461855286\\
-35.3590521980698	60.8044080841969\\
-37.4622811396456	39.2877865860583\\
-39.8770623483742	22.3084666776452\\
-36.0344904942684	53.7743108355357\\
-40.0258095303973	21.5407390551593\\
-32.9240737430791	77.1674587065132\\
-39.4837572938755	24.4660524470951\\
-39.7182234865812	23.1567969959555\\
-9.3842841700594	88.5587373359836\\
-40.0811069804453	21.2616663862124\\
-38.4744854472882	30.9966619233646\\
-33.2810070244342	75.7123388212849\\
-38.2464753540496	32.7009762780514\\
-34.0322772560431	71.7109984649821\\
-36.6104860955641	47.6621228806058\\
-39.0673229358711	26.9736348692644\\
-38.572719949421	30.2902485963232\\
-39.5095811731233	24.3183914878462\\
-39.5360311186428	24.1680555591401\\
-38.0295560008468	34.4089627280394\\
-37.4386976793572	39.5035329639839\\
-35.4971620049957	59.4140569750991\\
-38.4119146725443	31.4552687770997\\
-38.1253631600541	33.6440065917086\\
-38.7325162382167	29.1758023523534\\
-38.1612483762703	33.3618143942112\\
-39.4617982029872	24.5923064862472\\
-32.8391518954244	77.4788441887845\\
-35.1281877548394	63.0432220259435\\
-34.8388622868693	65.6664667507763\\
-39.1234937378509	26.6209722730844\\
-30.7436550927979	82.3942579890017\\
-37.921571474812	35.2914649946493\\
};
\addplot[scatter, only marks, mark=*, mark size=5.25pt, scatter src=explicit, 
scatter/use mapped color={mark options={}, draw=black, fill=mapped color},
nodes near coords*={\lbl},
visualization depends on={value \thisrow{label} \as \lbl},
every node near coord/.style={anchor=center, font=\normalsize\bfseries, text=blue!50!white},
forget plot] table[row sep=crcr, meta=color]{%
x	y	color	label\\
37.7668423671815	-29.9423050138247	1.15299337946187	1\\
48.6950324490743	-69.7645275237754	6.20347511361886	2\\
-44.7615820447793	-42.6060384406171	0.07350846498328	3\\
49.6051027366823	-66.060826815162	1.8352473358286	4\\
15.8831095470491	-57.2215006161576	0.278875658928503	5\\
-48.2951514000708	-28.0739086317017	0.107031334285864	6\\
-26.5802137359542	-30.1411392153783	0.0359226627796311	7\\
5.62578230459806	-40.6791707652455	0.16419449808473	8\\
54.9008202521157	-26.4051109118662	0.181789830779549	9\\
55.7866242239132	-69.0199817599323	1.65177592229638	10\\
-41.0864301986942	-36.1556298212871	0.108469995360417	11\\
56.4711338112739	-38.0803530296104	1.20098611676113	12\\
54.8600337891535	-28.9505767671532	0.209124603070388	13\\
-1.75492215325905	-28.49130456038	0.0353844029981799	14\\
36.033656266656	-48.2301649180406	4.69713182714106	15\\
-42.9736393647342	-34.6675349149033	0.0800493418378402	16\\
-9.388646084847	-35.9447130169263	0.10103544398732	17\\
49.888263022688	-31.0495392279534	0.405460882767637	18\\
35.0648795471465	-29.8842348248309	1.45237885182066	19\\
55.1390911671484	-29.1237983691689	0.225588519906515	20\\
18.6888838987904	-42.6523993143263	0.327672488149874	21\\
-55.7145985710973	-30.4156854858531	0.0954740351452351	22\\
41.8955167042533	-30.8789029755063	0.827276346293124	23\\
52.0791897309061	-50.2001497515931	7.02786845283029	24\\
21.4482185829328	-54.5214598118171	0.478903960388335	25\\
30.9288156694	-34.4346089565159	1.0620814476809	26\\
29.1758961749899	-26.2915051902991	0.627838662487498	27\\
-12.9327576558998	-39.6785136795441	0.112230162050778	28\\
18.6573468213068	-32.3197625786399	0.272847077119772	29\\
-39.4575974626126	-45.6532354648834	0.0900529690103519	30\\
};
\addplot[only marks, mark=square*, mark options={}, mark size=5pt, draw=white!20!black, fill=white!35!black, forget plot] table[row sep=crcr]{%
x	y\\
0	0\\
};

\addlegendimage{legend image code/.code={%
  \draw[red, line width=3.0pt] plot[mark=x, mark size=5.0pt] coordinates {(0.3cm,0cm)};
  \draw[violet, line width=2.0pt] plot[mark=x, mark size=4.0pt] coordinates {(0.3cm,0cm)};
}}
\addlegendentry{Satellites}

\addlegendimage{only marks, mark=*, mark size=4.0pt, draw=black, fill=black}
\addlegendentry{UEs}

\addlegendimage{only marks, mark=square*, mark size=3.5pt, draw=white!20!black, fill=white!35!black}
\addlegendentry{gNB}

\end{axis}
\end{tikzpicture}%

%% file: fig/Fig7_AngularPositions_gNB1_NoSats40_Nt16x16_lambda1_snap1.tex
\definecolor{mycolor1}{rgb}{0.06600,0.44300,0.74500}%
\definecolor{mycolor2}{rgb}{0.86600,0.32900,0.00000}%
\definecolor{mycolor3}{rgb}{0.20000,0.55000,1.00000}%
\definecolor{mycolor4}{rgb}{0.81900,0.01500,0.54500}%
\definecolor{mycolor5}{rgb}{0.12941,0.12941,0.12941}%
\begin{tikzpicture}

\tikzset{centered/.style={anchor=center}}

\begin{axis}[%
width=4in,
height=3in,
tick label style={font=\fontsize{20pt}{26pt}\selectfont},
at={(1.254in,0.802in)},
scale only axis,
point meta min=0,
point meta max=2.57941551638494,
colormap={mymap}{[1pt] rgb(0pt)=(0.0104,0.0000,0.0000); rgb(1pt)=(0.0208,0.0000,0.0000); rgb(2pt)=(0.0312,0.0000,0.0000); rgb(3pt)=(0.0417,0.0000,0.0000); rgb(4pt)=(0.0521,0.0000,0.0000); rgb(5pt)=(0.0625,0.0000,0.0000); rgb(6pt)=(0.0729,0.0000,0.0000); rgb(7pt)=(0.0833,0.0000,0.0000); rgb(8pt)=(0.0938,0.0000,0.0000); rgb(9pt)=(0.1042,0.0000,0.0000); rgb(10pt)=(0.1146,0.0000,0.0000); rgb(11pt)=(0.1250,0.0000,0.0000); rgb(12pt)=(0.1354,0.0000,0.0000); rgb(13pt)=(0.1458,0.0000,0.0000); rgb(14pt)=(0.1562,0.0000,0.0000); rgb(15pt)=(0.1667,0.0000,0.0000); rgb(16pt)=(0.1771,0.0000,0.0000); rgb(17pt)=(0.1875,0.0000,0.0000); rgb(18pt)=(0.1979,0.0000,0.0000); rgb(19pt)=(0.2083,0.0000,0.0000); rgb(20pt)=(0.2188,0.0000,0.0000); rgb(21pt)=(0.2292,0.0000,0.0000); rgb(22pt)=(0.2396,0.0000,0.0000); rgb(23pt)=(0.2500,0.0000,0.0000); rgb(24pt)=(0.2604,0.0000,0.0000); rgb(25pt)=(0.2708,0.0000,0.0000); rgb(26pt)=(0.2812,0.0000,0.0000); rgb(27pt)=(0.2917,0.0000,0.0000); rgb(28pt)=(0.3021,0.0000,0.0000); rgb(29pt)=(0.3125,0.0000,0.0000); rgb(30pt)=(0.3229,0.0000,0.0000); rgb(31pt)=(0.3333,0.0000,0.0000); rgb(32pt)=(0.3438,0.0000,0.0000); rgb(33pt)=(0.3542,0.0000,0.0000); rgb(34pt)=(0.3646,0.0000,0.0000); rgb(35pt)=(0.3750,0.0000,0.0000); rgb(36pt)=(0.3854,0.0000,0.0000); rgb(37pt)=(0.3958,0.0000,0.0000); rgb(38pt)=(0.4062,0.0000,0.0000); rgb(39pt)=(0.4167,0.0000,0.0000); rgb(40pt)=(0.4271,0.0000,0.0000); rgb(41pt)=(0.4375,0.0000,0.0000); rgb(42pt)=(0.4479,0.0000,0.0000); rgb(43pt)=(0.4583,0.0000,0.0000); rgb(44pt)=(0.4688,0.0000,0.0000); rgb(45pt)=(0.4792,0.0000,0.0000); rgb(46pt)=(0.4896,0.0000,0.0000); rgb(47pt)=(0.5000,0.0000,0.0000); rgb(48pt)=(0.5104,0.0000,0.0000); rgb(49pt)=(0.5208,0.0000,0.0000); rgb(50pt)=(0.5312,0.0000,0.0000); rgb(51pt)=(0.5417,0.0000,0.0000); rgb(52pt)=(0.5521,0.0000,0.0000); rgb(53pt)=(0.5625,0.0000,0.0000); rgb(54pt)=(0.5729,0.0000,0.0000); rgb(55pt)=(0.5833,0.0000,0.0000); rgb(56pt)=(0.5938,0.0000,0.0000); rgb(57pt)=(0.6042,0.0000,0.0000); rgb(58pt)=(0.6146,0.0000,0.0000); rgb(59pt)=(0.6250,0.0000,0.0000); rgb(60pt)=(0.6354,0.0000,0.0000); rgb(61pt)=(0.6458,0.0000,0.0000); rgb(62pt)=(0.6562,0.0000,0.0000); rgb(63pt)=(0.6667,0.0000,0.0000); rgb(64pt)=(0.6771,0.0000,0.0000); rgb(65pt)=(0.6875,0.0000,0.0000); rgb(66pt)=(0.6979,0.0000,0.0000); rgb(67pt)=(0.7083,0.0000,0.0000); rgb(68pt)=(0.7188,0.0000,0.0000); rgb(69pt)=(0.7292,0.0000,0.0000); rgb(70pt)=(0.7396,0.0000,0.0000); rgb(71pt)=(0.7500,0.0000,0.0000); rgb(72pt)=(0.7604,0.0000,0.0000); rgb(73pt)=(0.7708,0.0000,0.0000); rgb(74pt)=(0.7812,0.0000,0.0000); rgb(75pt)=(0.7917,0.0000,0.0000); rgb(76pt)=(0.8021,0.0000,0.0000); rgb(77pt)=(0.8125,0.0000,0.0000); rgb(78pt)=(0.8229,0.0000,0.0000); rgb(79pt)=(0.8333,0.0000,0.0000); rgb(80pt)=(0.8438,0.0000,0.0000); rgb(81pt)=(0.8542,0.0000,0.0000); rgb(82pt)=(0.8646,0.0000,0.0000); rgb(83pt)=(0.8750,0.0000,0.0000); rgb(84pt)=(0.8854,0.0000,0.0000); rgb(85pt)=(0.8958,0.0000,0.0000); rgb(86pt)=(0.9062,0.0000,0.0000); rgb(87pt)=(0.9167,0.0000,0.0000); rgb(88pt)=(0.9271,0.0000,0.0000); rgb(89pt)=(0.9375,0.0000,0.0000); rgb(90pt)=(0.9479,0.0000,0.0000); rgb(91pt)=(0.9583,0.0000,0.0000); rgb(92pt)=(0.9688,0.0000,0.0000); rgb(93pt)=(0.9792,0.0000,0.0000); rgb(94pt)=(0.9896,0.0000,0.0000); rgb(95pt)=(1.0000,0.0000,0.0000); rgb(96pt)=(1.0000,0.0104,0.0000); rgb(97pt)=(1.0000,0.0208,0.0000); rgb(98pt)=(1.0000,0.0312,0.0000); rgb(99pt)=(1.0000,0.0417,0.0000); rgb(100pt)=(1.0000,0.0521,0.0000); rgb(101pt)=(1.0000,0.0625,0.0000); rgb(102pt)=(1.0000,0.0729,0.0000); rgb(103pt)=(1.0000,0.0833,0.0000); rgb(104pt)=(1.0000,0.0938,0.0000); rgb(105pt)=(1.0000,0.1042,0.0000); rgb(106pt)=(1.0000,0.1146,0.0000); rgb(107pt)=(1.0000,0.1250,0.0000); rgb(108pt)=(1.0000,0.1354,0.0000); rgb(109pt)=(1.0000,0.1458,0.0000); rgb(110pt)=(1.0000,0.1562,0.0000); rgb(111pt)=(1.0000,0.1667,0.0000); rgb(112pt)=(1.0000,0.1771,0.0000); rgb(113pt)=(1.0000,0.1875,0.0000); rgb(114pt)=(1.0000,0.1979,0.0000); rgb(115pt)=(1.0000,0.2083,0.0000); rgb(116pt)=(1.0000,0.2188,0.0000); rgb(117pt)=(1.0000,0.2292,0.0000); rgb(118pt)=(1.0000,0.2396,0.0000); rgb(119pt)=(1.0000,0.2500,0.0000); rgb(120pt)=(1.0000,0.2604,0.0000); rgb(121pt)=(1.0000,0.2708,0.0000); rgb(122pt)=(1.0000,0.2812,0.0000); rgb(123pt)=(1.0000,0.2917,0.0000); rgb(124pt)=(1.0000,0.3021,0.0000); rgb(125pt)=(1.0000,0.3125,0.0000); rgb(126pt)=(1.0000,0.3229,0.0000); rgb(127pt)=(1.0000,0.3333,0.0000); rgb(128pt)=(1.0000,0.3438,0.0000); rgb(129pt)=(1.0000,0.3542,0.0000); rgb(130pt)=(1.0000,0.3646,0.0000); rgb(131pt)=(1.0000,0.3750,0.0000); rgb(132pt)=(1.0000,0.3854,0.0000); rgb(133pt)=(1.0000,0.3958,0.0000); rgb(134pt)=(1.0000,0.4062,0.0000); rgb(135pt)=(1.0000,0.4167,0.0000); rgb(136pt)=(1.0000,0.4271,0.0000); rgb(137pt)=(1.0000,0.4375,0.0000); rgb(138pt)=(1.0000,0.4479,0.0000); rgb(139pt)=(1.0000,0.4583,0.0000); rgb(140pt)=(1.0000,0.4688,0.0000); rgb(141pt)=(1.0000,0.4792,0.0000); rgb(142pt)=(1.0000,0.4896,0.0000); rgb(143pt)=(1.0000,0.5000,0.0000); rgb(144pt)=(1.0000,0.5104,0.0000); rgb(145pt)=(1.0000,0.5208,0.0000); rgb(146pt)=(1.0000,0.5312,0.0000); rgb(147pt)=(1.0000,0.5417,0.0000); rgb(148pt)=(1.0000,0.5521,0.0000); rgb(149pt)=(1.0000,0.5625,0.0000); rgb(150pt)=(1.0000,0.5729,0.0000); rgb(151pt)=(1.0000,0.5833,0.0000); rgb(152pt)=(1.0000,0.5938,0.0000); rgb(153pt)=(1.0000,0.6042,0.0000); rgb(154pt)=(1.0000,0.6146,0.0000); rgb(155pt)=(1.0000,0.6250,0.0000); rgb(156pt)=(1.0000,0.6354,0.0000); rgb(157pt)=(1.0000,0.6458,0.0000); rgb(158pt)=(1.0000,0.6562,0.0000); rgb(159pt)=(1.0000,0.6667,0.0000); rgb(160pt)=(1.0000,0.6771,0.0000); rgb(161pt)=(1.0000,0.6875,0.0000); rgb(162pt)=(1.0000,0.6979,0.0000); rgb(163pt)=(1.0000,0.7083,0.0000); rgb(164pt)=(1.0000,0.7188,0.0000); rgb(165pt)=(1.0000,0.7292,0.0000); rgb(166pt)=(1.0000,0.7396,0.0000); rgb(167pt)=(1.0000,0.7500,0.0000); rgb(168pt)=(1.0000,0.7604,0.0000); rgb(169pt)=(1.0000,0.7708,0.0000); rgb(170pt)=(1.0000,0.7812,0.0000); rgb(171pt)=(1.0000,0.7917,0.0000); rgb(172pt)=(1.0000,0.8021,0.0000); rgb(173pt)=(1.0000,0.8125,0.0000); rgb(174pt)=(1.0000,0.8229,0.0000); rgb(175pt)=(1.0000,0.8333,0.0000); rgb(176pt)=(1.0000,0.8438,0.0000); rgb(177pt)=(1.0000,0.8542,0.0000); rgb(178pt)=(1.0000,0.8646,0.0000); rgb(179pt)=(1.0000,0.8750,0.0000); rgb(180pt)=(1.0000,0.8854,0.0000); rgb(181pt)=(1.0000,0.8958,0.0000); rgb(182pt)=(1.0000,0.9062,0.0000); rgb(183pt)=(1.0000,0.9167,0.0000); rgb(184pt)=(1.0000,0.9271,0.0000); rgb(185pt)=(1.0000,0.9375,0.0000); rgb(186pt)=(1.0000,0.9479,0.0000); rgb(187pt)=(1.0000,0.9583,0.0000); rgb(188pt)=(1.0000,0.9688,0.0000); rgb(189pt)=(1.0000,0.9792,0.0000); rgb(190pt)=(1.0000,0.9896,0.0000); rgb(191pt)=(1.0000,1.0000,0.0000); rgb(192pt)=(1.0000,1.0000,0.0156); rgb(193pt)=(1.0000,1.0000,0.0312); rgb(194pt)=(1.0000,1.0000,0.0469); rgb(195pt)=(1.0000,1.0000,0.0625); rgb(196pt)=(1.0000,1.0000,0.0781); rgb(197pt)=(1.0000,1.0000,0.0938); rgb(198pt)=(1.0000,1.0000,0.1094); rgb(199pt)=(1.0000,1.0000,0.1250); rgb(200pt)=(1.0000,1.0000,0.1406); rgb(201pt)=(1.0000,1.0000,0.1562); rgb(202pt)=(1.0000,1.0000,0.1719); rgb(203pt)=(1.0000,1.0000,0.1875); rgb(204pt)=(1.0000,1.0000,0.2031); rgb(205pt)=(1.0000,1.0000,0.2188); rgb(206pt)=(1.0000,1.0000,0.2344); rgb(207pt)=(1.0000,1.0000,0.2500); rgb(208pt)=(1.0000,1.0000,0.2656); rgb(209pt)=(1.0000,1.0000,0.2812); rgb(210pt)=(1.0000,1.0000,0.2969); rgb(211pt)=(1.0000,1.0000,0.3125); rgb(212pt)=(1.0000,1.0000,0.3281); rgb(213pt)=(1.0000,1.0000,0.3438); rgb(214pt)=(1.0000,1.0000,0.3594); rgb(215pt)=(1.0000,1.0000,0.3750); rgb(216pt)=(1.0000,1.0000,0.3906); rgb(217pt)=(1.0000,1.0000,0.4062); rgb(218pt)=(1.0000,1.0000,0.4219); rgb(219pt)=(1.0000,1.0000,0.4375); rgb(220pt)=(1.0000,1.0000,0.4531); rgb(221pt)=(1.0000,1.0000,0.4688); rgb(222pt)=(1.0000,1.0000,0.4844); rgb(223pt)=(1.0000,1.0000,0.5000); rgb(224pt)=(1.0000,1.0000,0.5156); rgb(225pt)=(1.0000,1.0000,0.5312); rgb(226pt)=(1.0000,1.0000,0.5469); rgb(227pt)=(1.0000,1.0000,0.5625); rgb(228pt)=(1.0000,1.0000,0.5781); rgb(229pt)=(1.0000,1.0000,0.5938); rgb(230pt)=(1.0000,1.0000,0.6094); rgb(231pt)=(1.0000,1.0000,0.6250); rgb(232pt)=(1.0000,1.0000,0.6406); rgb(233pt)=(1.0000,1.0000,0.6562); rgb(234pt)=(1.0000,1.0000,0.6719); rgb(235pt)=(1.0000,1.0000,0.6875); rgb(236pt)=(1.0000,1.0000,0.7031); rgb(237pt)=(1.0000,1.0000,0.7188); rgb(238pt)=(1.0000,1.0000,0.7344); rgb(239pt)=(1.0000,1.0000,0.7500); rgb(240pt)=(1.0000,1.0000,0.7656); rgb(241pt)=(1.0000,1.0000,0.7812); rgb(242pt)=(1.0000,1.0000,0.7969); rgb(243pt)=(1.0000,1.0000,0.8125); rgb(244pt)=(1.0000,1.0000,0.8281); rgb(245pt)=(1.0000,1.0000,0.8438); rgb(246pt)=(1.0000,1.0000,0.8594); rgb(247pt)=(1.0000,1.0000,0.8750); rgb(248pt)=(1.0000,1.0000,0.8906); rgb(249pt)=(1.0000,1.0000,0.9062); rgb(250pt)=(1.0000,1.0000,0.9219); rgb(251pt)=(1.0000,1.0000,0.9375); rgb(252pt)=(1.0000,1.0000,0.9531); rgb(253pt)=(1.0000,1.0000,0.9688); rgb(254pt)=(1.0000,1.0000,0.9844); rgb(255pt)=(1.0000,1.0000,1.0000)},
unbounded coords=jump,
xmin=-60,
xmax=60,
xlabel style={font=\fontsize{48}{50}\selectfont},
xlabel={Azimuth angle ($\phi$) [$^\circ$]},
ymin=-80,
ymax=90,
ylabel style={font=\fontsize{48}{50}\selectfont},
ylabel={Elevation angle ($\theta$) [$^\circ$]},
axis background/.style={fill=white},
axis x line*=bottom,
axis y line*=left,
axis lines=box,
xmajorgrids,
ymajorgrids,
legend style={legend cell align=left, align=left, font=\fontsize{24pt}{30pt}\selectfont},
colorbar,
colorbar style={ylabel style={font=\Huge}, ylabel={RSS Degradation [dB]}},
ylabel style={font=\Huge},
xlabel style={font=\Huge}
]
\addplot[only marks, mark=x, mark options={}, mark size=3.7081pt, draw=red, forget plot] table[row sep=crcr]{%
x	y\\
-29.913045582444	83.467765732328\\
-37.331931452409	40.4928270911073\\
-37.3857422966607	39.9916557051464\\
-35.9439569233537	54.7399846405927\\
-37.5269899201166	38.701041726755\\
-34.482955942135	68.571425380883\\
-21.7903138594012	87.3036461855286\\
-35.3590521980698	60.8044080841969\\
-37.4622811396456	39.2877865860583\\
-39.8770623483742	22.3084666776452\\
-36.0344904942684	53.7743108355357\\
-40.0258095303973	21.5407390551593\\
-32.9240737430791	77.1674587065132\\
-39.4837572938755	24.4660524470951\\
-39.7182234865812	23.1567969959555\\
-9.3842841700594	88.5587373359836\\
-40.0811069804453	21.2616663862124\\
-38.4744854472882	30.9966619233646\\
-33.2810070244342	75.7123388212849\\
-38.2464753540496	32.7009762780514\\
-34.0322772560431	71.7109984649821\\
-36.6104860955641	47.6621228806058\\
-39.0673229358711	26.9736348692644\\
-38.572719949421	30.2902485963232\\
-39.5095811731233	24.3183914878462\\
-39.5360311186428	24.1680555591401\\
-38.0295560008468	34.4089627280394\\
-37.4386976793572	39.5035329639839\\
-35.4971620049957	59.4140569750991\\
-38.4119146725443	31.4552687770997\\
-38.1253631600541	33.6440065917086\\
-38.7325162382167	29.1758023523534\\
-38.1612483762703	33.3618143942112\\
-39.4617982029872	24.5923064862472\\
-32.8391518954244	77.4788441887845\\
-35.1281877548394	63.0432220259435\\
-34.8388622868693	65.6664667507763\\
-39.1234937378509	26.6209722730844\\
-30.7436550927979	82.3942579890017\\
-37.921571474812	35.2914649946493\\
};
\addplot[only marks, mark=x, mark options={}, mark size=2.9580pt, draw=violet, forget plot] table[row sep=crcr]{%
x	y\\
-29.913045582444	83.467765732328\\
-37.331931452409	40.4928270911073\\
-37.3857422966607	39.9916557051464\\
-35.9439569233537	54.7399846405927\\
-37.5269899201166	38.701041726755\\
-34.482955942135	68.571425380883\\
-21.7903138594012	87.3036461855286\\
-35.3590521980698	60.8044080841969\\
-37.4622811396456	39.2877865860583\\
-39.8770623483742	22.3084666776452\\
-36.0344904942684	53.7743108355357\\
-40.0258095303973	21.5407390551593\\
-32.9240737430791	77.1674587065132\\
-39.4837572938755	24.4660524470951\\
-39.7182234865812	23.1567969959555\\
-9.3842841700594	88.5587373359836\\
-40.0811069804453	21.2616663862124\\
-38.4744854472882	30.9966619233646\\
-33.2810070244342	75.7123388212849\\
-38.2464753540496	32.7009762780514\\
-34.0322772560431	71.7109984649821\\
-36.6104860955641	47.6621228806058\\
-39.0673229358711	26.9736348692644\\
-38.572719949421	30.2902485963232\\
-39.5095811731233	24.3183914878462\\
-39.5360311186428	24.1680555591401\\
-38.0295560008468	34.4089627280394\\
-37.4386976793572	39.5035329639839\\
-35.4971620049957	59.4140569750991\\
-38.4119146725443	31.4552687770997\\
-38.1253631600541	33.6440065917086\\
-38.7325162382167	29.1758023523534\\
-38.1612483762703	33.3618143942112\\
-39.4617982029872	24.5923064862472\\
-32.8391518954244	77.4788441887845\\
-35.1281877548394	63.0432220259435\\
-34.8388622868693	65.6664667507763\\
-39.1234937378509	26.6209722730844\\
-30.7436550927979	82.3942579890017\\
-37.921571474812	35.2914649946493\\
};
\addplot[scatter, only marks, mark=*, mark size=5.25pt, scatter src=explicit, 
scatter/use mapped color={mark options={}, draw=black, fill=mapped color},
nodes near coords*={\lbl},
visualization depends on={value \thisrow{label} \as \lbl},
every node near coord/.style={anchor=center, font=\normalsize\bfseries, text=blue!50!white},
forget plot] table[row sep=crcr, meta=color]{%
x	y	color	label\\
37.7668423671815	-29.9423050138247	0.23705493753529	1\\
48.6950324490743	-69.7645275237754	1.229093606269	2\\
-44.7615820447793	-42.6060384406171	0.0156075045102558	3\\
49.6051027366823	-66.060826815162	1.10487999618769	4\\
15.8831095470491	-57.2215006161576	0.0568562800707415	5\\
-48.2951514000708	-28.0739086317017	0.0522213624640664	6\\
-26.5802137359542	-30.1411392153783	0.00363246326950658	7\\
5.62578230459806	-40.6791707652455	0.00852528028812588	8\\
54.9008202521157	-26.4051109118662	0.0430453267660335	9\\
55.7866242239132	-69.0199817599323	0.751462851012748	10\\
-41.0864301986942	-36.1556298212871	0.0164385381008975	11\\
56.4711338112739	-38.0803530296104	0.243200362593324	12\\
54.8600337891535	-28.9505767671532	0.0427576074958065	13\\
-1.75492215325905	-28.49130456038	0.00824580286197361	14\\
36.033656266656	-48.2301649180406	0.848642473196207	15\\
-42.9736393647342	-34.6675349149033	0.00600739494782726	16\\
-9.388646084847	-35.9447130169263	0.0148508753473707	17\\
49.888263022688	-31.0495392279534	0.0486133723410983	18\\
35.0648795471465	-29.8842348248309	0.363070990237134	19\\
55.1390911671484	-29.1237983691689	0.0464891499235339	20\\
18.6888838987904	-42.6523993143263	0.026524633025901	21\\
-55.7145985710973	-30.4156854858531	0.0545972765655344	22\\
41.8955167042533	-30.8789029755063	0.0895500797894202	23\\
52.0791897309061	-50.2001497515931	2.57941551638494	24\\
21.4482185829328	-54.5214598118171	0.0703160428447652	25\\
30.9288156694	-34.4346089565159	0.328477666441394	26\\
29.1758961749899	-26.2915051902991	0.260910637355968	27\\
-12.9327576558998	-39.6785136795441	0.0193455629858627	28\\
18.6573468213068	-32.3197625786399	0.102088187939013	29\\
-39.4575974626126	-45.6532354648834	0.0109564843782947	30\\
};
\addplot[only marks, mark=square*, mark options={}, mark size=5pt, draw=white!20!black, fill=white!35!black, forget plot] table[row sep=crcr]{%
x	y\\
0	0\\
};

\addlegendimage{legend image code/.code={%
  \draw[red, line width=3.0pt] plot[mark=x, mark size=5.0pt] coordinates {(0.3cm,0cm)};
  \draw[violet, line width=2.0pt] plot[mark=x, mark size=4.0pt] coordinates {(0.3cm,0cm)};
}}
\addlegendentry{Satellites}

\addlegendimage{only marks, mark=*, mark size=4.0pt, draw=black, fill=black}
\addlegendentry{UEs}

\addlegendimage{only marks, mark=square*, mark size=3.5pt, draw=white!20!black, fill=white!35!black}
\addlegendentry{gNB}

\end{axis}
\end{tikzpicture}%

%% file: fig/Fig7_AngularPositions_gNB1_NoSats40_Nt32x32_lambda1_snap1.tex
\definecolor{mycolor1}{rgb}{0.06600,0.44300,0.74500}%
\definecolor{mycolor2}{rgb}{0.86600,0.32900,0.00000}%
\definecolor{mycolor3}{rgb}{0.20000,0.55000,1.00000}%
\definecolor{mycolor4}{rgb}{0.81900,0.01500,0.54500}%
\definecolor{mycolor5}{rgb}{0.12941,0.12941,0.12941}%
\begin{tikzpicture}

\tikzset{centered/.style={anchor=center}}

\begin{axis}[%
width=4in,
height=3in,
tick label style={font=\fontsize{20pt}{26pt}\selectfont},
at={(1.254in,0.802in)},
scale only axis,
point meta min=0,
point meta max=0.500959309984684,
colormap={mymap}{[1pt] rgb(0pt)=(0.0104,0.0000,0.0000); rgb(1pt)=(0.0208,0.0000,0.0000); rgb(2pt)=(0.0312,0.0000,0.0000); rgb(3pt)=(0.0417,0.0000,0.0000); rgb(4pt)=(0.0521,0.0000,0.0000); rgb(5pt)=(0.0625,0.0000,0.0000); rgb(6pt)=(0.0729,0.0000,0.0000); rgb(7pt)=(0.0833,0.0000,0.0000); rgb(8pt)=(0.0938,0.0000,0.0000); rgb(9pt)=(0.1042,0.0000,0.0000); rgb(10pt)=(0.1146,0.0000,0.0000); rgb(11pt)=(0.1250,0.0000,0.0000); rgb(12pt)=(0.1354,0.0000,0.0000); rgb(13pt)=(0.1458,0.0000,0.0000); rgb(14pt)=(0.1562,0.0000,0.0000); rgb(15pt)=(0.1667,0.0000,0.0000); rgb(16pt)=(0.1771,0.0000,0.0000); rgb(17pt)=(0.1875,0.0000,0.0000); rgb(18pt)=(0.1979,0.0000,0.0000); rgb(19pt)=(0.2083,0.0000,0.0000); rgb(20pt)=(0.2188,0.0000,0.0000); rgb(21pt)=(0.2292,0.0000,0.0000); rgb(22pt)=(0.2396,0.0000,0.0000); rgb(23pt)=(0.2500,0.0000,0.0000); rgb(24pt)=(0.2604,0.0000,0.0000); rgb(25pt)=(0.2708,0.0000,0.0000); rgb(26pt)=(0.2812,0.0000,0.0000); rgb(27pt)=(0.2917,0.0000,0.0000); rgb(28pt)=(0.3021,0.0000,0.0000); rgb(29pt)=(0.3125,0.0000,0.0000); rgb(30pt)=(0.3229,0.0000,0.0000); rgb(31pt)=(0.3333,0.0000,0.0000); rgb(32pt)=(0.3438,0.0000,0.0000); rgb(33pt)=(0.3542,0.0000,0.0000); rgb(34pt)=(0.3646,0.0000,0.0000); rgb(35pt)=(0.3750,0.0000,0.0000); rgb(36pt)=(0.3854,0.0000,0.0000); rgb(37pt)=(0.3958,0.0000,0.0000); rgb(38pt)=(0.4062,0.0000,0.0000); rgb(39pt)=(0.4167,0.0000,0.0000); rgb(40pt)=(0.4271,0.0000,0.0000); rgb(41pt)=(0.4375,0.0000,0.0000); rgb(42pt)=(0.4479,0.0000,0.0000); rgb(43pt)=(0.4583,0.0000,0.0000); rgb(44pt)=(0.4688,0.0000,0.0000); rgb(45pt)=(0.4792,0.0000,0.0000); rgb(46pt)=(0.4896,0.0000,0.0000); rgb(47pt)=(0.5000,0.0000,0.0000); rgb(48pt)=(0.5104,0.0000,0.0000); rgb(49pt)=(0.5208,0.0000,0.0000); rgb(50pt)=(0.5312,0.0000,0.0000); rgb(51pt)=(0.5417,0.0000,0.0000); rgb(52pt)=(0.5521,0.0000,0.0000); rgb(53pt)=(0.5625,0.0000,0.0000); rgb(54pt)=(0.5729,0.0000,0.0000); rgb(55pt)=(0.5833,0.0000,0.0000); rgb(56pt)=(0.5938,0.0000,0.0000); rgb(57pt)=(0.6042,0.0000,0.0000); rgb(58pt)=(0.6146,0.0000,0.0000); rgb(59pt)=(0.6250,0.0000,0.0000); rgb(60pt)=(0.6354,0.0000,0.0000); rgb(61pt)=(0.6458,0.0000,0.0000); rgb(62pt)=(0.6562,0.0000,0.0000); rgb(63pt)=(0.6667,0.0000,0.0000); rgb(64pt)=(0.6771,0.0000,0.0000); rgb(65pt)=(0.6875,0.0000,0.0000); rgb(66pt)=(0.6979,0.0000,0.0000); rgb(67pt)=(0.7083,0.0000,0.0000); rgb(68pt)=(0.7188,0.0000,0.0000); rgb(69pt)=(0.7292,0.0000,0.0000); rgb(70pt)=(0.7396,0.0000,0.0000); rgb(71pt)=(0.7500,0.0000,0.0000); rgb(72pt)=(0.7604,0.0000,0.0000); rgb(73pt)=(0.7708,0.0000,0.0000); rgb(74pt)=(0.7812,0.0000,0.0000); rgb(75pt)=(0.7917,0.0000,0.0000); rgb(76pt)=(0.8021,0.0000,0.0000); rgb(77pt)=(0.8125,0.0000,0.0000); rgb(78pt)=(0.8229,0.0000,0.0000); rgb(79pt)=(0.8333,0.0000,0.0000); rgb(80pt)=(0.8438,0.0000,0.0000); rgb(81pt)=(0.8542,0.0000,0.0000); rgb(82pt)=(0.8646,0.0000,0.0000); rgb(83pt)=(0.8750,0.0000,0.0000); rgb(84pt)=(0.8854,0.0000,0.0000); rgb(85pt)=(0.8958,0.0000,0.0000); rgb(86pt)=(0.9062,0.0000,0.0000); rgb(87pt)=(0.9167,0.0000,0.0000); rgb(88pt)=(0.9271,0.0000,0.0000); rgb(89pt)=(0.9375,0.0000,0.0000); rgb(90pt)=(0.9479,0.0000,0.0000); rgb(91pt)=(0.9583,0.0000,0.0000); rgb(92pt)=(0.9688,0.0000,0.0000); rgb(93pt)=(0.9792,0.0000,0.0000); rgb(94pt)=(0.9896,0.0000,0.0000); rgb(95pt)=(1.0000,0.0000,0.0000); rgb(96pt)=(1.0000,0.0104,0.0000); rgb(97pt)=(1.0000,0.0208,0.0000); rgb(98pt)=(1.0000,0.0312,0.0000); rgb(99pt)=(1.0000,0.0417,0.0000); rgb(100pt)=(1.0000,0.0521,0.0000); rgb(101pt)=(1.0000,0.0625,0.0000); rgb(102pt)=(1.0000,0.0729,0.0000); rgb(103pt)=(1.0000,0.0833,0.0000); rgb(104pt)=(1.0000,0.0938,0.0000); rgb(105pt)=(1.0000,0.1042,0.0000); rgb(106pt)=(1.0000,0.1146,0.0000); rgb(107pt)=(1.0000,0.1250,0.0000); rgb(108pt)=(1.0000,0.1354,0.0000); rgb(109pt)=(1.0000,0.1458,0.0000); rgb(110pt)=(1.0000,0.1562,0.0000); rgb(111pt)=(1.0000,0.1667,0.0000); rgb(112pt)=(1.0000,0.1771,0.0000); rgb(113pt)=(1.0000,0.1875,0.0000); rgb(114pt)=(1.0000,0.1979,0.0000); rgb(115pt)=(1.0000,0.2083,0.0000); rgb(116pt)=(1.0000,0.2188,0.0000); rgb(117pt)=(1.0000,0.2292,0.0000); rgb(118pt)=(1.0000,0.2396,0.0000); rgb(119pt)=(1.0000,0.2500,0.0000); rgb(120pt)=(1.0000,0.2604,0.0000); rgb(121pt)=(1.0000,0.2708,0.0000); rgb(122pt)=(1.0000,0.2812,0.0000); rgb(123pt)=(1.0000,0.2917,0.0000); rgb(124pt)=(1.0000,0.3021,0.0000); rgb(125pt)=(1.0000,0.3125,0.0000); rgb(126pt)=(1.0000,0.3229,0.0000); rgb(127pt)=(1.0000,0.3333,0.0000); rgb(128pt)=(1.0000,0.3438,0.0000); rgb(129pt)=(1.0000,0.3542,0.0000); rgb(130pt)=(1.0000,0.3646,0.0000); rgb(131pt)=(1.0000,0.3750,0.0000); rgb(132pt)=(1.0000,0.3854,0.0000); rgb(133pt)=(1.0000,0.3958,0.0000); rgb(134pt)=(1.0000,0.4062,0.0000); rgb(135pt)=(1.0000,0.4167,0.0000); rgb(136pt)=(1.0000,0.4271,0.0000); rgb(137pt)=(1.0000,0.4375,0.0000); rgb(138pt)=(1.0000,0.4479,0.0000); rgb(139pt)=(1.0000,0.4583,0.0000); rgb(140pt)=(1.0000,0.4688,0.0000); rgb(141pt)=(1.0000,0.4792,0.0000); rgb(142pt)=(1.0000,0.4896,0.0000); rgb(143pt)=(1.0000,0.5000,0.0000); rgb(144pt)=(1.0000,0.5104,0.0000); rgb(145pt)=(1.0000,0.5208,0.0000); rgb(146pt)=(1.0000,0.5312,0.0000); rgb(147pt)=(1.0000,0.5417,0.0000); rgb(148pt)=(1.0000,0.5521,0.0000); rgb(149pt)=(1.0000,0.5625,0.0000); rgb(150pt)=(1.0000,0.5729,0.0000); rgb(151pt)=(1.0000,0.5833,0.0000); rgb(152pt)=(1.0000,0.5938,0.0000); rgb(153pt)=(1.0000,0.6042,0.0000); rgb(154pt)=(1.0000,0.6146,0.0000); rgb(155pt)=(1.0000,0.6250,0.0000); rgb(156pt)=(1.0000,0.6354,0.0000); rgb(157pt)=(1.0000,0.6458,0.0000); rgb(158pt)=(1.0000,0.6562,0.0000); rgb(159pt)=(1.0000,0.6667,0.0000); rgb(160pt)=(1.0000,0.6771,0.0000); rgb(161pt)=(1.0000,0.6875,0.0000); rgb(162pt)=(1.0000,0.6979,0.0000); rgb(163pt)=(1.0000,0.7083,0.0000); rgb(164pt)=(1.0000,0.7188,0.0000); rgb(165pt)=(1.0000,0.7292,0.0000); rgb(166pt)=(1.0000,0.7396,0.0000); rgb(167pt)=(1.0000,0.7500,0.0000); rgb(168pt)=(1.0000,0.7604,0.0000); rgb(169pt)=(1.0000,0.7708,0.0000); rgb(170pt)=(1.0000,0.7812,0.0000); rgb(171pt)=(1.0000,0.7917,0.0000); rgb(172pt)=(1.0000,0.8021,0.0000); rgb(173pt)=(1.0000,0.8125,0.0000); rgb(174pt)=(1.0000,0.8229,0.0000); rgb(175pt)=(1.0000,0.8333,0.0000); rgb(176pt)=(1.0000,0.8438,0.0000); rgb(177pt)=(1.0000,0.8542,0.0000); rgb(178pt)=(1.0000,0.8646,0.0000); rgb(179pt)=(1.0000,0.8750,0.0000); rgb(180pt)=(1.0000,0.8854,0.0000); rgb(181pt)=(1.0000,0.8958,0.0000); rgb(182pt)=(1.0000,0.9062,0.0000); rgb(183pt)=(1.0000,0.9167,0.0000); rgb(184pt)=(1.0000,0.9271,0.0000); rgb(185pt)=(1.0000,0.9375,0.0000); rgb(186pt)=(1.0000,0.9479,0.0000); rgb(187pt)=(1.0000,0.9583,0.0000); rgb(188pt)=(1.0000,0.9688,0.0000); rgb(189pt)=(1.0000,0.9792,0.0000); rgb(190pt)=(1.0000,0.9896,0.0000); rgb(191pt)=(1.0000,1.0000,0.0000); rgb(192pt)=(1.0000,1.0000,0.0156); rgb(193pt)=(1.0000,1.0000,0.0312); rgb(194pt)=(1.0000,1.0000,0.0469); rgb(195pt)=(1.0000,1.0000,0.0625); rgb(196pt)=(1.0000,1.0000,0.0781); rgb(197pt)=(1.0000,1.0000,0.0938); rgb(198pt)=(1.0000,1.0000,0.1094); rgb(199pt)=(1.0000,1.0000,0.1250); rgb(200pt)=(1.0000,1.0000,0.1406); rgb(201pt)=(1.0000,1.0000,0.1562); rgb(202pt)=(1.0000,1.0000,0.1719); rgb(203pt)=(1.0000,1.0000,0.1875); rgb(204pt)=(1.0000,1.0000,0.2031); rgb(205pt)=(1.0000,1.0000,0.2188); rgb(206pt)=(1.0000,1.0000,0.2344); rgb(207pt)=(1.0000,1.0000,0.2500); rgb(208pt)=(1.0000,1.0000,0.2656); rgb(209pt)=(1.0000,1.0000,0.2812); rgb(210pt)=(1.0000,1.0000,0.2969); rgb(211pt)=(1.0000,1.0000,0.3125); rgb(212pt)=(1.0000,1.0000,0.3281); rgb(213pt)=(1.0000,1.0000,0.3438); rgb(214pt)=(1.0000,1.0000,0.3594); rgb(215pt)=(1.0000,1.0000,0.3750); rgb(216pt)=(1.0000,1.0000,0.3906); rgb(217pt)=(1.0000,1.0000,0.4062); rgb(218pt)=(1.0000,1.0000,0.4219); rgb(219pt)=(1.0000,1.0000,0.4375); rgb(220pt)=(1.0000,1.0000,0.4531); rgb(221pt)=(1.0000,1.0000,0.4688); rgb(222pt)=(1.0000,1.0000,0.4844); rgb(223pt)=(1.0000,1.0000,0.5000); rgb(224pt)=(1.0000,1.0000,0.5156); rgb(225pt)=(1.0000,1.0000,0.5312); rgb(226pt)=(1.0000,1.0000,0.5469); rgb(227pt)=(1.0000,1.0000,0.5625); rgb(228pt)=(1.0000,1.0000,0.5781); rgb(229pt)=(1.0000,1.0000,0.5938); rgb(230pt)=(1.0000,1.0000,0.6094); rgb(231pt)=(1.0000,1.0000,0.6250); rgb(232pt)=(1.0000,1.0000,0.6406); rgb(233pt)=(1.0000,1.0000,0.6562); rgb(234pt)=(1.0000,1.0000,0.6719); rgb(235pt)=(1.0000,1.0000,0.6875); rgb(236pt)=(1.0000,1.0000,0.7031); rgb(237pt)=(1.0000,1.0000,0.7188); rgb(238pt)=(1.0000,1.0000,0.7344); rgb(239pt)=(1.0000,1.0000,0.7500); rgb(240pt)=(1.0000,1.0000,0.7656); rgb(241pt)=(1.0000,1.0000,0.7812); rgb(242pt)=(1.0000,1.0000,0.7969); rgb(243pt)=(1.0000,1.0000,0.8125); rgb(244pt)=(1.0000,1.0000,0.8281); rgb(245pt)=(1.0000,1.0000,0.8438); rgb(246pt)=(1.0000,1.0000,0.8594); rgb(247pt)=(1.0000,1.0000,0.8750); rgb(248pt)=(1.0000,1.0000,0.8906); rgb(249pt)=(1.0000,1.0000,0.9062); rgb(250pt)=(1.0000,1.0000,0.9219); rgb(251pt)=(1.0000,1.0000,0.9375); rgb(252pt)=(1.0000,1.0000,0.9531); rgb(253pt)=(1.0000,1.0000,0.9688); rgb(254pt)=(1.0000,1.0000,0.9844); rgb(255pt)=(1.0000,1.0000,1.0000)},
unbounded coords=jump,
xmin=-60,
xmax=60,
xlabel style={font=\fontsize{48}{50}\selectfont},
xlabel={Azimuth angle ($\phi$) [$^\circ$]},
ymin=-80,
ymax=90,
ylabel style={font=\fontsize{48}{50}\selectfont},
ylabel={Elevation angle ($\theta$)  [$^\circ$]},
axis background/.style={fill=white},
axis x line*=bottom,
axis y line*=left,
axis lines=box,
xmajorgrids,
ymajorgrids,
legend style={legend cell align=left, align=left, font=\fontsize{24pt}{30pt}\selectfont},
colorbar,
colorbar style={ylabel style={font=\Huge}, ylabel={RSS Degradation [dB]}},
ylabel style={font=\Huge},
xlabel style={font=\Huge}
]
\addplot[only marks, mark=x, mark options={}, mark size=3.7081pt, draw=red, forget plot] table[row sep=crcr]{%
x	y\\
-29.913045582444	83.467765732328\\
-37.331931452409	40.4928270911073\\
-37.3857422966607	39.9916557051464\\
-35.9439569233537	54.7399846405927\\
-37.5269899201166	38.701041726755\\
-34.482955942135	68.571425380883\\
-21.7903138594012	87.3036461855286\\
-35.3590521980698	60.8044080841969\\
-37.4622811396456	39.2877865860583\\
-39.8770623483742	22.3084666776452\\
-36.0344904942684	53.7743108355357\\
-40.0258095303973	21.5407390551593\\
-32.9240737430791	77.1674587065132\\
-39.4837572938755	24.4660524470951\\
-39.7182234865812	23.1567969959555\\
-9.3842841700594	88.5587373359836\\
-40.0811069804453	21.2616663862124\\
-38.4744854472882	30.9966619233646\\
-33.2810070244342	75.7123388212849\\
-38.2464753540496	32.7009762780514\\
-34.0322772560431	71.7109984649821\\
-36.6104860955641	47.6621228806058\\
-39.0673229358711	26.9736348692644\\
-38.572719949421	30.2902485963232\\
-39.5095811731233	24.3183914878462\\
-39.5360311186428	24.1680555591401\\
-38.0295560008468	34.4089627280394\\
-37.4386976793572	39.5035329639839\\
-35.4971620049957	59.4140569750991\\
-38.4119146725443	31.4552687770997\\
-38.1253631600541	33.6440065917086\\
-38.7325162382167	29.1758023523534\\
-38.1612483762703	33.3618143942112\\
-39.4617982029872	24.5923064862472\\
-32.8391518954244	77.4788441887845\\
-35.1281877548394	63.0432220259435\\
-34.8388622868693	65.6664667507763\\
-39.1234937378509	26.6209722730844\\
-30.7436550927979	82.3942579890017\\
-37.921571474812	35.2914649946493\\
};
\addplot[only marks, mark=x, mark options={}, mark size=2.9580pt, draw=violet, forget plot] table[row sep=crcr]{%
x	y\\
-29.913045582444	83.467765732328\\
-37.331931452409	40.4928270911073\\
-37.3857422966607	39.9916557051464\\
-35.9439569233537	54.7399846405927\\
-37.5269899201166	38.701041726755\\
-34.482955942135	68.571425380883\\
-21.7903138594012	87.3036461855286\\
-35.3590521980698	60.8044080841969\\
-37.4622811396456	39.2877865860583\\
-39.8770623483742	22.3084666776452\\
-36.0344904942684	53.7743108355357\\
-40.0258095303973	21.5407390551593\\
-32.9240737430791	77.1674587065132\\
-39.4837572938755	24.4660524470951\\
-39.7182234865812	23.1567969959555\\
-9.3842841700594	88.5587373359836\\
-40.0811069804453	21.2616663862124\\
-38.4744854472882	30.9966619233646\\
-33.2810070244342	75.7123388212849\\
-38.2464753540496	32.7009762780514\\
-34.0322772560431	71.7109984649821\\
-36.6104860955641	47.6621228806058\\
-39.0673229358711	26.9736348692644\\
-38.572719949421	30.2902485963232\\
-39.5095811731233	24.3183914878462\\
-39.5360311186428	24.1680555591401\\
-38.0295560008468	34.4089627280394\\
-37.4386976793572	39.5035329639839\\
-35.4971620049957	59.4140569750991\\
-38.4119146725443	31.4552687770997\\
-38.1253631600541	33.6440065917086\\
-38.7325162382167	29.1758023523534\\
-38.1612483762703	33.3618143942112\\
-39.4617982029872	24.5923064862472\\
-32.8391518954244	77.4788441887845\\
-35.1281877548394	63.0432220259435\\
-34.8388622868693	65.6664667507763\\
-39.1234937378509	26.6209722730844\\
-30.7436550927979	82.3942579890017\\
-37.921571474812	35.2914649946493\\
};
\addplot[scatter, only marks, mark=*, mark size=5.25pt, scatter src=explicit, 
scatter/use mapped color={mark options={}, draw=black, fill=mapped color},
nodes near coords*={\lbl},
visualization depends on={value \thisrow{label} \as \lbl},
every node near coord/.style={anchor=center, font=\normalsize\bfseries, text=blue!50!white},
forget plot] table[row sep=crcr, meta=color]{%
x	y	color	label\\
37.7668423671815	-29.9423050138247	0.0185532357502995	1\\
48.6950324490743	-69.7645275237754	0.500959309984684	2\\
-44.7615820447793	-42.6060384406171	0.00269770036058625	3\\
49.6051027366823	-66.060826815162	0.29916944065558	4\\
15.8831095470491	-57.2215006161576	0.0175182167026889	5\\
-48.2951514000708	-28.0739086317017	0.00832471853217265	6\\
-26.5802137359542	-30.1411392153783	0.000536046061422676	7\\
5.62578230459806	-40.6791707652455	0.00254223303964139	8\\
54.9008202521157	-26.4051109118662	0.00923760990168379	9\\
55.7866242239132	-69.0199817599323	0.426713421045724	10\\
-41.0864301986942	-36.1556298212871	0.0137457542058322	11\\
56.4711338112739	-38.0803530296104	0.0213282133482858	12\\
54.8600337891535	-28.9505767671532	0.011770012095346	13\\
-1.75492215325905	-28.49130456038	0.00230512435902154	14\\
36.033656266656	-48.2301649180406	0.0440242160841977	15\\
-42.9736393647342	-34.6675349149033	0.00215064372876506	16\\
-9.388646084847	-35.9447130169263	0.00227824012898996	17\\
49.888263022688	-31.0495392279534	0.0191846790322414	18\\
35.0648795471465	-29.8842348248309	0.0252851676630078	19\\
55.1390911671484	-29.1237983691689	0.0130320635061217	20\\
18.6888838987904	-42.6523993143263	0.00469700475910507	21\\
-55.7145985710973	-30.4156854858531	0.0311061385114971	22\\
41.8955167042533	-30.8789029755063	0.0234641454562929	23\\
52.0791897309061	-50.2001497515931	0.290989445818877	24\\
21.4482185829328	-54.5214598118171	0.00797671810265604	25\\
30.9288156694	-34.4346089565159	0.0111108826964875	26\\
29.1758961749899	-26.2915051902991	0.0213266323616282	27\\
-12.9327576558998	-39.6785136795441	0.00448547062847266	28\\
18.6573468213068	-32.3197625786399	0.0498812081392137	29\\
-39.4575974626126	-45.6532354648834	0.00422339864309909	30\\
};
\addplot[only marks, mark=square*, mark options={}, mark size=5pt, draw=white!20!black, fill=white!35!black, forget plot] table[row sep=crcr]{%
x	y\\
0	0\\
};

\addlegendimage{legend image code/.code={%
  \draw[red, line width=3.0pt] plot[mark=x, mark size=5.0pt] coordinates {(0.3cm,0cm)};
  \draw[violet, line width=2.0pt] plot[mark=x, mark size=4.0pt] coordinates {(0.3cm,0cm)};
}}
\addlegendentry{Satellites}

\addlegendimage{only marks, mark=*, mark size=4.0pt, draw=black, fill=black}
\addlegendentry{UEs}

\addlegendimage{only marks, mark=square*, mark size=3.5pt, draw=white!20!black, fill=white!35!black}
\addlegendentry{gNB}

\end{axis}
\end{tikzpicture}%

%% file: tex/conclusion.tex
% !TEX root = ../thesis_dissertation template.tex
% conclusion.tex

\chapter{Conclusions}
\label{chap:conclusion}

This dissertation has focused on the algorithmic design of network management solutions at the system level for the optimization of the \gls{gnb}'s scheduling operations, ranging from classical optimization and game theory to \gls{ml} and \gls{ai} on and for the \gls{ran}. 

In detail, we have focused on the design of \gls{drl} solutions for slicing and scheduling control, leveraged at the network's edge. We demonstrated the feasibility of the deployment of \gls{ai}/\gls{ml} algorithms that can on-demand fine-tune the network's resources and functionalities, enabling \gls{ai} on and for the \gls{ran}. More importantly, we contributed to the research body by performing the largest evaluation of xApps ($23$) in the literature, spanning various design control choices and \gls{ml} architectures, providing the research community, telco operators, and xApp developers a framework that automates \gls{drl} agent training, xApp on-boarding, and large-scale evaluation and testing with Open~\gls{ran} within Colosseum, the world's largest wireless network emulator with hardware-in-the-loop. Our work demonstrated the benefits of leveraging \gls{ai} solutions on the \gls{ran} that can coexist with minimal conflicts using realistic protocol stacks, yielding practical lessons and offering key insights for the potential deployment of \gls{ai} on commercial \glspl{ran}.

Next, building on the principles that \gls{6g} research will emerge from simulation, we have designed an online \gls{rl} solution for link adaptation. We have designed the algorithmic framework within a \gls{dt}, showcasing how \gls{ai} on the \gls{ran}, deployed at the \gls{gnb}'s scheduler, can optimize performance by yielding higher spectral efficiency compared to industry standard and state-of-the-art algorithmic designs. Finally, we bridged the connection between simulation and real-world deployment, demonstrating the suitability of \gls{rl} for link adaptation on \gls{5g} data collected \gls{ota} from $24$~h operating \gls{5g} networks deployed at Northeastern University.

Apart from treating the wireless environment as a static entity, we also treated the \gls{rf} propagation as a quantity that can be optimized. Through intelligent beamforming via \glspl{ris}, we proposed a hierarchical game-theoretic approach for an energy-efficient power control mechanism that can be deployed in both \gls{fr1} and \gls{fr2}, providing enhanced energy efficiency without compromising \gls{qos}. Our work demonstrated the enhanced power saving for the latter band, pinpointing the necessity for the deployment of small panels, avoiding the complexity of larger panel arrays. Taking a step further, we deployed the \gls{ris}-enabled channels within the Colosseum channel emulator, demonstrating for the first time the emulation of realistic \gls{ris}-enabled channels within a massive channel emulator. Finally, this work is one of the first of its kind, discussing the bridging between Open~\gls{ran} and \gls{ris}. By deploying an xApp at the Near-RT \gls{ric} within Colosseum, we demonstrated the decreased levels of latency introduced by \glspl{ris} for the \gls{urllc} slice, indicating that both a \gls{ris} and the proper fine-tuning of the scheduling algorithm within the \gls{gnb} can enhance network performance in terms of spectral efficiency and utilization within resource-constrained environments for \glspl{ue} allocated to various network slices.

Finally, we have discussed the coexistence of cellular and \glspl{ntn}, a critical issue alongside \gls{ris} and \gls{ai}-\gls{ran} for \gls{6g} research. We evaluated state-of-the-art beamforming and power control, and proposed a joint approach combining both methods that strikes a balance between interference reduction towards the satellites while not compromising terrestrial \gls{qos}. Our work, in contrast to the plethora of research, focuses on dense satellite deployments, providing for the first time realistic insights for the design of optimization solutions for enhanced cellular-satellite coexistence. These insights are potentially valuable for the deployment of commercial \gls{nextg} networks that coexist with incumbents.

Overall, this work has focused on the design of novel algorithmic frameworks for advanced \gls{gnb} control, focusing on \gls{6g} and beyond-ready networks encompassing \gls{ai}, \gls{ris}, and coexistence mechanisms. This work tries to bridge the demands of both industry and academia for novel algorithmic designs and their successful deployment on realistic platforms for advanced wireless research, paving the way for commercial deployment by providing insights and lessons learned.

% \section{Section Title}
% \label{sec:conclusion:section1}

% This is where the section text goes.

% --- EOF ---

%% file: tex/acknowledgments.tex
% !TEX root = ../thesis_dissertation template.tex
% acknowledgments.tex:

\begin{acknowledgements}
The completion of my Ph.D. brings back so many memories of hope, perseverance, joy, effort, and determination to succeed. The past five years have tremendously shaped my character, making this journey one of the most intense and transformative experiences of my life so far. First and foremost, I would like to sincerely thank my academic advisor, Prof. Tommaso Melodia, for offering me the opportunity to pursue a Ph.D. at one of the world’s leading academic institutions, Northeastern University in Boston. I am forever grateful to him for investing in me and giving me the opportunity to conduct cutting-edge research in the United States of America. Knowing that he believed in me was my greatest motivation to persevere and, most importantly, to strive to produce high-quality research. Alongside Prof. Tommaso Melodia, I would also like to thank the other members of my Ph.D. dissertation committee, Prof. Josep-Miquel Jornet and Prof. Athanasios D. Panagopoulos, for all the knowledge and support they have offered me. I am beyond grateful to have worked with all of you.

I would also like to express my heartfelt thanks to the U.S. National Spectrum Consortium (NSC), and specifically to its Executive Director Joe Kochan, Chief Program Officer Mari Silbey, and Ericsson’s Chief Technology Officer and NSC board member Dr. Sheryl Genco, for recognizing me as one of the Women in Spectrum Scholars. I am deeply grateful for their support and belief in my abilities, as well as for offering me valuable opportunities, such as a work experience at Nokia Illinois.
% , where I gained an eye-opening, lifelong experience. 
Furthermore, I would like to thank Nokia for inviting me as one of the panelists at the Brooklyn 6G Summit in 2025 and for nominating me to participate in the wireless game show (students vs. professors) and, ultimately, for providing me with a chance to brag that we, the students, beat Prof. Tommaso Melodia and the other top professors in the game show. I promised my academic advisor I would never let him forget it, and here I am.

In addition, I would like to thank the Ph.D. students of the Institute for Intelligent Networked Systems for the countless hours of shared workspace, jokes, and laughter. I would also like to thank Prof. Diomedes and Faye Logothetis, Chris and Georgia Lagadinos, Christos Kazilas and Aimilia Tsolia, Fr. Sotiri and Presvytera Stephanie Malamis, Fr. Anthony Tandilyan and Despoina Lioliou, Anthony Ladas, Anastasia Asimacopoulos, Dawn (my first Airbnb host in the United States), SooMi Lee, Samar Elmaadawy, Neagin Neasamoni Santhi, Duschia Bodet, Sherif Badran, Ravis Shirkhani, Ali Saeizadeh, Albert Diez Comas, Stefano Maxenti, and Matteo Bordin for their friendship and for offering a helping hand whenever I needed it.

Last but not least, I would like to thank the most important people in my life so far, my family and closest friends. I want to thank Fr. Athanasios Polyzos for all his support and for endlessly listening to my worries over the phone, as well as my dearest friends, Athanasia Psorogianni and Maria Fragkou. Most importantly, I would like to thank my mother, Kostoula, for being my greatest strength and my driving force to keep moving forward throughout these years. I am beyond thankful to you, Mom, for all your sacrifices. Thank you for teaching me that the people who ultimately succeed are not those who get every response they hope for; instead, they are the ones who keep moving forward with their heads held high and with noble dignity. You inspire me to be as strong, yet as humble, as you are.
\end{acknowledgements}

%% file: bib/thesis.bib
@INPROCEEDINGS{10437367,
  author={Tsampazi, Maria and D'Oro, Salvatore and Polese, Michele and Bonati, Leonardo and Poitau, Gwenael and Healy, Michael and Melodia, Tommaso},
  booktitle={GLOBECOM 2023 - 2023 IEEE Global Communications Conference}, 
  title="{A Comparative Analysis of Deep Reinforcement Learning-Based xApps in O-RAN}", 
  year={2023},
  volume={},
  number={},
  pages={1638-1643},
  doi={10.1109/GLOBECOM54140.2023.10437367}}

@ARTICLE{11007770,
  author={Brach del Prever, Pietro and D’Oro, Salvatore and Bonati, Leonardo and Polese, Michele and Tsampazi, Maria and Lehmann, Heiko and Melodia, Tommaso},
  journal={IEEE Transactions on Mobile Computing}, 
  title="{PACIFISTA: Conflict Evaluation and Management in Open RAN}", 
  year={2025},
  volume={24},
  number={10},
  pages={10590-10605},
  doi={10.1109/TMC.2025.3570632}}

@ARTICLE{10973151,
  author={Tsampazi, Maria and Melodia, Tommaso},
  journal={IEEE Transactions on Vehicular Technology}, 
  title="{System-Level Experimental Evaluation of Reconfigurable Intelligent Surfaces for NextG Communication Systems}", 
  year={2025},
  volume={74},
  number={9},
  pages={14138-14153},
  doi={10.1109/TVT.2025.3563184}}

@INPROCEEDINGS{11152828,
  author={Tsampazi, Maria and Polese, Michele and Dressler, Falko and Melodia, Tommaso},
  booktitle={IEEE INFOCOM 2025 - IEEE Conference on Computer Communications Workshops (INFOCOM WKSHPS)}, 
  title="{O-RIS-ing: Evaluating RIS-Assisted NextG Open RAN}", 
  year={2025},
  volume={},
  number={},
  pages={1-6},
  doi={10.1109/INFOCOMWKSHPS65812.2025.11152828},
  ISSN={2833-0587},
  month={May},}

@ARTICLE{10766614,
  author={Tsampazi, Maria and D'Oro, Salvatore and Polese, Michele and Bonati, Leonardo and Poitau, Gwenael and Healy, Michael and Alavirad, Mohammad and Melodia, Tommaso},
  journal={IEEE Transactions on Mobile Computing},
  title={{PandORA}: Automated Design and Comprehensive Evaluation of Deep Reinforcement Learning Agents for Open {RAN}},
  year={2025},
  volume={24},
  number={4},
  pages={3223-3240},
  doi={10.1109/TMC.2024.3505781}
}

@article{tsampazi2026satellite,
  title="{Satellite-Terrestrial Spectrum Sharing in FR3 through QoS-Aware Power Control and Spatial Nulling}",
  author={Tsampazi, Maria and Testolina, Paolo and Polese, Michele and Melodia, Tommaso},
  journal={IEEE International Symposium on Dynamic Spectrum Access (DySPAN)},
  year={2026}
}

@ARTICLE{10643670,
  author={Polese, Michele and Bonati, Leonardo and D'Oro, Salvatore and Johari, Pedram and Villa, Davide and Velumani, Sakthivel and Gangula, Rajeev and Tsampazi, Maria and Paul Robinson, Clifton and Gemmi, Gabriele and Lacava, Andrea and Maxenti, Stefano and Cheng, Hai and Melodia, Tommaso},
  journal={IEEE Open Journal of the Communications Society}, 
  title="{Colosseum: The Open RAN Digital Twin}", 
  year={2024},
  volume={5},
  number={},
  pages={5452-5466},
  doi={10.1109/OJCOMS.2024.3447472}}

@article{wang2022self,
  title="{Self-play learning strategies for resource assignment in Open-RAN networks}",
  author={Wang, Xiaoyang and Thomas, Jonathan D and Piechocki, Robert J and Kapoor, Shipra and Santos-Rodr{\'\i}guez, Ra{\'u}l and Parekh, Arjun},
  journal={Computer Networks},
  volume={206},
  pages={108682},
  year={2022},
  publisher={Elsevier}
}

@inproceedings{kouchaki2022actor,
  title="{Actor-Critic Network for O-RAN Resource Allocation: xApp Design, Deployment, and Analysis}",
  author={Kouchaki, Mohammadreza and Marojevic, Vuk},
  booktitle={IEEE Globecom Workshops (GC Wkshps)},
  pages={968--973},
  year={2022},
  organization={}
}

@inproceedings{johnson2022nexran,
  title="{NexRAN: Closed-loop RAN slicing in POWDER-A top-to-bottom open-source open-RAN use case}",
  author={Johnson, David and Maas, Dustin and Van Der Merwe, Jacobus},
  booktitle={Proceedings of the 15th ACM Workshop on Wireless Network Testbeds, Experimental evaluation \& CHaracterization},
  pages={17--23},
  year={2022}
}

@article{filali2023communication,
  title="{Communication and computation O-RAN resource slicing for URLLC services using deep reinforcement learning}",
  author={Filali, Abderrahime and Nour, Boubakr and Cherkaoui, Soumaya and Kobbane, Abdellatif},
  journal={IEEE Communications Standards Magazine},
  volume={7},
  number={1},
  pages={66--73},
  year={2023},
  publisher={IEEE}
}

@inproceedings{zhang2022federated,
  title="{Federated deep reinforcement learning for resource allocation in O-RAN slicing}",
  author={{Zhang}, Han and Zhou, Hao and Erol-Kantarci, Melike},
  booktitle={IEEE Global Communications Conference (GLOBECOM)},
  pages={958--963},
  year={2022}
}

@article{iturria2022multi,
author = {Iturria Rivera, Pedro and Zhang, Han and Zhou, Hao and Mollahasani, Shahram and Erol Kantarci, Melike},
year = {2022},
month = {07},
pages = {5375},
title = {Multi-Agent Team Learning in Virtualized Open Radio Access Networks (O-RAN)},
volume = {22},
journal = {Sensors},
doi = {10.3390/s22145375}
}

@inproceedings{villa2022cast,
  title="{CaST: a toolchain for creating and characterizing realistic wireless network emulation scenarios}",
  author={Villa, Davide and Tehrani-Moayyed, Miead and Johari, Pedram and Basagni, Stefano and Melodia, Tommaso},
  booktitle={Proceedings of the 16th ACM Workshop on Wireless Network Testbeds, Experimental evaluation \& CHaracterization},
  pages={45--52},
  year={2022}
}

@article{diamanti2021prospect,
  title="{The prospect of reconfigurable intelligent surfaces in integrated access and backhaul networks}",
  author={Diamanti, Maria and Charatsaris, Panagiotis and Tsiropoulou, Eirini Eleni and Papavassiliou, Symeon},
  journal={IEEE Transactions on Green Communications and Networking},
  volume={6},
  number={2},
  pages={859--872},
  year={2021},
  publisher={IEEE}
}

@techreport{ETSI_RIS001,
  title = "{Reconfigurable Intelligent Surfaces (RIS); Use Cases, Deployment Scenarios and Requirements}",
  author = {ETSI},
  institution = {ETSI},
  type = {ETSI Group Report},
  number = {RIS 001 V1.1.1},
  year = {2023},
  month = {April},
  note = {[Available Online]: \url{https://www.etsi.org/deliver/etsi_gr/RIS/001_099/001/01.01.01_60/gr_RIS001v010101p.pdf}}
}

@article{liaskos2022software,
  title={Software-defined reconfigurable intelligent surfaces: From theory to end-to-end implementation},
  author={Liaskos, Christos and Mamatas, Lefteris and Pourdamghani, Arash and Tsioliaridou, Ageliki and Ioannidis, Sotiris and Pitsillides, Andreas and Schmid, Stefan and Akyildiz, Ian F},
  journal={Proceedings of the IEEE},
  volume={110},
  number={9},
  pages={1466--1493},
  year={2022},
  publisher={IEEE}
}

@inproceedings{gomez2016srslte,
  title="{{srsLTE}: An Open-Source Platform for {LTE} Evolution and Experimentation}",
  author={Gomez-Miguelez, Ismael and Garcia-Saavedra, Andres and Sutton, Paul D and Serrano, Pablo and Cano, Cristina and Leith, Doug J},
  booktitle={Proceedings of the Tenth ACM International Workshop on Wireless Network Testbeds, Experimental Evaluation, and Characterization},
  pages={25--32},
  year={2016}
}

@misc{mgen,
    title = {{U.S. Naval Research Laboratory, "Multi-Generator ({MGEN}) Network Test Tool"}},
    howpublished={[Available Online]: \url{https://www.nrl.navy.mil/itd/ncs/products/mgen}},
    note={2019}
}

@techreport{jaeckel2019f,
  title = "{F. Burkhardt, E. Eberlein,“QuaDRiGa-Quasi Deterministic Radio Channel Generator,” User Manual and Documentation}",
  author = {Jaeckel, S and Raschkowski, L and Boerner, K and Thiele, L},
  year = {2019},
  institution = {Tech. Rep. v2. 2.0, Fraunhofer Heinrich Hertz Institute},
  note = {[Available Online]: \url{https://github.com/fraunhoferhhi/QuaDRiGa/blob/main/quadriga_documentation_v2.8.1-0.pdf}}
}

@article{jaeckel2014quadriga,
  title="{QuaDRiGa: A 3-D multi-cell channel model with time evolution for enabling virtual field trials}",
  author={Jaeckel, Stephan and Raschkowski, Leszek and B{\"o}rner, Kai and Thiele, Lars},
  journal={IEEE transactions on antennas and propagation},
  volume={62},
  number={6},
  pages={3242--3256},
  year={2014}
}

@article{li2020reconfigurable,
  title="{Reconfigurable intelligent surface assisted UAV communication: Joint trajectory design and passive beamforming}",
  author={Li, Sixian and Duo, Bin and Yuan, Xiaojun and Liang, Ying-Chang and Di Renzo, Marco},
  journal={Wireless Communications Letters},
  volume={9},
  number={5},
  pages={716--720},
  year={2020},
  publisher={IEEE}
}

@inproceedings{diamanti2021energy,
  title="{Energy efficient multi-user communications aided by reconfigurable intelligent surfaces and UAVs}",
  author={Diamanti, Maria and Tsampazi, Maria and Tsiropoulou, Eirini Eleni and Papavassiliou, Symeon},
  booktitle={IEEE International Conference on Smart Computing (SMARTCOMP)},
  pages={371--376},
  year={2021}
}

@ARTICLE{8970580,  author={Zheng, Beixiong and Wu, Qingqing and Zhang, Rui},  journal={IEEE Communications Letters},   title="{Intelligent Reflecting Surface-Assisted Multiple Access With User Pairing: NOMA or OMA?}",   year={2020},  volume={24},  number={4},  pages={753-757},  doi={10.1109/LCOMM.2020.2969870}}

@inproceedings{angjo2023side,
    author = {Angjo, Joana and Zubow, Anatolij and Dressler, Falko},
    doi = {10.1109/GCWkshps58843.2023.10464696},
    title = {{Side Effects of IRS: On the Need for Coordination in 6G Multi-Operator IRS-assisted Networks}},
    pages = {1380--1385},
    publisher = {IEEE},
    booktitle = {IEEE Global Communications Conference (GLOBECOM), 4th Workshop on Emerging Topics in 6G Communications (6GComm)},
    month = {12},
    year = {2023},
   }

@misc{AIRANAlliance2024,
  author       = "{AI-RAN Alliance}",
  title        = "{AI-RAN Alliance -
Vision and Mission White Paper}",
  howpublished = {[Available Online]: \url{https://ai-ran.org/wp-content/uploads/2024/12/AI-RAN_Alliance_Whitepaper.pdf}},
  year         = "2024",
  month        = "December"
}

@techreport{3gpp38213,
     author={{3GPP TS 38.213}},                                                                                                   
     title={{3GPP TS 38.213: NR; Physical layer procedures for control}},
     institution={3rd Generation Partnership Project (3GPP)},                                                                                                  
     year={2023},
     type={Technical Specification}                                                                                                                            
   }

@techreport{3gpp38901,
   author={3GPP TR 38.901},
   title="{{Study on channel model for frequencies from 0.5 to 100 GHz}}",
   institution={3GPP},
   year={2023},
   type={Technical Report},
  
 }

@inproceedings{kang2024terrestrial,
  title="{Terrestrial-satellite spectrum sharing in the upper mid-band with interference nulling}",
  author={Kang, Seongjoon and Geraci, Giovanni and Mezzavilla, Marco and Rangan, Sundeep},
  booktitle={IEEE International Conference on Communications},
  pages={},
  year={2024}
}

@techreport{3GPP38821,
  author = {{3GPP TR 38.821}},
  title = "{Solutions for {NR} to support non-terrestrial networks ({NTN})}",
  institution = {3GPP},
  type = {Technical Report},
  year = {2023},
  month = mar,
  version = {V16.2.0},
  note = {Release 16}
}

@techreport{ITU-R-RS.2017-0,
  author = {{ITU-R RS.2017-0}},
  title = "{Performance and interference criteria for satellite passive remote sensing}",
  institution = {International Telecommunication Union, Radiocommunication Sector},
  type = {Recommendation},
  year = {2012},
}

@article{tsiropoulou2015combined,
  title="{Combined power and rate allocation in self-optimized multi-service two-tier femtocell networks}",
  author={Tsiropoulou, Eirini Eleni and Vamvakas, Panagiotis and Katsinis, Georgios K and Papavassiliou, Symeon},
  journal={Computer Communications},
  volume={72},
  year={2015}
}

@inproceedings{wadaskar2025satellite,
  title="{Satellite-Terrestrial Coexistence in FR3 Band via Hybrid True-Time-Delay Array-based Nulling}",
  author={Wadaskar, Aditya and Cabric, Danijela},
  booktitle={IEEE International Symposium on Dynamic Spectrum Access Networks},
  pages={},
  organization={},
  year={2025}
}

@article{jain1984quantitative,
  title="{A quantitative measure of fairness and discrimination}",
  author={Jain, Rajendra K and Chiu, Dah-Ming W and Hawe, William R and others},
  journal={Eastern Research Laboratory, Digital Equipment Corporation, Hudson, MA},
  volume={21},
  number={1},
  pages={},
  year={1984}
}

@misc{MATLAB:2025,
  author = {MathWorks},
  title = "{MATLAB (R2025a)}",
  year = {2025},
  howpublished = {[Available Online]: \url{https://www.mathworks.com}}
}

@inproceedings{jaeckel20225g,
  title="{A 5G-NR satellite extension for the QuaDRiGa channel model}",
  author={Jaeckel, Stephan and Raschkowski, Leszek and Thieley, Lars},
  booktitle={IEEE Joint European Conference on Networks and Communications \& 6G Summit},
  pages={},
  year={2022},
  organization={}
}

@article{li2002analytical,
  title="{An analytical model to predict the probability density function of elevation angles for LEO satellite systems}",
  author={Li, Sheng-Yi and Liu, CH},
  journal={IEEE Communications Letters},
  volume={6},
  number={4},
  pages={},
  year={2002},
  publisher={}
}

@book{balanis2016antenna,
  title="{Antenna theory: analysis and design}",
  author={Balanis, Constantine A},
  year={2016},
  publisher={John Wiley \& Sons}
}

@book{van2002optimum,
  title="{Optimum array processing: Part IV of detection, estimation, and modulation theory}",
  author={Van Trees, Harry L},
  year={2002},
  publisher={John Wiley \& Sons}
}

@article{bazzi2025upper,
  title="{Upper mid-band spectrum for 6G: Vision, opportunity and challenges}",
  author={Bazzi, Ahmad and Bomfin, Roberto and Mezzavilla, Marco and Rangan, Sundeep and Rappaport, Theodore and Chafii, Marwa},
  journal={arXiv preprint arXiv:2502.17914},
  year={2025}
}

@article{agarwal2023coexistence,
  title="{Coexistence assessment and interference mitigation for 5G and Fixed Satellite Stations in C-band in India}",
  author={Agarwal, Avinash},
  journal={arXiv preprint arXiv:2312.16079},
  year={2023}
}

@inproceedings{niloy2024ascent,
  title="{ASCENT: A Context-Aware Spectrum Coexistence Design and Implementation Toolset for Policymakers in Satellite Bands}",
  author={Niloy, Ta-Seen Reaz and Kumar, Saurav and Hore, Aniruddha and Hassan, Zoheb and Dietrich, Carl and Burger, Eric W and Reed, Jeffrey H and Shah, Vijay K},
  booktitle={International Symposium on Dynamic Spectrum Access Networks (DySPAN)},
  pages={240--248},
  year={2024},
  organization={IEEE}
}

@article{niloy2023interference,
  title="{Interference Analysis of Coexisting 5G Networks and NGSO FSS Receivers in the 12-GHz Band}",
  author={Niloy, Ta-Seen Reaz and Hassan, Zoheb and Stephenson, Nathan and Shah, Vijay K},
  journal={IEEE Wireless Communications Letters},
  volume={12},
  number={9},
  pages={1528--1532},
  year={2023},
  publisher={IEEE}
}

@article{wu2023space,
  title="{Space-Ground Multicast Group Control for Multiuser LEO Satellite Networks}",
  author={Wu, Dapeng and Qin, Chen and Cui, Yaping and He, Peng and Wang, Ruyan},
  journal={IEEE Transactions on Wireless Communications},
  year={2023},
  publisher={IEEE}
}

@article{ma2023resource,
  title="{Resource Scheduling for High-Capacity Multicast Service in Ultra-Dense LEO Satellite Networks}",
  author={Ma, Ting and Qian, Bo and Qin, Xiaohan and Zhang, Xin and Cai, Lin X and Zhou, Haibo},
  journal={IEEE Transactions on Vehicular Technology},
  year={2023},
  publisher={IEEE}
}

@article{heydarishahreza2024spectrum,
  title="{Spectrum sharing and interference management for 6G LEO satellite-terrestrial network integration}",
  author={Heydarishahreza, Navid and Han, Tao and Ansari, Nirwan},
  journal={Communications Surveys \& Tutorials},
  year={2024},
  publisher={IEEE}
}

@misc{nokia_ran_dt2026,
  author = {{Nokia}},
  title = "{Nokia launches {Nokia RAN Digital Twin} to turbo-charge {AI}-native {6G}, powered by {NVIDIA Aerial Omniverse Digital Twin}}",
  howpublished = {Nokia Corporate Blog},
  month = feb,
  year = {2026},
  note = {[Available Online]: \url{https://www.nokia.com/blog/nokia-launches-nokia-ran-digital-twin-to-turbo-charge-ai-native-6g-powered-by-nvidia-aerial-omniverse-digital-twin/}}
}

@misc{nvidia_nokia2025,
  author = {{NVIDIA Corporation}},
  title = "{NVIDIA and Nokia to Pioneer the {AI} Platform for {6G} --- Powering America's Return to Telecommunications Leadership}",
  howpublished = {NVIDIA Newsroom, Press Release},
  month = oct,
  year = {2025},
  note = {[Available Online]: \url{https://nvidianews.nvidia.com/news/nvidia-nokia-ai-telecommunications}}
}

@article{sionna2022,
  author = {Hoydis, Jakob and Cammerer, Sebastian and Ait Aoudia, Fay\c{c}al and Nimier-David, Matthieu and Keller, Alexander and Vem, Avinash and Stark, Matthias and O'Shea, Timothy},
  title = "{Sionna: An Open-Source Library for Next-Generation Physical Layer Research}",
  journal = {arXiv preprint arXiv:2203.11854},
  year = {2022}
}

@misc{3gpp38214,
  author = {{3rd Generation Partnership Project}},
  title = "{NR; Physical layer procedures for data (3GPP TS 38.214)}",
  year = {2023}
}

@article{kochan2017ppo,
  title="{PPO: Platforms for Advanced Wireless Research (PAWR) Project Office}",
  author={Kochan, Joseph},
  journal={NSF Award Number 1719547. Directorate for Computer and Information Science and Engineering},
  volume={17},
  number={1719547},
  pages={19547},
  year={2017}
}

@article{maggi2026sinr,
  title = "{SINR Estimation under Limited Feedback via Online Convex Optimization}",
  author = {Maggi, Lorenzo and Bonev, Boris and Wiesmayr, Reinhard and Cammerer, Sebastian and Keller, Alexander},
  journal = {arXiv preprint arXiv:2603.02061},
  year = {2026}
}

@article{wiesmayr2025salad,
  title = "{SALAD: Self-adaptive link adaptation}",
  author = {Wiesmayr, Reinhard and Maggi, Lorenzo and Cammerer, Sebastian and Hoydis, Jakob and Aoudia, Fay{\c{c}}al A{\"\i}t and Keller, Alexander},
  journal = {arXiv preprint arXiv:2510.05784},
  year = {2025}
}

@inproceedings{pedersen2007frequency,
  title = "{Frequency domain scheduling for OFDMA with limited and noisy channel feedback}",
  author = {Pedersen, Klaus I and Monghal, Guillaume and Kovacs, Istvan Z and Kolding, Troels E and Pokhariyal, Akhilesh and Frederiksen, Frank and Mogensen, Preben},
  booktitle = {IEEE 66th Vehicular Technology Conference},
  pages = {1792--1796},
  year = {2007}
}

@misc{schulman2017proximal,
  author = {Schulman, John and Wolski, Filip and Dhariwal, Prafulla and Radford, Alec and Klimov, Oleg},
  title = "{Proximal Policy Optimization Algorithms}",
  year = {2017},
  note = {arXiv:1707.06347}
}

@misc{stable-baselines3,
  author = {Raffin, Antonin and Hill, Ashley and Gleave, Adam and Kanervisto, Anssi and Ernestus, Maxim and Dormann, Noah},
  title = "{Stable-Baselines3: Reliable Reinforcement Learning Implementations}",
  howpublished = {GitHub Repository},
  year = {2021},
  note = {[Available Online]: \url{https://github.com/DLR-RM/stable-baselines3}}
}

@misc{gymnasium2023,
  author = {{Farama Foundation}},
  title = "{Gymnasium: A Standard Interface for Reinforcement Learning Environments}",
  howpublished = {GitHub Repository},
  year = {2023},
  note = {[Available Online]: \url{https://github.com/Farama-Foundation/Gymnasium}}
}

@article{quinlan1986induction,
  title = "{Induction of decision trees}",
  author = {Quinlan, J. Ross},
  journal = {Machine Learning},
  volume = {1},
  number = {1},
  pages = {81--106},
  year = {1986},
  publisher = {Springer}
}

@article{kalman1960,
  author = {Kalman, Rudolf E.},
  title = "{A New Approach to Linear Filtering and Prediction Problems}",
  journal = {Transactions of the ASME--Journal of Basic Engineering},
  volume = {82},
  pages = {35--45},
  year = {1960}
}

@article{breiman2001rf,
  author = {Breiman, Leo},
  title = "{Random Forests}",
  journal = {Machine Learning},
  volume = {45},
  year = {2001}
}

@article{ernst2005tree,
  title = "{Tree-based batch mode reinforcement learning}",
  author = {Ernst, Damien and Geurts, Pierre and Wehenkel, Louis},
  journal = {Journal of Machine Learning Research},
  volume = {6},
  year = {2005},
  publisher = {Microtome Publishing}
}

@article{villa2025x5g,
  title = "{X5G: An open, programmable, multi-vendor, end-to-end, private 5G O-RAN testbed with NVIDIA ARC and OpenAirInterface}",
  author = {Villa, Davide and Khan, Imran and Kaltenberger, Florian and Hedberg, Nicholas and da Silva, R{\'u}ben Soares and Maxenti, Stefano and Bonati, Leonardo and Kelkar, Anupa and Dick, Chris and Baena, Eduardo and others},
  journal = {IEEE Transactions on Mobile Computing},
  year = {2025},
  publisher = {IEEE}
}

@inproceedings{zubow2021grgym,
  title = "{GrGym: When GNU radio goes to (AI) gym}",
  author = {Zubow, Anatolij and R{\"o}sler, Sascha and Gaw{\l}owicz, Piotr and Dressler, Falko},
  booktitle = {Proc. of the 22nd International Workshop on Mobile Computing Systems and Applications},
  pages = {8--14},
  year = {2021}
}

@inproceedings{peri2025offline,
  title = "{Offline reinforcement learning and sequence modeling for downlink link adaptation}",
  author = {Peri, Samuele and Russo, Alessio and Fodor, Gabor and Soldati, Pablo},
  booktitle = {International Conference on Machine Learning for Communication and Networking (ICMLCN)},
  pages = {1--7},
  year = {2025},
  organization = {IEEE}
}

@article{pulliyakode2017reinforcement,
  title = "{Reinforcement learning techniques for outer loop link adaptation in 4G/5G systems}",
  author = {Pulliyakode, Saishankar Katri and Kalyani, Sheetal},
  journal = {arXiv preprint arXiv:1708.00994},
  year = {2017}
}

@inproceedings{leite2012flexible,
  title = "{A flexible framework based on reinforcement learning for adaptive modulation and coding in OFDM wireless systems}",
  author = {Leite, Joao P and de Carvalho, Paulo Henrique P and Vieira, Robson D},
  booktitle = {Wireless Communications and Networking Conference (WCNC)},
  pages = {809--814},
  year = {2012},
  organization = {IEEE}
}

@inproceedings{khedhri2025adaptive,
  title = "{Adaptive Modulation Selection in Wireless Communications: A Comparative Study of Reinforcement Learning, Deep Learning, Deep Reinforcement Learning, and Traditional Policies}",
  author = {Khedhri, Nader and Najar, Monia},
  booktitle = {International Wireless Communications and Mobile Computing (IWCMC)},
  pages = {1622--1625},
  year = {2025},
  organization = {IEEE}
}

@inproceedings{saxena2019contextual,
  title = "{Contextual multi-armed bandits for link adaptation in cellular networks}",
  author = {Saxena, Vidit and Jald{\'e}n, Joakim and Gonzalez, Joseph E and Bengtsson, Mats and Tullberg, Hugo and Stoica, Ion},
  booktitle = {Workshop on Network Meets AI \& ML},
  pages = {44--49},
  year = {2019}
}

@inproceedings{kela2022reinforcement,
  title = "{Reinforcement learning for delay sensitive uplink outer-loop link adaptation}",
  author = {Kela, Petteri and H{\"o}hne, Thomas and Veijalainen, Teemu and Abdulrahman, Hussein},
  booktitle = {Joint European Conference on Networks and Communications \& 6G Summit},
  pages = {59--64},
  year = {2022},
  organization = {IEEE}
}

@book{sutton2018reinforcement,
  title = "{Reinforcement learning: An introduction}",
  author = {Sutton, Richard S and Barto, Andrew G},
  year = {2018},
  publisher = {MIT press}
}

@inproceedings{wiebusch2023towards,
  title="{Towards open 6g: Experimental o-ran framework for predictive uplink slicing}",
  author={Wiebusch, Robin and Wagner, Niklas A and Overbeck, Dennis and Kurtz, Fabian and Wietfeld, Christian},
  booktitle={IEEE International Conference on Communications},
  pages={4834--4839},
  year={2023},
  organization={IEEE}
}

@article{yeh2023deep,
  title="{Deep Learning for Intelligent and Automated Network Slicing in 5G Open RAN (ORAN) Deployment}",
  author={Yeh, Shu-ping and Bhattacharya, Sonia and Sharma, Rashika and Moustafa, Hassnaa},
  journal={Open Journal of the Communications Society},
  year={2023},
  publisher={IEEE}
}

@ARTICLE{10329913,
  author={Zangooei, Mohammad and Golkarifard, Morteza and Rouili, Mohamed and Saha, Niloy and Boutaba, Raouf},
  journal={IEEE Journal on Selected Areas in Communications}, 
  title="{Flexible RAN Slicing in Open RAN With Constrained Multi-Agent Reinforcement Learning}", 
  year={2024},
  volume={42},
  number={2},
  pages={280-294},
  doi={10.1109/JSAC.2023.3336156}}

@article{sun2019dynamic,
  title="{Dynamic reservation and deep reinforcement learning based autonomous resource slicing for virtualized radio access networks}",
  author={Sun, Guolin and Gebrekidan, Zemuy Tesfay and Boateng, Gordon Owusu and Ayepah-Mensah, Daniel and Jiang, Wei},
  journal={Ieee Access},
  volume={7},
  pages={45758--45772},
  year={2019},
  publisher={IEEE}
}

@ARTICLE{9933014,
  author={Rezazadeh, Farhad and Zanzi, Lanfranco and Devoti, Francesco and Chergui, Hatim and Costa-Pérez, Xavier and Verikoukis, Christos},
  journal={IEEE Transactions on Vehicular Technology}, 
  title="{On the Specialization of FDRL Agents for Scalable and Distributed 6G RAN Slicing Orchestration}", 
  year={2023},
  volume={72},
  number={3},
  pages={3473-3487},
  doi={10.1109/TVT.2022.3218158}}

@inproceedings{bonati2021colosseum,
  title={{C}olosseum: {L}arge-scale {W}ireless {E}xperimentation through {H}ardware-in-the-{L}oop {N}etwork {E}mulation},
  author={Bonati, Leonardo and Johari, Pedram and Polese, Michele and D’Oro, Salvatore and Mohanti, Subhramoy and Tehrani-Moayyed, Miead and Villa, Davide and Shrivastava, Shweta and Tassie, Chinenye and Yoder, Kurt and others},
  booktitle={IEEE International Symposium on Dynamic Spectrum Access Networks (DySPAN)},
  pages={105--113},
  year={2021},
  organization={}
}

@inproceedings{bonati2021scope,
  title="{{SCOPE}: An Open and Softwarized Prototyping Platform for NextG Systems}",
  author={Bonati, Leonardo and D'Oro, Salvatore and Basagni, Stefano and Melodia, Tommaso},
  booktitle={Proceedings of the 19th Annual International Conference on Mobile Systems, Applications, and Services},
  pages={415--426},
  year={2021}
}

@article{bonati2023openran,
  title="{{OpenRAN Gym}: AI/ML development, data collection, and testing for O-RAN on PAWR platforms}",
  author={Bonati, Leonardo and Polese, Michele and D’Oro, Salvatore and Basagni, Stefano and Melodia, Tommaso},
  journal={Computer Networks},
  volume={220},
  pages={109502},
  year={2023},
  publisher={Elsevier}
}

@article{mnih2013playing,
  title="{Playing atari with deep reinforcement learning}",
  author={Mnih, Volodymyr and Kavukcuoglu, Koray and Silver, David and Graves, Alex and Antonoglou, Ioannis and Wierstra, Daan and Riedmiller, Martin},
  journal={arXiv preprint arXiv:1312.5602},
  year={2013}
}

@inproceedings{kakade2002approximately,
  title="{Approximately optimal approximate reinforcement learning}",
  author={Kakade, Sham and Langford, John},
  booktitle={Proceedings of the Nineteenth International Conference on Machine Learning},
  pages={267--274},
  year={2002}
}

@inproceedings{kozlica2023deep,
  title="{Deep q-learning versus proximal policy optimization: Performance comparison in a material sorting task}",
  author={Kozlica, Reuf and Wegenkittl, Stefan and Hir{\"a}nder, Simon},
  booktitle={IEEE 32nd International Symposium on Industrial Electronics (ISIE)},
  pages={1--6},
  year={2023},
  organization={}
}

@article{kim2022adaptive,
  title="{Adaptive Discount Factor for Deep Reinforcement Learning in Continuing Tasks with Uncertainty}",
  author={Kim, MyeongSeop and Kim, Jung-Su and Choi, Myoung-Su and Park, Jae-Han},
  journal={Sensors},
  volume={22},
  number={19},
  pages={7266},
  year={2022},
  publisher={MDPI}
}

@article{zhang2021minibatch,
  title="{Minibatch Recursive Least Squares Q-Learning}",
  author={Zhang, Chunyuan and Song, Qi and Meng, Zeng and others},
  journal={Computational Intelligence and Neuroscience},
  year={2021},
  publisher={Hindawi}
}

@article{lillicrap2015continuous,
  title="{Continuous control with deep reinforcement learning}",
  author={Lillicrap, Timothy P and Hunt, Jonathan J and Pritzel, Alexander and Heess, Nicolas and Erez, Tom and Tassa, Yuval and Silver, David and Wierstra, Daan},
  journal={arXiv preprint arXiv:1509.02971},
  year={2015}
}

@article{mnih2015human,
  title="{Human-level control through deep reinforcement learning}",
  author={Mnih, Volodymyr and Kavukcuoglu, Koray and Silver, David and Rusu, Andrei A and Veness, Joel and Bellemare, Marc G and Graves, Alex and Riedmiller, Martin and Fidjeland, Andreas K and Ostrovski, Georg and others},
  journal={Nature},
  volume={518},
  number={7540},
  pages={529--533},
  year={2015},
  publisher={Nature Publishing Group}
}

@inproceedings{hessel2018rainbow,
  title="{Rainbow: Combining improvements in deep reinforcement learning}",
  author={Hessel, Matteo and Modayil, Joseph and Van Hasselt, Hado and Schaul, Tom and Ostrovski, Georg and Dabney, Will and Horgan, Dan and Piot, Bilal and Azar, Mohammad and Silver, David},
  booktitle={Proceedings of the AAAI conference on artificial intelligence},
  volume={32},
  number={1},
  year={2018}
}

@article{fortunato2017noisy,
  title="{Noisy networks for exploration}",
  author={Fortunato, Meire and Azar, Mohammad Gheshlaghi and Piot, Bilal and Menick, Jacob and Osband, Ian and Graves, Alex and Mnih, Vlad and Munos, Remi and Hassabis, Demis and Pietquin, Olivier and others},
  journal={arXiv preprint arXiv:1706.10295},
  year={2017}
}

@inproceedings{schulman2015trust,
  title="{Trust region policy optimization}",
  author={Schulman, John and Levine, Sergey and Abbeel, Pieter and Jordan, Michael and Moritz, Philipp},
  booktitle={International conference on machine learning},
  pages={1889--1897},
  year={2015},
  organization={PMLR}
}

@inproceedings{fujimoto2019off,
  title="{Off-policy deep reinforcement learning without exploration}",
  author={Fujimoto, Scott and Meger, David and Precup, Doina},
  booktitle={International conference on machine learning},
  pages={2052--2062},
  year={2019},
  organization={PMLR}
}

@article{kak2023hexran,
  title="{{HexRAN}: A Programmable Multi-RAT Platform for Network Slicing in the Open RAN Ecosystem}",
  author={Kak, Ahan and Pham, Van-Quan and Thieu, Huu-Trung and Choi, Nakjung},
  journal={arXiv preprint arXiv:2304.12560},
  year={2023}
}

@conference{EURECOM+7416,
  author = {Chen, Chieh-Chun and  Chang, Chia-Yu and  Nikaein, Navid},
  title = {FlexSlice: Flexible and real-time programmable RAN slicing framework},
  booktitle = {IEEE Global Communications Conference},
  year = {2023},
  address = {Kuala Lumpur},
}

@misc{tensorflow2015-whitepaper,
title="{{TensorFlow}: Large-Scale Machine Learning on Heterogeneous Systems}",
note = {Software available from tensorflow.org. [Available Online]: \url{https://www.tensorflow.org/}},
author={
    Mart\'{i}n~Abadi and
    Ashish~Agarwal and
    Paul~Barham and
    Eugene~Brevdo and
    Zhifeng~Chen and
    Craig~Citro and
    Greg~S.~Corrado and
    Andy~Davis and
    Jeffrey~Dean and
    Matthieu~Devin and
    Sanjay~Ghemawat and
    Ian~Goodfellow and
    Andrew~Harp and
    Geoffrey~Irving and
    Michael~Isard and
    Yangqing Jia and
    Rafal~Jozefowicz and
    Lukasz~Kaiser and
    Manjunath~Kudlur and
    Josh~Levenberg and
    Dandelion~Man\'{e} and
    Rajat~Monga and
    Sherry~Moore and
    Derek~Murray and
    Chris~Olah and
    Mike~Schuster and
    Jonathon~Shlens and
    Benoit~Steiner and
    Ilya~Sutskever and
    Kunal~Talwar and
    Paul~Tucker and
    Vincent~Vanhoucke and
    Vijay~Vasudevan and
    Fernanda~Vi\'{e}gas and
    Oriol~Vinyals and
    Pete~Warden and
    Martin~Wattenberg and
    Martin~Wicke and
    Yuan~Yu and
    Xiaoqiang~Zheng},
  year={2015},
}

@inproceedings{zhang2022team,
  title="{Team learning-based resource allocation for open radio access network (O-RAN)}",
  author={Zhang, Han and Zhou, Hao and Erol-Kantarci, Melike},
  booktitle={IEEE International Conference on Communications},
  pages={4938--4943},
  year={2022},
  organization={IEEE}
}

@article{suh2022deep,
  title="{Deep reinforcement learning-based network slicing for beyond 5G}",
  author={Suh, Kyungjoo and Kim, Sunwoo and Ahn, Yongjun and Kim, Seungnyun and Ju, Hyungyu and Shim, Byonghyo},
  journal={IEEE Access},
  volume={10},
  pages={7384--7395},
  year={2022},
  publisher={IEEE}
}

@article{yan2023deep,
  title="{Deep Reinforcement Learning Based Resource Allocation for Network Slicing with Massive MIMO}",
  author={Yan, Dandan and Ng, Benjamin K and Ke, Wei and Lam, Chan-Tong},
  journal={IEEE Access},
  year={2023},
  publisher={IEEE}
}

@article{naderializadeh2021resource,
  title="{Resource management in wireless networks via multi-agent deep reinforcement learning}",
  author={Naderializadeh, Navid and Sydir, Jaroslaw J and Simsek, Meryem and Nikopour, Hosein},
  journal={IEEE Transactions on Wireless Communications},
  volume={20},
  number={6},
  pages={3507--3523},
  year={2021},
  publisher={IEEE}
}

@article{azimi2021energy,
  title="{Energy-efficient deep reinforcement learning assisted resource allocation for 5G-RAN slicing}",
  author={Azimi, Yaser and Yousefi, Saleh and Kalbkhani, Hashem and Kunz, Thomas},
  journal={IEEE Transactions on Vehicular Technology},
  volume={71},
  number={1},
  pages={856--871},
  year={2021},
  publisher={IEEE}
}

@inproceedings{liu2020deepslicing,
  title="{DeepSlicing: Deep reinforcement learning assisted resource allocation for network slicing}",
  author={Liu, Qiang and Han, Tao and Zhang, Ning and Wang, Ye},
  booktitle={GLOBECOM 2020-2020 IEEE Global Communications Conference},
  pages={1--6},
  year={2020},
  organization={IEEE}
}

@article{rahimi2022novel,
  title="{A novel approach to efficient resource allocation in load-balanced cellular networks using hierarchical DRL}",
  author={Rahimi, A Mirzaei and Ziaeddini, Amir and Gonglee, Shu},
  journal={Journal of Ambient Intelligence and Humanized Computing},
  volume={13},
  number={5},
  pages={2887--2901},
  year={2022},
  publisher={Springer}
}

@article{luong2019applications,
  title="{Applications of deep reinforcement learning in communications and networking: A survey}",
  author={Luong, Nguyen Cong and Hoang, Dinh Thai and Gong, Shimin and Niyato, Dusit and Wang, Ping and Liang, Ying-Chang and Kim, Dong In},
  journal={IEEE Communications Surveys \& Tutorials},
  volume={21},
  number={4},
  pages={3133--3174},
  year={2019},
  publisher={IEEE}
}

@article{alwarafy2021deep,
  title="{Deep reinforcement learning for radio resource allocation and management in next generation heterogeneous wireless networks: A survey}",
  author={Alwarafy, Abdulmalik and Abdallah, Mohamed and Ciftler, Bekir Sait and Al-Fuqaha, Ala and Hamdi, Mounir},
  journal={arXiv preprint arXiv:2106.00574},
  year={2021}
}

@misc{TFAgents,
  title = "{{TF-Agents}: A library for Reinforcement Learning in TensorFlow}",
  author = {Sergio Guadarrama and Anoop Korattikara and Oscar Ramirez and
     Pablo Castro and Ethan Holly and Sam Fishman and Ke Wang and
     Ekaterina Gonina and Neal Wu and Efi Kokiopoulou and Luciano Sbaiz and
     Jamie Smith and Gábor Bartók and Jesse Berent and Chris Harris and
     Vincent Vanhoucke and Eugene Brevdo},
  howpublished = {[Available Online]: \url{https://github.com/tensorflow/agents}},
  year = 2018
}

@article{wang2023resource,
  title="{Resource allocation based on Radio Intelligence Controller for Open RAN towards 6G}",
  author={Wang, Qingtian and Liu, Yang and Wang, Yanchao and Xiong, Xiong and Zong, Jiaying and Wang, Jianxiu and Chen, Peng},
  journal={IEEE Access},
  year={2023},
  publisher={IEEE}
}

@article{brik2022deep,
  title="{Deep learning for B5G open radio access network: Evolution, survey, case studies, and challenges}",
  author={Brik, Bouziane and Boutiba, Karim and Ksentini, Adlen},
  journal={IEEE Open Journal of the Communications Society},
  volume={3},
  pages={228--250},
  year={2022},
  publisher={IEEE}
}

@misc{kaloxylos2021ai,
  title="{AI and ML--Enablers for beyond 5G Networks}",
  author={Kaloxylos, Alexandros and Gavras, Anastasius and Camps, Daniel and Ghoraishi, Mir and Hrasnica, Halid},
  howpublished = {[Available Online]: \url{https://5g-ppp.eu/wp-content/uploads/2021/05/AI-MLforNetworks-v1-0.pdf}},
  year={2021}
}

@article{yang2019application,
  title="{Application of reinforcement learning in UAV cluster task scheduling}",
  author={Yang, Jun and You, Xinghui and Wu, Gaoxiang and Hassan, Mohammad Mehedi and Almogren, Ahmad and Guna, Joze},
  journal={Future generation computer systems},
  volume={95},
  pages={140--148},
  year={2019},
  publisher={Elsevier}
}

@article{alliance2021ran,
  title="{O-RAN Working Group 2:“O-RAN AI/ML workflow description and requirements 1.03}",
  author={Alliance, ORAN},
  journal={O-RAN. WG2. AIML-v01. 03 Technical Specification},
  year={2021}
}

@INPROCEEDINGS{9068634,
  author={Xiao, Yikai and Zhang, Qixia and Liu, Fangming and Wang, Jia and Zhao, Miao and Zhang, Zhongxing and Zhang, Jiaxing},
  booktitle={IEEE/ACM International Symposium on Quality of Service}, 
  title="{NFVdeep: Adaptive Online Service Function Chain Deployment with Deep Reinforcement Learning}", 
  year={2019},
  volume={},
  number={},
  pages={1-10},
  doi={10.1145/3326285.3329056}}

@inproceedings{villa2023twinning,
  title="{Twinning Commercial Radio Waveforms in the Colosseum Wireless Network Emulator}",
  author={Villa, Davide and Uvaydov, Daniel and Bonati, Leonardo and Johari, Pedram and Jornet, Josep Miquel and Melodia, Tommaso},
  booktitle={Proceedings of the 17th ACM Workshop on Wireless Network Testbeds, Experimental evaluation \& Characterization},
  pages={33--40},
  year={2023}
}

@article{bjornson2020reconfigurable,
  title="{Reconfigurable intelligent surfaces: Three myths and two critical questions}",
  author={Bj{\"o}rnson, Emil and {\"O}zdogan, {\"O}zgecan and Larsson, Erik G},
  journal={IEEE Communications Magazine},
  volume={58},
  number={12},
  pages={90--96},
  year={2020},
  publisher={IEEE}
}

@article{siddiqi2022reconfigurable,
  title="{Reconfigurable intelligent surface-aided wireless communications: An overview}",
  author={Siddiqi, Muhammad Zain and Mir, Talha},
  journal={Intelligent and Converged Networks},
  volume={3},
  number={1},
  pages={33--63},
  year={2022},
  publisher={TUP}
}

@article{pang2022investigation,
  title="{Investigation and Comparison of 5G Channel Models: From QuaDRiGa, NYUSIM, and MG5G Perspectives}",
  author={Pang, Lihua and Zhang, Jin and Zhang, Yang and Huang, Xinyi and Chen, Yijian and LI, Jiandong},
  journal={Chinese Journal of Electronics},
  volume={31},
  number={1},
  pages={1--17},
  year={2022},
  publisher={Wiley Online Library}
}

@article{kyosti20074,
  title="{IST-4-027756 WINNER II D1. 1.2 V1. 2: WINNER II channel models}",
  author={Kyosti, P and Meinila, J and Hentila, L and Zhao, Xiongwen and Jamsa, T and Schneider, Christian and Narandzic, M and Milojevic, M and Hong, A and Ylitalo, J and others},
  journal={Available: www. ist-winner. org},
  year={2007}
}

@article{mondal20153d,
  title="{3D channel model in 3GPP}",
  author={Mondal, Bishwarup and Thomas, Timothy A and Visotsky, Eugene and Vook, Frederick W and Ghosh, Amitava and Nam, Young-Han and Li, Yang and Zhang, Jianzhong and Zhang, Min and Luo, Qinglin and others},
  journal={IEEE Communications Magazine},
  volume={53},
  number={3},
  pages={16--23},
  year={2015},
  publisher={IEEE}
}

@inproceedings{he2020investigation,
  title="{Investigation and Comparison of QuaDRiGa, NYUSIM and MG5G Channel Models for 5G Wireless Communications}",
  author={He, Yaping and Zhang, Yang and Zhang, Jin and Pang, Lihua and Chen, Yijian and Ren, Guangliang},
  booktitle={IEEE 92nd Vehicular Technology Conference (VTC2020-Fall)},
  pages={1--5},
  year={2020}
}

@inproceedings{giordani2019path,
  title="{Path loss models for V2V mmWave communication: performance evaluation and open challenges}",
  author={Giordani, Marco and Shimizu, Takayuki and Zanella, Andrea and Higuchi, Takamasa and Altintas, Onur and Zorzi, Michele},
  booktitle={IEEE Connected and Automated Vehicles Symposium (CAVS)},
  pages={1--5},
  year={2019},
  organization={IEEE}
}

@electronic{etsi5g,
    title="{ETSI TR 138 901 V14.3.0 (2018-01)}",
    howpublished={[Available Online]: \url{https://www.etsi.org/deliver/etsi_tr/138900_138999/138901/14.03.00_60/tr_138901v140300p.pdf}}
}

@article{arunachalaperumal2018enhanced,
  title="{Enhanced 3D MIMO channel for urban macro environment}",
  author={Arunachalaperumal, C and Dhilipkumar, S and Abija, G},
  journal={International Journal of Pure and Applied Mathematics, Special Issue},
  volume={118},
  number={10},
  pages={259--269},
  year={2018}
}

@article{ademaj20163gpp,
  title="{3GPP 3D MIMO channel model: A holistic implementation guideline for open source simulation tools}",
  author={Ademaj, Fjolla and Taranetz, Martin and Rupp, Markus},
  journal={EURASIP Journal on Wireless Communications and Networking},
  number={1},
  pages={1--14},
  year={2016},
  publisher={SpringerOpen}
}

@inproceedings{ju2019millimeter,
  title="{A millimeter-wave channel simulator NYUSIM with spatial consistency and human blockage}",
  author={Ju, Shihao and Kanhere, Ojas and Xing, Yunchou and Rappaport, Theodore S},
  booktitle={IEEE global communications conference (GLOBECOM)},
  pages={1--6},
  year={2019}
}

@electronic{etsi19deliv,
    title="{Intelligent Transport Systems (ITS); Access Layer; Part 1: Channel Models for the 5,9 GHz frequency band}",
    note = {Sophia Antipolis Cedex, FRANCE, ETSI TR 103 257-1 V1.1.1, Tech. Rep., May, 2019. [Available Online]: \url{https://www.etsi.org/deliver/etsi_tr/103200_103299/10325701/01.01.01_60/tr_10325701v010101p.pdf}}
}

@article{ranjkesh2015optimized,
  title="{The Optimized QuaDRiGa Wi-Fi Channel Model}",
  author={Ranjkesh, Mohammad E and Khazaei, Ali Akbar},
  journal={International Journal of Science and Engineering Investigations},
  volume={4},
  number={9},
  pages={51--61},
  year={2015}
}

@electronic{quadrigacode,
    title="{QuaDRiGa Source Code GitHub repository}",
    howpublished={[Available Online]: \url{https://github.com/fraunhoferhhi/QuaDRiGa}}
}

@article{mecklenbrauker2011vehicular,
  title="{Vehicular channel characterization and its implications for wireless system design and performance}",
  author={Mecklenbrauker, Christoph F and Molisch, Andreas F and Karedal, Johan and Tufvesson, Fredrik and Paier, Alexander and Bernad{\'o}, Laura and Zemen, Thomas and Klemp, Oliver and Czink, Nicolai},
  journal={Proceedings of the IEEE},
  volume={99},
  number={7},
  pages={1189--1212},
  year={2011},
  publisher={IEEE}
}

@manual{MATLAB:2021a,
  address = {Natick, Massachusetts},
  organization = {The Mathworks, Inc.},
  title = "{MATLAB R2021a}",
  year = {2021}
}

@misc{googlemaps,
  title = "{Google Maps}",
  howpublished = {[Available Online]: \url{https://www.google.com/maps}}
}

@article{bjornson2022reconfigurable,
  title="{Reconfigurable intelligent surfaces: A signal processing perspective with wireless applications}",
  author={Bj{\"o}rnson, Emil and Wymeersch, Henk and Matthiesen, Bho and Popovski, Petar and Sanguinetti, Luca and de Carvalho, Elisabeth},
  journal={IEEE Signal Processing Magazine},
  volume={39},
  number={2},
  pages={135--158},
  year={2022}
}

@misc{etsi2022,
  title = "{5G; NR; User Equipment (UE) radio transmission and reception; Part 1: Range 1 Standalone (3GPP TS 38.101-1 version 17.5.0 Release 17)}",
  author = {{ETSI TS 138 101-1}},
  howpublished = {[Available Online]: \url{https://www.etsi.org/deliver/etsi_ts/138100_138199/13810101/17.05.00_60/ts_13810101v170500p.pdf}},
  note = {ETSI TS 138 101-1 V17.5.0 (2022-05)},
  year = {2022}
}

@misc{etsi2017,
  title = "{LTE; Evolved Universal Terrestrial Radio Access (E-UTRA); Physical layer procedures (3GPP TS 36.213 version 14.2.0 Release 14)}",
  author = {{ETSI TS 136 213}},
  howpublished = {[Available Online]: \url{https://www.etsi.org/deliver/etsi_ts/136200_136299/136213/14.02.00_60/ts_136213v140200p.pdf}},
  note = {ETSI TS 136 213 V14.2.0 (2017-04)},
  year = {2017}
}

@techreport{etsi2022b,
  title = "{5G; NR; User Equipment (UE) radio transmission and reception; Part 2: Range 2 Standalone (3GPP TS 38.101-2 version 17.5.0 Release 17)}",
  author = {{ETSI TS 138 101-2}},
  institution = {ETSI},
  number = {ETSI TS 138 101-2 V17.5.0},
  year = {2022}
}

@article{dang2021geometry,
  title="{A geometry-based stochastic channel model and its application for intelligent reflecting surface assisted wireless communication}",
  author={Dang, Jian and Gao, Shicheng and Zhu, Yongdong and Guo, Rongbin and Jiang, Hao and Zhang, Zaichen and Wu, Liang and Zhu, Bingcheng and Wang, Lei},
  journal={IET Communications},
  volume={15},
  number={3},
  pages={421--434},
  year={2021},
  publisher={Wiley Online Library}
}

@inproceedings{rossanese2022designing,
  title="{Designing, building, and characterizing RF switch-based reconfigurable intelligent surfaces}",
  author={Rossanese, Marco and Mursia, Placido and Garcia-Saavedra, Andres and Sciancalepore, Vincenzo and Asadi, Arash and Costa-Perez, Xavier},
  booktitle={Proceedings of the 16th ACM Workshop on Wireless Network Testbeds, Experimental evaluation \& CHaracterization},
  pages={69--76},
  year={2022}
}

@inproceedings{tehrani2021creating,
  title="{Creating {RF} Scenarios for Large-Scale, Real-Time Wireless Channel Emulators}",
  author={Tehrani-Moayyed, Miead and Bonati, Leonardo and Johari, Pedram and Melodia, Tommaso and Basagni, Stefano},
  booktitle={Proceedings of IEEE MedComNet},
  year={2021}
}

@electronic{srsran,
    title = {{srsRAN website}},
    howpublished = {[Available Online]: \url{https://www.srslte.com/}},
    note = {Accessed September 2022}
}

@article{da2023varactor,
  title="{A varactor-based 1024-element RIS design for mm-waves}",
  author={da Silva, Luis G and Chu, Zheng and Xiao, Pei and Cerqueira S Jr, Arismar},
  journal={Frontiers in Communications and Networks},
  volume={4},
  pages={1086011},
  year={2023}
}

@article{bloessl2017performance,
  title="{Performance assessment of IEEE 802.11p with an open source SDR-based prototype}",
  author={Bloessl, Bastian and Segata, Michele and Sommer, Christoph and Dressler, Falko},
  journal={IEEE transactions on mobile computing},
  volume={17},
  number={5},
  pages={1162--1175},
  year={2017}
}

@article{tapio2021survey,
  title="{Survey on reconfigurable intelligent surfaces below 10 GHz}",
  author={Tapio, Visa and Hemadeh, Ibrahim and Mourad, Alain and Shojaeifard, Arman and Juntti, Markku},
  journal={EURASIP Journal on Wireless Communications and Networking},
  pages={1--18},
  year={2021},
  publisher={Springer}
}

@article{khan2022vehicle,
  title="{Vehicle to everything (V2X) and edge computing: A secure lifecycle for UAV-assisted vehicle network and offloading with blockchain}",
  author={Khan, Abdullah Ayub and Laghari, Asif Ali and Shafiq, Muhammad and Awan, Shafique Ahmed and Gu, Zhaoquan},
  journal={Drones},
  volume={6},
  number={12},
  pages={377},
  year={2022},
  publisher={MDPI}
}

@article{attar20225g,
  title="{5G System Overview for Ongoing Smart Applications: Structure, Requirements, and Specifications}",
  author={Attar, Hani and Issa, Haitham and Ababneh, Jafar and Abbasi, Mahdi and Solyman, Ahmed AA and Khosravi, Mohammad and Said Agieb, Ramy and others},
  journal={Computational Intelligence and Neuroscience},
  volume={2022},
  year={2022},
  publisher={Hindawi}
}

@article{shehab20215g,
  title="{5G networks towards smart and sustainable cities: A review of recent developments, applications and future perspectives}",
  author={Shehab, Muhammad J and Kassem, Ihab and Kutty, Adeeb A and Kucukvar, Murat and Onat, Nuri and Khattab, Tamer},
  journal={IEEE Access},
  volume={10},
  pages={2987--3006},
  year={2021}
}

@article{sobhi2020energy,
  title="{Energy-efficient power allocation and user selection for mmWave-NOMA transmission in M2M communications underlaying cellular heterogeneous networks}",
  author={Sobhi-Givi, Sima and Shayesteh, Mahrokh G and Kalbkhani, Hashem},
  journal={IEEE Transactions on Vehicular Technology},
  volume={69},
  number={9},
  pages={9866--9881},
  year={2020}
}

@article{zhang2017network,
  title="{Network slicing based 5G and future mobile networks: Mobility, resource management, and challenges}",
  author={Zhang, Haijun and Liu, Na and Chu, Xiaoli and Long, Keping and Aghvami, Abdol-Hamid and Leung, Victor CM},
  journal={IEEE communications magazine},
  volume={55},
  number={8},
  pages={138--145},
  year={2017}
}

@article{wang2018survey,
  title="{A survey of 5G channel measurements and models}",
  author={Wang, Cheng-Xiang and Bian, Ji and Sun, Jian and Zhang, Wensheng and Zhang, Minggao},
  journal={IEEE Communications Surveys \& Tutorials},
  volume={20},
  number={4},
  pages={3142--3168},
  year={2018}
}

@inproceedings{lopes20235g,
  title="{How 5G Will Transform Smart Cities: A Literature Review}",
  author={Lopes, Isabel and Guarda, Teresa and Fernandes, AJG and Ribeiro, Maria Isabel},
  booktitle={International Conference on Computational Science and Its Applications},
  pages={70--81},
  year={2023},
  organization={Springer}
}

@article{liu2021promoting,
  title="{Promoting smart cities into the 5G era with multi-field Internet of Things (IoT) applications powered with advanced mechanical energy harvesters}",
  author={Liu, Long and Guo, Xinge and Lee, Chengkuo},
  journal={Nano Energy},
  volume={88},
  pages={106304},
  year={2021},
  publisher={Elsevier}
}

@article{sharif2019compact,
  title="{Compact base station antenna based on image theory for UWB/5G RTLS embraced smart parking of driverless cars}",
  author={Sharif, Abubakar and Guo, Jinhao and Ouyang, Jun and Sun, Sheng and Arshad, Kamran and Imran, Muhammad Ali and Abbasi, Qammer H},
  journal={IEEE Access},
  volume={7},
  pages={180898--180909},
  year={2019}
}

@article{marabissi2018real,
  title="{A real case of implementation of the future 5G city}",
  author={Marabissi, Dania and Mucchi, Lorenzo and Fantacci, Romano and Spada, Maria Rita and Massimiani, Fabio and Fratini, Andrea and Cau, Giorgio and Yunpeng, Jia and Fedele, Lucio},
  journal={Future Internet},
  volume={11},
  number={1},
  pages={4},
  year={2018},
  publisher={MDPI}
}

@article{jiang2019smart,
  title="{Smart urban living: Enabling emotion-guided interaction with next generation sensing fabric}",
  author={Jiang, Yingying and Xiao, Wenjing and Wang, Rui and Barnawi, Ahmed},
  journal={IEEE Access},
  volume={8},
  pages={28395--28402},
  year={2019}
}

@article{zhao2021nanogenerators,
  title="{Nanogenerators for smart cities in the era of 5G and Internet of Things}",
  author={Zhao, Xun and Askari, Hassan and Chen, Jun},
  journal={Joule},
  volume={5},
  number={6},
  pages={1391--1431},
  year={2021},
  publisher={Elsevier}
}

@article{sanchez2021review,
  title="{Review of methods to reduce energy consumption in a smart city based on IoT and 5G technology}",
  author={S{\'a}nchez-Cano, Julieta Evangelina and Garc{\'\i}a-Quilachamin, Washington Xavier and P{\'e}rez-V{\'e}liz, Jonny and Herrera-Tapia, Jorge and Fuentes, Kelvin Atiencia},
  journal={iJOE},
  volume={17},
  number={08},
  pages={5},
  year={2021}
}

@article{rana2023review,
  title="{Review paper on hardware of reconfigurable intelligent surfaces}",
  author={Rana, Biswarup and Cho, Sung-Sil and Hong, Ic-Pyo},
  journal={IEEE Access},
  year={2023}
}

@inproceedings{zivuku2022maximizing,
  title="{Maximizing the Number of Served Users in a Smart City using Reconfigurable Intelligent Surfaces}",
  author={Zivuku, Progress and Kisseleff, Steven and Nguyen, Van-Dinh and Ntontin, Konstantinos and Martins, Wallace A and Chatzinotas, Symeon and Ottersten, Bj{\"o}rn},
  booktitle={IEEE Wireless Communications and Networking Conference (WCNC)},
  pages={494--499},
  year={2022}
}

@article{al2023emerging,
  title="{Emerging 6G/B6G wireless communication for the power infrastructure in smart cities: Innovations, challenges, and future perspectives}",
  author={Al Amin, Ahmed and Hong, Junho and Bui, Van-Hai and Su, Wencong},
  journal={Algorithms},
  volume={16},
  number={10},
  pages={474},
  year={2023},
  publisher={MDPI}
}

@article{kisseleff2020reconfigurable,
  title="{Reconfigurable intelligent surfaces for smart cities: Research challenges and opportunities}",
  author={Kisseleff, Steven and Martins, Wallace A and Al-Hraishawi, Hayder and Chatzinotas, Symeon and Ottersten, Bj{\"o}rn},
  journal={IEEE Open Journal of the Communications Society},
  volume={1},
  pages={1781--1797},
  year={2020}
}

@article{liu2023integrated,
  title="{Integrated sensing and communication with reconfigurable intelligent surfaces: Opportunities, applications, and future directions}",
  author={Liu, Rang and Li, Ming and Luo, Honghao and Liu, Qian and Swindlehurst, A Lee},
  journal={IEEE Wireless Communications},
  volume={30},
  number={1},
  pages={50--57},
  year={2023}
}

@article{salah2022paving,
  title="{Paving the way for economically accepted and technically pronounced smart radio environment}",
  author={Salah, Mostafa and Pitsillides, Andreas and Mubarak, Ahmed S},
  journal={China Communications},
  volume={19},
  number={8},
  pages={247--266},
  year={2022},
  publisher={IEEE}
}

@article{makarfi2020reconfigurable,
  title="{Reconfigurable intelligent surfaces-enabled vehicular networks: A physical layer security perspective}",
  author={Makarfi, Abubakar U and Rabie, Khaled M and Kaiwartya, Omprakash and Adhikari, Kabita and Li, Xingwang and Quiroz-Castellanos, Marcela and Kharel, Rupak},
  journal={arXiv preprint arXiv:2004.11288},
  year={2020}
}

@article{kamruzzaman2022key,
  title="{Key technologies, applications and trends of internet of things for energy-efficient 6G wireless communication in smart cities}",
  author={Kamruzzaman, MM},
  journal={Energies},
  volume={15},
  number={15},
  pages={5608},
  year={2022},
  publisher={MDPI}
}

@article{bariah2023digital,
  title="{Digital twin-empowered smart cities: A new frontier of wireless networks}",
  author={Bariah, Lina and Sari, Hikmet and Debbah, Merouane},
  journal={Authorea Preprints},
  year={2023},
  publisher={Authorea}
}

@inproceedings{dagiuklas2023journey,
  title="{The journey from 5G towards 6G}",
  author={Dagiuklas, Tasos},
  booktitle={8th International Symposium on Electrical and Electronics Engineering (ISEEE)},
  pages={14--18},
  year={2023},
  organization={IEEE}
}

@article{li2023liquid,
  title="{From Liquid Crystal on Silicon and Liquid Crystal Reflectarray to Reconfigurable Intelligent Surfaces for Post-5G Networks}",
  author={Li, Jinfeng},
  journal={Applied Sciences},
  volume={13},
  number={13},
  pages={7407},
  year={2023},
  publisher={MDPI}
}

@incollection{mishra20236g,
  title="{6G-IoT Framework for Sustainable Smart City: Vision and Challenges}",
  author={Mishra, Priyanka and Singh, Ghanshyam},
  booktitle={Sustainable Smart Cities: Enabling Technologies, Energy Trends and Potential Applications},
  pages={97--117},
  year={2023},
  publisher={Springer}
}

@article{masouros2023guest,
  title="{Guest Editorial: Integrated Sensing and Communications for 6G}",
  author={Masouros, Christos and Zhang, J Andrew and Liu, Fan and Zheng, Le and Wymeersch, Henk and Di Renzo, Marco},
  journal={IEEE Wireless Communications},
  volume={30},
  number={1},
  pages={14--15},
  year={2023}
}

@article{renzo2019smart,
  title="{Smart radio environments empowered by reconfigurable AI meta-surfaces: An idea whose time has come}",
  author={Renzo, Marco Di and Debbah, Merouane and Phan-Huy, Dinh-Thuy and Zappone, Alessio and Alouini, Mohamed-Slim and Yuen, Chau and Sciancalepore, Vincenzo and Alexandropoulos, George C and Hoydis, Jakob and Gacanin, Haris and others},
  journal={EURASIP Journal on Wireless Communications and Networking},
  volume={2019},
  number={1},
  pages={1--20},
  year={2019},
  publisher={Springer}
}

@article{hou2020reconfigurable,
  title="{Reconfigurable intelligent surface aided NOMA networks}",
  author={Hou, Tianwei and Liu, Yuanwei and Song, Zhengyu and Sun, Xin and Chen, Yue and Hanzo, Lajos},
  journal={IEEE Journal on Selected Areas in Communications},
  volume={38},
  number={11},
  pages={2575--2588},
  year={2020}
}

@ARTICLE{9079457,
  author="{Zhou, Shaoqing and Xu, Wei and Wang, Kezhi and Di Renzo, Marco and Alouini, Mohamed-Slim}",
  journal={IEEE Wireless Communications Letters}, 
  title={Spectral and Energy Efficiency of IRS-Assisted MISO Communication With Hardware Impairments}, 
  year={2020},
  volume={9},
  number={9},
  pages={1366-1369},
  doi={10.1109/LWC.2020.2990431}}

@article{guan2022irs,
  title="{IRS-Enabled Spectrum Sharing: Interference Modeling, Channel Estimation and Robust Passive Beamforming}",
  author={GUAN, Xinrong and WU, Qingqing},
  journal={ZTE Communications},
  volume={20},
  number={1},
  pages={28--35},
  year={2022}
}

@article{ji2022reconfigurable,
  title="{Reconfigurable intelligent surface aided cellular networks with device-to-device users}",
  author={Ji, Zelin and Qin, Zhijin and Parini, Clive G},
  journal={IEEE Transactions on Communications},
  volume={70},
  number={3},
  pages={1808--1819},
  year={2022}
}

@inproceedings{banchs2021network,
  title="{Network intelligence in 6G: Challenges and opportunities}",
  author={Banchs, Albert and Fiore, Marco and Garcia-Saavedra, Andres and Gramaglia, Marco},
  booktitle={Proceedings of the 16th ACM Workshop on Mobility in the Evolving Internet Architecture},
  pages={7--12},
  year={2021}
}

@article{basharat2022exploring,
  title="{Exploring reconfigurable intelligent surfaces for 6G: State-of-the-art and the road ahead}",
  author={Basharat, Sarah and Khan, Maryam and Iqbal, Muhammad and Hashmi, Umair Sajid and Zaidi, Syed Ali Raza and Robertson, Ian},
  journal={IET Communications},
  volume={16},
  number={13},
  pages={1458--1474},
  year={2022},
  publisher={Wiley Online Library}
}

@article{liu2021reconfigurable,
  title="{Reconfigurable intelligent surfaces: Principles and opportunities}",
  author={Liu, Yuanwei and Liu, Xiao and Mu, Xidong and Hou, Tianwei and Xu, Jiaqi and Di Renzo, Marco and Al-Dhahir, Naofal},
  journal={IEEE communications surveys \& tutorials},
  volume={23},
  number={3},
  pages={1546--1577},
  year={2021}
}

@article{mu2021capacity,
  title="{Capacity and optimal resource allocation for IRS-assisted multi-user communication systems}",
  author={Mu, Xidong and Liu, Yuanwei and Guo, Li and Lin, Jiaru and Al-Dhahir, Naofal},
  journal={IEEE Transactions on Communications},
  volume={69},
  number={6},
  pages={3771--3786},
  year={2021}
}

@article{elzanaty2021reconfigurable,
  title="{Reconfigurable intelligent surfaces for localization: Position and orientation error bounds}",
  author={Elzanaty, Ahmed and Guerra, Anna and Guidi, Francesco and Alouini, Mohamed-Slim},
  journal={IEEE Transactions on Signal Processing},
  volume={69},
  pages={5386--5402},
  year={2021}
}

@article{yang2020intelligent,
  title="{Intelligent reflecting surface assisted anti-jamming communications: A fast reinforcement learning approach}",
  author={Yang, Helin and Xiong, Zehui and Zhao, Jun and Niyato, Dusit and Wu, Qingqing and Poor, H Vincent and Tornatore, Massimo},
  journal={IEEE transactions on wireless communications},
  volume={20},
  number={3},
  pages={1963--1974},
  year={2020},
  publisher={IEEE}
}

@article{salih2023enhancing,
  title="{Enhancing UAV communication links with Reconfigurable intelligent surfaces}",
  author={Salih, Nameer Mufeed and Aldababsa, Mahmoud and Yahya, Khalid},
  journal={AEU-International Journal of Electronics and Communications},
  volume={171},
  pages={154933},
  year={2023},
  publisher={Elsevier}
}

@article{basharat2021reconfigurable,
  title="{Reconfigurable intelligent surfaces: Potentials, applications, and challenges for 6G wireless networks}",
  author={Basharat, Sarah and Hassan, Syed Ali and Pervaiz, Haris and Mahmood, Aamir and Ding, Zhiguo and Gidlund, Mikael},
  journal={Wireless Communications},
  volume={28},
  number={6},
  pages={184--191},
  year={2021},
  publisher={IEEE}
}

@article{ning2023intelligent,
  title="{Intelligent-Reflecting-Surface-Assisted UAV Communications for 6G Networks}",
  author={Ning, Zhaolong and Li, Tengfeng and Wu, Yu and Wang, Xiaojie and Wu, Qingqing and Yu, Fei Richard and Guo, Song},
  journal={arXiv preprint arXiv:2310.20242},
  year={2023}
}

@article{yu2022fair,
  title="{Fair downlink communications for RIS-UAV enabled mobile vehicles}",
  author={Yu, Yingfeng and Liu, Xin and Leung, Victor CM},
  journal={Wireless Communications Letters},
  volume={11},
  number={5},
  pages={1042--1046},
  year={2022},
  publisher={IEEE}
}

@article{basar2021reconfigurable,
  title="{Reconfigurable intelligent surfaces for future wireless networks: A channel modeling perspective}",
  author={Basar, Ertugrul and Yildirim, Ibrahim},
  journal={IEEE Wireless Communications},
  volume={28},
  number={3},
  pages={108--114},
  year={2021}
}

@inproceedings{basar2020simris,
  title="{SimRIS channel simulator for reconfigurable intelligent surface-empowered communication systems}",
  author="{E. Basar, and I. Yildirim}",
  booktitle={Latin-American Conference on Communications (LATINCOM)},
  pages={1--6},
  year={2020},
  organization={IEEE}
}

@article{sihlbom2022reconfigurable,
  title="{Reconfigurable intelligent surfaces: Performance assessment through a system-level simulator}",
  author={Sihlbom, Bjorn and Poulakis, Marios I and Di Renzo, Marco},
  journal={Wireless Communications},
  year={2022},
  publisher={IEEE}
}

@article{sun20213d,
  title="{A 3D non-stationary channel model for 6G wireless systems employing intelligent reflecting surfaces with practical phase shifts}",
  author={Sun, Yingzhuo and Wang, Cheng-Xiang and Huang, Jie and Wang, Jun},
  journal={IEEE Transactions on Cognitive Communications and Networking},
  volume={7},
  number={2},
  pages={496--510},
  year={2021}}

@article{dajer2022reconfigurable,
  title="{Reconfigurable intelligent surface: Design the channel--A new opportunity for future wireless networks}",
  author={Dajer, Miguel and Ma, Zhengxiang and Piazzi, Leonard and Prasad, Narayan and Qi, Xiao-Feng and Sheen, Baoling and Yang, Jin and Yue, Guosen},
  journal={Digital Communications and Networks},
  volume={8},
  number={2},
  pages={87--104},
  year={2022},
  publisher={Elsevier}
}

@article{zhou2023survey,
  title={A survey on model-based, heuristic, and machine learning optimization approaches in RIS-aided wireless networks},
  author={Zhou, Hao and Erol-Kantarci, Melike and Liu, Yuanwei and Poor, H Vincent},
  journal={IEEE Communications Surveys \& Tutorials},
  year={2023},
  publisher={IEEE}
}

@inproceedings{kong2021channel,
  title="{Channel modeling and analysis of reconfigurable intelligent surfaces assisted vehicular networks}",
  author={Kong, Long and He, Jiguang and Ai, Yun and Chatzinotas, Symeon and Ottersten, Bj{\"o}rn},
  booktitle={International Conference on Communications Workshops (ICC Workshops)},
  pages={1--6},
  year={2021},
  organization={IEEE}
}

@inproceedings{tian2023reconfigurable,
  title="{Reconfigurable intelligent surface-aided spectrum sharing coexisting with multiple primary networks}",
  author={Tian, Zhong and Chen, Zhengchuan and Wang, Min and Jia, Yunjian and Wen, Wanli},
  booktitle={IEEE Wireless Communications and Networking Conference (WCNC)},
  pages={1--6},
  year={2023}
}

@article{chen2021qos,
  title="{QoS-driven spectrum sharing for reconfigurable intelligent surfaces (RISs) aided vehicular networks}",
  author={Chen, Yuanbin and Wang, Ying and Zhang, Jiayi and Di Renzo, Marco},
  journal={IEEE Transactions on Wireless Communications},
  volume={20},
  number={9},
  pages={5969--5985},
  year={2021}
}

@article{boulogeorgos2021coverage,
  title="{Coverage analysis of reconfigurable intelligent surface assisted THz wireless systems}",
  author={Boulogeorgos, Alexandros-Apostolos A and Alexiou, Angeliki},
  journal={IEEE Open Journal of Vehicular Technology},
  volume={2},
  pages={94--110},
  year={2021}
}

@article{yu2021smart,
  title={Smart and reconfigurable wireless communications: From IRS modeling to algorithm design},
  author={Yu, Xianghao and Jamali, Vahid and Xu, Dongfang and Ng, Derrick Wing Kwan and Schober, Robert},
  journal={Wireless Communications},
  volume={28},
  number={6},
  pages={118--125},
  year={2021},
  publisher={IEEE}
}

@article{xu2023exploiting,
  title="{Exploiting STAR-RISs in near-field communications}",
  author={Xu, Jiaqi and Mu, Xidong and Liu, Yuanwei},
  journal={IEEE Transactions on Wireless Communications},
  year={2023}
}

@article{tang2021channel,
  title="{On channel reciprocity in reconfigurable intelligent surface assisted wireless networks}",
  author={Tang, Wankai and Chen, Xiangyu and Chen, Ming Zheng and Dai, Jun Yan and Han, Yu and Jin, Shi and Cheng, Qiang and Li, Geoffrey Ye and Cui, Tie Jun},
  journal={Wireless Communications},
  volume={28},
  number={6},
  pages={94--101},
  year={2021},
  publisher={IEEE}
}

@article{liu2022reconfigurable,
  title="{Reconfigurable Intelligent Surface Physical Model in Channel Modeling}",
  author={Liu, Yiping and Dou, Jianwu and Cui, Yijun and Chen, Yijian and Yang, Jun and Qin, Fan and Wang, Yuxin},
  journal={Electronics},
  volume={11},
  number={17},
  pages={2798},
  year={2022},
  publisher={MDPI}
}

@article{liu2022simulation,
  title={Simulation and field trial results of reconfigurable intelligent surfaces in 5G networks},
  author={Liu, Ruiqi and Dou, Jianwu and Li, Ping and Wu, Jianjun and Cui, Yijun},
  journal={IEEE Access},
  volume={10},
  pages={122786--122795},
  year={2022}
}

@ARTICLE{8741198,
  author={Huang, Chongwen and Zappone, Alessio and Alexandropoulos, George C. and Debbah, Mérouane and Yuen, Chau},
  journal={IEEE Transactions on Wireless Communications}, 
  title="{Reconfigurable Intelligent Surfaces for Energy Efficiency in Wireless Communication}", 
  year={2019},
  volume={18},
  number={8},
  pages={4157-4170},
  doi={10.1109/TWC.2019.2922609}}

@ARTICLE{9950621,
  author={Yu, Xiangbin and Wang, Guangying and Huang, Xu and Wang, Kezhi and Xu, Weiye and Rui, Yun},
  journal={IEEE Transactions on Mobile Computing}, 
  title="{Energy Efficient Resource Allocation for Uplink RIS-Aided Millimeter-Wave Networks With NOMA}", 
  year={2024},
  volume={23},
  number={1},
  pages={423-436},
  doi={10.1109/TMC.2022.3222392}}

@INPROCEEDINGS{9810453,
  author={Deshpande, Anay Ajit and Vaca-Rubio, Cristian J. and Mohebi, Salman and Salami, Dariush and de Carvalho, Elisabeth and Popovski, Petar and Sigg, Stephan and Zorzi, Michele and Zanella, Andrea},
  booktitle={2022 20th Mediterranean Communication and Computer Networking Conference (MedComNet)}, 
  title={Energy-Efficient Design for RIS-assisted UAV communications in beyond-5G Networks}, 
  year={2022},
  volume={},
  number={},
  pages={158-165},
  doi={10.1109/MedComNet55087.2022.9810453}}

@INPROCEEDINGS{9628234,
  author={Ahsan, Muhammad and Jamil, Sonain and Ejaz, Muhammad Tauseef and Abbas, Muhammad Sohail},
  booktitle={2021 International Conference on Computing, Electronic and Electrical Engineering (ICE Cube)}, 
  title={Energy Efficiency Maximization in RIS-assisted Wireless Networks}, 
  year={2021},
  volume={},
  number={},
  pages={1-6},
  doi={10.1109/ICECube53880.2021.9628234}}

@INPROCEEDINGS{10096617,
  author={Fotock, Robert K. and Zappone, Alessio and Renzo, Marco Di},
  booktitle={ICASSP 2023 - 2023 IEEE International Conference on Acoustics, Speech and Signal Processing (ICASSP)}, 
  title="{Energy Efficiency Maximization in RIS-aided Networks with Global Reflection Constraints}", 
  year={2023},
  volume={},
  number={},
  pages={1-5},
  doi={10.1109/ICASSP49357.2023.10096617}}

@ARTICLE{10107766,
  author={Zhang, Yichi and Lu, Yang and Zhang, Ruichen and Ai, Bo and Niyato, Dusit},
  journal={IEEE Transactions on Vehicular Technology}, 
  title={Deep Reinforcement Learning for Secrecy Energy Efficiency Maximization in RIS-Assisted Networks}, 
  year={2023},
  volume={72},
  number={9},
  pages={12413-12418},
  doi={10.1109/TVT.2023.3269805}}

@ARTICLE{9964251,
  author={Guo, Yi and Fang, Fang and Cai, Donghong and Ding, Zhiguo},
  journal={IEEE Transactions on Vehicular Technology}, 
  title="{Energy-Efficient Design for a NOMA Assisted STAR-RIS Network With Deep Reinforcement Learning}", 
  year={2023},
  volume={72},
  number={4},
  pages={5424-5428},
  doi={10.1109/TVT.2022.3224926}}

@inproceedings{rusca2023mobile,
  title="{Mobile RF Scenario Design for Massive-Scale Wireless Channel Emulators}",
  author={Rusca, Riccardo and Raviglione, Francesco and Casetti, Claudio and Giaccone, Paolo and Restuccia, Francesco},
  booktitle={Joint European Conference on Networks and Communications \& 6G Summit (EuCNC/6G Summit)},
  pages={675--680},
  year={2023},
  organization={IEEE}
}

@article{lasaulce2009introducing,
  title="{Introducing hierarchy in energy games}",
  author={Lasaulce, Samson and Hayel, Yezekael and El Azouzi, Rachid and Debbah, M{\'e}rouane},
  journal={IEEE Transactions on Wireless Communications},
  volume={8},
  number={7},
  pages={3833--3843},
  year={2009},
  publisher={IEEE}
}

@inproceedings{he2011stackelberg,
  title="{Stackelberg games for energy-efficient power control in wireless networks}",
  author={He, Gaoning and Lasaulce, Samson and Hayel, Yezekael},
  booktitle={2011 Proceedings IEEE INFOCOM},
  pages={591--595},
  year={2011},
  organization={IEEE}
}

@article{paul2022omni,
  title="{An Omni-Directional Wideband Patch Antenna with Parasitic Elements for Sub-6 GHz Band Applications}",
  author={Paul, Liton Chandra and Saha, Himel Kumar and Rani, Tithi and Mahmud, Md Zulfiker and Roy, Tushar Kanti and Lee, Wang-Sang and others},
  journal={International Journal of Antennas and Propagation},
  year={2022},
  publisher={Hindawi}
}

@article{azim2021multi,
  title="{A multi-slotted antenna for LTE/5G Sub-6 GHz wireless communication applications}",
  author={Azim, Rezaul and Meaze, AKM Moinul H and Affandi, Adnan and Alam, Md Mottahir and Aktar, Rumi and Mia, Md S and Alam, Touhidul and Samsuzzaman, Md and Islam, Mohammad T},
  journal={International Journal of Microwave and Wireless Technologies},
  volume={13},
  number={5},
  pages={486--496},
  year={2021},
  publisher={Cambridge University Press}
}

@inproceedings{rani2022development,
  title="{Development of a Broadband Antenna for 5G Sub-6 GHz Cellular and IIoT Smart Automation Applications}",
  author={Rani, Tithi and Das, Sajeeb Chandra and Hossen, Md Shakhawat and Paul, Liton Chandra and Roy, Tushar Kanti},
  booktitle={12th International Conference on Electrical and Computer Engineering (ICECE)},
  pages={465--468},
  year={2022},
  organization={IEEE}
}

@article{sharawi2017two,
  title="{A two concentric slot loop based connected array MIMO antenna system for 4G/5G terminals}",
  author={Sharawi, Mohammad S and Ikram, Muhammad and Shamim, Atif},
  journal={IEEE Transactions on antennas and propagation},
  volume={65},
  number={12},
  pages={6679--6686},
  year={2017},
  publisher={IEEE}
}

@article{de2021radio,
  title="{Radio channel modeling in a ship hull: Path loss at 868 MHz and 2.4, 5.25, and 60 GHz}",
  author={De Beelde, Brecht and Tanghe, Emmeric and Yusuf, Marwan and Plets, David and Joseph, Wout},
  journal={IEEE Antennas and Wireless Propagation Letters},
  volume={20},
  number={4},
  pages={597--601},
  year={2021},
  publisher={IEEE}
}

@article{zaidi2020wide,
  title="{A Wide and Tri-band Flexible Antennas with Independently Controllable Notch Bands for Sub-6-GHz Communication System.}",
  author={Zaidi, Abir and Awan, Wahaj Abbas and Hussain, Niamat and Baghdad, Abdennaceur},
  journal={Radioengineering},
  volume={29},
  number={1},
  year={2020}
}

@article{mao2018planar,
  title="{Planar sub-millimeter-wave array antenna with enhanced gain and reduced sidelobes for 5G broadcast applications}",
  author={Mao, Chun-Xu and Khalily, Mohsen and Xiao, Pei and Brown, Tim WC and Gao, Steven},
  journal={IEEE Transactions on Antennas and Propagation},
  volume={67},
  number={1},
  pages={160--168},
  year={2018},
  publisher={IEEE}
}

@inproceedings{abirami2017review,
  title="{A review of patch antenna design for 5G}",
  author={Abirami, M},
  booktitle={IEEE International Conference on Electrical, Instrumentation and Communication Engineering (ICEICE)},
  pages={1--3},
  year={2017},
  organization={IEEE}
}

@article{ranvier2008low,
  title="{Low-cost planar omnidirectional antenna for mm-wave applications}",
  author={Ranvier, Sylvain and Dudorov, Sergey and Kyro, Mikko and Luxey, Cyril and Icheln, Clemens and Staraj, Robert and Vainikainen, Pertti},
  journal={IEEE Antennas and Wireless Propagation Letters},
  volume={7},
  pages={521--523},
  year={2008},
  publisher={IEEE}
}

@inproceedings{sun2015synthesizing,
  title="{Synthesizing omnidirectional antenna patterns, received power and path loss from directional antennas for 5G millimeter-wave communications}",
  author={Sun, Shu and MacCartney, George R and Samimi, Mathew K and Rappaport, Theodore S},
  booktitle={IEEE Global Communications Conference (GLOBECOM)},
  pages={1--7},
  year={2015},
  organization={IEEE}
}

@article{fan2018wideband,
  title="{Wideband horizontally polarized omnidirectional antenna with a conical beam for millimeter-wave applications}",
  author={Fan, Kuikui and Hao, Zhang-Cheng and Yuan, Quan and Hu, Jun and Luo, Guo Qing and Hong, Wei},
  journal={IEEE Transactions on Antennas and Propagation},
  volume={66},
  number={9},
  pages={4437--4448},
  year={2018},
  publisher={IEEE}
}

@article{maccartney2015millimeter,
  title="{Millimeter-wave omnidirectional path loss data for small cell 5G channel modeling}",
  author={Maccartney, George R and Rappaport, Theodore S and Samimi, Mathew K and Sun, Shu},
  journal={IEEE access},
  volume={3},
  pages={1573--1580},
  year={2015},
  publisher={IEEE}
}

@article{hasan2019dual,
  title="{Dual band omnidirectional millimeter wave antenna for 5G communications}",
  author={Hasan, Md Nazmul and Bashir, Shahid and Chu, Son},
  journal={Journal of Electromagnetic Waves and Applications},
  volume={33},
  number={12},
  pages={1581--1590},
  year={2019},
  publisher={Taylor \& Francis}
}

@article{niu2015survey,
  title="{A survey of millimeter wave communications (mmWave) for 5G: opportunities and challenges}",
  author={Niu, Yong and Li, Yong and Jin, Depeng and Su, Li and Vasilakos, Athanasios V},
  journal={Wireless networks},
  volume={21},
  pages={2657--2676},
  year={2015},
  publisher={Springer}
}

@inproceedings{pan2011dual,
  title="{Dual-polarized Mm-wave phased array antenna for multi-Gb/s 60GHz communication}",
  author={Pan, Helen K},
  booktitle={IEEE International Symposium on Antennas and Propagation (APSURSI)},
  pages={3279--3282},
  year={2011},
  organization={IEEE}
}

@article{mao2020compact,
  title="{Compact patch antenna with vertical polarization and omnidirectional radiation characteristics}",
  author={Mao, Chun Xu and Khalily, Mohsen and Zhang, Long and Xiao, Pei and Sun, Yuhang and Werner, Douglas H},
  journal={IEEE Transactions on Antennas and Propagation},
  volume={69},
  number={2},
  pages={1158--1161},
  year={2020},
  publisher={IEEE}
}

@article{HU2023342,
title = "{NeXT: Architecture, prototyping and measurement of a software-defined testing framework for integrated RF network simulation, experimentation and optimization}",
journal = {Computer Communications},
volume = {210},
pages = {342-355},
year = {2023},
issn = {0140-3664},
doi = {https://doi.org/10.1016/j.comcom.2023.08.018},
note = {[Available Online]: \url{https://www.sciencedirect.com/science/article/pii/S014036642300302X}},
author = {Jiangqi Hu and Zhiyuan Zhao and Maxwell McManus and Sabarish Krishna Moorthy and Yuqing Cui and Nicholas Mastronarde and Elizabeth Serena Bentley and Michael Medley and Zhangyu Guan},
}

@article{MCMANUS2023110000,
title = "{Digital twin-enabled domain adaptation for zero-touch UAV networks: Survey and challenges}",
journal = {Computer Networks},
volume = {236},
pages = {110000},
year = {2023},
issn = {1389-1286},
doi = {https://doi.org/10.1016/j.comnet.2023.110000},
note = {[Available Online]: \url{https://www.sciencedirect.com/science/article/pii/S1389128623004450}},
author = {Maxwell McManus and Yuqing Cui and Josh (Zhaoxi) Zhang and Jiangqi Hu and Sabarish Krishna Moorthy and Nicholas Mastronarde and Elizabeth Serena Bentley and Michael Medley and Zhangyu Guan},
}

@ARTICLE{abdalla2022toward,
  author={Abdalla, Aly S. and Upadhyaya, Pratheek S. and Shah, Vijay K. and Marojevic, Vuk},
  journal={IEEE Network}, 
  title="{Toward Next Generation Open Radio Access Networks: What O-RAN Can and Cannot Do!}", 
  year={2022},
  volume={36},
  number={6},
  pages={206-213},
  doi={10.1109/MNET.108.2100659},
  ISSN={1558-156X},
  month={November},}

@ARTICLE{alkhateeb2023real,
  author={Alkhateeb, Ahmed and Jiang, Shuaifeng and Charan, Gouranga},
  journal={IEEE Communications Magazine}, 
  title="{Real-Time Digital Twins: Vision and Research Directions for 6G and Beyond}", 
  year={2023},
  volume={61},
  number={11},
  pages={128-134},
  month={November},}

@inbook{bahl2023accelerating,
author = {Bahl, Paramvir and Balkwill, Matthew and Foukas, Xenofon and Kalia, Anuj and Kim, Daehyeok and Kotaru, Manikanta and Lai, Zhihua and Mehrotra, Sanjeev and Radunovic, Bozidar and Saroiu, Stefan and Settle, Connor and Verma, Ankit and Wolman, Alec and Yan, Francis Y. and Zhang, Yongguang},
title = "{Accelerating Open RAN Research Through an Enterprise-scale 5G Testbed}",
year = {2023},
publisher = {Association for Computing Machinery},
booktitle = {Proceedings of the 29th Annual International Conference on Mobile Computing and Networking},
articleno = {138},
numpages = {3}
}

@ARTICLE{balasubramanian2021ric,
  author={Balasubramanian, Bharath and Daniels, E. Scott and Hiltunen, Matti and Jana, Rittwik and Joshi, Kaustubh and Sivaraj, Rajarajan and Tran, Tuyen X. and Wang, Chengwei},
  journal={IEEE Internet Computing}, 
  title="{RIC: A RAN Intelligent Controller Platform for AI-Enabled Cellular Networks}", 
  year={2021},
  volume={25},
  number={2},
  pages={7-17},
  doi={10.1109/MIC.2021.3062487},
  ISSN={1941-0131},
  month={March},}

@INPROCEEDINGS{baranda2021demo,
  author={Baranda, J. and Mangues-Bafalluy, J. and Zeydan, E. and Casetti, C. and Chiasserini, C. F. and Malinverno, M. and Puligheddu, C. and Groshev, M. and Guimarães, C. and Tomakh, K. and Kucherenko, D. and Kolodiazhnyi, O.},
  booktitle={IEEE Conference on Computer Communications Workshops (INFOCOM WKSHPS)}, 
  title="{Demo: AIML-as-a-Service for SLA management of a Digital Twin Virtual Network Service}", 
  year={2021},
  volume={},
  number={},
  pages={1-2},
  month={May},}

@inproceedings{breen2020powder,
title="{POWDER: Platform for Open Wireless Data-driven Experimental Research}",
author={Breen, Joe and Buffmire, Andrew and Duerig, Jonathon and Dutt,
Kevin and Eide, Eric and Hibler, Mike and Johnson, David and Kumar
Kasera, Sneha and Lewis, Earl and Maas, Dustin and Orange, Alex and
Patwari, Neal and Reading, Daniel and Ricci, Robert and Schurig, David
and B. Stoller, Leigh and Van der Merwe, Kobus and Webb, Kirk and
Wong, Gary},
booktitle={Proceedings of ACM WiNTECH},
address = {London, United Kingdom},
month={Sept.},
year={2020}}

@ARTICLE{challita2020when,
  author={Challita, Ursula and Ryden, Henrik and Tullberg, Hugo},
  journal={IEEE Communications Magazine}, 
  title={When Machine Learning Meets Wireless Cellular Networks: Deployment, Challenges, and Applications}, 
  year={2020},
  volume={58},
  number={6},
  pages={12-18},
  doi={10.1109/MCOM.001.1900664},
  ISSN={1558-1896},
  month={June},}

@INPROCEEDINGS{chaudhari2018scalable,
  author={Chaudhari, Ashish and Braun, Martin},
  booktitle={2018 13th International Symposium on Reconfigurable Communication-centric Systems-on-Chip (ReCoSoC)}, 
  title="{A Scalable FPGA Architecture for Flexible, Large-Scale, Real-Time RF Channel Emulation}", 
  year={2018},
  volume={},
  number={},
  pages={1-8},
  ISSN={},
  month={July},}

@misc{colosseumImages,
author="{Colosseum Website}",
title="{Colosseum Software}",
note = {Accessed on March 12, 2024. [Available Online]: \url{https://www.northeastern.edu/colosseum/colosseum-software/}}}

@incollection{corici2023digital,
  title="{Digital Twin for 5G Networks}",
  author={Corici, Marius and Magedanz, Thomas},
  booktitle={The Digital Twin},
  pages={433--446},
  year={2023},
  publisher={Springer}
}

@misc{dellOroRAN,
author={Stefan Pongratz},
title="{Open RAN Market Opportunity and Risks}",
year={2021},
note = {[Available Online]: \url{https://www.delloro.com/wp-content/uploads/2021/11/DellOro-Group-Article-Open-RAN-Market-Opportunity-and-Risks.pdf}},
howpublished = {Dell'Oro Group report},}

@INPROCEEDINGS{deng2021digital,
  author={Deng, Juan and Zheng, Qingbi and Liu, Guangyi and Bai, Jielin and Tian, Kaicong and Sun, Changhao and Yan, Yujie and Liu, Yitong},
  booktitle={IEEE Wireless Communications and Networking Conference Workshops (WCNCW)}, 
  title="{A Digital Twin Approach for Self-optimization of Mobile Networks}", 
  year={2021},
  volume={},
  number={},
  pages={1-6},
  month={March},}

@inproceedings{jagannath2022digital,
author = {Jagannath, Jithin and Ramezanpour, Keyvan and Jagannath, Anu},
title = "{Digital Twin Virtualization with Machine Learning for IoT and Beyond 5G Networks: Research Directions for Security and Optimal Control}",
year = {2022},
isbn = {9781450392778},
publisher = {Association for Computing Machinery},
address = {New York, NY, USA},
note = {[Available Online]: \url{https://doi.org/10.1145/3522783.3529519}},
booktitle = {Proceedings of the 2022 ACM Workshop on Wireless Security and Machine Learning},
pages = {81–86},
numpages = {6},
location = {San Antonio, TX, USA},
series = {WiseML '22}
}

@article{kaltenberger2020openairinterface,
    author = {Kaltenberger, Florian and P. Silva, Aloizio and Gosain, Abhimanyu and Wang, Luhan and Nguyen, Tien-Thinh},
    title = {{OpenAirInterface}: Democratizing Innovation in the {5G} Era},
    journal = {Computer Networks},
    number = {107284},
    month = {May},
    year = {2020}
}

@inproceedings{koumaras20185genesis,
  author={Koumaras, Harilaos and Tsolkas, Dimitris and Gardikis, Georgios and Gomez, Pedro Merino and Frascolla, Valerio and Triantafyllopoulou, Dionysia and Emmelmann, Marc and Koumaras, Vaios and Osma, Maria L. Garcia and Munaretto, Daniele and Atxutegi, Eneko and Puga, Jara Suárez de and Alay, Ozgu and Brunstrom, Anna and Bosneag, Anne Marie Cristina},
  title={{5GENESIS: The Genesis of a Flexible 5G Facility}},
  booktitle={Proceedings of IEEE CAMAD},
  address={Barcelona, Spain},
  month={September},
  year={2018}
}

@ARTICLE{li2022rlops,
  author={Li, Peizheng and Thomas, Jonathan and Wang, Xiaoyang and Khalil, Ahmed and Ahmad, Abdelrahim and Inacio, Rui and Kapoor, Shipra and Parekh, Arjun and Doufexi, Angela and Shojaeifard, Arman and Piechocki, Robert J.},
  journal={IEEE Access}, 
  title="{RLOps: Development Life-Cycle of Reinforcement Learning Aided Open RAN}", 
  year={2022},
  volume={10},
  number={},
  pages={113808-113826},
  month={},}

@ARTICLE{lin20236g,
  author={Lin, Xingqin and Kundu, Lopamudra and Dick, Chris and Obiodu, Emeka and Mostak, Todd and Flaxman, Mike},
  journal={IEEE Communications Magazine}, 
  title="{6G Digital Twin Networks: From Theory to Practice}", 
  year={2023},
  volume={61},
  number={11},
  pages={72-78},
  month={November},}

@ARTICLE{marojevic2020advanced,
  author={Marojevic, Vuk and Guvenc, Ismail and Dutta, Rudra and Sichitiu, Mihail L. and Floyd, Brian A.},
  journal={IEEE Vehicular Technology Magazine}, 
  title="{Advanced Wireless for Unmanned Aerial Systems: 5G Standardization, Research Challenges, and AERPAW Architecture}", 
  year={2020},
  volume={15},
  number={2},
  pages={22-30},
  doi={10.1109/MVT.2020.2979494},
  ISSN={1556-6080},
  month={June},}

@ARTICLE{masaracchia2023digital,
  author={Masaracchia, Antonino and Sharma, Vishal and Fahim, Muhammad and Dobre, Octavia A. and Duong, Trung Q.},
  journal={IEEE Communications Magazine}, 
  title="{Digital Twin for Open RAN: Toward Intelligent and Resilient 6G Radio Access Networks}", 
  year={2023},
  volume={61},
  number={11},
  pages={112-118},
  month={November},}

@article{mckeown2008openflow,
    author={McKeown, Nick and Anderson, Tom and Balakrishnan, Hari and Parulkar, Guru and Peterson, Larry and Rexford, Jennifer and Shenker, Scott and Turner, Jonathan},
    title={{{OpenFlow}: Enabling Innovation in Campus Networks}},
    year={2008},
    publisher={Association for Computing Machinery},
    volume={38},
    number={2},
    journal={ACM SIGCOMM Computer Communication Review},
    month={March},
    pages={69--74}
}

@article{mirzaei2023network,
  title="{Network Digital Twin for Open RAN: The Key Enablers, Standardization, and Use Cases}",
  author={Mirzaei, Javad and Abualhaol, Ibrahim and Poitau, Gwenael},
  journal={arXiv preprint arXiv:2308.02644},
  year={2023}
}

@INPROCEEDINGS{moorthy2022middleware,
  author={Moorthy, Sabarish Krishna and Harindranath, Ankush and McManus, Maxwell and Guan, Zhangyu and Mastronarde, Nicholas and Bentley, Elizabeth Serena and Medley, Michael},
  booktitle={18th International Conference on Distributed Computing in Sensor Systems (DCOSS)}, 
  title="{A Middleware for Digital Twin-Enabled Flying Network Simulations Using UBSim and UB-ANC}", 
  year={2022},
  volume={},
  number={},
  pages={322-327},
  month={May},}

@INPROCEEDINGS{ndikumana2023digital,
  author={Ndikumana, Anselme and Nguyen, Kim Khoa and Cheriet, Mohamed},
  booktitle={IEEE Global Communications Conference (GLOBECOM)}, 
  title="{Digital Twin Assisted Closed-Loops for Energy-Efficient Open RAN-Based Fixed Wireless Access Provisioning in Rural Areas}", 
  year={2023},
  volume={},
  number={},
  pages={6285-6290}
}

@ARTICLE{nguyen2021digital,
  author={Nguyen, Huan X. and Trestian, Ramona and To, Duc and Tatipamula, Mallik},
  journal={IEEE Communications Magazine}, 
  title="{Digital Twin for 5G and Beyond}", 
  year={2021},
  volume={59},
  number={2},
  pages={10-15},
  month={February},}

@misc{oran-wg3-ricarch,
    author={{O-RAN Working Group 3}},
    title={{O-RAN Near-RT RAN Intelligent Controller Near-RT RIC Architecture 2.00}},
    howpublished={O-RAN.WG3.RICARCH-v02.00},
    month = {March},
    year={2021}
}

@electronic{pawr,
    author={{\acrfull{pawr}}},
    title={\url{https://www.advancedwireless.org}},
    note={Accessed February 2024}
}

@article{polese2021colo,
author={Polese, Michele and Bonati, Leonardo and D'Oro, Salvatore and Basagni, Stefano and Melodia, Tommaso},
  journal={IEEE Transactions on Mobile Computing}, 
  title="{ColO-RAN: Developing Machine Learning-Based xApps for Open RAN Closed-Loop Control on Programmable Experimental Platforms}", 
  year={2023},
  volume={22},
  number={10},
  pages={5787-5800},
  month={Oct},}

@article{polese2022understanding,
author={Polese, Michele and Bonati, Leonardo and D’Oro, Salvatore and Basagni, Stefano and Melodia, Tommaso},
  journal={IEEE Communications Surveys \& Tutorials}, 
  title="{Understanding O-RAN: Architecture, Interfaces, Algorithms, Security, and Research Challenges}", 
  year={2023},
  volume={25},
  number={2},
  pages={1376-1411},
  month={Second quarter},}

@inproceedings{raychaudhuri2020cosmos,
    author={Raychaudhuri, D. and Seskar, I. and Zussman, G. and Korakis, T. and Kilper, D. and Chen, T. and Kolodziejski, J. and Sherman, M. and Kostic, Z. and Gu, X. and Krishnaswamy, H. and Maheshwari, S. and Skrimponis, P. and Gutterman, C.},
    title={Challenge: {COSMOS}: A City-Scale Programmable Testbed for Experimentation with Advanced Wireless},
    booktitle={Proceedings of ACM MobiCom},
    address={London, United Kingdom},
    month={September},
    year={2020}
}

@ARTICLE{ren2023end,
  author={Ren, Yinlin and Guo, Shaoyong and Cao, Bin and Qiu, Xuesong},
  journal={IEEE Transactions on Mobile Computing}, 
  title="{End-to-End Network SLA Quality Assurance for C-RAN: A Closed-Loop Management Method Based on Digital Twin Network}", 
  year={2023},
  volume={},
  number={},
  pages={1-18},
  month={},}

@article{upadhyaya2022prototyping,
  title="{Prototyping next-generation O-RAN research testbeds with SDRs}",
  author={Upadhyaya, Pratheek S and Abdalla, Aly S and Marojevic, Vuk and Reed, Jeffrey H and Shah, Vijay K},
  journal={arXiv preprint arXiv:2205.13178},
  year={2022}
}

@ARTICLE{upadhyaya2023open,
  author={Upadhyaya, Pratheek S. and Tripathi, Nishith and Gaeddert, Joseph and Reed, Jeffrey H.},
  journal={IEEE Network}, 
  title="{Open AI Cellular (OAIC): An Open Source 5G O-RAN Testbed for Design and Testing of AI-Based RAN Management Algorithms}", 
  year={2023},
  volume={37},
  number={5},
  pages={7-15},
  month={Sep.},}

@article{vila2023design,
  title={On the design of a network digital twin for the radio access network in 5G and beyond},
  author={Vil{\`a}, Irene and Sallent, Oriol and P{\'e}rez-Romero, Jordi},
  journal={Sensors},
  volume={23},
  number={3},
  pages={1197},
  year={2023},
  publisher={MDPI}
}

@article{villa2024dt,
author = {Villa, Davide and Tehrani-Moayyed, Miead and Robinson, Clifton P. and Bonati, Leonardo and Johari, Pedram and Polese, Michele and Melodia, Tommaso},
title = {{Colosseum as a Digital Twin: Bridging Real-World Experimentation and Wireless Network Emulation}},
year = {2024},
month = {In press,},
journal = {IEEE Transactions on Mobile Computing},
pages = {1--17}
}

@inproceedings{zhang2021ara,
  author = {Zhang, Hongwei and Guan, Yong and Kamal, Ahmed and Qiao, Daji and Zheng, Mai and Arora, Anish and Boyraz, Ozdal and Cox, Brian and Daniels, Thomas and Darr, Matthew and Jacobson, Doug and Khokhar, Ashfaq and Kim, Sang and Koltes, James and Liu, Jia and Luby, Mike and Nadolny, Larysa and Peschel, Joshua and Schnable, Patrick and Sharma, Anuj and Somani, Arun and Tang, Lie},
  title = {{ARA: A Wireless Living Lab Vision for Smart and Connected Rural Communities}},
  booktitle = {Proceedings of ACM WiNTECH},
  address = {New Orleans, LA, USA},
  month = {October},
  year = {2021}
}

@electronic{gnuradio,
    author={\relax{Eric Blossom and GNU Radio Project}},
    title={{\relax{GNU Radio}}},
    url_={https://www.gnuradio.org},
    howpublished={[Available Online]: \url{https://www.gnuradio.org}},
    note={Accessed February 2023}
}

@article{chen2023static,
  title         = "{Static Standing Balance With Musculoskeletal Models Using PPO With Reward Shaping}",
  author        = {Chen, Wenqian and Chen, Yaru and Wang, Yongxuan and Liu, Rong},
  journal       = {Procedia Computer Science},
  volume        = {226},
  pages         = {78--84},
  year          = {2023},
  publisher     = {Elsevier}
}

@misc{oran-wg1-use-cases,
  author        = {{O-RAN Working Group 1}},
  title         = {{{O-RAN} Use Cases Detailed Specification 6.0}},
  howpublished  = {O-RAN.WG1.Use-Cases-Detailed-Specification-v06.00 Technical Specification},
  month         = {July},
  year          = {2021}
}

@misc{oran-wg3-con-mit,
  author        = {{O-RAN Working Group 3}},
  title         = {{Conflict Mitigation}},
  howpublished  = {O-RAN.WG3.TR.ConMit-R004-v01.00 Technical Specification},
  month         = {October},
  year          = {2024}
}

@inproceedings{adamczyk2023challenges,
  title         = {{Challenges for Conflict Mitigation in {O-RAN} 's {RAN} Intelligent Controllers}},
  author        = {Adamczyk, Cezary},
  booktitle     = {Proceedings of IEEE SoftCOM},
  pages_        = {1--6},
  year          = {2023}
}

@article{adamczyk2023conflict,
  title         = {{Conflict Mitigation Framework and Conflict Detection in O-RAN Near-RT RIC}},
  author        = {Adamczyk, Cezary and Kliks, Adrian},
  journal       = {IEEE Communications Magazine},
  year          = {2023},
  publisher     = {IEEE}
}

@article{wadud2024qacm,
  title         = {{QACM: QoS-Aware xApp Conflict Mitigation in Open RAN}},
  author        = {Wadud, Abdul and Golpayegani, Fatemeh and Afraz, Nima},
  journal       = {arXiv preprint arXiv:2405.07324},
  year          = {2024}
}

@article{hazra2016biostatistics,
  title         = {{Biostatistics series module 4: comparing groups--categorical variables}},
  author        = {Hazra, Avijit and Gogtay, Nithya},
  journal       = {Indian Journal of Dermatology},
  volume        = {61},
  number        = {4},
  pages         = {385--392},
  year          = {2016},
  publisher     = {Medknow}
}

@techreport{3gpp22261,
     author={{3GPP TS 22.261}},                       
     title={{3GPP TS 22.261: Service requirements for the 5G system}},                 
     institution={3rd Generation Partnership Project (3GPP)},                        
     year={2022},                       type={Technical Specification},
     note={Release 17, v17.11.00}                    
  }

@techreport{3gpp23501,
     author={{3GPP TS 23.501}},
     title={{3GPP TS 23.501: System architecture for the 5G system (5GS)}},                            institution={3rd Generation Partnership Project (3GPP)},        year={2023},                
     type={Technical Specification},    note={Release 17}                  
  }

@misc{tsampazi2026ariadne,
  author       = {M. Tsampazi and N. N. Santhi and N. Perrotta and F. Dressler and T. Melodia},
  title        = {{ARIADNE}: AI-RAN Informed Link Adaptation in Digital Twin Network Environments},
  year         = {2026},
  note         = {European Wireless}
}
